%% file: frg_review.tex
\documentclass[3p,times]{elsarticle}

\usepackage{graphicx,amsfonts,amssymb,amsmath} 
\usepackage{natbib}
\usepackage{textcomp}
\usepackage{esint}
\usepackage{soul}
\usepackage{graphics}
\usepackage{subfigure}
\usepackage{bm}
\usepackage{bbm}
\usepackage[colorlinks,linktocpage,bookmarks=false,citecolor=blue,linkcolor=red,urlcolor=blue]{hyperref}

\newcommand{\p}{\partial} 
\newcommand{\vx}{{\bf x}}
\newcommand{\vq}{{\bf q}}
\newcommand{\vp}{{\bf p}}
\newcommand{\vv}{{\bf v}}
\newcommand{\vf}{{\bf f}}

\newcommand{\vnabla}{{\bm{\nabla}}}

\newcommand{\rgk}{k}
\newcommand{\pre}{Phys. Rev. E }
\newcommand{\prb}{Phys. Rev. B }
\newcommand{\prl}{Phys. Rev. Lett. }

\allowdisplaybreaks

\begin{document}

\title{The nonperturbative functional renormalization group and its applications}

\author[paris]{N. Dupuis}

\author[grenoble]{L. Canet}

\author[CP3,heidelberg]{A. Eichhorn}

\author[MPI]{W. Metzner}
 
\author[heidelberg,darmstadt]{J.~M. Pawlowski}

\author[paris]{M. Tissier}

\author[MV]{N. Wschebor}

\address[paris]{Sorbonne Universit\'e, CNRS, Laboratoire de Physique Th\'eorique de la Mati\`ere Condens\'ee, LPTMC, F-75005 Paris, France}
\address[grenoble]{Universit\'e Grenoble Alpes and CNRS, Laboratoire de Physique et Mod\'elisation des Milieux Condens\'es, LPMMC, 38000 Grenoble, France}
\address[CP3]{CP3-Origins, University of Southern Denmark, Campusvej 55, DK-5230 Odense M, Denmark} 
\address[heidelberg]{Institut f\"ur Theoretische Physik, Universit\"at Heidelberg, Philosophenweg 16, 69120 Heidelberg, Germany}
\address[MPI]{Max-Planck-Institute for Solid State Research,
	Heisenbergstra{\ss}e 1, D-70569 Stuttgart, Germany} 
\address[darmstadt]{ExtreMe Matter Institute EMMI, GSI, Planckstr. 1, D-64291 Darmstadt, Germany}
\address[MV]{Instituto de F\'isica, Facultad de Ingenier\'ia, Universidad de la Rep\'ublica,
J.H.y Reissig 565, 11000 Montevideo, Uruguay}

\date{\today}

\begin{abstract}
The renormalization group plays an essential role in many areas of physics, both conceptually and as a practical tool to determine the long-distance low-energy properties of many systems on the one hand and on the other hand search for viable ultraviolet completions in fundamental physics. It provides us with a natural framework to study theoretical models where degrees of freedom are correlated over long distances and that may exhibit very distinct behavior on different energy scales. The nonperturbative functional renormalization-group (FRG) approach is a modern implementation of Wilson's RG, which allows one to set up nonperturbative approximation schemes that go beyond the standard perturbative RG approaches. The FRG is based on an exact functional flow equation of a coarse-grained effective action (or Gibbs free energy in the language of statistical mechanics). We review the main approximation schemes that are commonly used to solve this flow equation and discuss applications in equilibrium and out-of-equilibrium statistical physics, quantum many-particle systems, high-energy physics and quantum gravity. 
\end{abstract}

\maketitle
\tableofcontents

\input{./SEC_INTRO/intro.tex}

\graphicspath{{./SEC_FRG/}}
\input{./SEC_FRG/frg.tex}

\graphicspath{{./SEC_STATMECH/}}

\input{./SEC_STATMECH/statmech_Nico.tex}

\input{./SEC_STATMECH/statmech_Matth.tex} 
\input{./SEC_STATMECH/stat_mech_Leo.tex}

\input{./SEC_HighEnergy/noneqquantum.tex}

\graphicspath{{./SEC_FB/}}
\input{./SEC_FB/fermions.tex}

\graphicspath{{./SEC_HighEnergy/}}
\input{./SEC_HighEnergy/frg_HighEnergy.tex}

\graphicspath{{./SEC_GRAVITY/}}

\input{./SEC_GRAVITY/frg_review-gravity.tex}

\section{Acknowledgments}

We have benefited from collaborations and discussions with many people. In particular we are grateful to Sabine Andergassen, Ivan Balog, Jean-Paul Blaizot, Federico Benitez, Jens Braun, Anton Cyrol, Bertrand Delamotte, Gonzalo De Polsi, Sebastian Diehl, Andreas Eberlein, Wei-jie Fu, Thomas Gasenzer, Holger Gies, Aaron Held, Carsten Honerkamp, Yuji Igarashi, Katsumi Itoh, Pawel Jakubczyk, Thomas Kloss, Tim Koslowski, Daniel Litim, Volker Meden, Mario Mitter, Dominique Mouhanna, Carlo Pagani, Marcela Pel\`aez, Roberto Percacci, Antonio Pereira, Alessia Platania, Adam Ran\c{c}on, Manuel Reichert, Fabian Rennecke, Martin Reuter, F\'elix Rose, Manfred Salmhofer, Frank Saueressig, Bernd-Jochen Schaefer, Michael Scherer, Kurt Sch\"onhammer, Lorenz von Smekal, Davide Squizzato, Nils Strodthoff, Gilles Tarjus, Malo Tarpin, Arno Tripolt, Demetrio Vilardi, Christof Wetterich, Nicolas Wink, Masatoshi Yamada and Hiroyuki Yamase. We also thank Bertrand Delamotte and D. Benedetti, the organizers of the conference ERG2018, who suggested the writing of this review. 

L.~C. acknowledges support from Institut Universitaire de France and from ANR through the project NeqFluids (grant ANR-18-CE92-0019). 
A.~E. is supported by the DFG under grant no.~Ei-1037/1, by a research grant (29405) from VILLUM FONDEN, and also partially supported by a visiting fellowship at the Perimeter Institute for Theoretical Physics. 
L.~C., M.~T. and N.~W. acknowledge support from the program ECOS Sud U17E01 and from IRP "Institut Franco-Uruguayen de Physique". N.~W. is supported by Grant I+D number 412 of the CSIC (Ude-laR) Commission and Programa de Desarrollo de las Ciencias B\'asicas (PEDECIBA). J.M.P.\ acknowledges support by the DFG under Germany's Excellence Strategy EXC 2181/1 - 390900948 (the Heidelberg STRUCTURES Excellence Cluster), the DFG Collaborative Research Centre SFB 1225 (ISOQUANT), and the BMBF grant 05P18VHFCA.

\graphicspath{{./SEC_APP/}}
\input{./SEC_APP/appendix.tex}

\input{./frg_review_final.bbl}
\bibliographystyle{elsarticle-num}
\biboptions{sort&compress} 
\journal{Physics Reports}

%\bibliography{../BIB/BIBnd/nprg.bib,../BIB/BIBnd/bosons.bib,../BIB/BIBnd/disorder.bib,../BIB/BIBnd/impurity.bib,../BIB/BIBnd/book.bib,../BIB/BIBnd/notes.bib,../BIB/BIBnd/stat_phys.bib,../BIB/BIBnd/bcsbec.bib,../BIB/BIBnd/cold_atoms.bib,../BIB/BIBnd/fermions.bib,../BIB/BIBnd/fermions_rg.bib,../BIB/BIBnd/supra.bib,../BIB/BIBnd/magnetism.bib,../BIB/BIBnd/stages_theses.bib,../BIB/bib_master.bib,../BIB/statmech_Nico.bib,../BIB/statmech.bib,../BIB/biblc.bib,../BIB/fermions.bib}

\end{document}

%% file: SEC_INTRO/intro.tex
\section{Introduction} 

Linking descriptions of physics at various scales and relating the macroscopic physical properties of systems to the microscopic interactions and degrees of freedom is the primary goal of research in many areas of physics, from condensed matter and cold atoms to high-energy physics and quantum gravity. The aim of this review is to describe a theoretical approach, the nonperturbative functional renormalization group (FRG), which provides us with an efficient and versatile tool to bridge the gap between micro- and macroscopic scales and thus determine the physical properties of a wide variety of systems. 
 
Strongly correlated systems, such as electrons in solids interacting {\it via} the Coulomb interaction or quarks in nucleons subjected to the strong interaction, although different in some respects share nevertheless a number of properties. When external parameters (temperature, density, etc.) are varied, they often exhibit rich phase diagrams due to competing collective phenomena. The theoretical study of their properties faces two major difficulties. First, there is often no small parameter that would allow for a systematic perturbative expansion. Second, whenever the degrees of freedom are correlated over distances much larger than the microscopic scales, collective effects become important at low energies. This implies that these systems may exhibit very distinct behavior on different energy scales and the relevant degrees of freedom which permit a simple formulation of the low-energy (macroscopic) properties may be different from the microscopic ones. This diversity of scales explains the difficulty of a straightforward numerical solution of microscopic models, since the interesting phenomena emerge only at low energies and in large-size systems. It is also responsible for the fact that perturbation theories are often plagued with infrared divergences and may be inapplicable even at weak coupling. Conversely, in fundamental physics one is often confronted with the opposite problem: While the low-energy description is known, one searches for a consistent underlying microphysics. Many of the technical challenges and conceptual insights -- maybe surprisingly -- resemble those of the previous examples.

\subsection{Wilson's renormalization group}

The renormalization group\footnote{The RG was pioneered in high-energy physics~\cite{Petermann:1953wpa, GellMann:1954fq, Bogolyubov:1983gp, Symanzik:1970rt, Callan:1970yg}.} (RG) is a natural framework to study systems with many degrees of freedom correlated over long distances. In Wilson's modern formulation, fluctuations at short distances and high energies are progressively integrated out  to obtain an effective (coarse-grained) description at long distances and low energies~\cite{Kadanoff66,Wilson71,Wilson71a,Wilson74,Wegner73,Polchinski84,Fisher98,Bagnuls01}. The RG not only gives us an explanation of cooperative behavior and universality but also provides us with a practical tool to study systems where correlations and fluctuations play an important role, the prime example being systems in the vicinity of a second-order phase transition~\cite{Wilson74,Guida98,Pelissetto02}. In high-energy physics, the RG provides us with a powerful conceptual understanding of fundamental interactions, as it allows to distinguish effective theories (which break down in the UV limit, e.g., due to the triviality problem) from fundamental theories which hold over an infinite range of scales due to asymptotic freedom or safety.

In its standard formulation however, the RG usually relies on perturbation theory and is therefore restricted to weakly interacting systems where a small expansion parameter allows one to systematically compute the effects of fluctuations beyond the noninteracting limit or the mean-field theory.\footnote{High-order perturbative expansions are usually obtained within the field theoretical RG approach; see, e.g., \cite{Guida98,Pelissetto02}.\label{sec_intro:footnote1}}$^{,}$\footnote{The RG has also been used as a mathematical tool for a rigorous non-perturbative construction of field theories~\cite{Gawedzki1985,Balaban:1988rr,Brydges1987,Feldman:1987zq,rivasseau_book}, and as a non-perturbative computational tool, such as Wilson's numerical RG for the Kondo problem~\cite{Wilson:1974mb} or the density matrix renormalization group (DMRG) for one-dimensional lattice models~\cite{Schollwock:2005zz}.} For example in one of the most studied models, the O($N$) model ($\varphi^4$ theory with O($N$) symmetry), the critical exponents are computed either from a Ginzburg-Landau-Wilson functional in an $\epsilon=4-d$ expansion~\cite{Wilson72,Wilson74} or from the nonlinear sigma model in an $\epsilon=d-2$ expansion~\cite{Migdal75,Polyakov75,Brezin76a,Brezin76b,Nelson77} (with $d$ being the space dimension). In the latter case the perturbative RG series is usually considered as useless due to the lack of Borel summability. Thus even high-order perturbative RG expansions cannot relate the two expansions, except in the large-$N$ limit~\cite{Moshe03}. This is not crucial for the O($N$) model where the critical behavior does not change qualitatively for $2<d<4$ but in other models forbids a completely coherent picture of the physics between $d=2$ and $d=4$ (or, more generally, between the lower and upper critical dimensions). 

Moreover, even when the field theoretical perturbative RG$^{\ref{sec_intro:footnote1}}$ yields an accurate determination of universal quantities (e.g., the critical exponents or universal scaling functions), it is often not clear how to compute nonuniversal quantities such as a transition temperature, the phase diagram of equilibrium and out-of-equilibrium systems, the spectrum of bound states in strongly correlated theories, etc., since these properties depend on the underlying microscopic models. 

Last, the perturbative RG is useless for genuinely nonperturbative problems such as the Berezinskii-Kosterliz-Thouless (BKT) transition~\cite{Berezinskii71,Berezinskii72,Kosterlitz73,Kosterlitz74} in the two-dimensional O(2) model,\footnote{The BKT transition is often studied in the framework of the Coulomb gas, Villain or sine-Gordon models for which the perturbative RG is useful; see, e.g.,~\cite{Chaikin_book}. A direct study in the O(2) model, i.e., without introducing explicitly the vortices, is much more challenging.} the growth of stochastically growing interfaces, turbulent flows, the confinement of quarks in quantum chromodynamics (QCD), gravity at the Planck scale, etc. In the latter case, a perturbative analysis indicates perturbative non-renormalizability, rendering non-perturbative tools necessary to describe quantum gravity beyond the Planck scale.

\subsection{The functional renormalization group}

Although various systems may be characterized by different microscopic energy scales (e.g., $10^{-7}$~meV for a ultracold atomic gas, 1~eV for conduction electrons in a solid, 1~GeV for QCD, 125~GeV for the Higgs particle, $10^{19}$~GeV for the Planck scale in quantum gravity), from a theoretical point of view there is no fundamental difference between relativistic quantum field theory (that describes elementary particles and their interactions) and statistical field theory (that describes the statistical properties of quantum or classical systems where the degrees of freedom are represented by fields).\footnote{In statistical field theory, the UV cutoff $\Lambda$ of the theory has usually a well-defined physical meaning (e.g., the inverse of the lattice spacing of the original model) and the interactions at that scale are known from experiments or {\it ab initio} calculations based on microscopic (realistic) models. In quantum field theory, $\Lambda$ stands for the highest momentum scale where the theory is valid.} In a modern language, both are formulated as functional integrals in $d$ space dimensions or $d+1$ spacetime dimensions. The primary goal of these functional approaches is to compute correlation functions as well as the free energy of the system. 

The FRG combines the functional approach with the Wilson RG idea of integrating out fluctuations not all at once but progressively from high- to low-energy scales. The expression {\it functional} RG stems from the fact that one naturally deals with (possibly singular) functions of the field rather than a finite number of coupling constants. In the literature, the (nonperturbative) FRG is sometimes merely referred to as the nonperturbative RG. Both aspects (nonperturbative and functional) are actually crucial features of the method presented in this review. Note however that the RG approach can be functional and perturbative, or nonperturbative and nonfunctional. The FRG is also referred to as the Exact RG due to its one-loop exact (closed) functional form. This epithet does not imply that  the flow equation can be solved exactly (except for models that can be solved more easily with other methods): Most nontrivial applications rely on an approximate solution.

Different versions of exact functional RG equations, such as the functional Callan-Symanzik, Wilson-Polchinski and Wegner-Houghton formulations, have already a long history~\cite{Symanzik:1970rt,Wegner73,Wilson74,Polchinski84,Hasenfratz86,Chang92,Parola95,Nicoll74,Nicoll77,Parola84}. However the application of these methods has been hindered for a long time by the complexity of functional differential equations and the difficulty to devise nonperturbative and reliable approximation schemes. Early works~\cite{Wegner73,Nicoll74,Nicoll76,Newman84,Newman84a,Golner86,Hasenfratz86,Hasenfratz88,Zumbach93,Zumbach94a,Zumbach94b} were mainly based on the so-called local potential approximation (see Sec.~\ref{sec_frg} for a detailed discussion) and neglected the momentum dependence of the interaction vertices, with apparently no possibility of a systematic improvement. The FRG has nevertheless been useful in its perturbative formulation for the study of disordered systems~\cite{Fisher85,Narayan92,Nattermann92,Chauve01,Ledoussal04,Tarjus04}. The necessity of a functional approach in this context is due to an infinite number of operators being marginal at the upper or lower (whichever the case of interest) critical dimension. It is therefore not possible, in some disordered systems, to restrict oneself to a finite number of coupling constants and one must consider functions of the field (Sec.~\ref{sec_disordered}). 

The formulation of the FRG based on a formally exact flow equation for a scale-dependent ``effective action'' $\Gamma_k[\phi]$ (or Gibbs free energy in the language of statistical physics), the generating functional of one-particle irreducible vertices, has proven successful in devising nonperturbative approximation schemes~\cite{Ringwald90,Wetterich91,Wetterich93a,Wetterich93b,Wetterich93,Berges:2000ew,Ellwanger94,Morris94,Bonini93}. The functional $\Gamma_k[\phi]$ may be seen as a coarse-grained free energy that includes only fluctuations with momenta or energies larger than a scale $k$.\footnote{$\Gamma_k[\phi]$ is a scale-dependent generating functional of 1PI vertices (see Sec.~\ref{sec_frg})} The field $\phi({\bf r})$ or $\phi({\bf r},t)$ represents the relevant microscopic degrees of freedom, e.g., the local magnetization in a solid.\footnote{The field $\phi({\bf r},t)$ depends on time in out-of-equilibrium classical and quantum (be them at equilibrium or not) systems. In the Euclidean (quantum) formalism $t=-i\tau$ is an imaginary time. $\phi$ is an anticommuting Grassmann variable if the degrees of freedom are fermionic.} If necessary, one may include additional fields corresponding to emerging low-energy (collective) degrees of freedom: a pairing field in a superconductor, a meson field in QCD, etc. The functional $\Gamma_{k=0}[\phi]$ includes fluctuations on all scales, and allows us to obtain the free energy and the one-particle irreducible vertices. In principle, all correlation functions can be deduced from $\Gamma_{k=0}[\phi]$. Thus the FRG replaces the difficult determination of $\Gamma_{k=0}[\phi]$ from a direct calculation of the functional integral (that defines the partition function) by the solving of a functional differential equation $\partial_k\Gamma_k[\phi]$ with an initial condition $\Gamma_\Lambda[\phi]$, which often (but not always) corresponds to the bare action or the mean-field solution of the model, at some microscopic scale $\Lambda$. The flow equation $\partial_k\Gamma_k[\phi]$ closely resembles a renormalization-group improved one-loop equation, but is exact. This close connection to perturbation theory, for which we have an intuitive understanding, is an important key for devising meaningful nonperturbative approximations.

\subsection{Scope of the review}

The aim of the review is to give an up-to-date nontechnical presentation of the nonperturbative FRG approach that emphasizes its applications in various fields of physics.\footnote{For previous general reviews on the nonperturbative FRG, see~\cite{Aoki00,Bagnuls01,Berges:2000ew,Polonyi:2001se,Delamotte04,Pawlowski:2005xe,Rosten:2010vm,Kopietz_book,Braun:2011pp,Delamotte12,Gies12}.} Section~\ref{sec_frg} is devoted to a general presentation of the FRG. We discuss the properties of the scale-dependent effective action $\Gamma_k[\phi]$ and the main two approximations used for the solution of the exact flow equation: the derivative expansion and the vertex expansion. We also emphasize the applicability of the FRG to microscopic, classical or quantum, models. In the following sections, we show how the method can be used in practice in statistical mechanics, quantum many-body physics, high-energy physics and quantum gravity (Secs.~\ref{sec_sm}-\ref{sec_gr}). Although an exhaustive account of all applications is obviously impossible given the broad scope of the review, we have tried nevertheless to cover most subjects. A more technical presentation of the nonperturbative FRG approach, focusing on the derivative expansion, can be found in the Appendices. 

We set $\hbar=k_B=c=1$ throughout the paper.

%% file: SEC_FRG/frg.tex
\section{The FRG in a nutshell} 
\label{sec_frg} 

While we wish to stress the general concepts of the FRG rather than its application to a particular model, we shall base our discussion in this section on the $d$-dimensional O($N$) model (or $\varphi^4$ theory). Its partition function
\begin{equation}
{\cal Z}[{\bf J}] = \int {\cal D}[\boldsymbol{\varphi}] \, e^{-S[\boldsymbol{\varphi}] + \int_{\bf r} {\bf J}\cdot\boldsymbol{\varphi} } 
\label{sec_frg:Zdef}
\end{equation} 
can be written as a functional integral, with the action 
\begin{equation}
S[\boldsymbol{\varphi}] = \int_{\bf r} \left\lbrace \frac{1}{2} (\boldsymbol{\nabla}\boldsymbol{\varphi})^2 + \frac{r_0}{2} \boldsymbol{\varphi}^2 + \frac{u_0}{4!} {(\boldsymbol{\varphi}^2)}^2 \right\rbrace ,
\label{sec_frg:Smicro}
\end{equation} 
where $\boldsymbol{\varphi}=(\varphi_1\cdots\varphi_N)$ is an $N$-component real field, ${\bf r}$ a $d$-dimensional coordinate and $\int_{{\bf r}}=\int d^dr$. We shall mostly use the language of classical statistical mechanics where $S[\boldsymbol{\varphi}]$ is simply the Hamiltonian $H[\boldsymbol{\varphi}]$ multiplied by the inverse temperature $\beta=1/T$. ${\bf J}$ is an external ``source'' (e.g. a magnetic field for a magnetic system) which couples linearly to the field. The model is regularized by a UV momentum cutoff $\Lambda$ which can be thought of as the inverse lattice spacing if the continuum model~({\ref{sec_frg:Smicro}) is derived from a lattice model. We refer to the action~({\ref{sec_frg:Smicro}) as the ``microscopic" action, i.e., the action describing the physics at length scales $\sim\Lambda^{-1}$. Assuming the value of $u_0$ fixed, the O($N$) model exhibits a second-order phase transition between a disordered phase ($r_0>r_{0c}$) and an ordered phase ($r_0<r_{0c}$) where $\langle\boldsymbol{\varphi}({\bf r})\rangle\neq 0$ and the O($N$) symmetry is spontaneously broken. $r_0$ is naturally related to the temperature by setting $r_0\equiv\bar r_0(T-T_0)$; $r_{0c}=\bar r_0(T_c-T_0)$ then defines the critical temperature $T_c$ while $T_0$ is the mean-field transition temperature. 
 
We are typically interested in the Helmholtz free energy $F[{\bf J}] = -T\ln {\cal Z}[{\bf J}]$ (usually for a vanishing source, ${\bf J}=0$) and the correlation functions such as the two-point one (or propagator)
\begin{equation} 
G_{ij}({\bf r}-{\bf r}')=\langle \varphi_i({\bf r})\varphi_j({\bf r}')\rangle_c= \frac{\delta^2\ln{\cal Z}[{\bf J}]}{\delta J_i({\bf r})\delta J_j({\bf r}')}\biggl|_{{\bf J}=0} ,
\label{sec_frg:Gdef}
\end{equation}
where $\langle\varphi_i\varphi_j\rangle_c\equiv\langle\varphi_i\varphi_j\rangle-\langle\varphi_i\rangle\langle\varphi_j\rangle$.

\subsection{The scale-dependent effective action}
\label{sec_frg:subsec_gammak}

The main idea of Wilson's RG is to compute the partition function~(\ref{sec_frg:Zdef}) by progressively integrating out short-distance (or high-energy) degrees of freedom~\cite{Kadanoff66,Wilson71,Wilson71a,Wilson74,Wegner73,Polchinski84,Fisher98,Bagnuls01}. In the FRG approach, one builds a family of models indexed by a momentum scale $k$ such that fluctuations are smoothly taken into account as $k$ is lowered from some initial scale $k_{\rm in}\geq \Lambda$ down to 0. In practice this is achieved by adding to the action $S[\boldsymbol{\varphi}$] a quadratic term $\Delta S_k[\boldsymbol{\varphi}]$ defined by 
\begin{equation}
\Delta S_k[\boldsymbol{\varphi}] = \frac{1}{2} \int_{{\bf p}} \sum_{i=1}^N \varphi_i(-{\bf p}) R_k({\bf p}) \varphi_i({\bf p}) 
\label{sec_frg:DeltaS}
\end{equation}
with $\int_{\bf p}=\int d^d p/(2{\pi})^d$. The typical shape of the regulator function $R_k({\bf p})$ is shown in Fig.~\ref{sec_frg:fig_Rk}; it is strongly suppressed for $|{\bf p}|\gg k$ and of order $k^2$ for $|{\bf p}|\ll k$. In the language of high-energy physics, $R_k({\bf p})$ can be interpreted as a momentum-dependent mass-like term that gives a mass of order $k^2$ to the low-energy modes and thus suppresses their fluctuations. The regulator function is discussed in more detail below. 

Rather than considering the (now $k$-dependent) Helmholtz free energy $F_k[{\bf J}]=-T\ln{\cal Z}_k[{\bf J}]$, one introduces the scale-dependent ``effective action'' (aka average effective action), or ``Gibbs free energy'' in the language of statistical mechanics,
\begin{equation}
\Gamma_k[\boldsymbol{\phi}] = - \ln {\cal Z}_k[{\bf J}] + \int_{\bf r} {\bf J}\cdot\boldsymbol{\phi} - \Delta S_k[\boldsymbol{\phi}] ,
\label{sec_frg:gammak}
\end{equation} 
defined as a (slightly modified) Legendre transform which includes the subtraction of $\Delta S_k[\boldsymbol{\phi}]$~\cite{Ringwald90,Wetterich91,Wetterich93a,Wetterich93b,Wetterich93,Berges:2000ew}. Here $\boldsymbol{\phi}=\langle\boldsymbol{\varphi}\rangle\equiv \boldsymbol{\phi}_k[{\bf J}]$ is the order-parameter field and ${\bf J}\equiv{\bf J}_k[\boldsymbol{\phi}]$ in~(\ref{sec_frg:gammak}) should be understood as a functional of $\boldsymbol{\phi}$ obtained by inverting $\boldsymbol{\phi}_k[{\bf J}]$.  

Thermodynamic properties can be obtained from the effective potential 
\begin{equation}
U_k(\rho) = \frac{1}{V} \Gamma_k[\boldsymbol{\phi}] \Bigl|_{\boldsymbol{\phi}\;\rm unif.} 
\label{sec_frg:Udef} 
\end{equation} 
($V$ is the volume of the system) which is proportional to $\Gamma_k[\boldsymbol{\phi}]$ evaluated in a uniform field configuration $\boldsymbol{\phi}({\bf r})=\boldsymbol{\phi}$. Because of the O($N$) symmetry of the model, $U_k$ is a function of the O($N$) invariant $\rho=\boldsymbol{\phi}^2/2$. $U_k(\rho)$ may exhibit a minimum at $\rho_{0,k}$. Spontaneous breaking of the O($N$) symmetry is characterized by a nonvanishing expectation value of the field $\langle\boldsymbol{\varphi}\rangle_{{\bf J}\to 0^+}$ in the thermodynamic limit and occurs if $\lim_{k\to 0}\rho_{0,k}=\rho_0>0$. 

On the other hand correlation functions can be related to the one-particle irreducible vertices $\Gamma_k^{(n)}[\boldsymbol{\phi}]$ defined as the $n$th-order functional derivatives of $\Gamma_k[\boldsymbol{\phi}]$~\cite{Zinn_book}. In particular the propagator defined by~(\ref{sec_frg:Gdef}), $G_k[\boldsymbol{\phi}]=(\Gamma_k^{(2)}[\boldsymbol{\phi}]+R_k)^{-1}$ (written here in a matrix form), is simply related to the two-point vertex $\Gamma_k^{(2)}$.\footnote{We refer to~\ref{appEA} for a brief introduction to the effective action formalism.}

\subsubsection{The regulator function $R_k$} 
\label{sec_frg:subsubsec_Rk} 

The regulator function $R_k$ is chosen such that $\Gamma_k$ smoothly interpolates between the microscopic action $S$ for $k=k_{\rm in}$ and the effective action of the original model~(\ref{sec_frg:Smicro}) for $k=0$. It must therefore satisfy the following properties: 
\smallskip

i) At $k=k_{\rm in}$, $R_{k_{\rm in}}({\bf p})=\infty$. All fluctuations are then frozen and $\Gamma_{k_{\rm in}}[\boldsymbol{\phi}]=S[\boldsymbol{\phi}]$~\cite{Berges:2000ew} as in Landau's mean-field theory of phase transitions. In practice, it is sufficient to choose $k_{\rm in}\gg\Lambda$ to ensure that $R_{k_{\rm in}}({\bf p})$ is much larger than all microscopic mass scales in the problem.
\smallskip

ii) At $k=0$, $R_{k=0}({\bf p})=0$ so that $\Delta S_{k=0}=0$. All fluctuations are taken into account and the effective action $\Gamma_{k=0}[\boldsymbol{\phi}]\equiv\Gamma[\boldsymbol{\phi}]$ coincides with the effective action of the original model. 
\smallskip

iii) For $0<k<k_{\rm in}$, $R_k({\bf p})$ must suppress fluctuations with momenta below the scale $k$ but leave unchanged those with momenta larger than $k$. In general one chooses a ``soft'' regulator (as opposed to the sharp regulator commonly used in the weak-coupling momentum-shell RG), see Fig.~\ref{sec_frg:fig_Rk}. Two popular choices are the exponential regulator $R_k({\bf p})=\alpha{\bf p}^2/(e^{{\bf p}^2/k^2}-1)$ and the theta regulator $R_k({\bf p})=\alpha(k^2-{\bf p}^2)\Theta(k^2-{\bf p}^2)$~\cite{Litim01} (with $\alpha$ a constant of order unity). The latter is a not a smooth function of ${\bf p}$ and cannot be used beyond the second order of the derivative expansion (see Sec.~\ref{sec_frg:subsec_de})~\cite{Morris05}. The generic form of the regulator is $R_k({\bf p})={\bf p}^2 r({\bf p}^2/k^2)$ with $r(y)$ satisfying $r(y\to 0)\sim 1/y$ and $r(y\gg 1)\ll 1$. 

\begin{figure}
\centerline{\includegraphics[width=5.5cm]{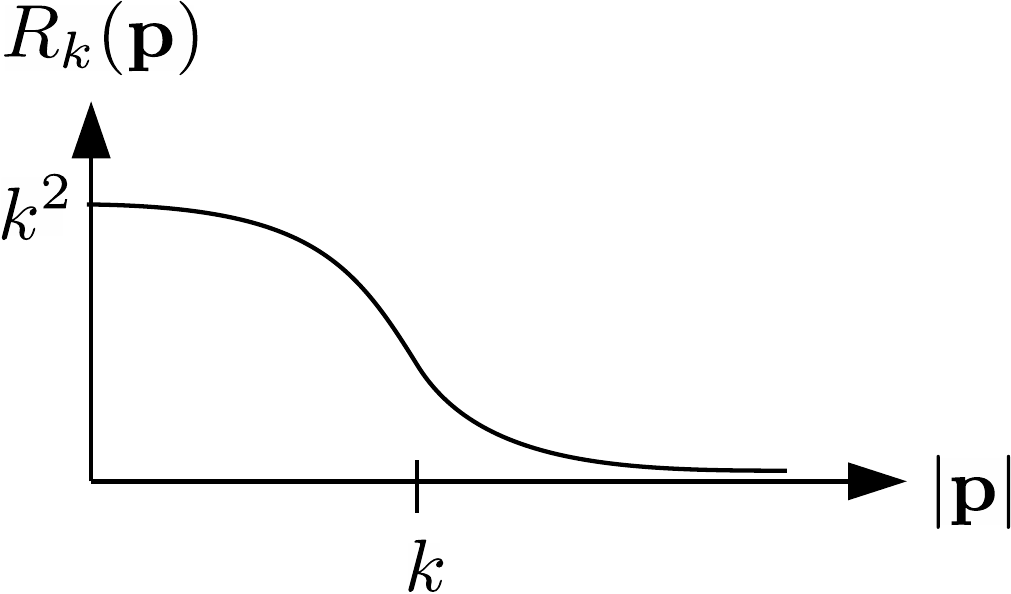}}
\caption{Typical shape of the regulator function $R_k({\bf p})$.}
\label{sec_frg:fig_Rk}
\end{figure}

\subsubsection{General properties of the effective action $\Gamma_k$}
\label{sec_frg:subsubsec_gammak_general}

i) The condition $R_{k_{\rm in}}=\infty$, which ensures that $\Gamma_{k_{\rm in}}=S$, can often be relaxed in particular when one is interested in universal properties of a model (e.g. the critical exponents or the universal scaling functions associated with a phase transition). In that case the microscopic physics can be directly parametrized by $\Gamma_\Lambda$ (with no need to specify the microscopic action) and one may simply choose $k_{\rm in}=\Lambda$. This is the most common situation and in the following, unless stated otherwise, we will assume $k_{\rm in}=\Lambda$ and $\Gamma_\Lambda=S$. However, in lattice models (and whenever one starts from a well-defined microscopic action) it is important to treat the initial condition at $k=k_{\rm in}\gg\Lambda$ carefully if one wants to compute nonuniversal quantities as a function of the microscopic parameters (see Sec.~\ref{sec_frg:subsec_lattice}).  
\smallskip

ii) Since $k$ acts as an infrared regulator, somewhat similar to a box of finite size $\sim k^{-1}$, 
the critical fluctuations are cut off by the $R_k$ term and
the effective action $\Gamma_k$ is analytic for $k>0$; there may be, however, some exceptions, e.g. in fermion systems (Sec.~\ref{sec_fb}) or disordered systems (Sec.~\ref{sec_disordered}).
The singularities associated with critical behavior therefore arise only for $k=0$. This implies in particular that
the vertices $\Gamma_{k,i_1\cdots i_n}^{(n)}[{\bf p}_1,\cdots,{\bf p}_n;\boldsymbol{\phi}]$ are smooth functions of the momenta and can 
be expanded in powers of ${\bf p}_i^2/k^2$ or ${\bf p}_i^2/m^2$, whichever
is the smallest, where $m=\xi^{-1}$ is the smallest ``mass'' of the problem and $\xi$ the correlation length. For the same reason, 
the effective action $\Gamma_k[\boldsymbol{\phi}]$ itself 
can be expanded in derivatives if one is interested only in the physics at length scales larger than either $k^{-1}$
or $\xi$. This property of the effective action and the $n$-point vertices is crucial as it underlies 
both the derivative expansion (Sec.~\ref{sec_frg:subsec_de}) and the Blaizot--M\'endez-Galain--Wschebor approximation (Sec.~\ref{sec_frg:sec_vertexp}). 
\smallskip

iii) All linear symmetries of the model that are respected by the infrared regulator $\Delta S_k$ are automatically symmetries of $\Gamma_k$. As a consequence, $\Gamma_k$ can be expanded in terms of invariants of these symmetries. 
\smallskip

iv) The effective action $\Gamma_k[\boldsymbol{\phi}]$ and the Wilsonian effective action $S^{\rm W}_k[\boldsymbol{\varphi}]$ are related but carry different physical meanings~\cite{Bonini93,Ellwanger94,Morris94,Morris05}.
%There are important conceptual differences between the effective action $\Gamma_k[\boldsymbol{\phi}]$ and the Wilsonian effective action $S^{\rm W}_k[\boldsymbol{\varphi}]$~\cite{Bonini93} (although they essentially coincide at the level of the local potential approximation (Sec.~\ref{sec_frg:subsubsec_lpa})~\cite{Morris05}). 
In the Wilson approach $k$ plays the role of a UV cutoff for the low-energy modes that remain to be integrated out~\cite{Wilson74,Wegner73,Polchinski84}. $S^{\rm W}_k[\boldsymbol{\varphi}]$ describes a set of different actions, parametrized by $k$, for the same model. The correlation functions are independent of $k$ and have to be computed from $S^{\rm W}_k[\boldsymbol{\varphi}]$ by functional integration. Information about correlation functions with momenta above $k$ is lost. 
In contrast $k$ acts as an infrared regulator in the effective action method. 
Moreover, $\Gamma_k$ is the effective action for a set of different models. The $n$-point correlation functions depend on $k$ and can be obtained from the $n$-point vertices $\Gamma_k^{(n)}$. The latter are defined for any value of external momenta.

\subsection{The exact flow equation}
\label{sec_frg:subsec_exacteq}

The FRG approach aims at relating the physics at different scales, e.g., in many cases of interest determining $\Gamma[\boldsymbol{\phi}]\equiv\Gamma_{k=0}[\boldsymbol{\phi}]$ from $\Gamma_{\Lambda}[\boldsymbol{\phi}]$ using Wetterich's equation~\cite{Wetterich93,Ellwanger94,Morris94,Bonini93}
\begin{equation}
\partial_k \Gamma_k[\boldsymbol{\phi}] = \frac{1}{2} {\rm Tr} \left\lbrace \partial_k R_k \bigl(\Gamma^{(2)}_k[\boldsymbol{\phi}]+R_k \bigr)^{-1} \right\rbrace ,
\label{sec_frg:eqwet}
\end{equation}   
where ${\rm Tr}$ denotes a trace wrt space and the O($N$) index of the field.\footnote{For earlier versions of Eq.~(\ref{sec_frg:eqwet}), see~\cite{Symanzik:1970rt,Nicoll77,Parola84,Chang92,Parola95}.} By taking successive functional derivatives of~(\ref{sec_frg:eqwet}), one obtains an infinite hierarchy of equations for the 1PI vertices, the first two of which are shown in Fig.~\ref{sec_frg:fig_eqwet}. It is sometimes convenient to introduce the (negative) RG ``time'' $t=\ln(k/\Lambda)$ and consider the equation $\partial_t \Gamma_k[\boldsymbol{\phi}]=k\partial_k\Gamma[\boldsymbol{\phi}]$. 

\begin{figure}
\centerline{\includegraphics[width=7.5cm]{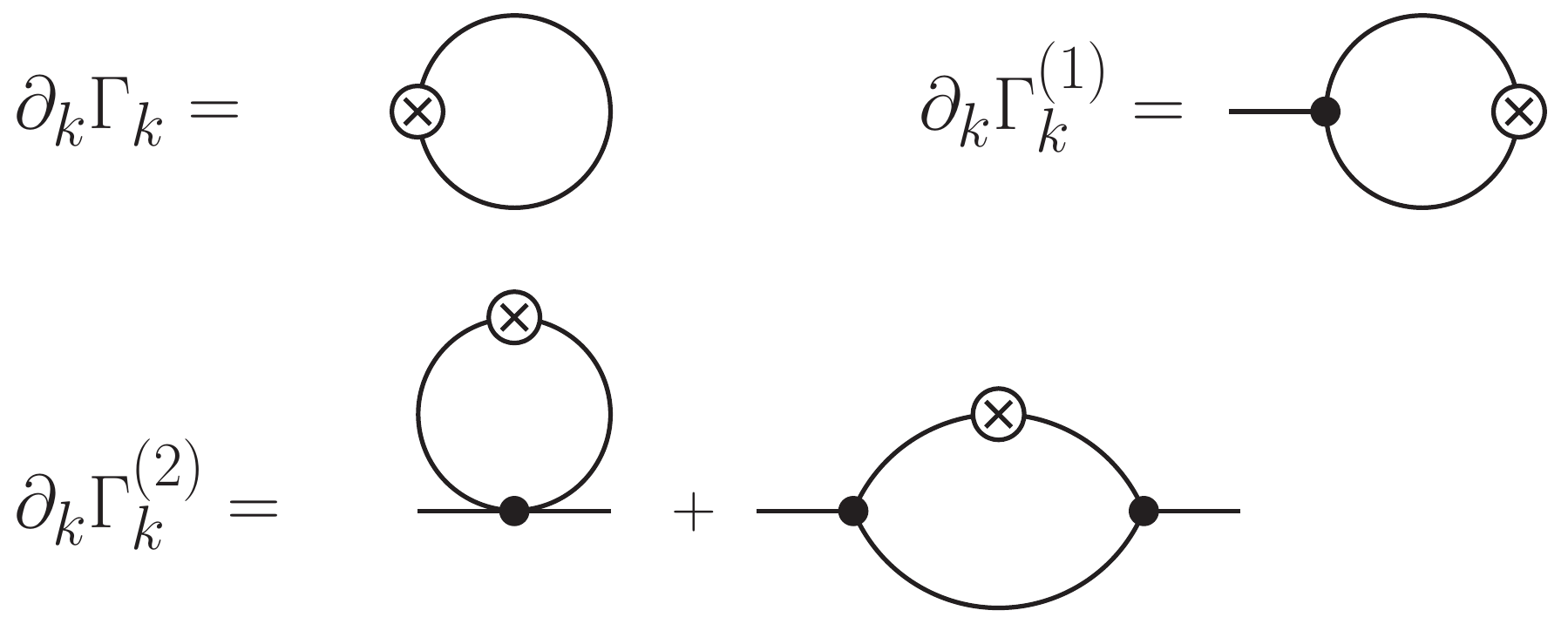}}
\caption{Diagrammatic representation of the RG equations satisfied by the effective action [Eq.~(\ref{sec_frg:eqwet})] and the vertices $\Gamma_k^{(1)}$ and $\Gamma_k^{(2)}$. The solid line stands for the propagator $G_k$, the cross for $\partial_k R_k$ and the dot with $n$ legs for $\Gamma_k^{(n)}$. (Signs and symmetry factors are not shown explicitly.)}
\label{sec_frg:fig_eqwet}
\end{figure}

Let us point out important properties satisfied by the flow equation~(\ref{sec_frg:eqwet}): 
\smallskip

i) The standard perturbative expansion about the Gaussian model can be retrieved from Eq.~(\ref{sec_frg:eqwet})~\cite{Papenbrock95,Bonini97,Morris99,Kopietz:2000bh}. 
\smallskip

ii) The flow equations for the $\Gamma_k^{(n)}$'s look very much like one-loop equations but where the vertices are the exact ones, $\Gamma_k^{(n)}[\boldsymbol{\phi}]$ (see Fig.~\ref{sec_frg:fig_eqwet}). Substitution of $\Gamma^{(2)}_k[\boldsymbol{\phi}]$ by $S^{(2)}[\boldsymbol{\phi}]$ in~(\ref{sec_frg:eqwet}) gives a flow equation which can be easily integrated out and yields the one-loop correction to the mean-field result $\Gamma_{\rm MF}[\boldsymbol{\phi}]=S[\boldsymbol{\phi}]$. The one-loop structure of the flow equation is important in practice as it implies that a single $d$-dimensional momentum integration has to be carried out in contrast to standard perturbation theory where $l$-loop diagrams require $l$-momentum integrals.  
\smallskip

iii) Since the one-loop approximation is recovered by approximating $\Gamma^{(2)}_k[\boldsymbol{\phi}]$ by $S^{(2)}[\boldsymbol{\phi}]$ in~(\ref{sec_frg:eqwet}), any sensible approximation of the flow equation will be one-loop exact (in the sense that it encompasses the one-loop result when expanded in the coupling constants). This implies that all results obtained from a one-loop approximation must be recovered from the nonperturbative flow equation. This includes for example the computation of the critical exponents to ${\cal O}(\epsilon=4-d)$ near four dimensions (the upper critical dimension of the O($N$) model) (see Sec.~\ref{sec_frg:subsec_de}).  
\smallskip

iv) The presence of $\partial_k R_k({\bf q})$ in the trace of Eq.~(\ref{sec_frg:eqwet}) implies that only momenta ${\bf q}$ of order $k$ or less contribute to the flow at scale $k$ (provided that $R_k({\bf q})$ decays sufficiently fast for $|{\bf q}|\gg k$),  which implements Wilson's idea of momentum shell integration of fluctuations (with a soft separation between fast and slow modes). 
This, in particular, ensures that the momentum integration is 
UV finite. Furthermore, the $R_k$ term appearing in the propagator 
$G_k=(\Gamma_k^{(2)}+R_k)^{-1}$ acts as an infrared regulator and ensures that the momentum integration 
in~(\ref{sec_frg:eqwet}) is free of infrared divergences. This makes the formulation well-suited 
to deal with theories that are plagued with infrared problems in perturbation theory, e.g. in the vicinity
of a second-order phase transition.   
\smallskip

\begin{figure}
\centerline{\includegraphics[width=4cm]{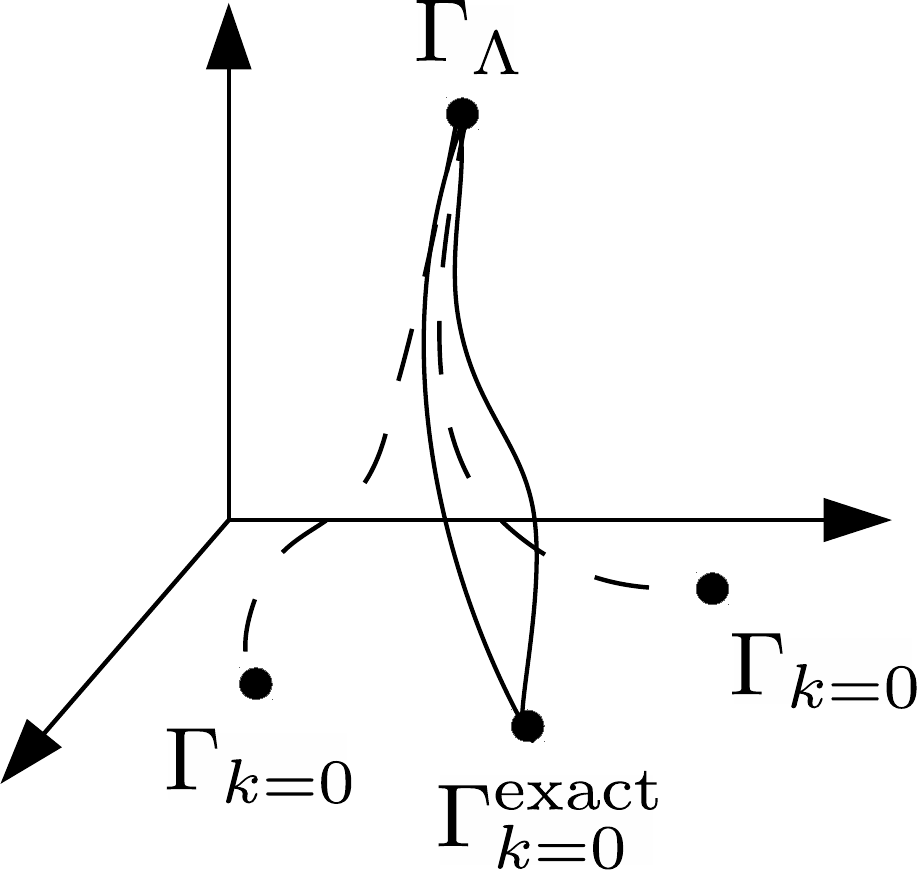}}
\caption{RG flow in the parameter space of the effective action. The solid lines show the exact RG flows obtained with two different regulator functions $R_k$. The dashed lines show the RG flows obtained by solving the RG equation with the same approximation and two different regulator functions.}
\label{sec_frg:fig_flow_gammak} 
\end{figure}

v) Different choices of the regulator function $R_k$ correspond to different trajectories 
in the space of effective actions. If no approximation were made on the flow equation, the final point 
$\Gamma_{k=0}$ would be the same for all trajectories. However, once approximations are made, 
$\Gamma_{k=0}$  acquires a dependence on the precise shape of $R_k$ (Fig.~\ref{sec_frg:fig_flow_gammak}).\footnote{For a study in the framework of general Wilsonian RG flows, see~\cite{Latorre:2000qc,Latorre01}. For a discussion at one- and two-loop order in the Wilson-Polchinski and effective-action formulations, see~\cite{Arnone:2002yh,Arnone:2003pa}.}$^{,}$\footnote{The $R_k$ dependence is similar to the scheme dependence in perturbative RG (physical results depend on the RG prescription, e.g. $\overline{\mbox{MS}}$ scheme, massive zero-momentum scheme, etc.), see e.g.~\cite{Litim:1996nw, Pernici:1997ie, Ellwanger:1997tp, Pernici:1998tp, Latorre:2000qc, Arnone:2002yh, Arnone:2003pa, Pawlowski:2005xe, Rosten:2006pd, Codello:2013bra}.} This dependence can be used to study the robustness of the approximations used to solve~(\ref{sec_frg:eqwet})~\cite{Schnoerr13}.  
\smallskip

vi) The flow equation~(\ref{sec_frg:eqwet}) is a complicated functional integro-differential equation, 
which cannot (except in trivial cases) be solved exactly. Two main types of approximations have 
been designed: the derivative expansion, which is based on an ansatz for $\Gamma_k$ involving a 
finite number of derivatives of the field (Sec.~\ref{sec_frg:subsec_de}), and the vertex expansion, which is based on a 
truncation of the infinite hierarchy of equations satisfied by the $\Gamma_k^{(n)}$'s (Sec.~\ref{sec_frg:sec_vertexp}).  
\smallskip

vii) The flow equation -- unlike the path-integral expression for $\Gamma_k$ -- no longer depends on the microscopic action $S$. This enables a search for a consistent microscopic dynamics through a fixed-point search of Eq.~(\ref{sec_frg:eqwet}). This property, which will be
discussed in more detail in Secs.~\ref{sec_hep} and \ref{sec_gr}, is key for the application
in quantum gravity as well as in many beyond-Standard-Model settings.

\subsection[The derivative expansion (DE)]{The derivative expansion\footnote{See~\ref{appDE} for a further discussion of the DE.} (DE)}
\label{sec_frg:subsec_de}

The DE is based on the regularity of the scale-dependent effective action at small momentum scales, $|{\bf p}|\leq {\rm max}(k,\xi^{-1})$ (Sec.~\ref{sec_frg:subsubsec_gammak_general}). Since the $\partial_k R_k$ term in Eq.~(\ref{sec_frg:eqwet}) implies that the integral over the internal loop momentum ${\bf q}$ is dominated by $|{\bf q}|\leq k$, an expansion in internal and external momenta of the vertices, corresponding to a DE of the effective action, makes sense and can be used to obtain the thermodynamics and the long-distance behavior of the
system. As will be explained in detail in Sec.~\ref{sec_frg:validityDE} and Appendix~\ref{appDE}, long-distance quantities computed within the
DE converge very quickly to their exact physical values.

\subsubsection{Local-potential approximation: LPA and LPA$'$} 
\label{sec_frg:subsubsec_lpa}

To lowest order of the DE, the local potential approximation (LPA), the effective action 
\begin{equation}
\Gamma_k^{\rm LPA}[\boldsymbol{\phi}] = \int_{\bf r} \left\lbrace \frac{1}{2} (\boldsymbol{\nabla}\boldsymbol{\phi})^2 + U_k(\rho) \right\rbrace 
\label{sec_frg:gammaklpa}
\end{equation}
is entirely determined by the effective potential $U_k(\rho)$ whereas the derivative term keeps its bare (unrenormalized) form~\cite{Reuter93,Morris94b}. Despite its simplicity $\Gamma_k^{\rm LPA}$ is highly nontrivial from a perturbation theory point of view since it includes vertices to all orders: $\Gamma_k^{{\rm LPA}(n)}\sim \partial_\phi^n U_k$ for $n\geq 3$. The effective potential satisfies the exact equation
\begin{equation}
\partial_k U_k(\rho) = \frac{1}{2} \int_{\bf q} \partial_k R_k({\bf q}) [ G_{k,\rm L}({\bf q},\rho) + (N-1) G_{k,\rm T}({\bf q},\rho) ] 
\label{sec_frg:eqLPA} 
\end{equation} 
with initial condition $U_\Lambda(\rho)=r_0\rho+(u_0/6)\rho^2$, where 
$G_{k,\alpha}({\bf q},\rho) = [\Gamma^{(2)}_{k,\alpha}({\bf q},\rho) + R_k({\bf q})]^{-1}$ ($\alpha={\rm L,T}$) are the longitudinal and transverse parts (wrt the order parameter $\boldsymbol{\phi}$) of the propagator $G_k({\bf q},\boldsymbol{\phi})$ evaluated in the uniform field configuration $\boldsymbol{\phi}({\bf r})=\boldsymbol{\phi}$. The contribution of the transverse propagator appears with a factor $N-1$ corresponding to the number of transverse modes when $\rho$ is nonzero. Within the LPA, one has 
\begin{equation} 
\begin{split}
G_{k,\rm L}({\bf q},\rho) &= [{\bf q}^2+U'_k(\rho)+2\rho U_k''(\rho)+R_k({\bf q})]^{-1} , \\  
G_{k,\rm T}({\bf q},\rho) &= [{\bf q}^2+U'_k(\rho)+R_k({\bf q})]^{-1} .
\end{split}
\label{sec_frg:GLTlpa}
\end{equation}

A negative $r_0$ corresponds to a system which is in the ordered phase at the mean-field level and the potential $U_\Lambda(\rho)$ then exhibits a minimum at $\rho_{0,\Lambda}=-3r_0/u_0>0$. When $r_0$ is smaller than a critical value $r_{0c}<0$, fluctuations are not sufficiently strong to fully suppress long-range order and the system is in the ordered phase, i.e., $0<\rho_0<\rho_{0,\Lambda}$ where $\rho_0=\lim_{k\to 0}\rho_{0,k}$.  In that case the effective potential $U_{k=0}(\rho)$ is flat for $\rho<\rho_0$. The convexity of the effective potential $U_{k=0}$, which in the exact solution is a consequence of its definition as a Legendre transform, can be ensured in the LPA by a proper choice of the regulator. The approach to convexity of $\Gamma_k[\boldsymbol{\phi}]$ (which, being not a pure Legendre transform for $k>0$, is not necessarily convex) is discussed in~\cite{Tetradis92,Tetradis96,Berges:2000ew,Pelaez16,Litim:2006nn}. For $r_0>r_{0c}$ the system is in the disordered phase, $\rho_0=0$, with a finite correlation length $\xi$.  

At criticality ($r_0=r_{0c}$) the scale invariance due to the infinite correlation length can be made manifest by expressing all quantities in unit of the running momentum scale $k$. This amounts to defining dimensionless variables (coordinate, field and potential) as 
\begin{equation}
\tilde {\bf r} = k {\bf r} , \quad \tilde \rho = k^{-(d-2)} \rho, \quad \tilde U_k({\tilde\rho}) = k^{-d} U_k(\rho) 
\label{sec_frg:adim1}
\end{equation} 
(this is equivalent to the usual momentum and field rescaling in the standard formulation of the Wilsonian RG). 
The dimensionless effective potential $\tilde U_k({\tilde\rho})$ of the critical system flows to a fixed point $\tilde U^*({\tilde\rho})$ of the flow equation $\partial_k \tilde U_k({\tilde\rho})$. The fixed-point equation and its numerical solution are discussed in~\cite{Morris94a,Morris94b,Bagnuls01,Ball95,Comellas98,Morris98,Zumbach94}. Linearizing the flow about the fixed-point value $\tilde U^*$ gives the correlation-length exponent $\nu$ and the correction-to-scaling exponent $\omega$. The LPA can also be used to study the ordered phase ($r_0<r_{0c}$) and the results~\cite{Defenu15} agree with the Mermin-Wagner theorem forbidding spontaneous broken symmetry when $N\geq 2$ and $d\leq 2$~\cite{Mermin66,Hohenberg67,Coleman73}. The solution of the fixed-point equation for $\tilde{U}_k$, which is independent of $\tilde{U}_{\Lambda}$, is a paradigmatic example showcasing the power of the FRG in the search for scale-invariance that is central in the study of asymptotic safety in a high-energy context, see, e.g., \cite{Labus:2015ska} for the generalization of Eq.~(\ref{sec_frg:eqLPA}) to the case with quantum gravity and the gravitationally dressed Wilson-Fisher fixed point.

An important limitation of the LPA is the absence of anomalous dimension, since $G^{\rm LPA}_{k=0}({\bf p},\boldsymbol{\phi}=0)=1/|{\bf p}|^2$ at criticality, whereas one expects $\sim 1/|{\bf p}|^{2-\eta}$ with $\eta>0$ when $d<4$. This form can never be obtained in the DE (to all orders) since $\Gamma_k^{(2)}({\bf p},\boldsymbol{\phi})$ is a regular function of ${\bf p}$ when $|{\bf p}|\ll k$ (the domain of validity of the DE). The anomalous dimension $\eta$ can nevertheless be obtained from a slight improvement of the LPA effective action, 
\begin{equation}
\Gamma_k^{{\rm LPA}'}[\boldsymbol{\phi}] = \int_{\bf r} \left\lbrace \frac{Z_k}{2} (\boldsymbol{\nabla}\boldsymbol{\phi})^2 + U_k(\rho) \right\rbrace ,
\label{sec_frg:gammalpap}
\end{equation}
which includes a field renormalization factor $Z_k$.
To allow for a scaling solution of the flow equations and therefore a fixed point at criticality, the regulator function must be defined as $R_k({\bf p})=Z_k{\bf p}^2 r({\bf p}^2/k^2)$.\footnote{The prefactor $Z_k$ in the definition of $R_k$ is required by the Ward identities associated with scale invariance~\cite{Delamotte16a}. It ensures that no intrinsic scale is introduced in the renormalized inverse propagator $\Gamma^{(2)}_k({\bf p})/Z_k={\bf p}^2(1+r({\bf p}^2/k^2))$ when $U_k=0$.} 
One can then define a ``running'' anomalous dimension $\eta_k=-k\partial_k\ln Z_k$. At criticality $\lim_{k\to 0}\eta_k\equiv\eta>0$ and $Z_k$ diverges as $k^{-\eta}$. The two-point vertex $\Gamma^{(2)}_{k}({\bf p},\boldsymbol{\phi})-\Gamma^{(2)}_{k}(0,\boldsymbol{\phi})\simeq Z_k{\bf p}^2$ is a regular function of ${\bf p}$ when $|{\bf p}|\ll k$ as ensured by the infrared regulator $R_k$. For $|{\bf p}|\gg k$, a momentum range which is outside the domain of validity of the DE, one expects the singular behavior $\Gamma^{(2)}_{k}({\bf p},\boldsymbol{\phi})-\Gamma^{(2)}_{k}(0,\boldsymbol{\phi})\sim |{\bf p}|^{2-\eta}$ for $|{\bf p}|$ smaller than the Ginzburg momentum scale $p_G\sim u_0^{1/(4-d)}$. Since $Z_k\sim k^{-\eta}$ these two limiting forms (for $|{\bf p}|\ll k$ and $|{\bf p}|\gg k$) match for $|{\bf p}|\sim k$. The fact that the anomalous dimension controls the divergence of $Z_k$, and can therefore be obtained from the LPA$'$, can be shown rigorously~\cite{Blaizot06a}. 

It is possible to simplify the LPA$'$ by expanding the effective potential to lowest (nontrivial) order about $\rho_{0,k}$, 
\begin{equation}
U_k(\rho) = U_{0,k} + \delta_k (\rho-\rho_{0,k}) + \frac{\lambda_k}{2} (\rho-\rho_{0,k})^2 , 
\label{sec_frg:Utroncated} 
\end{equation}
where $\delta_k=0$ if $\rho_{0,k}>0$. This gives coupled ordinary differential equations for the coupling constants $\rho_{0,k}$, $\delta_k$, $\lambda_k$ and $Z_k$ (and a separate one for the free energy $U_{0,k}$). The truncated LPA$'$ is still nonperturbative to the extent that $\partial_k \delta_k$, $\partial_k\lambda_k$, etc., are nonpolynomial functions of the coupling constants. This simple truncation retains the main features of the LPA$'$ flow and turns out to be sufficient to recover the critical exponents to leading order in the large-$N$ limit as well as to ${\cal O}(\epsilon)$ near four dimensions ($d=4-\epsilon$) and two dimensions ($d=2+\epsilon$) as obtained from perturbative RG in the O($N$) nonlinear sigma model. 

The LPA$'$ is not reliable for a precise estimate of the critical exponents but it shows that the FRG, even with a very simple truncation of the effective action, interpolates smoothly between two and four dimensions and suggests that with more involved truncations one can reliably explore the behavior of the system in any dimension and in particular $d=3$~\cite{Codello12,Codello13,Codello15} (see Secs.~\ref{sec_frg:subsubsec_DE2}, \ref{sec_frg:sec_vertexp} and \ref{appDE}).

\subsubsection{Second order of the DE}
\label{sec_frg:subsubsec_DE2}

To obtain reliable estimates of the critical exponents, it is necessary at least to consider the second order of the DE where the effective action, 
\begin{equation}
\Gamma_k^{{\rm DE}_2}[\boldsymbol{\phi}] = \int_{\bf r} \left\lbrace \frac{1}{2} Z_k(\rho) (\boldsymbol{\nabla}\boldsymbol{\phi})^2 + \frac{1}{4} Y_k(\rho) (\boldsymbol{\nabla}\rho)^2 + U_k(\rho) \right\rbrace, 
\label{sec_frg:gammaDE2}
\end{equation}
is defined by three functions of $\rho$~\cite{Wetterich93a,Tetradis94,Aoki98,Morris97,Morris98,Morris99,Berges:2000ew}. In addition to the effective potential, there are two derivative terms, reflecting the fact that transverse and longitudinal fluctuations (wrt the local order parameter $\boldsymbol{\phi}({\bf r})$) have different stiffness. For $N=1$ the term $Y_k(\rho)(\boldsymbol{\nabla}\rho)^2$ should be omitted since it can be put in the form $Z_k(\rho)(\boldsymbol{\nabla}\boldsymbol{\phi})^2$. As in the LPA it is convenient to use dimensionless variables, 
\begin{equation}
 \tilde U_k({\tilde\rho}) = k^{-d} U_k(\rho), \quad \tilde Z_k({\tilde\rho}) = Z_k^{-1} Z_k(\rho) , \quad 
 \tilde Y_k({\tilde\rho}) = Z_k^{-2} k^{d-2} Y_k(\rho) , 
\label{sec_frg:adim2}
\end{equation}
where ${\tilde\rho}=Z_k k^{-(d-2)}\rho$. The field renormalization factor $Z_k$ is defined by imposing the condition $\tilde Z_k({\tilde\rho}_{\rm r})=1$, where ${\tilde\rho}_{\rm r}$ is an arbitrary renormalization point (e.g. ${\tilde\rho}_r=0$) and, as mentioned in the discussion of the LPA$'$, must appear as a prefactor in the definition of the regulator function $R_k$. 
We note that only if the factor $Z_k$ is introduced in the regulator does the fixed-point condition become identical to the Ward identity for scale invariance (in presence of the infrared regulator)~\cite{Delamotte16a}. This implements Wilson's original idea identifying RG fixed point and scale invariance. By inserting the ansatz~(\ref{sec_frg:gammaDE2}) into the flow equation~(\ref{sec_frg:eqwet}) one obtains four coupled differential equations for the three functions $\tilde U_k$, $\tilde Z_k$ and $\tilde Y_k$ and the running anomalous dimension $\eta_k=-k\partial_k \ln Z_k$.

\begin{table}[t]
	\small
	\centering
	\caption{Critical exponents $\nu$, $\eta$ and $\omega$ for the
		three-dimensional O($N$) universality class obtained in the FRG approach
		from DE to second~\cite{Seide99,Gersdorff01}, fourth~\cite{DePolsi20}
		and sixth~\cite{Balog19} orders, LPA$''$~\cite{Hasselmann12,Rose18} and 
		BMW approximation~\cite{Benitez09,Benitez12}, compared to Monte Carlo
		(MC) simulations~\cite{Hasenbusch10,Campostrini06,Campostrini02,Hasenbusch:2019jkj,Clisby16,Clisby_2017},
		$d=3$ perturbative RG (PT)~\cite{Guida98},
		$\epsilon$-expansion at order $\epsilon^6$ ($\epsilon$-exp)~\cite{Kompaniets:2017yct}
		and conformal bootstrap (CB)~\cite{Shimada:2015gda,Kos16,Simmons-Duffin:2016wlq,Echeverri2016,chester2019carving}  (when several estimates are available in the literature, we show the one with the smallest error bar).}
	\vspace{0.25cm}
	\begin{tabular}{l@{\hskip5.5pt}l@{\hskip5.5pt}l@{\hskip5.5pt}l@{\hskip5.5pt}l@{\hskip5.5pt}l@{\hskip5.5pt}l@{\hskip5.5pt}l@{\hskip5.5pt}l@{\hskip5.5pt}l@{\hskip5.5pt}l@{\hskip5.5pt}l}
		\hline \hline
		& \multicolumn{9}{c}{Correlation-length exponent $\nu$}  \\
		\multicolumn{1}{c}{$N$}         &\multicolumn{1}{c}{LPA}
		&\multicolumn{1}{c}{DE$_2$}&\multicolumn{1}{c}{DE$_4$}&
		\multicolumn{1}{c}{DE$_6$} & \multicolumn{1}{c}{LPA$''$} &
		\multicolumn{1}{c}{BMW} & \multicolumn{1}{c}{MC} &
		\multicolumn{1}{c}{PT} &
		\multicolumn{1}{c}{$\epsilon$-exp} &
		\multicolumn{1}{c}{CB} \\
		\hline
		$0$    & 0.5925  & 0.5879(13)   & 0.5876(2) & \multicolumn{1}{c}{--} & \multicolumn{1}{c}{--}  & \multicolumn{1}{c}{0.589} & 0.58759700(40) & 0.5882(11) & 0.5874(3) & 0.5876(12) \\
		$1$    & $0.650$ & $0.6308(27)$ & $0.62989(25)$ &  $0.63012(16)$ & \multicolumn{1}{c}{$0.631$} & \multicolumn{1}{c}{$0.632$}  &
		$0.63002(10)$ & $0.6304(13)$ & $0.6292(5)$ & $0.629971(4)$ \\
		$2$    & 0.7090  & 0.6725(52) & 0.6716(6) & \multicolumn{1}{c}{--} & \multicolumn{1}{c}{$0.679$}  & \multicolumn{1}{c}{$0.674$} & 0.67169(7)   & 
		0.6703(15) & 0.6690(10) & 0.6718(1) \\
		$3$    & 0.7620 & 0.7125(71) &  0.7114(9) &\multicolumn{1}{c}{--} & \multicolumn{1}{c}{$0.725$} & \multicolumn{1}{c}{$0.715$} & 0.7112(5)    & 
		0.7073(35) & 0.7059(20) &  0.7120(23) \\
		$4$    & 0.805  & 0.749(8)   & 0.7478(9)  & \multicolumn{1}{c}{--} & \multicolumn{1}{c}{0.765}  & \multicolumn{1}{c}{0.754} & 0.7477(8)    &
		0.741(6)& 0.7397(35) & 0.7472(87)\\
		\hline
	\end{tabular}
	
	\vspace{0.5cm}
	\begin{tabular}{l@{\hskip5.5pt}l@{\hskip5.5pt}l@{\hskip5.5pt}l@{\hskip5.5pt}l@{\hskip5.5pt}l@{\hskip5.5pt}l@{\hskip5.5pt}l@{\hskip5.5pt}l@{\hskip5.5pt}l}
		\hline \hline
		& \multicolumn{8}{c}{Anomalous dimension $\eta$}  \\
		\multicolumn{1}{c}{$N$}          &\multicolumn{1}{c}{DE$_2$}
		&\multicolumn{1}{c}{DE$_4$} & \multicolumn{1}{c}{DE$_6$} &
		\multicolumn{1}{c}{LPA$''$} & \multicolumn{1}{c}{BMW} &
		\multicolumn{1}{c}{MC} & \multicolumn{1}{c}{PT} &
		\multicolumn{1}{c}{$\epsilon$-exp} & \multicolumn{1}{c}{CB} \\
		\hline
		$0$        & 0.0326(47) & 0.0312(9) & \multicolumn{1}{c}{--} & \multicolumn{1}{c}{--}   & \multicolumn{1}{c}{0.034} & 0.0310434(30) & 0.0284(25) & 0.0310(7) & 0.0282(4) \\
		$1$        & $0.0387(55)$ & $0.0362(12)$ & $0.0361(11)$ & $0.0506$ & \multicolumn{1}{c}{$0.039$} &
		$0.03627(10)$ & $0.0335(25)$ & $0.0362(6)$ & $0.0362978(20)$  \\
		$2$        & 0.0410(59) & 0.0380(13) & \multicolumn{1}{c}{--} & $0.0491$ & \multicolumn{1}{c}{$0.041$} & 0.03810(8)   &
		0.0354(25) & 0.0380(6) & 0.03818(4) \\
		$3$        & 0.0408(58)& 0.0376(13)  & \multicolumn{1}{c}{--} & $0.0459$ & \multicolumn{1}{c}{$0.040$} & 0.0375(5)   &
		0.0355(25) & 0.0378(5) & 0.0385(13) \\
		$4$        & 0.0389(56)& 0.0360(12)  & \multicolumn{1}{c}{--} & 0.0420     & \multicolumn{1}{c}{0.038} & 0.0360(4)   &
		0.0350(45) & 0.0366(4) & 0.0378(32)\\
		\hline
	\end{tabular}
	
	\vspace{0.5cm}
	\begin{tabular}{l@{\hskip5.5pt}l@{\hskip5.5pt}l@{\hskip5.5pt}l@{\hskip5.5pt}l@{\hskip5.5pt}@{\hskip5.5pt}l@{\hskip5.5pt}l@{\hskip5.5pt}l@{\hskip5.5pt}l}
		\hline \hline
		& \multicolumn{8}{c}{Correction-to-scaling exponent $\omega$}  \\
		\multicolumn{1}{c}{$N$}         &\multicolumn{1}{c}{LPA}
		&\multicolumn{1}{c}{DE$_2$}&\multicolumn{1}{c}{DE$_4$}&  \multicolumn{1}{c}{BMW} &
		\multicolumn{1}{c}{MC} & \multicolumn{1}{c}{PT} &
		\multicolumn{1}{c}{$\epsilon$-exp}  & \multicolumn{1}{c}{CB} \\
		\hline
		$0$ & 0.66    & 1.00(19)    & 0.901(24) & \multicolumn{1}{c}{0.83} & 0.899(14)& 0.812(16) & 0.841(13)& \multicolumn{1}{c}{--} \\
		$1$ & $0.654$ & $0.870(55)$ & $0.832(14)$ & \multicolumn{1}{c}{0.78} & 0.832(6) & 0.799(11) & 0.820(7) & 0.82968(23) \\
		$2$ & 0.672   & 0.798(34)   & 0.791(8)    & \multicolumn{1}{c}{0.75} & 0.789(4) & 0.789(11) & 0.804(3) & 0.794(8) \\
		$3$ & 0.702   & 0.754(34)   & 0.769(11)   & \multicolumn{1}{c}{0.73} & 0.773 & 0.782(13) & 0.795(7) & 0.791(22)  \\
		$4$ & 0.737   & 0.731(34)   & 0.761(12)   & \multicolumn{1}{c}{0.72} & 0.765    &  0.774(20)  & 0.794(9) & 0.817(30) \\
		\hline
	\end{tabular}
	\label{sec_frg:table_critexp}
	\normalsize
\end{table}

The choice of the regulator function $R_k$ is crucial when looking for accurate estimates of critical exponents. One usually considers a family of functions depending on one or more parameters $\{\alpha_i\}$ (see, e.g., the exponential and theta regulators defined in Sec.~\ref{sec_frg:subsubsec_Rk} which depend on a single parameter $\alpha$). We determine the optimal value of $\{\alpha_i\}$ from the principle of minimal sensitivity, that is by demanding that locally critical exponents be independent of $\{\alpha_i\}$, e.g. $d\nu/d\alpha_i=0$ for the correlation-length exponent. The renormalization point ${\tilde\rho}_{\rm r}$ is usually taken fixed (for numerical convenience) and, provided that the fixed point exists, a change in ${\tilde\rho}_{\rm r}$ is equivalent to a change in the amplitude of $R_k$ (which is usually one of the $\alpha_i$'s) so that the critical exponents are independent of ${\tilde\rho}_{\rm r}$~\cite{Balog19}. The optimization of the regulator choice is discussed in~\cite{Litim00,Litim01,Litim01b,Litim02,Litim05,Liao00,Canet03a,Canet03b,Canet05,Morris05,Pawlowski:2005xe,Nandori14,Codello:2013bra,Pawlowski15b,Balog19,DePolsi20}.

Results for the critical exponents of the three-dimensional O($N$) universality class obtained from the LPA and the DE to second, fourth~\cite{Canet03b,Litim11,DePolsi20} and sixth~\cite{Balog19} orders are shown in Table~\ref{sec_frg:table_critexp} for $N=0,1,2,3,4$ and compared to Monte Carlo simulations, fixed-dimension perturbative RG, $\epsilon$-expansion and conformal bootstrap (the two-dimensional O(1) model (Ising universality class) is discussed in~\cite{Morris95,Ballhausen04,Defenu18}). In the large-$N$ limit the DE to second order becomes exact for the critical exponents and the functions $U_k(\rho)$ and $Z_k(\rho)$~\cite{Reuter93,Tetradis94,Tetradis96,DAttanasio97,Rose18}.\footnote{{\it Stricto sensu} this is true only if the longitudinal propagator $G_{k,\rm L}({\bf q},\rho)$ is ${\cal O}(N^0)$ for any value of $\rho$ (the $N\to\infty$ limit of $U_k(\rho)$ is then a regular function), which is the case for the Wilson-Fisher fixed point but not for all fixed points~\cite{Yabunaka17,Yabunaka18,Katsis:2018bvc}.\label{sec_frg:footnote1}}

Since the DE is {\it a priori} valid in all dimensions and for all $N$, it can be applied to the two-dimensional O(2) model where the transition, as predicted by BKT~\cite{Berezinskii71,Berezinskii72,Kosterlitz73,Kosterlitz74}, is driven by topological defects (vortices). The FRG approach requires a fine tuning of the regulator function $R_k$ in order to reproduce, {\it stricto sensu}, the line of fixed points in the low-temperature phase but otherwise recovers most universal features of the BKT transition~\cite{Graeter95,Gersdorff01,Jakubczyk14,Jakubczyk16,Rancon17,Jakubczyk17,Jakubczyk17a,Defenu17,Krieg17a}. This significantly differs from more traditional studies, based on the Coulomb gas or Villain models~\cite{Kosterlitz74,Villain75}, where vortices are introduced explicitly. 

An important feature of the DE is that all linear symmetries can be implemented at the level of the effective action with the result that physical quantities satisfy these symmetries if the regulator does. The latter condition is trivially realized in the O($N$) model but is more difficult to satisfy in other cases, e.g. in gauge theories (see Secs.~\ref{sec_hep} and \ref{sec_gr}).

The numerical solution of the flow equations is an important part of the FRG approach (for both the DE and the methods described in Sec.~\ref{sec_frg:sec_vertexp}). Differential equations can be solved using the explicit Euler or Runge-Kutta methods (with a discretized RG time $t=\ln(k/\Lambda)$) while momentum integrals can be computed using standard techniques. The variable $\tilde\rho$ is  often discretized but it is also possible to use pseudospectral methods (e.g. based on Chebyshev polynomials) for the field-dependent functions~\cite{Fischer:2004uk,Borchardt16,Borchardt15,Rose16a}, or discontinuous Galerkin methods, that combine the strength of finite-volume methods and pseudo-spectral methods, see~\cite{Grossi:2019urj}. The DE can be simplified by truncating the effective action in powers of the field (as in Eq.~(\ref{sec_frg:Utroncated})); the convergence of this expansion is discussed in~\cite{Liao00,Canet03a,Canet03b,Litim01a,Litim02,Aoki98}.

\subsubsection{Validity of the DE}
\label{sec_frg:validityDE}

The success of the DE can be partially understood by the functional form of the flow equations which makes the expansion nonperturbative in the coupling constants even to lowest order (LPA). But the convergence and the existence of a small parameter that would make the DE a fully controlled approximation is not {\it a priori} obvious.

In the previous sections the DE was (loosely) justified by the scale-dependent effective action $\Gamma_k[\boldsymbol{\phi}]$ being regular at small momentum scales $|{\bf p}|\ll {\rm max}(k,\xi^{-1})$ and the fact that its flow equation is insensitive to momenta larger than $k$ due to the presence of $\partial_k R_k({\bf p})$ in the momentum integrals. More precisely the convergence of the DE requires two properties: i) the expansion in ${\bf p}^2/k^2$ of the effective action $\Gamma_k[\boldsymbol{\phi}]$ must have a nonzero radius of convergence and ii) the momentum cutoff $|{\bf p}|\lesssim p_{\rm max}$ due to $\partial_k R_k({\bf p})$ must be sufficiently efficient for the parameter $p^2_{\rm max}/k^2$ to be significantly smaller than the radius of convergence. 

These two conditions are likely to be satisfied in all unitary theories (i.e., Euclidean theories whose analytic continuation in Minkowski space is unitary) for which the structure of nonanalyticities of correlation functions is known. The radius of convergence of the momentum expansion of a given correlation function is determined by the singularity closest to the origin ${\bf p}=0$. For the two-point correlation function, this singularity is located in the complex plane at ${\bf p}^2=-m^2$ where $m$ is the mass (the inverse correlation length). The next singularity, which shows up in all correlation functions, is located at ${\bf p}^2=-9m^2$ and $-4m^2$ in the disordered and ordered phases, respectively, and corresponds to the threshold of the two-particle excitation continuum.

Consider now the regulated model defined by the action $S+\Delta S_k$. Because of the regulator function $R_k$, at criticality the two-point correlation function $G_k=(\Gamma_k^{(2)}+R_k)^{-1}$ exhibits a mass $m_k\equiv k$ for $k\to 0$ due to the regulator $R_k$ (one can always redefine the running momentum scale such that $m_k=k$). We therefore expect the Taylor expansion in ${\bf p}$ in that model to have the same radius of convergence as in the Ising model with a mass $m\equiv k$. The singularity at ${\bf p}^2=-m^2\equiv -k^2$ determines the radius of convergence of $G_k$ but not that of $\Gamma_k^{(2)}$ and higher-order vertices. The singularity closest to the origin in $\Gamma_k^{(2)}$ corresponds to the threshold of the two-particle continuum and is expected to be in the range $[-9k^2,-4k^2]$. This implies a radius of convergence for the ${\bf p}^2/k^2$ expansion of the effective action in the range $[4,9]$ and ensures that the condition (i) defined above is satisfied. The condition (ii) is then easily fulfilled by choosing a regulator function $R_k$ which cuts off momentum integrals (at least) exponentially for $|{\bf p}|\gtrsim k$ (which is the case of most regulators used in practice).\footnote{One could expect that it is possible to devise a function $R_k$ which yields a mass $m_k=k$ and is (at least) exponentially suppressed for momenta above $p_{\rm max}\ll k$. Such a regulator would however introduce a singularity near the origin (implying large high-order derivatives) and spoil the analytic structure of the correlation functions and thus the convergence of the DE.\label{sec_frg:footnote2}} Finally one must also add that the corrections to second order and higher in the DE are suppressed by a factor $\eta$~\cite{Balog19,DePolsi20}, the anomalous dimension, which is a small number in the $\varphi^4$ theory. The convergence of the DE has been nicely illustrated by 4th- and 6th-order calculations in the three-dimensional O($N$)-model universality class (see Sec.~\ref{sec_frg:subsubsec_DE2} and Table~\ref{sec_frg:table_critexp} as well as \ref{appDE}).

This reasoning also explains why the DE exhibits poor convergence properties in the Wilson-Polchinski formulation, i.e., without performing the Legendre transform~\cite{Morris99,Kubyshin02}, even if the LPA gives satisfactory results~\cite{Litim05,Morris05}. The leading singularity being located at ${\bf p}^2=-k^2$ for the two-point correlation function, the radius of convergence in ${\bf p}^2/k^2$ is of order one and thus of the same order as $p^2_{\rm max}/k^2\sim 1$,$^{\ref{sec_frg:footnote2}}$ so that there is no small parameter.

\subsubsection[Further results obtained from the DE]{Further results obtained from the DE\footnote{In this section we mainly focus on O($N$)-like models. Many other applications of the DE are described in Secs.~\ref{sec_sm}-\ref{sec_gr}.}}
%\subsubsection{Further results obtained from the DE}

In Sec.~\ref{sec_frg:subsubsec_DE2} we have emphasized the computation of critical exponents but the DE also allows one to compute the scaling functions determining the universal equation of state in the vicinity of a second-order phase transition both in classical and quantum systems~\cite{Berges96,Rancon13a,Rancon13b,Rancon16}. The DE can be used to study the high-temperature disordered phase~\cite{Morris97} as well as the ordered phase~\cite{Tetradis92,Tetradis96,Berges:2000ew,Dupuis11,Caillol12a,Pelaez16} of the O($N$) model where for $2<d\leq 4$ and $N\geq 2$ the longitudinal susceptibility diverges due its coupling to transverse fluctuations, a general phenomenon in systems with a continuous broken symmetry~\cite{Patasinskij73,Zwerger04}. 

Many authors have considered multicritical points in dimensions $d\le 4$~\cite{Morris95,Yabunaka17,Katsis:2018bvc,Yabunaka18,Codello12,Codello13,Codello15,Hellwig:2015woa,Eichhorn:2013zza,Codello:2017hhh,Marchais:2017jqc} and the more speculative existence of critical fixed points for $4<d<6$~\cite{Eichhorn16,Percacci:2014tfa,Mati:2014xma}. Some multicritical fixed points of the O($N$) model for $d<4$ show singularities in the form of cusps at $N=\infty$ in their effective potential that become a boundary layer at finite $N$ and are therefore overlooked in the standard $1/N$ expansion~\cite{Yabunaka17,Yabunaka18,Katsis:2018bvc}. 

Let us also mention the following studies: $O(N)$ models with long-range interactions~\cite{Defenu:2014bea,Defenu:2017utb,Goll:2018vdj,Defenu:2020umv}; fixed points 
with imaginary couplings~\cite{Litim:2016hlb,Juttner:2017cpr}; non-polynomial perturbations to fixed points~\cite{Halpern95,Halpern96,Morris:1996nx,Gies:2000xr,Bridle:2016nsu}; O($N$) models in finite geometries and critical Casimir forces~\cite{Jakubczyk:2012iza,Rancon16}; nonlinear sigma models~\cite{Codello:2008qq,Flore:2012ma,Percacci:2013jpa}; the Potts model~\cite{Zinati18}.

The DE has been used to study the sine-Gordon model with emphasis on the BKT transition~\cite{Nagy09,Pangon12,Pangon11} or the central charge and the $c$-function~\cite{Bacso15,Oak17}. It yields an accurate estimate of the (exactly known) soliton and soliton-antisoliton bound state masses in the massive phase of this model~\cite{Daviet19} and strongly supports the Lukyanov-Zamolodchikov conjecture~\cite{Lukyanov97} regarding the amplitude of the field fluctuations.

\subsection{Computing momentum-dependent correlation functions}
\label{sec_frg:sec_vertexp}

To obtain the full momentum dependence of correlation functions it is necessary to go beyond the DE, since the latter is restricted to the momentum range $|{\bf p}|\lesssim {\rm max}(k,\xi^{-1})$. Momentum-dependent correlation functions can be computed by means of a vertex expansion. 

The flow equation~(\ref{sec_frg:eqwet}) yields an infinite hierarchy of equations satisfied by the vertices $\Gamma_k^{(n)}$. The vertex expansion, in its simplest formulation, amounts to truncating this hierarchy by retaining a finite number of low-order vertices. This leads to a closed system of equations that can be solved. Retaining the momentum dependence of the vertices allows one to obtain that of the correlation functions ~\cite{Blaizot:2004qa,Blaizot06a,Blaizot06b,Blaizot07,Ledowski04,Hasselmann07,Sinner08,Kopietz_book}. Systematic vertex expansion schemes with full momentum dependence have been also used in condensed matter systems, QCD and gravity, for more details see Secs.~\ref{sec_fb}, \ref{sec_hep} and \ref{sec_gr}.

Keeping only a finite number of vertices is however not always sufficient. In some problems, it is necessary to keep both the momentum dependence of low-order vertices and the full set of vertices in the zero-momentum sector (which amounts to considering the full effective potential). A possible approximation~\cite{Guerra07,Hasselmann12,Rose18}, inspired by the LPA$'$, is defined by the effective action 
\begin{equation}
\Gamma_k^{{\rm LPA}''}[\boldsymbol{\phi}] = 
\int_{\bf r} \biggl\lbrace  \frac{1}{2} (\partial_\mu\boldsymbol{\phi}) \cdot Z_k(-\boldsymbol{\nabla}^2) (\partial_\mu\boldsymbol{\phi}) 
 + \frac{1}{4} (\partial_\mu \rho) Y_k(-\boldsymbol{\nabla}^2) (\partial_\mu \rho) + U_k(\rho) \biggr\rbrace ,
\label{sec_frg:gammalpapp} 
\end{equation}
with a sum over $\mu=1\cdots d$. The full momentum dependence of the propagator $G_k[\boldsymbol{\phi}]$ is preserved by virtue of the nonlocal two- and four-point vertices. The anomalous dimension can now be deduced from $Z_{k=0}({\bf p})\sim |{\bf p}|^{-\eta}$ for $|{\bf p}|\ll p_G$ when the system is critical. The value of the critical exponents $\nu$ and $\eta$ is shown in Table~\ref{sec_frg:table_critexp}. This approximation scheme, which is sometimes referred to as the LPA$''$, is not numerically more costly than the DE and usually allows for an easy implementation of the symmetries, and for these reasons has been used in various contexts, see e.g.~\cite{Hasselmann11,Mathey14,Canet11b,Canet16sm,Feldmann:2017ooy}.

A more elaborate approximation scheme, which also keeps all vertices in the zero-momentum sector, has been proposed by Blaizot, M\'endez-Galain and Wschebor (BMW)~\cite{Blaizot06,Benitez08,Benitez09,Benitez12,Rose15} and, in the context of liquid theory, by Parola and Reatto \cite{Parola84,Parola95}. The flow equation of the two-point vertex $\Gamma^{(2)}_k({\bf p},\boldsymbol{\phi})$ in a uniform field $\boldsymbol{\phi}$ involves $\Gamma_k^{(4)}({\bf p},-{\bf p},{\bf q},-{\bf q},\boldsymbol{\phi})$ and $\Gamma_k^{(3)}({\bf p},-{\bf q},-{\bf p}-{\bf q},\boldsymbol{\phi})$ as well as $\partial_k R_k({\bf q})$ (see Fig.~\ref{sec_frg:fig_eqwet}). Because of the latter term,  which restricts the integral over the loop momentum to $|{\bf q}|\lesssim k$, to leading order one can set ${\bf q}=0$ in $\Gamma_k^{(4)}$ and $\Gamma_k^{(3)}$. Since $\Gamma^{(3)}_{k,ilj}({\bf p},0,-{\bf p},\boldsymbol{\phi})=\partial\Gamma^{(2)}_{k,ij}({\bf p},\boldsymbol{\phi})/\partial\phi_l$ (and a similar relation for $\Gamma^{(4)}_{k,ijlm}({\bf p},-{\bf p},0,0,\boldsymbol{\phi})$), one obtains a closed equation for $\Gamma_k^{(2)}({\bf p},\boldsymbol{\phi})$ which must be solved together with the exact flow equation~(\ref{sec_frg:eqLPA}) of the effective potential.
The BMW approximation scheme is numerically more involved than the DE and the LPA$''$ since $\Gamma_k^{(2)}({\bf p},\boldsymbol{\phi})$ is a two-variable ($|{\bf p}|$ and $\rho$) function. Symmetries may also sometimes be difficult to implement.

The critical exponents obtained from BMW compare favorably with those derived from the DE to second order or the LPA$''$. $\nu$ and $\eta$ are within 0.41\% and 7.45\%, respectively, of the conformal bootstrap results for the three-dimensional O($N$) universality class and $N=1,2,3$ (Table~\ref{sec_frg:table_critexp}). For the two-dimensional Ising model, BMW gives $\nu\simeq 1.00$ and $\eta\simeq 0.254$~\cite{Benitez12}, to be compared with the exact values $\nu=1$, $\eta=1/4$~\cite{Pogorelov07}. The BMW approximation becomes exact in the large-$N$ limit~\cite{Blaizot06,Rose15}.$^{\ref{sec_frg:footnote1}}$ The small parameter of the order of 1/4-1/9 that has been mentioned in the context of the DE (Sec.~\ref{sec_frg:validityDE}) applies since the BMW approximation also corresponds to an expansion in momenta (although the internal ones).

The BMW approximation (and simplified versions of it~\cite{Guerra07,Blaizot07,Benitez08}) has been applied to the quantum O($N$) model and interacting bosons (Sec.~\ref{sec_sm}), as well as the Kardar-Parisi-Zhang and Navier-Stokes equations~\cite{Canet11b,Canet16sm}.

\subsection{Lattice models and realistic microscopic actions} 
\label{sec_frg:subsec_lattice}

The FRG approach to continuum models described in the preceding sections can be straightforwardly extended to lattice models~\cite{Dupuis08}.\footnote{Application of the FRG to lattice models is common in fermion systems, see Sec.~\ref{sec_fb}.}  The lattice is taken into account by replacing the ${\bf p}^2$ dispersion in the bare propagator by the actual lattice dispersion $\epsilon_0({\bf p})$ and restricting the momentum to the first Brillouin zone. Alternatively, one can start from an initial condition of the RG flow corresponding to the local limit of decoupled sites~\cite{Machado10}. The flow equation then implements an expansion about the single-site limit and is reminiscent, to some extent, of Kadanoff's idea of block spins~\cite{Kadanoff66}. This lattice FRG captures both local and critical fluctuations and therefore enables us to compute nonuniversal quantities such as transition temperatures. It has been applied to the O($N$) model defined on a lattice~\cite{Machado10,Caillol12b,Caillol13,Banerjee:2018hqr,Banerjee:2018pkt} and to classical~\cite{Machado10} and quantum~\cite{Rancon14a,Krieg19} spin models as well as the superfluid-Mott transition in the Bose-Hubbard model~\cite{Rancon11a,Rancon11b,Rancon12a,Rancon12d,Rancon13b}.   

The possibility to start from an initial condition that already includes short-range fluctuations has been used in the Hierarchical Reference Theory of fluids~\cite{Parola95}, an approach which bears many similarities with the lattice FRG. More recently, similar ideas have appeared in the RG approach to interacting fermions~\cite{Reuther14,Wentzell15,Taranto14}.

\subsection{Quantum models} 
\label{sec_frg:subsec_quantum}

There is no difficulty to extend the FRG approach to quantum bosonic models. In the Euclidean (Matsubara) formalism, the latter map onto $(d+1)$-dimensional classical field theories with a finite extension $\beta=1/T$ in the $(d+1)$th (imaginary time) direction~\cite{Sachdev_book}; the effective action $\Gamma_k$ becomes a functional of a space- and time-dependent bosonic field $\boldsymbol{\phi}({\bf r},\tau)$ where $\tau\in [0,\beta]$~\cite{Tetradis93,Reuter93,Litim06a}. 
Galilean-invariant bosons and the quantum O($N$) model, the simplest quantum generalization of Eq.~(\ref{sec_frg:Smicro}) with space-time Lorentz invariance, are discussed in Sec.~\ref{sec:IVA}. 

Fermionic models are more difficult to deal with since the field $\boldsymbol{\phi}({\bf r},\tau)=\langle\boldsymbol{\varphi}({\bf r},\tau)\rangle$ in that case is an anticommuting Grassmann variable. Functionals of Grassmann variables make sense only {\it via} their Taylor expansions and, for example, the very concept of an effective potential with a well-defined minimum is lost. Thus the only {\it a priori} available method for fermionic models is a vertex expansion where one retains a finite number of (momentum-dependent) vertices evaluated at $\boldsymbol{\phi}=0$. It is however possible to introduce, {\it via} Hubbard-Stratonovich transformations, collective bosonic fields which can be treated nonperturbatively using the methods discussed in the previous sections (see Secs.~\ref{sec_fb}-\ref{sec_hep}).

A well-known difficulty in the study of quantum systems is the computation of real-frequency correlation functions from numerical data obtained in the Euclidean formalism, in particular at finite temperatures. At zero temperatures, the resonances-via-Pad\'e method~\cite{Schlessinger68,Vidberg77,Tripolt19} and other, Bayesian, reconstruction methods have been used in several works~\cite{Dupuis09b,Sinner10,Schmidt11,Rose15,Rose16a,Rose17a,Rose18,Tripolt17}. Alternative methods, where the analytic continuation is performed at the level of the flow equations, have been proposed~\cite{Rohe05,Jakobs10a,Floerchinger:2011sc,Tripolt14,Tripolt:2013jra,Kamikado:2013sia,Haas:2013hpa,Christiansen:2014ypa,Wambach14,Pawlowski:2015mia,Pawlowski18,Khedri18,Cyrol:2018xeq}

%% file: SEC_STATMECH/statmech_Nico.tex
%\newpage
\section{Statistical mechanics} 
\label{sec_sm}
\subsection{Why and when is the FRG useful?}

Statistical mechanics is a natural area of application of the FRG since it aims at calculating macroscopic properties of a system from a microscopical model. This is the reason why the general presentation of Sec.~\ref{sec_frg} was framed in the language of equilibrium statistical 
mechanics by using examples drawn from the Ising and $O(N)$ models.

Many models (including the $O(N)$ model) can be treated by traditional
means such as perturbation theory (in the critical regime, improved with
perturbative RG). When does the FRG become useful compared to
perturbative RG? There is no general answer but some general remarks can be made. First, the FRG is
well suited for approximations where the {\it functional} form of
various terms of the scale-dependent effective action plays a major role. This functional aspect becomes
unavoidable when the considered functions develop
nonanalyticities, such as ``cusps'' (see, e.g.,
Sec.~\ref{sec_disordered} below). Second, in some cases, the physics
to be studied is beyond the reach of approximations based on
perturbation theory. This, for example, occurs in the study of the
Kardar-Parisi-Zhang equation (Sec.~\ref{sec_NEQ}).
The main advantage of the FRG in this respect is that its flow equations are dominated by a small shell
of momenta, making it extremely robust and flexible when employing
approximations going beyond perturbation theory (see Sec.~\ref{sec_frg}). This is at odds with
other nonperturbative formulations such as Schwinger-Dyson equations
which involve integrals over a large region of momenta.  This
``locality in momentum'' is at the origin of the success of the
expansion schemes described in Sec.~\ref{sec_frg}. In particular, the decoupling of different momenta explains {\it a
  posteriori} the apparent convergence of the DE and the accurate results
obtained by this method, even though a small expansion parameter has only recently been identified (Sec.~\ref{sec_frg:validityDE}). There is a third reason why the FRG
proves useful: it can lead to the determination of nonuniversal
properties, such as a phase diagram (see, e.g., Sec.~\ref{sec_NEQ-RD}), something which is often challenging in other RG approaches.
Finally, as pointed out in Sec.~\ref{sec_frg}, in the FRG framework it is particularly simple to vary the dimension of the
theory. Even within very simple approximations, the behavior near the upper and lower dimensions is reproduced and one can therefore obtain results at
intermediate dimensions that {\it interpolate} between controlled limiting cases. This typically makes the results much more robust than
when only an {\it extrapolation} from the upper critical dimension is done.

In this section, we present applications of the FRG to classical statistical physics. Section~\ref{secequilibrium} is devoted to classical
equilibrium statistical mechanics, Sec.~\ref{sec_disordered} to disordered systems, and 
out-of-equilibrium systems are discussed in Sec.~\ref{sec_NEQ}. The subject has
reached a mature level and it is therefore not possible to present a
full account of all the topics that have been addressed within the FRG
method. Some paradigmatic examples are nevertheless presented in detail and a
survey of other works is briefly given.

\subsection{Equilibrium statistical mechanics}
\label{secequilibrium}

%\subsubsection{General considerations}
The implementation of the FRG procedure relies on a microscopic
Hamiltonian appropriate to describe the system under
consideration. When studying universal features, it is sufficient
to consider a general enough low-energy effective Hamiltonian respecting the symmetries of
the problem and including the main infrared degrees of freedom.\footnote{In order to determine nonuniversal properties
with the FRG, it is necessary to consider a more realistic microscopic
Hamiltonian as discussed in Sec.~\ref{non-universal}.} The derivation of the low-energy effective theory may be nontrivial and in some cases not even known. To illustrate this procedure in a concrete example, in the following section we discuss a paradigmatic problem at
equilibrium: the nature of the phase transition in Stacked Triangular Antiferromagnets (STA), an important class of frustrated magnets.

\subsubsection{Frustrated magnets as an example} 
\label{frustrated}
\begin{figure}
\centerline{\includegraphics[width=4.5cm]{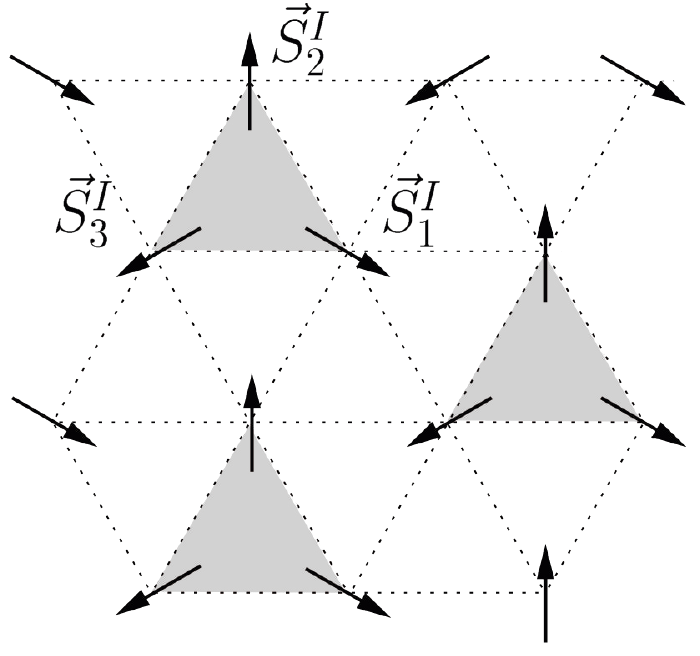}}
 \caption{Configuration minimizing the energy in the STA model (reprinted from Ref.~\cite{Delamotte04}).}
\label{frustrationfigure}
\end{figure}

{\it The model.} The STA model has been
proposed to describe several frustrated magnets (we refer to
\cite{Delamotte04} for a review). It describes $N$-component classical spins with antiferromagnetic nearest-neighbor
interactions. The spins are located at the lattice sites of a $d$-dimensional lattice consisting of stacked two-dimensional triangular lattices. The corresponding Hamiltonian is
$O(N)$-invariant:
\begin{equation}
H=J \sum_{\langle i,j\rangle}{\bf S}_i\cdot {\bf S}_j \quad
\mathrm{with}\quad J>0.
\end{equation}
The triangular planar structure induces {\it frustration} (it is not
possible to minimize simultaneously all nearest-neighbor
interactions).  This makes the standard $O(N)$ Ginzburg-Landau
model unsuited for the description of the long-distance properties of STA.
The derivation of a field theory that correctly describes the
critical physics of a given model relies on the knowledge of the
ground state, which is often nontrivial in presence of
frustration.
In the case of STA, the configuration which minimizes the energy is known and takes 
the form shown in Fig.~\ref{frustrationfigure} where the three spins of
  each triangular plaquette point 120$^\circ$ one from another. 
The
degree of freedom corresponding to the magnetization of each triangular block is therefore frozen and 
exhibits gapped excitations (even at the transition). Accordingly, in the critical regime, the degree of freedom associated with the block magnetization can be
integrated out and the orientation of the spins on a plaquette can
then be described by {\it two} $N$-component
vectors that are not colinear. Indeed it is sufficient to consider one of the two spins and the projection of the second one
in the direction orthogonal to the first one. As such, the block variables can be chosen to be two orthogonal vectors $\boldsymbol{\varphi}_1$ and $\boldsymbol{\varphi}_2$, of unit norm. In terms of these block variables, the
Hamiltonian becomes {\it ferromagnetic}. As a consequence, the effective Hamiltonian takes the form
\begin{equation}
H=-J \sum_{\langle I,J\rangle}\Big(\boldsymbol{\varphi}_1^I\cdot \boldsymbol{\varphi}_1^J+\boldsymbol{\varphi}_2^I\cdot \boldsymbol{\varphi}_2^J\Big) ,
\end{equation}
where the sum now runs over nearest-neighbor blocks. The resulting Hamiltonian has, in addition to the original $O(N)$ symmetry,
an $O(2)$ symmetry corresponding to a rotation in the $1-2$ block variable plane. Having obtained a ferromagnetic effective
Hamiltonian, one can introduce an
associated Ginzburg-Landau model with an $O(N)\times O(2)$ symmetry, where now the block variables are not constrained to
have unit norm or being orthogonal. 
However, in order to be equivalent to the spin system one requires that the Ginzburg-Landau
potential has its minimum in a configuration where the two vectors
$\boldsymbol{\varphi}_1$ and
$\boldsymbol{\varphi}_2$ are orthogonal and of the same norm. For that purpose it is convenient to introduce the $2\times N$ matrix
$\Phi=\big(\boldsymbol{\varphi}_1,\boldsymbol{\varphi}_2\big)$, in terms of which the Ginzburg-Landau Hamiltonian reads
\begin{equation}
H_{\rm GL}=\int_{\bf r} \bigg\{ \frac 1 2 {\rm Tr} (\nabla \Phi^t \nabla \Phi ) +\frac r 2 
\rho+\frac{\lambda}{16}\rho^2+\frac{\mu}{4}\tau\bigg\},
\label{sec_sm:hamSTA} 
\end{equation}
with $\rho={\rm Tr} (\Phi^t \Phi)$ and $\tau=\frac 1 2 {\rm Tr} [(\Phi^t \Phi-\frac{\rho}{2} 
1_2)^2]$. % I have replaced \openone by 1 (Nicolas D, 14/01/2002) 
The coupling constants $\mu$ and $\lambda$ are chosen
positive in order to ensure, as stated above, that the minimum of the potential
corresponds to a configuration where the vectors $\boldsymbol{\varphi}_1$ and
$\boldsymbol{\varphi}_2$ are mutually orthogonal and of the same norm. One can choose the coordinates such that the
vacuum state takes the form
\begin{equation}
 \Phi_0\propto \left(
 \begin{array}{cc}
  1 & 0 \\
  0 & 1 \\
  0 & 0 \\
  \vdots & \vdots
 \end{array}
 \right) .
\end{equation}
This corresponds to a breaking of the
$O(N)\times O(2)$ symmetry to $O(N-2)\times O(2)_{diag}$ in the ordered phase. 
The Hamiltonian~(\ref{sec_sm:hamSTA}) also describes spin-one bosons~\cite{Debelhoir16a,Debelhoir16b}.\\

{\it Main open problems in STA}. The STA model has been studied in perturbation theory ($d=4-\epsilon$, 
$d=2+\epsilon$, fixed $d$), large-$N$, Monte-Carlo simulations, and within the FRG. There are also many experimental studies of materials expected to be in the STA universality class. The main results from these theoretical and experimental studies appear to be conflicting. 

\begin{figure}
\centerline{\includegraphics[width=4.2cm]{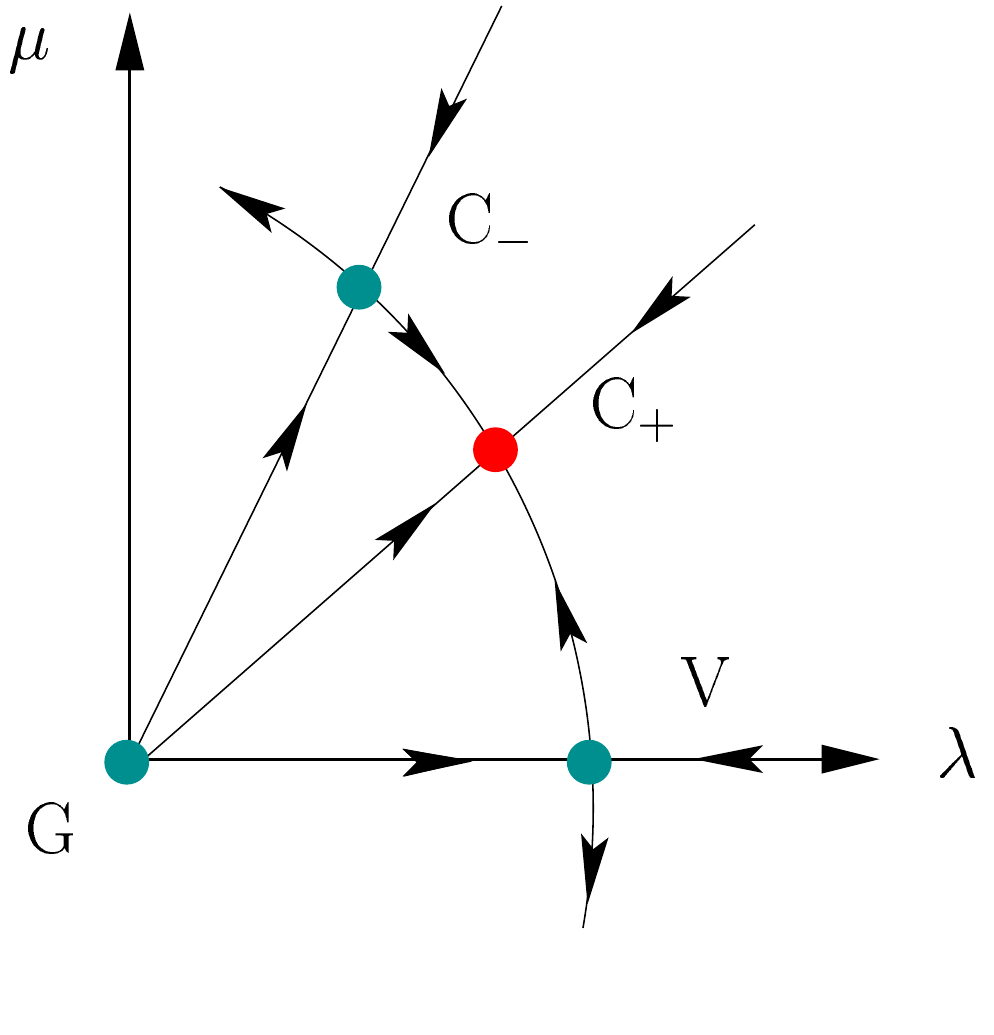}
	\hspace{1cm}
 \includegraphics[width=4.2cm]{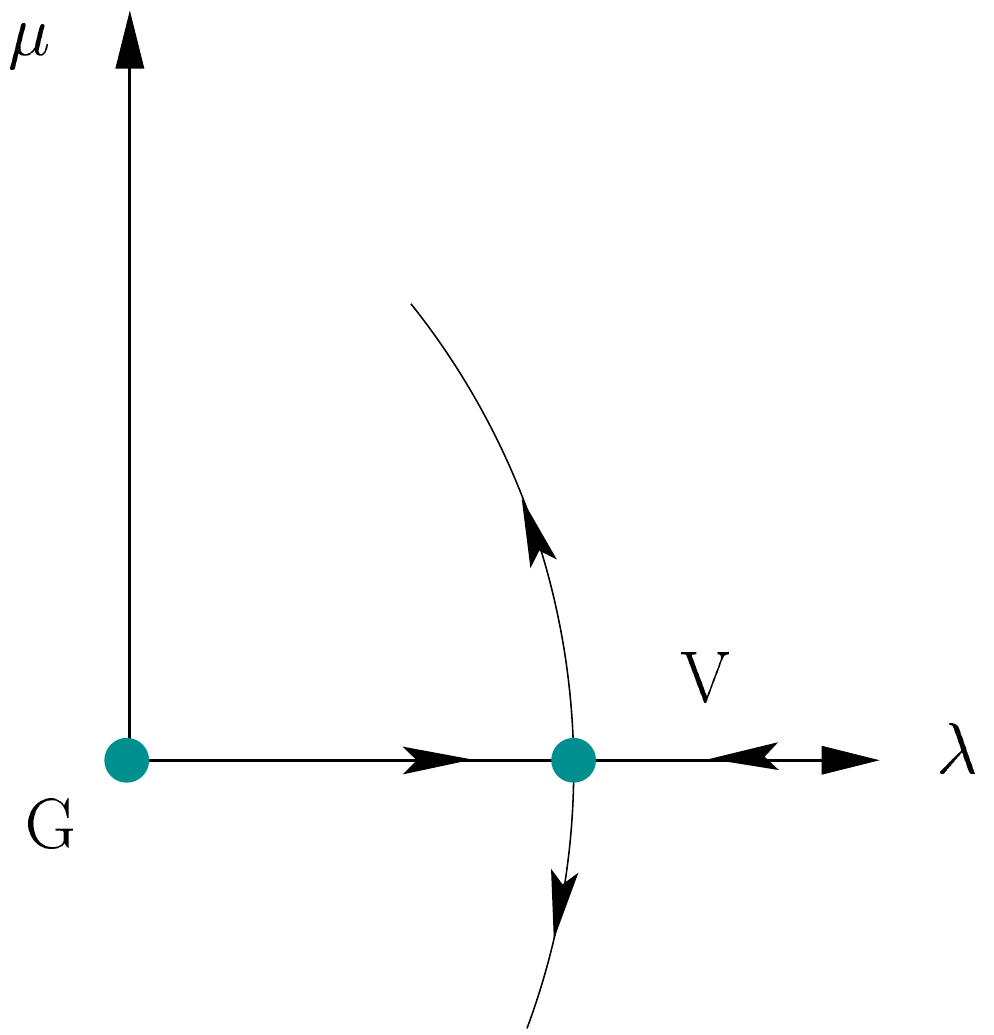}}
 \caption{\label{critsufSTA} Critical RG flow for $N>N_c(d)$ (left) and $N<N_c(d)$ (right)  (reprinted from Ref.~\cite{Delamotte04}).}
\end{figure}

Results from perturbation theory near the upper critical dimension $d_c=4$ are schematically shown in Fig.~\ref{critsufSTA}.
One observes two different cases. For values of 
$N$ above a certain critical value $N_c(d)$, the critical regime is controlled 
by a nontrivial fixed point (denoted by $C_+$ in Fig.~\ref{critsufSTA}). Below $N_c(d)$ there is no stable fixed point anymore in the would-be critical surface and the 
transition is first-order (the $O(2N)$-invariant fixed point denoted by V in Fig.~\ref{critsufSTA} has an unstable direction).
The main problem is then to  properly determine the function $N_c(d)$. Near $d=4$, $N_c(d)$ can be determined from the
$\epsilon$-expansion \cite{Antonenko95a,Calabrese03sm}:
\begin{equation}
N_c(d)=21.80-23.43 \epsilon+7.09 \epsilon^2-0.03 \epsilon^3+4.26 \epsilon^4 + 
\mathcal{O}(\epsilon^5)  .
\label{sec_sm:Nceps}
\end{equation} 
For $d=3$, Eq.~(\ref{sec_sm:Nceps}) gives results that strongly oscillate with the order of the $\epsilon$ expansion. Moreover, resummation techniques~\cite{Pelissetto:2000ne,Calabrese02} yield very unstable predictions (at odds with the three-dimensional $O(N)$ model).
The model has also been studied in the fixed $d=3$ resumed perturbative expansion at six loops, giving $N_c(d=3)=6.4(4)$
\cite{Pelissetto:2000ne}. However, the $d=3$ perturbative expansion
finds that the transition becomes again of second order for $N < 5.7(3)$,
which disagrees with the conclusion of the nonperturbative FRG analysis (see below).
The STA have also been studied within the conformal bootstrap program
\cite{Nakayama:2014sba,Stergiou:2019dcv,Henriksson:2020fqi} but this approach, even if it has been very successful in the calculation of
critical exponents in cases where the transition is second order~\cite{Poland19},
relies on the assumption that the transition is continuous and is therefore of no use for determining whether the transition is actually first or second order.

On the experimental and numerical side, scaling is mostly observed. However, the 
exponents seem to be {\it nonuniversal} (see \cite{Delamotte04} for details). 
Large-scale simulations \cite{Itakura01,Ngo08} indicate a {\it weakly first-order transition}.

{\it FRG approach to STA}. The phase transition in the O($N$)$\times$O(2) model can be studied within the FRG approach. Given that the anomalous dimension is small, the DE to second order, DE$_2$, is expected to be a good approximation. To this order, the most general effective action compatible with symmetries reads \cite{Delamotte:1998ay,Tissier:1999hv,Tissier:2000tz,Tissier:2001uk,Delamotte:2016acs}.
\begin{align}
&\Gamma_k[\boldsymbol{\phi}_1,\boldsymbol{\phi}_2]=\int_{\bf r} \Big\{ U_k(\rho,\tau)+
\frac{Z_k(\rho,\tau)}{2} \big[ ({\bf \nabla} \boldsymbol{\phi}_1)^2+ ({\bf 
\nabla} \boldsymbol{\phi}_2)^2 \big]
+ \frac{ Y^{(1)}_k(\rho,\tau)}{4} (\boldsymbol{\phi}_1\cdot{\bf \nabla} \boldsymbol{\phi}_2- 
\boldsymbol{\phi}_2\cdot{\bf \nabla}
\boldsymbol{\phi}_1 )^2 \nonumber \\
+&\frac{Y^{(2)}_k(\rho,\tau)}{4} (
\boldsymbol{\phi}_1\cdot{\bf \nabla}\boldsymbol{\phi}_1+ \boldsymbol{\phi}_2\cdot{\bf \nabla}
\boldsymbol{\phi}_2 )^2
+ \frac 14 Y^{(3)}_k(\rho,\tau) \bigl[
(\boldsymbol{\phi}_1\cdot{\bf \nabla} \boldsymbol{\phi}_1- \boldsymbol{\phi}_2\cdot{\bf \nabla}
\boldsymbol{\phi}_2 )^2+ (\boldsymbol{\phi}_1\cdot{\bf \nabla} \boldsymbol{\phi}_2+
\boldsymbol{\phi}_2\cdot{\bf \nabla} \boldsymbol{\phi}_1)^2\bigr] \Big\} \ .
\end{align}
To reduce the numerical cost of solving flow equations for two-variable functions, it is possible to expand $U_k(\rho,\tau)$, $Z_k(\rho,\tau)$ and $Y^{(i)}_k(\rho,\tau)$ about $\tau=0$ and $\rho=\rho_{0,k}$, which corresponds to the minimum of the effective potential \cite{Delamotte:1998ay,Tissier:1999hv,Tissier:2000tz,Tissier:2001uk}.
As in the case of the O($N$) model, the DE$_2$ approximation with a field expansion around the minimum of the potential reproduces the leading behavior
near $d=4$ and $d=2$ and in the large-$N$ limit. More recently, this truncation has been improved by expanding in the invariant $\tau$ but treating the full $\rho$-dependence of the various functions~\cite{Delamotte:2016acs}. 

\begin{figure}
	\centerline{\includegraphics[width=8.4cm]{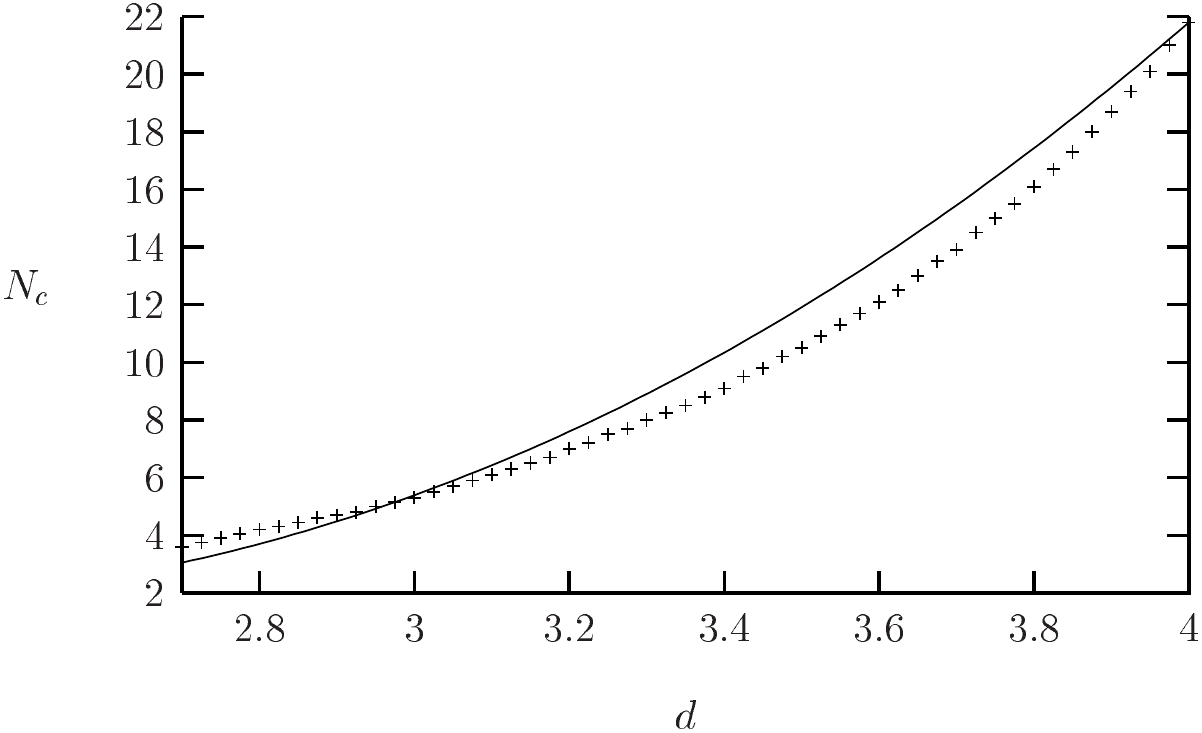}}
	\caption{Critical value $N_c(d)$ below which the transition becomes first order in the O($N$)$\times$O(2) model.
		Solid line: results obtained from the three-loop calculation improved by the constraint $N_c(d=2)=2$~\cite{Pelissetto:2001fi}.
		Crosses: FRG results~\cite{Delamotte04}. 
		\label{Nc-FRG}}
\end{figure}

Figure~\ref{Nc-FRG} shows $N_c(d)$ obtained from DE$_2$ and three-loop calculation in the $\epsilon=4-d$ expansion improved by the exactly known condition $N_c(d=2)=2$ \cite{Pelissetto:2001fi}. Both curves are qualitatively similar. In both cases, $N_c(d=3)$ is larger than 3.  Thus the FRG predicts a {\it first-order} phase transition in the O(3)$\times$O(2) and O(2)$\times$O(2) models in agreement with Monte-Carlo simulations \cite{Itakura01,Ngo08} but in disagreement with fixed-dimension RG studies~\cite{Pelissetto:2001fi}.

\begin{table}[b] %[htbp]
	\centering
	\begin{tabular}{lllllll}
		\hline \hline
		\multicolumn{1}{l}{Method} & \multicolumn{1}{c}{$\alpha$} & \multicolumn{1}{c}{$\beta$} & \multicolumn{1}{c}{$\gamma$} & \multicolumn{1}{c}{$\nu$} & \multicolumn{1}{c}{$\eta$} \\
		\hline
		FRG&  0.38&0.29&1.04&0.54&0.072\\
		6-loop& 0.35(9)&0.30(2)&1.06(5)&0.55(3)&0.08\\
		\hline
	\end{tabular}
	\caption{Typical values of the pseudo-critical exponents associated with the weakly first-order transition in
		STA (obtained from the FRG for $N=3$ and $d=3$)~\cite{Delamotte04}, compared with critical exponents of resumed
		perturbative series at six loops~\cite{Pelissetto:2000ne}.\label{tableSTAexp}}
\end{table}

The FRG also explains the pseudo-scaling observed in numerical simulations and experiments. When $N>N_c(d)$, the phase transition is second order and scaling (with universal critical exponents) is associated with the critical fixed point $C_+$. The two fixed points
$C_+$ and $C_-$ disappear by merging when $N=N_c(d)$, and for $N<N_c(d)$ two fixed points with complex coordinates appear. The RG flow becomes very slow in the vicinity of these two (unphysical) fixed points; hence a large, although finite,
correlation length that varies as $\xi\sim (T-T_c)^\nu$ over a large temperature range with a (nonuniversal) pseudocritical
exponent $\nu$.

Being nonuniversal, the value of the pseudo-exponent $\nu$ depends on the initial condition of the flow. However, as explained above, there is a finite region of couplings that shows a quasi-fixed point behavior. As a consequence, if the bare couplings take natural values (that is, are of order one in units of the lattice spacing)
the exponents lie within a small range of values. The typical value of these pseudo-critical exponents compares well with the results obtained in the six-loop
	perturbative expansion (even if the interpretation is quite
	different because the $d=3$ perturbative expansion predicts a genuine second order transition \cite{Pelissetto:2000ne}), see Table~\ref{tableSTAexp}.

An FRG study of a lattice O(2)$\times$O(2) model~\cite{Debelhoir16a,Debelhoir16b} has confirmed the existence of a weak first-order transition and shown that the correlation length is larger than the typical size of systems studied in Monte Carlo simulations, thus explaining why the true nature of the phase transition may have been missed in some numerical calculations~\cite{Calabrese04a}.

FRG studies of frustrated magnets are reviewed in~\cite{Delamotte04}. Related works on phase transitions with matrix order parameters can be found in Refs.~\cite{Berges:1996ja,Delamotte:1998ay,Tissier:1999hv,Tissier:2000tz,
Kindermann:2000vj,Tissier:2001uk,Delamotte:2016acs}.

\subsubsection{Critical phenomena and universal long distance 
regime}
\label{criticalphen}

We now review some universal long-distance properties obtained from the DE in various models
(and that were not already discussed in Sec.~\ref{sec_frg}).
 
The critical regime of models with cubic symmetry and multiple scalar fields has been considered in detail in Refs.~\cite{Bornholdt:1994rf,Bornholdt:1995rn,Bornholdt:1996ir,Tissier:2002zz,Chlebicki:2019yks}. Other studies of classical models include some particular Potts models~\cite{Zinati18}, the Yang-Lee model~\cite{An:2016lni,Zambelli:2016cbw}, and the spontaneous 
breaking of space rotational invariance~\cite{Lauscher:2000ux}.
In some clock models, the DE has given the striking result that
the exponents can be different on the two sides of the transition~\cite{Leonard15}. A discussion of the relation between the LPA and Dyson's Hierarchical Model can be found in~\cite{Meurice:2007zg,Litim:2007jb}.

The analysis of the existence of nontrivial critical points has been extended to 
very general models with a rich field content or symmetries 
\cite{Eichhorn:2014asa,Eichhorn:2015woa,Boettcher:2015pja,Borchardt:2016kco}.

FRG equations have also been employed to study various aspects of 
polymerized membranes, including the crumpling phase transition and their flat phase~\cite{Kownacki09,Essafi11,Essafi14a,Essafi14b,Coquand2016,Coquand2017}, as well as the wetting transition~\cite{PhysRevE.84.021124,PhysRevB.86.075142,Jakubczyk_2015}.

Critical properties have been studied within the DE not only in scalar theories 
but also in fermionic models (as well as models presenting both fermions and scalars)~\cite{Hofling:2002hj,Jaeckel:2002rm,Wetterich:2002ky,Gies:2009da,Gies:2010st,
Braun:2010tt,Scherer:2012nn,Janssen:2012pq,Scherer:2013pda,Strack:2013jta,
Janssen:2014gea,Bauer:2015qwa,Gehring:2015vja,Knorr:2016sfs,Janssen:2016xvc,Ihrig:2018hho}. This 
includes the analysis of some properties of graphene 
\cite{Janssen:2014gea,Bauer:2015qwa}.

\subsubsection{Nonuniversal properties}
\label{non-universal}

The FRG approach allows one to compute nonuniversal properties, in particular in the case of first-order phase transitions (including the problem of bubble
nucleation)~\cite{Litim:1996nw,Berges97,Berges:1996ja,Seide99,Strumia99,Strumia99a,Strumia99b,Strumia99c,Tetradis98,Munster:2000kk,Delamotte04,Tissier:1999hv,Tissier00a,Delamotte:2016acs,Debelhoir16a,Debelhoir16b,Qin:2018xox} as discussed in detail for frustrated magnets in Sec.~\ref{frustrated}.

A related topic is the description of nontrivial 
field configurations such as instantons. Even if the DE 
can reproduce in some cases the main consequences of the existence of topological excitations (e.g. vortices in the two-dimensional XY model or solitons in the sine-Gordon model, see Sec.~\ref{sec_frg:subsec_de}), it does not seem to yield in general a proper
quantitative description of nontrivial classical field configurations. For instance, it fails to reproduce all the features of the quantum-mechanical tunneling~\cite{Kapoyannis:2000sp,Zappala:2001nv}; this would likely require to retain the full momentum dependence of 
correlation functions.\footnote{In fact even sophisticated
momentum-dependent approximations, such as the BMW approach, seem to have difficulties
to quantitatively reproduce the quantum-mechanical tunneling~\cite{Rulquin15}.}

We pointed out in Sec.~\ref{sec_frg:subsec_lattice} that the FRG approach can predict phase diagrams and transition temperatures provided that the precise form of the microscopic action is used as the initial condition of the flow equations.\footnote{The precise knowledge of the microscopic action is not always necessary. In dilute Bose gases, for instance, the inter-particle interaction potential enters the low-temperature equation of state only through the $s$-wave scattering length so that the determination of the equation of state can be based on a simple interaction potential with a trivial UV behavior (e.g. a delta potential with a UV regularization).} This idea was used to determine the equation of state (and, possibly, the transition temperature to the low-temperature phase) of various classical and quantum fluids~\cite{Parola95,Bergerhoff:1995zq,Bergerhoff:1995zm,Seide99,Caillol06,Tarjus11,
Boettcher:2013kia,Rancon12b,Rancon12a,Rancon12d}.

\subsubsection{Conformal invariance and c-theorem}
\label{conformal}

Most FRG studies of critical phenomena do not consider the possible role of the conformal invariance in the critical
regime. In recent years, this issue has been the subject of many publications. The implementation of conformal symmetry in the
framework of FRG equations, together with the modified Ward identities due to the regularization of the theory in the
infrared~\cite{Ellwanger:1994iz}, has been considered in Refs.~\cite{Codello:2012sn,Codello:2015ana,Delamotte16a,Rosten:2016zap,Codello:2017hhh,Pagani:2017tdr,DePolsi:2018vxc,Morris:2018zgy,Rosten:2014oja,Rosten:2016nmc,Sonoda:2017zgl,Sonoda:2015pva,DePolsi:2019owi}.
In particular, this has led to a proof that the critical point in the Ising- and O($N$)-model universality classes is conformal invariant~\cite{Delamotte16a,DePolsi:2019owi}. Closely related to the conformal symmetry, the trace of the energy-momentum tensor has been analyzed in the context of FRG regularizations~\cite{Morris:2018zgy}. The structure of the Operator Product Expansion in the FRG context has been studied in Ref.~\cite{Pagani:2020ejb}.

The global structure of the flow has also been studied by constructing 
quantities that behave as C-functions \cite{Zomolodchikov66} in various approximations of FRG equations~\cite{Generowicz:1997he,Codello14,Bacso15}.

%%% Local Variables:
%%% mode: latex
%%% TeX-master: t
%%% End:

%% file: SEC_STATMECH/statmech_Matth.tex
\subsection{Disordered systems}
\label{sec_disordered}

Impurities, defects and other types of imperfections are ubiquitous in realistic many-body systems. Such a random disorder can prevent phases of matter from forming, change the critical properties of a phase transition, etc. However disorder is also responsible for a variety of novel phenomena that often have no counterparts in clean systems: metastability, non-ergodicity, pinning, avalanches, chaotic behavior, slow dynamics and aging, etc.

\subsubsection{Replica formalism and FRG}
\label{sec_d_observables}

When the impurities have a relaxation time much larger than the typical time of the experiment, the disorder is said to be quenched and can be modeled by a static random field $h(x)$ coupling to the density of particles or the order parameter field, etc.\footnote{Depending on the context, $h$ is referred to as a random field, a random potential, etc.} $\mathcal W_h[J]=\ln Z[J;h]$ becomes a functional of both the external source $J$ and the random field $h$ ($Z[J;h]$ is the partition function for a given random field $h$). The disorder averaged value of many observables (e.g. the order parameter or the susceptibility) depends on the mean free energy $W_1[J]=\overline{\mathcal W_h[J]}$ (an overline indicates disorder averaging). However to fully characterize the disorder effect, in particular in the absence of self-averaging~\cite{Aharony96}, one also needs to consider higher-order cumulants $W_i$ of the random functional $\mathcal W_h[J]$ and in particular the second-order one,
\begin{equation}
W_2[J_a,J_b] = \overline {\mathcal W_h[J_a]\mathcal
		W_h[J_b]}-\overline {\mathcal W_h[J_a]}\ \overline{\mathcal W_h[J_b]} .
\end{equation} 
Note that the $i$th-order cumulant depends on $i$ distinct sources: $J_{a_1}\cdots J_{a_i}$. This allows one to obtain all physical observables of interest by taking functional derivatives of the $W_i$'s and then setting the sources equal to zero. 

The standard method to compute the cumulants $W_i$ is to consider $n$ copies (or replicas) of the system with the same random field $h$ but with distinct sources $J_{a}$. The partition function then reads 
\begin{equation}
e^{W[\{J_a\}]} = \overline{\prod_{a=1}^n Z[J_a;h]} = \overline{\exp\left(\sum_{a=1}^n \mathcal
		W_h[J_a]\right)} ,
\label{sec_d_Wdef}
\end{equation}
where the functional $W[\{J_a\}]$ can be expanded in free replica sums,
\begin{equation}
W[\{J_a\}] = \sum_{a=1}^n W_1[J_a]+\frac 12 \sum_{a,b=1}^n W_2[J_a,J_b]+\cdots 
\end{equation}
the $W_i$'s being the cumulants introduced above.

In the FRG formalism, the quantity of interest is the effective action $\Gamma[\{\phi_a\}]$ defined as the Legendre transform of $W[\{J_a\}]$ where $\phi_a(x)=\delta W[J]/\delta J_a(x)$~\cite{tissier12-2}. $\Gamma[\{\phi_a\}]$ can be expanded in free replica sums, 
\begin{equation}
\Gamma[\{\phi_a\}]=\sum_{a=1}^n\Gamma_1[\phi_a]-\frac 12 \sum_{a,b=1}^n\Gamma_2[\phi_a,\phi_b]+\cdots 
\end{equation}
$\Gamma_1$ is the Legendre transform of $W_1$ and contains all information on the average order parameter, susceptibility, etc. $\Gamma_2$ is simply related to $W_2$ and encodes the effective disorder correlator $\overline{h(x)h(x')}$. 

An essential feature of disordered systems is that the disorder correlator, $W_2[J_a,J_b]$ or $\Gamma_2[\phi_a,\phi_b]$, may assume a nonanalytic (cuspy) functional form that encodes the existence of metastable states and the ensuing glassy properties of the system~\cite{Balents96}. In order to take into account such nonanalyticities, it is necessary to implement a RG approach that retains the functional form of the ``cumulants'' $\Gamma_i[\phi_{a_1}\cdots\phi_{a_i}]$. Perturbative implementations of the FRG in disordered systems have a long history~\cite{Fisher85,Narayan92,Nattermann92,Balents_1993,Chauve01,Tarjus04,PhysRevLett.56.1964,PhysRevB.48.5949,PhysRevB.46.11520,PhysRevB.62.6241,Ledoussal04,PhysRevLett.88.177202,Ledoussal10}. In the following section, we focus on the three-dimensional random-field Ising model, a system that is far from the upper critical dimension $d_{\rm uc}=6$ and where nonperturbative effects are crucial (see~\cite{Tarjus19} for a recent review).

\subsubsection{The Random Field Ising Model}
\label{sec_RFIM}

\begin{figure}
\centerline{\includegraphics[width=6cm]{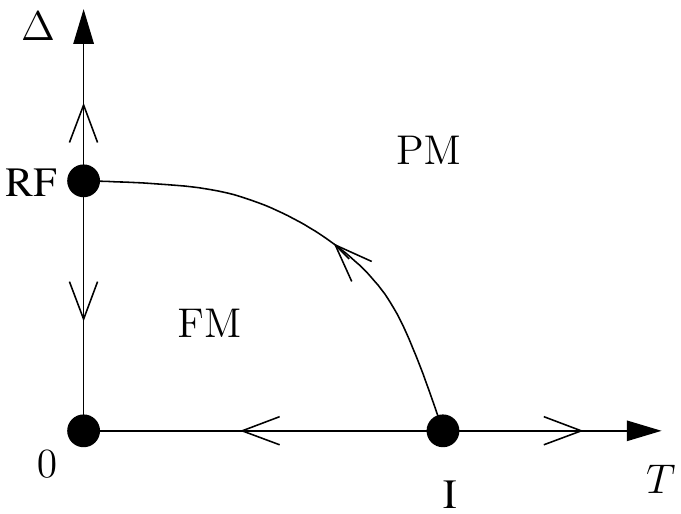}}
\caption{Schematic phase diagram of the RFIM in the disorder
	strength $\Delta$-temperature $T$ plane above the lower critical dimension
	$d_{\rm lc}=2$. At low disorder and low temperature, the system is ferromagnetic, and is paramagnetic otherwise. The arrows
	describe how the renormalized parameters evolve under the RG
	flow at long distance, and I and RF denote the critical fixed points
	of the pure and random-field Ising models, respectively.}
\label{sec_d:fig_rfim_dia} 
\end{figure}

The random-field Ising model (RFIM)~\cite{imry75} is defined by the Hamiltonian
\begin{equation}
  \label{eq_hamiltonianrfim}
  H=-J\sum_{\langle ij\rangle}S_iS_j-\sum_iS_ih_i \qquad (J>0) , 
\end{equation}
where the quenched disorder appears through a random magnetic field
$h_i$ which is Gaussian distributed, with zero mean and variance
$\Delta$. This model is used to describe, among others,
diluted antiferromagnets in a homogeneous field~\cite{1984PhRvB..29..505C} and the critical point of a fluid adsorbed
in a porous matrix~\cite{brochard83}. It is characterized by
two control parameters: the temperature (or equivalently the
exchange $J$) and the variance of the disorder. The phase diagram is shown in Fig.~\ref{sec_d:fig_rfim_dia}. The temperature is
known to be a dangerously irrelevant term, which flows to zero like a
power law at the fixed point describing the critical physics of the
model. This indicates that sample-to-sample fluctuations dominate the
thermal fluctuations. As a consequence, the transition can be studied
at vanishing temperature, by changing the variance of the disorder.

The critical behavior of the phase transition in the RFIM can also be studied from the field theory defined by the action 
\begin{equation}
S[\varphi] = \int d^dx \left\lbrace \frac{1}{2} (\partial_\mu\varphi)^2 + \frac{r_0}{2} \varphi^2 + \frac{u_0}{4!} \varphi^4 - h \varphi \right\rbrace ,
\label{sec_d:FT}
\end{equation}
where $h(x)$ is a Gaussian distributed random field. Perturbative calculations performed in the 70's on the RFIM unveiled
the very surprising property of dimensional reduction: The critical
exponents of the RFIM are identical, to all orders of perturbation
theory, to those of the pure Ising model in $2$ dimensions less~\cite{grinstein76,aharony76,young77}. This result was first derived by
comparing Feynman diagrams in the two theories. It was later understood that
dimensional reduction is a consequence of a hidden
(super)symmetry~\cite{parisi79}.

Surprisingly for a result true to all orders in perturbation theory,
dimensional reduction is invalid in three dimensions~\cite{imbrie84,bricmont87,bricmont88}. The failure of dimensional
reduction was expected to be related to the existence of many
metastable states, which were overlooked in the supersymmetric
construction of~\cite{parisi79}. Moreover, in several other disordered
systems, such as random manifolds pinned by impurities, it was shown
that the existence of these many metastable states can lead to
nonanalycities in the vertex functions~\cite{PhysRevLett.56.1964,Nattermann92,PhysRevB.46.11520,PhysRevLett.88.177202,Balents96}. This
possibility was not considered in the diagrammatic derivation
performed in~\cite{grinstein76,aharony76sm,young77} and is probably at
the origin of the invalid prediction of dimensional reduction.

The strategy followed to study the RFIM in the nonperturbative FRG approach is
standard~\cite{tissier06,tissier06b,tissier08a,tissier08,tissier11b-2,tissier12b-2,Tarjus_2013}. We first introduce a regulator in the construction described
in Sec.~\ref{sec_d_observables} and obtain the exact
flow equation for $\Gamma_k[\{\phi_a\}]$. We then propose an ansatz
for this quantity and derive the approximate flow equations. Before embarking in this discussion,
we have to ensure that supersymmetry, which is responsible for
dimensional reduction is not explicitly broken. In this respect,
dimensional regularization which is often used in perturbative
approaches is very powerful because it satisfies automatically many
symmetries, may they be explicit or hidden. On the contrary, the
regulator introduced in the FRG may break some symmetries and we have to
be cautious about that.\footnote{In the first calculations of RFIM
  critical properties performed with the nonperturbative FRG~\cite{tissier06,tissier06b,tissier08a,tissier08}, the regulator term
  was breaking explicitly the supersymmetry. It incorrectly led to
  critical exponents that never fulfilled dimensional reduction.}

In Refs.~\cite{tissier12-2,tissier12b-2}, it was shown that the nonperturbative FRG can indeed reproduce
supersymmetry and dimensional reduction, if appropriate truncations
are performed. Supersymmetry shows up as relations between $\Gamma_i$
and $\Gamma_{i+1}$, which are preserved by the renormalization-group
flow, under the assumptions that the vertex functions are sufficiently regular. The simplest of these relations reads
\begin{equation}
  \Gamma_2^{(1,1)}[q^2;\phi,\phi]=\Delta_{\text B}\partial_{q^2}\Gamma_1^{(2)}[q^2;\phi]
\end{equation}
where the superscripts indicate functional derivatives. A
similar property must be fulfilled by the regulating term $\Delta S_k$
[Eq.~(\ref{sec_frg:DeltaS})]. This
implies that we need to introduce a regulator $\hat R$ in the
1-replica part and a related regulator $\tilde R$ in the 2-replica
part.

The minimal truncation that preserves this symmetry consists in
neglecting $\Gamma_i$ when $i>2$ and assuming
\begin{align}
  \Gamma_{1,k}[\phi]&=\int d^d x \biggl\{ U_k(\phi)+\frac 12
  Z_k(\phi)(\partial_\mu\phi)^2 \biggr\} , \\
  \Gamma_{2,k}[\phi_1,\phi_2]&=\int d^d x\ V_k(\phi_1,\phi_2) ,
\end{align}
which amounts to implementing the DE to second order in $\Gamma_{1,k}$ and to LPA order in $\Gamma_{2,k}$.
The flow equations for the three functions appearing in this
truncation have been integrated numerically. For space dimensions
larger than $d_{\text{dr}}\sim 5.13$, the flow attains a fixed point
after fine tuning the initial condition. During the whole flow, the
functions $U_k(\phi)$ and $Z_k(\phi)$ remain equal to the associated
functions in the pure Ising model in 2 dimensions less, computed at
the same level of truncation (up to unimportant multiplicative
factors). The critical exponents fulfill the dimensional reduction
property. For lower dimensions, a nonanalyticity appears at a finite RG
scale, called the Larkin length. Below this scale,
$V^{(11)}_k(\phi_a,\phi_b)$ develops a cuspy
behavior $\propto |\phi_a-\phi_b|$, the flow of $U_k$ and $Z_k$ depart from
those of the pure Ising model in two dimension less and the fixed
point does not present dimensional reduction. The appearance of such a
cusp is common to many systems governed by a zero-temperature fixed
point. It is associated with the presence of avalanches, or shocks,
which correspond to reorganizations of the ground state on large scales
when the external parameters are slightly modified.

\begin{figure}
  \centerline{\includegraphics[width=6.5cm]{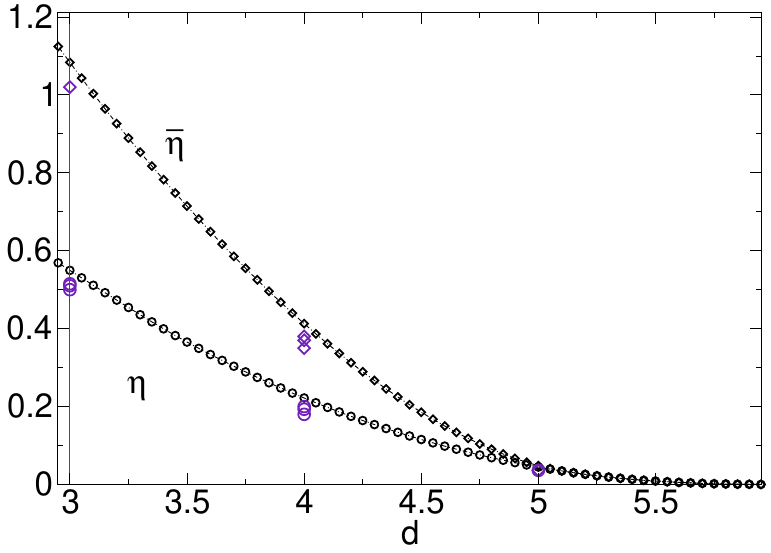}}
  \caption{Anomalous dimensions $\eta$ and $\bar \eta$ of the RFIM as
a function of space dimension. The large symbols for integer dimensions are lattice results from Refs.~\cite{PhysRevB.64.214419,PhysRevB.65.134411,PhysRevLett.110.227201,PhysRevE.95.042117}.}
  \label{fig_dr}
\end{figure}

This change of behavior in dimension is depicted in Fig.~\ref{fig_dr},
where the two anomalous dimensions
$\eta$ and $\bar\eta$, associated with the long-distance power-law
decay of the correlation functions,
\begin{align}
  \overline{\langle\varphi(x)\varphi(y)\rangle}-\overline{\langle\varphi(x)\rangle\langle\varphi(y)\rangle}
                        &\sim\frac
 1{|x-y|^{d-2+\eta}} , \\
  \overline{\langle\varphi(x)\rangle\langle\varphi(y)\rangle}-\overline{\langle\varphi(x)\rangle}\,\overline{\langle\varphi(y)\rangle}&\sim\frac1{|x-y|^{d-4+\overline\eta}} ,
\end{align}
are shown as a function of the dimension. The agreement with
lattice simulations is very good.

The picture emerging from the nonperturbative FRG approach is that
dimensional reduction is broken in low enough dimensions but is
realized in high dimensions, in particular in the vicinity of the
upper critical dimension 6. This may seem surprising at first sight
because there is no qualitative difference between these two regimes, in
what concerns the metastable states which are responsible for the
breaking of supersymmetry and dimensional reduction. A criterion was
derived in~\cite{PhysRevLett.110.135703} to quantify under which
conditions metastable states destabilize the dimensional reduction
property.

Other works in the same line aimed at studying the universality class
of the RFIM in presence of long
range interactions and/or correlations of the disorder~\cite{Balog_2014,PhysRevB.88.014204}. The RFIM presents also
interesting dynamical properties, which were studied both at
equilibrium~\cite{PhysRevB.91.214201} and out-of-equilibrium~\cite{PhysRevB.89.104201,PhysRevB.97.094204}.

\subsubsection{Other disordered systems}
\label{sec_d_other}

The perturbative FRG has been used to study many disordered systems, in particular random manifolds pinned by
impurities, at and out of equilibrium~\cite{PhysRevLett.56.1964,PhysRevB.48.5949,Nattermann92,Balents_1993,PhysRevB.46.11520,PhysRevB.62.6241,Ledoussal04,PhysRevLett.88.177202,Balents96}.

Other works used the nonperturbative aspect of Wetterich's equation~(\ref{sec_frg:eqwet}). The random elastic manifold model in and out of equilibrium is studied in~\cite{Balog19a}. The authors of Ref.~\cite{Tissier:2002zz} considered the random bond
model where the disorder appears as an
inhomogeneity of the exchange term (or, equivalently, of the $\varphi^2$
term in the continuum field theory~(\ref{sec_d:FT})). In
Ref.~\cite{SciPostPhys.1.1.007}, a first attempt to study the
spin-glass to paramagnetic transition was performed. The authors of Ref.~\cite{Coquand2017} studied the influence of disorder on graphene and
crystalline membranes and a glass phase was identified. The disordered Bose fluid is discussed in~\cite{Dupuis19,Dupuis20,Dupuis20a}. %\red{\bf Y a-t-il d'autres refs a rajouter?} 

%%% Local Variables:
%%% mode: latex
%%% TeX-master: "frg_review"
%%% End:

%% file: SEC_STATMECH/stat_mech_Leo.tex
\subsection{Classical non-equilibrium systems and FRG}
\label{sec_NEQ}

In this section, we mostly consider {\it classical}  systems out of equilibrium. 
The FRG formalism has also been developed for nonequilibrium {\it quantum} systems where it has led to fruitful applications (see {\it e.g.}~\cite{Koenig99, Jakobs07, Gasenzer:2007za, Pietroni:2008jx, Berges:2008sr, Berges12, Sieberer13,Chiocchetta17}). We also refer the reader to Secs.~\ref{sec:QuantumNonEq} and \ref{sec:IVB4} of this review.
In the following,  we focus on long-time stationary dynamics, but the FRG formalism can also be used to study the universal short-time dynamics of systems, for instance following a quench in the vicinity of a critical point \cite{Chiocchetta16, Chiocchetta17}.

In a classical system, thermodynamics and dynamics are decoupled, which means 
 that the dynamics has to be specified through an equation of motion. In statistical mechanics,
 there are two standard ways of defining
  a (stochastic) dynamics: a microscopic description via a master equation or an effective approach based on a Langevin type 
  of equation. Both descriptions can be recast into a path integral formulation,  adapted to field-theoretical treatments. 
  Systems of both classes have been studied within the FRG framework.
  In the following, we put emphasis on the four most studied and paradigmatic examples, which we present in details,
  while other works are only briefly reviewed.

\subsubsection{Langevin stochastic dynamics}

 A Langevin equation (in a broad sense) describes the time evolution of a mesoscopic (coarse-grained) field $\phi(t,\vx)$,
  subjected to some microscopic noise.
 For simplicity, we consider a scalar field, the generalization to more complex situations is straightforward. 
 The Langevin equation has the generic form
\begin{equation}
\p_t \phi(t,\vx) = -{\cal F}[\phi(t,\vx)] + \eta (t,\vx) \, ,
\label{sec_kpz:eqlangevin}
\end{equation}
where the functional ${\cal F}[\phi]$ encompasses the deterministic part of the dynamics and $\eta$
 models the randomness. This noise  is in general  chosen with a Gaussian probability distribution, of zero mean  $\langle \eta \rangle=0$ and variance 
\begin{equation}
 \langle \eta(t,\vx)\eta(t',\vx\,')\rangle= 2 D[\phi(t,\vx)] \delta(t-t') \delta^d(\vx-\vx\,')\,
\end{equation}
where $D$ is a (possibly field-dependent) diffusion coefficient\footnote{The Langevin equation is written here using the Ito discretization for time derivatives.}.
This form can be generalized to  include  spatial or temporal correlations of the microscopic noise.
 
The Langevin equation (\ref{sec_kpz:eqlangevin}) can be cast into a field theory using the Martin-Siggia-Rose-Janssen-de Dominicis (MSRJD) formalism \cite{Martin73,Janssen76,Dominicis76}. This procedure involves the introduction
 of a response field $\bar\phi$, and yields an action with the general structure
\begin{equation}
{\cal S}[\phi,\bar\phi] = \int_{t,\vx} \Big\{\bar\phi\big(\p_t \phi + {\cal F}[\phi]\big)
 - \bar\phi D[\phi]\bar\phi\Big\} \, ,\label{sec_kpz:actionGen}
\end{equation}
 where the deterministic part of the dynamics is linear in $\bar\phi$ and the noise correlator
  is encoded in the quadratic term in $\bar\phi$.

\vspace{0.2cm}
\noindent{\it a- FRG formalism for Langevin stochastic dynamics}
\vspace{0.2cm}

The equilibrium FRG formalism can be simply extended to study classical nonequilibrium systems, by including  the additional response fields, and taking into account
 Ito's prescription and causality issues \cite{Canet04a,Canet11b,Benitez13}.
 The minimal FRG scheme consists in implementing only a spatial coarse-graining, which amounts  to adding to the action a quadratic  scale-dependent term which depends on momentum but not on frequency, of the form 
\begin{equation}
 \Delta {\cal S}_\rgk = \frac 1 2\int_{\omega,\vq} \Phi_i [R_\rgk]_{i j}(\vq) \Phi_j
\label{sec_kpz:deltaSk}
\end{equation}
where $\Phi$ has component $\Phi_1=\phi$ and $\Phi_2=\bar\phi$.
In most cases,  the integrals in the frequency domain are convergent, and this scheme suffices to achieve
 the progressive integration of the fluctuation modes and to regularize the theory.
A frequency regularization has been implemented up to now only in  \cite{Duclut17}.
 The effective average action $\Gamma_k$ is then defined as usual and its exact flow equation reads
\begin{equation}
 \p_s \Gamma_k = \frac 1 2 {\rm Tr} \int_{\omega,\vq} \p_s R_\rgk \cdot \big[\Gamma_\rgk^{(2)} + R_\rgk\big]^{-1}.
\label{sec_kpz:dkgam}
\end{equation}
 with $\partial_s\equiv \rgk\partial_\rgk$ \footnote{The notation $s$ is used in this section for the RG ``time'' since $t$ denotes the physical time.}, and where $R_\rgk$ and $\Gamma_\rgk^{(2)}$ are $2\times 2$ matrices.

This formalism has been used to study the dynamics of spin models  (Model A, reviewed below,  and Model C \cite{Mesterhazy13} 
    in the classification by  Hohenberg and Halperin \cite{Hohenberg77}), and famous non-linear partial differential equations
 such as the Kardar-Parisi-Zhang equation and
 the  stochastic Navier-Stokes equation (see below), or Burgers equation  \cite{Mathey14}.

\vspace{0.2cm}
\noindent{\it b- Critical Dynamics: Model A}
\vspace{0.2cm}

Model A describes the purely dissipative relaxation towards equilibrium of a non-conserved scalar order parameter $\phi(t,\vx)$ with Ising symmetry.
 %It corresponds at the microscopic level to Glauber dynamics (independent random spin flips).
 The model is defined by the Langevin equation (\ref{sec_kpz:eqlangevin}) with a simple additive noise 
 $D[\phi]\equiv D$. The functional  ${\cal F}\equiv D\frac {\delta {\cal H}}{\delta \phi}$ derives from  the  equilibrium Ising  Hamiltonian ${\cal H}$ 
 where the coefficient $D$  is identical to the amplitude of the noise in order to fulfill  Einstein's relation, and ensure that the system evolves towards the equilibrium state.
  According to Eq.~(\ref{sec_kpz:actionGen}), the action of  model A is then given by
\begin{align}
{\cal S}[\phi,\bar\phi] &= \int_{t,\vx} \Big\{\bar\phi\left(\partial_t\phi - D \vnabla^2\phi+ D V'(\phi)\right) - 
 D\bar\phi^2 \Big\} ,
\label{sec_kpz:eqactionA}
\end{align}
where $V$ is the $\phi^4$ potential.
 When approaching the continuous phase transition of the model, 
the relaxation time of the order parameter  diverges. This behavior reflects  the critical slowing down of the dynamics near the critical point, characterized  by the dynamical exponent $z$
  which relates the divergence of the
  relaxation time $\tau$ to the divergence  of the correlation length $\xi$ in the vicinity 
 of  the critical point as $\tau\sim \xi^z\sim |T-T_c|^{-z\nu}$.
  The value of the exponent $z$ depends on the precise relaxation mechanism 
 and differs in the different models A, B, C \dots. For model A, its mean field value is $z=2$. 
 
  The classical action (\ref{sec_kpz:eqactionA}) possesses 
  a time-reversal symmetry (in the long-time limit), that can be  expressed 
 as an invariance of the action (\ref{sec_kpz:eqactionA}) under the following field transformation \cite{Andreanov06,Canet07b}
\begin{equation}
\left\{
\begin{array}{l l l}
t & \to& -t\\
\phi & \to & \phi \\
\bar\phi & \to & \bar\phi - \frac 1 D\p_t\phi.
\end{array}
\right.
\label{sec_kpz:eqtrs}
\end{equation}
 The time-reversal symmetry can also be expressed as a supersymmetry \cite{Zinn_book}, and the corresponding  supersymmetric FRG formalism for Model A is expounded  in Ref.~\cite{Canet11b}.
 The consequence of this time-reversal symmetry (or supersymmetry) is the existence of a fluctuation-dissipation theorem,
  which yields the decoupling of the statics and of the dynamics. The static critical exponents $\nu$ and $\eta$ for Model A are thus those of the equilibrium Ising model.

This model has been studied within the second order of DE (DE$_2$).
The minimal regularization, which is a straightforward extension
 of the static case,  consists in introducing only  off-diagonal 
 and frequency-independent cutoff terms $[R_\rgk]_{12}(\vq)=[R_\rgk]_{21}(\vq)$ in (\ref{sec_kpz:deltaSk}).
 If one also  implements a frequency regularization, 
  then a diagonal cutoff term, satisfying  $[R_\rgk]_{12}(t,\vx)=2 \theta(-t)\p_t[R_\rgk]_{22}(t,\vx)$ 
   must be included in order to preserve the time-reversal symmetry (\ref{sec_kpz:eqtrs}) and causality \cite{Duclut17}.
 The symmetry (\ref{sec_kpz:eqtrs}) also imposes constraints on the ansatz for
 $\Gamma_\rgk$, which must have the following structure  \cite{Canet07b}
\begin{align}
\Gamma_\rgk[\varphi,\bar\varphi]&=\int_{\vx,t} \,\Bigg\{
 X_\rgk(\varphi)\left(\bar\varphi\,\p_t \varphi - \bar\varphi^2\right) +\bar\varphi\,\Big(U_\rgk'(\varphi)  - 
Z_\rgk(\varphi)\,\vnabla^2\,\varphi -\frac{1}{2}\,\p_\varphi Z_\rgk(\varphi) (\vnabla \varphi)^2 \Big)\Bigg\}.
\label{sec_kpz:eqanzA}
\end{align}
where the time derivative and quadratic term in $\bar\varphi$ are renormalised in the same way %as a consequence of 
 %the time-reversal symmetry (\ref{sec_kpz:eqtrs}),
  and the rest derives from
 the standard ansatz for the equilibrium Ising model at DE$_2$. At the bare level,  $X_\Lambda=1/D$.
The flow equations for Model A have been integrated numerically at different orders for the dynamical parts, referred to as LPA, LPA$'$ and DE$_2$,  which all correspond to the complete
 DE$_2$ for the static part, and respectively $X_\rgk=1$ (LPA), $X_\rgk(\rho)=X_\rgk$ (LPA$'$), and $X_\rgk(\rho)$ (DE$_2$).
 At DE$_2$, the integration has been performed both with and without frequency regularization (using different frequency regulators) \cite{Duclut17}.
 As an illustration,  the results obtained for the critical exponents
 in $d=3$  in the different schemes are reported in Table  \ref{sec_kpz:tabmodA1},
   and compared with the best estimates available in the literature.  The agreement is satisfactory, and the frequency regularization turns out 
 to improve the results.
\begin{table}
\begin{center}
\begin{tabular}{l c c c }
\hline
\hline
 ${d=3}$       &  $\nu$ & $\eta$ & $z$ \\
\hline
 LPA     & 0.65  & 0.11 & 2.05  \\
  LPA$'$     &  0.63 &   0.05     & 2.14  \\
 DE$_2$  & 0.628   &  0.0443 &  2.032$^{a}$  2.024$^{b}$\\
\hline
FT & 0.6304(13) & 0.0335(25)& 2.0237(55) \\
MC     & 0.63002(10) & 0.03627(10)& 2.032(4)  2.055(10)  \\
\hline
\end{tabular}
\caption{Critical exponents of Model A in $d=3$ from the different 
 FRG approximations (LPA, LPA$'$ \cite{Canet07b} and DE$_2$ without$^{a}$ or with$^{b}$ frequency regularization \cite{Duclut17}), 
 compared with results from  other field theoretical
 methods (FT) \cite{Guida98,Prudnikov06} and Monte Carlo simulations (MC) \cite{Hasenbusch10,Ito00,Grassberger95}.}
\label{sec_kpz:tabmodA1}
\end{center}
\end{table}

 The critical exponents of Model A have also been obtained in $d=2$ and for $N=2$ and 3 \cite{Duclut17}. 
 Moreover,  the initial-slip exponent, which characterizes the nonequilibrium universal short-time behavior of both the order parameter and correlation functions after a quench of the temperature of the system, has been computed for model A in $2\leq d \leq 4$  \cite{Chiocchetta16}.
 The relaxational critical dynamics of a nonconserved order parameter coupled to a conserved scalar density, 
 which corresponds to Model C, has been studied in \cite{Mesterhazy13}. The properties of the phase diagram
  for the dynamic critical behavior are determined. In this work, different scaling regimes corresponding to 
  different critical exponents are identified and  characterized.

\vspace{0.2cm}
\noindent{\it c- Kardar-Parisi-Zhang equation}
\vspace{0.2cm}

A prominent example of non-equilibrium and non-linear  Langevin dynamics is 
 the Kardar-Parisi-Zhang (KPZ) equation.  It was originally derived to
  describe the critical roughening of stochastically growing interfaces,
  whose dynamics was modeled by \cite{kardar86}
\begin{equation}
\frac{\p h(t,\vx)}{\p t} = \nu\,\vnabla^2 h(t,\vx) \, + 
\,\frac{\lambda}{2}\,\big(\vnabla h(t,\vx)\big)^2 \,+\,\eta(t,\vx)
\label{sec_kpz:eqkpz}
\end{equation}
where the diffusion term provides  the interface
 with a smoothening mechanism, and the non-linear term encompasses an enhanced growth
  along the local normal to the surface, and leads to critical roughening.
  Note that the
  deterministic part of Eq. (\ref{sec_kpz:eqkpz}) is not the functional derivative of some Hamiltonian,
  and the amplitude of the noise $D$ is no longer related to the diffusion part, 
  such that this dynamics leads to intrinsically  non-equilibrium steady-states.
In fact, a remarkable feature of the KPZ equation is generic scale invariance, also termed 
self-organized criticality, in any dimension. For $d\leq 2$, the interface always becomes rough (and critical)
 as it grows, without  fine-tuning any control parameter,
 contrary to usual equilibrium phase transitions. The rough interface is characterized
  by two universal critical exponents, the roughening exponent $\chi$ and the dynamical exponent $z$.

Due to its simplicity,  KPZ universality class arises in connection with  an extremely large class of nonequilibrium or
disordered systems, such as randomly stirred fluids (Burgers equation) \cite{Forster77},
  directed polymers in random media \cite{Kardar87},  or even driven-dissipative Bose-Einstein condensation \cite{Sieberer13,Squizzato18}, to cite only a few. 
  The KPZ equation has thereby emerged as one of the 
 fundamental theoretical models to investigate  nonequilibrium scaling phenomena
 and phase transitions \cite{Halpin-Healy95}, and has witnessed an intense renewal of interests in recent years. 
Indeed, a considerable breakthrough has been achieved in the last decade regarding
 the characterization of the KPZ universality class in 1+1 dimensions, 
 sustained by a wealth of exact results \cite{Corwin12},
 and by high-precision measurements in turbulent convection of liquid crystals  \cite{Takeuchi10sm,Takeuchi12}.
 %A particularly striking feature is the discovery of universality sub-classes
 % for the distribution of height fluctuations, determined by the nature of the initial conditions
 %(flat, sharp-wedge, or stochastic), which has revealed a 
 % deep connection with random matrix theory  \cite{Corwin12}.
   %\cite{Calabrese11,Amir11,Sasamoto10a,Calabrese12,Imamura12}. 
  % Moreover, experiments in liquid crystals provided the first set-up to
  % allow for quantitative measurements of KPZ properties, and 
 %they confirmed with extreme accuracy the theoretical results \cite{Takeuchi10sm,Takeuchi12}.
   However, in higher dimensions, or in the presence of additional ingredients
   such as non-delta correlations of the microscopic noise,
    the integrability of the KPZ equation is broken, and controlled analytical
     methods to describe the rough phase are scarse. In this context,
      the FRG has been particularly useful since it is the only method which can access 
     in a controlled way the strong-coupling KPZ fixed-point in any dimension \cite{Canet10sm}.

Following the MSRJD  procedure, the field theory associated with the KPZ equation reads  
\begin{equation}
{\cal S}[h,\bar h]  \!\! = \!\!\! \int_{t,\vx}  \left\{ \bar h\left(\p_t h -\nu \,\vnabla^2 h - 
\frac{\lambda}{2}\,({\vnabla} h)^2 \right) - D\, \bar h^2  \right\}.
\label{sec_kpz:eqftkpz}
\end{equation}
This action is invariant under  shifts of the height field $h \to h + c$ where $c$ is an arbitrary constant. 
It is also invariant under  a Galilean transformation for  Burgers' velocity $\vv(t) \propto \vnabla h$, which
 corresponds for the height field to an infinitesimal tilt of the interface.
  This symmetry enforces the exact identity between critical exponents $z+\chi=2$ in all dimensions \cite{Halpin-Healy95}.

In fact, both
the Galilean and shift symmetries admit a stronger form,  gauged in time \cite{Lebedev94,Canet11asm}.
The corresponding  transformations read
\begin{eqnarray}
%{{\cal G}(\vv(t))}  \,: \,
&& \left\{
\begin{array}{l}
h'(t,\vx)=\vx \cdot \p_t \vv(t) + h(t,\vx+ \lambda \vv(t))\label{sec_kpz:eqgalg}\\
\bar h'(t,\vx)=\bar h(t,\vx+ \lambda \vv(t))
\end{array}
\right.\\
%{\cal C}(c(t)) \, :\, 
&& \;\;\;\;
  h'(t,\vx)=h(t,\vx)+c(t). \label{sec_kpz:eqtimeg}
\end{eqnarray}
where $c(t)$ and $\vv(t)$ are  infinitesimal time-dependent quantities\footnote{
 The transformation with $\vv(t) = \vv\times t$ corresponds to the usual (non-gauged) Galilean symmetry.}. 
More precisely, these time-gauged forms are extended symmetries, 
 in the sense that the KPZ action is not strictly invariant under these 
   transformations, but its variation is linear in the fields,
 and this  entails important non-renormalization theorems and general Ward identities  \cite{Canet11asm}.
  Extended symmetries proved as an extremely powerful tool in the context of FRG, as
   illustrated below.
 
 Furthermore, for a one-dimensional interface, the KPZ equation satisfies  an `accidental' 
time-reversal symmetry  which, as  shown in \cite{Canet05b,Canet11asm},  can be encoded in the discrete transformation 
\begin{equation}
 \left\{
\begin{array}{l}
h'(t,\vx)=- h(-t,\vx) \\
\bar h'(t,\vx)=\bar h(-t,\vx) +\frac{\nu}{D} \vnabla^2 h(-t,\vx).
\end{array}
\right. \label{sec_kpz:eqfdt}
\end{equation}
and which fixes the exponents exactly to $\chi=1/2$ and $z=3/2$ in $d=1$.

  The KPZ field theory was studied within FRG using different approximations \cite{Canet05b,Canet10sm,Canet11asm}.
 One first needs to specify the cutoff matrix in (\ref{sec_kpz:deltaSk}). An appropriate
  choice compatible with the symmetries is $[R_\rgk]_{12}(\vq) = [R_\rgk]_{21}(\vq) = \nu_\rgk q^2 r(q/\rgk)$
   and $[R_\rgk]_{22}(\vq) = -2D_\rgk r(q/\rgk)$ where $\nu_\rgk$ and $D_\rgk$ are two running coefficients defined
   below. Note that the Galilean invariance precludes for implementing a frequency regularization, at least in a simply manageable way.
  The KPZ equation is an interesting example for which the LPA$'$ already goes beyond all-order perturbative
   RG  but where a simple DE scheme is not enough to yield accurate results.
   Indeed, a rescaling of the time and the height leads to a single KPZ coupling   $g=\lambda^2 D/\nu^3$.
   Its $\beta$-function  was
    computed to all orders in perturbation \cite{Wiese97} and it fails to capture the strong-coupling fixed point describing the rough phase in $d\ge 2$,
    whereas the LPA$'$ already contains this fixed point in any dimension.
  However,  the critical exponents in $d>1$ are poorly determined, even at DE$_4$ \cite{Canet05b}. 
  This is very unusual in the FRG context, and probably originates in the derivative nature of the  KPZ vertex. 
   Hence, one needs to resort to an alternative approximation scheme, such as the vertex expansion, in order 
   to better account for the momemtum dependence. However, the application of the BMW approximation scheme,
 as presented in Sec. \ref{sec_frg:sec_vertexp}, is hindered  by the symmetries for the KPZ problem, and in particular the Galilean one.
  Indeed,  the associated Ward identities relate in a nontrivial way
the momentum and frequency dependencies of vertex functions of different orders,
 preventing the direct BMW  expansions of the vertices. 

 This difficulty has been successfully 
circumvented  \cite{Canet10sm,Canet11asm}. The strategy is to build an ansatz which automatically encodes the symmetries 
while leaving an arbitrary momentum and frequency dependence of the two-point functions.
This is achieved by combining together the basic Galilean scalars of the theory --
 which are $\bar{h}$, $\nabla_i\nabla_j h$ and $D_t h = \partial_t h-\frac{\lambda}{2} (\vnabla h)^2$, and arbitrary powers of their gradients
and covariant time derivatives $\tilde{D}_t= \p_t -\lambda \vnabla h\cdot\vnabla$.
Truncated at Second Order in the response field, the corresponding SO ansatz reads
%\begin{widetext}
\begin{align}
\Gamma_\rgk[\varphi,\bar \varphi]&= \int_{t,\vx} 
\left\{ \bar \varphi f_\rgk^\lambda D_t\varphi - 
\bar \varphi f_\rgk^{\tiny D}\bar \varphi %\right. \nonumber\\&  \left.
- \frac{\nu}{D}% \left[\vnabla^2 \varphi f_\rgk^\nu \bar \varphi + 
 \bar \varphi f_\rgk^\nu  \vnabla^2 \varphi
%\right]
 \right\},
\label{sec_kpz:eqanzso}
\end{align}
%\end{widetext}
where the three functions $f_\rgk^X$, $X=\nu,D,\lambda$, are functions of $\tilde D_t$ and $\nabla$,
 i.e $f_\rgk^X\equiv f_\rgk^X(-\tilde D_t^2,-\vnabla^2)$
 and with $\varphi=\langle h\rangle$ and $\bar \varphi=\langle \bar h\rangle$.
Note that arbitrary powers of the field itself are included {\it via} the functional dependence 
in $\varphi$ of the covariant derivative $\tilde D_t$.
The gauged-shift symmetry  furthermore imposes that $f_\rgk^\lambda(\omega^2,\vp^{\,2}=0) \equiv 1$.
One defines two running coefficients through the normalization conditions $f_\rgk^\nu(0,0)=\nu_\rgk$ and 
 $f_\rgk^D(0,0)=D_\rgk$, which entail the anomalous dimensions $\eta^\nu_\rgk=-\p_s \ln \nu_\rgk$
  and $\eta^D_\rgk=-\p_s \ln D_\rgk$ whose fixed point values are related to the critical exponents
   as $\chi = (2-d-\eta_*^\nu+\eta_*^D)/2$  and $z = 2-\eta_*^\nu$.

\begin{figure}[t]
	\begin{center}
		\includegraphics[height=5cm]{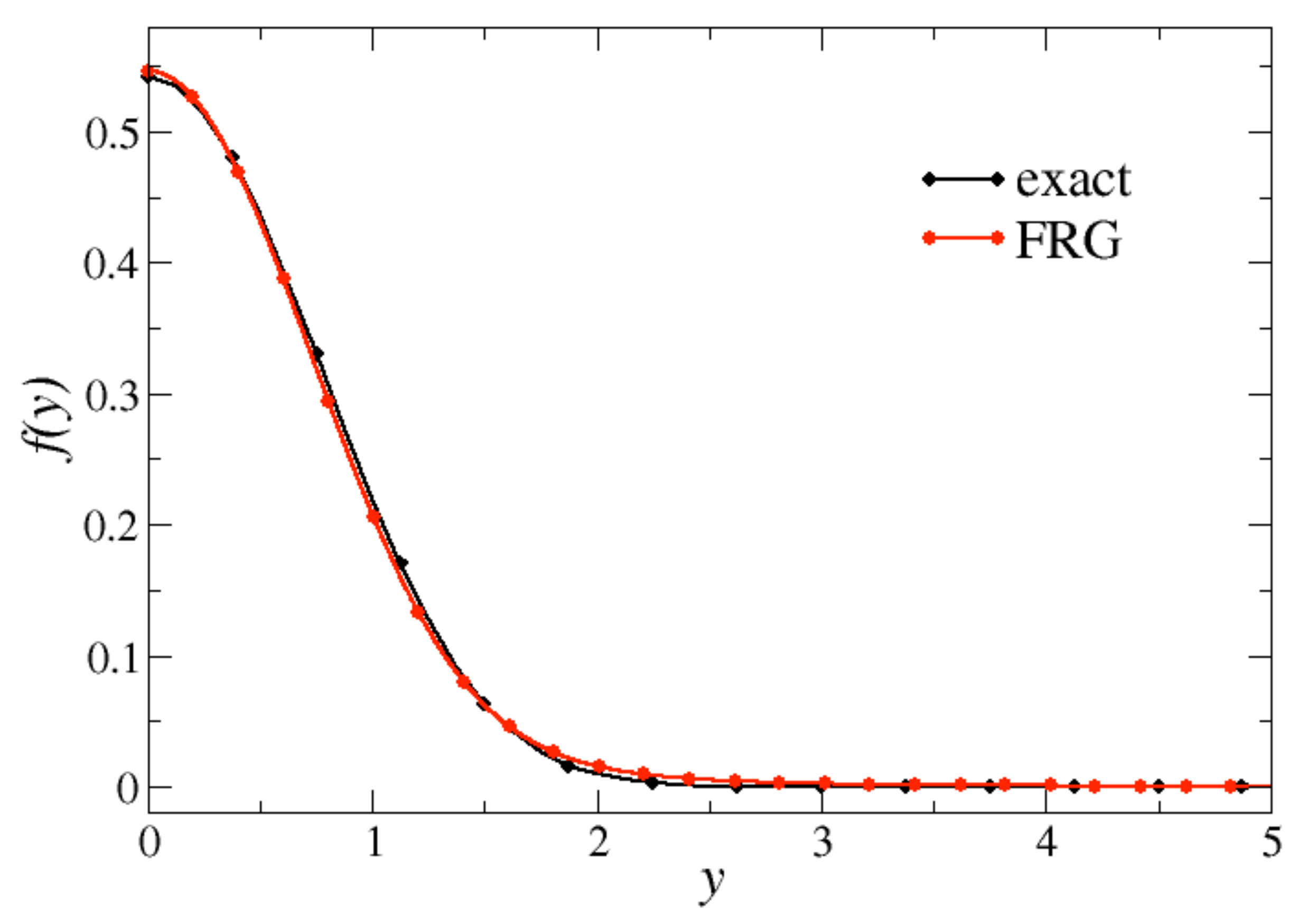}\hspace{1cm}\includegraphics[height=5cm]{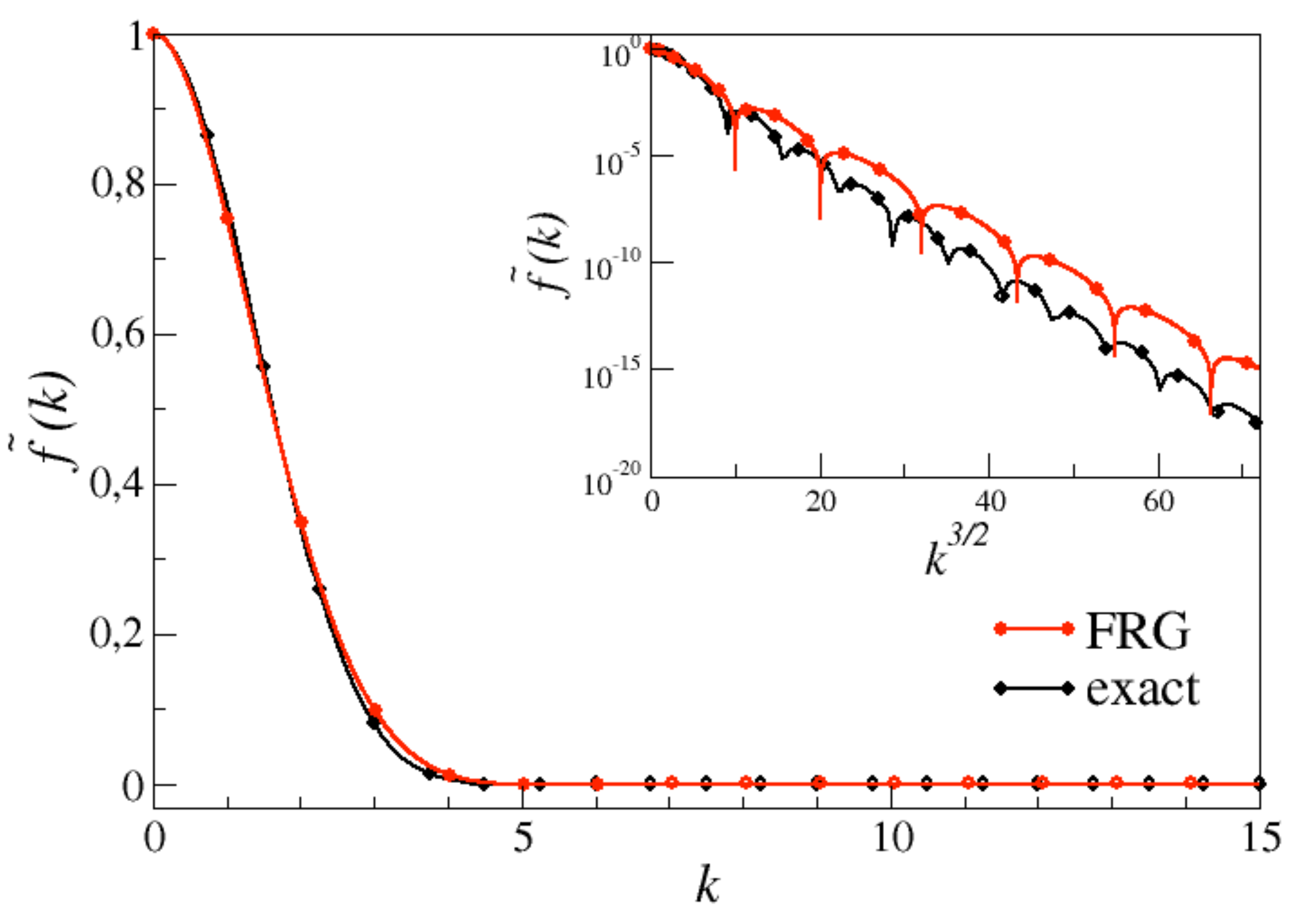}
	\end{center}
	\caption{Comparison (without any fitting parameters) of the scaling functions $f(y)$ (left panel) and $\tilde f(k)$ (right panel)
		obtained from FRG (red curves with dots) \cite{Canet11asm,Canet12Err} with the exact ones  (black curves with squares) \cite{Praehofer04}.
		The inset on the right panel shows the stretched exponential behavior of the tail with the superimposed oscillations, 
		developing on the same scale $k^{3/2}$. Note the vertical scale:  this behavior develops with amplitudes
		below typically $10^{-6}$.}
	\label{sec_kpz:figfhat}
\end{figure}

 This ansatz yields extremely accurate results for a one-dimensional interface \cite{Canet11asm}. 
 The numerical integration of the flow equations
  leads to a fully attractive, non-Gaussian fixed point, which means that the interface always become
   rough, and critical, as physically observed. As a consequence, one can demonstrate analytically,
   from the fixed point equation and the decoupling of the non-linear terms of the flow equations of the two-point functions at large
    momentum and frequency, the existence of generic scaling.
  In particular, one can show that the two-point correlation function endows a scaling form
  $C(\omega,\vp) = p^{d+2+\chi} F(\omega/p^z)$, and the scaling function $F$ can be compared
   to exact results that were obtained in \cite{Praehofer04},
     for $\tilde f(k) = \int_{0}^\infty d\tau \cos(\tau k^{3/2}) F(\tau)$
    and for its Fourier transform $f(y)$. The comparison, which does not involve  any free parameter,  is excellent, as shown in Fig.~\ref{sec_kpz:figfhat}.
    The FRG even captures
     very fine details of the tail of $\tilde f(k)$ featuring a stretched exponential with superimposed oscillations.

%%%%%%%%%%%%%%%%%%%%%%%%%%%%%%%%%%%%%%%%%%%%%%%%%%%%%%%%%%%%%%%%%%%%%%%%%%%%%%%%%%%%%%%%%%
%%%%%%%%%%%%%%%%%%%%%%%%%%%%%%%%%%%%%%%%%%%%%%%%%%%%%%%%%%%%%%%%%%%%%%%%%%%%%%%%%%%%%%%%%% 
 
Moreover, the FRG approach can be applied in all non-integrable cases. It has been used in particular
 in a simplified form for interfaces of $d >1$ dimensions, where the critical exponents and scaling functions
 were calculated, and provided their first analytical estimates.
It also yielded predictions for dimensionless ratios in $d=2$ and $3$ \cite{Kloss12} which 
were later accurately confirmed by large-scale numerical simulations \cite{Halpin-Healy13,Halpin-Healy13Err}.
  This framework was extended to study the influence of anisotropy \cite{Kloss14b}, and  of spatial correlations
   in the microscopic noise,  following a power-law \cite{Kloss14a} or with a finite length-scale \cite{Mathey17sm}, 
    and also the influence of temporal correlations of the noise \cite{Strack15,Squizzato19}. 

   Let us emphasize that temporal correlations break at the microscopic level the Galilean invariance (and the time-reversal symmetry in 1D) which are fundamental symmetries of the KPZ equation. Thus, it is not clear a priori whether the KPZ universality should survive even an infinitesimal breaking. This raises the  important issue of its realisability in experimental systems since delta-correlations are an idealization and any real physical source of noise does possess some  finite time correlation.
     This issue has been the subject of an active debate and conflicting answers have been
      given (see \cite{Squizzato19} and references therein).
      The FRG analysis brought convincing evidence to settle this debate by showing that in fact, as long as the temporal correlations of the  noise are not too long-ranged, the symmetries are dynamically restored along the flow, and the KPZ universal properties emerge at long distances and long times \cite{Squizzato19}.

\vspace{0.2cm}
\noindent{\it d- Stochastic Navier-Stokes equation}
\vspace{0.2cm}

Turbulence remains one of the major unsolved problems of classical physics. Whereas the constitutive model for fluid dynamics
 is well-known, no one has succeeded yet in deriving from it the actual statistical properties of the turbulent state \cite{Frisch95}. 
 The dynamics of an incompressible fluid is given by the Navier-Stokes (NS) equation
 \begin{equation}
 \p_t \vv+ \vv \cdot \vnabla \vv=-\frac 1\rho 
\vnabla p +\nu \nabla^2 \vv +\vf \, , \quad\hbox{with}\quad\vnabla \cdot \vv =0
\label{eq:ns}
 \end{equation}
where $\nu$ is the kinematic viscosity, 
$\rho$ the density of the fluid, $p$ the pressure field and $\vf$ the forcing necessary to maintain a stationary turbulent state. 
We here  focus on isotropic and homogeneous  turbulence.
To describe its universal properties, it is convenient to consider a stochastic forcing, with a Gaussian distribution 
 and concentrated in Fourier space at a large scale $L$, called the integral scale, where energy is injected
 \begin{equation}
 \langle f_\alpha(t,\vx)f_\beta(t',\vx\,')\rangle=2 \delta_{\alpha\beta}\delta(t-t')N_{L^{\text{-}1}}(|\vx-\vx\,'|) \, .
\end{equation}

 The phenomenology of turbulence is well-known from  many experiments and Direct Numerical Simulations (DNS) \cite{Frisch95}.
 One striking feature is the emergence of very robust power-law behaviors, observed over a wide range of scales called the inertial range,
 and irrespective of the nature of the fluid (liquids, gas, or even quantum fluids). This universal behavior
 was first explained by the pioneering statistical theory of turbulence proposed by Kolmogorov in 1941 \cite{Kolmogorov41a,Kolmogorov41c}.
 However, the power-laws observed in turbulence do not fit in the framework of  standard scale invariance, 
 but rather involve multi-scaling or multi-fractality, and this anomalous behavior is generically called intermittency. Another peculiar signature of intermittency in turbulence is 
 the enhancement of extreme events at small scales, generally associated with quasi-singularities of the vorticity field
 at small-scales \cite{Frisch95}. The calculation of the multi-fractal spectrum of anomalous exponents from first principles is an open issue. 
  
  The idea of applying RG methods to turbulence dates back to the late seventies and has a long history which is not reviewed here,  but it has essentially failed to bridge this gap \cite{Adzhemyan99,Zhou10sm}. One difficulty is the absence of upper critical dimension, and the necessary introduction of an expansion parameter through a power-law forcing
 profile, whose extrapolation to physical large-scale forcing turns out to be extremely problematic. 
 This has thus motivated several authors to revisit turbulence using FRG \cite{Tomassini97,Monasterio12,Canet16sm}. In these works,  it was shown that
 the turbulent state corresponds to a fixed point, for a physical large-scale forcing (no need to artificially introduce a power-law profile).
   More interestingly, the FRG has proven to be a very powerful tool in this context, since  it allows for the derivation
  of  exact results for the time-dependence (in the stationary state, {\it i.e.} the dependence in time delays) of any generic $n$-point correlation function
     in the limit of large wave-numbers, both in 3D \cite{Canet17,Tarpin18} and 2D \cite{Tarpin19}. This is all the more remarkable that exact results are scarce in turbulence.
     From a FRG perspective, turbulence also stands as a unique example where the RG flow equation for an arbitrary 
     $n$-point correlation function can be closed
       exactly in the limit of large wave-numbers, without any further approximations. 
  This is a compelling illustration of the importance of extended symmetries, which are the key ingredients in this derivation, within the FRG framework.
  
 Upon considering a stochastic forcing, the NS equation (\ref{eq:ns})  formally stands as a Langevin equation,  and the MSRJD procedure yields 
 %the generating functional 
%\begin{align}
% {\cal Z} =  &\int  \mathcal{D} \vv \mathcal{D}p \mathcal{D}\bar{\vv} \mathcal{D}\bar p\,
% \, e^{-{\cal S} + \int_{t,\vx}\{ \vJ\cdot \vv+\bar{\vJ}\cdot \bar{\vv}+K p+\bar K\bar p\}} 
% \, ,\label{eq:Z}
%\end{align}
 the NS action \citep{Canet16sm}
\begin{align}
 {\cal S} &= \int_{t,\vx}\Big\{ \bar v_\alpha\Big[ \p_t 
v_\alpha -\nu \nabla^2 v_\alpha +v_\beta\p_\beta v_\alpha +\frac 1\rho \p_\alpha p  \Big] +\bar p\,\p_\alpha v_\alpha \Big\} -\int_{t,\vx,\vx'}\bar v_\alpha N_{L^{\text{-}1}}(|\vx-\vx'|)\bar v_\alpha\, .
\label{eq:NSaction}
\end{align}
Similarly to the KPZ equation, the NS equation admits an extended (time-gauged) form of Galilean invariance,
 corresponding to the infinitesimal transformation:
\begin{align} 
 \delta v_\alpha(t,\vx) =-\dot{\epsilon}_\alpha(t)+\epsilon_\beta(t) \p_\beta v_\alpha(t,\vx)\,,\quad\quad
    \delta \varphi(t,\vx) =\epsilon_\beta(t) \p_\beta \varphi(t,\vx)\,, 
\label{eq:defGal}
\end{align}
where $\varphi$ denotes any of the other fields $\bar{v}_\alpha$, $p$ and $\bar{p}$.
From this symmetry one can derive Ward identities which fix the value of any vertex function 
 $\Gamma^{(n)}$ with one vanishing momentum carried by a velocity field in terms of lower-order ones.
Another extended symmetry, unveiled in \cite{Canet15sm}, corresponds to a time-gauged shift of the response fields as
\begin{equation}
 \delta \bar v_\alpha(\vx) =\bar \epsilon_\alpha(t)\,,\quad\quad
 \delta \bar p(\vx)= v_\beta(\vx) \bar \epsilon_\beta(t) \, .\label{eq:defShift}
\end{equation}
Interestingly, this symmetry yields Ward identities which impose that any 
 vertex function with one zero-momentum carried by a response velocity
 is vanishing. Due to the time dependence of the parameter of these symmetries, both set of Ward
  identities hold for arbitrary frequencies.  
  
  To implememt FRG, the forcing term in (\ref{eq:NSaction}) can be promoted to a regulator by 
  replacing $L^{-1}$ by the running RG scale $\rgk$, and is then interpreted as an effective forcing.
  The usual off-diagonal regulator term can also be added (and is required in $d=2$), and plays the role of an effective
   friction. For the NS field theory, the BMW scheme (introduced in Sec. \ref{sec_frg:sec_vertexp}) allows one to exactly close the flow equation
  of any $n$-point correlation function in the limit of large momentum.
 Indeed, consider the flow equation of a vertex function. In the regime where all external wave-numbers are large compared to $\rgk$
  (which ultimately means large compared to $L^{-1}$), then the internal (loop) momentum is negligible in all the vertex functions,
 which can  therefore be expanded around $\vq\simeq 0$. At leading order, i.e. keeping only the zeroth order term of this expansion,
 all the vertex functions with one (or two) zero momentum are given exactly in terms of lower order ones thanks to the Ward identities. Compared to the usual BWM scheme,
 it is not necessary to keep a dependence in a background field to express these vertices. 
 Moreover, the peculiarity of the NS flow equations is that the non-linear part of the flow is not negligible compared to the linear one in the limit of large momentum. This implies that there is no decoupling of the scales, and it yields  a violation of standard scale invariance pertaining to intermittency.  The physical origin of this non-decoupling of scales roots in what is known phenomenologically as the random sweeping effect, which is the random advection of small-(length)scale velocities by the large-(length)scale vortices of the flow \cite{Frisch95}.
 
 Furthermore, it turns out that due to the structure of the Ward identities, the
 resulting flow equations endow a very simple  form when expressed in terms of the correlation functions $G^{(n)}$ rather than
 in terms of the vertex functions  $\Gamma^{(n)}$ (even if in the calculation the $\vq$ expansion is applied only to vertex functions).
 As a result, the flow equation for an arbitrary $n$-point generalized correlation function (correlation and response) can be solved analytically at the fixed point, in both regimes of small 
 and large time delays in the stationary state. This solution provides the exact 
 time dependence of correlation function in turbulence at leading order in wave-numbers \cite{Canet17,Tarpin18,Tarpin19}.

 For the 2-point correlation function,
  the FRG result yields a  Gaussian decay in the variable $tk$
    where $k$ is the wave-number and $t$
   a small time delay in the stationary state. This form can be physically understood as the effect of sweeping \cite{Frisch95}. 
     However, the FRG result shows that
      the Gaussian form is only valid at small time delays, and it predicts that the temporal decay crosses over to a simple exponential at large time delays, which was not known. Similar explicit expressions were obtained for the temporal dependence of any multi-point correlation. 
 The related predictions for the two-point function have been compared to DNS of the NS equation \cite{Canet17} and experiments \cite{Debue18,Gorbunova19},
  and the agreement is very accurate.

\subsubsection{Master equations: reaction-diffusion processes}
\label{sec_NEQ-RD}

 Reaction-diffusion processes are simple models which describe one or several species of particles, $A$, $B$, $\cdots$,
  which diffuse on a lattice and can undergo some reactions at a certain rate when they meet  ({\it e.g.} $A+B \to B+ B$). Owing to their simplicity, these models are widely used in physics (percolation or growth processes, cellular automata,\dots), chemistry (chemical reactions),
   biology (dynamics of population or disease), and also economics (stockmarket evolution). 
   This dynamics can be described by a master equation, which gives the time evolution of the probability 
  distribution of the micro-states of the system (number of particles of each species on each lattice site).
    When the transition probabilities do not satisfy the detailed balance condition,
  the system reaches genuinely nonthermal (nonequilibrium) steady-states,
  whose probability distribution is not known in general.

 Reaction-diffusion systems exhibit phase transitions between these nonequilibrium steady-states.
 The most common ones are absorbing phase 
 transitions, which separate an active fluctuating steady-state from one, or several, absorbing states, with no 
 fluctuations \cite{Hinrichsen00,Odor04,Henkel08}. An important challenge is to obtain
  a full classification of the associated non-equilibrium universality classes.
  When there is a unique absorbing state, and when the phase transition is characterized by a scalar order parameter 
  (the mean density for instance), it generically belongs to the Directed Percolation (DP) universality class.
  % as conjectured by Janssen and Grassberger \cite{janssen81,grassberger82}.
    Besides this  prominent class,
  other universality classes can emerge, for instance in the presence of additional symmetries (see below).

 Let us focus on one-species  reaction-diffusion processes. They are generically defined as Branching and Annihilating 
 Random Walks (BARW)
  which consist of identical particles $A$, with  particle creation $A\xrightarrow{\sigma_{m}}  (m+1)A$ and destruction $nA\xrightarrow{\lambda_{n}} \varnothing$
   \cite{Bramson85,Cardy96,Cardy98,Tauber05}. 
  The standard procedure to derive a field theory from such  microscopic rules  is the Doi-Peliti formalism
 \cite{doi76,peliti84}. It consists in introducing creation and annihilation operators, in analogy with  second quantification, to turn the associated master  equation  into an imaginary-time Shr\"odinger equation. 
 One then  resorts to the coherent state formalism to cast it into a field theory. 
 Details of this formalism can be found for instance in \cite{Cardy98,Tauber05,Tauber14}.
 The resulting action is formulated in terms of two fields $\phi$ and $\bar\phi$, as in the Langevin case, 
 which are originally complex conjugate, but often generalized to independent fields. It has the following structure
\begin{equation}
 {\cal S}[\phi,\bar\phi] = \int_{t,\vx}  \Big\{ \bar \phi\,\big(\p_t  - D\, \vnabla^2 \big)\,\phi + U[\phi,\bar\phi]\Big\}
\end{equation}
where the quadratic dynamical part encodes the diffusion and the potential $U$ encompasses all the microscopic reactions.
 
\vspace{0.2cm}
\noindent{\it a- FRG formalism for reaction-diffusion processes}
\vspace{0.2cm}

The FRG formalism to study reaction-diffusion processes is very similar to the Langevin case Eqs. (\ref{sec_kpz:deltaSk},\ref{sec_kpz:dkgam}). The simplest approximation to study these processes is the DE.
 The  generic ansatz for $\Gamma_\rgk$ at LPA$'$  can be written as
\begin{equation}
 \Gamma_\rgk[\varphi,\bar\varphi] = \int_{t,\vx}  \Big\{ \bar \varphi\,\big( X_\rgk \,\p_t  - Z_\rgk\, \vnabla^2 \big)\,\varphi + U_\rgk[\varphi,\bar\varphi]\Big\}
\label{sec_kpz:anzreacdif}
\end{equation}
where $\varphi_i=\langle\phi_i \rangle$, and 
with the two running coefficients $Z_\rgk$ and $X_\rgk$ leading to the  anomalous dimensions $\eta_\rgk^Z = -\p_s \ln Z_\rgk$ and  $\eta_\rgk^X = -\p_s \ln X_\rgk$. 
 The effective running potential $U_\rgk$ must encode the symmetries of the specific reactions under study. 
 For generic processes, the anomalous dimension $\eta$ of the fields and the dynamic exponent $z$
 are defined such that at criticality,
 $\varphi\bar\varphi \sim \rgk^{d+\eta}$ and $\omega\sim \rgk^z$, respectively \cite{Wijland98}, thus one has $\eta =\eta_*^X$ and $z= 2 + \eta_*^X -\eta_*^Z$.
 
\vspace{0.2cm}
\noindent{\it b- Directed Percolation}
\vspace{0.2cm}
 
 The directed percolation universality class can be represented by the set of reactions
 \begin{equation}
 \label{sec_kpz:DPrules}
  A \xrightarrow{\sigma} 2A \, , \quad  2A \xrightarrow{\lambda}\varnothing
  \end{equation}
 which corresponds to the effective potential
 %%%%%%%%%%%%%%%%%%%%%%%%%%%%%%%%%%%%%%%%%%%%%%%%%%%%%%%%%%%%%%%%%
\begin{equation}
U_\rgk[\phi,\bar\phi]  =   -\sigma \bar \phi\,\phi 
+ 2\lambda \, \bar \phi \phi^2 -\sigma \phi\bar\phi^2 
 +\lambda\,\big(\bar\phi\phi \big)^2. \label{sec_kpz:actionDP}
\end{equation}
%%%%%%%%%%%%%%%%%%%%%%%%%%%%%%%%%%%%%%%%%%%%%%%%%%%%%%%%%%%%%%%%%
The fundamental symmetry of the DP process is the so-called rapidity symmetry
which corresponds to the transformation $\phi(t) \to -\frac{\sigma}{2\lambda}\bar\phi(-t)$, $\bar\phi(t) \to -\frac{2\lambda}{\sigma}\phi(-t)$.
  Hence, the effective potential $U_\rgk$
 can be parametrized in terms of the 
%All possible polynomials of the fields compatible with the rapidity symmetry can be parametrized in terms 
  two invariants $\rho \equiv \bar\varphi \varphi$ and $\xi \equiv \varphi -\frac{\sigma}{2\lambda}\bar\varphi$ \cite{Canet04a}.
    
 The critical exponents of the DP
  class in dimensions 1, 2 and 3, have been computed within the FRG at both LPA and LPA$'$, and they compare accurately with results from other  methods \cite{Canet04a,Buchhold16}.
 The DP exponents have also been obtained using FRG in the context of the Reggeon Field Theory,  whose critical behavior
  belongs to the DP class \cite{Bartels16}.
 %%%%%%%%%%%%%%%%%%%%%%%%%%%%%%%%%%%%%%%%%%%%%%%%%%%%%%%%%%%%%%%%%%%%%%%%%%%%%%%%%%%%%%%%% 
 \begin{figure}[h]
 \begin{center}
\includegraphics[width=7cm]{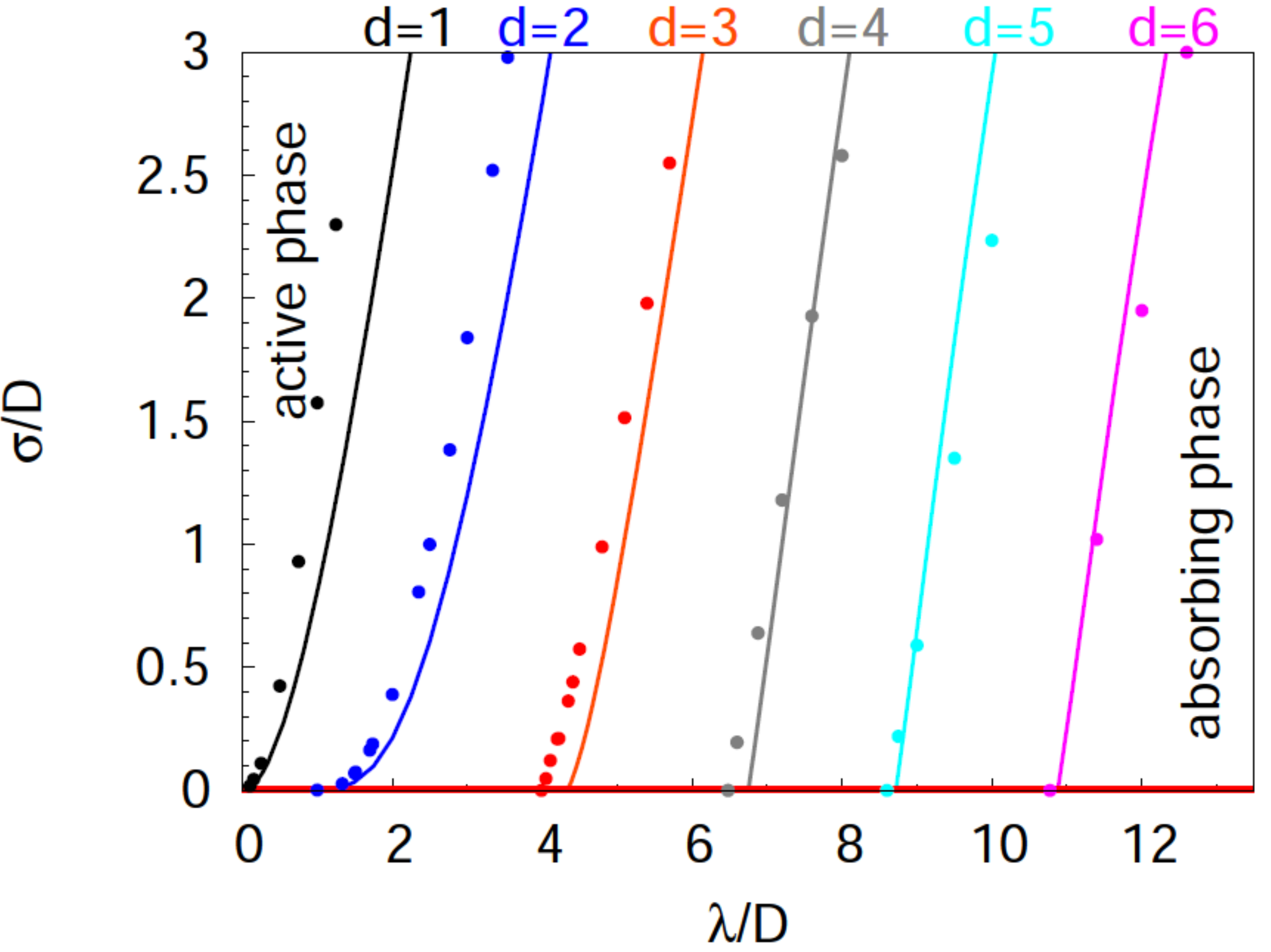}
\end{center}
\caption{Phase diagram of the BARW model (\ref{sec_kpz:DPrules}) in dimensions 1 to 6 from FRG calculations 
(solid lines) and numerical simulations (symbols) from \cite{Canet04b}. 
For each dimension, the active phase lies on the left of the transition line, 
and the absorbing phase on its right.}
\label{sec_kpz:figDP}
\end{figure}
%%%%%%%%%%%%%%%%%%%%%%%%%%%%%%%%%%%%%%%%%%%%%%%%%%%%%%%%%%%%%%%%%%%%%%%%%%%%%%%%%%%%%%%%%%
  Besides the value of the critical exponents, the FRG has been particularly useful to study nonuniversal 
  aspects. For the DP process  (\ref{sec_kpz:DPrules}), mean field theory predicts that the active phase is always stable, and thus that there is no transition in this model. Perturbative RG studies have shown that fluctuations  
   could  induce an absorbing transition in low dimensions $d \leq 2$  \cite{Cardy96,Cardy98}. 
  % In the ($\lambda/D$,$\sigma/D$) plane, the transition line found perturbatively is quadratic in $d=1$,
  %  given by $(\sigma/D)_c= (\lambda/(2\pi D))^2$, exponential in $d=2$, given by $(\sigma/D)_c=\exp(-4\pi D/\lambda)$,
  %   and no longer exists in $d>2$. 
    The model was re-examined  using FRG in \cite{Canet04a,Canet04b}. The results are in agreement with the perturbative
   ones in their region of validity, {\it i.e.} near vanishing reaction rates. However, a transition was found in $d=3$,
    and also in all higher dimensions investigated up to $d=10$. This result was confirmed by Monte Carlo
     simulations \cite{Canet04b,Odor04b}, and a remarkable agreement was found for the phase diagrams,
     as illustrated in Fig.~\ref{sec_kpz:figDP},
  which is noticeable given that the critical rates are nonuniversal quantities, similar to critical temperatures.

 The failure of perturbative RG to predict the transition in $d>2$ is due to the existence of a finite  
 threshold value $(\lambda/D)_{th}$  for the absorbing state to emerge at $\sigma/D=0$, which is not accessible
  at any order in perturbation theory
 (performed around  vanishing rates). These threshold values were observed to grow linearly with the dimension 
 in the FRG analysis. This result was  supported analytically by 
  a single-site approximation of the model, which becomes exact in the large $d$ limit \cite{Canet06b}.
  The model was eventually exactly
solved at  small $\sigma$ but arbitrary $\lambda$ in Ref.~\cite{Benitez12b,Benitez13}. The authors established  that for hypercubic lattices,
 the thresholds  grow as $(\lambda/D)_{th}\sim 2 d a^{d-2}$, where $a$ is the lattice spacing, when $d\to \infty$, confirming the FRG result.
 The mean field prediction of absence of phase transition  for this DP model is hence only valid in infinite dimension.

\vspace{0.2cm}
\noindent{\it c- Other processes}
\vspace{0.2cm}
 
 In the presence of an additional symmetry, which is the conservation of the parity of the number of particles, another universality class emerges. It can be represented by the simple BARW processes
 $A \xrightarrow{\sigma} 2A$ and   $3A \xrightarrow{\lambda}\varnothing$.
Within a perturbative RG analysis \cite{Cardy96,Cardy98}, the existence of the absorbing phase transition has been 
associated with a  change of stability of the Pure Annihilation (PA) fixed point (same model with $\sigma=0$) occuring
 in an emerging critical dimension $d_c=4/3$.    For $d>d_c$, the branching
 $\sigma$ is a relevant perturbation and only an active phase exists, whereas for $d<d_c$, $\sigma$ is irrelevant and
  an absorbing phase, which  long-distance properties are controlled by the PA fixed point, is expected for small $\sigma$.
 
\begin{figure}[t]
  \begin{center}
\includegraphics[width=5cm,angle=-90]{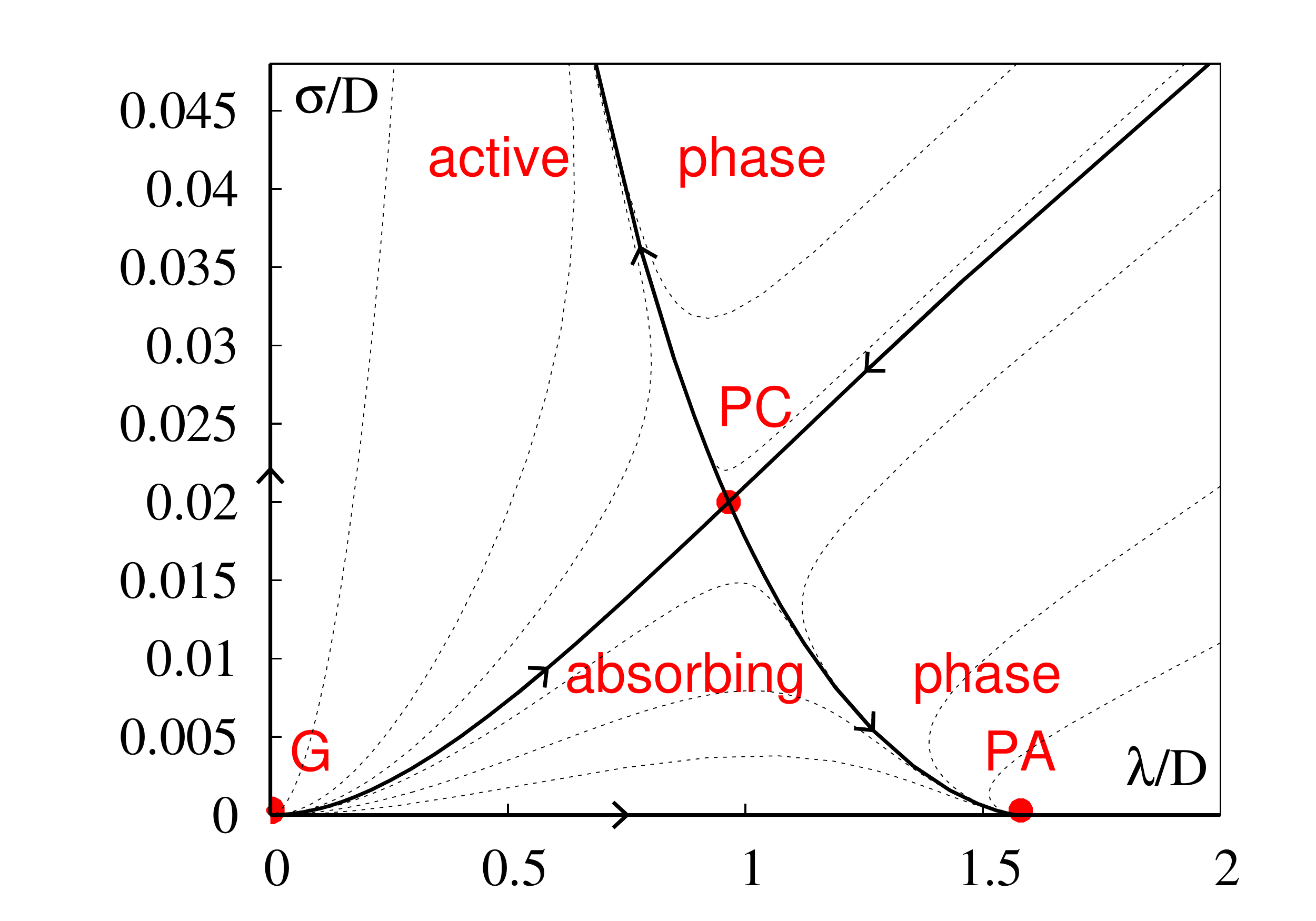}
\end{center}
\caption{Flow diagram of the BARW model $A \xrightarrow{\sigma} 2A$ and   $3A \xrightarrow{\lambda}\varnothing$ in $d=1$ 
from FRG at LPA \cite{Canet05}. The PC fixed point drives a transition between 
the active state and the absorbing one controlled by the PA fixed point.}
\label{sec_kpz:figPC}
\end{figure}

 This model was investigated using FRG within LPA in \cite{Canet05sm}. A genuinely nonperturbative fixed point 
  (not connected to the Gaussian fixed point in any dimension) is found in $d=1$ for finite $\sigma/D$ 
  and $\lambda/D$, driving an active-to-absorbing phase transition, as illustrated on Fig.~\ref{sec_kpz:figPC}.
   The  associated critical exponents differ from the DP ones, and are compatible (at least for $\nu$ at LPA) with 
   numerical simulations \cite{Canet05sm}.  Within LPA, this fixed  point  annihilates with the PA fixed point in $d\simeq 4/3$,   thereby changing its stability. The LPA analysis hence confirms  the perturbative scenario.
 However, an exact calculation at small $\sigma$ (that is to first order in $\sigma$ around the exactly solved PA fixed point, rather than around the Gaussian one) showed that the stability of the PA fixed point with respect to a perturbation 
  in $\sigma$ does not change between one and two dimensions \cite{Benitez12b,Benitez13}.
  An alternative scenario with the existence of two fixed points was proposed to reconcile existing results, but
   is yet to be elucidated.  

 The last important one-species reaction-diffusion process which has been studied using FRG is the Pair
  Contact Process with Diffusion. It differs from the previous BARW models in that branching requires 
  two particles to meet $2A \xrightarrow{\sigma} 3A$,
  supplemented with pair annihilation $2A \xrightarrow{\lambda}  \varnothing$. This process has particularly
   resisted perturbative analysis \cite{Henkel04}. In fact, FRG studies, within the LPA approximation and using a semifunctional treatment of the effective potential $U_\rgk[\varphi,\bar\varphi]$, revealed that
   $U_\rgk$ develops a nonanalyticity at a finite RG scale $\rgk_c$:
 linear terms, perturbatively forbidden, are dynamically generated. This scale corresponds to the scale where 
 perturbative RG flows blow up. This is reminiscent of the appearance of a cusp in Random Field Ising 
 Model and other disordered models (see Sec. \ref{sec_RFIM}).
 Let us finally mention that a two-species reaction-diffusion process called Diffusive Epidemic
  Process was studied using FRG in \cite{Tarpin17}.

%%% Local Variables:
%%% mode: latex
%%% TeX-master: "frg_review"
%%% End:

%% file: SEC_HighEnergy/noneqquantum.tex
\subsubsection{Non-equilibrium quantum systems} \label{sec:QuantumNonEq}

The understanding of non-equilibrium quantum phenomena is amongst the most pressing questions of modern physics. Potential applications ranging from the physics of the early universe over the early evolution of the fireball in heavy-ion collisions, the dynamics of the pair-creation in strong electromagnetic fields to the far-from-equilibrium dynamics of ultracold atomic systems. Non-equilibrium versions of the FRG may also help to understand the propagation of information, the generation of (entanglement-) entropy and the equilibration process. 

A generic non-equilibrium process is often initiated with an over-occupied initial state with large occupancies $n(t_0,{\bf x})\gg 1$ encoded in the initial density matrix $\rho(t_0)$. The statistical propagator $\langle \{\varphi(t_0,{\bf x})\,,\, \varphi(t_0,{\bf 0})\}\rangle \propto n(t_0,{\bf x})+1/2$  is first dominated by the large occupancies and the respective physics is well described by classical-statistical approximations to the full system. In $O(N)$-theories this classical-statistical regime is captured well by next-to-leading order computations in $1/N$-expansions ($s$-channel resummations) of 2PI-hierarchies, for reviews see e.g.~\cite{Berges:2015kfa, Schmied:2018mte}. 

For occupancies $n(t_0,{\bf x})\approx 1$ the quantum $1/2$ in $ n(t_0,{\bf x})+1/2$ gets sizable. The system enters a regime with non-equilibrium quantum dynamics which requires a non-equilibrium quantum field theoretical description. This can be done within the Schwinger-Keldysh formalism on a closed time path $\cal C$ with an initial density matrix $\rho(t_0)$ at the initial time $t_0$. Then, the real-time FRG follows straightforwardly with the standard  cutoff term quadratic in the fields. In the scalar case, the generating functional for non-equilibrium correlation functions is given by 
\begin{align}
\label{eq:ClosedTimePath}
Z_k[J;\rho] = \textrm{Tr}\, \rho(t_0)\,  \exp \left\{ i \int_{\cal C} \varphi(x) J(x) + \frac{i}{2}\int \varphi (x) R_k(x,y) \varphi(y)\right\}\,, 
\end{align}
where $x, y$ live on the Keldysh contour $\cal C$. For an introduction to the formulation with a standard spatial-momentum cutoff see \cite{Berges12}. A manifestly causal cutoff is obtained if closing the Keldysh contour at a finite time $t=\tau$. Then the cutoff parameter is simply the maximal time $k=\tau$. The respective flow equation does not integrate-out momentum-shells but simply propagates the system in time, see \cite{Gasenzer:2007za, Pietroni:2008jx, Gasenzer:2010rq, Corell:2019jxh}. The regularisation can also be introduced via deformations of the occupancies $n(t, {\bf x})\to n_k(t, {\bf x})$, see e.g.~\cite{Koenig99, Jakobs07} for applications in low-dimensional systems. 

The Schwinger-Keldysh formulation of the FRG has been applied to the time evolution and turbulent cascades at non-thermal fixed points in $O(N)$-models, \cite{Gasenzer:2007za, Berges09, Corell:2019jxh}, to the classification of dynamical critical phenomena in models with and without driving force, \cite{Sieberer13, Mesterhazy13, Sieberer14, Mesterhazy:2015uja}, and to the cosmological power spectrum, \cite{Pietroni:2008jx, Lesgourgues:2009am, Bartolo:2009rb, Audren:2011ne, Juergens12, Vollmer:2014pma}, see also \cite{Floerchinger:2016hja}. For further applications in low-dimensional systems see also Sec.~\ref{sec:IVB4}.

%% file: SEC_FB/fermions.tex
\section{Quantum many-particle systems}
\label{sec_fb}

Quantum many-particle theory was developed to describe interacting particles
in condensed matter, and is now relevant for ultracold atomic gases, too.
According to the two types of quantum statistics (leaving anyons aside),
this broad field is naturally divided in the theory of interacting bosons
and interacting fermions.
Applications of the FRG to quantum many-particle problems have already been
summarized in two extensive review articles published in 2012, one in
relation to cold atoms \cite{Boettcher:2012cm}, the other with a focus on
interacting fermions in condensed matter \cite{Metzner12}.

We note in passing that the FRG can also be applied to {\em few}-body systems.
A beautiful example is the Efimov effect related to the formation of bound states
of three or four interacting bosons or fermions, which is associated with a limit
cycle in the renormalization group flow \cite{Floerchinger:2011yv}.

%%%%%%%%%%%%%%%%%%%%%%%%%%%%%%%%%%%%%%%%%%%%%%%%%%%%%%%%%%%%%%%%%%%%%%%%%

\subsection{Bosons} \label{sec:IVA} %\label{sec_fb:subsec_bose} 

The first realization of an interacting quantum Bose system was liquid $^4$He.
Nowadays, ultracold bosonic atoms provide a vast additional playground for
exploring interaction effects in Bose systems. Moreover, interacting
Bose systems also emerge in effective low-energy theories.
In the following we review a selection of systems where the FRG has made
a major impact.

\subsubsection{Superfluidity in a dilute Bose gas}

The Euclidean action of a dilute Bose gas reads 
\begin{equation}
S = \int_0^\beta d\tau \int d^dr \left\{ \psi^*\left(\partial_\tau-\mu - \frac{\boldsymbol{\nabla}^2}{2m}
\right) \psi + \frac{g}{2} (\psi^*\psi)^2 \right\} ,
\label{sec_fb:bose1} 
\end{equation}
where the complex field $\psi$ satisfies periodic boundary conditions, $\psi({\bf r},\tau)=\psi({\bf r},\tau+\beta)$, and $\mu$ is the chemical potential. Most physical low-energy properties depend only on the boson mass $m$ and the $s$-wave scattering length $a\equiv a(g,\Lambda)$, which is a function of the repulsive interaction $g$ and the upper momentum cutoff $\Lambda$ in the model defined by~(\ref{sec_fb:bose1}). For any finite density ($\mu>0$), the ground state is superfluid and characterized by spontaneously broken U(1) symmetry: $\langle \psi({\bf r},\tau)\rangle\neq 0$. 

In three dimensions (and at zero temperatures in two dimensions) most physical properties of the superfluid phase, including the equation of state and the sound mode with linear dispersion, can be understood from the Bogoliubov theory~\cite{Bogoliubov47}, based on a Gaussian-fluctuation approximation about the mean-field solution. Nevertheless, perturbation theory beyond the Bogoliubov approximation is plagued with infrared divergences~\cite{Beliaev58a,Beliaev58b,Hugenholtz59,Gavoret64}. Although these divergences cancel out in gauge-invariant quantities (pressure, sound velocity, etc.), they have a definite physical origin since they reflect the divergence of the longitudinal susceptibility, a phenomenon which also occurs in the low-temperature phase of the O($N\geq 2$) model (Sec.~\ref{sec_frg:subsubsec_DE2}) but is not accounted for in the Bogoliubov theory. Popov has proposed an approach to superfluidity, based on an amplitude-phase representation of the boson field, which is free of infrared divergences and yields the correct low-energy behavior of the correlation functions but is restricted to the (low-momentum) hydrodynamic regime~\cite{Popov79,Popov_book2}. 

Field-theoretical diagrammatic methods~\cite{Nepomnyashchii75,Nepomnyashchii78,Nepomnyashchii83} or perturbative RG constrained by Ward identities~\cite{Castellani97,Pistolesi04} can be used to handle the infrared divergences of perturbative theory. The FRG provides us with a simple alternative method, free of any divergences, which encompasses both the Bogoliubov theory (valid, {\it stricto sensu}, only at momenta larger than a ``Ginzburg'' scale) and Popov's hydrodynamic approach~\cite{Dupuis09a}. It has been used to understand various properties of three- and two-dimensional superfluid Bose systems, from the equation of state to the excitation spectrum and the damping of quasi-particles~\cite{Wetterich08,Dupuis07,Dupuis09b,Sinner09,Sinner10,Dupuis11,Floerchinger08,Floerchinger09a,Floerchinger09b,Eichler09,Rancon12b,Krieg17,Isaule18,Isaule20}.   

Another important issue regarding dilute Bose gases is the determination of the superfluid transition temperature. In the absence of interactions ($a=0$), the bosons undergo a Bose-Einstein condensation at the temperature $T_c^0 = (2\pi/m)[n/\zeta(3/2)]^{2/3}$ in three dimensions where $n$ is the mean density. For weak interactions, the shift $\Delta T_c/T_c^0=(T_c-T_c^0)/T_c^0$ in the transition temperature is a universal function of $na^3$. The dependence of $\Delta T_c/T_c^0$ on $a$ has remained a controversial issue for a long time and even the sign of the effect has been debated~\cite{Lee58,Toyoda82,Huang99}. It is now understood that $\Delta T_c/T_c^0=c(an^{1/3})$ increases linearly with $a$~\cite{Gruter97,Holzmann99,Holzmann99a,Baym99,Baym01} and the proportionality coefficient $c$ has been estimated from various approaches. The difficulty in getting a precise estimate of $c$ comes from the fact that it requires the knowledge of the one-particle propagator at $T_c$ in a large momentum range including the crossover region between the critical and noncritical regimes. The determination of $c$ is therefore a nonperturbative problem which can be dealt with the FRG. Since the full knowledge of the momentum dependence of the propagator is required, the DE is not sufficient. The BMW approximation (Sec.~\ref{sec_frg:sec_vertexp}) gives $c\simeq 1.37$~\cite{Blaizot05,Benitez09,Benitez12} in reasonable agreement with lattice results ($c=1.32(2)$~\cite{Arnold01} and $c=1.29(5)$~\cite{Kashurnikov01}) and seven-loop resummed calculations ($c=1.27(10)$~\cite{Kastening04}). Another FRG calculation, based on a vertex expansion, gives $c = 1.23$~\cite{Hasselmann04,Ledowski04}. In the case of a model with $N$-component real fields (Eq.~(\ref{sec_fb:bose1}) corresponds to $N=2$), the value of $c$ obtained from FRG is remarkably accurate for $N\geq 3$, being within the error bars of lattice simulations and seven-loop calculations~\cite{Benitez12}.

\subsubsection{Superfluid to Mott-insulator transition}
\label{sec_fb:subsubsec_BHM}

\begin{figure}
	\centerline{\includegraphics[width=6.cm]{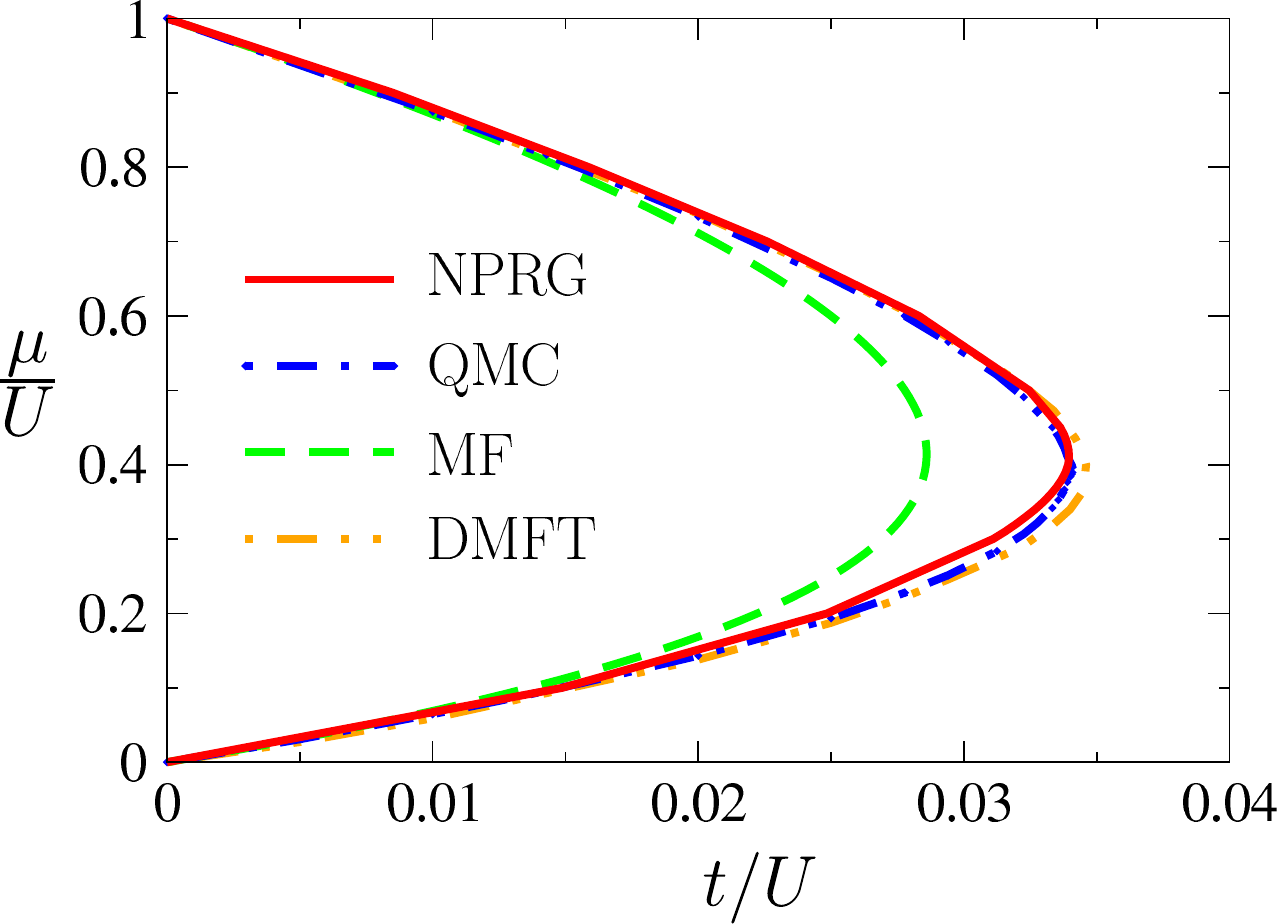} \hspace{0.5cm}	
		\includegraphics[width=6.cm]{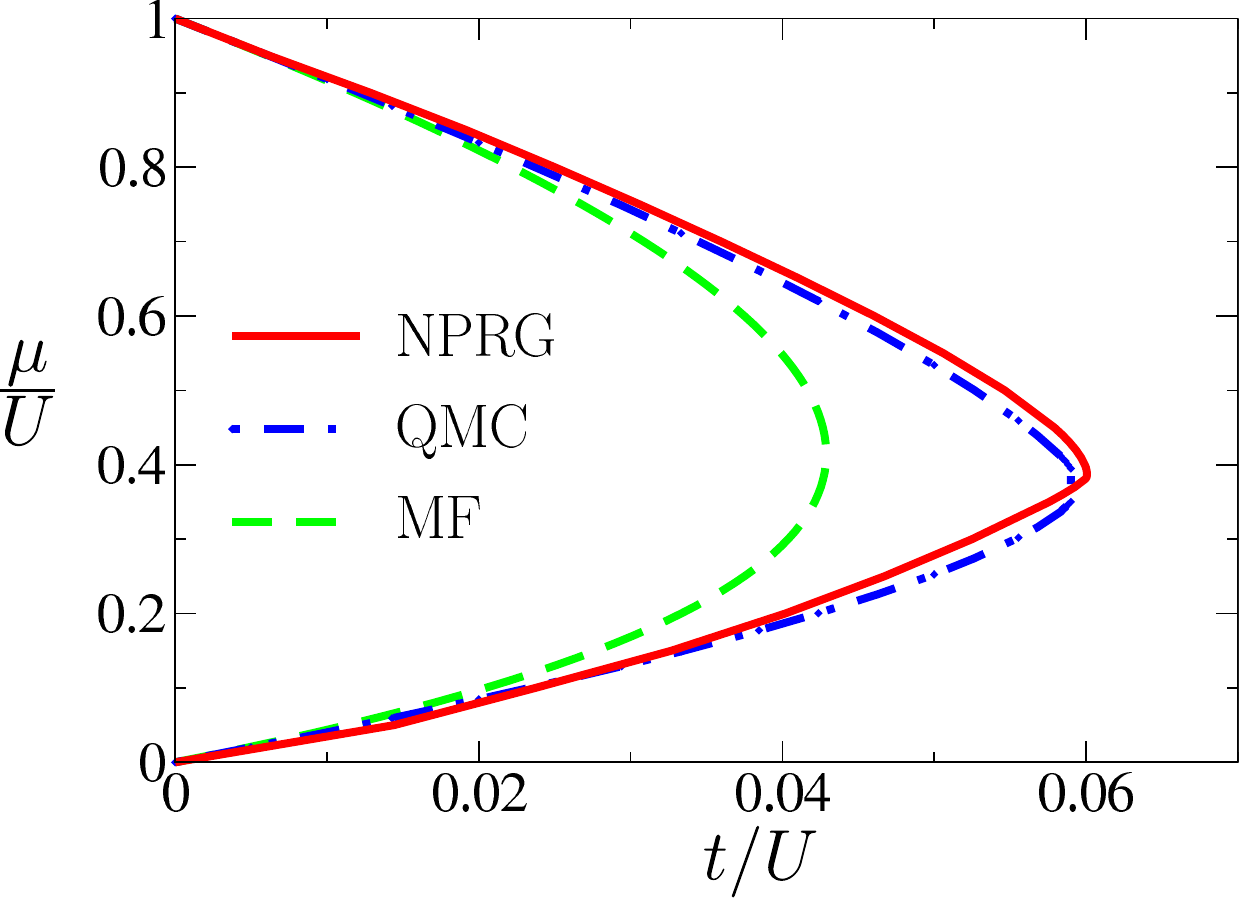}}
	\caption{(left) Phase diagram of the 3D Bose-Hubbard model showing the first Mott lobe (with a density $n=1$) and the surrounding superfluid phase. The solid (red) line shows the FRG result and the (green) dashed line the mean-field one. The QMC and DMFT data are obtained from Refs.~\cite{Capogrosso07} and~\cite{Anders11}, respectively . (right) Phase diagram of the 2D Bose-Hubbard model. The QMC data are obtained from Ref.~\cite{Capogrosso08}. Reprinted from Ref.~\cite{Rancon11b}.}
	\label{secIV:fig_BHM}
\end{figure}

The simplest model describing bosons moving in a $d$-dimensional hypercubic lattice is the one-band Bose-Hubbard model defined by the action~\cite{Fisher89} 
\begin{equation}
S = \int_0^\beta d\tau \biggl\{ 
\sum_{\bf r} \left[ \psi^*_{\bf r} (\partial_\tau - \mu) \psi_{\bf r} + \frac{U}{2}  n_{\bf r} ( n_{\bf r}-1)  \right] 
- t \sum_{\langle {\bf r},{\bf r}'\rangle} \left( \psi^*_{\bf r} \psi_{{\bf r}'} + {\rm h.c.} \right) \biggr\}, 
\end{equation}
where $ n_{\bf r}=\psi^*_{\bf r} \psi_{\bf r}$, $U$ is the on-site repulsion, $t$ the intersite hopping amplitude, and $\langle {\bf r},{\bf r}'\rangle$ denotes nearest-neighbor sites. When the average number $n$ of bosons per site is integer and $U\gg t$, the ground state is a Mott insulator with $n$ bosons localized at each lattice site, a vanishing compressibility $\kappa=\partial n/\partial\mu$ and a gap in the excitation spectrum. A transition to a superfluid ground state can be induced by either decreasing the ratio $U/t$ or varying $n$. In the first case the transition is in the universality class of the $(d+1)$-dimensional XY (or O(2)) model with an upper critical dimension $d_c^+=3$. In the second one, the transition is in the universality class of the dilute Bose gas and is mean-field-like above two dimensions ($d_c^+=2$). 

The three- and two-dimensional Bose-Hubbard models have been studied within the FRG approach using the lattice formulation (Sec.~\ref{sec_frg:subsec_lattice}) where the initial condition corresponds to the limit of decoupled sites~\cite{Rancon11a,Rancon11b}. The FRG yields a phase diagram in very good agreement with the (numerically exact) quantum Monte Carlo simulations~\cite{Capogrosso07,Capogrosso08} and with an accuracy similar to that obtained from dynamical-mean-field theory~\cite{Anders11,Panas15}; see Fig.~\ref{secIV:fig_BHM}. It also allows one to compute nonuniversal quantities such as the velocity (or the effective mass when the transition is induced by a density change) of the critical fluctuations. It reproduces the two universality classes of the superfluid--Mott-insulator transition and gives an accurate determination of the universal scaling functions associated with the equation of state near the quantum critical point~\cite{Rancon12a,Rancon12d,Rancon13b}.

\subsubsection{Relativistic bosons and quantum O($N$) model}  

\begin{figure}
	\centerline{
		\includegraphics[width=5.5cm]{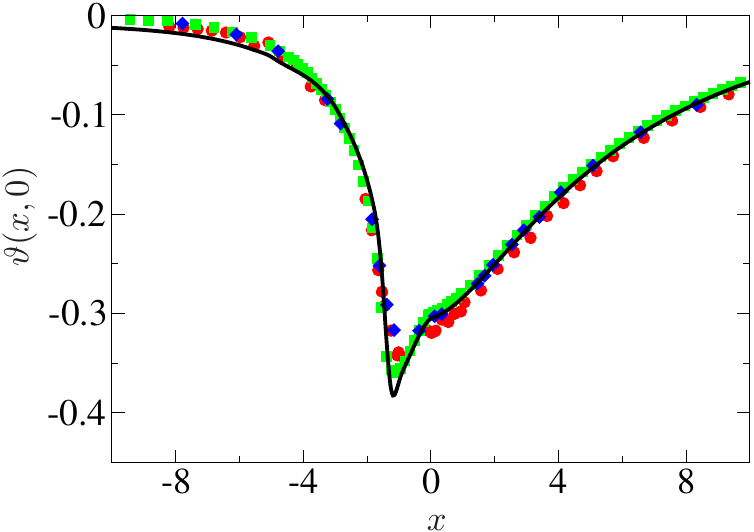}
		\includegraphics[width=5.1cm]{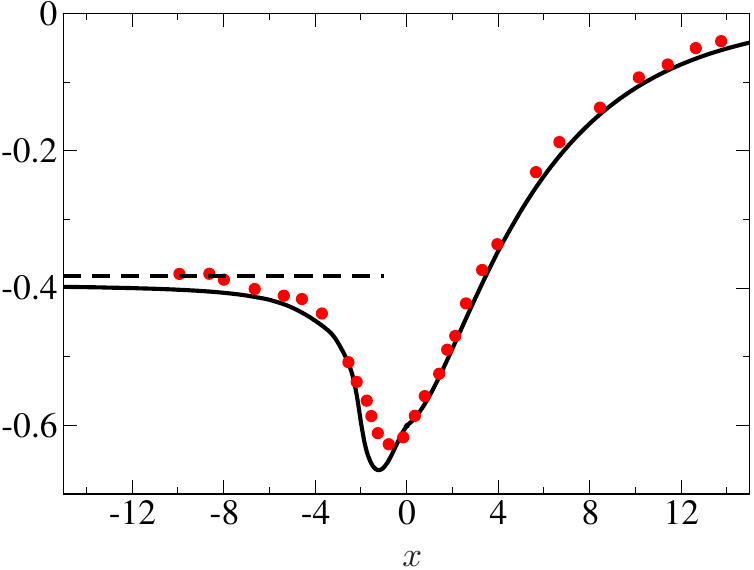}
		\includegraphics[width=5.1cm]{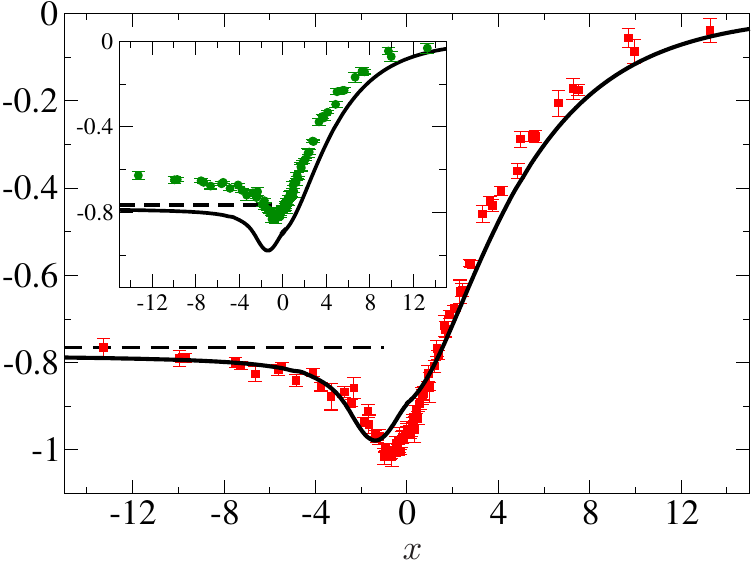}}
	\caption{Scaling function $\vartheta(x,0)$ determining the singular part of the internal energy density, $\epsilon_{\rm sing}(\delta,T)=\vartheta(x,0) (k_BT)^3/(\hbar c)^2$ ($c$ is the velocity of the critical fluctuations), in the two-dimensional quantum O($N$) model near its quantum critical point (full lines); $x\propto \delta/T^{1/\nu}$ where $\delta$ measures the distance to the quantum critical point and $\nu$ is the correlation-length critical exponent of the $T=0$ phase transition. The symbols show the results obtained from Monte Carlo simulations of three-dimensional classical spin models in a finite geometry where $\vartheta(x,0)$ determines the universal Casimir force.   
		The horizontal dashed line shows the (exact) limit $-2(N$$-$$1$$)\zeta(3)/2\pi$ for $x\to-\infty$. (left panel)  Ising ($N$$=$$1$) universality class. Monte Carlo simulations are from Ref.~\cite{Vasilyev09} (blue diamonds), Ref.~\cite{Hucht11} (green squares) and Ref.~\cite{LopesCardozo15} (red circles). (middle panel)  XY universality class ($N$$=$$2$). The Monte Carlo data  are from Ref.~\cite{Vasilyev09}. (right panel) Heisenberg universality class ($N$$=$$3$). The Monte Carlo  data~\cite{Dantchev04} have been rescaled so as to satisfy the correct asymptotic value for $x$$\to$$-\infty$, the bare data are shown in the inset. Reprinted from Ref.~\cite{Rancon16}.}
	\label{secIV:fig_N1N2N3}
\end{figure} 

The simplest quantum generalization of the O($N$) model discussed in Sec.~\ref{sec_frg}, with space-time Lorentz invariance, is defined by the Euclidean (imaginary-time) action 
\begin{equation}
S[\boldsymbol{\varphi}] = \int_0^{\beta} d\tau \int d^dr \biggl\lbrace \frac{1}{2}  (\boldsymbol{\nabla}\boldsymbol{\varphi})^2 + \frac{1}{2c^2} (\partial_\tau \boldsymbol{\varphi})^2  + \frac{r_0}{2} \boldsymbol{\varphi}^2 + \frac{u_0}{4!} {(\boldsymbol{\varphi}^2)}^2 \biggr\rbrace ,
\end{equation}
where the $N$-component real field satisfies periodic boundary conditions: $\boldsymbol{\varphi}({\bf r},\tau)=\boldsymbol{\varphi}({\bf r},\tau+\beta)$. $r_0$ and $u_0$ are temperature-independent coupling constants and $c$ is the (bare) velocity of the excitations. The quantum O(2) model describes relativistic bosons and is relevant for the superfluid--Mott-insulator transition (Sec.~\ref{sec_fb:subsubsec_BHM}); the $N=3$ model applies to quantum antiferromagnets, the $N=4$ to QCD, etc. At zero temperature the quantum model is equivalent to its $(d+1)$-dimensional classical counterpart and therefore exhibits a quantum phase transition between a disordered phase and an ordered phase where the O($N$) symmetry is spontaneously broken. At nonzero temperature the equivalent classical model has a finite size $L_\tau=\beta$ in the $(d+1)$th dimension. The finite-temperature thermodynamic has been studied within the FRG approach~\cite{Blaizot07a,Blaizot11,Rancon13a}. In the vicinity of the quantum critical point, the scaling functions determining the universal equation of state~\cite{Rancon16} are in striking agreement with Monte Carlo simulations of $(d+1)$-dimensional classical $N$-component spin models in a finite geometry (these models are in the same universality class as the two-dimensional quantum O($N$)  model)~\cite{Vasilyev09,Hucht11,Dantchev04,LopesCardozo15}; see Fig.~\ref{secIV:fig_N1N2N3}. 

Momentum-frequency dependent correlation functions and spectral functions of the quantum O($N$) model have also been obtained from the FRG. In the context of QCD, where the O(4) model is frequently used as a chiral effective model, spectral functions of sigma mesons and pions have been  computed~\cite{Tripolt:2013jra,Kamikado:2013sia,Wambach14,Pawlowski18}. In statistical physics, the FRG approach has lead to the confirmation of the existence of a bound state in the ordered phase of the two-dimensional quantum O(1) model~\cite{Rose16a}, with a mass within 1\% of previous Monte Carlo simulations and numerical diagonalization values~\cite{Agostini97,Caselle99,Nishiyama14}. In the ordered phase of the two-dimensional O(2) and O(3) models, the ``Higgs'' amplitude mode, whose existence and visibility in the vicinity of the quantum critical point have been overlooked for a long time~\cite{Podolsky11}, is well described by FRG~\cite{Rancon14,Rose15}, in quantitative agreement with Monte Carlo simulations~\cite{Gazit13}. The estimate of the ``Higgs'' mass in the quantum O(3) model from the BMW approach has been confirmed by subsequent exact diagonalizations~\cite{Nishiyama15,Nishiyama16} and quantum Monte Carlo simulations of quantum spin models~\cite{Lohofer15}. At zero-temperatures, the frequency-dependent conductivity and its universal features have also been studied~\cite{Rose17,Rose17a,Rose18}. 

%%%%%%%%%%%%%%%%%%%%%%%%%%%%%%%%%%%%%%%%%%%%%%%%%%%%%%%%%%%%%%%%%%%%%%%%%

\subsection{Fermions} \label{sec:IVB}

Interacting fermion systems display a particularly rich variety of phases and phenomena. The most important representatives are electrons in solids, liquid $^3$He, and ultracold fermionic atoms. In this section we focus on non-relativistic systems with spin-$\frac{1}{2}$ fermions.

For fermions moving in a continuum, the bare Euclidean action has the form \cite{Negele_book}
\begin{equation}
 S[\psi,\psi^*] = \int_0^\beta \! d\tau \int \! d^dr \, \psi^*({\bf r},\tau) \, 
 \Big( \partial_\tau - \mu - \frac{1}{2m} \nabla_{\bf r}^2 \Big) \,
 \psi({\bf r},\tau) + V[\psi,\psi^*] \, ,
\end{equation}
where $\psi = (\psi_\uparrow,\psi_\downarrow)$ and $\psi^* = (\psi_\uparrow^*,\psi_\downarrow^*)$ are anticommuting spinor fields with antiperiodic boundary conditions in $\tau$, and $V[\psi,\psi^*]$ is an arbitrary interaction term. For two-particle interactions, $V[\psi,\psi^*]$ is quartic in the fields. Note that $\psi$ and $\psi^*$ are independent variables.
For lattice fermions, the space variable $\bf r$ is replaced by a discrete lattice site variable $\bf j$, and the gradient term by hopping amplitudes between sites $\bf j$ and $\bf j'$.

The  Fermi-Dirac statistics of the particles and the existence of a 
Fermi surface at nonzero density leads to a number of difficulties that 
are not present in the bosonic case discussed in the previous section. 
First, the fermionic effective action $\Gamma_k[\psi,\psi^*]$ defined 
below is a functional of Grassmann variables. As pointed out in 
Sec.~\ref{sec_frg:subsec_quantum}, one must therefore truncate 
$\Gamma_k[\psi^*,\psi]$ in an expansion about $\psi = \psi^* = 0$, which 
amounts to retaining only a finite number of vertices and performing a 
loop expansion of the flow equations. This makes the fermionic FRG 
fundamentally perturbative. Second, since the low-energy fermion states 
live near the Fermi surface, it is not 
possible to neglect the momentum dependence of the vertices $\Gamma_k^{(n)}$
or perform a derivative expansion, i.e., an expansion about momentum
${\bf p}=0$. Thus the vertices are necessarily functionals of the 
momenta.\footnote{In the fermionic case, the terminology {\em functional}
RG originally stems from the momentum dependence of the vertices and 
not their field dependence.} Third, the Grassmannian field is not an 
order parameter; order parameters are defined by composite fields: 
$\langle\psi^*\psi\rangle$ for a charge- or spin-density wave, 
$\langle\psi\psi\rangle$ for a superconductor. A phase transition is 
signaled by a divergence of the associated susceptibility and the 
two-particle vertex $\Gamma_k^{(4)}$. The growth of $\Gamma_k^{(4)}$ 
implies a breakdown of the (perturbative) RG and prevents, at least in a 
straightforward way, to continue the flow into the ordered phase. 
Furthermore, the fluctuations of the bosonic fields $\psi^*\psi$ and 
$\psi\psi$  being not properly taken care of, the flow becomes 
uncontrolled whenever strong collective fluctuations with a large 
correlation length set in. We shall discuss below how one can, at least 
partially, overcome these difficulties, e.g., by introducing bosonic 
fields to take into account collective fluctuations or starting the FRG 
flow from a nonperturbative initial condition.

The FRG for interacting fermion systems and various applications have already been reviewed in Ref.~\cite{Metzner12}. That review also contains a short history of the renormalization group for fermion systems which shall not be repeated here. In the meantime, there have been substantial methodological developments, and the range of applications has broadened considerably. A concise review covering also rigorous mathematical work has been published recently by Salmhofer \cite{Salmhofer19}.

%%%%%%%%%%%%%%%%%%%%%%%%%%%%%%%%%%%%%%%%%%%%%%%%%%%%%%%%%%%%%%%%%%%%%%%%%%%%%%%%%%%

\subsubsection{Fermion flow equation} \label{sec:IVB1}

The scale dependent effective action and its flow equation are obtained in close analogy to the classical $O(N)$ model described in Sec.~II. The flow is defined by adding a scale dependent regulator term
$\Delta S_k = - (\psi^*, R_k \psi) = - \int dx \int dx' \psi^*(x') R_k(x',x) \psi(x)$
to the bare action, where $x$ and $x'$ include space and time variables (or, alternatively, momentum and frequency variables), and $k$ is the flow parameter.\footnote{In the literature on interacting fermion systems the greek letter $\Lambda$ is the most commonly used symbol for the flow parameter. However, for the sake of a consistent notation within this review, we denote the flow parameter by $k$.}
Here and in the following $(.,.)$ is a shorthand notation for the summation (or integration) over all fermion field indices in an arbitrary basis.
The regulator term modifies the bare propagator $G_0$ to $G_{0,k} = (G_0^{-1} + R_k)^{-1}$.
The effective action is defined by the Legendre transform of the free energy in the presence of Grassmann source fields
\begin{equation}
 \Gamma_k[\psi,\psi^*] = - \ln {\cal Z}_k[\eta_k,\eta_k^*] +
 (\eta_k^*,\psi) + (\psi^*,\eta_k) - \Delta S_k[\psi,\psi^*] \, , 
\end{equation}
where $\eta_k$ and $\eta_k^*$ are $k$-dependent functions of $\psi$ and $\psi^*$, as determined by the equations
$\psi = \partial_{\eta^*} \ln {\cal Z}_k[\eta,\eta^*]$ and
$\psi^* = - \partial_\eta \ln {\cal Z}_k[\eta,\eta^*]$.

The fermionic version of Wetterich's exact flow equation for the effective action reads \cite{Berges:2000ew,Metzner12}
\begin{equation} \label{eq:SecIV:flowexact}
 \partial_k \Gamma_k[\psi,\psi^*] = - \frac{1}{2} {\rm Tr} \left\{
 \partial_k {\bf R}_k \big( {\bf\Gamma}_k^{(2)}[\psi,\psi^*] +
 {\bf R}_k \big)^{-1} \right\} \, ,
\end{equation}
where
\begin{equation} \label{eq:secIV:Gamma2}
 {\bf\Gamma}_k^{(2)}[\psi,\psi^*](x',x) = 
 \left( \begin{array}{cc} 
 \frac{\partial^2\Gamma_k}{\partial\psi^*(x')\partial\psi(x)} &
 \frac{\partial^2\Gamma_k}{\partial\psi^*(x')\partial\psi^*(x)} \\[3mm]
 \frac{\partial^2\Gamma_k}{\partial\psi(x')\partial\psi(x)} &
 \frac{\partial^2\Gamma_k}{\partial\psi(x')\partial\psi^*(x)}
 \end{array} \right) \, ,
\end{equation}
and ${\bf R}_k(x',x) = {\rm diag}[R_k(x',x),-R_k(x,x')]$. Note that $\Gamma_k^{(2)}$ is a functional of the fields and a function of the field indices.
The matrix structure of the two-point function generated by the conjugated variables $\psi$ and $\psi^*$ gives rise to particle-particle and particle-hole channels in fermion systems.

\begin{figure} [tb]
\centerline{\includegraphics[width = 9cm]{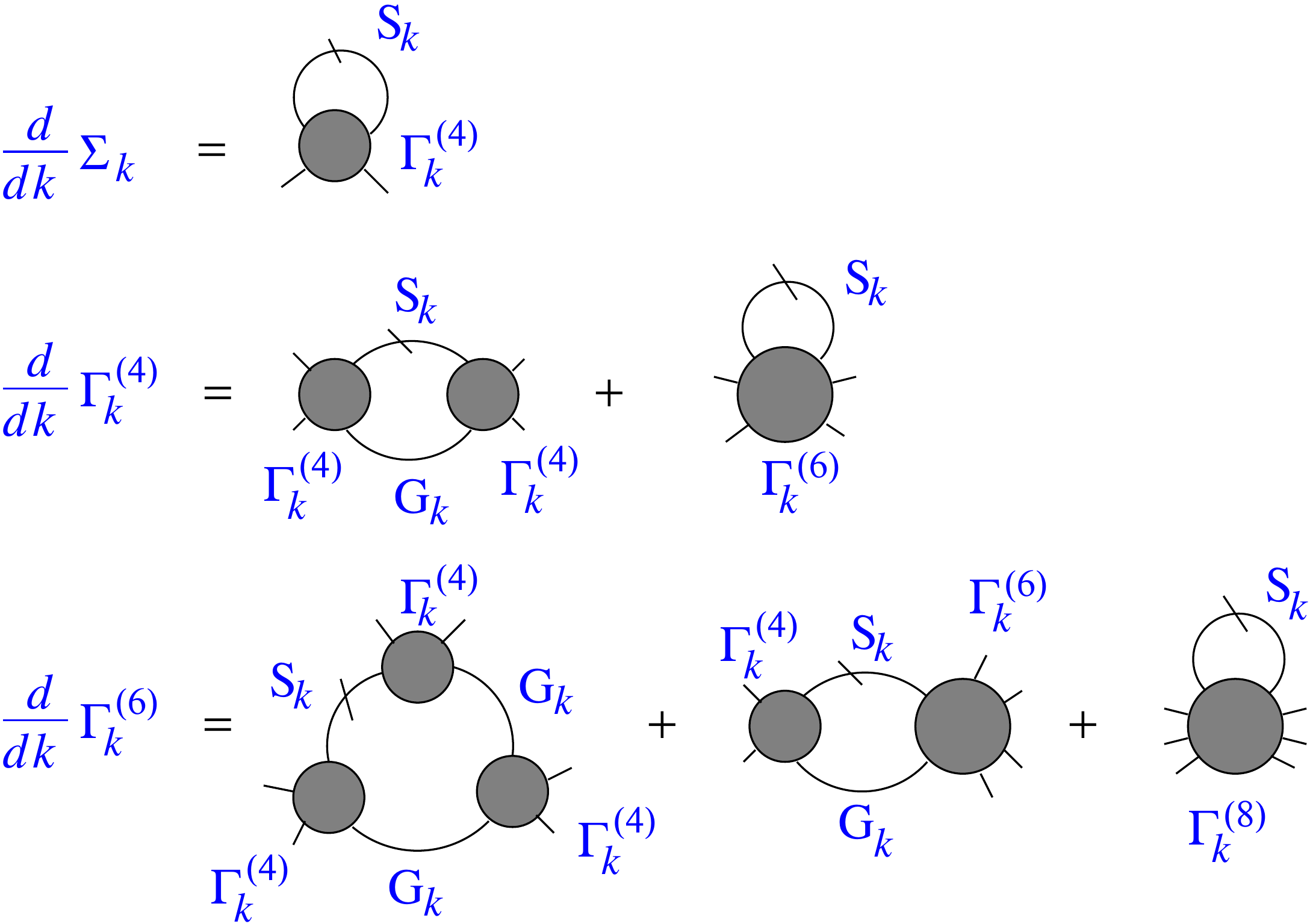}}
\caption{Diagrammatic representation of the flow equations for
 the self-energy $\Sigma_k$, the two-particle vertex
 $\Gamma_k^{(4)}$, and the three-particle vertex $\Gamma_k^{(6)}$
 in the one-particle irreducible version of the functional RG. 
 Lines with a dash correspond to the single scale propagator 
 $S_k$ (shown as a line with a cross in Fig.~\ref{sec_frg:fig_eqwet}), 
 the other lines correspond to the full propagator $G_k$.}
\label{fig:secIV:vertexexp}
\end{figure}
Expanding the functional flow equation (\ref{eq:SecIV:flowexact}) in powers of the fields, one obtains an exact hierarchy of flow equations for the vertex functions \cite{Berges:2000ew,Metzner12}.
For fermions, there are only vertex functions of even order in the fields.
In Fig.~\ref{fig:secIV:vertexexp} we show the Feynman diagrams describing the first three equations in the hierarchy. One of the internal lines corresponds to the single-scale propagator
\begin{equation}
 S_k = - G_k (\partial_k R_k) G_k =
 \partial_k G_k \big|_{\Sigma_k \, \mbox{const}} \, ,
\end{equation}
where $G_k = (G_{0,k}^{-1} - \Sigma_k)^{-1}$ is the full scale-dependent propagator.

The non-interacting propagator in a translation and spin-rotation invariant system has the spin-independent and momentum-diagonal form
\begin{equation}
 G_0({\bf p},i\omega_n) = \big[ i\omega_n - (\epsilon_{\bf p} - \mu) \big]^{-1} \, ,
\end{equation}
where $\epsilon_{\bf p}$ is the energy of single-particle states with momentum $\bf p$, and $\omega_n = (2n + 1) \pi T$ with integer $n$ is the fermionic Matsubara frequency.
At $T=0$ the propagator diverges for $\omega_n = 0$ and $\epsilon_{\bf p} = \mu$, that is, for momenta on the Fermi surface. Hence, in fermion systems with dimensionality $d > 1$ the propagator exhibits an extended singularity, not just a divergence at a point in momentum space.

\begin{figure}[tb]
\centerline{\includegraphics[width = 4.5cm]{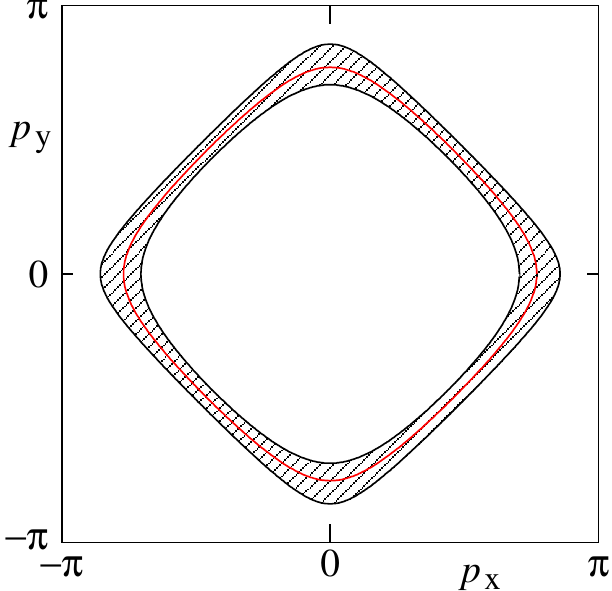}}
\caption{Momentum space region around the Fermi surface excluded by a sharp momentum cutoff
 for fermions with a tight-binding dispersion on a two-dimensional square lattice
 (lattice constant $a=1$).}
\label{fig:secIV:cutoff}
\end{figure}
There are various possibilities to regularize the Fermi surface singularity in the FRG flow. One is to choose a momentum dependent regulator that suppresses contributions with momenta close to the Fermi surface. An extreme choice is a sharp momentum cutoff, where contributions from momenta $\bf p$ with $|\epsilon_{\bf p} - \mu| < k$ are completely suppressed (see Fig.~\ref{fig:secIV:cutoff}). This corresponds to a regularized bare propagator of the form $G_{0,k} = \Theta(|\epsilon_{\bf p} - \mu| - k) G_0$. Of course one may also choose a smooth momentum cutoff. In any case, for fermion systems momentum cutoffs have two major drawbacks. First, the Fermi surface can actually be deformed by interactions, so that the regulator cannot be fixed a priori, if these deformations are taken into account. Second, the limit of vanishing momentum transfer ${\bf q} \to 0$ in interaction vertices does not commute with the limit $k \to 0$ for momentum cutoffs \cite{Metzner98}.
Both of these problems can be avoided by choosing a frequency cutoff \cite{Husemann09}, which is therefore the most popular choice nowadays. A frequency cutoff can also be used in systems without translation invariance \cite{Andergassen04}.
Alternatively one may use the interaction \cite{Honerkamp04} or temperature \cite{Honerkamp01b} as flow parameters. The interaction flow is computationally convenient, but it does not regularize the Fermi surface singularity.

%%%%%%%%%%%%%%%%%%%%%%%%%%%%%%%%%%%%%%%%%%%%%%%%%%%%%%%%%%%%%%%%%%%%%%%%%%%%%%%%%%

\subsubsection{Competing instabilities} \label{sec:IVB2}

At low temperatures, the normal metallic state of interacting fermion systems is usually unstable toward magnetism, superconductivity, or more exotic forms of order. For systems with weak or moderate interaction strengths, the FRG has turned out to be a valuable tool for the unbiased detection of instabilities, especially in cases where several instabilities of distinct nature compete with each other.

\begin{figure}[tb]
\centerline{\includegraphics[width = 10cm]{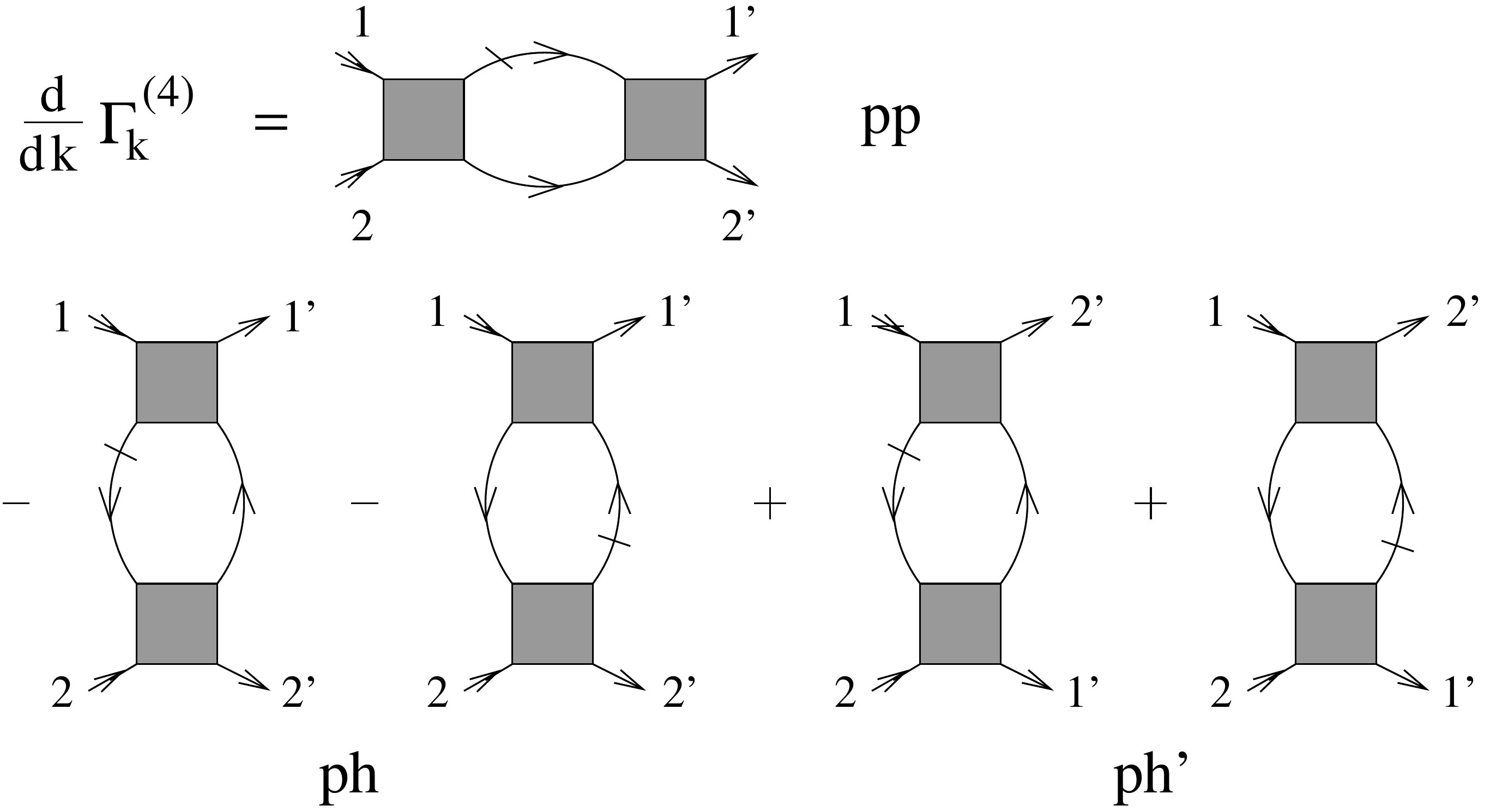}}
\caption{Contributions to the flow of the two-particle vertex with particle-particle (pp) and particle-hole channels (ph and ph') written explicitly, without the contribution from $\Gamma_k^{(6)}$.}
\label{fig:secIV:gamma4}
\end{figure}

Most instabilities are signalled by a divergence of the two-particle vertex $\Gamma_k^{(4)}$. For weak interactions, the flow of $\Gamma_k^{(4)}$ can be truncated at second order, that is, contributions from the self-energy and the three-particle vertex can be neglected such that only the first Feynman diagram in the second line of Fig.~\ref{fig:secIV:vertexexp} needs to be computed. Writing out the various terms arising from the matrix structure of $\Gamma_k^{(2)}[\psi,\psi^*]$ in Eq.~(\ref{eq:secIV:Gamma2}), one obtains three types of contributions to the flow of $\Gamma_k^{(4)}$, known as particle-particle, direct particle-hole, and crossed particle-hole channel, respectively (see Fig.~\ref{fig:secIV:gamma4}). 

The two-particle vertex is a function of four momenta and four Matsubara frequencies. The latter are constrained by energy conservation to three independent frequencies, and for translation invariant systems momentum conservation also reduces the number of independent momenta to three. The flow needs to be computed numerically, and a suitable parametrization of the two-particle vertex is a major issue. In weak coupling power counting, the frequency dependence and the momentum dependence perpendicular to the Fermi surface appear to be irrelevant \cite{Shankar94}. Hence, in most concrete evaluations of the vertex flow the frequency dependence of the vertex was neglected, and the momentum dependence was discretized by a partition of momentum space in patches, paying particular attention to an accurate resolution of the dependences along the Fermi surface. With $N$ patches one thus obtains ${\cal O}(N^3)$ ``running couplings'', where discrete symmetries can be exploited to reduce the precise number to a certain fraction of $N^3$.

The nature of the instabilities can be read off from the momentum and spin dependence of the diverging two-particle vertex. For example, pairing instabilities are signalled by a divergence in the Cooper channel, that is, for a vanishing total ingoing and outgoing momentum. To identify the instabilities it can also be helpful to analyze the corresponding susceptibilities, for which flow equations can be derived by adding external symmetry breaking fields to the bare action \cite{Metzner12}.

%%%%%%%%%%%%%%%%%%%%%%%%%%%%%%%%%%%%%%%%%%%%%%%%%%%%%%%%%%%%%%%%%%%%%%%%%%%%%%%%%%%%

{\em Two-dimensional Hubbard model:}
The FRG stability analysis was originally developed for the two-dimensional one-band Hubbard model, describing tight-binding fermions with a local interaction on a square lattice. The Hamiltonian is given by
\begin{equation}
 H = \sum_{{\bf j},{\bf j'}} \sum_\sigma t_{\bf jj'}
 c_{{\bf j'}\sigma}^\dagger c_{{\bf j}\sigma}^{\phantom\dagger} +
 U \sum_{\bf j} n_{{\bf j} \uparrow} n_{{\bf j}\downarrow}
\end{equation}
in standard second quantization notation, where $t_{\bf jj'}$ are hopping amplitudes between sites $\bf j$ and $\bf j'$ and $U$ is a local interaction. The two-dimensional Hubbard model was proposed by Anderson \cite{Anderson87} as an effective model for electrons moving in the copper-oxygen planes of cuprate high temperature superconductors. At half-filling, that is, for an average fermion density $n=1$, the ground state of the 2D Hubbard model is an antiferromagnetic insulator for sufficiently large $U$. The critical $U$ depends on the structure of the hopping matrix. For pure nearest neighbor hopping any positive $U$ leads to an antiferromagnetic ground state.
While the magnetic instability in the Hubbard model is revealed already by conventional mean-field theory, pairing is fluctuation-driven and hence more difficult to capture. Simple qualitative arguments suggesting $d$-wave pairing driven by magnetic fluctuations were corroborated by the fluctuation exchange approximation, that is, a resummation of a certain class of Feynman diagrams in perturbation theory \cite{Scalapino95}. However, only the unbiased stability analysis made possible by the FRG provided conclusive evidence for the existence of $d$-wave superconductivity with a sizable energy gap at moderate interaction strength, since the FRG treats all fluctuation channels on equal footing.
The first FRG results for the Hubbard model were actually obtained from three distinct FRG versions, namely Polchinski's flow equation \cite{Zanchi98,Zanchi00}, a Wick ordered flow equation \cite{Halboth00,Halboth00a}, and the Wetterich equation \cite{Honerkamp01}.

\begin{figure}[tb]
\begin{minipage}{7.5cm}
\centerline{\includegraphics[width = 6.5cm]{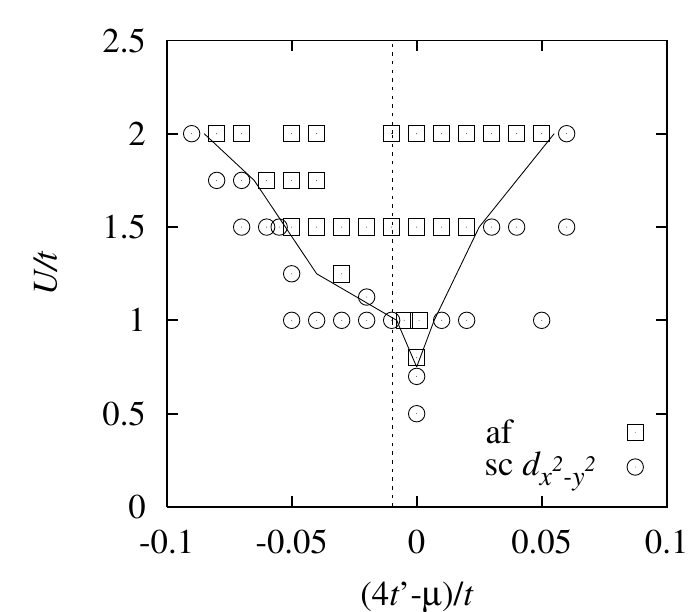}}
\end{minipage}
\begin{minipage}{7.5cm}
\centerline{\includegraphics[width = 6cm]{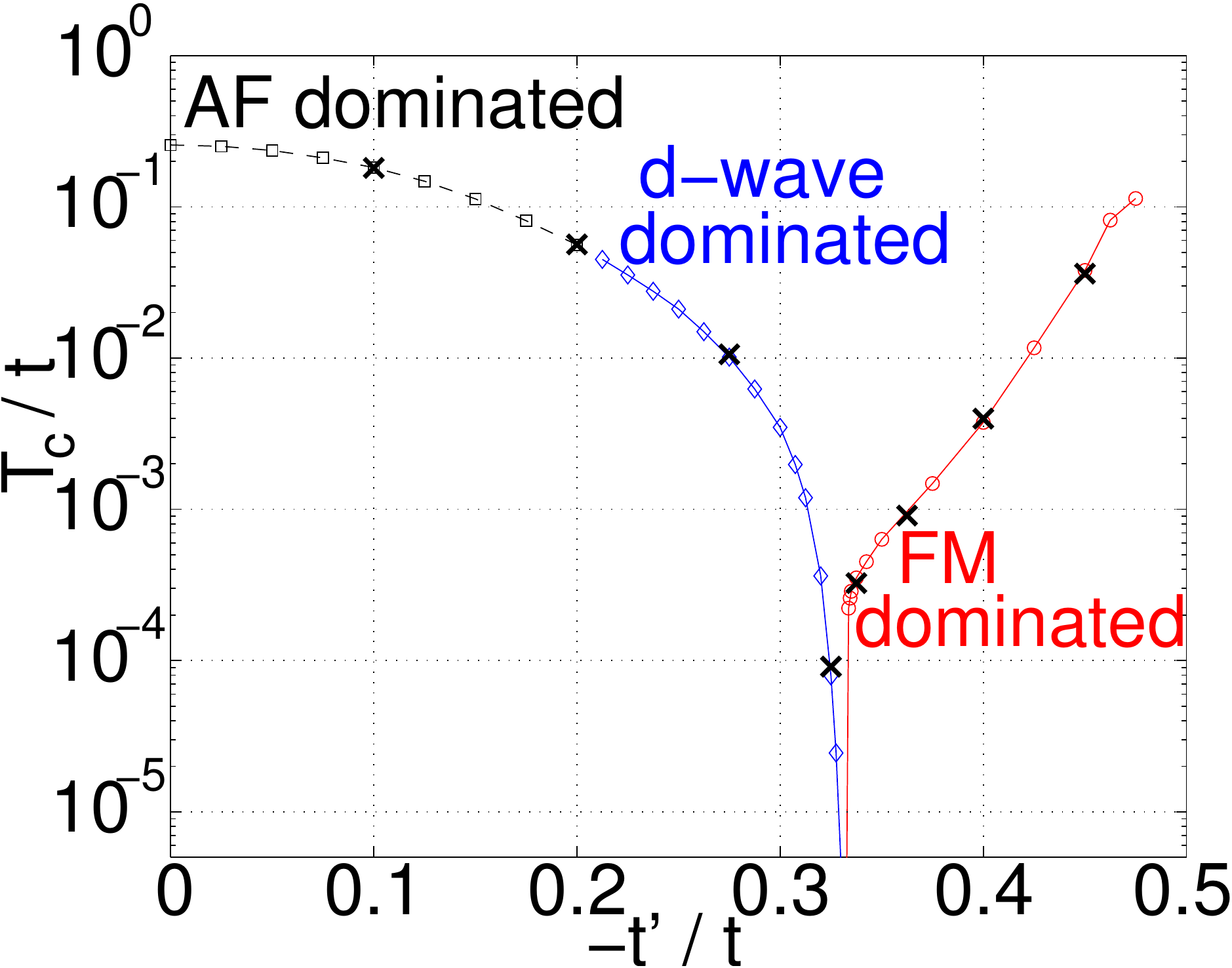}}
\end{minipage}
\caption{Left: Ground state phase diagram of the 2D Hubbard model near half-filling (marked
 by dashed line) at weak coupling ($U/t \leq 2$) for fixed $t' = -0.01t$ and variable
 $\mu$. The symbols indicate whether the dominant instability is magnetic (squares)
 or superconducting (circles); the solid line separates the magnetic from the pairing
 regime (from Ref.~\cite{Halboth00a}).
 Right: Pseudocritical temperature obtained from the temperature flow for the 2D
 Hubbard model at van Hove filling as a function of $t'/t$ for $U=3t$ (from Ref.~\cite{Honerkamp01a}).}
\label{fig:secIV:hubbard}
\end{figure}

In Fig.~\ref{fig:secIV:hubbard} (left panel) we show the ground state phase diagram of the 2D Hubbard model for a nearest neighbor hopping amplitude $t$ and a next-to-nearest neighbor hopping amplitude $t'=-0.01t$ in the weak coupling regime, as obtained from the FRG flow with a momentum cutoff \cite{Halboth00a}. Each symbol in the plot corresponds to a choice of $\mu$ and $U$ for which the dominant instability was identified from divergences of the two-particle vertex and susceptibilities at a critical cutoff $k_c$.
In Fig.~\ref{fig:secIV:hubbard} (right panel) we show the pseudocritical temperature, at which the two-particle vertex diverges in a temperature flow \cite{Honerkamp01a}. Here a fixed moderate coupling $U=3t$ was chosen, and the chemical potential was adjusted such that the Fermi surface touches the van Hove saddle points at $(\pi,0)$ and $(0,\pi)$ for each $t'/t$. The leading instability is antiferromagnetic for small $|t'/t|$, then $d$-wave pairing, and finally ferromagnetic for $t' < - t/3$. The temperature $T_c$ in this plot is not the true critical temperature, but rather the temperature scale at which strong magnetic or pairing correlations are formed. Order parameter fluctuations suppressing the actual transition temperature, especially in two dimensions, are not captured by the second order truncation of the flow. Remarkable is the drastic suppression of the pseudocritical scale at the boundary between the pairing and the ferromagnetic regime, which indicates a quantum critical point separating two symmetry-broken phases with distinct order parameters. Within Landau theory, such a behavior could be obtained only by tuning an additional parameter.
The self-energy at this quantum critical point obeys an unconventional powerlaw as a function of frequency with an exponent near $0.74$, for momenta on one of the van Hove points \cite{Giering12}. This implies that Landau quasi-particles are destroyed and Fermi liquid theory breaks down at this point.

The static approximation of the two-particle vertex turned out to be insufficient beyond the weak coupling regime. Improved parametrizations are based on a channel decomposition, where the fluctuation contributions to the vertex are decomposed in charge, magnetic, and pairing channels \cite{Karrasch08,Husemann09}. In the charge and magnetic channels dependences on the momentum and energy transfer variable are usually stronger than those on the remaining variables, while in the pairing channel the total momentum and energy dependences require the highest resolution. Using the channel decomposition it was shown that the frequency (= energy) dependence of the vertex is actually important already at moderate coupling \cite{Husemann12}. Neglecting it leads, in particular, to an overestimation of the energy scale for pairing in the 2D Hubbard model. Moreover, the frequency dependence is not even separable, that is, each term in the channel decomposition exhibits a substantial dependence on all three frequency variables already for moderate coupling strengths \cite{Vilardi17}. 

In view of the numerical difficulties posed already by the leading (second order) truncation of the vertex flow described above, it is clear that a complete inclusion of the three-particle or even higher order vertices is not feasible. However, an approximate evaluation of contributions from higher order vertices to the flow of the two-particle vertex was devised, and it was shown that the resulting multi-loop expansion sums up all parquet diagrams \cite{Kugler18,Kugler18a}. Employing an accurate and economic parametrization of momentum and frequency dependences \cite{Tagliavini19} it could be shown that the multi-loop expansion really (also ``in practice'') converges to the parquet approximation for the two-dimensional Hubbard model. The results agree very well with numerically exact Quantum Monte Carlo results for $U=2t$, and still decently for a moderate interaction $U=3t$ \cite{Hille20}.
To extend the application range of the FRG to the strong coupling regime, one needs to start the flow from a non-perturbative starting point, as described in Sec.~\ref{sec:IVB5}. 

%%%%%%%%%%%%%%%%%%%%%%%%%%%%%%%%%%%%%%%%%%%%%%%%%%%%%%%%%%%%%%%%%%%%%%%%%%%%%%%%%%

{\em Other models:} Competing instabilities and fluctuation induced superconductivity have been analyzed via FRG flows of the two-particle vertex for numerous other models. For multi-orbital systems, such as pnictide superconductors, the unitary transformation from orbital to band operators leads to momentum dependences already in the bare vertex, even for local (Hubbard-type) interactions \cite{Graser09,Maier09}. FRG flows have provided valuable information on instabilities of various compounds, especially of those where electron-electron interactions are weak or moderate. It is beyond the scope of this review to list all these applications here. Some of the earlier applications have been described in Ref.~\cite{Metzner12}.  A more extensive review of work on multi-orbital systems performed before 2013, in particular on pnictides and compounds with hexagonal lattices (triangular, honeycomb, and Kagome), has been published by Platt {\em et al.} \cite{Platt13}.
Since then the range of applications has significantly broadened, often following experimental discoveries, including systems with spin-orbit coupling \cite{Schober16}, new Dirac nodal-line materials \cite{Scherer18} and, most recently, moir\'e heterostructures \cite{Classen19}, to name just a few.

%%%%%%%%%%%%%%%%%%%%%%%%%%%%%%%%%%%%%%%%%%%%%%%%%%%%%%%%%%%%%%%%%%%%%%%%%%%%%%%%%%

\subsubsection{Spontaneous symmetry breaking} \label{sec:IVB3}

In most interacting fermion systems a symmetry of the Hamiltonian is spontaneously broken at sufficiently low temperatures, or at least in the ground state. In the FRG flow, common types of symmetry breaking such as magnetic order or superconductivity are associated with a divergence of the two-particle vertex at a finite scale $k_c > 0$. One may encounter a divergence even in cases where ultimately only quasi-long range order or two-particle bound states are formed.
To continue the flow beyond this critical scale, an order parameter needs to be implemented. There are two distinct ways of doing this.

\begin{figure} [tb]
\centerline{\includegraphics[width = 9.5cm]{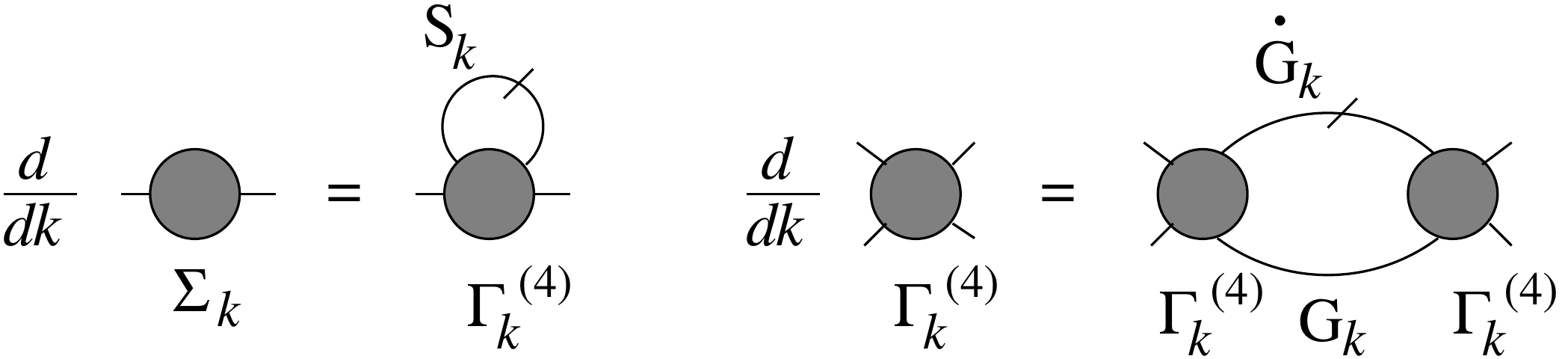}}
\caption{Katanin truncation for the coupled flow of self-energy and
 two-particle vertex.}
\label{fig:secIV:katanin}
\end{figure}
{\em Fermionic flows:}
In one approach the FRG flow is computed in the presence of a small (ideally infinitesimal) symmetry breaking field added to the bare action, which develops into a finite order parameter below the scale $k_c$ \cite{Salmhofer04}. The external field regularizes the divergence of the vertex at $k_c$. This field can be endowed with a scale-dependence, so that it can be removed completely at the end of the flow \cite{Eberlein13a}.
One then needs to apply a manageable truncation of the exact flow equations which does not break down at the critical scale where the two-particle vertex becomes large even for a weak bare interaction. A minimal requirement is that a truncation should at least provide a decent solution for mean-field models, such as the reduced BCS model. The simplest approximation satisfying this requirement is the Katanin truncation \cite{Katanin04} shown in Fig.~\ref{fig:secIV:katanin}. Here the single-scale propagator $S_k$ in the leading contribution to the flow of the two-particle vertex (cf.\ Fig.~\ref{fig:secIV:vertexexp}) has been replaced by $\partial_k G_k$. This takes self-energy corrections generated by the three-particle vertex into account, while other contributions from the three-particle vertex are neglected. Spontaneous symmetry breaking leads to anomalous contributions to both the self-energy and the two-particle vertex.
The flow obtained from Katanin's truncation solves the reduced BCS model and similar mean-field models for other types of order exactly \cite{Salmhofer04,Gersch05}. In spite of the diverging two-particle interaction the neglected contributions do not contribute because of a zero-measure momentum integration region.

The Katanin truncation has been used as an approximation to compute ground state order parameters for the two-dimensional Hubbard model with attractive and repulsive interactions. For the attractive Hubbard model, the $s$-wave pairing gap obtained from the FRG flow \cite{Gersch08,Eberlein13a} is substantially suppressed compared to mean-field theory even at weak coupling, in agreement with results from other methods.
For the repulsive Hubbard model, the $d$-wave pairing gap was computed for weak and moderate interactions, and it was shown that a sizable next-to-nearest neighbor hopping amplitude is beneficial for $d$-wave superconductivity \cite{Eberlein14fb}.
The Ward identity protecting gapless Goldstone modes in phases with a broken continuous symmetry is not automatically satisfied by the Katanin truncation. However, one may enforce it by a suitable projection procedure \cite{Eberlein13a}.

{\em Flows with order-parameter fields:}
Spontaneous symmetry breaking in interacting fermion systems can also be treated by introducing collective bosonic order parameter fields via a Hubbard-Stratonovich decoupling of the two-fermion interaction \cite{Popov87}, and applying the FRG to the resulting coupled boson-fermion action. This route to symmetry breaking was first used to treat the formation of an antiferromagnetic state in the repulsive 2D Hubbard model at half-filling \cite{Baier04}. The correct low-temperature behavior (usually described by a nonlinear sigma model) was recovered from from a simple truncation of the exact flow equation. Spontaneous symmetry breaking sets in at a critical scale $k_c$, where the boson mass vanishes, but at finite temperatures the symmetry is gradually restored by bosonic fluctuations at lower scales, in agreement with the Mermin-Wagner theorem.

Subsequently, the coupled flow of fermions and bosonic order parameter fluctuations was applied to the superfluid phase of fermion systems with attractive interactions, using $U(1)$-symmetric truncations of the effective fermion-boson action. A simple truncation, where the bosonic part of the effective action has the form of a Bose gas with a local $\phi^4$ interaction, yields already sensible results for the pairing gap and the transition temperature in three dimensions in the entire range from BCS superfluidity at weak coupling to Bose-Einstein condensation of tightly bound pairs at strong coupling \cite{Birse05fb,Diehl07fb,Diehl:2007ri,Krippa07}.
To include effects from particle-hole fluctuations, which tend to reduce the pairing gap, one has to take effective two-fermion interactions into account. Decoupling these at each step in the flow by a scale-dependent ``dynamical bosonisation'' \cite{Gies:2001nw,Gies:2002hq}, one obtains a substantial quantitative improvement \cite{Floerchinger:2008qc}.
Instead of using a U(1)-symmetric ansatz for the effective action, it is also possible to start from the hierarchy of flow equations for the vertex functions and implement the U(1) symmetry via Ward identities \cite{Bartosch09fb}.

To distinguish the Goldstone mode from longitudinal order parameter fluctuations in a $U(1)$-symmetric ansatz, one needs to include quartic gradient terms in the bosonic part of the effective action \cite{Tetradis94}. A relatively simple truncation of the complete effective action (for fermions and bosons) yields the correct low-energy behavior of the superfluid ground state in dimensions $d \geq 2$ \cite{Obert13}.

The interplay of antiferromagnetism and superconductivity in the two-dimensional repulsive Hubbard model was also analyzed by FRG flows with order-parameter fields.
It was clarified how $d$-wave pairing mediated by antiferromagnetic fluctuations emerges from the coupled fermion-boson flow \cite{Krahl09a}, and a phase diagram with magnetic and superconducting order was computed at weak coupling \cite{Friederich10,Friederich11}.

{\em Renormalized mean-field theory:}
In case of competing instabilities and possible coexistence of two or more order parameters, both routes to symmetry breaking described above become quite involved. For a computation of order parameters in the ground state one may neglect low-energy fluctuations and combine flow equations at high scales with a mean-field approximation at low scales \cite{Reiss07,Wang14}. In this approach the flow of the two-particle vertex $\Gamma_k^{(4)}$ is stopped at a scale $k_{\rm mf}$ slightly above the critical scale $k_c$, that is, before entering the symmetry-broken regime. The remaining low-energy degrees of freedom are treated in mean-field approximation, with a reduced effective interaction extracted from  $\Gamma_{k_{\rm mf}}^{(4)}$.
An application of this renormalized mean-field theory revealed broad doping regions with coexistence of $d$-wave superconductivity and N\'eel or incommensurate antiferromagnetic order in the ground state of the 2D Hubbard model \cite{Reiss07,Wang14,Yamase16}. 

{\em FRG in the two-particle irreducible (2PI) formalism:}
Although the 2PI formalism in the framework of the FRG has been considered in
various contexts~\citep{Nagy11,Polonyi05,Pawlowski:2005xe,Blaizot11b}, few works
have focused on interacting fermions. In the 2PI 
approach~\cite{Luttinger60b,Baym61,Baym62,Dedominicis64a,Dedominicis64b,Cornwall74}, 
the scale-dependent effective action $\Gamma_k[G]$ is a functional of 
the one-particle propagator and the Wetterich equation becomes a flow 
equation for the scale-dependent Luttinger-Ward functional 
$\Phi_k[G]$~\cite{Wetterich:2002ky,Dupuis:2005ij,Dupuis14,Rentrop15}. The 
``classical'' variable is thus bosonic in nature and is itself an order 
parameter. There are no conceptual difficulties to describe phases with 
spontaneously broken symmetries and one can even start the RG flow in a 
broken-symmetry phase (with the Hartree-Fock-RPA theory as initial 
condition). The 2PI formalism has been proposed as a possible approach 
to the Hubbard model~\cite{Dupuis14}, in particular in combination with 
dynamical mean-field theory~\cite{Katanin19} (see Sec.~\ref{sec:IVB5}). 
It has also been applied to quantum impurity models~\cite{Rentrop16}.
Finally, the 2PI FRG has been used in connection with density functional
theory to study many-body systems in condensed or nuclear matter~\cite{Polonyi02fb,Kemler13,Yokota19,Yokota19a,Yokota19b}.

%%%%%%%%%%%%%%%%%%%%%%%%%%%%%%%%%%%%%%%%%%%%%%%%%%%%%%%%%%%%%%%%%%%%%%%%%%%%%%%%%%

\subsubsection{Quantum transport} \label{sec:IVB4}

Electronic transport through nanostructured devices such as quantum wires and quantum dots is an important area in modern condensed matter physics. The typical setup is a mesoscopic region coupled to two or more leads. While the electrons in the leads move more or less independently, interesting correlation effects due to the restricted geometry can occur in the mesoscopic region \cite{Hanson07}.

In general, the calculation of transport properties in interacting systems requires the calculation of two-particle quantities such as current-current correlation functions \cite{Mahan_book}. Under certain assumptions, the electronic transport through a mesoscopic region can be viewed as a scattering problem, and the electric conductance can be written in the Landauer-B\"uttiker form \cite{Landauer57,Buettiker86}
\begin{equation} \label{eq:secIV:conductance}
 C = - \frac{e^2}{2\pi} \sum_\zeta
 \int d\epsilon \, {\cal T}_\zeta(\epsilon) f'(\epsilon) \, ,
\end{equation}
where $\zeta$ labels the scattering channels, ${\cal T}_\zeta(\epsilon)$ is the transmission probability for each channel, and $f(\epsilon)$ is the Fermi function. Note that we use natural units where $\hbar=1$ such that the conductance quantum is $e^2/(2\pi)$.
For a frequency-independent self-energy, Eq.~(\ref{eq:secIV:conductance}) is actually exact, and the transmission probability can be expressed by the one-particle propagator at the contacts connecting the interacting region to the leads \cite{Oguri01}.

An important quantum transport problem involving correlated electrons is posed by electrons moving through a one-dimensional metallic wire with one or a few impurities \cite{Kane92,Matveev93}.
Interacting electrons in one-dimensional metals form a Luttinger liquid \cite{Giamarchi_book}, which, unlike Fermi liquids, has no resemblance to a Fermi gas.
While perturbation theory diverges at any interaction strength for these systems, a strikingly simple truncation of the FRG flow equations captures the most important effects related to non-magnetic impurities in one-dimensional metals \cite{Meden02,Andergassen04}.
Indeed, already the first equation in the flow equation hierarchy (see Fig.~\ref{fig:secIV:vertexexp}), with the bare instead of the effective two-particle vertex, yields several power-laws describing the low-energy behavior, where the exponents are correct to leading order in the interaction strength.
Since the self-energy is frequency independent in this first order truncation, the conductance can be computed without further approximations from the one-particle propagator \cite{Enss05}. The conductance through a wire with a single impurity vanishes with a power-law as a function of temperature as expected \cite{Kane92}.
In case of two impurities a rich behavior involving resonant tunneling and crossovers between distinct power-laws could be obtained \cite{Enss05,Meden05}.

The same FRG approach turned out to be equally fruitful in a computation of persistent currents in a mesoscopic Luttinger liquid ring \cite{Meden03,Meden03a}, in a study of quantum transport through a Y-junction of three Luttinger liquid wires \cite{Barnabe05}, and to describe transport
through a correlated quantum dot setup dominated by charge fluctuations (absence of the Kondo effect) \cite{Karrasch10}.

The FRG was also extended to steady-state nonequilibrium transport through quantum wires and quantum dots in the presence of a finite voltage \cite{Gezzi07,Jakobs07,Karrasch10}.
This extension makes use of real-time or real-frequency Green functions on the Keldysh contour \cite{Rammer86}. Compared to the equilibrium imaginary frequency FRG, in the steady state the only technical complication is an additional Keldysh index $\pm$.
An important issue is to define a flow that does not violate causality and Kubo-Martin-Schwinger (KMS) relations after truncation of the flow equation hierarchy \cite{Jakobs10}.
Suitable choices are a judiciously designed imaginary frequency cutoff \cite{Jakobs07,Jakobs10} or scale dependent local auxiliary terms, which are removed at the end of the flow \cite{Jakobs10,Jakobs10a}.
The non-equilibrium FRG was applied to simple models of quantum dots \cite{Karrasch10} and quantum wires \cite{Jakobs07}.
It is even possible to compute time-dependent non-equilibrium transport of setups with correlated mesoscopic regions using Keldysh-FRG. This approach was used to study the quench dynamics of quantum dots \cite{Kennes12} and quantum wires \cite{Kennes13}, as well as the dynamics of periodically driven systems \cite{Eissing16}.
The FRG on the Keldysh contour can also be used to compute real frequency quantities in equilibrium, avoiding thus the numerically unstable analytic continuation from imaginary to real frequencies \cite{Jakobs10a,Khedri18}.

%%%%%%%%%%%%%%%%%%%%%%%%%%%%%%%%%%%%%%%%%%%%%%%%%%%%%%%%%%%%%%%%%%%%%%%%%%%%%%%%%%

\subsubsection{Leap to strong coupling} \label{sec:IVB5}

A truncation of the FRG hierarchy of flow equations at some finite order can be justified only for weak interactions \cite{Salmhofer01}, with the exception of mean-field models where phase space restrictions suppress higher order contributions \cite{Salmhofer04}.
Although bare interactions are usually two-particle interactions, effective $m$-particle interactions with $m>2$ are generated by the flow and affect the effective two-particle interaction and the self-energy, see Fig.~\ref{fig:secIV:vertexexp}.
For strong bare interactions, these contributions become important already at high energy scales far above the critical scales for instabilities.
One of the most important strong coupling phenomena in correlated lattice electron systems like the Hubbard model is the Mott metal-insulator transition. Cuprate high-temperature superconductors are doped Mott insulators \cite{Anderson87}. Truncated FRG flow equations starting from the bare action do not capture the Mott transition.

For interacting fermion systems there are no non-perturbative approximations such as the derivative expansion of the exact flow equation for bosons. However, one may choose a suitable non-perturbative approximation as a starting point for the flow. The Mott transition is essentially a consequence of strong {\em local} correlations. As such, it is well described by the dynamical mean-field theory (DMFT), which treats local correlations non-perturbatively \cite{Metzner89,Georges92,Georges96}.
The DMFT and the FRG can be consistently merged in the socalled DMF$^2$RG \cite{Taranto14,Wentzell15}. In this approach the FRG flow does not start from the bare action of the system, but rather from the DMFT solution. Strong local correlations and the Mott physics are thus captured via the DMFT starting point, while the nonlocal correlations are generated by the FRG flow. While an obvious small expansion parameter is still lacking, the weaker nonlocal correlations may be captured by a manageable truncation of the exact FRG hierarchy. The flow can be defined such that it is generated exclusively by non-local correlations \cite{Vilardi19}.

Concrete DMF$^2$RG calculations have so far been performed with a truncation of the flow equations at the two-particle level \cite{Taranto14,Vilardi19}, involving thus the two-particle vertex and the self-energy. For strong interactions, the two-particle vertex exhibits strong frequency dependencies which cannot be reduced to one frequency per interaction channel. Nevertheless, numerical solutions of the DMF$^2$RG flow equations have been obtained for the two-dimensional Hubbard model in the strong coupling regime, and evidence for $d$-wave pairing at the expected (for cuprates) temperature scale has been found \cite{Vilardi19}.

Recently, a combination of the extended DMFT for non-local interactions with the 2PI version of the FRG has been proposed \cite{Katanin19}. Encouraging first benchmarks were obtained for the conventional and the extended Hubbard model at half-filling.

%%%%%%%%%%%%%%%%%%%%%%%%%%%%%%%%%%%%%%%%%%%%%%%%%%%%%%%%%%%%%%%%%%%%%%%%%%%%%%%%%%%%%%%%%

\subsubsection{Omissions} \label{sec:IVB6}

The range of applications of the FRG has become too broad to be covered in a single review. Here we briefly mention two important devopments related to fermion systems which we could not review in detail.

{\em Spin systems:}
Using an auxiliary fermion representation of spin degrees of freedom, the FRG can also be applied to quantum spin systems \cite{Reuther10,Reuther11}. The pseudofermions carry no kinetic energy, and the interactions depend only on one space or momentum variable. Using the Katanin truncation \cite{Katanin04} of the flow equations and keeping the full frequency dependence of the self-energy and interaction vertex leads to an approximation scheme that works remarkably well for a variety of spin systems. This success of a perturbative truncatation for a strongly coupled system can be partially understood since leading orders in $1/N$ and $1/S$ expansions are captured by the approximation. The accuracy can be systematically improved by using decoupled clusters instead of single pseudo-fermions as starting point for an expansion \cite{Reuther14}.
Alternative ways of treating quantum spin systems with the FRG have been developed by mapping spins on hard core bosons \cite{Rancon14} or working directly with spin operators \cite{Krieg19}.

{\em Quantum criticality:}
Numerous quantum many-body systems undergo a continuous transition between ground states with distinct symmetry or topology upon tuning a non-thermal control parameter. In the vicinity of the quantum critical point separating the two phases, quantum fluctuations are particularly important and lead to unconventional properties not only in the ground states, but also at low finite temperatures \cite{Sachdev99}. In metallic systems quantum critical fluctuations prevent the existence of Landau quasi-particles. Fermi liquid theory thus breaks down and unconventional, so-called non-Fermi liquid behavior is observed \cite{Loehneysen07}. 
A purely bosonic order parameter theory for quantum critical metals as developed by Hertz \cite{Hertz76} and Millis \cite{Millis93} is problematic since gapless fermionic excitations lead to singular interactions between the order parameter fluctuations. This problem is particularly serious in low-dimensional systems. Hence, for a complete and controlled treatment a coupled theory treating fermions (with a Fermi surface) and critical order parameter fluctuations on equal footing is required.

The FRG is an ideal tool to study quantum critical systems, and quantum critical fermion systems in particular. Some of the FRG work on quantum critical metals and semi-metals done before 2012 has been reviewed in Ref.~\cite{Metzner12}. Hertz-Millis theory can be formulated as a simple truncation of the bosonic FRG with few running couplings \cite{Metzner12}.
Some studies of metallic quantum critical points have been performed by using perturbative truncations for the coupled flow of fermions and order parameter fluctuations. This procedure helped clarifying, in particular, the intriguing case of quantum criticality at the onset of antiferromagnetic order in two dimensions \cite{Maier16}.
Since order parameter fluctuations are described by bosonic fields, their action is amenable to non-perturbative truncations. Non-perturbative truncations are required, in particular, to establish fluctuation-induced continuous quantum phase transitions in certain systems which exhibit only first order transitions on the mean-field level. Illustrative examples are the nematic phase transition in two-dimensional metals \cite{Jakubczyk09}, and gapless Dirac fermions coupled to a $Z_3$ symmetric order parameter within a two-dimensional Gross-Neveu model \cite{Classen17}. More references, especially on relativistic critical fermion systems, are provided at the end of Sec.~\ref{criticalphen}.

%% file: SEC_HighEnergy/frg_HighEnergy.tex
%%%%%%%%%%%%%%%%%%%% commands Jan %%%%%%%%%%%%%%%%%

\newcommand{\imag}{\mathrm{i}}
\def\dr{{D\!\llap{/}}\,}
\def\Dr{{D\!\llap{/}}\,} 

%\newpage

\section{High-energy physics}
\label{sec_hep}
\subsection{Introduction: High-energy physics}\label{sec_hep-intro}

High energy physics is very successfully described by the Standard Model (SM) of particle physics. Its different parts, QCD and the electroweak gauge theory with the Higgs sector give rise to many fascinating phenomena that either quantitatively or even qualitatively can only be described with nonperturbative methods. Most notably, together with quantum gravity discussed in Sec.~\ref{sec_gr}, high energy physics is described with Abelian and non-Abelian gauge theories whose properties are responsible for quite some of its nonperturbative physics. 

A paradigmatic example is QCD, and most of the conceptual investigations within the FRG as well as the physics results have been achieved here. Accordingly, we mostly concentrate on QCD in the following. At large momentum scales, QCD is asymptotically free, and the running strong coupling $\alpha_s(p)=g_s^2(p)/(4 \pi)$ decays logarithmically with the momentum scale $p^2$. The converse of this behaviour is an increasing strong coupling in the low energy (infrared) regime of QCD. In this regime, QCD also exhibits spontaneous chiral symmetry breaking and confinement as well as the formation of a rich bound-state spectrum. While an  increasing coupling does not necessarily entail the breakdown of perturbation theory and is in particular not tantamount to confinement, a full understanding of the latter as well as of spontaneous chiral symmetry breaking certainly requires nonperturbative techniques. Moreover, while high-energy QCD is well described with perturbation theory for weakly-interacting quarks and gluons due to asymptotic freedom, low-energy QCD is well-described by low-energy effective theories with weakly-interacting hadrons. This dynamical change of the relevant degrees of freedoms as well as the access to the emerging hadron spectrum are further challenges to be met. 

For the physics of the early universe as well as that of heavy-ion collisions ('little big bangs') we also require an understanding of the rich phase structure of QCD at finite temperature $T$ and density $n$ or quark chemical potential $\mu_q$: At large temperatures the theory is deconfined and chirally symmetric (except for the  current quark masses induced by the Higgs mechanism). Most of its properties are well-described by thermal perturbation theory of quarks and gluons.\footnote{Note that high-temperature QCD still exhibits spatial confinement, so not all its properties are captured by perturbation theory.} At low temperatures we enter the hadronic low-energy regime with chiral symmetry breaking and confinement. At vanishing density and physical quark masses the chiral transition is a crossover, while at larger density the chiral crossover may turn into a 1st-order phase transition at a 2nd-order critical end point (CEP). Whether or not this CEP exists, as well as its location and further QCD properties at large densities such as possible inhomogeneous phases and so-called condensed matter phases with color superconductivity are pressing open questions. Their resolution gives e.g.~access to the QCD equation of state at large densities which is required for the physics of neutron stars. 

Finally, the physics of a heavy ion collision in its early stages as well as the QCD phase transition in the early universe are genuinely non-equilibrium processes. Their understanding necessitates the computation of real-time (in and out-of equilibrium) correlation functions. These correlation functions are relevant, e.g., for transport and hydrodynamic phases in the above-mentioned non-equilibrium evolutions. Moreover, time-like correlation functions are already required for unravelling the rich hadronic structure of QCD. More details can be found in Sec.~\ref{sec_hep-QCD}. 

The treatment of gauge theories within the FRG as required for high energy physics and quantum gravity is discussed in Sec.~\ref{sec:FRGGauge}. Readers, who are either already familiar with the setup or are more interested in the physics of QCD and high energy physics in general may skip this part for a first reading and jump to Sec.~\ref{sec_hep-QCD} and beyond.

\subsection{The functional renormalization group for gauge theories}\label{sec:FRGGauge}

In this section we briefly discuss some important features of the FRG-formulation of gauge theories at the example of an SU(N) gauge theory with the classical Euclidean Yang-Mills action, 
\begin{align}
\label{sec_he:SYM}
	S_\text{YM}[A]
	= \frac{1}{2} \int_x \mathrm{tr}\ F_{\mu\nu}F_{\mu\nu} 
	= \frac{1}{4} \int_x F^a_{\mu\nu}F^a_{\mu\nu}\,,\qquad \textrm{with} \qquad F_{\mu\nu}= F^a_{\mu\nu} t^a\,, 
\end{align}
with $\int_x =\int \textrm{d}^4 x$. The fieldstrength tensor in \eqref{sec_he:SYM} is the curvature tensor $F_{\mu\nu}= \imag/g_s [D_\mu\,,\,D_\nu]$, where the covariant derivative  and the generators $t^a$ of the group are given by 
\begin{align}
\label{eq:Cov+Group} 
D_\mu=\partial_\mu -\imag g_s A_\mu\,,\qquad \textrm{with}\quad A_\mu=A_\mu^a t^a\,, \qquad \textrm{and} \qquad [t^a, t^b] = \imag f^{abc} t^c\,,\quad \mathrm{tr}\, t^a t^b= \frac12  \delta^{ab}\, .
\end{align}
The traces in \eqref{sec_he:SYM}, \eqref{eq:Cov+Group} are taken in the fundamental representation. The non-Abelian gauge field in \eqref{eq:Cov+Group} has component fields $A_\mu^a$ with $a=1,...,\text{N}^2-1$, and lives in the Lie algebra $\text{su(N)}$ of the gauge group $\text{SU(N)}$. The generators $t^a$ are normalized according to \eqref{eq:Cov+Group} and satisfy the $\text{su(N)}$-Lie algebra in  \eqref{eq:Cov+Group}. The components of the fieldstrength tensor $F_{\mu\nu}$ used in \eqref{sec_he:SYM} are given by 
\begin{align}
\label{sec_he:Fmunu}
  F_{\mu\nu}^a = \partial_\mu A^a_\nu - \partial_\nu A^a_\mu
  + g_s f^{abc} A_\mu^b A_\nu^c 
\, . 
\end{align}
The gauge theories introduced above encompass the Standard-Model with the gauge group $U(1)\times SU(2)\times SU(3)$.  There, the coupling to matter is given by Dirac actions for the fermionic matter fields, leptons and quarks, in the fundamental representation. For the present example  we restrict ourselves to QCD with the coupling of the gluon $A_\mu\in su(3)$ to the quarks $(q_i)^A_\xi$. Here $A=1,2,3$ is the color index in the fundamental representation and $i=1,...,N_f$ counts the quark flavors. The quarks are Dirac fermions with the Dirac index $\xi=1,...,4$. The Dirac term in Euclidean spacetime is given by 
\begin{align}\label{sec_he:SDirac}
S_{\textrm{Dirac}}[q,\bar q,A]=\int_x \bar q\left( i \,\slash{\hspace{-2.2mm} D}+m_q+\gamma_0\,\mu_q\right) q \,,\qquad \textrm{with} \qquad \slash{\hspace{-2.2mm} D}=\gamma_\mu D_\mu\,,\quad \textrm{and}\quad  \{\gamma_\mu\,,\,\gamma_\nu\}=2 \delta_{\mu\nu}\,, 
\end{align}
In \eqref{sec_he:SDirac} we have suppressed gauge group, Dirac and flavor indices, and introduced a mass matrix $m_q$ in flavor space as well as a quark chemical potential. The Yang-Mills action \eqref{sec_he:SYM} and the matter term in \eqref{sec_he:SDirac} are invariant under gauge transformations with $S_\text{YM}[A^\mathcal{U} ]+S_{\textrm{Dirac}}[q,\bar q,A^\mathcal{U} ]=S_\text{YM}[A]+S_{\textrm{Dirac}}[q,\bar q,A]$. The gauge group element is given by $\mathcal{U}(x) = e^{\imag\, \omega(x)}$ with $\omega\in\text{su(N)}$, and the gauge transformation reads 
\begin{align}
\label{sec_he:gauge_trafo_A}
  A_\mu \rightarrow A_\mu^\mathcal{U} 
  = \frac{\imag}{g}  \mathcal{U} \left( D_\mu \mathcal{U}^\dagger\right) =
  \mathcal{U} A_\mu \mathcal{U}^\dagger +
  \frac{\imag}{g} \mathcal{U} (\partial_\mu \mathcal{U}^\dagger)
  \, ,\qquad \qquad  q\to \mathcal{U}q,\quad \bar q \to \bar q \mathcal{U}^\dagger\,.
\end{align}
Naively, we would simply adopt the general formulation of the FRG discussed in Sec.~\ref{sec_frg} for bosons and in Sec.~\ref{sec:IVB} for fermions, and substitute the scalar field with the gauge field. However, the standard formulation of the functional renormalization group rests on a momentum cutoff with a cutoff term that is quadratic in the field, see \eqref{sec_frg:DeltaS}. Both properties are tightly related to the practical accessibility of the flow equation for the scale dependent effective action, \eqref{sec_frg:eqwet}, and giving them up comes at a high price. For this reason 
the commonly-used FRG-approach to gauge theories utilises a quadratic momentum cutoff term in the gauge field, which breaks gauge invariance for finite cutoff scales $k\neq 0$. Moreover, the right hand side of \eqref{sec_frg:eqwet} depends on the full propagator of the theory, and in particular its gauge field component 
\begin{align} \label{sec_he:prop} G_{A A}(x_1,x_2) =
  \left(\frac{1}{\Gamma^{(2)}+R_k}\right)_{AA}(x_1,x_2)\,. 
\end{align}
Accordingly, the flow equation requires the existence of the inverse of the full kinetic operator of the theory in the presence of the infrared regulator, $ \Gamma^{(2)}+R_k$.

\subsubsection{Gauge-fixed flows and modified Slavnov-Taylor identities}

The inverse in \eqref{sec_he:prop} is only defined with a gauge fixing due to gauge redundancies carried by the gauge field. Hence the standard FRG-formulation of gauge theories requires gauge-fixing. The most common choice in nonperturbative functional approaches (FRG, Dyson-Schwinger equations, $n$-particle irreducible approaches) is the Landau gauge, in which the covariant gauge is strictly implemented: The respective gauge-fixing term in the presence of a background field reads 
\begin{align}\label{sec_he:eq:LandauGauge} 
  S_\textrm{gf}[\bar A, a]= \frac{1}{2 \xi}\int_x (\bar D_\mu a_\mu)^2\,,
  \quad \bar D_\mu
  =\partial_\mu - \imag\,g_s \bar A_\mu\,, \qquad \textrm{where}  \qquad A_\mu=\bar A_\mu + a_\mu\,.
\end{align}
In \eqref{sec_he:eq:LandauGauge} we have introduced the split of the full gauge field into the sum of an auxiliary background field $\bar A_\mu$ and a dynamical fluctuation field $a_\mu$. The strict implementation of the gauge \eqref{sec_he:eq:LandauGauge} is achieved for $\xi=0$. Then, only configurations with $\bar D_\mu a_\mu=0$ survive in the path integral. With a non-vanishing background, \eqref{sec_he:eq:LandauGauge} with $\xi=0$ is also called the Landau-DeWitt gauge.\footnote{The gauge \eqref{sec_he:eq:LandauGauge} and many other sufficiently smooth gauges are subject to the Gribov problem: the solution of \eqref{sec_he:eq:LandauGauge} is not unique, there are Gribov copies, and their treatment is relevant for the infrared regime of correlation functions. The discussion of this highly interesting issue goes far beyond the scope of the present work, for reviews and recent work see e.g.~\cite{Maas:2011se, Vandersickel:2012tz, Capri:2017bfd}.}

In non-Abelian gauge theories or non-linear gauges in Abelian theories the transformation from an unrestricted integration over the gauge field to one over gauge-fixed configurations has a gauge-field dependent Jacobi-determinant  $\Delta_{\cal F}[\bar A,a]$ (Faddeev-Popov determinant) for a given gauge-fixing condition ${\cal F}[A]=0$. For Lorentz gauges we have ${\cal F}[A]=\partial_\mu A_\mu$. Typically, it is taken into account as a Grassmann integral over auxiliary fermionic fields, the Faddeev-Popov ghosts. For the gauge fixing \eqref{sec_he:eq:LandauGauge} the ghost action reads 
\begin{align} \label{sec_he:eq:Sghost}
S_\textrm{gh}[\bar A, a, c,\bar c]= -\int_x \bar c^a (\bar D_\mu D_\mu)^{ab} c^b\,, 
\end{align}
with a negative quadratic dispersion for the ghost field. The appeal of the background-field approach is the existence of a gauge-invariant effective action, the background field effective action $\Gamma[A] = \Gamma[\bar A=A,a=0]$ that can be defined at vanishing fluctuation gauge field $a=0$. Here, for the sake of simplicity we have suppressed the dependence of $\Gamma$ on all the other fields $c,\bar c, q, \bar q,...$. Note that the background field effective action $\Gamma[A]$ inherits gauge invariance from gauge transformations of the auxiliary background field $\bar A_\mu$. The equivalence of this auxiliary  symmetry with physical gauge invariance of the theory follows from (onshell) background-independence of the approach and the Slavnov-Taylor identities (STIs). The latter encode the gauge-transformation properties or BRST-symmetries of the dynamical fluctuation field. Background independence, typically encoded in Nielsen identities, and the STIs can also be used to show, that the background field correlation functions are related to S-matrix elements similarly to the fluctuation correlation functions, see e.g.~\ \cite{Abbott:1983zw}. 

The background effective action $\Gamma[A] $ can be expanded in gauge-invariant operators leading to easily accessible approximation schemes. Consequently, many of the early applications have been performed in the background field approach, ranging from computations of the (cutoff-) running of the QCD and Yang-Mills couplings towards the infrared to computations of the infrared effective potential of $\textrm{tr} \, F_{\mu\nu}^2$. For applications to Yang-Mills theories see \cite{Reuter:1993kw, Reuter:1994zn, Wetterich:1996kf, Reuter:1997gx, Gies:2002af, Gies:2003ic, Codello:2013wxa}, for QED and scalar QED (Abelian Higgs model) see \cite{Gies:2004hy} and \cite{Reuter:1992uk, Reuter:1994sg, Bergerhoff:1995zm, Bergerhoff:1995zq, Freire:2000sx} respectively. The electroweak sector of the Standard Model was discussed in \cite{Reuter:1993nn}.

\begin{figure}[t]
	\begin{center}
	\includegraphics[width=0.6\columnwidth]{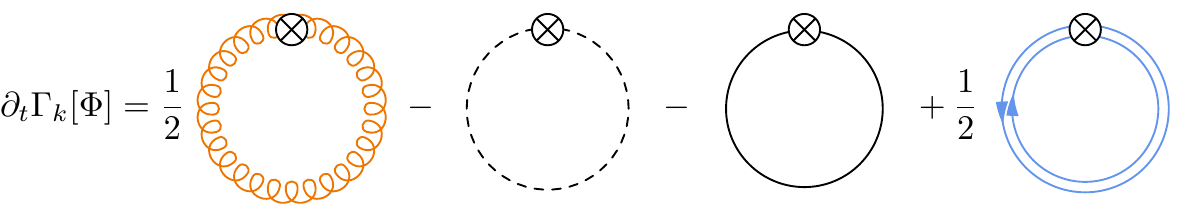}
	\caption{ Flow of the effective action of QCD. The first three
		diagrams arise from the gluon, ghost, and the quark degrees of
		freedom respectively. The last diagram is that of the mesonic
		contribution. The double line with the up-down arrows indicates
		the nature of the mesons as quark-antiquark composites. The
		crossed circles indicate the regulator insertion in the flow
		equation.  }\label{sec_he:QCD_equation}
	\end{center}
\end{figure}

In the gauge-fixed approach with gauge fixing \eqref{sec_he:eq:LandauGauge} and ghost action \eqref{sec_he:eq:Sghost}, the flow equation for the effective action of gauge theories with and without background field is based on a quadratic cutoff term for both the dynamical gauge field $A_\mu$ and the ghost field. These cutoff terms read   
\begin{align}\label{eq:sec_he:A-cutoff}
  \Delta S_{A,k}[A] = \int_p A^a_\mu(-p) \, R_{k,\mu\nu}^{ab}(p)\, A^b_\nu(p)\,,\qquad \qquad  \Delta S_{\textrm{gh},k}[c,\bar c] = \int_p  \bar c^a(-p) \, R_k^{ab}(p)\, c^b(p)\,, 
\end{align}
and the corresponding flows are given in the first two terms in Fig.~\ref{sec_he:QCD_equation}. There we depict the flow equation for QCD with  dynamical quarks and mesonic low-energy degrees of freedom as an illustrative example and for later use. The implementation of the latter low-energy degrees of freedom in first principles QCD is discussed in Sec.~\ref{sec_he:VacuumQCD}. Both cutoff terms in \eqref{eq:sec_he:A-cutoff} explicitly break gauge invariance, as do the cutoff terms for the colored matter fields. This entails that the STIs are modified. Moreover, while background gauge invariance can be maintained with the substitution $p^2\to -\bar D^2$ in the cutoff terms, background independence is lost, also manifested in modified Nielsen identities. Finally, the propagator of the gauge field in the flow equation for the background field effective action $\Gamma[A]$ is that of the fluctuation field $a_\mu$, since the flow $\partial_t \Gamma_k[A]$ is simply $\partial_t \Gamma_k[A, a=0]$. Here we have used the (inverse) RG-time $t=\ln k/\Lambda$ with some reference scale $\Lambda$. In summary, for a finite cutoff scale the underlying gauge invariance is broken by the cutoff terms and the gauge invariance of the background field effective action ceases to be physical. 

In the limit $k\to 0$ background independence as well as the standard STIs are recovered in a controlled way: First we note that for a non-linear gauge symmetry the breaking of gauge invariance is necessarily (at least) quadratic in the gauge field. This leads to loop terms in symmetry identities such as the Slavnov-Taylor identities. A derivation and discussion is beyond the scope of the present review, here we simply quote it in a very convenient form, the \textit{modified Quantum Master Equation} (mQME) or \textit{ modified Slavnov Taylor identity} (mSTI)  
\begin{align}
\label{sec_he:eq:mMaster}
  \int_x \frac{\delta \Gamma_k}{\delta \mathcal{Q}^i(x)}
  \, \frac{\delta \Gamma_k}{\delta \Phi_i(x)}
  =
\textrm{Tr} \,  R^{ij} G_{jl}  \frac{\delta^2 \Gamma_k[\Phi,
  \mathcal{Q}]}{\delta\Phi_l \delta Q^i}
  \, , \qquad \qquad \Phi=(A_\mu,c,\bar c,....)\,. 
\end{align}
In \eqref{sec_he:eq:mMaster}, the superfield $\Phi$ contains all fields of the given theory including auxiliary fields. The indices $i,j,l$ comprise all internal  (group, flavor) and Lorentz indices as well as the species of fields and are summed over.  The trace on the right hand side sums over spacetime or momentum space. This condensed notation is similar to that used in \cite{Pawlowski:2005xe} (a variation of DeWitt's condensed notation). The left hand side of \eqref{sec_he:eq:mMaster} is nothing but the Zinn-Justin equation in a concise form: In order to arrive at this form one has to augment the fields with an additional auxiliary field, the Nakanishi-Lautrup field $B$. In QCD this leads to $\Phi=(A_\mu, c,\bar c, B, q,\bar q)$.  Moreover, the source term $\int_x J^i \Phi_i$ in the path integral is amended by sources for BRST-transformations $\int_x Q^i \mathfrak{s}\Phi_i$. Here $Q^i$ are the BRST-sources of the fields $\Phi_i$, and$(\delta\Gamma/\delta Q^i)\delta /\delta\Phi_i$ generates (quantum) BRST-transformations. 

The right hand side is the modification of the STI in the presence of the cutoff term: it is proportional to the regulator function $R_k$ and the full propagator $G_{jl}$ of the fields $\Phi_i,\Phi_l$. This term resembles the flow equation itself. There, the loop carries the violation of the scaling relations of the underlying theory by the cutoff term, here it is the violation of BRST-symmetry.  

This allows us to discuss the physical limit $k\to0$: Naturally, the right hand side of \eqref{sec_he:eq:mMaster} vanishes for $R_k\equiv 0$. Note however that this is not enough to guarantee a well-defined limit in the sense that we arrive at the original massless gauge theory. Another important property is the fact that the breaking of BRST-invariance in \eqref{sec_he:eq:mMaster} is local in momentum space: the operator in the trace decays for large momenta due to the regulator. It is this important property that signifies the breaking of gauge invariance as a local perturbation. In summary this leads to a well-defined and smooth limit of the gauge theory within a gauge symmetry-breaking regularization to the gauge theory at $k=0$. The existence of this limit can be studied in perturbation theory, where an iterative solution of the flow equation in terms of loop orders is nothing but the regularization and renormalization of the gauge theory with a momentum cutoff and the respective gauge-variant counter terms: a generalized BPHZ-scheme (Bogoliubov-Parasiuk-Hepp-Zimmermann--scheme). 

Eq.~\eqref{sec_he:eq:mMaster} guarantees the standard Master equation for the effective action at $k=0$. This implies BRST-invariance of the effective action as well as the gauge invariance of observables. It is left to guarantee the gauge-independence of the approach. This property is encoded in Nielsen identities, which take a convenient form in the background gauge approach with the gauge fixing \eqref{sec_he:eq:LandauGauge}: the background effective action $\Gamma[\bar A,a]$ carries an explicit dependence on the background field $\bar A$, which vanishes onshell on the equations of motion. There the source term in the path integral vanishes and the Fadeev-Popov trick can be undone. The Nielsen identity (NI) monitors the difference of fluctuation field derivatives and background field derivatives. In the presence of the cutoff term the Nielsen identity gets modified similarly to the mSTI. The modified Nilsen identity (mNI) reads schematically 
\begin{align}\label{eq:Nielsen} 
\frac{\delta \Gamma_k}{\delta a} -\frac{\delta \Gamma_k}{\delta \bar A} 
-\left\langle \frac{\delta (S_\textrm{gf}+S_\textrm{gh})}{\delta a} -
\frac{\delta  (S_\textrm{gf}+S_\textrm{gh})}{\delta \bar A} \right\rangle =  \frac12 \textrm{Tr} \,  \frac{\delta R^{ij}_k}{\delta \bar A}\, G_{ji} \,,
\end{align} 
within the same condensed notation used for the mSTI,  \eqref{sec_he:eq:mMaster}. At vanishing cutoff the right hand side of \eqref{eq:Nielsen}  vanishes and we are left with the standard NI. The fluctuation field derivative of the effective action vanishes onshell (vanishing currents). This entails that the Fadeev-Popov trick can be undon, and the expectation value on the left hand side vanishes. Consequently also the background field derivative vanishes. This is how the NI encodes the onshell background-independence of the background-field approach. In summary, the modified symmetry relations \eqref{sec_he:eq:mMaster} and \eqref{eq:Nielsen} monitor the violation of gauge invariance and gauge independence for $k\neq 0$ and guarantee their reinstatement at $k=0$. Background independence is also of also of crucial importance in quantum gravity, where the Nielsen identity is also called split Ward identity, see \eqref{sec_gr:sWI}, the respective discussion in Sec.~\ref{sec_gr}, and the recent review \cite{Pawlowski:2020qer}. 

Monitoring global and local symmetries, and in particular BRST-symmetry and gauge independence, and hence the physical gauge symmetry of the theory, is essential for the FRG-approach to gauge theories. Hence this topic has received much interest. Modified STIs and Master equations for Abelian, non-Abelian gauge theories and gravity have been discussed and applied in \cite{Bonini:1993sj, Bonini:1994kp, Bonini:1994dz, Bonini:1995tx, Becchi:1996an, Ellwanger:1994iz, DAttanasio:1996tzp, Reuter:1997gx, Freire:2000bq, Igarashi:1999rm, Igarashi:2000vf, Igarashi:2001mf, Igarashi:2001ey, Igarashi:2001cv, Pawlowski:2003sk, Pawlowski:2005xe, Sonoda:2007dj, Igarashi:2007fw, Igarashi:2008bb, Igarashi:2009tj, Donkin:2012ud, Lavrov:2012xz, Sonoda:2013dwa, Safari:2015dva, Safari:2016dwj, Safari:2016gtj, Igarashi:2016gcf, Asnafi:2018pre, Igarashi:2019gkm, Barra:2019rhz, Lavrov:2019agp, Lavrov:2020exa, Pawlowski:2020qer}. For gravity applications see also Sec.~\ref{sec_gr}. The related fate of anomalies and topological terms in gauge theories have been discussed in \cite{Bonini:1994xj, Reuter:1996be, Pawlowski:1996ch}. 

We close this section with two remarks: 

In the absence of the regulator term on the right hand side in \eqref{sec_he:eq:mMaster} one can use powerful cohomological methods for an iterative construction of the \textit{perturbative} effective action. Beyond perturbation theory, the mQME or mSTI \eqref{sec_he:eq:mMaster} at $R_k=0$ at least provides \textit{algebraic} relations between longitudinal and transverse correlation functions which are discussed in more detail in the next section Sec.~\eqref{sec_hep-locality}. This has triggered some investigations of reformulations of gauge theories and the mSTI without the modification of \eqref{sec_he:eq:mMaster} on the right hand side, see e.g.~\cite{Igarashi:1999rm,  Pawlowski:2005xe, Lavrov:2012xz, Igarashi:2019gkm}. Loosely speaking these attempts either work with non-linear gauges, non-linear field transformations or further auxiliary degrees of freedom. In the formulation with the mSTI \eqref{sec_he:eq:mMaster} this amounts to absorbing its right hand side in the symmetry transformations themselves. Typically then the generator of the symmetry looses its ultralocality, a well-studied problem for chiral symmetry transformations as well as supersymmetry in lattice formulation, for a point of view close to the FRG see \cite{Bergner:2012nu}. Note that symmetries, and in particular the definition of (Noether) charges require locality (exponential decay in spacetime) of the generators of this symmetry. A well-studied example is chiral symmetry on the lattice: while chiral symmetry for Wilson fermions with explicit mass terms for the fermionic doublers  can be formulated, the respective generator of chiral symmetry only has an algebraic decay in spacetime, see \cite{Bergner:2012nu}. This leads to a slow  convergence towards the continuum limit of observables with chiral properties. In turn, the generator of chiral symmetry for Ginsparg-Wilson fermions is local, though not ultralocal, and the continuum limit of observables with chiral properties is more rapid in comparison. In summary, in most of the FRG-formulations for gauge theories without the symmetry-breaking term on the right hand side of \eqref{sec_he:eq:mMaster} it can be shown that this locality is lost. For the remaining formulations this issue has not been studied yet. 

Most of the applications of the FRG to gauge theories have been performed in the Landau gauge or the Landau-DeWitt gauge \eqref{sec_he:eq:LandauGauge} given its technical and numerical advantages. In turn, so called \textit{physical} gauges as well as dual formulations are advantageous for approaching the underlying conceptual problems in strongly-correlated gauge theories. This has led to progress within Abelian gauges, \cite{Ellwanger:1997wv}, Axial gauges, \cite{ Litim:1998qi, Simionato:1998iz, Simionato:1998te, Litim:2002ce}, the Polyakov gauge, \cite{Marhauser:2008fz, Kondo:2010ts}, within the Cho-Faddeev-Niemi decomposition, \cite{Gies:2001hk}, the use of the Landau gauge with fieldstrength formulation for the correlator of the fieldstrength in \cite{Ellwanger:1998th}, the dual fieldstrength in \cite{Ellwanger:1999vc}, as well as the formulation of massive Yang-Mills theory in Abelian gauges in \cite{Ellwanger:2002sj} and in the Abelian theory in \cite{Ellwanger:2002xa}. Hamiltonian FRG flows have been derived and applied to  the infrared sector of Coulomb gauge Yang-Mills theory in \cite{Leder:2010ji, Leder:2011yc}.

%
%%%%%%%%%%%%%%%%%%%%%%%%%%%%%%%%%%%%%%%%%%%%%%%%%%%%%%%%%%%%%%%%%%%%%
\begin{figure}[t]
	\begin{center}
				\subfigure[FRG Yang-Mills Gluon propagator $G_A(p^2)$, \cite{Cyrol:2016tym}, 
		in comparison to lattice data, \cite{Sternbeck:2006cg}. The shaded area for small momenta potentially signals different infrared solutions of Landau gauge Yang-Mills theory, for more details see \cite{Cyrol:2016tym}.]{\hspace{-0.4cm}\label{fig:GluonpropYM}
			\includegraphics[scale=0.38]{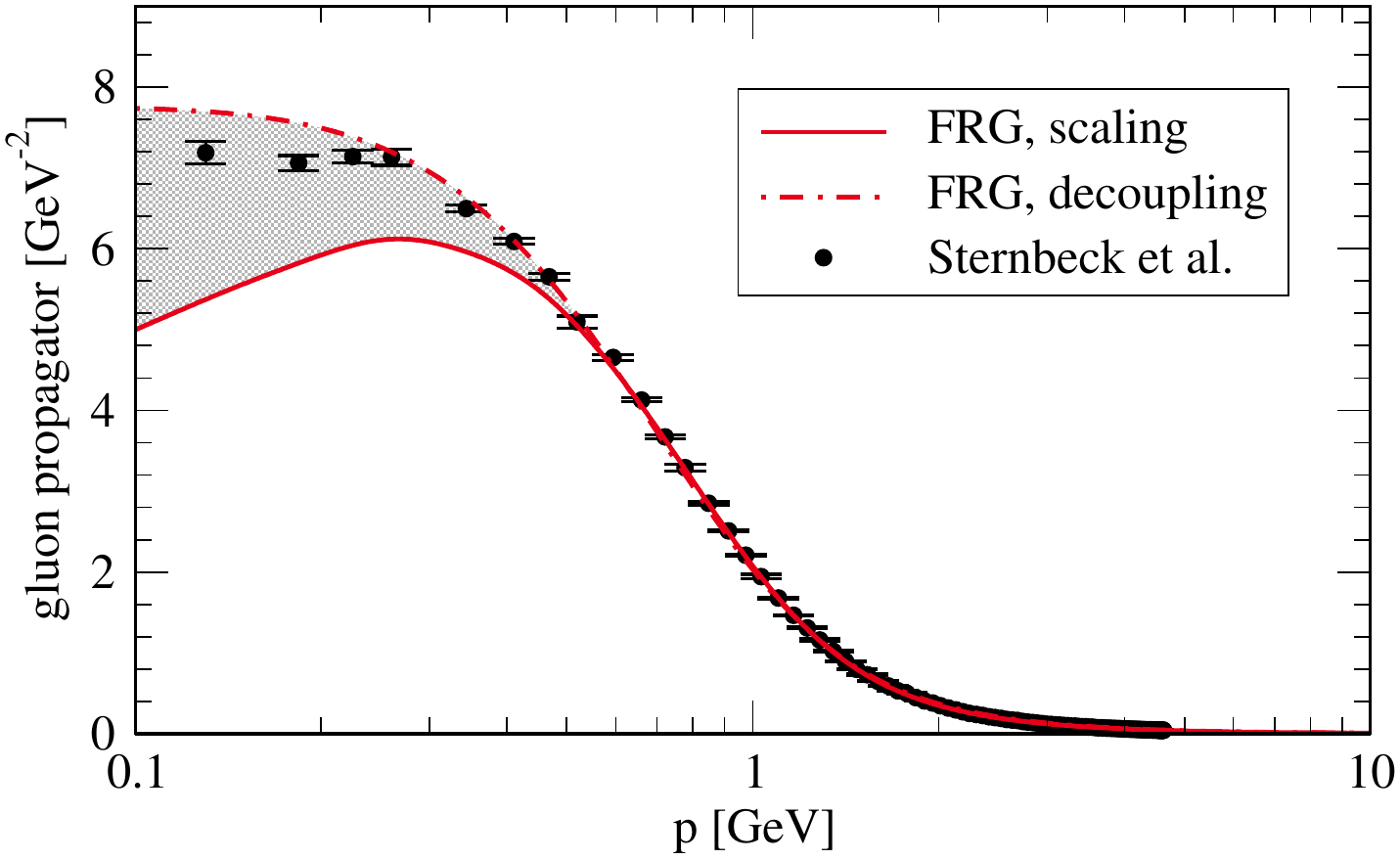}}\hspace{0.1cm}
\subfigure[Effective 
running couplings, \cite{Cyrol:2016tym}, as obtained
from different Yang-Mills vertices: three-gluon vertex: $\alpha_{A^3}$, four-gluon vertex: $\alpha_{A^^4}$, ghost-gluon vertex: $\alpha_{A\bar c c}$. At large momenta the couplings agree due to universality.]{\hspace{-0.0cm}	\label{fig:runningcouplings}
	\includegraphics[scale=0.38]{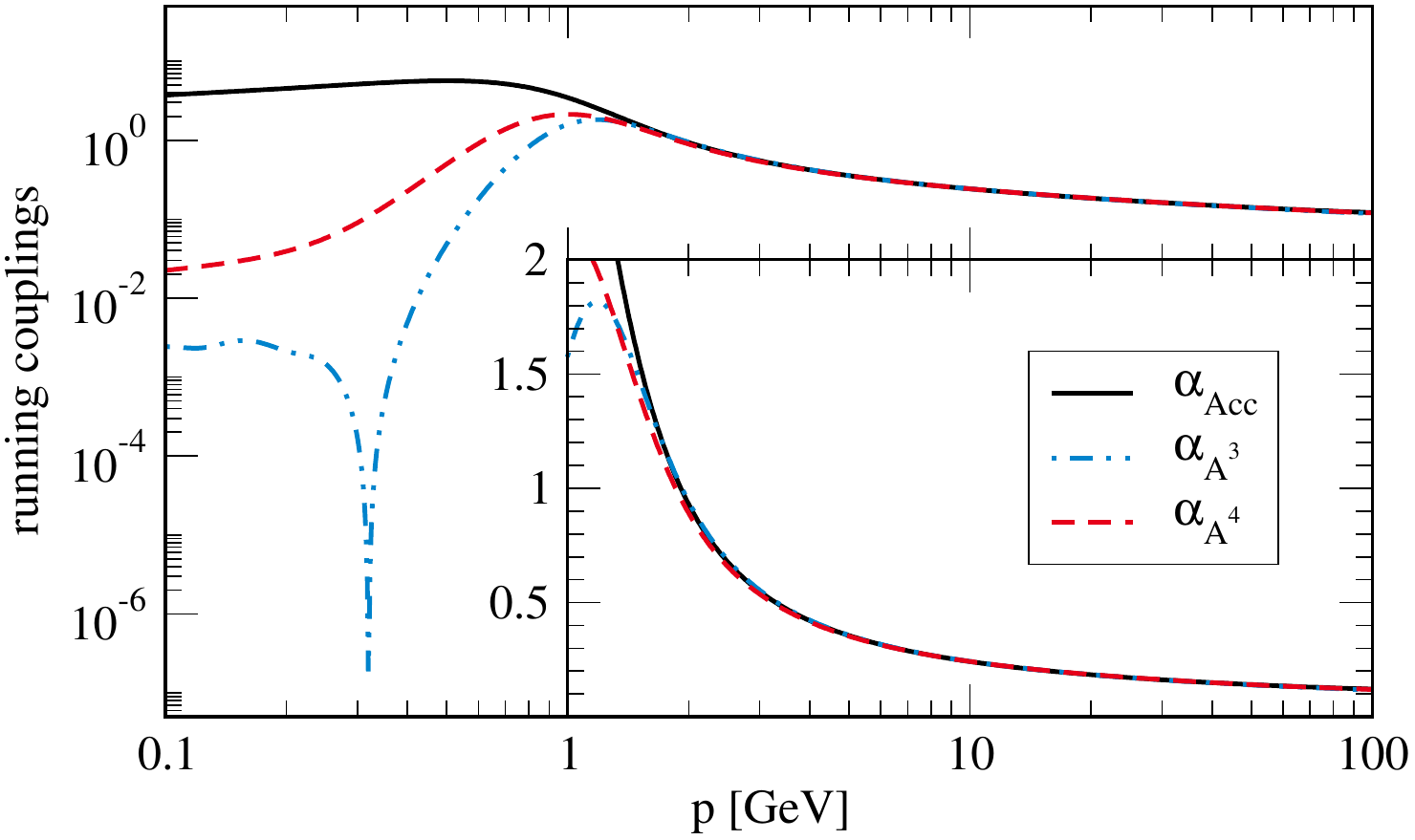}}	\hspace{0.1cm}
		\subfigure[$2-$ and $2+1$-flavor gluon dressing functions
			$1/Z_{A}(p^2)= p^2\,G_A(p^2)$ with the gluon propagator $G_A(p^2)$.
			 FRG: $2$-flavor \cite{Cyrol:2017ewj}, $2+1$-flavor \cite{Fu:2019hdw}.  
		Lattice: $2$-flavor \cite{Sternbeck:2012qs},  
			$2+1$-flavor \cite{Zafeiropoulos:2019flq, Boucaud:2018xup}.]{\hspace{-0.0cm}	\label{fig:inZA_lattice2-2+1}
			\includegraphics[scale=0.42]{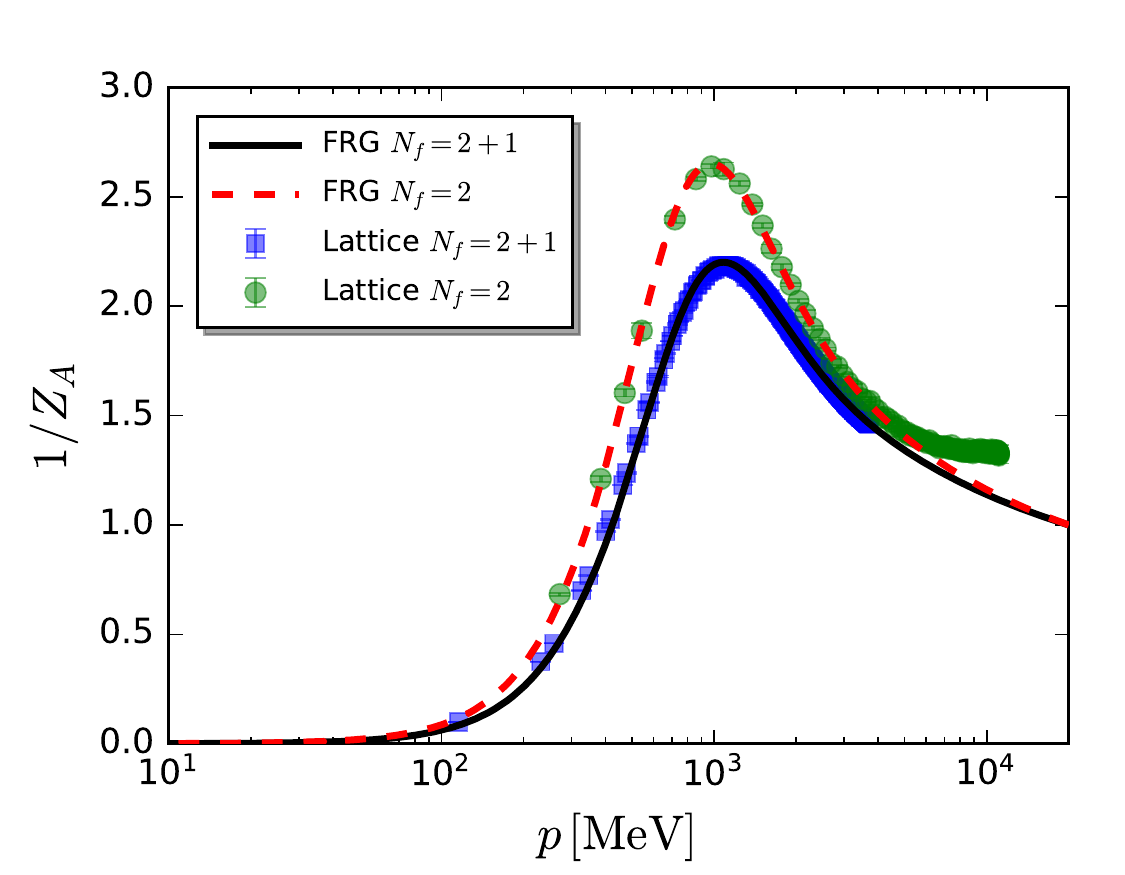}\hspace{0.0cm}}	
	\end{center}
	\caption{YM gluon propagator $G_A(p^2)$ (right) and couplings (middle) and $2-$ and $2+1$-flavor gluon dressing function $p^2 G_A(p^2)$ (right) in comparison to lattice data. The results show the (apparent) convergence of the vertex expansion scheme, for more details see in particular \cite{Cyrol:2016tym, Cyrol:2017ewj}.} 
	\label{fig:YMprops}
\end{figure}
%%%%%%%%%%%%%%%%%%%%%%%%%%%%%%%%%%%%%%%%%%%%%%%%%%%%%%%%%%%%%%%%%%%%%%
%
\subsubsection{Gauge invariance, locality \& confinement} \label{sec_hep-locality} 

The functional setup developed in the last sections in terms of gauge-fixed correlation functions allows us to discuss some important aspects of functional approaches to gauge theories related to gauge invariance, locality as well as the regularity of correlation functions. These aspects are also relevant for the access to confinement in Yang-Mills theory and QCD in these approaches, mostly done in the Landau(-DeWitt) gauge, $\xi=0$ in \eqref{sec_he:eq:LandauGauge}. 

The current argument follows the analysis for Yang-Mills theory in \cite{Cyrol:2016tym} within the FRG approach. This work, together with corresponding ones for quenched QCD \cite{Mitter:2014wpa}, unquenched QCD, \cite{Cyrol:2017ewj} as well finite temperature Yang-Mills theory, \cite{Cyrol:2017qkl, Hajizadeh:2019qrj} and three-dimensional Yang-Mills theory \cite{Corell:2018yil} constitute the technically most advanced FRG-QCD computations including the nonperturbative infrared regime of QCD. This line of works has been initiated in \cite{Ellwanger:1994iz, Ellwanger:1995qf, Ellwanger:1996wy, Ellwanger:1997tp, Bergerhoff:1997cv}. While the results in \cite{Mitter:2014wpa, Cyrol:2016tym, Cyrol:2017ewj, Cyrol:2017qkl} also include vertex dressings, we refer to the original works for the details. In Fig.~\ref{fig:YMprops} we only depict a selection of results for the gluon propagator as well as the running couplings in vacuum Yang-Mills theory and QCD. For momenta $p^2\gtrsim \Lambda_\textrm{QCD}^2$ the FRG-results are in quantitative agreement with the corresponding lattice results. For these results as well as results for Landau-gauge QCD with other functional methods we refer the reader to some reviews and works, see e.g.~\cite{Alkofer:2000wg, Fischer:2006ub, Sternbeck:2006cg, Fischer:2008uz, Aguilar:2008xm, Binosi:2009qm, Maas:2011se, Boucaud:2011ug, Cucchieri:2011ig, Sternbeck:2012qs, Pelaez:2013cpa, Aguilar:2013vaa, Reinosa:2017qtf, Huber:2018ned, Oliveira:2018lln, Boucaud:2018xup, Zafeiropoulos:2019flq, Gracey:2019xom, Li:2019hyv, Hajizadeh:2019qrj,  Aguilar:2019uob, Huber:2020keu} and references therein. 

The differences in the deep infrared are related to the Gribov ambiguity. The Gribov problem in gauge-fixed approaches is yet not fully resolved, for reviews and recent work see e.g.~\cite{Maas:2011se, Vandersickel:2012tz, Capri:2017bfd}. In the works \cite{Mitter:2014wpa, Cyrol:2016tym, Cyrol:2017ewj, Cyrol:2017qkl} a bootstrap approach for its resolution has been used: the BRST-symmetry underlying the mSTI and mNI in \eqref{sec_he:eq:mMaster} and \eqref{eq:Nielsen} is standard Landau gauge BRST-symmetry without infrared modifications as in the Gribov-Zwanziger approach. In this setup the Kugo-Ojima confinement scenario, \cite{Kugo:1979gm, Kugo:1995km}, applies if the respective BRST-charge is well-defined. Note that the STI only covers the invariance under infinitesimal BRST-transformations and not the existence of the BRST-charge, see \cite{Fischer:2008uz}. Under the \textit{assumption} that the BRST-charge exists, the ghost and gluon propagators are required to have a non-trivial infrared scaling. Moreover, this scaling is unique, \cite{Fischer:2006vf, Fischer:2009tn, Alkofer:2008jy}. The initial conditions for the (classical) vertices and propagators in \cite{Mitter:2014wpa, Cyrol:2016tym, Cyrol:2017ewj, Cyrol:2017qkl} at the ultraviolet initial scale $k=\Lambda_\textrm{UV}$ have been fine-tuned such that this infrared scaling behaviour is obtained. A detailed discussion can be found in \cite{Cyrol:2017qkl}. This construction has passed several consistency checks: first the existence of such a scaling solution with the correct perturbative UV behaviour is by not trival. Second it has been shown that the STI consistence gets worse, if moving away from the scaling solution. Note however, that this is a bootstrap resolution based on the assumption of the existence of BRST-charges in the Landau gauge. We emphasise that a satisfactory resolution of the non-perturbative Gribov problem in gauge-fixed approaches has yet to be obtained not only for the FRG-approach, but for functional approaches in general as well as on the lattice. 

Evidently, the results for the correlation functions depicted in Fig.~\ref{fig:YMprops} are gauge-dependent. We still have to compute observables from these correlation functions. Moreover, we have to check the underlying gauge-consistency of the numerical results obtained within approximations to the full hierarchy of flow equations. In perturbation theory the STIs simply provide checks for the computations and facilitate the relation of  the computed gauge-dependent correlation functions to physical observables such as S-matrix elements. A well-known textbook example is the identification of different renormalized strong couplings $\alpha_s$ extracted from different renormalization functions of primitively divergent (classical) vertices and propagators.  Its momentum-dependent version follows from the STIs. Strictly speaking the respective running couplings are those of longitudinal projections of classical vertices since the STIs relate the dressings of longitudinal vertices to a combination of dressings of longitudinal and transverse correlation functions. Accordingly, an underlying implicit assumption within the STI construction of the running coupling is the identification of transverse and longitudinal dressings of the respective vertices. In the following we will call this the \textit{regularity} assumption. A relevant example for a vertex that does violate the regularity assumption is the ghost-gluon vertex, $[\Gamma^{(3)}_{A\bar c c}]_\mu^{abc}(|p,|q|,\cos\theta)$. Here $p$ is the gluon momentum, $q$ is the anti-ghost momentum $q$, and $\theta$ is the angle between them. If the vertex still depends on $\theta$ in the limit $|p|\to 0$, it is irregular. Such a vertex does not allow for a simple identification of different running couplings as described above. While in perturbation theory the vertices can shown to be regular, irregularities are required for confinement in the Landau gauge, for a detailed discussion see \cite{Cyrol:2016tym}. 

As has been discussed in the last section, in nonperturbative functional approaches such as the FRG or Dyson-Schwinger equations, the STIs also provide non-trivial relations between correlation functions that can be used in order to maintain or monitor gauge invariance during the flow. This has been discussed for non-Abelian gauge theories in \cite{Fischer:2008uz}, for gravity see \cite{Denz:2016qks,Christiansen:2017bsy}. In particular it should be noted, that gauge invariance alone is not necessarily the hallmark of a good approximation. This statement will now be discussed within the example of Landau gauge flows, and will lead to a refined criterion: 

First we note that in the Landau gauge the propagator of the gauge field is transverse. Note also that this property is present in any approximation within the Landau gauge. Hence, \textit{internal} gauge field legs of vertices are transversally projected as they are contracted with the transverse propagator of the gauge field. If additionally applying transverse projection operators to all \textit{external} gauge field legs of the flow equations for a given correlation function $\Gamma^{(n)}$, we are led to flow equations with completely transverse vertices $\Gamma^{(n)}_\bot$: all gauge field legs, internal and external ones are transversally projected. In summary, this leads us to a closed set of flow equations for the completely transverse correlation functions $\Gamma^{(n)}_\bot$. Schematically this reads
\begin{align} \label{sec_he:eq:TransverseFlows} \partial_t
  \Gamma^{(n)}_\bot = \textrm{Flow}^{(n)}_\bot[\{ \Gamma^{(m)}_\bot\}]\,, 
\end{align}
where $m\leq n+2$ follows from the one-loop structure of the Wetterich equation. Accordingly, we can solve the flows \eqref{sec_he:eq:TransverseFlows} without the knowledge of the correlation functions with at least one longitudinal leg, in the following denoted with $\Gamma_L^{(n)}$. Note that some of the legs in $\Gamma_L^{(n)}$ can still be transverse. This analysis entails that the full dynamics of the theory is carried by the transverse correlation functions and hence by \eqref{sec_he:eq:TransverseFlows}. 

We are led to the peculiar situation that we can always arrange for a BRST-invariant solution from a given set of transverse correlation functions computed from \eqref{sec_he:eq:TransverseFlows}: we simply use the mSTIs for the correlation functions that follow from the functional mSTI \eqref{sec_he:eq:mMaster} to compute the corresponding longitudinal correlation functions. This construction of BRST-invariant solutions works even for transverse correlation functions, that even qualitatively do not have the correct behaviour. This already entails that BRST-invariance alone is no hallmark of a good approximation. 

For collecting further information on the quality of the approximation we evaluate both, the flows of correlation functions $\Gamma_L^{(n)}$ with at least one longitudinal gauge field leg as well as the corresponding mSTIs or mQMEs. Both sets of relations are not closed but require the transverse ones as input. Schematically this reads
\begin{subequations}
\begin{align}\label{sec_he:eq:LongFlows}
    \partial_t \Gamma^{(n)}_L =
    &\, 
      \textrm{Flow}^{(n)}_L[\{ \Gamma^{(m)}_L\}\,,\,\{
      \Gamma^{(m)}_\bot\}]\,, \\[1ex]
    \Gamma^{(n)}_L =
    &\,\textrm{mSTI}^{(n)}[\{ \Gamma^{(m)}_L\}\,,\,\{
      \Gamma^{(m)}_\bot\}]\, .
      \label{sec_he:eq:LongmSTIs}
\end{align}
\label{sec_he:eq:LongG}
\end{subequations}
The above equations relate (flows of) longitudinal correlation functions or vertices $\Gamma^{(n)}_L$ on the left hand side to loops with sets of both, longitudinal correlation functions, $\{\Gamma^{(m)}_L\}$ and transverse ones,  $\{\Gamma^{(m)}_\bot\}$. Hence, in contradistinction to the closed relations for the completely transverse correlation functions, the relations for the longitudinal correlation functions can only be solved for a given input of transverse correlation functions.  

As mentioned above, BRST-invariance within a given approximation is easily achieved by solving the mSTIs \eqref{sec_he:eq:LongmSTIs} with the given input of a set of transverse correlation functions $\Gamma_\bot^{(n)}$ computed from \eqref{sec_he:eq:TransverseFlows}. This leads to a gauge-invariant closure of the effective action. One may be tempted to argue that one should only consider approximations where the longitudinal correlation functions satisfy both relations in \eqref{sec_he:eq:LongG}. However, the comparison between the correlation functions from the mSTIs \eqref{sec_he:eq:LongmSTIs} and the longitudinal flows \eqref{sec_he:eq:LongFlows} only indirectly carries some information about \textrm{physical} gauge invariance. Instead, differences are to be expected in nonperturbative approximations. For example, if comparing the approximate solution of a set of flow equations with, e.g.~the DSE or nPI solution for the same approximation, they will only agree if the given approximation leads to the same resummation of -- nonperturbative -- diagrams. This rarely happens, the only known examples are perturbation theory and self-consistent nPI schemes, \cite{Wetterich:2002ky, Dupuis:2005ij, Pawlowski:2005xe, Blaizot11b, Carrington:2012ea, Carrington:2014lba, Carrington:2016zlc, Carrington:2017lry, Carrington:2019fwp, Alexander:2019cgw, Alexander:2019quf}. Moreover, even in the known examples, perturbation theory and nPI schemes, a comparison is only possible if taking into account the generically different RG scheme in the functional RG, for related work see \cite{Litim:1996nw, Pernici:1997ie, Ellwanger:1997tp, Pernici:1998tp, Latorre:2000qc, Arnone:2002yh, Arnone:2003pa, Pawlowski:2005xe, Rosten:2006pd, Codello:2013bra}.

This leaves us with the yet only partially solved important question of how to evaluate the preservation of gauge invariance and the reliability within a given approximate solution for transverse correlation functions. Nonetheless, while the analysis is not fully conclusive, it provides us with the refined criterion promised above: Keeping in mind, that also for the definition of observables gauge invariance is not enough but also \textit{locality} is required\footnote{If one gives up locality, even the Landau-gauge propagator is an observable as we can define a gauge-invariant gauge field $A_\textrm{gi}$ with $A_\textrm{gi}[A] =A^{{\cal U}_L[A]}$, where ${\cal U}_L[A]$ is the gauge transformation that implements the Landau gauge on $A$. Then $\langle A_\textrm{gi} A_\textrm{gi}\rangle$ is gauge-invariant and equals the Landau gauge propagator. However, this 'observable' is not local.}, this is a sensible but only implicit criterium for judging the quality of given results: the results for the expectation values of \textit{local} gauge invariant operator computed from the gauge-fixed correlation functions have to be \textit{local}. 

We close this section with a remark on irregularity of vertices and confinement. We first note that regularity implies, that the dressing of the longitudinal projection of a correlation function has to agree with that of the transverse projection for vanishing projection momentum. In particular, the regularity assumption together with the STI implies the vanishing of the inverse gluon propagator at vanishing cutoff, $k=0$, and momentum: $\Gamma^{(2)}_{\bot}(p=0)=0$. However, confinement requires a gluon mass gap in covariant gauges, see \cite{Braun:2007bx, Fister:2013bh} and the discussion in Sec.~\ref{sec_QCD-Conf}. Consequently, this mass gap can only be generated with irregularities in the vertices, for a discussion within the FRG approach see \cite{Cyrol:2016tym}. We conclude that confinement also necessitates the irregularity of vertices, which has been already pointed out in \cite{Cornwall:1981zr}. In summary, the seemingly very formal (ir)regularity property has direct consequences for the confinement phenomenon. This property is also tightly related to the above discussion of locality and gauge invariance, as irregularities introduce non-localities. While the gauge-fixed correlation functions are not required to be local, they have to cancel out in local observables that are computed from these correlation functions.

\subsubsection{Gauge-invariant flows \& the quest for simplicity}
\label{sec_hep-GaugeInv} 

The above intricacies have triggered many attempts to derive gauge-invariant flow equations. Such a framework with a gauge-invariant effective action has not only the appeal of a direct physics interpretation of the respective correlation functions but also that of technical simplicity: A gauge-invariant effective action can be expanded in gauge-invariant operators, e.g.~powers of the fieldstrength $F_{\mu\nu}$ leading to $\int_x V(F)$. This potential is the analogue of the effective potential in a scalar theory and the respective expansion in gauge theories is the covariant analogue of the derivative expansion in scalar theories. Parts of it, $\int_x W(\theta)$ with $\theta=\textrm{tr}\,F^2$,  have been studied in the literature, see e.g.~\cite{Reuter:1994zn, Reuter:1997gx, Gies:2002af, Eichhorn:2010zc}. Further terms are generalised kinetic terms such as $\int_x \textrm{tr} F_{\mu\nu} f_{\mu\nu\rho\sigma}(-D) F_{\rho\sigma}$, see \cite{Ellwanger:1995qf, Ellwanger:1996wy, Bergerhoff:1997cv}, or gauge-invariant matter terms such as $\int_x \bar q \dr^n q$, see \cite{Mitter:2014wpa, Cyrol:2017ewj}.

The above examples have all been employed in a gauge-fixed settings (either Landau gauge or Landau-DeWitt gauge) with the implicit assumption that the full effective action 
\begin{subequations}
	\label{eq:BackApprox}
\begin{align}
\label{eq:BackApproxG}
\Gamma[\bar A, a]\approx \Gamma_\textrm{Dyn}[A]+ \Gamma_\textrm{gf}[\bar A,a]\,,
\end{align}
can be split into a dynamical gauge-invariant part $\Gamma_\textrm{Dyn}[A]$ and a nearly classical gauge fixing part 
\begin{align}\label{eq:Ggf=Sgf}
\Gamma_\textrm{gf}\approx S_\textrm{gf}[\bar A,a] +S_\textrm{ghost}[\bar A,
a,c,\bar c]\,. 
\end{align}
\end{subequations}
While this underlying assumption has been checked to hold quantitatively in QCD for scales $k\gtrsim 1$\,GeV for the above mentioned matter terms in \cite{Mitter:2014wpa, Cyrol:2017ewj}, respective checks for the pure gauge field dynamics are still lacking, and are highly warranted.

These complications could be avoided within fully gauge-invariant formulations of the FRG-approach to gauge theories. Most of the respective constructions can be linked to the background field approach. A direct cousin is the \textrm{geometrical} or Vilkovisky-DeWitt effective action. This uses the geometry of the configuration space for defining gauge-invariant fields as tangent vectors of geodesics with respect to the Vilkovisky connection $\Gamma_V(\bar A)$ from a given background $\bar A$ to a general configuration $A$. The Vilkovisky connection is constructed with the demand of a maximal disentanglement of gauge degrees of freedom $a_\textrm{gauge}$ (gauge fibre) and the dynamical gauge-invariant ones $a_\textrm{dyn}$ (base space), with $\bar D\cdot a_\textrm{dyn}=0$. It can be shown that the effective action only depends on the gauge-invariant field $a_\textrm{dyn}$ and not on $a_\textrm{gauge}$. 

Naturally, the construction is highly non-local and $\Gamma_V \propto (1/\bar D^2) \bar D$. Its linear approximation is simply the standard \textit{linear split} in the background field approach, $A=\bar A+a$. Without such an approximation it involves an infinite series, schematically $A=\bar A+a-1/2 \Gamma_V \cdot a^2 +O(a^3)$. In perturbation theory it is of limited use, and on the technical level it boils down to the standard background field computations in the Landau-DeWitt gauge. In the FRG-approach it allows for a gauge-invariant regularisation, as the dynamical field $a_\textrm{dyn}$ itself is gauge-invariant. Seemingly the geometrical approach depends on two fields, $\bar A$ and $a$, but it can be shown with Nielsen or split Ward identities (NIs, sWI) that the respective field derivatives are linearly related, in contradistinction to the standard background field approach. This property of the geometrical approach originates in its reparameterisation invariance. As already mentioned before, these attractive features come at the price, that the construction is highly non-local, both in the configuration space of the gauge field as well as in momentum space. We have already discussed and emphasised the importance of locality (and symmetry identities) at the end of the last Sec.~\ref{sec_hep-locality}. In the present framework it is carried by the Nielsen identities and the \textit{locality} of the gauge-invariant effective action. As in the gauge-fixed approaches, the assessment of these properties is highly non-trivial. 

The FRG-approach to the geometrical effective action has been studied in non-Abelian theories and gravity in \cite{Branchina:2003ek, Pawlowski:2003sk, Pawlowski:2005xe, Donkin:2012ud, Demmel:2014hla, Safari:2015dva}. In the presence of the cutoff term, the Nielsen identities acquire a non-linear term similar to the right hand side of \eqref{sec_he:eq:mMaster}, see \cite{Pawlowski:2003sk, Pawlowski:2005xe, Safari:2015dva, Safari:2016dwj, Safari:2016gtj}. This additional term mirrors a similar one in the Nielsen or split Ward identity in the standard background field approach, see \cite{Litim:2002ce, Litim:2002hj, Litim:2002xm, Pawlowski:2005xe, Folkerts:2011jz, Bridle:2013sra, Dietz:2015owa, Labus:2016lkh}. There, however, it adds to a non-linear term already present due to the gauge fixing. Importantly, in the geometrical approach gauge invariance is kept trivially at all stages of the construction thanks to the gauge invariance of the fields. The latter is the core ingredient of the construction: a parametrisation of the theory, that allows the distinction between gauge degrees of freedom and the dynamical ones without mixing terms. So far, the only nonperturbative application has been a diffeomorphism-invariant computation of UV-IR phase structure of asymptotically safe quantum gravity with the ultraviolet Reuter fixed point in \cite{Donkin:2012ud}. 

Related approaches based on such a disentanglement have also been put forward in \cite{Wetterich:2016ewc, Wetterich:2017aoy, Pawlowski:2018ixd, Wetterich:2019zdo}, based on a so called \textit{physical} gauge fixing that allows to use the dynamical (mean) field in the regulator. This splitting is also at the root of a gauge-fixed approach with standard STIs put forward in \cite{Asnafi:2018pre}. While these formulation can be embedded or related to the geometrical approach discussed above, they have the advantage of (relative) technical simplicity. In the case of the gauge-invariant flow \cite{Wetterich:2016ewc} this has already led to first results in quantum gravity, see also Sec.~\ref{sec_gr}.

Another promising approach has been put forward in \cite{Morris:1999px, Morris:2000fs, Arnone:2001iy, Arnone:2002cs, Arnone:2005fb, Arnone:2005vd, Morris:2005tv, Morris:2006in, Rosten:2006qx, Rosten:2006tk, Rosten:2006pd, Rosten:2008zp}, the \textit{manifestly gauge-invariant FRG}. This approach utilizes a specific property of the Polchinski-RG for the Wilson effective action (or Schwinger functional), the generating functional for amputated connected correlation functions: the flow admits field-dependent regulators. It has been mostly used for gauge-invariant perturbative computations of $\beta$-functions, in particular the two-loop $\beta$-function of
Yang-Mills theory, \cite{Morris:2005tv}. It also has been shown, how to set-up the approach for a computation of Wilson loop expectation values, see \cite{Rosten:2006qx}. Variants of such a flow have also been discussed more recently in \cite{deAlwis:2017ysy, Bonanno:2019ukb}. In summary, this approach may give direct access to the flow of observables such as the confinement-deconfinement order parameter. 

Very recently, a gauge-invariant, background- and scheme-independent FRG-approach for the Wilson effective action has suggested in \cite{Falls:2020tmj}. It combines the geometrical approach to the FRG with the gauge-invariant FRG discussed in the previous two paragraphs. This promising approach may circumvent all the above mentioned obstacles. 

A final possibility is to take more general regularisations than the standard momentum cutoff. This idea has been pursued at finite temperature in \cite{DAttanasio:1996psq,Comelli:1997ru}. In these works the temporal boundary conditions at finite temperature have been modified via the thermal distribution function. This does not lead to
modifications of the Slavnov-Taylor identities. More generally, regularisations can be implemented in position space. This has been implemented within real-time approaches in \cite{Gasenzer:2007za, Gasenzer:2010rq, Corell:2019jxh} and \cite{Pietroni:2008jx, Lesgourgues:2009am, Bartolo:2009rb}. If applied to gauge theories, the standard STIs would be maintained in these approaches.

In summary, there are various very promising options that allow to either keep the standard STIs or to obtain gauge-invariant flows. Still, we close this Section with a word of caution: the geometry of the configuration space in theories with non-linear symmetries structurally does not allow for a global linear disentanglement of the gauge degrees of freedom. In turn, any such disentanglement is inherently non-linear, if not non-local. This remark re-iterated the importance of locality which cannot be emphasized enough. Hence it remains to be seen how far these approaches can be pushed in the necessary quest for quantitative precision.

\subsection{QCD}\label{sec_hep-QCD}

We have already mentioned in the introduction Sec.~\ref{sec_hep-intro}, that QCD is a paradigmatic example for a gauge theory with many nonperturbative phenomena. As discussed there, QCD physics questions range from the confinement mechanism and that of chiral symmetry breaking, the determination of the bound state and resonance structure of QCD, to the phase structure of QCD including the condensed matter phases at high densities, as well as the real-time dynamics in and out of equilibrium. The resolution of these questions is highly relevant for an understanding of the QCD phase transition in the early universe, for nuclear astrophysics with intriguing links to gravitational-wave physics, the formation of matter around us as well as that of heavy-ion collisions. Therefore, QCD has been studied extensively with the FRG since the beginning of the 90ties. Apart from a plethora of research works this also has led to a number of QCD-related reviews that cover some or many of the relevant physics aspects, see \cite{Jungnickel:1997ke, Litim:1998nf, Berges:2000ew, Polonyi:2001se, Pawlowski:2005xe, Schaefer:2006sr, Gies:2006wv, Sonoda:2007av, Rosten:2010vm, Braun:2011pp, Schaefer:2011pn, Pawlowski:2014aha, Strodthoff:2016dip, Klein:2017shl}. Due to its paradigmatic nature we have already shown some results for Yang-Mills and QCD correlation functions in Fig.~\ref{fig:YMprops}. Here we give a brief overview of the progress in describing QCD  physics with the functional RG while also reporting on further technical advances within the FRG.

\subsubsection{Confinement and the flow of composite operators}\label{sec_QCD-Conf} 

We start our investigation with a discussion of confinement, which is the peculiar property in QCD that the potential $V^\textrm{\tiny{static}}_{q\bar q}(r)$ between a \textit{static} quark--anti-quark pair at a large spatial distance $r$ grows linearly with that distance, 
\begin{align}
\label{eq:Vqbarq} 
V^\textrm{\tiny{static}}_{q\bar q}(r\to\infty )\simeq \sigma r\,,\qquad \qquad V^\textrm{\tiny{static}}_{q\bar q}(r\to 0)\simeq \frac{\alpha_s(r)}{r}\,.
\end{align}
For small differences, it resembles a Coulomb potential with a logarithmically decreasing coupling $\alpha_s(r\to 0) \to 0$ owing to asymptotic freedom. The proportionality factor $\sigma$ is called the string tension and has the dimension of an inverse area. The linear rise in \eqref{eq:Vqbarq}  entails that the energy between the quark and the anti-quark is stored in a flux tube between them. In QCD with dynamical quarks this asymptotic linear potential is not to be seen, since for sufficiently large distances $r\gtrsim r_\textrm{sb}$, larger than the string-breaking scale $r_\textrm{sb}$, the flux-tube energy is large enough to allow for the generation of a quark--anti-quark pair that shields the original $q\bar q$-pair. Accordingly, for distances $r> r_\textrm{sb}$, the quark--anti-quark potential is that of a color-dipole pair and levels off. 

String breaking cannot happen in quenched QCD and pure Yang-Mills theory, and the potential or free energy of a static $q\bar q$-pair indeed shows an asymptotic linear growth. The operator, that describes the generation of a static quark--anti-quark pair at a time $t=0$ and distance $r=\vec x-\vec y$ and its annihilation at time $t$, is the traced Wilson loop. The Wilson loop is the path-ordered exponential of the gauge field with the path being the rectangular wordline ${\cal C}_{t,r}$ of the quark--antiquark pair, see \eqref{eq:Wilson-Polyakov}. The free energy or potential in Euclidean spacetime is proportional to the logarithm of the traced Wilson loop  expectation value, see \eqref{eq:Wilson-Polyakov}, $V_{q\bar q}(r) \propto \log \left \langle W(t,r)\right\rangle $. 

Accordingly, confinement in Euclidean spacetime is signalled by an area law of '$\log \left \langle W(t,r)\right\rangle$', and also implies a linear rise of this observable in both $r$ and $t$. For the sake of simplicity we have restricted ourselves to rectangular paths of the static situation described above. Let us now also assume for the time being, that our Euclidean spacetime is a box  $T^4=\prod_{i=0}^4 [0,L_i]$ with periodic boundary conditions in all directions (four-dimensional torus). This is the spacetime manifold one considers on the lattice. Then, for large temporal distances $t\to  L_0$ the Wilson loop is proportional to the two-point correlation function of two Polyakov loops, the latter winding around the full temporal extent, to wit, 
\begin{align}
\label{eq:Wilson-Polyakov}
\lim_{t\to L_0}\left \langle W(t,r)\right\rangle \propto \langle L(\vec x)  L(\vec y) \rangle\,, \qquad \textrm{with}\qquad   W(t,r) = \frac{1}{N_c} \textrm{tr} \, {\cal P }  e^{ - \imag g_s  \int_{ {\cal C}_{t,r}} (x,y) 
	A_\mu(z) d z_\mu} \,, 
\end{align} 
where $\cal P$ stands for path ordering. The traced Polyakov loop $L(\vec x)$ is the traced Wilson loop, that winds around the full temporal extent of the torus $T^4$,  
\begin{align}
\label{eq:Polyakov} 
L(\vec x) = \frac{1}{N_c}\textrm{tr} \, P(\vec x) \qquad \textrm{with}\qquad P(\vec x)=
  {\cal P} e^{-\imag g_s  \int_0^{L_0} A_0(\tau,\vec x) \,d\tau}\,, 
\end{align}
The proportionality constant in the relation \eqref{eq:Wilson-Polyakov} tends to zero in the limit $L_0\to \infty$. In \eqref{eq:Wilson-Polyakov}, \eqref{eq:Polyakov} the traces are taken in the fundamental representation, since the quarks carry this representation. The relation \eqref{eq:Wilson-Polyakov} maps the Wilson loop to the two-point correlation function of a gauge-invariant operator that is local in position space. The area law for the logarithm of the Wilson loop expectation value is in one-to-one correspondence to a linear dependence on the distance $r$ for the logarithm of the correlator of Polyakov loops. 

However, as the Polyakov loops span the whole temporal extent of the manifold, in the vacuum one typically uses the area law for the Wilson loop as a signature for confinement. Still, with a proper normalization or limit the Polyakov loop correlator signals confinement in the vacuum, and it is the confinement signature that has been studied in the FRG-literature. While it is not commonly used in the vacuum, the Polyakov loop correlation is typically used as an order parameter for the confinement-deconfinement phase transition at finite temperature $T$, with the identification $T=1/L_0$.  There, for large distances we have declustering,  
\begin{align}
\lim_{r\to\infty}\langle L(\vec x)  L(\vec y) \rangle \to \langle L(\vec x) \rangle \langle  L(\vec y) \rangle\,, 
\end{align}
and confinement entails the vanishing of the Polyakov loop expectation value itself, $\langle L(\vec x) \rangle=0$. This happens for low temperatures and for $T=0$ it is tantamount to confinement in the vacuum. In turn, in the deconfined phase for $T\to\infty$ with small fluctuations of the temporal gauge field $A_0$, the traced Polyakov loop is finite and tends to unity. The Polyakov loop also allows to easily identify the symmetry behind the confinement-deconfinement symmetry in Yang-Mills theory: the center symmetry of the gauge group. While the action is invariant under center transformations, the traced Polyakov loop is not. In SU($N_c$) we have $P(\vec x) \to z\,P(\vec x)$ with $z\in Z_{N_c}$. The elements of the center group $Z_{N_c}$ are the $N_c$th roots of unity, $z^{N_c}=\mathbbm{1}$.  Accordingly, the symmetry group of the phase transition is the center $Z_{N_c}$, and we can deduce the universality class of the phase transitions for different $N_c$. For SU(2) the center is $Z_2$ and the transition is a second order phase transition and lies in the Ising universality class. For SU(3) and all higher $N_c$ the phase transition is of first order.  \footnote{This analysis also implies that strictly speaking the full breaking of center symmetry underlying confinement is signalled by $\langle \textrm{tr} \,{\cal P}^n(\vec x)\rangle =0$ for $n=1,...,N_c^2-1$, see e.g. \cite{Ford:1998bt, vanBaal:2000zc, Reinosa:2014ooa, Herbst:2015ona} and references therein.} 

A challenge specific for the FRG-approach to QCD is the fact, that the above observables are infinite-order correlation function in the gauge field and the FRG-approach is based on the computation of finite-order correlation function.  Indeed, the pivotal ones are the propagators, the two-point correlation functions of the theory. There are several ways out, all of which have been undertaken in the FRG:  

A first option is the reformulation of QCD in terms of Wilson lines along the lines of \cite{Morris:1999px}. These reparameterisations of theories are very naturally implemented in the FRG as it is typically formulated such that it is quadratic in the degree of freedom that is regularised, see \cite{Pawlowski:2005xe} for a detailed discussion. Another possibility is the reparameterisation in terms of a suitable gauge fixing that simplifies the relation between the Wilson and Polyakov loops and the gauge field. In finite temperature gauge theories a natural choice is the Polyakov gauge, that has been studied in \cite{Marhauser:2008fz, Kondo:2010ts} for $SU(2)$. For this choice the flow equation for the effective potential $V_\textrm{eff}[A_0]$ resembles that of the effective potential of a real scalar field\footnote{In contradistinction to the effective potential of a the real scalar field, $V_\textrm{eff}[A_0]$ is singular at the maximally center-symmetry breaking points with $|L[A_0]|=1$. These singluarities are already present at one-loop. Consequently the effective potential is convex within the Weyl chambers. This complicates its numerical solution for $SU(N>2)$ and requires advanced numerical techniques such as put forward in \cite{Grossi:2019urj}.}, and the second order phase transition with Ising universality class in $SU(2)$ Yang-Mills theory has been found. 

The second option is to define variants of the standard order parameters, the Wilson loop $\langle W\rangle$ and the Polyakov loop $\langle L\rangle$, that can be defined in particular from correlation functions within standard covariant gauges \eqref{sec_he:eq:LandauGauge}, and in particular in the Landau(DeWitt) gauge. The latter is the best studied and numerically most advanced and successful formulation of QCD with functional methods, as discussed before. 

In the Landau-DeWitt gauge we can compute the gauge-invariant background-effective action $\Gamma[A ]$ from the flow equation of the effective action with fluctuation fields. At finite temperature, $\Gamma[A]$ includes a gauge-invariant effective potential $V_\textrm{eff}[A_0]$ of the temporal gauge field $A_0$. This potential can be understood as an order-parameter potential: it inherits  center symmetry from the effective action, but its minima $A_{0,\textrm{EoM}}$ may or may not break center symmetry. Indeed, the Polyakov loop of the solution of the EoM, $A_{0,\textrm{EoM}}$, vanishes at the center-symmetric points of $A_{0,\textrm{EoM}}$. Accordingly, it vanishes for temperatures below the confinement temperature $T_c$ 
and is non-vanishing above $T_c$,   
\begin{align}\label{sec_he:eq:PolCrit}
L[A_{0,\textrm{EoM}}] \  \left\{\begin{array}{lr} = 0 & T<T_c \\[1ex] 
\neq 0 & T>T_c
	\end{array}  \right.\,. 
\end{align}
This functional approach to confinement is based on corresponding studies of the high temperature regime in perturbation theory founded in \cite{Gross:1980br, Weiss:1980rj}. It has been put forward in \cite{Braun:2007bx} within the FRG and has been extended later also to general functional approaches, \cite{Fister:2013bh}. By now results have been obtained for generic $SU(N_c)$ gauge groups in Yang-Mills theory, \cite{Braun:2010cy}, and for full dynamical two-flavor QCD in the chiral limit, \cite{Braun:2009gm}. For results within this approach obtained with other functional methods see \cite{Fister:2013bh, Fischer:2013eca, Fischer:2014vxa, Fischer:2014ata, Reinosa:2014ooa, Reinosa:2014zta, Reinosa:2015oua, Reinosa:2015gxn, Reinosa:2016iml, Maelger:2017amh, Reinhardt:2012qe, Reinhardt:2013iia, Heffner:2015zna, Quandt:2016ykm}.

Moreover, it has been shown in \cite{Braun:2007bx}, that confinement in covariant gauges necessitates a relative gapping of the gluon propagator to that of the ghost: confinement implies the vanishing of \eqref{sec_he:eq:PolCrit} and the part of the Polyakov loop potential stemming from the dressed gluon loop in the flow equation is deconfining. In turn, the contribution from the ghost loop is confining, which originates from the relative fermionic minus sign of the ghost loop. For large temperatures (or in perturbation theory) the gluon loop dominates. At small temperatures the ghost loop has to take over in order to guarantee confinement. However, it only dominates if the gluons are gapped relative to the ghost. The details of this mechanism have been corroborated and extended in \cite{Braun:2010cy, Fister:2013bh}. The gluon and ghost propagators in covariant gauges have this property, for a detailed discussion see Sec.~\ref{sec_hep-locality}. Respective FRG-results and applications for the gapped gluon propagator in Yang-Mills theory and QCD in the Landau gauge are given in \cite{Pawlowski:2003sk, Fischer:2004uk, Fischer:2006vf,	Fischer:2008uz, Fischer:2009tn, Cyrol:2016tym, Cyrol:2017ewj, Corell:2018yil}, for related work see \cite{Ellwanger:1995qf, Ellwanger:1996wy, Bergerhoff:1997cv, Gies:2006nz, Eichhorn:2010zc}.

The results for \eqref{sec_he:eq:PolCrit} show a second order phase transition for $SU(2)$ Yang-Mills theory,  while the phase transition is of first order for higher $SU(N_c>2)$, and a  crossover\footnote{Dynamical quarks break the center symmetry of the gauge group as they live in the fundamental representation. Hence for sufficiently small quark masses the phase transition in QCD is a crossover.} for  two-flavor QCD. This agrees with the expected orders of the phase transitions, and the critical temperatures are in quantitative agreement with the lattice results. 

Finally, the results for $L[A_{0,\textrm{EoM}}]$ can be used within the flow of Wilson- and Polyakov-loop observables. The latter can be studied with the flow equation for general composite operators derived in \cite{Pawlowski:2005xe, Igarashi:2009tj, Pagani:2016pad}, 
\begin{align}
\label{eq:FlowComposite} 
\partial_t {\cal O}_k[\Phi] = -\frac12 \textrm{Tr} \, G_k[\Phi]\,\partial_t R_k\,G_k[\Phi] \, {\cal O}^{(2)}_k[\Phi]\,. 
\end{align}
In \eqref{eq:FlowComposite} the operator ${\cal O}^{(2)}_k$ is the second derivative of ${\cal O}_k$ with respect to the fields. This operator is contracted with $G_k\,\partial_t R_k\,G_k$ in the trace. The set of composite operators ${\cal O}_k[\Phi] $ with the flow \eqref{eq:FlowComposite} includes general correlation functions with their connected and disconnected parts as well as functions of the source $J[\Phi]$. For the latter case further terms enter \eqref{eq:FlowComposite}, see \cite{Pawlowski:2005xe}. An educative example  for the former case of general correlation functions and the  necessity of including the disconnected terms is the full two-point function $G_{\Phi_1\Phi_2} +\Phi_1 \Phi_2$. Eq.~\eqref{eq:FlowComposite} has been put to work in Yang-Mills theories \cite{Herbst:2015ona} for the traced Polyakov loop observables. In gravity it has been used in  
\cite{Pagani:2016dof, Becker:2019fhi, Houthoff:2020zqy, Kurov:2020csd} for the study of the renormalization and scaling of composite operators, see Sec.~\ref{sec_gr}. 

Evidently, the expectation value of the traced Polyakov loop falls into the class of general correlation functions ${\cal O}[\Phi]$ with the flow \eqref{eq:FlowComposite}, and its flow is given by 
\begin{align} \label{eq:PolFlow}
\partial_t \langle L \rangle [\Phi] = -\frac12 \textrm{Tr} \, G_k[\Phi]\,\partial_t R_k\,G_k[\Phi] \, \langle L\rangle^{(2)}[\Phi]\,.
\end{align}
The flow \eqref{eq:PolFlow} has been solved in a first approximation in an expansion of $\langle L\rangle[\Phi]$ about $L[A_{0,\textrm{EoM}}]$ in \cite{Herbst:2015ona} for $SU(3)$ Yang-Mills theory. The result is already in quantitative agreement with the lattice results for the Polyakov loop expectation value. 

In summary by now the FRG-approach offers a well-developed and quantitative access to quite some aspects of the exciting phenomenon of confinement and in particular to the confinement-deconfinement phase transition at finite temperature. Remaining challenges concern in particular the thermodynamics of gauge theories, see e.g. \cite{Fister:2011uw, Fister:2012lug, Fister:2013bh}. Its resolution may also require a quantitative access to the subleading momentum and frequency dependence of correlation functions, for first work in this direction as well as non-relativistic analogues see \cite{Boettcher:2012dh, Hajizadeh:2019qrj}.

\subsubsection{Chiral symmetry breaking}\label{sec_QCD-Chiral}

\begin{figure}[t]
	\begin{center}
		\includegraphics[width=0.45\columnwidth]{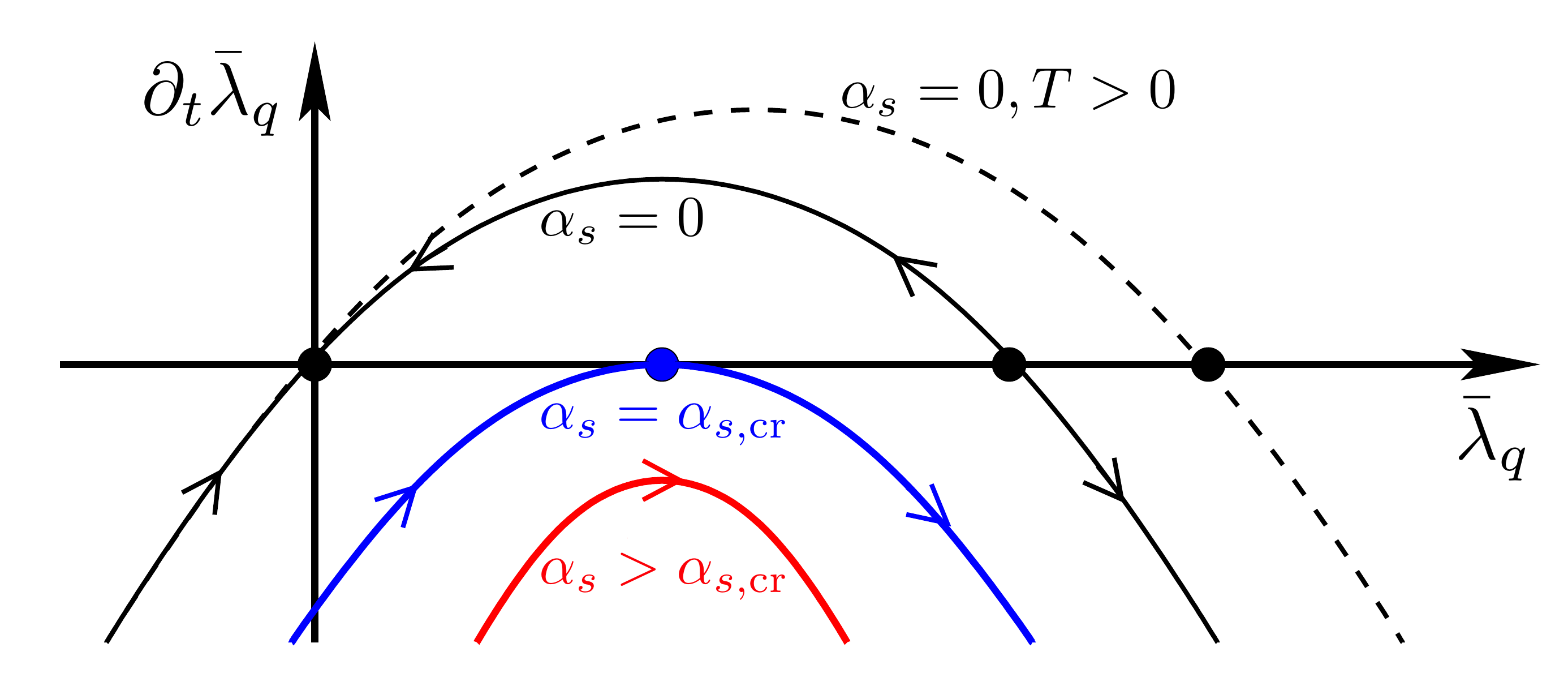}
		\caption{Flow $\partial_t\bar{ \lambda}_{q}$ of the dimensionless four-quark coupling $\bar\lambda_q=\lambda_q k^2$ in the scalar-pseudoscalar channel as a function of the four-quark coupling for different values of the strong couplings $\alpha_s$. If the strong coupling is large enough, $\alpha_s>\alpha_{s,\textrm{cr}}$ with the critical coupling $\alpha_{s,\textrm{cr}}$, the flow is negative for all $\bar{ \lambda}_q$. Then,  chiral symmetry breaking happens for all initial values of $\bar\lambda_q$, for more details see the discussion below \eqref{sec_he:4quarkflow}.}\label{sec_he:lambdaparabolaofgandT}
	\end{center}
\end{figure}

The ultraviolet behaviour of QCD is governed by \textit{asymptotic} freedom, the theory approaches the ultraviolet Gau\ss ian fixed point with vanishing strong coupling $\alpha_s(p\to\infty) =0$ with $\alpha_s = g_s^2 /(4 \pi)$. In turn, in the infrared the coupling grows strong and triggers chiral symmetry breaking. This is easily investigated from  the flow of the dimensionless four-quark interaction $\bar \lambda_q \propto  \lambda_q \,k^2$ in the scalar--pseudoscalar
channel:
$-\lambda_q/4 \int_x \left[(\bar q\, q)^2 - (\bar q\imag \gamma_5 \vec
  \tau \, q)^2 \right]$. Its flow reads schematically
\begin{align}\label{sec_he:4quarkflow}
  \partial_t \bar\lambda_q = 2 \bar\lambda_q - A_k\, \bar\lambda_q^2 -
  B_k\bar\lambda_q \alpha_s  - C_k\, \alpha_s^2 + \textrm{tadpole-terms}\,,
\end{align}
with positive constants $A_k, B_k, C_k$ that depend on other parameters of the theory, and in particular on the mass scales. It is depicted in Fig.~\ref{sec_he:lambdaparabolaofgandT}. In \eqref{sec_he:4quarkflow}, $2 \bar\lambda_q$ is the canonical scaling term and the other terms are quantum corrections computed from the respective diagrams in the flow. For $\alpha_s=0$ the flow \eqref{sec_he:4quarkflow} is that of the four-quark coupling in an NJL-type model. Then, for large enough initial coupling $\left.\bar\lambda^2_{q}\right|_{k=\Lambda}>2/A_\Lambda$ at the UV cutoff scale $\Lambda$, the $\beta$-function $\partial_t\bar\lambda_q$ is negative and the coupling grows towards the infrared. It finally diverges at a pole that signals chiral symmetry breaking. In turn, for $\bar\lambda_{q,\Lambda}<2/A_\Lambda$ the coupling weakens towards the infrared and runs into the Gau\ss ian fixed point without chiral symmetry breaking.

When switching on the strong coupling, the $\beta$-function of the four-quark coupling in \eqref{sec_he:4quarkflow} is deformed: Firstly, the canonical running gets an anomalous part with $2\bar\lambda_q\to (2-B_k \alpha_s)\bar\lambda_q$. More importantly the whole $\beta$-function is shifted down globally by $- C_k\, \alpha_s^2$, see Fig.~\ref{sec_he:lambdaparabolaofgandT}. This term originates from quark-gluon box
diagrams and is negative. For large enough $\alpha_s$ the $\beta$-function $\partial_t \bar{ \lambda}_q$ is negative  for all $\bar\lambda_q$, 
\begin{align}\label{sec_he:alpha*}
\partial_t \bar\lambda_q<0\quad \forall \bar \lambda\quad \textrm{if}\quad    \alpha_s > \alpha_{s,\textrm{cr}}\,,\qquad \textrm{with}\qquad
  \alpha_{s,\textrm{cr}}=\frac{2}{B_k + 2 \sqrt{A_k C_k}}\,, 
\end{align}
with the critical coupling $ \alpha_{s,\textrm{cr}}$, see Fig.~\ref{sec_he:lambdaparabolaofgandT}. Accordingly, if the growth of the strong coupling towards the infrared is unlimited, chiral symmetry breaking in QCD is always present, and is basically the converse of asymptotic freedom. More precisely it is the one-gluon exchange coupling in the quark-gluon box diagrams that has to satisfy \eqref{sec_he:alpha*} in the infrared. In this context it is important to mention that it is the gapping of the gluon related to confinement that stops the growth of the exchange couplings in the infrared and even leads to their decay for small momenta, see Fig.~\ref{fig:runningcouplings}. 

The simple relation between the size of the strong coupling and chiral symmetry breaking also provides a simple explanation for the restoration at finite temperature: the strong coupling melts down and finally we have $\alpha_s< \alpha_{s,\textrm{cr}}$ for all frequencies and momenta, and spontaneous chiral symmetry breaking cannot happen any more. 

We emphasise that while the above argument has been discussed within a relatively simple approximation, its structure carries over straightforwardly to the full system of flow equations in QCD. For FRG-literature in gauge theories on the many facets of chiral symmetry breaking in the vacuum, \cite{Pawlowski:1996ch, Aoki:1996fh, Aoki:1999dv, Aoki:1999dw, Aoki:2000dh, Meggiolaro:2000kp, Gies:2002hq, Aoki:2009zza, Aoki:2009zza, Aoki:2012mj, Aoki:2014ola, Mitter:2014wpa, Braun:2014ata, Rennecke:2015eba, Cyrol:2017ewj, Fu:2019hdw}.

In combination the results and mechanisms discussed in the last two sections lead to the interesting interplay between chiral symmetry breaking and confinement. While being short of a proof, the following picture unfolds: While a growing one-gluon exchange coupling leads to chiral symmetry breaking, confinement requires a gapped gluon which
stops this growth and even leads to exchange couplings that decay in the infrared. In turn, massless quarks lead to a massless dispersion in the gluon (for $N_f$ large enough) and hence no confinement. These properties relate to the {'t\phantom{-}Hooft} anomaly matching argument of \textit{'No confinement without	chiral symmetry breaking'} (for $N_f\geq 3$).

\subsubsection{Vacuum QCD and hadronic bound states} \label{sec_he:VacuumQCD}

The mechanisms of chiral symmetry breaking and confinement as described in such a gauge-fixed functional approach lead to a very robust access to both phenomena. In particular the above mechanism of chiral symmetry breaking yields very robust results for strong couplings $\alpha_s$ that far exceed the infrared bound $\alpha_s> \alpha_* $ in \eqref{sec_he:alpha*} for low cutoff scales. The strength of chiral symmetry breaking is measured in the chiral condensate $\Delta_q \propto \int_x \langle \bar q(x) q(x)\rangle$, which agrees well with that in full QCD as measured on the lattice or within quantitative computations with functional methods already in crude approximations. 

These approximations have to be improved for a reliable quantitative, and even for a qualitative, access to the hadron structure as well as the phase structure of QCD. In the ongoing and crucial quest for quantitative precision it has turned out that the chiral and confinement dynamics of QCD leads to gluon exchange couplings close to the bound \eqref{sec_he:alpha*}. In this regime the size of the chiral symmetry breaking, i.e.\ the size of the chiral condensate $\Delta_q$ strongly depends on the size of the running coupling: variations of $\alpha_s$ by a few percent lead to variations of $\Delta_q$ by factors $1/2$ to $2$ or more, see \cite{Mitter:2014wpa, Cyrol:2017ewj}. In short, \textit{QCD is living on the edge}. One may be tempted to classify this behaviour as a merely technical challenge, but it has far-reaching physics consequences: Even small changes of external parameters such as temperatures, density or (chromo-) electric and (chromo-) magnetic background fields potentially have a large impact on observables.

Accordingly, the qualitative and even more so quantitative access to strongly correlated infrared QCD from first principles (the only input being the fundamental parameters of QCD, the current quark masses) requires advanced approximation schemes.  Advanced numerical and computer algebraic tools for the derivation of large systems of flow equations, Dyson-Schwinger equations and Slavnov-Taylor identities, as well as their numerical solution have been developed in \cite{Huber:2011qr, Huber:2019dkb, Cyrol:2016zqb, github:FormTracer}. Related first principles work can be found in \cite{Mitter:2014wpa, Cyrol:2016tym, Cyrol:2017ewj, Corell:2018yil, Hajizadeh:2019qrj}. While being developed in the context of QCD, they also can be readily applied to generic relativistic and non-relativistic systems.

One of the chiefly interesting properties of QCD is the transformation of a weakly-interacting almost chiral theory of quarks and gluons for large momentum scales into a weakly-interacting theory of hadrons with large masses and the light pions at small momentum scales, governed by chiral perturbation theory. The formation of the hadronic bound states and the related dynamical change of the relevant degrees of freedom is very well captured with \textit{dynamical hadronization} \cite{Gies:2001nw, Gies:2002hq}, see also \cite{Ellwanger:1993mw,  Ellwanger:1994wy}. Further formal developments can be found in \cite{Pawlowski:2005xe, Floerchinger:2009uf, Fu:2019hdw}, applications to QCD are found in \cite{Mitter:2014wpa, Braun:2014ata,  Cyrol:2017ewj, Alkofer:2018guy, Fu:2019hdw}. It is reminiscent of a Hubbard-Stratonovich transformation, but does not share the potential double-counting problems of the latter that are well-known from applications in low-energy effective field theories. 

Here we explain this problem at the example of the scalar channel of the four-quark interaction already discussed in the last section Sec.~\ref{sec_QCD-Chiral} on chiral symmetry breaking. The Hubbard-Stratonovich transformation of the four-quark term is a reformulation in terms of effective exchange fields (here the scalar $\sigma$-field) and a Yukawa interaction,
\begin{align}\label{sec_he:HS}
  - \frac{\lambda_q}{4} (\bar q\, q)^2  =
  \left[   \frac12 m_\phi^2 \sigma^2 + \frac12 h_\sigma \sigma \bar q
  q\right]_{\sigma_\textrm{EoM}}\,, \qquad \textrm{with} \qquad 
  \frac{h_\sigma^2}{2 m_\phi^2}=\lambda_q \qquad \textrm{and} \quad \sigma_\textrm{EoM}(q,\bar q)= \frac{h_\sigma}{2 m_\phi^2} \bar q q\,,
\end{align}
where $\sigma_\textrm{EoM}(q,\bar q)$ is the solution of the $\sigma$- equation of motion (EoM). Note that the Yukawa coupling and meson mass function $m_\phi$ can also be chosen such that only a part of the four-quark interaction is captured by the right hand side. Double counting may arise due to the fact that the original four-quark interaction can now be represented both in a purely fermionic or a mixed fermion-scalar form. While this is easily disentangled on the classical level, it is hard to resolve on the quantum level for the quantum effective action.

\begin{figure}[t]
	\begin{center}
		\subfigure[Full four-quark interaction in the scalar-pseudoscalar channel as an effective scalar-pseudoscalar meson exchange and a residual four-quark interaction (square). The meson exchange diagram provides the scalar-pseudoscalar four-quark coupling at $(p_1+p_3)^2=(p_2+p_4)^2 =0$ ($s$-channel). Higher order
		contributions in the flow \eqref{sec_he:eq:flow4q} are dropped. ]{\hspace{-0.0cm}	
			\label{sec_he:fig:rebos_definition2}
			\includegraphics[scale=0.23]{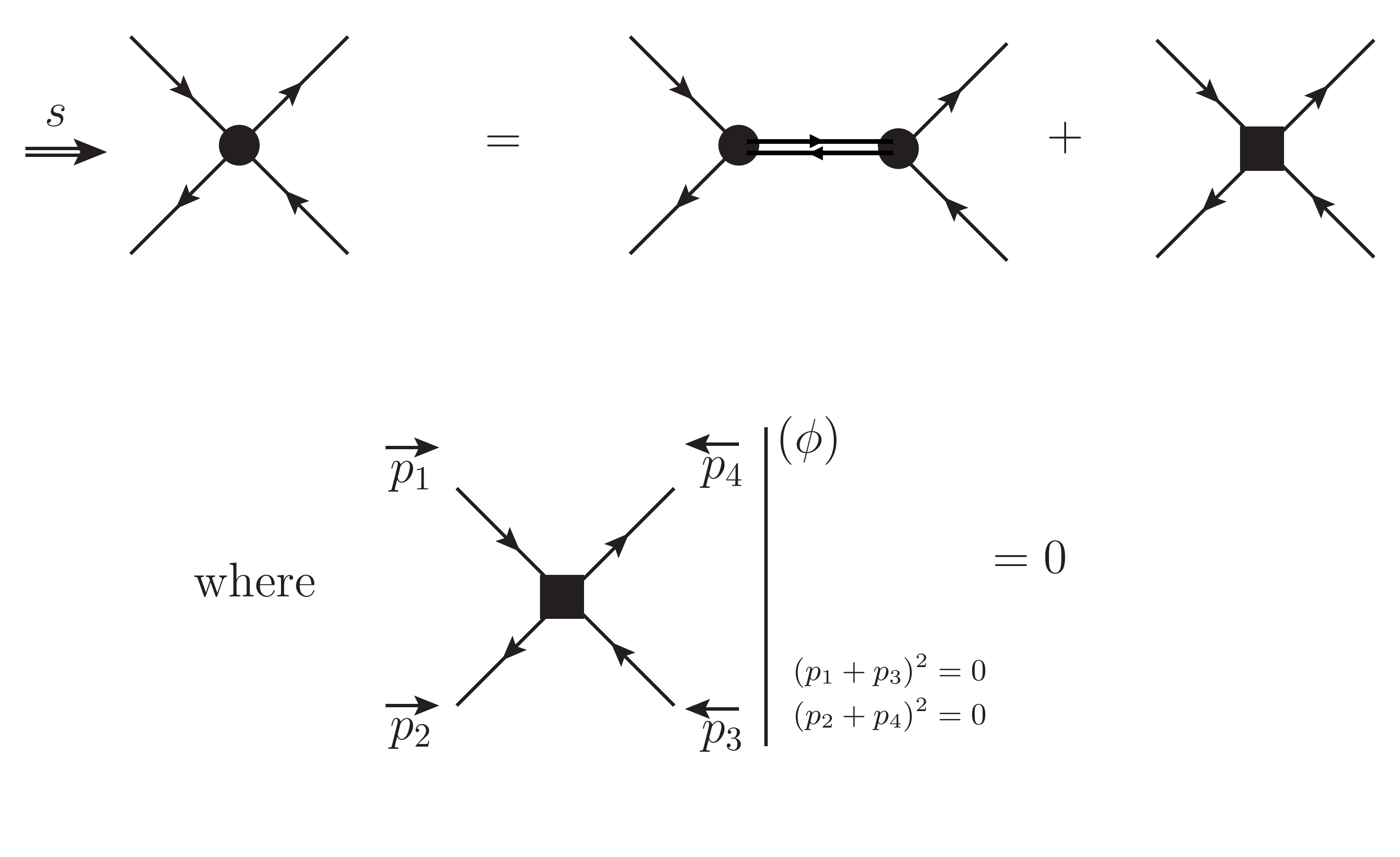}}	\hspace{0.5cm}
		\subfigure[Cutoff-dependence of gluon-exchange coupling 
		for the light $u,d$ quarks ($g^2_{\bar l A l}$) and the strange 
		quark ($g^2_{\bar s A s}$).  Meson-exchange
		couplings ($\bar h^2 /(1+\tilde m^2)$ with $\tilde m^2 = m^2/k^2$. Gluons,
		quarks, and mesons decouple sequentially from the matter
		dynamics. Figure taken from \cite{,Fu:2019hdw}.]{\hspace{-0.0cm}	\label{fig:SequentialDecoupling}
			\includegraphics[scale=0.54]{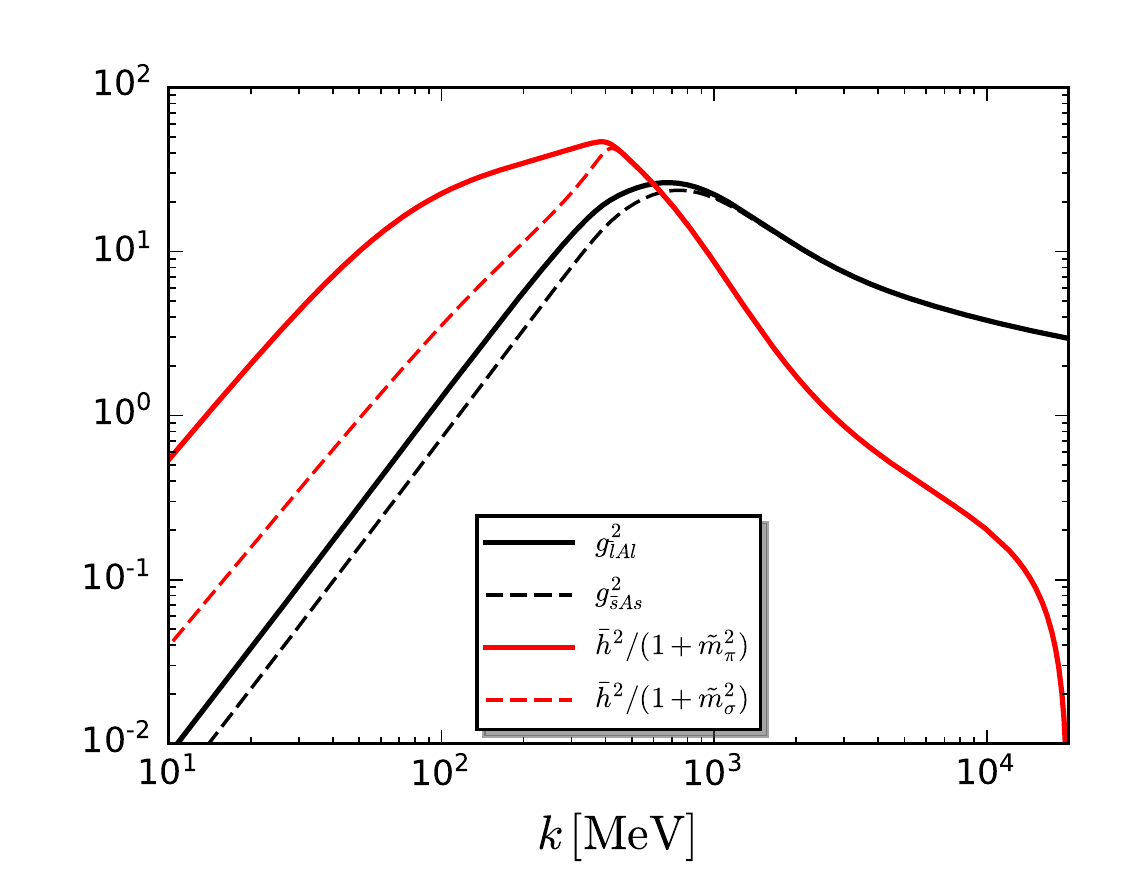}\hspace{0.0cm}}	
	\end{center}
	\caption{Four-quark coupling with dynamical hadronization (left), strength of exchange couplings in physical $N_f=2+1$ flavor QCD (right). } 
	\label{fig:DynHad}
\end{figure}

Within the FRG-approach this problem is avoided with \textit{dynamical hadronization}.  There, the dynamical re-distribution of the full scalar four-quark interaction in the effective action is described by the flow of the four-quark coupling, 
\begin{align}
  \partial_{t}\bar{\lambda}_{q} - 2\left( 1+\eta_q\right)
  \bar{\lambda}_{q}
  -\bar{h}_\sigma \,\dot{\bar A}
  =  \overline{\textrm{Flow}}^{(4)}_{(\bar q q)^2}\,. 
 \label{sec_he:eq:flow4q}
\end{align}
In \eqref{sec_he:eq:flow4q} the right hand side stands for the diagrams of the scalar four-quark flow. Now this flow also includes diagrams with the effective meson field, and the subscript ${}_{(\bar q q)^2}$ indicates the projection on the scalar part of the interaction. Moreover, for the time being we only consider vanishing momenta. The term proportional to $\dot{\bar A}$ is the analogue of the choice for $h_\sigma^2/(2 m_\phi^2)$ in the Hubbard-Stratonovich transformation. The typical choice for the dynamical hadronization parameter $\dot{\bar A}$ is such that the flow of the four-quark coupling vanishes identically, $\partial_{t}\bar{\lambda}_{q} \equiv 0$. This leads us to $\dot{\bar A} = - \overline{\textrm{Flow}}^{(4)}_{(\bar q q)^2}/h_\sigma$. Then, with the initial four-quark interaction $\Gamma^{(4)}_{(\bar q\bar q)^2,\Lambda}\equiv 0$ at a large initial cutoff scale $\Lambda$ the full scalar channel of the four-quark interaction is parametrised in terms of the exchange fields $\sigma$ and $\vec \pi$. For a diagrammatic depiction see  Fig.~\ref{sec_he:fig:rebos_definition2}, where we have dropped higher order loop terms. 

%
%%%%%%%%%%%%%%%%%%%%%%%%%%%%%%%%%%%%%%%%%%%%%%%%%%%%%%%%%%%%%%%%%%%%%
%%%%%%%%%%%%%%%%%%%%%%%%%%%%%%%%%%%%%%%%%%%%%%%%%%%%%%%%%%%%%%%%%%%%%%
%

In QCD the above simple example of a scalar four-quark coupling has to be embedded in the more complicated four-quark interaction structure of the theory. For example, in two-flavor QCD we have 10 momentum-independent four-quark tensor structures that respect the symmetries of the theory, and in 2+1-flavor QCD this number increases to 27 momentum-independent tensor structures.  Each tensor structure comes with a scalar dressing function depending on all momenta. The size of the full basis of tensor structures including momentum-dependent ones is even larger by an order of magnitude, see e.g.~\cite{Eichmann:2015nra}. The latter extension is important for quantitative studies of the hadron-resonance structure.  A fully reliable treatment of even the momentum-independent tensor structures necessitates to disentangle all of them  in order to avoid the Fierz ambiguity. This will be discussed further in Sec.~\ref{sec:QCD1st}, for a respective review see \cite{Braun:2011pp}. 

In QCD, dynamical hadronization has been used for the scalar-pseudoscalar ($\sigma-\pi$) channel of the four-quark interaction. On the level of the effective action this leads to an extension of the field content by the mesonic $O(4)$-field $\phi=(\sigma, \vec \pi)$.  We emphasize, that the effective action $\Gamma_\textrm{DynHad}$ with these additional mesonic fields is still an effective action of first principles QCD: The standard effective action of QCD, $\Gamma_\textrm{QCD}[A_\mu, c,\bar c, q, \bar q]$, can be obtained from $\Gamma_\textrm{DynHad}$ with 
\begin{align}\label{sec_he:eq:GQCD=GDynHad}
  \Gamma_\textrm{QCD}[A,c,\bar c,q,\bar q]=
  \Gamma_\textrm{DynHad}[A,c,\bar c,q,\bar q,\phi_\textrm{EoM}]\,, 
\end{align}
where the mesonic solution of the EoM depends on the fundamental fields, $\phi_\textrm{EoM}=\phi_\textrm{EoM}[A,c,\bar c,q,\bar q]$. This is reminiscent of going from the  2PI-effective action to the 1PI-effective action with the EoM for the propagator. Indeed, technically $\Gamma_\textrm{DynHad}$ is the 2PPI (two-point particle) effective action. For more details see in particular \cite{Fu:2019hdw}. 

Dynamical hadronization is nothing but a dynamical change of the field basis in the theory. Within an optimal use it allows to expand the dynamics in terms of weakly interacting degrees of freedom at all cutoff scales and hence speeds-up the convergence of a given expansion scheme. In particular, the low energy limit of higher order scatterings of resonant interaction channels is now described very efficiently in terms of an effective potential $V_\textrm{eff}(\rho)$ of the composite mesonic field $\phi$ with $\rho=\phi^2/2$. In the original formulation in terms of quarks the higher order terms $\rho^2, \rho^3,\rho^4,...$ are related to channels of multi-quark scattering vertices, $(\bar q q)^3,(\bar q q)^4,(\bar q q)^5, ...$ which are technically less accessible. Note however, that dynamical hadronization does not describe the complete four-quark interaction, but only a given momentum channel of a specific tensor structure. The remnant vertex has to be kept, in particular if discussing competing order effects, see for example \cite{FukushimaPawlowskiStrodthoff, Denz:2019ogb}. This is depicted in Fig.~\ref{sec_he:fig:rebos_definition2}. A complete bosonization in terms of a series expansion has been discussed in \cite{Jakovac:2018dkp, Jakovac:2019zzw}.  

In Fig.~\ref{fig:SequentialDecoupling} one also sees the sequential decoupling of the different degrees of freedom  within the flow equation for first principles QCD with dynamical hadronization of the $(\sigma-\pi)$ channel depicted in Fig.~\ref{sec_he:QCD_equation}:  The full field-dependent propagators of gluons, quarks and mesons and hence the respective exchange couplings are gapped with different mass scales. In the ultraviolet regime for $k\gtrsim 1$\, GeV, the quark-gluon dynamics is dominant as the gluon-exchange couplings dominate over the other couplings, see  Fig.~\ref{fig:SequentialDecoupling}. At about 600\,MeV, the quark-meson dynamics takes over and the gluon dynamics decouples due to the  gluon mass gap $m_\textrm{gap}$. The latter is the confinement scale in disguise, see also Sec.~\ref{sec_QCD-Conf}. Finally only the dynamics of the pseudo-Goldstone bosons, the pions, survives. Below the pion mass also the latter dynamics dies out. Note that the auxiliary ghost field is massless but only interacts with the matter fields via the gluon. Accordingly the ghost decouples together with the gluon from the matter sector. 

Further hadronic resonances (and heavier quarks) are far too heavy to give sizeable effects to the Euclidean vacuum dynamics of QCD. We rush to add that this decoupling only concerns the \textit{offshell} dynamics in the loops in the flow equation, see Fig.~\ref{sec_he:QCD_equation}. It does not entails the irrelevance of onshell hadrons for observables. However, the respective correlation functions of hadrons or more generally asymptotic states can be built from the offshell dynamics of ghosts, gluons, quarks and light mesons.

This sequential decoupling already explains the success and the natural emergence of chiral perturbation theory  from QCD. It also allows for the computation of its low energy parameters within the FRG, for recent works see \cite{Eser:2018jqo, Divotgey:2019xea}. The concept of dynamical hadronization straightforwardly carries over to gluonic excitations such as glueballs or to effective degrees of freedom inside hadrons such as the pomeron. For first FRG-steps in the latter direction see \cite{Bartels:2018pin, Bartels:2019qho}, for the description of baryonic degrees of freedom see \cite{Weyrich:2015hha}.  Hence, dynamical hadronization is a very natural way to describe the quark-hadron duality in QCD. 

Another very interesting link is that to bound-state equations of QCD, Bethe-Salpeter equations (BSEs), Faddeev equations (FEs), and higher ones for e.g.~tetraquarks, for a recent review see \cite{Eichmann:2016yit}. To see this link more clearly, we evaluate the pseudoscalar channel of the four-quark interaction close to the pion pole with $p^2 = -m_{\pi,\textrm{pole}}^2$. There, the pion propagator grows large and the higher order terms in Fig.~\ref{sec_he:fig:rebos_definition2} are suppressed. This leaves us with the exchange term depicted in Fig.~\ref{sec_he:fig:rebos_definition2}, which is given by the pion propagator close to the pole, $1/(p^2 +m_{\pi,\textrm{pole}}^2)\to\infty$, sandwiched by two Yukawa vertices. Note that these vertices are not simply $h_\sigma$ but carry all allowed tensor structures with momentum-dependent dressings. A comparison with the respective BSE expressions reveals that the Yukawa vertices $\Gamma^{(3)}_{\bar q\vec\pi q}$ can be identified with the Bethe-Salpeter wave function of the pion. The latter carries low energy constants of QCD such as the pion decay constant $f_\pi$. These relations carry over to generic BSE, FE and higher hadronic bound state equations if the respective channels are hadronized. 

In summary, dynamical hadronization allows for a natural implementation of bound-state equations of QCD within an effective action approach. Together with the recent developments for real-time flow equations, see Sec.~\ref{sec:QCDPhaseModelResults}, this opens a yet unexplored door to a uniform access to one of the major challenges in QCD, the description of the spectrum of hadron resonances. In particular this would encompass the higher states, also including decays and scattering processes. The setup is applicable to general theories and generic composite fields. It does not only allow for the hadronization of mesons, but also, e.g., baryons and tetraquarks. 

In the context of condensed matter and statistical physics systems the present setup is called \textit{dynamical condensation/pairing}. In particular ultracold atomic systems with their very exciting strongly correlated physics and pairing phenomena are ideally-suited for the application of the techniques described above. With the plethora of precise experimental results they also provide an ideal testbed for the techniques described above. For FRG-works on two-, three-, and four-body resonances with and without dynamical pairing, and related work on limit cycles in non-relativistic systems see e.g. \cite{Birse05fb, Diehl07fb, Diehl:2007ri, Diehl:2007xz, Floerchinger:2008qc, Diehl:2009ma, Krippa:2009vu, Floerchinger:2009pg, Braun:2011uq, Boettcher:2012dh, Moroz:2009nm, Floerchinger209, Floerchinger109, Schmidt:2009kq, Schmidt:2012yn, Avila:2013rda, Avila:2015iza}, for introductory reviews see \cite{Floerchinger:2011yv, Boettcher:2012cm}.

\subsection{Phase structure and dynamics of QCD}\label{sec_QCD-Phase}

A major challenge in QCD concerns its phase structure and dynamics. This is important for the explanation of heavy-ion collisions (HICs), the QCD phase transition in the early universe, and for neutron star physics.

The related physics offers several intricate and yet unsolved questions ranging from the equilibrium phase structure of QCD at finite density to the non-equilibrium dynamics at early times of a heavy ion collision. Specific questions concern in particular the existence and location of a critical end point (CEP) for the chiral and confinement-deconfinement crossovers at large density, the possible existence of mixed or inhomogeneous phases as well as the phase structure beyond nuclear densities. In the latter case most likely competing order dynamics has to be resolved, see also Sec.~\ref{sec:IVB2}. At these large densities we are also interested in the equation of state (EoS), which is chiefly important for a resolution of the mass range of neutron stars. Finally, for the description of HICs we need access to the dynamics of strongly correlated QCD from the first far from equilibrium phase over kinetic and hydrodynamical phases to transport close to equilibrium. It is in particular the non-equilibrium (real-time) physics as well as the high density regime that are at present not accessible with lattice simulations due to sign-problems. In turn, in the vacuum at $T,\mu=0$ and at finite temperature lattice simulations provide quantitative results that can be used for benchmarking computations and approximations within the FRG-approach to QCD.

\subsubsection{The unreasonable effectiveness of low-energy effective theories}\label{sec:UnreasonableEffectiveness}

The highly exciting phenomena mentioned above in the introduction of Sec.~\ref{sec_QCD-Phase} require an access to QCD at momentum and cutoff scales $p^2,k^2\lesssim 1$\, (GeV$)^2$. In this regime QCD is strongly correlated, and nonperturbative methods are required. In Sec.~\ref{sec_he:VacuumQCD} we have already discussed how dynamical low energy degrees of freedom of QCD emerge naturally in the FRG-approach to QCD with dynamical hadronization, including the sequential decoupling of the gluonic and quark degrees of freedom, see in particular Fig.~\ref{fig:DynHad}. 

Below the gluon-decoupling scale of about 600\,MeV the flow of the QCD effective action is only driven by diagrams with quarks and mesons. Note however, that this dynamics takes place in a gluonic background, the solution of the gluonic equations of motion. As discussed in Sec.~\ref{sec_QCD-Conf}, at finite temperature and density such a background, $A_{0,\textrm{EoM}}\neq 0$, gives rise to a non-trivial Polyakov-loop expectation value. This property of QCD can be included in low-energy effective theories (LEFTs) of QCD in terms of a Polyakov-loop background $L$, or that of a temporal gauge field $A_0$. In these Polyakov-loop enhanced models this background couples to the quark fields and the effective action of the LEFT is augmented with the effective potential in QCD, $V_{\textrm{Pol}}[L]$ or $V_\textrm{eff}[A_0]$.

Clearly the Polyakov-loop enhanced models are not confining themselves but the Polyakov-loop background carries over confinement information from QCD. This is usually called \textit{Statistical Confinement}. Note however, that the QCD-embedding of the models discussed above entails that, while it is not the only confining property to be taken into account, it carries a good deal of it. For a recent review on Polyakov loop-enhanced models see \cite{Fukushima:2017csk}.

Note also that one of the consequences of the QCD-embedding with and without dynamical hadronization is the formal equivalence of fermionic NJL-type models and models with quark and hadronic degrees of freedom such as the quark-meson model (QM). The renormalizability of the latter in comparison to the former models is a red herring: Firstly, both models have a physical ultraviolet cutoff: one due to non-renormalizability (NJL), the other due to a Landau pole (QM). Moreover, they cease to describe QCD even for scales smaller than the ultraviolet cutoff, see e.g.~\cite{Alkofer:2018guy}. Still one should keep in mind that models with hadronic degrees of freedom are better suited to describe the infrared dynamics of QCD, in particular that of pions.

In summary, the flow of full QCD naturally leads to an emergent low-energy effective theory of quarks and hadrons as well as a background temporal gluon or Polyakov loop in the infrared. These \textit{QCD-assisted} models can be improved within systematic expansion schemes towards QCD, for a more detailed description see \cite{Springer:2016cji, Braun:2018svj, Alkofer:2018guy, Fu:2019hdw} and references therein. This allows to study many phenomena in low energy QCD within these LEFTs. The latter allow for an easier access to the dynamics both in Euclidean spacetime and also with genuinely real-time FRG-methods. The LEFTs are also an ideal testbed for conceptual advances in the FRG to be used later also in QCD. 

 %
 %%%%%%%%%%%%%%%%%%%%%%%%%%%%%%%%%%%%%%%%%%%%%%%%%%%%%%%%%%%%%%%%%%%%%
 \begin{figure}[t]
 	\begin{center}
 		\includegraphics[scale=0.8]{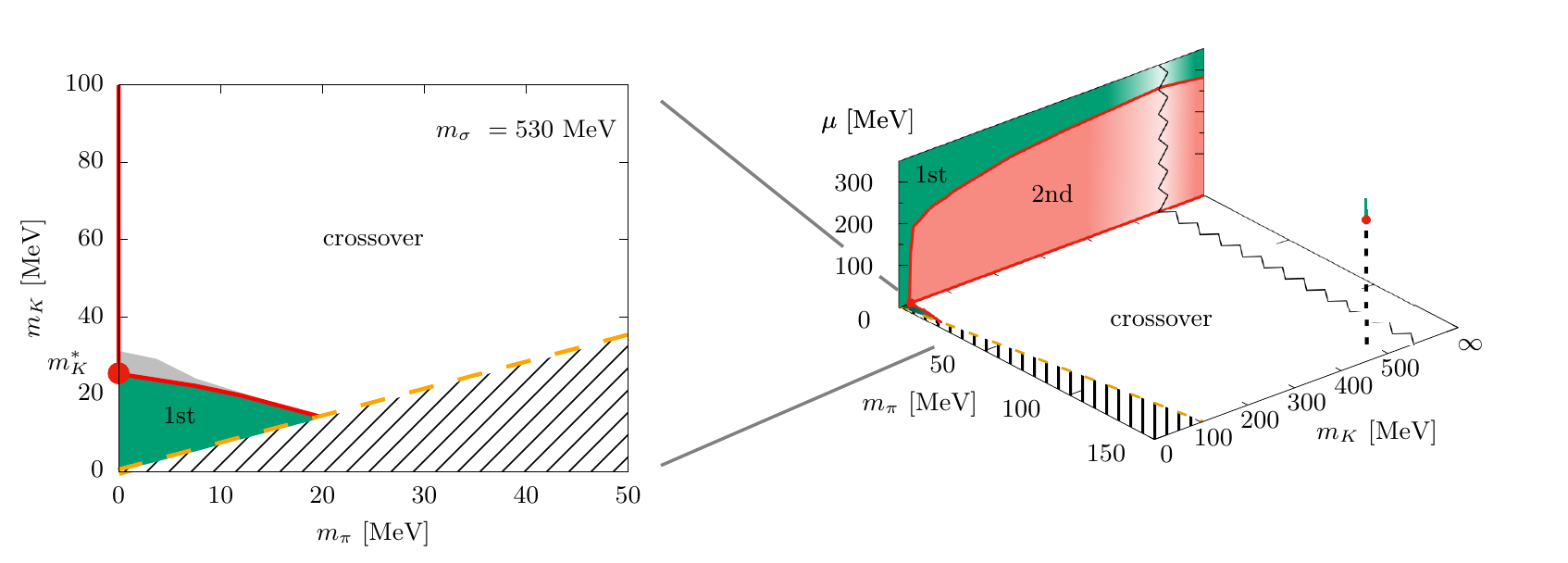}
 	\end{center}
 	\caption{Chiral phase structure for 2+1 flavor QCD in the $(m_\pi,m_K)$- and $(\mu,m_K)$-planes (Columbia plot) obtained in \cite{Resch:2017vjs}. At $\mu=0$ there is a small first-order region  around the chiral limit. The anomalous $U_A(1)$-breaking is chiefly important for this results, without the anomaly the phase structure is significantly changed, see \cite{Resch:2017vjs}.} 
 	\label{fig:columbiafull}
 \end{figure}
 %%%%%%%%%%%%%%%%%%%%%%%%%%%%%%%%%%%%%%%%%%%%%%%%%%%%%%%%%%%%%%%%%%%%%%
 %

\subsubsection{The phase structure of QCD from low-energy effective models}\label{sec:QCDPhaseModelResults} 

This has led to a plethora of FRG-works in low-energy effective models (LEFT) on the phase structure of QCD with NJL-type and quark-meson models (QM-model) with and without Polyakov-loop background. Below we provide a brief recollection of FRG-works on the phase structure of QCD, and single out only a few of the many highlights.  

The chiral phase structure has been studied in many works, starting with works at finite temperature and vanishing density, see \cite{Jungnickel:1995fp, Berges:1997eu, Berges:1998sd, Berges:1998ha, Papp:1999he, Bergerhoff:1999hr, Bohr:2000gp, Schaefer:2001cn, Braun:2009si, Fukushima:2010ji}. The chiral phase structure at finite density or baryon-chemical potential including the volume-dependence of the chiral phase structure has been considered in \cite{Berges:1998sd, Berges:1998ha, Schaefer:1998my, Schaefer:1999em, Braun:2003ii, Schaefer:2004en, Braun:2004yk, Braun:2005gy, Braun:2006vd, Braun:2005fj, Braun:2007td, Braun:2008sg, Fukushima:2010ji, Braun:2010vd, Schaefer:2011pn, Braun:2011iz, Jiang:2012wm, Kamikado:2012cp, Tripolt:2013zfa, Mitter:2013fxa, Drews:2013hha, Pawlowski:2014zaa, Aoki:2015hsa, Springer:2015kxa, Aoki:2015mqa, Wang:2015bky, Jiang:2015xqz, Aoki:2017rjl, Zhang:2017icm, Tripolt:2017zgc, Resch:2017vjs, Braun:2017srn, Braun:2018bik, Yin:2019ebz, Otto:2019zjy, CamaraPereira:2020xla, Otto:2020hoz, Braun:2020bhy, Connelly:2020pno, Connelly:2020gwa}, superfluid phases with gapless fermionic excitations at finite (isospin) density and their counterparts in non-relativistic systems have been studied in \cite{Boettcher:2014tfa, Boettcher:2014xna}. Here we want to single out one of the many important and interesting results, which concerns the chiral phase structure of QCD as a function of the masses of the pseudo-Goldstone bosons of chiral symmetry breaking, the pions and kaons, \cite{Resch:2017vjs}. Their masses can be varied by varying the current quark masses that originate in the Higgs mechanism. This analysis is specifically important as it also constrains the QCD phase structure finite density, for FRG-studies of the relevant Yang-Lee edge singularities and related work see \cite{An:2016lni, Litim:2016hlb, Zambelli:2016cbw, Connelly:2020pno, Connelly:2020gwa}. Moreover, the axial anomaly plays an important role for this investigation. It is a genuine nonperturbative problem, where a good grip on the scaling  behaviour of light or massless degrees of freedom in the chiral limit is crucial. The chiral limit of QCD is still a major challenge for lattice simulations, and hence this problem is specifically well-suited for a renormalization group study. For more details see the $2+1$-flavor investigation within the quark-meson model in \cite{Resch:2017vjs}. The respective phase structure is depicted in Fig.~\ref{fig:columbiafull}. 

Further investigations with the FRG include not only the chiral dynamics within the phase structure, but also statistical confinement in terms of a background Polyakov loop or a temporal background gluon. The phase structure of QCD and the interplay between confinement and chiral symmetry at finite temperature and density have been studied in \cite{Skokov:2010wb, Herbst:2010rf, Braun:2011fw, Morita:2011jva, Braun:2012zq, Herbst:2013ail, Herbst:2013ufa, Haas:2013qwp, Fu:2018wxq}. Note in this context that in the large density regime with $\mu_B/T\gtrsim 2-3$ with baryon chemical potential $\mu_B = 3 \mu_q$, regime lattice simulations are obstructed by the sign problem. In turn, functional approaches have to deal with the increasingly complex structure of the dynamical low energy degrees of freedom. In the presence of a  chemical potential the pole- and cut-structure in the complex frequency plane gets relevant, leading to exciting phenomena such as the Silver-Blaze property of QCD. For FRG-works see \cite{Strodthoff:2011tz, Kamikado:2012bt, Strodthoff:2013cua, Khan:2015puu}. In heavy ion collisions one often investigates fluctuation observables, for FRG-works see \cite{Schaefer:2006ds,  Nakano:2009ps,Skokov:2010uh, Skokov:2011rq, Skokov:2012ds, Morita:2013tu, Fu:2015naa, Fu:2015amv, Fu:2016tey, Rennecke:2016tkm, Almasi:2016zqf, Almasi:2017bhq, Sun:2018ozp, Fu:2018swz,Fu:2018qsk,Wen:2018nkn, Wen:2019ruz}. In heavy ion collisions we also expect strong (chromo-) magnetic and electric fields, FRG-work in this direction has been done in  \cite{Skokov:2011ib, Fukushima:2012xw, Braun:2014fua, Mueller:2015fka, Aoki:2015mqa, Fu:2017vvg, Li:2019nzj}. 

A very exciting development has taken place in the past decade concerning the access of real-time properties of QCD with the FRG. Naturally, the real-time approach to QCD correlation functions has not only applications to the QCD phase structure but also to the hadron spectrum, see the discussion at the end of Sec.~\ref{sec_he:VacuumQCD}.  Applications to heavy ion collisions range from dilepton rates, see e.g.~\cite{Tripolt:2018jre, Tripolt:2020dac} or computing transport processes and transport coefficients relevant for the hydrodynamical phase of the collision, see e.g.~\cite{Haas:2013hpa, Christiansen:2014ypa, Bluhm:2018qkf}.  

%
%%%%%%%%%%%%%%%%%%%%%%%%%%%%%%%%%%%%%%%%%%%%%%%%%%%%%%%%%%%%%%%%%%%%%
\begin{figure}[t]
	\begin{center}
		\subfigure[Vector meson spectral functions from \cite{Jung:2016yxl, Jung:2019nnr} with the analytically continued FRG (aFRG). The inlay shows the temperature dependence of the pole masses. The different scattering thresholds are clearly visible.  ]{\hspace{-0.0cm}	\label{fig:SpecFunBos}
			\hspace{+0.1cm}\includegraphics[scale=0.15]{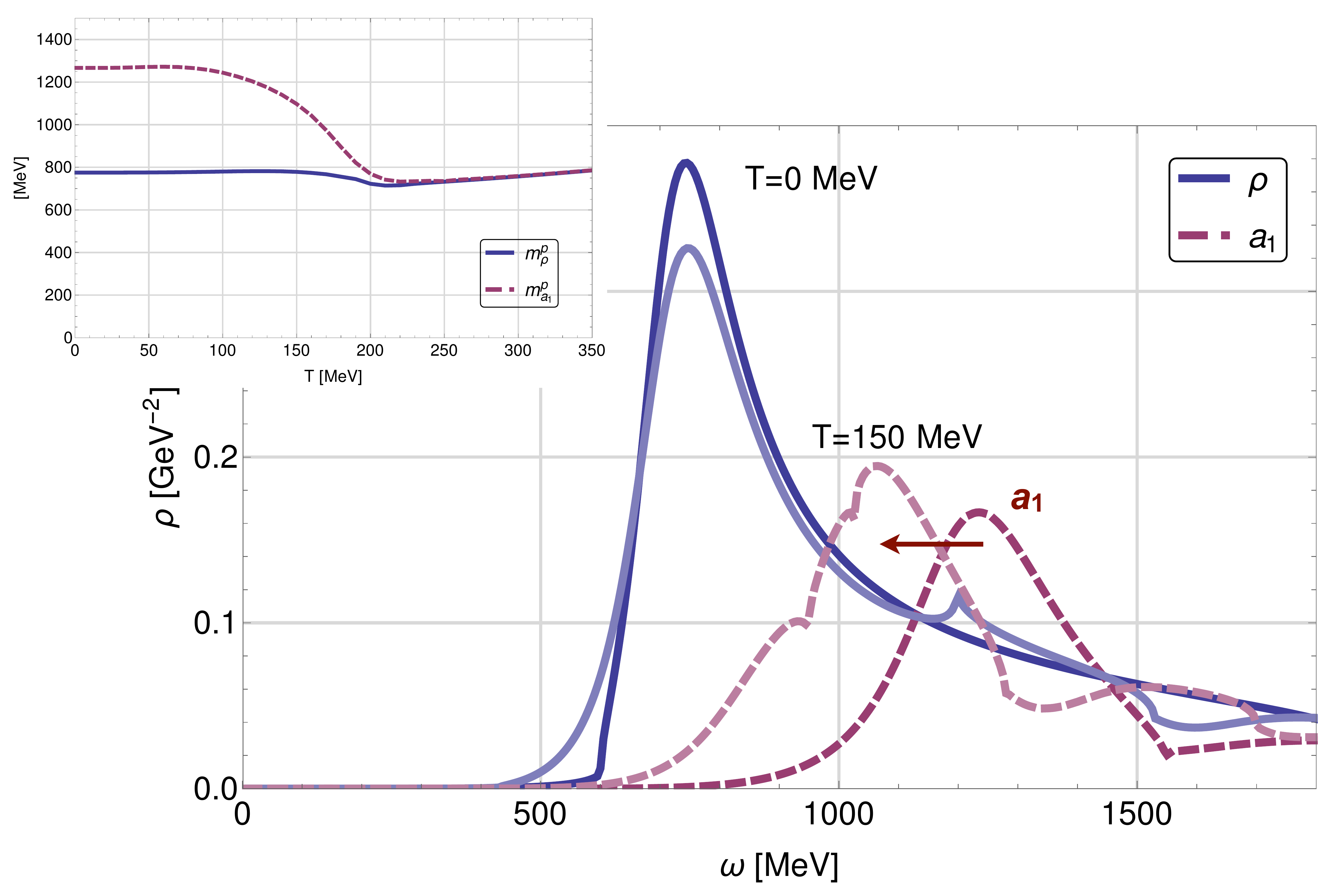}
		}	\hspace{.3cm}
		\subfigure[Quark spectral functions from \cite{Tripolt:2020irx} with the aFRG. The temperature is $T=180$\, MeV, close but above the crossover temperature, see also \cite{Tripolt:2018qvi, Wang:2018osm}.  The spectral function shows  quasi-particle and plasmino peaks as well as the ultrasoft mode. ]{\hspace{0cm}	\label{fig:SpecFunFerm}
			\includegraphics[scale=0.155]{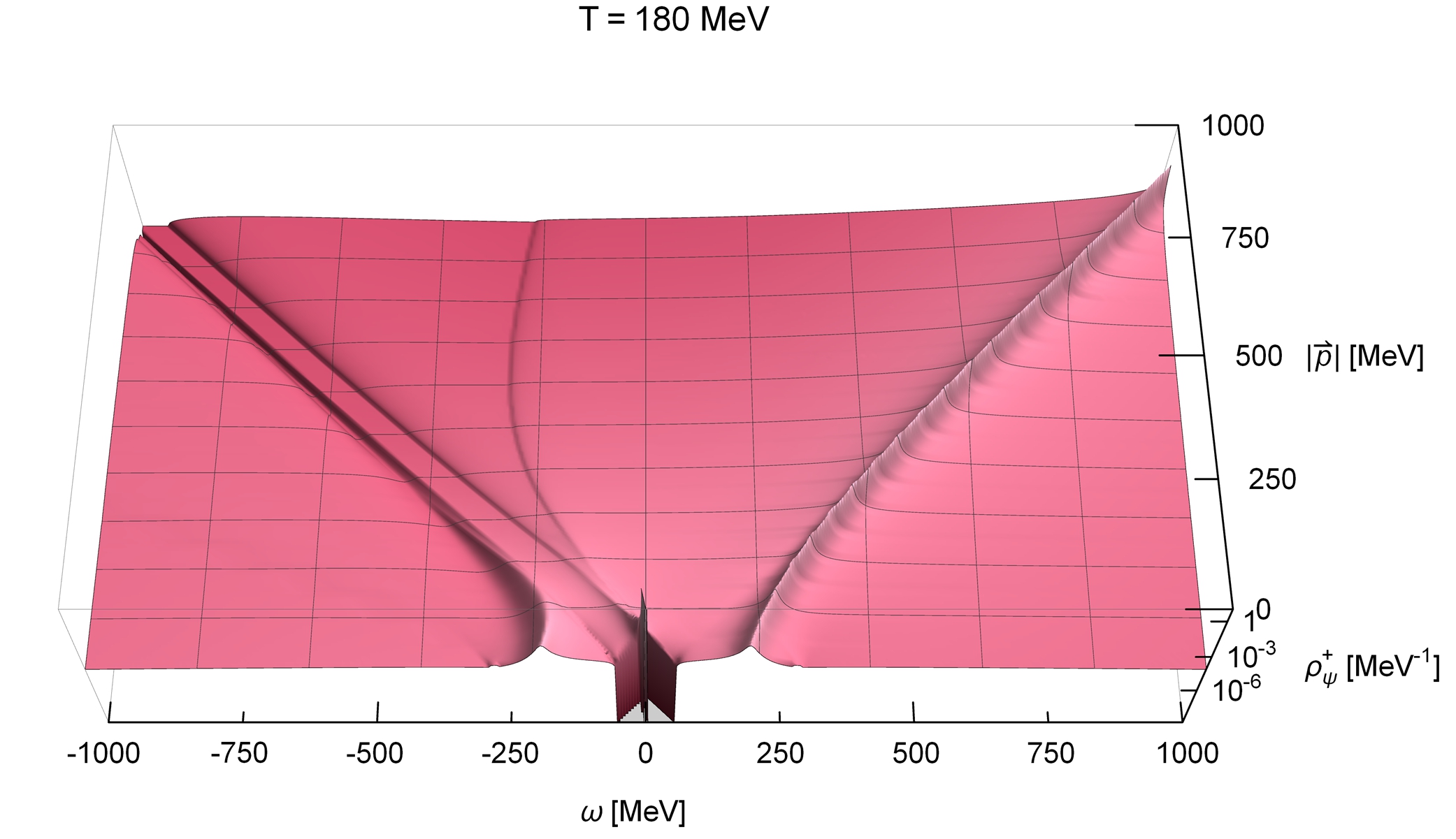}
		}	
	\end{center}
	\caption{Spectral functions in  the 2-flavor QM-model: Vector meson spectral functions (left), quark spectral functions (right).}
	\label{fig:Columbia}
\end{figure}
%%%%%%%%%%%%%%%%%%%%%%%%%%%%%%%%%%%%%%%%%%%%%%%%%%%%%%%%%%%%%%%%%%%%%%
%
Within the FRG, real-time correlation functions are either worked out with numerical analytic continuation, see e.g.~\cite{Cyrol:2018xeq,  Binosi:2019ecz, Li:2019hyv}, or computed directly. Real-time FRGs have been developed in different areas of physics, see e.g.~\cite{Gasenzer:2007za, Gasenzer:2010rq, Corell:2019jxh, Pietroni:2008jx, Lesgourgues:2009am, Bartolo:2009rb}, typically also using specific properties of the situation under consideration, for related work see Secs.~\ref{sec_sm}, \ref{sec_gr} .The following works are based on ideas developed in \cite{Floerchinger:2011sc, Kamikado:2013sia, Pawlowski:2015mia, Steib:2019xrv}. The promising  developments, that started with \cite{Kamikado:2013sia}, have been called aFRG (analytically-continued FRG), for works on meson spectral functions in quark-meson and meson effective theories in the vacuum and at finite temperatures and densities see \cite{ Floerchinger:2011sc, Kamikado:2013sia, Tripolt:2013jra, Helmboldt:2014iya, Tripolt:2014wra, Pawlowski:2015mia, Strodthoff:2016pxx, Yokota:2016tip, Yokota:2017uzu, Wang:2017vis, Jung:2016yxl, Jung:2019nnr}, for quark spectral functions see \cite{Tripolt:2018qvi, Wang:2018osm, Tripolt:2020irx}. In Figs.~\ref{fig:SpecFunBos}, \ref{fig:SpecFunFerm} we have depicted aFRG results for the nonperturbative spectral functions for vector mesons and quarks obtained in \cite{Jung:2016yxl, Tripolt:2020irx}. These real-time results allow to study the emergence and dissolution of the hadronic low-energy degrees of freedom onshell. This is important for an interpretation of HIC results. Moreover, within Euclidean numerical computations the respective physics information is diluted or even lost due to the necessity of analytically continuing numerical data with a non-trivial systematic error and finite numerical accuracy.

%%%%%%%%%%%%%%%%%%%%%%%%%%%%%%%%%%%%%%%%%%%%%%%%%%%%%%%%%%%%%%%%%%%%%
\begin{figure}[t]
	\begin{center}
		\subfigure[Chiral phase structure of QCD for 2+1-flavor QCD. The black dashed line is the chiral crossover line from the FRG-computation in \cite{Fu:2019hdw}. The blue dashed line is from an FRG-assisted DSE-computation in \cite{Gao:2020qsj, Gao:2020fbl}. The transition temperature is determined by the peak of the thermal susceptibility of the chiral condensate of the light quarks. The black circle indicates the critical end point of the crossover line, by now corroborated by the most recent DSE result, \cite{Gao:2020fbl}. The hatched red area depicts a regime with a sizable chiral condensate and a minimum of the meson dispersion at non-vanishing spatial momentum. This indicates a potentially inhomogeneous regime. We also provide lines for $\mu_B/T=2,3$ related to reliability bounds for lattice results, \cite{Bellwied:2015rza, Bazavov:2018mes, Braguta:2019yci}, and $\mu_B/T=3,4$ for results from functional methods, \cite{Fischer:2014ata, Gao:2015kea, Fu:2019hdw, Gao:2020fbl, Gao:2020qsj}. Freeze-out data:  \cite{Adamczyk:2017iwn, Alba:2014eba, Andronic:2017pug, Becattini:2016xct, Vovchenko:2015idt, Sagun:2017eye}. ]{\hspace{0.4cm}\label{fig:phasediagram}
			\hspace{-0.0cm}\includegraphics[width=0.43\linewidth]{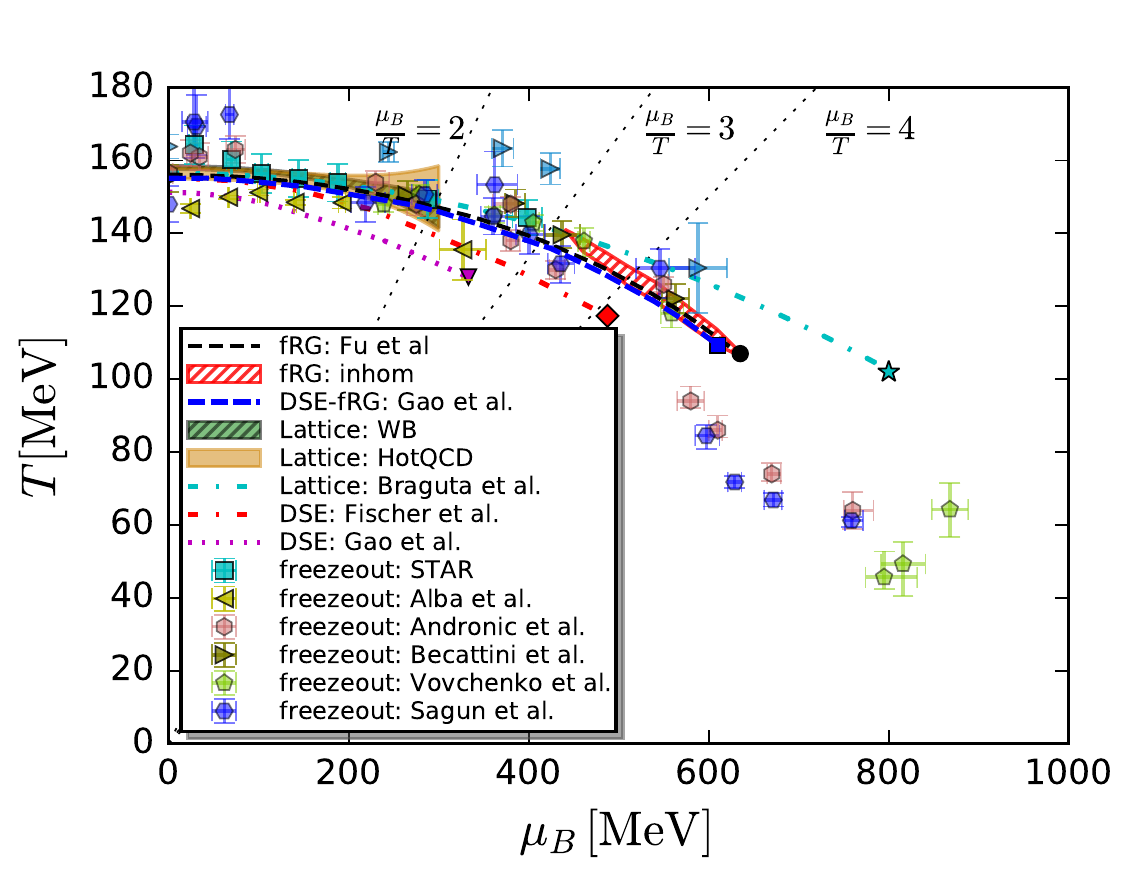}}\hspace{0.5cm}
		\subfigure[Strength of all four-quark couplings in 2-flavor QCD at the crossover temperature in terms of quark chemical potential $\mu_q$ over $T_0=T_c(\mu_q=0)$ from \cite{Braun:2019aow}. This Fierz-complete computation shows  dominance of the  ($\sigma-\pi)$-channel  (scalar-pseudoscalar) for small chemical potentials. The onset of the transition from dominance of the ($\sigma-\pi)$-channel  to that of the (csc)-channel (diquark), is given by $\mu_\textrm{q,trans}\approx  T_0$, for the details see \cite{Braun:2019aow}. Relative temperatures and chemical potentials agree well for 2- and 2+1-flavor QCD, \cite{Fu:2019hdw}, and for $2+1$-flavor QCD, $\mu_\textrm{q,trans}$ agrees well with the onset of the  potentially inhomogeneous regime at the crossover line: the intersection point of the hatched red area in Fig.~\ref{fig:phasediagram} with the crossover line. In this regime the FRG-computations so far lack full reliability and have to be systematically improved. This requires in particular a Fierz-complete basis combined with higher-order mesonic scatterings.   ]{\hspace{-0.0cm}	\label{fig:dominancepatternUA1}
		\includegraphics[width=0.48\linewidth]{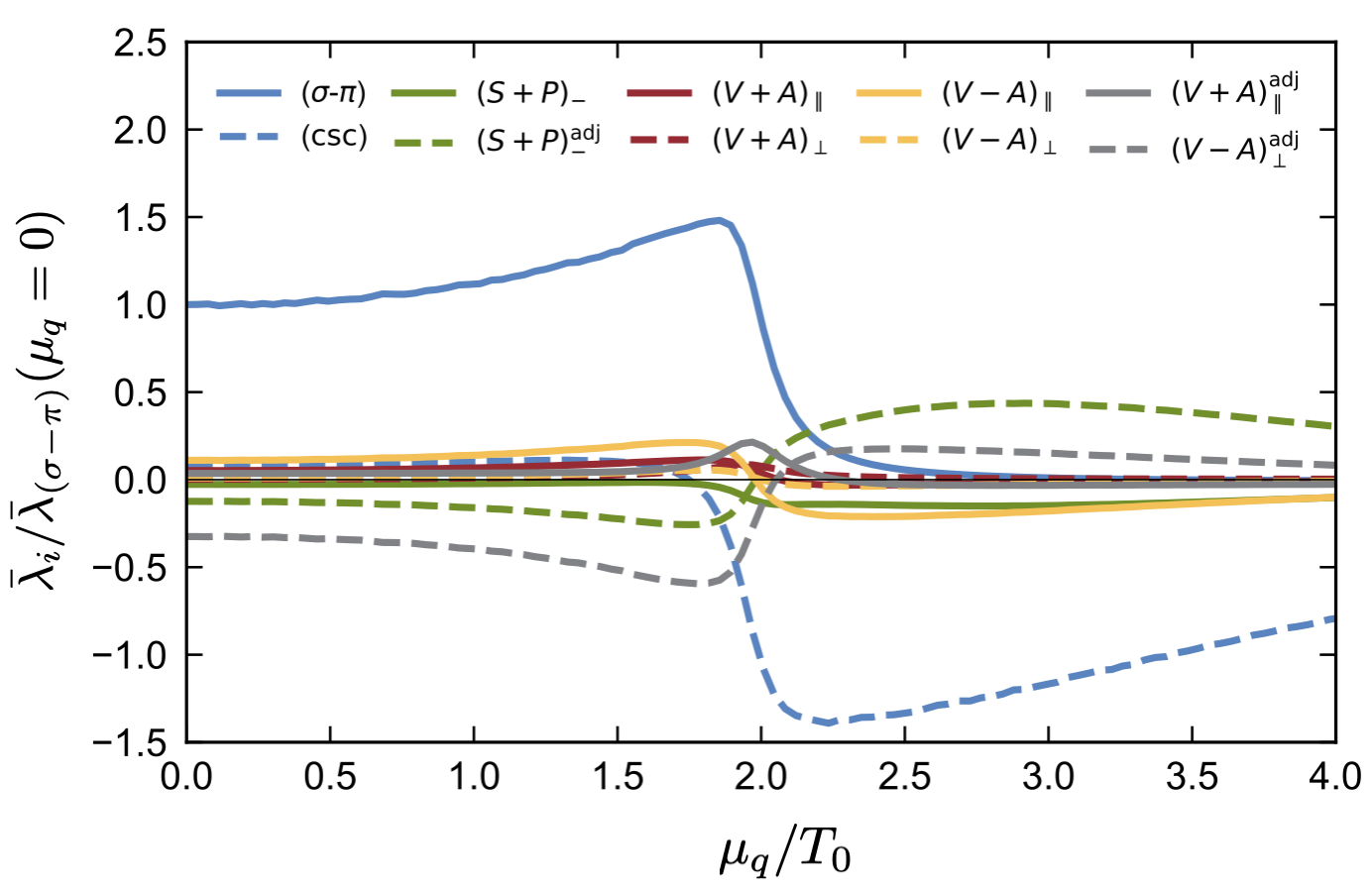}
}	\hspace{0.1cm}
	\end{center}
	\caption{QCD phase diagram from QCD flows, Data taken from \cite{Fu:2019hdw, Gao:2020qsj, Gao:2020fbl, Braun:2019aow}, more details can be found there. } 
	\label{fig:QCDPhaseStructure}
\end{figure}
%%%%%%%%%%%%%%%%%%%%%%%%%%%%%%%%%%%%%%%%%%%%%%%%%%%%%%%%%%%%%%%%%%%%%%
%
\subsubsection{Towards the QCD phase structure from first principles} \label{sec:QCD1st}

The advances in the FRG-approach to QCD concerning quantitative  computations in extended truncations in the vacuum, see Sections~\ref{sec_QCD-Conf}, \ref{sec_QCD-Chiral} and \ref{sec_he:VacuumQCD}, and the plethora of results for LEFTs for finite temperature and density discussed in the Sections~\ref{sec:UnreasonableEffectiveness} and \ref{sec:QCDPhaseModelResults}, prepare the stage for FRG-applications to QCD at finite temperature and density. While these investigations have not reached the quantitative level of the vacuum applications,  results on two-flavor QCD at finite temperature and imaginary chemical potential, \cite{Braun:2009gm}, as well as one-, two- and 2+1 flavor QCD at finite temperature and chemical potential have been obtained, \cite{Braun:2008pi, Fu:2019hdw, Braun:2019aow}, see Fig~\ref{fig:QCDPhaseStructure} for some results. The nuclear equation of state (EoS) at low temperatures and high densities has been studied in \cite{Leonhardt:2019fua}. In the recent work \cite{Braun:2020ada}, the pion mass dependence of the chiral phase transition temperature (magnetic EoS) has been studied. For lattice results we refer the reader to \cite{Bellwied:2015rza, Bazavov:2018mes, Braguta:2019yci, Borsanyi:2020fev, Ding:2020rtq} and references therein, for results from other functional methods we refer the reader to \cite{Fister:2013bh, Fischer:2013eca, Fischer:2014vxa, Fischer:2014ata, Eichmann:2015kfa, Gao:2015kea,  Isserstedt:2019pgx, Reinosa:2015oua, Maelger:2017amh, Maelger:2018vow, Maelger:2019cbk, Gao:2020qsj, Gao:2020fbl} and references therein, for a recent review see \cite{Fischer:2018sdj}. 

The phase boundary obtained in \cite{Fu:2019hdw} for $N_f=2+1$ flavors, see  Fig~\ref{fig:phasediagram}, agrees quantitatively with respective 2+1 flavor lattice results at small densities. It also agrees well with the most recent DSE-computations (FRG-assisted DSE), \cite{Gao:2020fbl, Gao:2020qsj}. In these works, the DSEs for the correlation functions are expanded about the quantitative $N_f=2$-flavour vacuum correlation functions obtained with the FRG in \cite{Cyrol:2017ewj}. In turn, at large densities no lattice results are present due to the notorious sign problem, and lattice result at large densities such as those in \cite{Braguta:2019yci} depicted in Fig~\ref{fig:phasediagram} are extrapolations. Interestingly,  \cite{Fu:2019hdw} predicts a critical end point at large baryon chemical potential $\mu_B$, as well as a potentially inhomogeneous regime. The location of the critical end is by now corroborated by the FRG-assisted DSE results in \cite{Gao:2020fbl}. However, for obtaining quantitative predictivity in this regime, functional approaches still require systematic improvements: The onset of the latter agrees well with the change of the dominant four-quark interaction channel from the scalar-pseudoscalar channel to a diquark channel found in \cite{Braun:2019aow}, see the discussion below  Fig~\ref{fig:dominancepatternUA1}. This coincidence is even more remarkable and non-trivial, as \cite{Fu:2019hdw} works with dynamical hadronization of the scalar-pseudoscalar channel which gives access to higher order scattering of mesonic degrees of freedom,  while \cite{Braun:2019aow} works with a Fierz-complete basis. This indicates the onset of new physics as well as additional resonant channels. Consequently, in this regime the respective systematic error of the present truncations grows large, and the results have to be corroborated within systematically improved truncations. For example, multi-scattering of potentially resonant diquark and density channels have to be taken into account. These systematic improvements pose solely technical challenges in comparison to the conceptual intricacies of the sign problem. We emphasize that the intricacies of the  interaction-structure of QCD at large densities may still pose an insurmountable challenge despite being of technical nature. Whether or not apparent convergence can be achieved in the large density region of QCD within a first principles FRG-study remains to be seen in the future. 

We close the QCD-part with a brief account of FRG-investigations in many-flavor QCD. This already connects QCD to high-energy physics at the high-energy frontier. Asymptotic freedom in QCD comes from the negative sign of the QCD $\beta$-function. At one loop it is given by $\beta_{\alpha_s} = -1/(8 \pi^2) ( 11/3 N_c  -2/3 N_f)$. For many flavors the sign  of the $\beta$-function changes and asymptotic safety is lost, at one loop this happens at  $N^\textrm{af}_f = 11/2 N_c$. The theory may also feature an infrared fixed point (triggered by the two-loop coefficient) for $N_f > N_f^\textrm{CBZ}$, the Caswell-Banks-Zaks fixed point with $N_f^\textrm{CBZ}< N^\textrm{af}_f$. Finally, QCD may also have a conformal regime for $N_f^\textrm{CBZ}\leq  N_f^\textrm{cr} < N_f <N^\textrm{af}_f$, for an FRG-discussion of all these features see \cite{Braun:2010qs}. 

These fixed point structures and scaling properties call for FRG-applications. Indeed, the smallness of the $\beta$-function in this regime supports the reliability of semi-perturbative approximations, while the nonperturbative FRG-setup is required and well-suited to unravel the fixed points and non-trivial scalings, for FRG-works see \cite{Gies:2005as, Braun:2005uj, Braun:2006jd, Terao:2007jm, Braun:2009ns, Braun:2010qs}, for a QED${}_3$ analogue see \cite{Braun:2014wja}. These works contain many results on the fixed point locations, but in particular on the scaling pattern in the regime, e.g.~the 'beyond-Miransky scaling' at the quantum critical point \cite{Braun:2010qs}.  

\subsection{Electroweak phase transition, BSM physics \& Supersymmetry}\label{sec_hep-ew-BSM-Susy}

The interesting regime of QCD for small number of flavors is the low energy limit. The high energy regime of the Standard Model and its ultraviolet closure also harbour many fascinating phenomena and conceptual questions, ranging from the details of the electroweak phase transition, see \cite{Reuter:1993nn, Bergerhoff:1994sj}, over potential beyond the Standard Model scenarios up to the question of a unification with (quantum) gravity, which is explained in more detail in Section~\ref{sec_gr}.  

An exciting link to the UV-completion of high energy physics including gravity are asymptotically safe matter systems, which includes asymptotically free systems as a specific case. Such systems may allow for a UV-completion of the Standard model in terms of stable interacting or free ultraviolet fixed points with a finite number of UV-relevant directions. In this setting the Standard Model emerges from a specific UV-IR trajectory. Note that such a embedding of the Standard Model in an asymptotically safe UV-completion of matter should not be seen as fundamental theory. Most likely quantum gravity effects deform or even change the UV completion beyond the Planck scale. This happens both within asymptotically safe gravity, see the review \cite{Eichhorn:2018yfc}, and beyond, see \cite{deAlwis:2019aud}. Still, already the existence of such a fixed point constrains predictions in the presence of a physical UV-cutoff beyond which the low energy theory is changed by new physics, \cite{Held:2020kze}.

Prior to the discovery of the Higgs particle, various aspects of Higgs physics were explored with FRG techniques in \cite{Ellwanger:1992us, Gies:2003dp}, with a particular focus on asymptotic safety. Asymptotic safety in nonlinear SU(N) sigma models without propagating Higgs mode was investigated in \cite{Percacci:2009fh, Fabbrichesi:2010xy, Bazzocchi:2011vr, Fabbrichesi:2011bx}. Asymptotic safety has also been searched for in simple and chiral Yukawa models \cite{Gies:2009hq, Gies:2009sv, Scherer:2009wu,Vacca:2015nta} and in gauged chiral Yukawa models in \cite{Gies:2013pma}. Recently, asymptotic safety in QED with a Pauli term was explored in \cite{Gies:2020xuh}, whereas a search for asymptotic safety in axion electrodynamics can be found in \cite{Eichhorn:2012uv}. Beyond the Standard Model, asymptotic safety in the dark-matter sector was explored in \cite{Eichhorn:2018vah}; for studies including gravity we refer to Sec.~\ref{sec_gr}.

Initiated in \cite{Litim:2014uca}, there is by now a growing class of asymptotically safe gauge-matter models, where the UV fixed point can be accessed perturbatively in the Veneziano limit, see e.g.~\ \cite{Litim:2014uca, Litim:2015iea, Mann:2017wzh, Bond:2018oco, Dondi:2019ivp, Bond:2019npq, Dondi:2020qfj}. The stability of these models in the presence of higher order couplings has been explored with the FRG for a gauge-Yukawa system in \cite{Buyukbese:2017ehm}.

The triviality problem in the Higgs sector was tackled in \cite{Gies:2015lia, Gies:2016kkk, Gies:2018vwk, Gies:2019nij}, where asymptotically free scaling solutions were uncovered by imposing generalized boundary conditions on the correlation functions. With the Higgs mass of about 125 GeV and the top mass of about 172 GeV, the Standard Model lies on the border of stability: Specifically, for the present central value of the top mass, the Higgs quartic coupling must be negative at the Planck scale. Moreover, it only crosses into the positive values at a momentum scale of about $10^{10}\, \rm GeV$. This is typically viewed as an indication of a metastable potential, i.e., the electroweak vacuum is only a local, but not a global minimum of the potential. The FRG is tailor-made for an investigation of this question, which has started in a simple Yukawa model \cite{Gies:2013fua}, also explored in \cite{Jakovac:2015kka, Sondenheimer:2017jin}. This model was upgraded to a chiral one with the two heaviest quark flavors in \cite{Gies:2014xha}. The effect of gauge interactions was added in \cite{Eichhorn:2015kea}. It was shown, that higher-order operators in the Higgs potential, as they arise from new physics, can lower the Higgs mass bound, see also \cite{Gies:2017zwf} for studies of higher-order Yukawa interactions. In  \cite{Borchardt:2016xju} the vacuum stability was studied with spectral methods, that allow to explore the global form of the potential. Studies of vacuum stability with new degrees of freedom, specifically scalar and fermionic dark matter, coupled through a Higgs portal, can be found in \cite{Eichhorn:2014qka,Held:2018cxd}. For a summary and references on this question with other techniques, see \cite{Gies:2017ajd}. 

In the context of vacuum stability, new physics at a relatively high scales is of central interest. In contrast, new physics at scales close to LHC-scales is in the focus of a study of a first-order electroweak phase transition \cite{Reichert:2017puo}. As a key advantage of FRG techniques, non-perturbative corrections to the Higgs potential, such as, e.g., operators of the form $\phi^4 {\rm exp}{-\Lambda^2/\phi^2}$, where $\Lambda$ is the scale of new physics, can be explored in a well-controlled way, using a numerical grid to evaluate the evolution of the potential as a function of field $\phi$, temperature $T$ and RG scale $k$. 

Even though the minimal supersymmetric extension of the Standard Model fails to meet experimental bounds, supersymmetric theories are still viable extensions of the Standard Model, and are specifically relevant in the context of low-energy effective theories of string theory.  While full supersymmetry is typically broken in the presence of an infrared regulator as used in the FRG, it can be monitored similarly to gauge symmetries by modified STIs. FRG-applications to supersymmetric models can be found in \cite{Synatschke:2008pv, Sonoda:2008dz, Rosten:2008ih, Sonoda:2009df, Synatschke:2009nm, Heilmann:2014iga, Feldmann:2017ooy, Hellwig:2015woa}, for supersymmetry at finite temperature see \cite{Synatschke:2010ub} ($N=1$ Wess-Zumino model). Supersymmetric gauge theories have been studied with the FRG in \cite{Granda:1997xk, Falkenberg:1998bg, Arnone:1998zc, Bonini:1998ec, Arnone:2000ij}. The critical behaviour and the phase structure of  supersymmetric $O(N)$-models is discussed in \cite{Litim:2011bf, Heilmann:2012yf}, and the breaking of supersymmetry as a phase transition has been described in \cite{Gies:2009az}, for emergent supersymmetry see \cite{Gies:2017tod}.

\subsection{Summary} 

We have discussed the FRG-setup for gauge theories, that are relevant for the formulation of Particle Physics and Gravity. By far the most FRG-applications are concerned with the strongly-correlated infrared sector of QCD at vanishing and finite temperature and density. These applications have been discussed in detail in Sec.~\ref{sec_hep-QCD} and Sec.~\ref{sec_QCD-Phase}. The phase structure of QCD features spontaneous symmetry breaking (chiral symmetry breaking and confinement), competing order effects (color superconducting phases at large densities), as well as the emergent dynamical degrees of freedom (from fundamental quarks \& gluons to hadrons). These applications as well as technical developments hence also works as the showcase example for applications in and beyond the Standard Model. The respective works have been briefly discussed in Section~\ref{sec_hep-ew-BSM-Susy}. There, we have listed the FRG-works on the question of vacuum stability of the Higgs sector, investigations of asymptotically safe matter systems, and that of supersymmetric theories. This part already is already tightly linked with the quantum gravity applications reviewed in Section~\ref{sec_gr}.

%% file: SEC_GRAVITY/frg_review-gravity.tex
\section{Gravity}
\label{sec_gr} 
\subsection{Introduction: Quantum gravity - why and what?}
A model of quantum gravity brings together the key aspects of quantum physics -- the superposition principle, quantum interference and the uncertainty principle -- with the insight that gravity is encoded in dynamical spacetime geometry. Thus, a quantum theory of gravity allows us to explore the consequences of quantum superpositions of spacetime. Why do we need such a theory? \\
A key motivation comes from the observation of gravitational waves by the LIGO/VIRGO collaboration \cite{Abbott:2016blz} and the first image of a black hole captured by the EHT collaboration \cite{Akiyama:2019cqa}. Within the uncertainties of the measurements, these  are in agreement with the predictions from General Relativity (GR). Nevertheless, these observations reinforce the need to go beyond GR. The reason is that within the theoretical description by General Relativity the observed objects, black holes, contain curvature singularities --  corresponding to diverging tidal forces -- where the curvature exceeds the Planck scale. Briefly reinstating $\hbar$ and $c$, the Planck scale reads
\begin{equation}
M_{\rm Planck}\! =\! \sqrt{\frac{\hbar \, c}{G_N}} \approx 10 ^{19}\, {\rm GeV},\,\, l_{\rm Planck}\! =\! \sqrt{\frac{\hbar \,G_N}{c^3}} \approx 10^{-35}\,{\rm m}.
\end{equation}
This scale is characteristic of a relativistic (due to $c$), gravitational  (due to $G_N$), quantum theory (due to $\hbar$). Hence, it is expected to be the typical scale of quantum gravity. The resolution of spacetime-singularities is thus expected to be one of the hallmarks of a successful model of quantum gravity. 

The key challenge of quantum gravity is to understand the consequences of quantum fluctuations of spacetime at all scales. An additional complication is that there is no fixed notion of scales, or more generally spacetime geometry, in quantum gravity. Nevertheless, (Wilsonian) RG concepts can be put to good use in this setting.

Due to the huge ratio of the Planck scale to experimentally accessible energy scales, e.g., at the LHC, $E_{\rm LHC}/M_{\rm Planck} \approx 10^{-15}$, direct experimental tests of quantum gravity remain out of reach (with the notable exception of tests of the breaking of Lorentz invariance at (trans)planckian energies \cite{AmelinoCamelia:1997gz,Abdo:2009zza,Ackermann:2009aa,Vasileiou:2013vra}). As a consequence, a diversity of theoretical ideas about Planck-scale physics is being explored, see, e.g., \cite{Oriti:2009zz,Ashtekar:2014kba,Carlip:2015asa} for overviews over several approaches. Several settings, e.g., asymptotically safe gravity, tensor models, dynamical triangulations, spin foams and causal sets are based on the gravitational path integral, schematically
\begin{equation}
Z = \int_{\rm spacetimes}e^{i\,S[{\rm spacetime}]}.\label{sec_gr:eq:Zgrav}
\end{equation}
With the FRG, one can tackle such path integrals both in the continuum as well as in discrete settings, see Subsec.~\ref{sec_gr:discrete}. As we will see, the notion of the RG scale is slightly different in these two settings. Nevertheless, a one-loop flow equation can be derived both in the continuum and in the discrete setting by introducing a partition of the configuration space into "microscopic" and "macroscopic" configurations.\\
Before introducing the FRG for quantum gravity, let us review the problem of quantum gravity from the perspective of a perturbative quantization of General Relativity:
To define what is meant by Eq.~\eqref{sec_gr:eq:Zgrav}, one needs to specify several things: Firstly, a configuration space to be summed over needs to be defined. The minimalistic choice is a summation over geometries at fixed topology and dimensionality, for instance expressed as a sum over metrics\footnote{Alternatively, additional geometric degrees of freedom, such as, e.g., spacetime torsion can be taken into account. There are several choices for the fields to be included in the path integral, many of which lead to the identical classical dynamics, but which are expected to lead to different quantum theories of gravity.}.  
Secondly, a microscopic dynamics is needed. The Einstein-Hilbert action
\begin{equation}
S_{\rm EH} = \frac{1}{16\pi G_N}\int d^dx\, \sqrt{-g}\left(-2\bar{\Lambda} + R \right),
\label{sec_gr:eq:EH}
\end{equation}
features the curvature scalar $R$, as well as a cosmological constant term, where $g$ is the determinant of the metric and $\bar{\Lambda}$ is the cosmological constant. This action is not a good starting point within perturbation theory: Loop divergences require additional curvature invariants beyond the Einstein-Hilbert action to be added as counterterms, e.g., in $d=4$, $\sqrt{-g}R^2$ and $\sqrt{-g}R_{\mu\nu}R^{\mu\nu}$ at the one-loop level with matter \cite{Deser:1974cy,Deser:1974cz,tHooft:1974toh}, and $\sqrt{-g}R_{\mu\nu\kappa\lambda}R^{\kappa\lambda}_{\,\,\,\,\rho\sigma}R^{\rho\sigma}_{\,\,\,\,\mu\nu}$ at the two-loop level \cite{Goroff:1985sz,vandeVen:1991gw}. This follows, as the action in Eq.~\eqref{sec_gr:eq:EH}, when expanded in fluctuations of the metric around a flat background metric, features derivative interactions with a coupling with negative mass dimension, $G_N$. This renders the theory power-counting nonrenormalizable and is expected to result in infinitely many counterterms at infinitely high loop order. Each counterterm comes with an independent coupling that is a free parameter, leading to a breakdown of predictivity of the perturbative approach based on the Einstein-Hilbert dynamics. As long as one focuses on energies sufficiently below the Planck scale, the additional terms are negligible, and predictions can be derived within an effective field theory approach to quantum gravity \cite{Donoghue:1993eb}. As one approaches the Planck scale, all higher-order terms become important and the effective field theory approach breaks down. Thus, a different microscopic dynamics and/or nonperturbative techniques appear indicated\footnote{An enhancement of the symmetry to maximal supersymmetry, i.e., $N=8$ supergravity, is also being explored as a way to reduce the number of ultraviolet divergences, see \cite{Bern:2018jmv} and references therein.}. 
 \subsection{Path integral for quantum gravity and asymptotic safety}
A Wilsonian approach to gravity, like, e.g., the FRG,  allows us to shift the focus away from a specific choice of $S$. Instead the theory space takes center stage. It is defined once the field content, i.e., the configuration space, and symmetries of the theory are specified. The theory space is spanned by the dimensionless counterparts of the couplings of all quasilocal (i.e., containing only non-negative powers of derivatives) symmetry invariants\footnote{The presence of nonlocalities of the effective action $\Gamma_{k\rightarrow 0}$ does not imply that one needs to include inverse powers of derivatives among the operators whose couplings span the theory space. The increasing canonical relevance of operators with an increasing negative power of derivatives makes it unlikely that such a theory space can contain a fixed point with a finite number of relevant couplings, see, e.g., \cite{Machado:2007ea}. Indeed, non-localities in $\Gamma_{k}$ of order $k^{-1}$ are expected to be parameterized by the couplings of quasi-local operators, see \cite{Codello:2010mj} for a specific example.}. For instance, for a diffeomorphism invariant configuration space of the metric, the theory space is spanned by the dimensionless versions of the couplings of the curvature operators, i.e., $\sqrt{g}$, $\sqrt{g}R$, $\sqrt{g}R^2$, $\sqrt{g}R_{\mu\nu}R^{\mu\nu}$, etc.~which are written in the Euclidean setting, as required for the application of the FRG. This expansion is closely related to the derivative expansion widely used within many FRG applications, as presented, e.g., in Sect.~\ref{sec_sm}. \footnote{To bridge the gap to quantum field theories on a flat background discussed in the previous sections, one can expand each invariant in perturbations of the metric about a flat background, i.e., in the graviton. Then, each term contributes to arbitrarily high $n$-point graviton vertices, and each successive power of the curvature generates two additional derivatives. Thus, all interactions of gravitons, except those from the cosmological-constant term $\Lambda \int d^dx\sqrt{g}$ are derivative interactions.}.

The search for asymptotic safety proceeds in dimensionless couplings, akin to the previously discussed fixed-point searches in statistical physics.
For instance, for the first two terms in the curvature expansion, the dimensionless couplings in $d$ spacetime dimensions are 
\begin{equation}
\lambda = \bar{\Lambda}k^{-2}, \quad G = G_N\, k^{d-2}.
\end{equation}

As discussed in Sec.~\ref{sec_hep}, the gauge fixing and the regularization require the presence of a background field, and a breaking of diffeomorphism symmetry. This leads to a considerable enlargement of the theory space by the corresponding couplings. We will get back to this question below.

The FRG provides a vector field in the space of all couplings, denoted $g_i$, which corresponds to the scale-derivative of the couplings at each point, i.e., the beta functions $\beta_i$. The integral curves of this vector field are the RG trajectories. This is in full analogy to the setting described in all previous sections. The key difference lies in the fact that in statistical physics, condensed matter and high-energy particle physics one  is typically interested in the RG flow starting from a given initial condition that describes the microscopic physics. In the asymptotic-safety approach to quantum gravity, the focus is different. Instead of using the Wetterich equation together with an initial condition to obtain $\Gamma_{k\rightarrow 0}$, first we are  interested in discovering whether UV complete trajectories exist. One can view this as the search for a consistent microscopic initial condition, i.e., the search for a microscopic action $S$ that will lead to a predictive theory\footnote{The relation between $\Gamma_{k\rightarrow \infty}$ and the bare action $S$ is addressed in \cite{Manrique:2008zw,Morris:2015oca}. Reconstructing the bare action is an intermediate step to bridge the gap to formulations of quantum gravity based on the Hamiltonian.}. The derivation of $\Gamma_{k\rightarrow 0}$ which encodes the full physics of the theory is then a second step, to be tackled once the existence of a well-defined microscopic starting point $\Gamma_{k\rightarrow \infty}$ for the dynamics has been found.  \\
In order to provide a predictive UV completion, a point in theory space has to satisfy two conditions: \\
Firstly, it must be a fixed point of the RG flow, i.e., a zero of the vector field\footnote{More exotic possibilities such as, e.g., limit cycles, see \cite{Litim:2012vz}, could also in principle be viable, or a fixed point at infinite value of the coupling, i.e., a flow that only leads to an infinite value of the coupling in the limit $k \rightarrow \infty$, see \cite{Eichhorn:2012uv}.}, i.e., $\beta_i=0\, \, \forall i$ at $g_i=g_{i\, \ast}$. This generates  quantum scale-invariance \cite{Wetterich:2019qzx} at this point, just as scale-invariance close to second-order phase transitions in statistical physics is associated to fixed points. Scale-invariance allows the RG flow to spend an infinite amount of RG ``time" at the fixed point before the flow away from it breaks scale-invariance, making new physics at some high scale an option instead of a theoretical necessity.\\
\begin{figure}[!t]
\begin{center}
\includegraphics[width=0.5\linewidth,clip=true, trim=2cm 2cm 13cm 9cm]{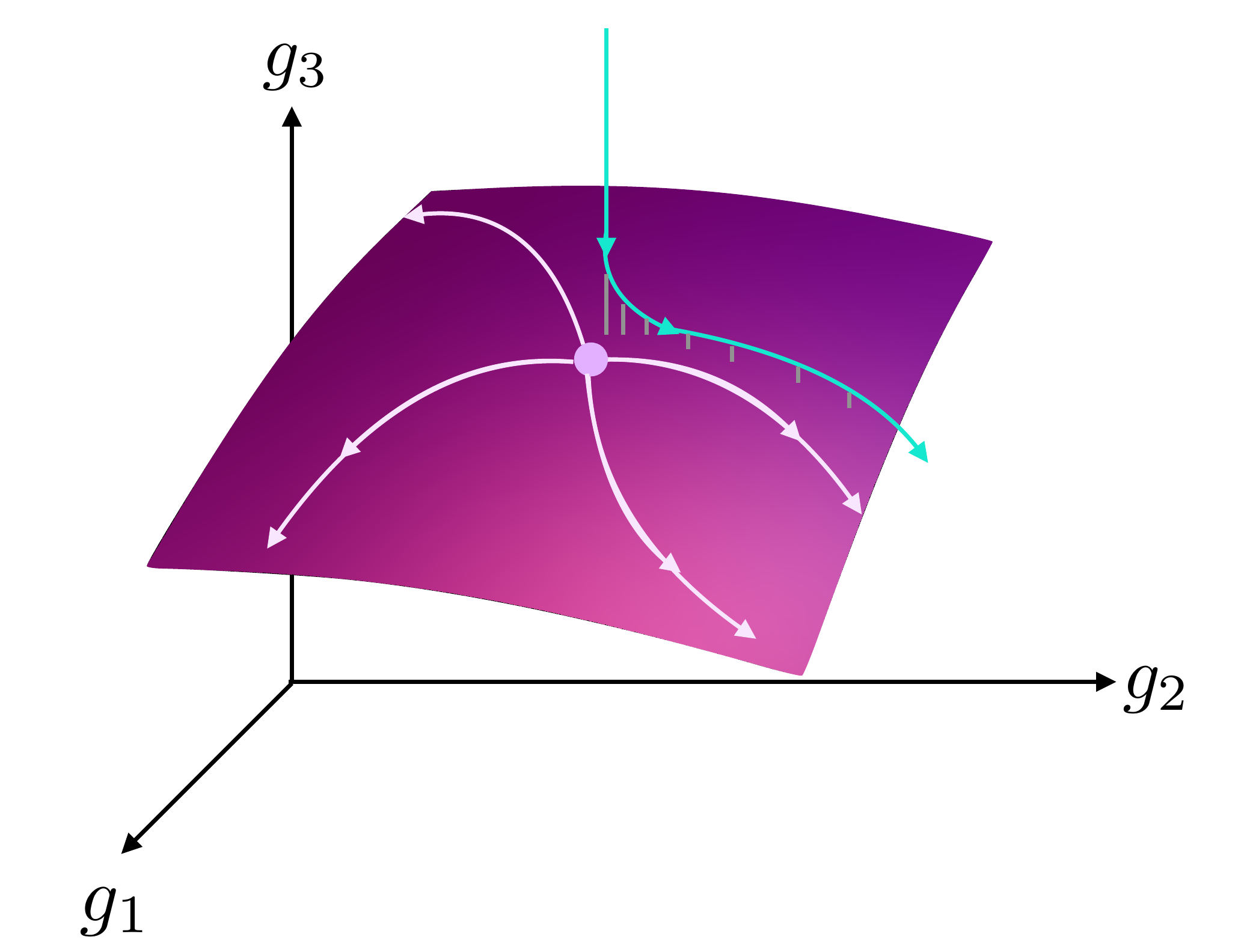}
\end{center}
\caption{\label{sec_gr:fig:ASFP}A fixed point (purple point) comes with a critical hypersurface, spanned by the relevant directions. RG trajectories emanating out of the fixed point must lie within the critical hypersurface, as it is IR attractive (cf.~cyan trajectory). This results in predictions for the values of irrelevant couplings; e.g., in this case, $g_3$ is determined at all scales once experimental input fixes one UV-complete RG trajectory by determining the values of $g_1$ and $g_2$ at some scale.}
\end{figure}

Secondly, the vector field $-\beta_i$ should only have finitely many  independent directions along which it points \emph{away} from the fixed point, i.e., a finite number of relevant directions, corresponding to positive critical exponents\footnote{We use the following convention for critical exponents: $\theta_I = -{\rm eig} \left(\frac{\partial \beta_i}{\partial g_j} \right)\Big|_{g_{\ast}}$. The opposite sign convention can also sometimes be found in the literature.}, cf.~Fig.~\ref{sec_gr:fig:ASFP}. These are the directions along which the RG flow towards the IR can leave the scale-invariant fixed-point regime. The low-energy values of the corresponding relevant couplings cannot be predicted and are the free parameters of the theory. In contrast, all IR-attractive, i.e., irrelevant couplings are determined at all scales on a trajectory that emanates from the fixed point. \\
If a theory space contains a fixed point with the above properties at nonzero values for (some of) the couplings, then a model with the corresponding field content and symmetries can be asymptotically safe. This idea was first put forward as a potential UV completion for gravity in \cite{Weinberg:1980gg}. For several fixed points in the same theory space, each one in principle defines a UV completion. Whether one of those is realized in nature can be tested by measuring whether the low-energy values of the couplings lie in its critical hypersurface.\\
Note the similarity to fixed points in statistical physics: The relevant directions determine the deviation from scale invariance. In statistical physics they can be translated into experimentally tunable quantities, such as, e.g., the temperature, which allow one to tune a system to criticality. In gravity and high-energy physics, the relevant parameters encode how far the low-energy dynamics has deviated from scale-invariance and constitute the quantities that need measurements and cannot be calculated within the theory.\\
To stress the similarity between fixed points in (often lower-dimensional) systems and gravity, let us highlight that
it is a misconception that a fixed point is either UV or IR: Unless it features only IR repulsive or only IR attractive directions, a fixed point can actually be both; depending on the choice of RG trajectory, the fixed point is approached by the trajectory in the IR or serves as a repulsor of the trajectory in the UV, cf.~the discussion in \cite{Rosten:2010vm}. Therefore, the fixed points of relevance in statistical physics and the tentative gravitational fixed point are actually structurally similar in many aspects and can in particular be searched for and investigated with the same set of tools.

\subsection{The FRG for quantum gravity: a brief manual}
The construction in Sec.~\ref{sec_frg} can be generalized to a gravitational setting, as pioneered by Martin Reuter \cite{Reuter:1996cp}, see also \cite{Dou:1997fg}. This entails working with Euclidean signature, i.e., the underlying path integral is for quantum \emph{space}, not spacetime. The analytical continuation to real time is an outstanding challenge which is far more complicated in gravity than in theories on a flat background, see, e.g., \cite{Demmel:2015zfa,Baldazzi:2018mtl}.\\
To set up the Wetterich equation, one must distinguish the ``fast" from the ``slow" modes, i.e., the UV from the IR. The natural generalization of the flat-space construction of the regulator would at a first glance appear to be of the form $g_{\mu\nu}R_k^{\mu\nu\kappa\lambda}(\Delta)g_{\kappa\lambda}$, where $\Delta= -g^{\mu\nu}D_{\mu}D_{\nu}$, and $D_{\mu}$ is the covariant derivative. Yet, this has two problems: Firstly, because of metric compatibility, $D_{\mu}g_{\kappa\lambda}=0$, and secondly, $D_{\mu}$ itself depends on the metric, thus rendering the regulator term higher-order in the field. Yet, the quadratic nature of the regulator is key to the one-loop structure of the flow equation. 
Accordingly, setting up the Wetterich equation for gravity typically requires the introduction of an \emph{auxiliary} background metric $\bar{g}_{\mu\nu}$ which is also used to set up a background-invariant gauge fixing, see also Sec.~\ref{sec_hep}; see also \cite{Falls:2020tmj} for an alternative development. The regulator depends on the background-covariant Laplacian $\bar{\Delta}=-\bar{D^2}$, taking the form
\begin{equation}
\Delta S_k[\bar{g}_{\mu\nu}, h_{\mu\nu}]= \frac{1}{2}\int d^dx\, \sqrt{\bar{g}}\,h_{\mu\nu}R^{\mu\nu\kappa\lambda}_k(\bar{\Delta})h_{\kappa\lambda},
\end{equation}
where
\begin{equation}
h_{\mu\nu} = g_{\mu\nu}- \bar{g}_{\mu\nu}\label{secgrav_eq:linearsplit}
\end{equation}
is the fluctuation field. Introducing the fluctuation field simply corresponds to a shift in the integration variable in the path integral; Eq.~\eqref{secgrav_eq:linearsplit} should not be read in a perturbative sense.
For the special choice $\bar{g}_{\mu\nu} = \delta_{\mu\nu}$, one recovers the construction familiar from previous sections, where the cutoff is implemented in momentum space.  
In the presence of a non-flat  background metric $\bar{g}_{\mu\nu}$, the momentum space is generalized, as the eigenfunctions of the Laplacian $\bar{\Delta}$ are no longer plane waves. Thus, the regulator in the Wetterich equation is set up with respect to the eigenvalues $\lambda_{\bar{\Delta}}$ of the Laplacian $\bar{\Delta}$: Eigenmodes with eigenvalues $\lambda_{\bar{\Delta}}>k^2$ are the UV-modes, and those with $\lambda_{\bar{\Delta}}<k^2$ are the IR-modes. Intuitively speaking, the background metric provides a notion of locality, allowing to define local patches in which the quantum fluctuations are averaged over, thus enabling a coarse-graining procedure.
However, there is an added complication in quantum gravity, as
 all configurations of spacetime are to be treated on an equal footing and no metric should play a distinguished role. The background metric is therefore not to be understood as a physical background, but as an auxiliary field. Accordingly, in addition to the flow equation, there is a Ward identity that relates the background-field dependence of the average effective action to its fluctuation-field dependence, also cf.~Sec.~\ref{sec_hep} for the discussion of such symmetry-identities and the flow equation.
  
The background field can also be used to gauge-fix the fluctuations, as their unregularized inverse propagator, $\Gamma_k^{(2)}$, is not invertible without a gauge-fixing. This is in direct analogy to the case of Yang-Mills theories discussed in the previous section, and entails similar technical challenges pertaining to the simultaneous solution of the Wetterich equation and symmetry-identities. The gauge fixing is often chosen to be of the form
\begin{eqnarray}
S_{\rm gf}&=& \frac{1}{2\alpha}\frac{1}{16\pi G_N}\int d^4x\sqrt{\bar{g}}\, \bar{g}^{\mu\nu}\left(\bar{D}^{\kappa}h_{\kappa \mu} - \frac{1+\beta}{4} \bar{D}_{\mu}h \right)\left(\bar{D}^{\rho}h_{\rho \nu} - \frac{1+\beta}{4} \bar{D}_{\nu}h \right),
\end{eqnarray}
where $h = h^{\mu}_{\mu}$, and indices are raised and lowered with the background metric. $\alpha$ and $\beta$ are gauge-parameters, and $\alpha\rightarrow 0$ implements the Landau-gauge limit of the gauge condition. The corresponding Faddeev-Popov ghost term also contributes to the flow.

The trace in the Wetterich equation \eqref{sec_frg:eqwet} can be evaluated by summing over the eigenvalues for backgrounds $\bar{g}_{\mu\nu}$ for which the full spectrum of the operator $\bar{\Delta}$ is known, e.g., \cite{Eichhorn:2009ah,Benedetti:2012dx,Alkofer:2018fxj}, or by using heat-kernel techniques, which only requires knowledge of the heat kernel instead of the full spectrum of the operator, see, e.g., \cite{Lauscher:2001ya,Reuter:2001ag,Codello:2008vh,Benedetti:2010nr,Kluth:2019vkg}.  This provides the machinery to extract beta functions for gravity from the Wetterich equation.

The last important aspect for applications of the Wetterich equation is the choice of truncation. Here, we will provide a brief overview of truncations that have mainly been used. 
It is generally assumed -- and by now supported by many explicit results -- that the canonical dimension is a robust guide to determine which couplings are relevant at the asymptotically safe fixed point. This is motivated firstly by the possibility that the Reuter fixed point could be connected continuously to the perturbative fixed point in $d=2+\epsilon$ dimensions \cite{Gastmans:1977ad,Christensen:1978sc}. Secondly, explicit studies of truncations of the form $\Gamma_k = \sum_{i=0}^n \int d^4x\, \sqrt{g}a_i R^i$, with $n=70$ \cite{Falls:2013bv,Falls:2014tra,Falls:2018ylp}, exhibit a near-canonical scaling for the higher-order operators, see also \cite{Lauscher:2002sq,Machado:2007ea,Codello:2008vh,Benedetti:2009rx,Ohta:2013uca,Falls:2013bv,Falls:2014tra,Ohta:2015fcu,Ohta:2015efa,Gies:2016con,Falls:2017lst,deBrito:2018jxt,Falls:2018ylp}; however note that a number of dimension-6 operators (and beyond) remain to be explored. Third, studies of the magnitude of diffeomorphism-symmetry breaking, expected to be large and connected to highly nontrivial Slavnov-Taylor identities in a nonperturbative regime, yield results compatible with a near-perturbative nature of the Reuter fixed point \cite{Eichhorn:2018akn,Eichhorn:2018ydy,Eichhorn:2018nda}. Fourth, there are also hints for the Reuter fixed point from perturbative studies \cite{Codello:2006in,Niedermaier:2009zz,Niedermaier:2010zz}. For such a fixed point, near-canonical scaling of the critical exponents is expected. Specifically, this means that shifts $\mathcal{O}(1)$ of the critical exponents, compared to the canonical scaling dimension, can be expected, but not shifts $\mathcal{O}(10)$.  This supports a truncation scheme based on the canonical dimension of couplings, where higher orders in the curvature and derivatives are expected to be irrelevant. It is worth to mention that, at odds with scalar theories where operators with high canonical dimension can be constructed without introducing high number of derivatives, in the quantum-gravity context the classification along canonical dimension in terms of diffeomorphism-invariant operators is more closely related to the derivative expansion.
 In principle, near-canonical scaling might even provide a small parameter with which to achieve systematic error estimates, namely the deviation of quantum scaling at the fixed point from canonical scaling, see also Sec.~\ref{sec_sm} for the discussion of a small parameter in the derivation expansion.

A considerable body of literature employs the single-metric approximation, in which the distinction between background metric and full metric is only made in the gauge-fixing, ghost and regulator term. Based on canonical power counting as a guiding principle, one then sets up truncations in a curvature expansion, i.e., 
\begin{equation}
\Gamma_k =\int d^4x\sqrt{g}\,\left( \frac{1}{16\pi\, G_N}\left(2 \bar{\Lambda }- R \right)+ a\, R^2 + b\, R_{\mu\nu}R^{\mu\nu}+... \right) + S_{\rm gf}+ S_{\rm gh}.
\end{equation}

The ghost-sector and gauge-fixing are typically only accounted for by including the corresponding classical actions $S_{\rm gf}$ and $S_{\rm gh}$. Explicit studies indicate that additional terms might be present for the ghost sector at a nontrivial fixed point \cite{Eichhorn:2013ug}.

Finally let us note that alternative gravitational theory spaces are also accessible to the FRG. Formulations of gravity which have been explored in this setup are based, e.g., on a unimodular theory space \cite{Eichhorn:2013xr,Eichhorn:2015bna,Benedetti:2015zsw} in which the determinant of the metric is held fixed such that the cosmological constant is no longer a term in the action and therefore not subject to the usual fine-tuning. Further studies focussed on a formulation based on the tetrad and the spin connection \cite{Daum:2010qt,Daum:2013fu,Harst:2014vca,Harst:2015eha}, the tetrad with a dependent connection \cite{Harst:2012ni}, or including dynamical torsion degrees of freedom \cite{Pagani:2015ema,Reuter:2015rta}.
Topologically massive 3d-(super)gravity has been studied in \cite{Percacci:2010yk,Percacci:2013ii}; for a study of the massive Pauli-Fierz action see \cite{Binder:2020ifl}. Conformal gravity has been explored in \cite{Ohta:2015zwa}. Intriguingly, results in truncations in many of these settings hint at asymptotically safe fixed points in these settings; in particular, the results in unimodular gravity are close to those for the Reuter fixed point, except for the fact that there is one relevant direction less in truncations, since the cosmological context is not a part of the theory space. This result would support the hypothesis that the spin-2 modes are most relevant to generate an asymptotically safe fixed point. Interestingly, the loosely speaking ``opposite" approximation, where only the conformal mode is kept, also features an asymptotically safe fixed point in the truncations explored in \cite{Reuter:2008qx,Reuter:2008wj,Daum:2008gr,Machado:2009ph,Bonanno:2012dg,Dietz:2015owa,Labus:2016lkh,Dietz:2016gzg}. 

 The development of a flow equation based on a single metric can be found in \cite{Wetterich:2016ewc} and a proper-time flow equation for gravity in \cite{Bonanno:2004sy}. The Polchinski equation has been applied to test the asymptotic-safety paradigm in quantum gravity in \cite{deAlwis:2017ysy,deAlwis:2018azs}.

\subsection{Status and open questions of asymptotically safe gravity}

\subsubsection{Indications for the Reuter fixed point}

\begin{figure}
\includegraphics[width=\linewidth, clip=true, trim=0cm 15cm 0cm 0cm]{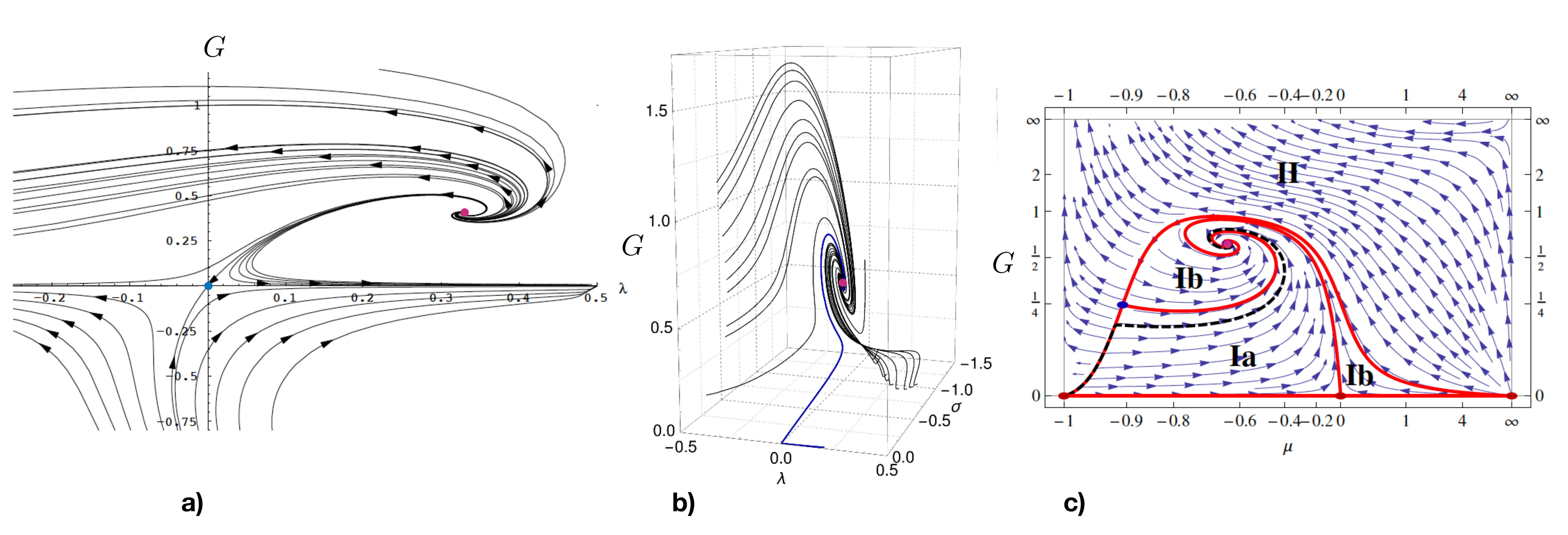}
\caption{\label{fig:gravity_flows} The RG flow in a) the Einstein-Hilbert truncation from \cite{Reuter:2001ag}, b) the Einstein-Hilbert truncation plus a curvature-cubed term from \cite{Gies:2016con} and c) in the reverse direction (i.e., flow trajectories shown towards the UV) in the Einstein-Hilbert truncation in the fluctuation-field approach (where $\mu = -2 \lambda$) from \cite{Christiansen:2014raa}. A UV fixed point that is IR repulsive in two directions and IR attractive in one direction (panel b)) is marked by a purple dot. Perturbative nonrenormalizability of GR is linked to the fact that the free fixed point (marked in blue in plot a)) is IR attractive in the Newton coupling.}
\end{figure}

\begin{figure}
\begin{center}
\includegraphics[width=0.5\linewidth]{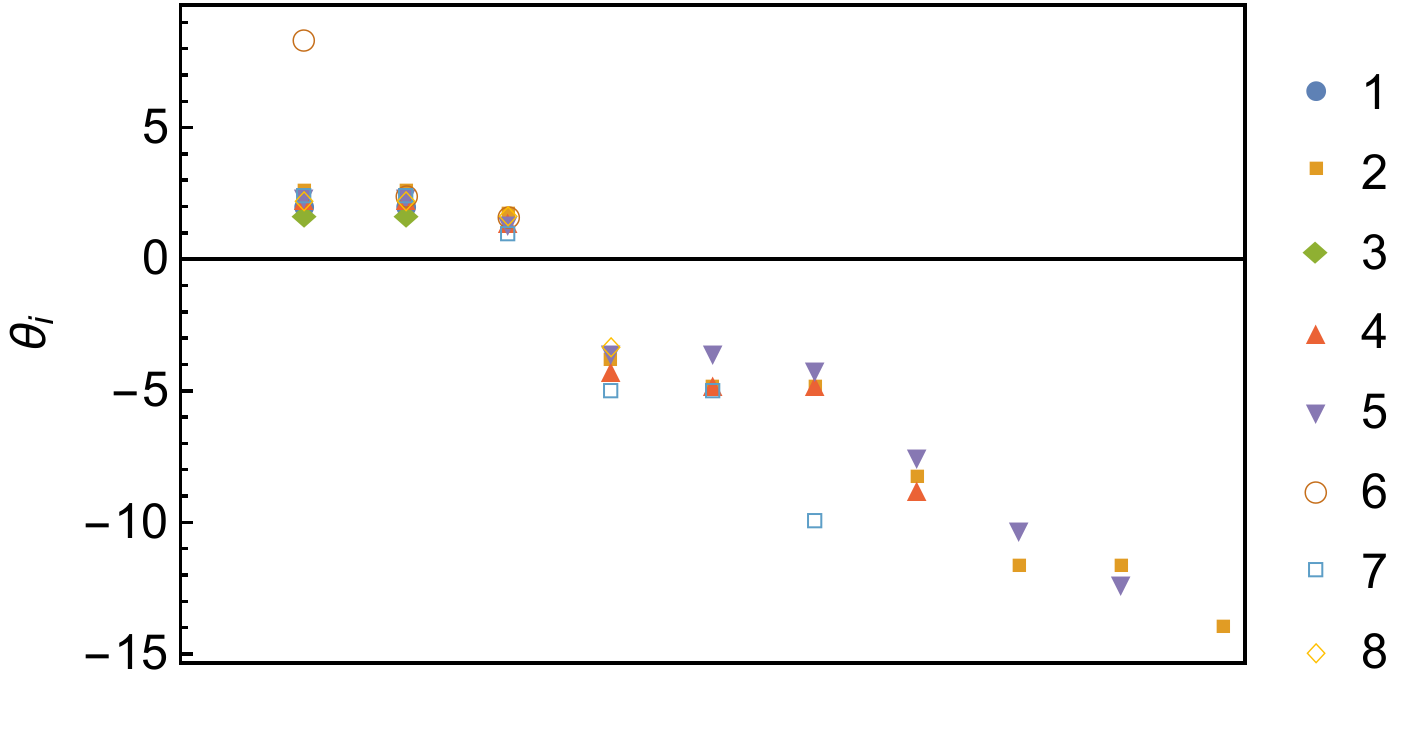}
\end{center}
\caption{\label{figgravtheta} Real part of critical exponents sorted from most relevant towards irrelevant from the following references in single-metric calculations: 1 = \cite{Reuter:2001ag}, 2=\cite{Falls:2013bv}, 3=\cite{Litim:2003vp},4 =\cite{Machado:2007ea}, 5=\cite{Codello:2008vh}, 6=\cite{Benedetti:2009rx}, 7=\cite{Falls:2017lst}, 8=\cite{Falls:2020qhj}. In \cite{Falls:2013bv,Falls:2017lst}, critical exponents are calculated to significantly higher orders in the expansion than is shown here and clearly show a near-Gaussian scaling. The results are consistent with a three-dimensional critical hypersurface.}
\end{figure}

The Reuter fixed point is the tentative interacting, predictive fixed point in gravity with full diffeomorphism invariance\footnote{It is thereby distinct from the asymptotically free fixed point in curvature squared gravity \cite{Avramidi:1985ki,deBerredoPeixoto:2003pj,deBerredoPeixoto:2004if} and the free fixed point in Horava gravity \cite{DOdorico:2014tyh,Barvinsky:2017kob}.}.
Indications for this fixed point have robustly been found in all quasilocal truncations in the literature, see Fig.~\ref{fig:gravity_flows} for an illustration, including studies of the Einstein-Hilbert truncation \cite{Souma:1999at,Lauscher:2001ya,Reuter:2001ag,Litim:2003vp,Groh:2010ta,Eichhorn:2010tb,Nagy:2013hka,Falls:2014zba,Gies:2015tca} and beyond \cite{Lauscher:2002sq,Machado:2007ea,Codello:2008vh,Benedetti:2009rx,Ohta:2013uca,Falls:2013bv,Falls:2014tra,Ohta:2015efa,Gies:2016con,Falls:2017lst,Nagy:2017zvc,Falls:2018ylp}, see \cite{Reuter:2012id,Percacci:2017fkn,Eichhorn:2018yfc,Reuter:2019byg,Pawlowski:2020qer} for reviews and \cite{Eichhorn:2020mte,Reichert:2020mja} for recent introductory lectures. 
Extensions to higher orders in $R^n$ can be found in \cite{Lauscher:2002sq,Machado:2007ea,Codello:2007bd,Codello:2008vh,Benedetti:2012dx,Demmel:2012ub,Dietz:2012ic,Dietz:2013sba,Falls:2013bv,Falls:2014tra,Demmel:2015oqa,Gonzalez-Martin:2017gza,Falls:2018ylp}, and inclusions of Ricci tensor invariants and Riemann tensor invariants in \cite{Benedetti:2009rx,Benedetti:2009gn,Gies:2016con,Falls:2017lst}. A complete study at order curvature-squared can be found in \cite{Falls:2020qhj}. These terms cannot consistently be set to zero at an interacting fixed point in the Einstein-Hilbert truncation; they are generated by the flow and it constitutes a nontrivial check of the fixed-point hypothesis to explore whether an extension of the fixed point in the Einstein-Hilbert truncation exists once these terms are included in the truncation.\\
On manifolds with boundaries, boundary terms have to be added to the action. First steps to explore their flows, focussing on the Gibbons-Hawking term, can be found in \cite{Becker:2012js,Falls:2017cze}.\\
A key difference to the perturbative renormalization of gravity is that higher-order terms, such as, e.g., $R_{\mu\nu\kappa\lambda}R^{\kappa \lambda}_{\, \, \, \, \rho \sigma}R^{\rho \sigma \mu\nu}$ add new free parameters in perturbation theory that cannot be neglected at the Planck scale. At the asymptotically safe fixed point, e.g., the so-called Goroff-Sagnotti term $R_{\mu\nu\kappa\lambda}R^{\kappa \lambda}_{\, \, \, \, \rho \sigma}R^{\rho \sigma \mu\nu}$ does not lead to the appearance of a new relevant direction within a truncation where it is coupled to the Einstein-Hilbert term \cite{Gies:2016con}. 
This constitutes an example of how the demand that RG trajectories emanate out of a UV fixed point in the flow towards the IR (and conversely, can be traced back into the fixed point if the flow is reversed, which can be done in any finite-dimensional truncation), is a strong principle inducing predictivity, i.e., fixing free parameters that are present in an effective-field-theory setting.\\
In the above truncations the Reuter fixed point features three relevant directions, although it is important to keep in mind that significant systematic uncertainties still affect the calculation of the critical exponents, cf.~Fig.~\ref{figgravtheta}.

As has been discussed, e.g., in Sec.~\ref{sec_sm}, the FRG can be applied with the dimensionality as a free parameter, allowing to track fixed points across different dimensionalities. This has been used to explore the Reuter universality class also away from four dimensions, see, e.g., \cite{Fischer:2006fz,Nink:2012vd,Falls:2017cze,Falls:2015qga,Biemans:2016rvp}. One should note that as $d$ increases, one expects the same truncation to perform worse than in low $d$, since increasingly many higher-order curvature invariants become canonically relevant. With this caveat in mind, results in the above references indicate that the fixed point in pure gravity can be extended beyond $d=4$, although it remains unclear whether there is an upper critical dimension.

The robustness of results is tested by i) extending the truncation and ii) studying the dependence of universal quantities such as critical exponents on unphysical parameters of the setup: Just as any other approximation scheme, including perturbation theory, physical quantities acquire a dependence on unphysical choices that define a scheme in a broad sense. In the FRG, these include the choice of regulator shape function, e.g., \cite{Groh:2010ta,Reuter:2001ag}, the choice of gauge fixing \cite{Gies:2015tca,Ohta:2016npm,Ohta:2016npm}, as well as the choice of parameterization \cite{Gies:2015tca,Ohta:2016npm,deBrito:2018jxt} linked to the introduction of the background field. 

Similarly to many systems discussed in previous sections, momentum-dependence can play an important role. Physically, this is clearly relevant for gravitational scattering processes, including graviton-mediated scattering processes as well as graviton-scattering. Despite the outstanding challenge to access such scattering amplitudes experimentally, their behavior at high energies is a useful theoretical check since it carries information on the unitarity of the theory.
To study the momentum-dependence in the gravitational case, one can for instance introduce  form-factors \cite{Bosma:2019aiu,Knorr:2019atm}, e.g., 
\begin{equation}
\Gamma_k = \Gamma_{\rm EH}+\int d^4x \sqrt{g} \left( R f_1 (-D^2) R + R_{\mu\nu}f_2(-D^2)R^{\mu\nu} + R_{\mu\nu\kappa \lambda}f_3(-D^2)R^{\mu\nu\kappa\lambda}+...  \right) + S_{\rm gf}+ S_{\rm gh}.
\end{equation}
This is simply a way of rearranging the couplings discussed previously into a different expansion; the couplings are given by the Taylor coefficients of the form factors. In a flat expansion, corresponding terms have been explored starting in \cite{Christiansen:2012rx,Christiansen:2015rva}.
Such form factors could become important to achieve a control of the behavior of the theory not only at high $k$, but at high, physical momentum scales; see also, e.g., \cite{Denz:2016qks,Knorr:2017fus,Christiansen:2017bsy,Eichhorn:2018nda} for studies of the momentum dependence of metric correlation functions. In particular, correlation functions for the metric can also be studied in a curvature-dependent fashion \cite{Knorr:2017fus,Christiansen:2017bsy,Burger:2019upn}.

\subsubsection{Background field and dynamical field}
The fixed point exists in the single-metric approximation, where one equates $g_{\mu\nu} = \bar{g}_{\mu\nu}$ after determining the form of $\Gamma_k^{(0,2)}[\bar{g}_{\mu\nu}; h_{\mu\nu}]$ in a given truncation. This approximation neglects the difference between background couplings and fluctuation couplings that exist due to the different appearance of the two fields in the regulator and gauge-fixing terms. 
Their difference is controlled by the shift Ward identity \cite{Reuter:1997gx,Litim:2002hj,Pawlowski:2005xe,Bridle:2013sra,Safari:2015dva}, 
a.~k.~a.~split Ward identity which takes the following form for the gravitational case
\begin{eqnarray}
&{}&\frac{\delta \Gamma_k}{\delta \bar{g}_{\mu\nu}}-\frac{\delta \Gamma_k}{\delta h_{\mu\nu}}
= \frac{1}{2}{\rm Tr}\Bigl[\frac{1}{\sqrt{\bar{g}}}\frac{\delta \sqrt{\bar{g}R_k[\bar{g}]}}{\delta \bar{g}_{\mu\nu}}\left(\Gamma_k^{(0,2)}[\bar{g}_{\mu\nu}; h_{\mu\nu}]+R_k[\bar{g}] \right)^{-1} \Bigr]+ \left< \frac{\delta S_{\rm gf}}{\delta \bar{g}_{\mu\nu}}-\frac{\delta S_{\rm gf}}{\delta h_{\mu\nu}}\right>+ \left< \frac{\delta S_{\rm gh}}{\delta \bar{g}_{\mu\nu}}-\frac{\delta S_{\rm gh}}{\delta h_{\mu\nu}}\right>.
\label{sec_gr:sWI}
\end{eqnarray}
The left hand-side is the difference between the background-field dependence and the fluctuation field dependence. The right-hand side encodes the sources of this difference, which are given by the background-field dependence of the regulator (first term on the lhs) as well as the gauge-fixing and ghost terms. Since fluctuations in $h_{\mu\nu}$ are gauge-fixed with respect to the background field, the gauge-fixing and ghost terms treat $\bar{g}_{\mu\nu}$ differently from $h_{\mu\nu}$ and must accordingly appear on the right-hand-side of this symmetry identity. Regarding the first term, one can see that it is structurally similar to the right-hand-side of the flow equation itself. This is a consequence of the fact that this term encodes the dependence of the regulator on an external field ($\bar{g}_{\mu\nu}$) or parameter ($k$); therefore, except for the substitution $\partial_k R_k \rightarrow  \delta \sqrt{\bar{g}R_k[\bar{g}]}/\delta \bar{g}_{\mu\nu}$, these terms must be the same.
The Ward-identity is explicitly studied in truncations in \cite{Donkin:2012ud,Dietz:2015owa,Labus:2016lkh,Morris:2016spn,Percacci:2016arh,Nieto:2017ddk,Ohta:2017dsq,Eichhorn:2018akn}. \\
Note that the physics of asymptotic safety is encoded in the full effective action at $g_{\mu\nu}= \bar{g}_{\mu\nu}$, i.e., $\Gamma_{k\rightarrow 0}[\bar{g}_{\mu\nu};0]$. Yet, in order to calculate this quantity, it is important to distinguish $\Gamma_k^{(0,2)}[\bar{g}_{\mu\nu}; h_{\mu\nu}]$ and $\Gamma_k^{(2,0)}[\bar{g}_{\mu\nu}; h_{\mu\nu}]$. It is the former, not the latter, that drives the flow; and they differ as described by Eq.~\eqref{sec_gr:sWI}. Accordingly, although the physics is encoded in background quantities, intermediate steps in their calculation require a clean distinction of background and fluctuation quantities. In particular, in \cite{Pagani:2019vfm}, the importance of keeping track of a self-consistent choice of background has been emphasized. The bimetric approximation distinguishes between the flow of background quantities and dynamical quantities. It can equivalently be implemented by tracking the dependence of the average effective action on $\bar{g}_{\mu\nu}$ and $g_{\mu\nu}$ or that on $\bar{g}_{\mu\nu}$ and $h_{\mu\nu}$. Studies in these various setups hint at the qualitative reliability of the single-metric approximation in pure gravity \cite{Manrique:2009uh,Manrique:2010am,Christiansen:2012rx,Becker:2014qya,Codello:2013fpa,Christiansen:2014raa,Christiansen:2015rva,Denz:2016qks,Knorr:2017fus,Knorr:2017mhu,Christiansen:2017bsy}, although the background cosmological constant appears to deviate significantly from its fluctuation counterpart under the impact of quantum fluctuations of matter \cite{Eichhorn:2018akn}. \\
Note that there are various ways to split the full metric $g_{\mu\nu}$ into a background $\bar{g}_{\mu\nu}$ and fluctuation field $h_{\mu\nu}$. Most studies to date employ a linear split as in Eq.~\eqref{secgrav_eq:linearsplit}, but alternative parameterizations have also been explored, see, e.g., \cite{Eichhorn:2013xr,Nink:2014yya,Eichhorn:2015bna,Falls:2015qga,Percacci:2015wwa,Gies:2015tca,Ohta:2015efa,Ohta:2016npm,deBrito:2018jxt}. A Vilkovisky-DeWitt approach in the FRG framework \cite{Pawlowski:2003sk} for gravity has been explored in \cite{Donkin:2012ud}. When changing from one parameterization to another, a Jacobian arises in the path integral. Further, the domain of integration can in general change, see, e.g., \cite{Nink:2014yya}. Note that the studies that compare the results within different  parameterizations  up to date neglect the Jacobian and change of domain of integration.

In summary, there are compelling indications for the existence of the Reuter fixed point in four-dimensional Euclidean gravity, although the implementation of background independence currently remains an outstanding problem, see also \cite{Bonanno:2020bil} for a discussion of current open questions in asymptotically safe gravity.

\subsubsection{Key challenges for asymptotically safe gravity}
Additional key points for future research include an understanding of Lorentzian gravity, see \cite{Manrique:2011jc}. Unlike in all other sections of this review, a Wick rotation is in general not available in quantum gravity \cite{Visser:2017atf,Baldazzi:2018mtl}. For instance, the existence of horizons, which are null hypersurfaces, is tied to the existence of causal structure, i.e., Lorentzian nature.
Further, the configuration space of Riemannian as compared to Lorentzian metrics has significantly different properties \cite{Demmel:2015zfa}.
Therefore a scale-invariant fixed-point regime in Euclidean gravity does not automatically imply the same for Lorentzian gravity, and Lorentzian gravity must be explored separately, see also \cite{Donoghue:2019clr} for a discussion of this point. Since a momentum cutoff is most straightforwardly imposed in Euclidean signature, the study of Lorentzian asymptotic safety is a challenge. 
The impact of a foliation structure on the Reuter fixed point has been explored in \cite{Biemans:2016rvp,Biemans:2017zca,Houthoff:2017oam}, also bringing the FRG studies closer to Monte Carlo simulations based on a configuration space where each configuration admits a foliation \cite{Ambjorn:2000dv,Ambjorn:2004qm,Ambjorn:2011cg,Ambjorn:2012jv}.

A further central open question is to determine which degrees of freedom propagate. This is not fixed uniquely once the field is specified. For instance, an action including $\int d^4x\, \sqrt{g}\,R^2$ in addition to the Einstein-Hilbert term propagates an additional scalar, as one can most easily see by performing a conformal transformation which isolates the kinetic term for the conformal mode. Given that the action in asymptotic safety is expected to contain higher-derivative terms,  there might be further modes that propagate beyond a massless spin-2 graviton. This is not only potentially phenomenologically relevant, as an additional scalar might drive an inflationary regime \cite{Bonanno:2015fga}, but such modes could threaten the consistency of the theory, as unitarity/kinematic stability could be violated \cite{Arici:2017whq,Becker:2017tcx}. In two-dimensional gravity, a unitary theory underlying asymptotic safety has been identified, see \cite{Nink:2015lmq}. Additionally,  it is worthwhile noting that Causal Dynamical Triangulations (CDTs) feature a self-adjoint and bounded transfer matrix. In CDTs, each configuration can be Wick-rotated, implying that a  well-defined Hamiltonian exists in the discrete setting \cite{Ambjorn:2000dv,Ambjorn:2001cv}. If this property persists in the continuum limit, and the universality class of that continuum limit is the Reuter fixed point, then asymptotically safe gravity would be expected to be unitary according to this notion of unitarity.\\
It should be noted that in general the question of unitarity is significantly more subtle in quantum gravity than in flat-space QFTs without gauge symmetry. These subtleties include but are not limited to the points that i) flat space does not necessarily have to be stable to agree with observations (since we live in a universe with a nonvanishing cosmological constant) and therefore (tachyonic) instabilities around flat space are not necessarily a problem for the viability of the theory; ii) the definition of an S-matrix on a generic non-trivial backgrounds is an outstanding challenge;  iii) spectral representations for the propagators in a gauge theory can feature negative parts without posing a problem for the unitarity of the theory -- in fact, Yang-Mills theory is an example, since the spectral function of the gluon features negative parts. Therefore the notion of unitarity that is used in flat-space QFTs without gauge symmetries might be applicable in quantum gravity in a perturbative regime, but might not suffice to analyze the theory in the fully non-perturbative regime -- or even in a pre-geometric one, where no nontrivial background metric is available.\\

\subsubsection{Understanding spacetime structure}
A central question of (asymptotically safe) quantum gravity concerns the properties of the effective quantum geometries that arise. Most importantly, these pertain to physically relevant spacetimes, such as that of black holes or in cosmology. Using the idea of RG-improvement, quantum-gravity inspired models have been explored. Their starting point is the RG scale dependence of the gravitational couplings: The dimensionful Newton coupling is constant in the classical gravity regime below the transition scale $1\gamma$ to the fixed-point regime. In the fixed-point regime, the dimensionless Newton coupling $G$ takes its constant fixed-point value, which is actually given by $G_{\ast}=1/\gamma$, such  that $G_N = G k^{-2} = 1/\gamma k^{-2}$ falls off like $k{-2}$.
The simplest form of the dimensionful  Newton coupling $G_N(k)$ that models this RG flow between the scale-invariant fixed-point regime and the classical regime in the infrared is given by 
\begin{equation}
G_N(k) = \frac{G_0}{1+\gamma\, G_0 k^2}, \label{secgravity_eq:GN}
\end{equation}
where $\gamma$ is the inverse dimensionless transition scale between the two regimes. $G_0$ is the value of the dimensionful Newton couplings at low scales, i.e., $G_N(k \rightarrow 0) \rightarrow G_0$, i.e., it is the value of $G_N$ that is achieved once all quantum fluctuations have been integrated out.
As one explores higher $k$, the fixed-point scaling stops the quadratic growth of the dimensionless Newton coupling $G = G_N k^2$, which instead levels off to a constant. Conversely, this implies that the dimensionful Newton coupling $G_N$ that is constant at low $k$, scales quadratically, $G_N \sim G_{\ast} k^{-2}$, at high $k$. Eq.~\eqref{secgravity_eq:GN} exhibits a relatively sharp transition between the classical and the fixed-point regime, which it shares with the scale dependence of $G_N(k)$ obtained by integrating the RG flow in a given truncation, see, e.g., \cite{Reuter:2001ag,Litim:2003vp,Denz:2016qks}.\\
One might be tempted to interpret this very directly in terms of the dynamics of gravity at high energy/ momentum/ curvature scales. Yet, one should be rather careful with an identification of $k$ with a physical scale, see, e.g., an example in the Lorentzian case \cite{Anber:2011ut}. Models of spacetime obtained from an identification of $k$ with physical scales should thus be considered as "quantum-gravity inspired" models that could provide a first qualitative understanding of potential quantum gravitational effects. With this in mind, the weakening of the dimensionfull Newton coupling towards high momentum scales could rise to the expectation that classically singular black-hole spacetimes should become regular. This has been explored using RG-improvement, i.e., upgrading the classical solution/dynamics by a scale-dependent coupling.
The Schwarzschild case, first analyzed in  \cite{Bonanno:1998ye,Bonanno:2000ep}, provides the
``blueprint" for later works \cite{Bonanno:2006eu,Cai:2010zh,Reuter:2010xb,Falls:2010he,Falls:2012nd,Torres:2013cya,Litim:2013gga,Koch:2013owa,Kofinas:2015sna,Torres:2017ygl,Pawlowski:2018swz,Adeifeoba:2018ydh}. For black holes, a ``natural" scale identification is to use the curvature scale, as this provides a coordinate-independent notion of scale. As black hole spacetimes are vacuum solutions of the Einstein equations, $R=0$ and $R_{\mu\nu}=0$. Instead, one uses the Kretschmann scalar, which is given by $K = R_{\mu\nu\kappa \lambda} R^{\mu\nu\kappa\lambda}=\frac{48G_0^2\, M^2}{r^6}$ for the spherically symmetric case, where $G_0$ is the classical value of the dimensionful Newton coupling, $M$ is the mass of the black hole and $r$ is the radial coordinate in Schwarzschild coordinates. Due to the spherical symmetry and static nature of the Schwarzschild spacetime, all other curvature invariants are proportional to the Kretschmann scalar and therefore yield the same result for the RG-improved metric.
Once $K^{1/4}$ exceeds the Planck mass, one expects quantum-gravity effects to kick in. This motivates the scale identification
\begin{equation}
k = \alpha\, K^{1/4},
\end{equation} 
where the power is fixed on dimensional grounds and $\alpha$ is a dimensionless number of order one.
Inserting $G_N(k) = G_N(K^{1/4})$ from Eq.~\eqref{secgravity_eq:GN} in the classical form of the metric leads to an upgraded metric, for which the Kretschmann scalar is no longer singular. This regular black-hole metric should be understood as an asymptotic-safety inspired model for a black hole, and has been explored further in the literature.
In \cite{Platania:2019kyx}, the RG improvement procedure is iterated using the Kretschmann scalar of the improved metric, in which case the RG improved black-hole spacetime converges to the so-called Dymnikova metric \cite{Dymnikova:1992ux}. In \cite{Held:2019xde}, the size and shape of black-hole shadows for the corresponding modified spacetimes are explored, see also \cite{Kumar:2019ohr}. RG improved  gravitational collapse has been explored in \cite{Casadio:2010fw,Fayos:2011zza,Torres:2014gta,Torres:2014pea,Bonanno:2016dyv,Bonanno:2017zen,Bonanno:2017kta} and RG improved stellar interiors have been studied in \cite{Bonanno:2019ilz}. It should be emphasized that the RG improvement procedure could be implemented at the level of a given spacetime geometry, at the level of the equations of motion, or at the level of the action, without a guarantee of agreement between the outcomes. \\
Besides black holes, another phenomenologically relevant class of spacetimes in which quantum gravity is expected to play a role is cosmological spacetimes. In particular, one expects effects of quantum gravity in the early universe. This provides a potential testing ground for asymptotic safety, as well as other models of quantum gravity. Various aspects of cosmology have been explored in the context of asymptotic safety, resulting in asymptotic-safety inspired cosmological models. For recent reviews of potential imprints of asymptotic safety in cosmology, as suggested by exploring RG improved models, see \cite{Bonanno:2017pkg,Platania:2020lqb}. \\
Cosmological singularity resolution and the suppression of anisotropies and inhomogeneities based on the presence of higher-order interactions in asymptotically safe gravity has been suggested in \cite{Lehners:2019ibe}. This work is based on an argument regarding the gravitational path integral, where early-universe spacetimes interfere destructively, if they are associated to a diverging action. This is the case for a curvature-squared action, as one might expect to play a role in asymptotic safety, when applied to early-universe spacetimes with anisotropies and inhomogeneities, which are disfavored also observationally.
Further, asymptotic safety might play a role in the context of inflation \cite{Weinberg:2009wa}. Here, the first main question is whether asymptotic safety could result in an inflationary phase without the need for an additional scalar field that is introduced ad-hoc. Given that, e.g., $f(R)$ theories come with an additional propagating scalar, as one can see by performing an appropriate conformal transformation on the metric, the inflaton might arise as an automatic consequence of asymptotically safe gravity. The second main question is whether an inflationary phase results in agreement with the observed spectrum of the cosmic microwave background. The deviation from scale invariance, tensor-to-scalar-ratio and energy scale associated to inflation, encoded in the amplitude of scalar fluctuations, are measured/constrained by Planck and therefore constitute potential checks of the quantum-gravity theory.
RG improvement has been used to derive quantum-gravity inspired models of inflation, see \cite{Bonanno:2001xi,Reuter:2005kb,Bonanno:2010bt,Bonanno:2010mk,Cai:2012qi,Bonanno:2012jy,Copeland:2013vva,Tronconi:2017wps,Bonanno:2018gck,Liu:2018hno,Platania:2019qvo} and the early universe \cite{Hindmarsh:2011hx,Kofinas:2016lcz}. Cosmological perturbation theory in an RG improved context has been studied in \cite{Bonanno:2002zb,Contillo:2010ju}.
Late-time modifications of cosmology that might arise due to infrared effects of quantum gravity in the context of RG fixed points, see \cite{Dou:1997fg,Bonanno:2001hi,Babic:2004ev,Ahn:2011qt,Bonanno:2011yx,Wetterich:2018qsl,Anagnostopoulos:2018jdq,Gubitosi:2018gsl}.  
 \\

Another central task for any approach to quantum gravity is to characterize the properties of quantum spacetime in the vicinity of the Planck scale. 
Indications for a reduction from $d=4$ to $d=2$ exist in the ultraviolet in the spectral dimension which is extracted from an RG improved diffusion equation \cite{Lauscher:2005qz,Reuter:2011ah,Rechenberger:2012pm,Calcagni:2013vsa}. The underlying idea is that a diffusion process, i.e., a random walk, probes the properties of the effective spacetime manifold at a given scale. For instance, the return probability to the starting point can give direct insight into the underlying dimensionality, providing what is known as the spectral dimension.
Here, some care is required when performing the RG improvement, as only the use of diffusion time as inverse RG scale results in a well-defined diffusion equation \cite{Calcagni:2013vsa}, highlighting the potential pitfalls of the RG improvement procedure.
The observed reduction in the spectral dimension does not imply a reduction in the topological dimension of the manifold: In common with fractals, dimensionality can become a resolution-scale dependent quantity that further depends on the ``prescription" with which it is measured. For instance, the effects of quantum fluctuations of spacetime that slow down the random walker underlying the measurement of the spectral dimension do not automatically impact the scaling of the volume of an $d$-sphere with its radius. The latter provides the Hausdorff dimension which actually stays constant at all scales in asymptotic safety \cite{Reuter:2011ah}.
A dimensional reduction is also familiar from some other quantum-gravity approaches, see \cite{Carlip:2017eud} for an overview.

To further characterize the geometry of spacetime, one can for instance analyze  the behavior of geodesic curves, or various hypersurfaces. Progress towards understanding these has been made in terms of a flow equation for composite operators \cite{Igarashi:2009tj,Pawlowski:2005xe,Pagani:2016dof,Pagani:2017tdr}, which provides hints for dimensional reduction from the scaling behavior of geodesics and areas \cite{Becker:2018quq,Houthoff:2020zqy,Kurov:2020csd}, see also \cite{Becker:2019fhi} for a comparison to higher-derivative gravity.

\subsubsection{Asymptotically safe gravity and particle physics}
Although a substantial part of quantum-gravity research is focused on a purely gravitational setting, it is neither necessary nor sufficient for a quantum description of spacetime to be viable in our universe to exist as a purely gravitational theory, since we know from observations that matter exists. Accordingly, the existence of a fixed point in gravity-matter systems is actually key for a phenomenologically viable asymptotic-safety paradigm. The exploration of asymptotically safe gravity-matter models started in \cite{Percacci:2002ie,Percacci:2003jz}.
Indications for the existence of the Reuter fixed point in the presence of minimally coupled Standard Model matter have been found in \cite{Dona:2013qba,Alkofer:2018fxj,Wetterich:2019zdo}, see \cite{Dona:2014pla} for an extension including gravitinos. In \cite{Dona:2013qba} and the follow-up works, an asymptotically safe fixed point has been found when the matter content of the Standard Model is added. This fixed point can be connected continuously to the pure-gravity fixed point, if the numbers of matter fields are treated as external parameters.
On the technical side, the treatment of fermions in the single-metric approximation has been discussed in  \cite{Dona:2012am} and the spin-base invariance formalism has been developed in \cite{Gies:2013noa,Gies:2015cka,Lippoldt:2015cea}. The impact of fermions, scalar and vectors has separately been studied in various truncations in \cite{Meibohm:2015twa,Dona:2015tnf,Eichhorn:2016vvy,Eichhorn:2017sok,Christiansen:2017cxa,Hamada:2017rvn,Eichhorn:2018akn,Eichhorn:2018nda,Burger:2019upn}, some of which go beyond minimal coupling. It should be stressed that extended truncations are required in order to settle whether the Reuter fixed point can be extended continuously to a large numbers of matter fields. Alternatively, gravity-matter systems with large numbers of matter fields might be a new universality class, or not feature a fixed point at all.

Conversely, the impact of quantum gravity on matter is key to generate an asymptotically safe model and bridge the gap from quantum gravity to phenomenology.
The Standard Model can be consistently extended all the way up to the Planck scale \footnote{Whether or not the electroweak vacuum is absolutely stable or metastable sensitively depends \cite{Bezrukov:2014ina} on the value of the top quark mass \cite{Aad:2019ntk,Sirunyan:2019zvx}.}, yet, beyond the Planck scale, the Landau pole problems from Abelian gauge theories and scalar theories are expected to render the Abelian hypercharge and the Higgs-Yukawa sector of the Standard Model inconsistent. Further, the Standard Model has 19 free parameters that lack a fundamental dynamical explanation. These include the ratio of the Higgs mass to the Planck mass, the tiny value of which is argued to pose a particularly severe problem. Yet, the values of the fermion masses, related to Yukawa couplings, as well as the strength of the Higgs self-interaction and the gauge interactions are also unexplained, with the Yukawa sector also containing a significant hierarchy. Moreover, the Standard Model only includes three of the four currently known fundamental interactions, i.e., lacks gravity.\\
The asymptotically safe perspective motivates the idea that all three challenges (UV completion, increase in predictivity, inclusion of gravity) can be tackled simultaneously, see \cite{Eichhorn:2017egq,Eichhorn:2018yfc} for reviews. 
Quantum-gravity effects have been explored on the scalar sector \cite{Narain:2009fy,Narain:2009gb,Eichhorn:2012va,Henz:2013oxa,Labus:2015ska,Percacci:2015wwa,Henz:2016aoh,Eichhorn:2017als,Eichhorn:2017sok,Pawlowski:2018ixd,Wetterich:2019rsn}, where the scalar mass parameter remains relevant unless quantum-gravity effects are very strong, and the quartic scalar coupling and higher-order couplings are irrelevant. Therefore, an asymptotically safe gravity-scalar fixed point results in predictive power in the scalar sector.
The Yukawa sector has been explored in \cite{Zanusso:2009bs,Vacca:2010mj,Oda:2015sma,Eichhorn:2016esv,Eichhorn:2017eht,Hamada:2017rvn,Eichhorn:2017ylw,deBrito:2019epw}, where a predictive fixed point for the Yukawa coupling appears to be possible in a restricted part of the gravitational parameter space. Additionally, fermions can remain light\footnote{These studies are based on the dynamics of four-fermion couplings. The latter actually provide a paradigmatic example of asymptotic safety in 3 dimensions \cite{Braun:2010tt,Gies:2010st,Gehring:2015vja,Classen:2015mar,Knorr:2016sfs}.} despite quantum-gravity fluctuations \cite{Eichhorn:2011pc,Meibohm:2016mkp,Gies:2018jnv} (in contrast to the analogous setting in non Abelian gauge theories, see, e.g., \cite{Braun:2005uj,Braun:2006jd,Braun:2011pp}). Finally, the 
 gauge sector has been studied in \cite{Daum:2009dn,Daum:2010bc,Harst:2011zx,Folkerts:2011jz,Christiansen:2017gtg,Eichhorn:2017lry,Christiansen:2017cxa,Eichhorn:2017muy}, finding indications that asymptotic freedom of non-Abelian gauge theories is compatible with an asymptotically safe fixed point, while Abelian gauge theories could be UV completed \cite{Harst:2011zx,Eichhorn:2017lry}.
For studies of these effects in a unimodular setting, see \cite{deBrito:2019umw}. Demanding UV complete matter sectors restricts the values of curvature-squared couplings in the unimodular setting, analogously to the standard setting \cite{Eichhorn:2017eht}.

\begin{figure}
\includegraphics[width=0.45\linewidth]{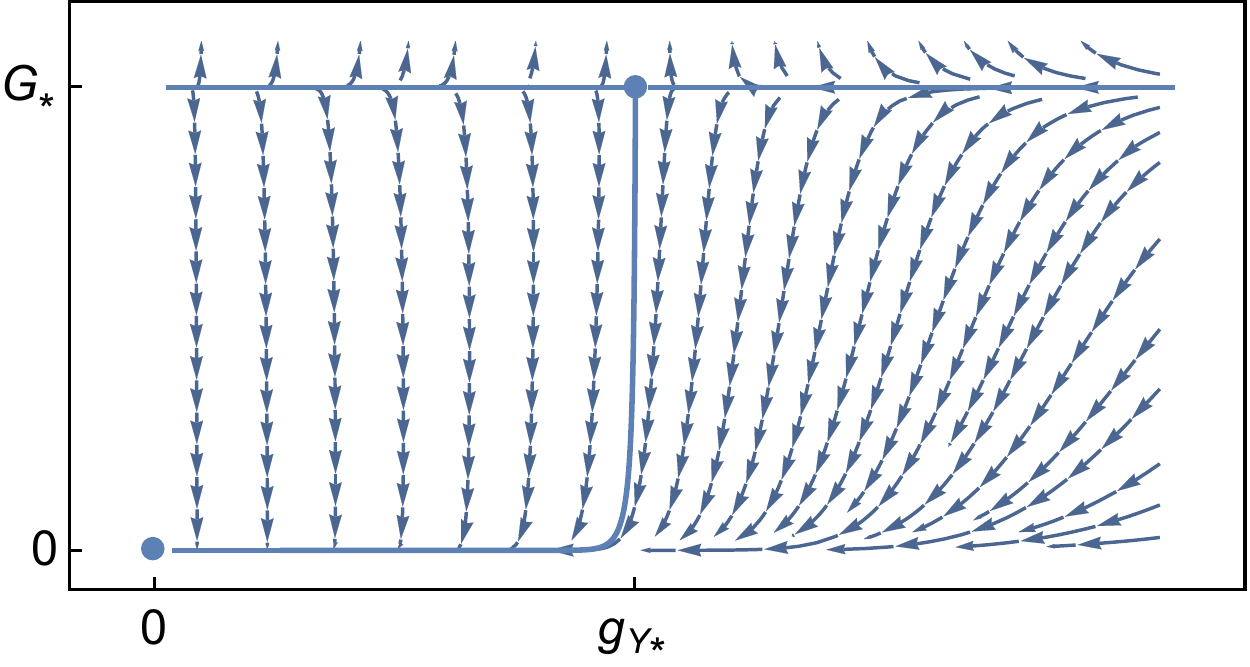}\quad \includegraphics[width=0.45\linewidth]{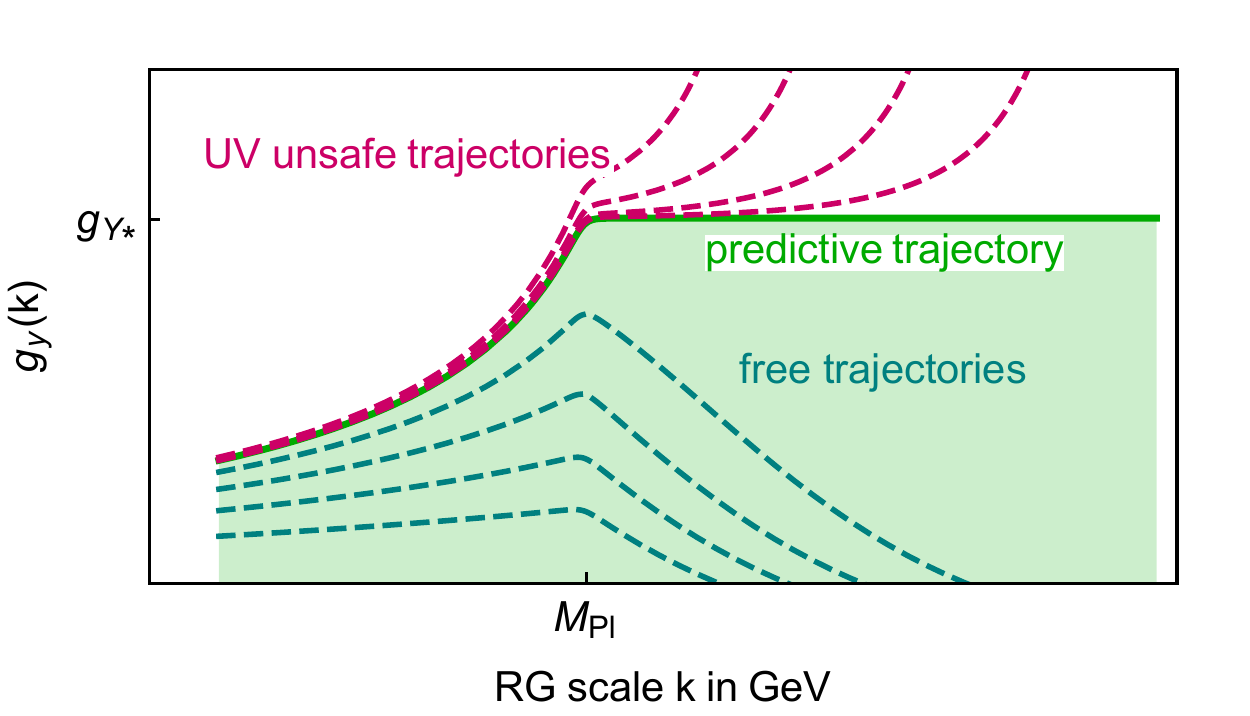}
\caption{\label{gravfig:upperbound} The RG flow in the plane spanned by the Newton coupling and an Abelian gauge coupling exhibits an interacting fixed point with one IR-attractive direction (left panel), cf.~\cite{Harst:2011zx,Eichhorn:2017lry,Eichhorn:2018whv}. The corresponding flows exhibit a transition between the regime with gravity at $k\geq M_{\rm Pl}$ and the classical regime (without gravity), where $G(k)\sim k^2$ and quantum-gravity effects are negligible. One single trajectory for the gauge coupling emanates from the interacting fixed point (right panel), resulting in a prediction of the IR value. }
\end{figure}

In truncations, which to date only include canonically relevant and marginal couplings, as well as a small subset of the canonically irrelevant couplings in some cases, the following is observed:
Quantum gravity triggers an asymptotically safe UV completion of Standard Model-like theories with novel, quantum-gravity generated interactions at high scales, \cite{Eichhorn:2017eht}. 
At the same time, marginally irrelevant couplings of the Standard Model, corresponding to the free parameters of  the Standard Model, could be irrelevant at a joint, interacting matter-gravity fixed point, cf.~Fig.~\ref{gravfig:upperbound}. As such, their values at all scales would be predictable. Once the Higgs vacuum expectation value, which remains a relevant parameter, unless quantum-gravity effects are very strong \cite{Wetterich:2016uxm,Eichhorn:2017als}, is set, the Higgs mass \cite{Shaposhnikov:2009pv}, top quark mass \cite{Eichhorn:2017ylw}, bottom quark mass \cite{Eichhorn:2018whv} and Abelian gauge coupling \cite{Harst:2011zx,Eichhorn:2017lry,Eichhorn:2018whv} become calculable within the toy models defined by the truncated RG flows of subsectors of the Standard Model. This is a first hint for an enhancement of predictive power in the parameters that determine the dynamics of the model.
Intriguingly, demanding that asymptotically safe gravity provides a consistent UV completion for Standard-Model like matter theories might potentially even set geometric parameters, and e.g., fix the spacetime dimensionality to 4 \cite{Eichhorn:2019yzm}.\\
If such findings persist in extended truncations and the universal quantities exhibit apparent convergence, this will allow to either obtain a model with a higher predictive power than the Standard Model which additionally is UV complete and includes gravity, or to rule out this model, as the reduced number of free parameters compared to the Standard Model could lead to a conflict with observational data. The latter point is noteworthy, as it implies that experimental constraints can be imposed on quantum gravity using data at scales much below the Planck scale, such as, e.g., the electroweak scale. 

Constraints that could arise on the number of fields in a dark sector have been discussed in \cite{Eichhorn:2017ylw} and the inclusion of a Higgs portal in asymptotic safety has been explored in \cite{Eichhorn:2017als,Reichert:2019car,Hamada:2020vnf,Eichhorn:2020kca}. Grand unified theories are expected to be constrained \cite{Eichhorn:2017muy,Eichhorn:2019dhg} if indeed they can be consistently coupled to asymptotically safe gravity. For studies of further BSM settings coupled to asymptotically safe quantum gravity, see  \cite{Kwapisz:2019wrl,Grabowski:2018fjj}. In summary, these exploratory first studies suggest that asymptotically safe gravity could have an enhanced predictive power in BSM settings, and therefore restrict the theoretically viable extensions of the Standard Model.

We briefly highlight that the gravitational contributions to beta functions in the matter sector is not universal, due to the nonvanishing canonical dimension of the Newton coupling. 
The universal statement therefore consists in the number of positive critical exponents that determines how many free parameters remain, and how many relations between various couplings can be imposed. While the actual flows are therefore expected to differ in different schemes, the values of actual physical observables in the IR only exhibit a scheme-dependence within approximations, but should converge to the same values as the approximation is improved. 

Finally, we highlight that an interacting RG fixed point in (matter)-gravity systems is also of interest in a setting where a UV cutoff $\Lambda_{\rm fund}$ exists, beyond which some more fundamental description of the degrees of freedom is applicable. In that setting, the RG fixed point can actually serve as an intermediate fixed point: The fundamental theory determines the values of all couplings in the effective description at $\Lambda_{\rm fund}$. From there, the RG flow can approach the interacting fixed point arbitrarily close, if the values of the relevant couplings at $\Lambda_{\rm fund}$ are appropriate. Then, the RG flow ``washes out" many of the microscopic details of the fundamental theory and the fixed point generates universality, in much the same way as it does in the previous sections, in particular for statistical physics. In particular, if the Reuter universality class is characterized by only a small number of free parameters, this high degree of predictivity is effectively imposed on the underlying, more fundamental theory as well.\\
For this scenario to be realistic, the RG trajectory cannot be a true fixed-point trajectory, i.e., it must not end in the Reuter fixed point in the IR, as this does not yield a classical gravity regime. Instead, the RG flow should pass very close to the Reuter fixed point and exhibit a long range of scales over which it is nearly scale-invariant. At a transition scale, the flow should then leave the near-fixed-point regime to flow towards a classical-gravity regime in the IR, as trajectories that emanate out of the Reuter fixed point exhibit \cite{Reuter:2001ag,Reuter:2005kb}.\\
For the first discussion of effective asymptotic safety in the gravitational context see \cite{Percacci:2010af}, for a proposal in the context of string theory see \cite{deAlwis:2019aud}, and for a quantitative measure of predictivity in effective asymptotic safety, see \cite{Held:2020kze}. \\

Let us also mention that the Wetterich equation can be applied to other forms of matter-gravity models, e.g., a study in the context of the spectral action can be found in \cite{Estrada:2012te}.

\subsection{The FRG in other approaches to quantum gravity}
Many quantum-gravity approaches face one or both of the following two questions:
\begin{itemize}
\item Given a proposal for a microscopic dynamics, is the emergent effective low-energy theory compatible with observations?
\item Defining the gravitational path integral through a discretization (typically not in metric variables), can the continuum limit be taken?
\end{itemize}
The FRG is a suitable tool to tackle both questions, and an example of each is introduced below.

\subsubsection{Lorentz-symmetry violating quantum gravity}
Lorentz symmetry is a cornerstone in the Standard Model of particle physics. This goes hand in hand with strong observational constraints on violations of Lorentz symmetry in high-energy physics. Yet, these constraints typically only cover a low-energy (in comparison to the Planck scale) range of scales. This motivates giving up Lorentz symmetry in quantum gravity, which in turn opens the door for perturbative renormalizability \cite{Horava:2009uw}, and in particular asymptotic freedom \cite{DOdorico:2014tyh,Barvinsky:2017kob}. A key question in this setting is whether violations of Lorentz symmetry at and beyond the Planck scale are compatible with the strong observational constraints at lower scales \cite{Yagi:2013qpa,Ramos:2018oku}. The FRG is a powerful tool to tackle this question \cite{Contillo:2013fua}. Its application to this setting, pioneered in \cite{Rechenberger:2012dt} relies on a 3+1 split of the metric, such that a (Euclidean) time-direction as well as spatial hypersurfaces can be distinguished. It allows to follow the RG flow in Horava-Lifshitz gravity \cite{Horava:2009uw} to low energies \cite{Contillo:2013fua,DOdorico:2015pil}
The flow of Lorentz-symmetry violating operators, such as, e.g., couplings of monomials build out of the extrinsic curvature of spatial hypersurfaces, has been explored, showing that such operators are RG relevant \cite{Knorr:2018fdu}. The proliferation of Lorentz-invariance violation from the gravitational to the matter sector with this tool has been explored in \cite{Eichhorn:2019ybe}.

\subsubsection{Discrete quantum-gravity models}\label{sec_gr:discrete}
\begin{figure}
\begin{center}
\includegraphics[width=0.5\linewidth]{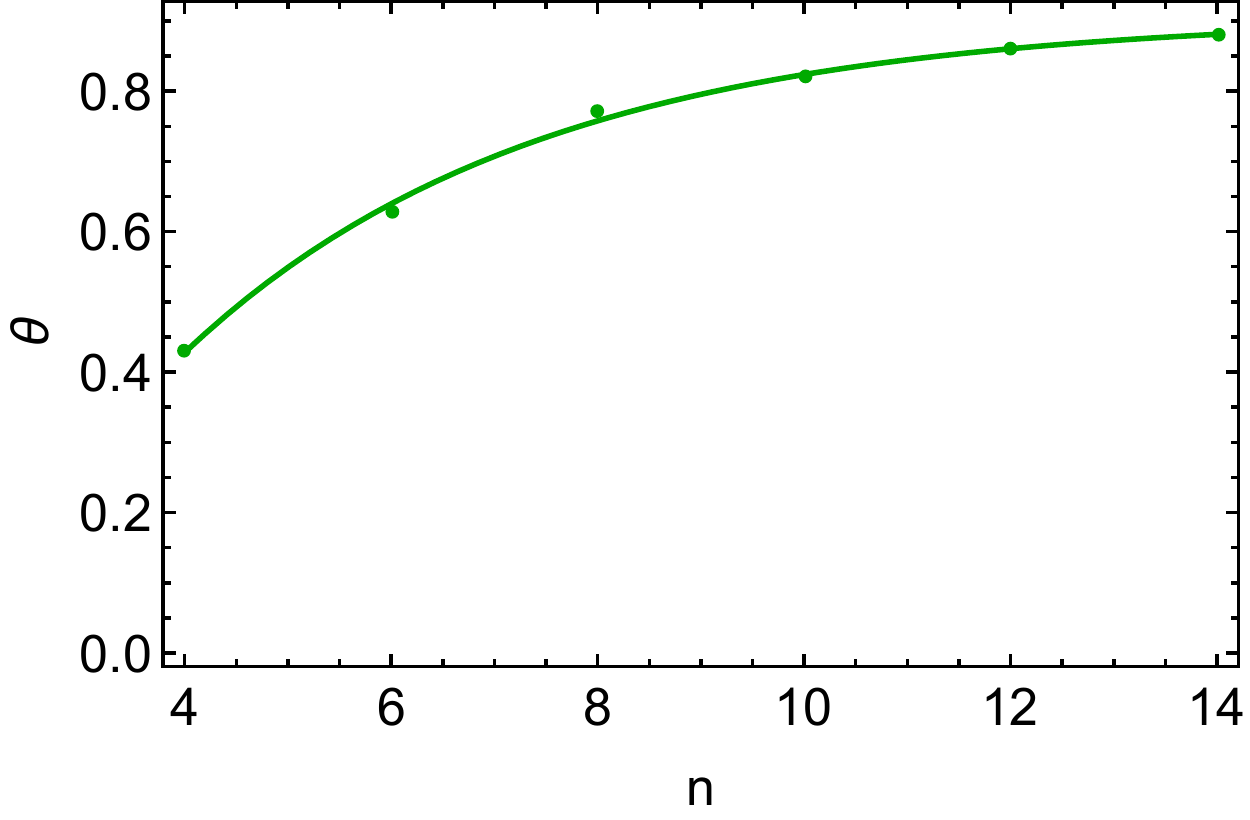}
\end{center}
\caption{\label{figgravmmtheta} Leading critical exponent in the hermitian matrix model for 2d gravity as a function of the order $n$ in an expansion $\Gamma_N = \sum_{i=1}^n g_i\, {\rm Tr}T^{i}$, as in \cite{Eichhorn:2018phj}. The critical exponent is evaluated at fixed anomalous dimension, and can be fit as $\theta=0.907 - 1.540\, {\rm exp}(-0.292\, n)$, extrapolating to $\theta =0.91$, compared to the exact result $\theta=0.8$.}
\end{figure}

In discrete quantum-gravity models, the FRG can be used to search for the existence of a continuum limit, as it is related to a second-order phase transition in the space of couplings, its universal properties in turn being encoded in a fixed point. A noteworthy aspect of these developments is a  reinterpretation of the coarse-graining scale. Formally, an analogous derivation to that of Eq.~\eqref{sec_frg:eqwet} goes through if the regulator is not a function of a momentum scale, but of some other external variable that allows to sort fluctuations in the path integral. This technical observation is the basis for a flow equation, derived in \cite{Eichhorn:2013isa} and following similar developments for fuzzy spaces \cite{Sfondrini:2010zm}, that takes the number  of degrees of freedom as a (non-local) coarse graining scale. In the UV, many degrees of freedom are present, whereas in the IR only few (effective) degrees of freedom are present. This reinterpretation of the notion of scale is particularly suited to fully background-independent quantum gravity models where a local coarse-graining procedure cannot be defined, as it relies on a background metric with respect to which one can define local ``patches" to be integrated over. Instead, these models are typically formulated as spacetime-free (or pre-geometric) matrix or tensor models, where the number of degrees of freedom is encoded in the tensor size $N$. Acccordingly, $N$ can be used as a cutoff scale and a background-independent version of the flow equation \eqref{sec_frg:eqwet} can be derived taking the form
\begin{eqnarray}
&{}&N\partial_N \Gamma_N[T_{a_1...a_d}]= \frac{1}{2}{\rm Tr}\left(\left[\frac{\delta^2\Gamma_N}{\delta T_{a_1...a_d}\delta T_{b_1...b_d}}+R_N(a_1,...,b_d)\right]^{-1}N\partial_N R_N \right).\label{sec_gr:eq:discflow}
\end{eqnarray}
 It encodes how the effective dynamics change, as the outermost ``layers" of a tensor (rows and columns in the case of matrix models) are integrated out, following the intuition of \cite{Brezin:1992yc}. For a version of the Polchinski equation following similar ideas see \cite{Krajewski:2015clk,Krajewski:2016svb}.
 This tool has been benchmarked for the case of 2d gravity, which corresponds to a matrix model. The leading critical exponent in that model is known to be $\theta=0.8$, enabling a check of the tool, cf.~Fig.~\ref{figgravmmtheta}.
 
 The connection to quantum gravity arises as the Feynman diagrams of such rank-$d$ models are dual to triangulations of $d$ dimensional space, see \cite{Ambjorn:1990ge,Sasakura:1990fs,Godfrey:1990dt,Gross:1991hx,Gurau:2010ba,Bonzom:2012hw}, see \cite{Rivasseau:2011hm,Gurau:2011xp,Rivasseau:2012yp,Rivasseau:2013uca,Rivasseau:2016zco,Rivasseau:2016wvy,Gurau:2016cjo} for reviews. To that end, the interactions of the model are tensor invariants, such that the identification of two indices of two tensors can be interpreted as dual to the gluing of two building blocks of $d-1$ dimensional space. For instance, $T_{a_1a_2a_3}T_{a_1b_2b_3}T_{b_1b_2a_3}T_{b_1a_2b_3}$ is dual to four triangles, one for each tensor, which are glued along adjacent edges to form a tetrahedron. Note that the combinatorial nonlocality (i.e., the indices are only contracted pairwise between tensors, there is no index common to all four tensors) of these interactions is crucial to allow for the dual interpretation in terms of building blocks of space.
 A universal continuum limit of these is related to a universal $N\rightarrow \infty$ limit --  which in turn is conjectured  to be discoverable as a fixed point of the background-independent flow equation. Tentative indications for such fixed-point candidates have been discovered in \cite{Eichhorn:2017xhy,Eichhorn:2018ylk,Eichhorn:2019hsa}, for reviews see \cite{Eichhorn:2018phj,Pereira:2019dbn}.  \\
In related models, so-called group-field theories, which are combinatorically nonlocal\footnote{This refers to a nonlocality of the interactions, which share the combinatorial structure with tensor models.} quantum field theories for quantum gravity defined on a group manifold, where an extension of the FRG-equation \eqref{sec_gr:eq:discflow} developed in \cite{Benedetti:2014qsa} has been used to search for infrared fixed points \cite{,Geloun:2015qfa,Benedetti:2015yaa,Geloun:2016qyb,Geloun:2016xep,Lahoche:2018vun,BenGeloun:2018ekd,Lahoche:2018oeo,Lahoche:2018ggd}, as well as to support indications for asymptotic safety \cite{Carrozza:2016tih}. Studies of the modified Ward-identity that arises due to symmetry-breaking by the regulator, cf.~Sec.~\ref{sec_hep} can be found in \cite{Lahoche:2019vzy,Lahoche:2018ggd,Lahoche:2018vun}; a first solution to the Ward-identity within a truncation has been put forward in \cite{Eichhorn:2014xaa}.

%
%\emph{Acknowledgments}
%%We are grateful for fruitful collaborations and/or discussions with...
%
%A.~E.~is supported by the DFG under grant no.~Ei-1037/1, by a research grant (29405) from VILLUM FONDEN, and also partially supported by a visiting fellowship at the Perimeter Institute for Theoretical Physics.
%
%
%%\bibliographystyle{elsarticle-num}
%%\bibliography{refs} A copy of this file is now in directory BIB (Nicolas) 
%
%\end{document}

%% file: SEC_APP/appendix.tex
\def\J{{\bf J}}
\def\varphibf{\boldsymbol{\varphi}}
\def\tvarphibf{\tilde{\boldsymbol{\varphi}}}
\def\phibf{\boldsymbol{\phi}}
\def\calD{{\cal D}}
\def\calO{{\cal O}}
\def\calR{{\cal R}}
\def\calZ{{\cal Z}}
\def\r{{\bf r}}
\def\p{{\bf p}}
\def\q{{\bf q}}
\def\u{{\bf u}}
\def\dt{\partial_t}
\def\half{\frac{1}{2}}
\def\Tr{{\rm Tr}} 
\def\nablabf{\boldsymbol{\nabla}}
\def\trho{\tilde\rho}
\def\tdelta{\tilde\delta}
\def\tlambda{\tilde\lambda}
\newcommand{\mean}[1]{\langle #1 \rangle}

\appendix

\section{The effective action formalism} 
\label{appEA} 

In this Appendix we briefly review the effective action formalism~\cite{Lebellac_book,Zinn_book}, commonly used in field theory and at the basis of the FRG approach. In statistical physics, the effective action is nothing but the Gibbs free energy and therefore contains all information about the thermodynamics of the system. It is also the generating functional of 1PI vertices, which are closely related to the correlation functions. For simplicity, we consider the O($N$) model defined by the action~(\ref{sec_frg:Smicro}).

\subsection{Effective action} 

In the presence of an external source $\J=(J_1,\cdots,J_N)$, the partition function reads
\begin{equation} 
\calZ[\J] =  \int \calD[\varphibf] \, e^{-S[\varphibf] + \int_\r J\cdot \varphibf } . 
\end{equation} 
$W[\J]=\ln\calZ[\J]$ is the generating functional of the connected $n$-point correlation functions (aka Green functions)
\begin{equation}
G^{(n)}_{\{i_j\}}[\{\r_j\};\J] = \frac{\delta^n W[\J]}{\delta J_{i_1}(\r_1)\cdots J_{i_n}(\r_n)}
= \mean{ \varphi_{i_1}(\r_1)\cdots \varphi_{i_n}(\r_n) }_c .
\end{equation}
For the lowest orders, one has
\begin{equation}
\begin{split}
G^{(1)}_i[\r;\J] &= \mean{\varphi_i(\r)} = \phi_i(\r) , \\ 
G^{(2)}_{i_1i_2}[\r_1,\r_2;\J] &= \mean{\varphi_{i_1}(\r_1)\varphi_{i_2}(\r_2)} - \mean{\varphi_{i_1}(\r_1)} \mean{\varphi_{i_2}(\r_2)} ,
\end{split}
\end{equation} 
where $\phi_i(\r)\equiv \phi_i[\r;\J]=\mean{\varphi_i(\r)}$ is the expectation value of the field $\varphi_i$ in the presence of the external source. A nonzero value of $\phibf(\r)$ in the limit $\J({\bf r})\to 0^+$ implies that the O($N$) symmetry of the action is spontaneously broken. This occurs if $d>1$ ($N=1$) or $d>2$ ($N\geq 2$) when $r_0$ is smaller than a critical value $r_{0c}=\bar r_0(T_c-T_0)$. 

The effective action $\Gamma[\phibf]$ is defined as the Legendre transform of $W[\J]$, 
\begin{equation}
\Gamma[\phibf] = - W[\J] + \int_\r  \J\cdot \phibf ,
\label{appEA:eq1}
\end{equation}
where $\J\equiv \J[\r;\phibf]$ must be considered as a functional of $\phibf$ obtained by inverting the relation $\phibf=\phibf[\J]$. The equation of state, relating the order parameter $\phibf$ to the external source, can now be written as 
\begin{equation}
\frac{\delta\Gamma[\phibf]}{\delta \phi_i(\r)} = J_i(\r) . 
\label{appEA:eq2}
\end{equation}
The effective potential  
\begin{equation}
U(\rho)= \frac{1}{V} \Gamma[\phibf]\bigl|_{\phibf\;{\rm unif.}} 
\label{appEA:eq1a}
\end{equation}
($V$ is the volume of the system) is obtained by evaluating the effective action in a uniform field configuration $\phibf(\r)=\phibf$ and, owing to the O($N$) symmetry, is a function of the O($N$) invariant $\rho=\phibf^2/2$. The field configurations $\phibf_{\rm eq}$ in the equilibrium state (corresponding in general to a vanishing external source\footnote{There are also physical situations corresponding to a nonzero source, e.g. a uniform external magnetic field ${\bf J}({\bf r})={\bf J}$ in a spin system.}) are thus obtained from the minimum of $U(\rho)$; they all have the same magnitude $|\phibf_{\rm eq}|=\sqrt{2\rho_0}$ (with $\rho_0$ the position of the minimum of $U$) and are related to one another by an O($N$) transformation.\footnote{When extending the 1PI formalism to out-of-equilibrium systems (Sec.~\ref{sec_NEQ}), the effective action may not be real but is nevertheless stationary in the absence of external sources.} The free energy of the system at zero external source is given by $F=-T\ln\calZ[\J=0]=TVU(\rho_0)$. 

\subsection{1PI vertices}

The effective action is the generating functional of the $n$-point 1PI vertices
\begin{equation}
\Gamma^{(n)}_{\{i_j\}}[\{\r_j\};\phibf] = \frac{\delta^n \Gamma[\phibf]}{\delta \phi_{i_1}(\r_1) \cdots \delta \phi_{i_n}(\r_n)} . 
\end{equation}
The correlation functions $G^{(n)}$ can be related to the $\Gamma^{(n)}$'s. Taking the functional derivative of~(\ref{appEA:eq2}) wrt the external source, one obtains 
\begin{align}
\delta_{i_1i_2} \delta(\r_1-\r_2) &= \frac{\delta}{\delta J_{i_2}(\r_2)} \frac{\delta\Gamma[\phibf]}{\delta \phi_{i_1}(\r_1)} \nonumber \\ 
&= \int_{\r_3} \frac{\delta^2\Gamma[\phibf]}{\delta \phi_{i_1}(\r_1)\delta \phi_{i_3}(\r_3)}
\frac{\delta \phi_{i_3}(\r_3)}{\delta J_{i_2}(\r_2)} \nonumber \\ 
&= \int_{\r_3} \frac{\delta^2\Gamma[\phibf]}{\delta \phi_{i_1}(\r_1)\delta \phi_{i_3}(\r_3)}
\frac{\delta^2 W[\J]}{\delta J_{i_3}(\r_3)\delta J_{i_2}(\r_2)}
\label{appEA:eq3}
\end{align} 
(a summation over repeated indices is implied) or, in matrix form, 
\begin{equation}
\Gamma^{(2)}[\phibf] = G[\J]^{-1} ,
\label{appEA:eq4}
\end{equation}
where we denote the 2-point correlation function (propagator) $G^{(2)}$ merely by $G$. The 2-point vertex $\Gamma^{(2)}[\phibf]=G_0^{-1}+\Sigma[\phibf]$ is often expressed in terms of the self-energy $\Sigma$ with $G_0(\p)=(\p^2+r_0)^{-1}$ the bare 2-point propagator. Equation~(\ref{appEA:eq4}) then becomes Dyson's equation $G=G_0-G_0\Sigma G$ (in matrix form). 

By taking an additional functional derivative in~(\ref{appEA:eq3}), one obtains a relation between the 3-point correlation function and the 3-point vertex, 
\begin{equation}
G^{(3)}_{i_1i_2i_3}(\r_1,\r_2,\r_3) = - \int_{\u_1,\u_2,\u_3} G_{i_1j_1}(\r_1,\u_1) G_{i_2j_2}(\r_2,\u_2) G_{i_3j_3}(\r_3,\u_3) \Gamma^{(3)}_{j_1j_2j_3}(\u_1,\u_2,\u_3) .
\label{appEA:eq6}
\end{equation}
To alleviate the notations, we do not write the functional dependence on $\J$ and $\phibf$. Similarly we can relate the 4-point correlation function $G^{(4)}$ to the 4- and 3-point vertices,
\begin{multline}
G^{(4)}_{i_1i_2i_3i_4}(\r_1,\r_2,\r_3,\r_4) = - \int_{\u_1,\u_2,\u_3,\u_4} G_{i_1j_1}(\r_1,\u_1) G_{i_2j_2}(\r_2,\u_2) G_{i_3j_3}(\r_3,\u_3) G_{i_4j_4}(\r_4,\u_4) \Gamma^{(4)}_{j_1j_2j_3j_4}(\u_1,\u_2,\u_3,\u_4) \\ 
 - \int_{\u_1,\u_2,\u_3} \bigl[ \Gamma^{(3)}_{j_1j_2j_3}(\u_1,\u_2,\u_3) G^{(3)}_{i_1j_1i_4}(\r_1,\u_1,\r_4)  G_{i_2j_2}(\r_2,\u_2) G_{i_3j_3}(\r_3,\u_3)
+ {\rm permutations} \bigr] .
\label{appEA:eq7}
\end{multline}
More generally the knowledge of the 1PI vertices is sufficient to reconstruct all connected correlation functions. In diagrammatic representation, the latter can be obtained as the sum of tree diagrams whose vertices are the $\Gamma^{(n)}$'s, as illustrated by Eqs.~(\ref{appEA:eq6}) and (\ref{appEA:eq7}). This provides us with a simple interpretation of the $\Gamma^{(n)}$'s as the effective interaction vertices; they are 1PI to the extent that they are represented by diagrams that cannot be split into two disconnected pieces by cutting only one line.

\subsection{Loop expansion} 

Let us write the partition function and the effective action as
\begin{equation} 
\calZ[\J] = \int \calD[\varphibf] \, e^{-\frac{1}{l}(S[\varphibf]-\int_\r \J\cdot \varphibf)}, \qquad 
\Gamma[\phibf] = - l \ln \calZ[\J] + \int_\r \J \cdot\phibf ,
\end{equation}
where $\phibf({\bf r})=l \delta\ln \calZ[\J]/\delta \J(\r)$. The real parameter $l$ (which will eventually be set to unity), as $\hbar$ in the path integral of a quantum-mechanical system, can be used to organize the perturbation expansion as a ``loop expansion''. 

In the limit $l\to 0$, the saddle-point approximation becomes exact,
\begin{equation}
\lim_{l\to 0} l \ln\calZ[\J] = - S[\varphibf_{\rm cl}] + \int_\r \J\cdot\varphibf_{\rm cl} ,
\end{equation}
where the ``classical' field $\varphibf_{\rm cl}$ is defined by $\delta S[\varphibf]/\delta \varphibf(\r)|_{\varphibf_{\rm cl}} - \J(\r) = 0$. This gives  
\begin{equation}
\lim_{l\to 0} \Gamma[\phibf] \equiv \Gamma_{\rm cl}[\phibf] = S[\phibf] , 
\label{appEA:eq5}
\end{equation} 
where $\Gamma_{\rm cl}$ is referred to as the classical (or mean-field) effective action. From~(\ref{appEA:eq5}) we deduce the 1PI vertices in a uniform field $\phibf(\r)=\phibf$ (for the O($N$) model), 
\begin{equation}
\begin{split}
\Gamma^{(1)}_{{\rm cl},i}(\p,\phibf) &= \sqrt{V} \delta_{\p,0} \phi_i \Bigl( r_0  + \frac{u_0}{6} \phibf^2  \Bigr) , \\ 
\Gamma^{(2)}_{{\rm cl},i_1i_2}(\p_1,\p_2,\phibf) &=  \delta_{\p_1+\p_2,0} \Bigl[ \delta_{i_1,i_2}  \Bigl(\p_1^2 + r_0 + \frac{u_0}{6}  \phibf^2 \Bigr) + \frac{u_0}{3} \phi_{i_1} \phi_{i_2} \Bigr] , \\ 
\Gamma^{(3)}_{{\rm cl},i_1i_2i_3}(\p_1,\p_2,\p_3,\phibf) &= \delta_{\p_1+\p_2+\p_3,0} \frac{u_0}{3\sqrt{V}} ( \delta_{i_1i_2} \phi_{i_3} +  \delta_{i_1i_3} \phi_{i_2} + \delta_{i_2i_3} \phi_{i_1} ) , \\ 
\Gamma^{(4)}_{{\rm cl},i_1i_2i_3i_4}(\p_1,\p_2,\p_3,\p_4,\phibf) &= \delta_{\p_1+\p_2+\p_3+\p_4,0} \frac{u_0}{3V}  ( \delta_{i_1i_2} \delta_{i_3i_4} +  \delta_{i_1i_3} \delta_{i_2i_4} + \delta_{i_1i_4} \delta_{i_2i_3} )  ,
\end{split}
\end{equation}
where $V$ is the volume of the system. The classical effective action reproduces the mean-field (Landau) theory. From the equation of state~(\ref{appEA:eq2}) with $\J=0$, we find 
\begin{equation}
\rho_0 = \frac{\phibf_{\rm eq}^2}{2} = \left\{ \begin{array}{lll} 0 & \mbox{if} & r_0\geq 0, \\ 
-3 \frac{r_0}{u_0} & \mbox{if} & r_0\leq 0 , \end{array}\right.
\label{appEA:eq5a}
\end{equation}
assuming a uniform field $\phibf_{\rm eq}$. In the broken-symmetry phase ($r_0<0$) the 2-point vertex 
\begin{equation} 
\Gamma^{(2)}_{{\rm cl},ij}(\p,\phibf_{\rm eq})= \frac{\phi_{{\rm eq},i}\phi_{{\rm eq},j}}{2\rho_0} \Gamma^{(2)}_{\rm cl,L}(\p,\phibf_{\rm eq}) + \biggl(\delta_{ij}-\frac{\phi_{{\rm eq},i}\phi_{{\rm eq},j}}{2\rho_0} \biggr) \Gamma^{(2)}_{\rm cl,T}(\p,\phibf_{\rm eq}) 
\end{equation}
is defined by its longitudinal and transverse parts, 
\begin{equation} 
\Gamma^{(2)}_{\rm cl,L}(\p,\phibf_{\rm eq}) = \p^2 + 2 |r_0|, \qquad 
\Gamma^{(2)}_{\rm cl,T}(\p,\phibf_{\rm eq}) = \p^2 ,
\end{equation}
while $\Gamma^{(2)}_{{\rm cl},ij}(\p,\phibf=0)=\delta_{ij}(\p^2+r_0)$ in the symmetric phase.  

It is possible to compute the correction to $\Gamma_{\rm cl}$ in a systematic expansion in $l$. To leading order, 
\begin{equation}
\Gamma[\phibf] = S[\phibf] + \frac{l}{2} \Tr\, \ln \Bigl(\Gamma^{(2)}_{\rm cl}[\phibf] \Bigr) + \calO(l^2) . 
\label{appEA:eq8}
\end{equation}
Taking functional derivatives of this equation, we can obtain the one-loop corrections to the 1PI vertices  $\Gamma^{(n)}_{\rm cl}$ (these corrections include all one-loop diagrams that are 1PI). We shall see in \ref{appDE:subsec_floweq} that the exact FRG flow equation~(\ref{sec_frg:eqwet}) is closely related to~(\ref{appEA:eq8}).

\section[The FRG at work: the case of the derivative expansion]{The FRG at work: the case of the derivative expansion\footnote{To make the discussion self-contained, some basic equations appearing in the main text are reproduced in the Appendix.}} 
\label{appDE}

The aim of this Appendix is to show how the FRG works in practice, considering the derivative expansion as an example and emphasizing technical aspects that were omitted or only briefly mentioned in Sec.~\ref{sec_frg:subsec_de}. We consider the O($N$) model defined by the action~(\ref{sec_frg:Smicro}).

\subsection{The exact flow equation} 
\label{appDE:subsec_floweq} 
 
In the presence of the infrared regulator term $\Delta S_k$ the scale-dependent partition function is defined by 
\begin{equation}
{\cal Z}_k[\J] = \int \calD[\varphibf] \, e^{-S[\varphibf] - \Delta S_k[\varphibf] + \int_\r J\cdot \varphibf } .
\label{appDE:eq1a}
\end{equation} 
One easily finds that the generating functional of connected correlation functions, $W_k[\J]=\ln \calZ_k[\J]$, satisfies the RG equation 
\begin{equation}
\dt W_k[\J] = - \half \int_{\r,\r'} \dt R_k(\r-\r') \left( 
\frac{\delta^2 W_k[\J]}{\delta J_i(\r) \delta J_i(\r')} 
+ \frac{\delta W_k[\J]}{\delta J_i(\r)} \frac{\delta W_k[\J]}{\delta J_i(\r')} \right) , 
\label{appDE:eq1}
\end{equation} 
where $t=\ln(k/\Lambda)$ is the (negative) RG time introduced in Sec.~\ref{sec_frg:subsec_exacteq} and a sum over repeated indices is implied. The derivative $\dt$ in~(\ref{appDE:eq1}) is taken at fixed external source $\J$. To obtain the flow equation of the scale-dependent effective action 
\begin{equation}
\Gamma_k[\phibf] = - W_k[\J] + \int_\r \J\cdot\phibf - \Delta S_k[\phibf]
\label{appDE:eq2}
\end{equation}
at fixed $\phibf$, we must consider $\J\equiv \J_k[\phibf]$ as a $k$-dependent functional of $\phibf$ defined by $\phi_{i,k}[\r;\J]=\delta W_k[\J]/\delta J_i(\r)$. Using 
\begin{align}
\dt W_k[\J] \bigl|_{\phibf} &= \dt W_k[\J] \bigl|_\J + \int_\r \frac{\delta W_k[\J]}{\delta J_i(\r)} \dt J_i(\r) |_{\phibf} \nonumber \\ &= \dt W_k[\J] \bigl|_\J + \int_\r \phi_i(\r)\dt J_i(\r) |_{\phibf} , 
\end{align}
we obtain Wetterich's equation~\cite{Wetterich93,Ellwanger94,Morris94,Bonini93}
\begin{align}
\dt \Gamma_k[\phibf] &= \half \int_{\r,\r'} \dt R_k(\r-\r') \frac{\delta^2 W_k[\J]}{\delta J_i(\r) \delta J_i(\r')} \nonumber \\ &= \half \Tr \bigl[ \dt R_k ( \Gamma^{(2)}[\phibf] + R_k)^{-1} \bigr] , 
\label{appDE:eq3}
\end{align}
where $\Tr$ denotes a trace wrt space and the O($N$) index of the field. The last result in~(\ref{appDE:eq3}) is obtained by noting that the propagator $G_{k,ij}[\r,\r';\J]=\delta^2 W_k[\J]/\delta J_i(\r) \delta J_j(\r')$ is the inverse (in a matrix sense) of $\Gamma_k^{(2)}[\phibf]+R_k$.\footnote{The relation $G^{-1}_k=\Gamma^{(2)}_k+R_k$ (rather than $G^{-1}_k=\Gamma^{(2)}_k$) is due to the fact that the true Legendre transform of $W_k[\J]$ is $\Gamma_k[\phibf]+\Delta S_k[\phibf]$ and not $\Gamma_k[\phibf]$; see Eq.~(\ref{appDE:eq2}).} Substitution of $\Gamma^{(2)}_k[\phibf]$ by $S^{(2)}[\phibf]$ in~(\ref{appDE:eq3}) gives a flow equation that can be easily integrated out and yields the one-loop result~(\ref{appEA:eq8}).

\subsection[The local potential approximation]{The local potential approximation (LPA)} 

In the LPA the effective action~(\ref{sec_frg:gammaklpa}) is entirely determined by the effective potential $U_k(\rho)$. To derive the flow equation $\dt U_k(\rho)$, it is sufficient to consider a uniform field configuration $\phibf(\r)=\phibf$. In that case the 2-point vertex is diagonal in Fourier space and, owing to the O($N$) symmetry, can be written as                                                                                                                         
\begin{equation}
\Gamma^{(2)}_{k,ij}(\p,\phibf) = \frac{\phi_i\phi_j}{2\rho} \Gamma^{(2)}_{k,\rm L}(\p,\rho) + \left( \delta_{i,j} - \frac{\phi_i\phi_j}{2\rho} \right) \Gamma^{(2)}_{k,\rm T}(\p,\rho) ,
\label{appDE:eq3a}
\end{equation}
where the longitudinal and transverse parts, $\Gamma^{(2)}_{k,\rm L}(\p,\rho)$ and $\Gamma^{(2)}_{k,\rm T}(\p,\rho)$, are functions of the $O(N)$ invariant $\rho$. Within the LPA, one has 
\begin{equation} 
\begin{split}
\Gamma^{(2)}_{k,\rm L}({\bf p},\rho) &= {\bf p}^2+U'_k(\rho)+2\rho U_k''(\rho)+R_k({\bf p}) , \\  
\Gamma^{(2)}_{k,\rm T}({\bf p},\rho) &= {\bf p}^2+U'_k(\rho)+R_k({\bf p}) .
\end{split}
\label{appDE:eq3b}
\end{equation}
An expression similar to~(\ref{appDE:eq3a}) holds for the propagator $G_{k,ij}(\p,\phibf)$, with 
$G_{k,\alpha}(\p,\rho) = 1/\Gamma^{(2)}_{k,\alpha}(\p,\rho)$ ($\alpha={\rm L,T}$).

From the definition~(\ref{appEA:eq1a}) of the effective potential and Wetterich's equation~(\ref{sec_frg:eqwet}), we then find the exact flow equation of the effective potential, 
\begin{align}
\dt U_k(\rho) &= \frac{1}{2} \int_{\p} \dt R_k(\p) G_{k,ii}(\p,\phibf) \nonumber \\ 
&= \frac{1}{2} \int_\p \dt R_k(\p) \left[ G_{k,\rm L}(\p,\rho) + (N-1) G_{k,\rm T}(\p,\rho) \right] .
\end{align}
Writing the regulator function as $R_k(\p)=\p^2 r(\p^2/k^2)$ and employing the LPA~(\ref{appDE:eq3b}) we finally obtain 
\begin{equation}
\dt U_k = -2 v_d k^d \int_0^\infty dy\, y^{d/2+1} r' \left[ \frac{1}{y(r+1)+k^{-2}(U_k'+2\rho U''_k)} + \frac{N-1}{y(r+1)+k^{-2}U_k'} \right] 
\label{appDE:eq4}
\end{equation}
(with $r\equiv r(y)$ and $r'\equiv r'(y)$), where we use the dimensionless variable $y=\p^2/k^2$ and $v_d^{-1}=2^{d+1}\pi^{d/2}\Gamma(d/2)$.\footnote{With the theta regulator $R_k(\q)=(k^2-\q^2)\Theta(k^2-\q^2)$, i.e., $r(y)=((1-y)/y)\Theta(1-y)$, the integral in the rhs of~(\ref{appDE:eq4}) can be carried out analytically~\cite{Litim01}.\label{appDE:footnote1}} 

\begin{figure}
\centerline{\includegraphics[height=4.6cm]{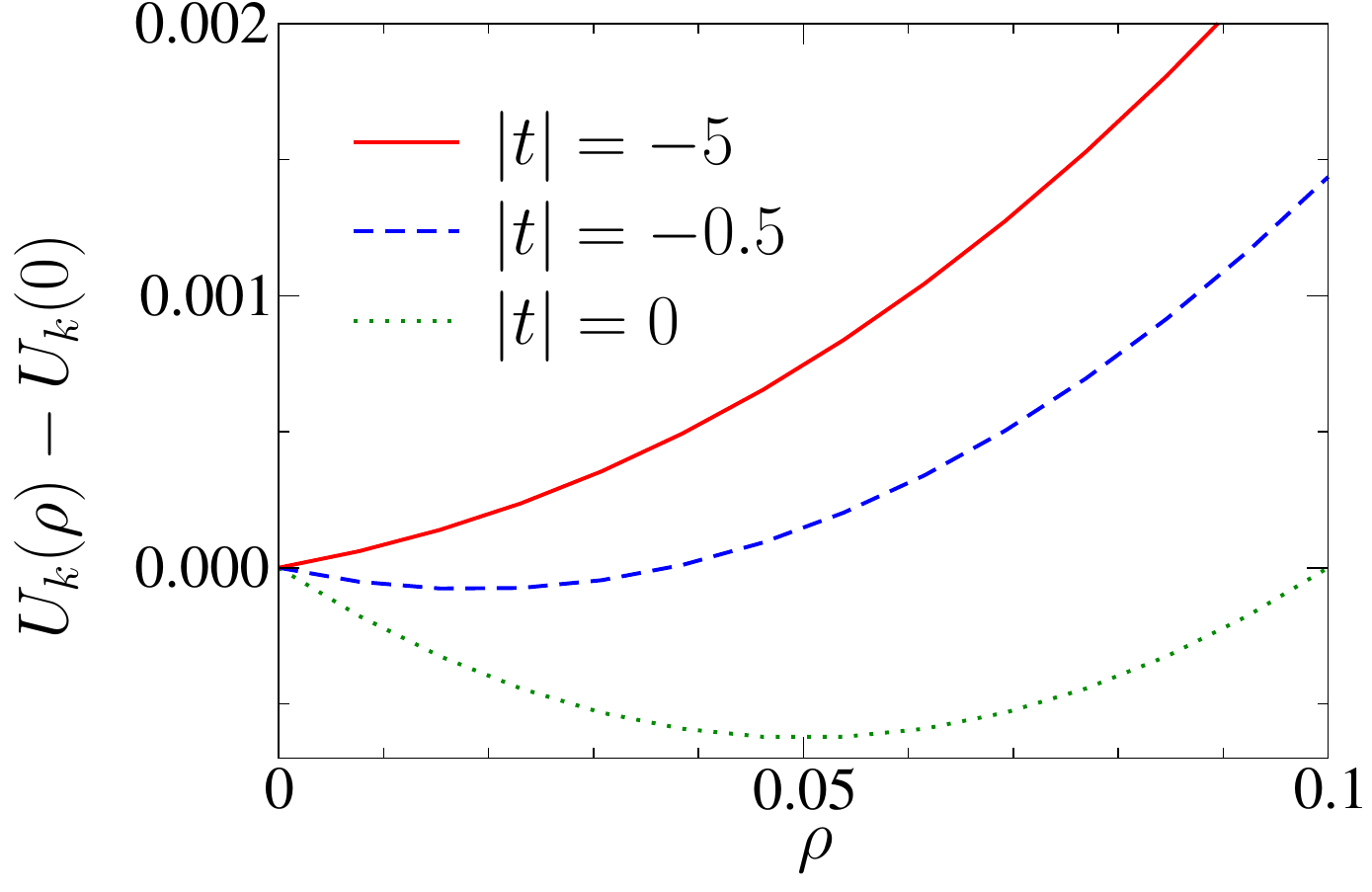}
	\hspace{0.2cm}
\includegraphics[height=4.5cm]{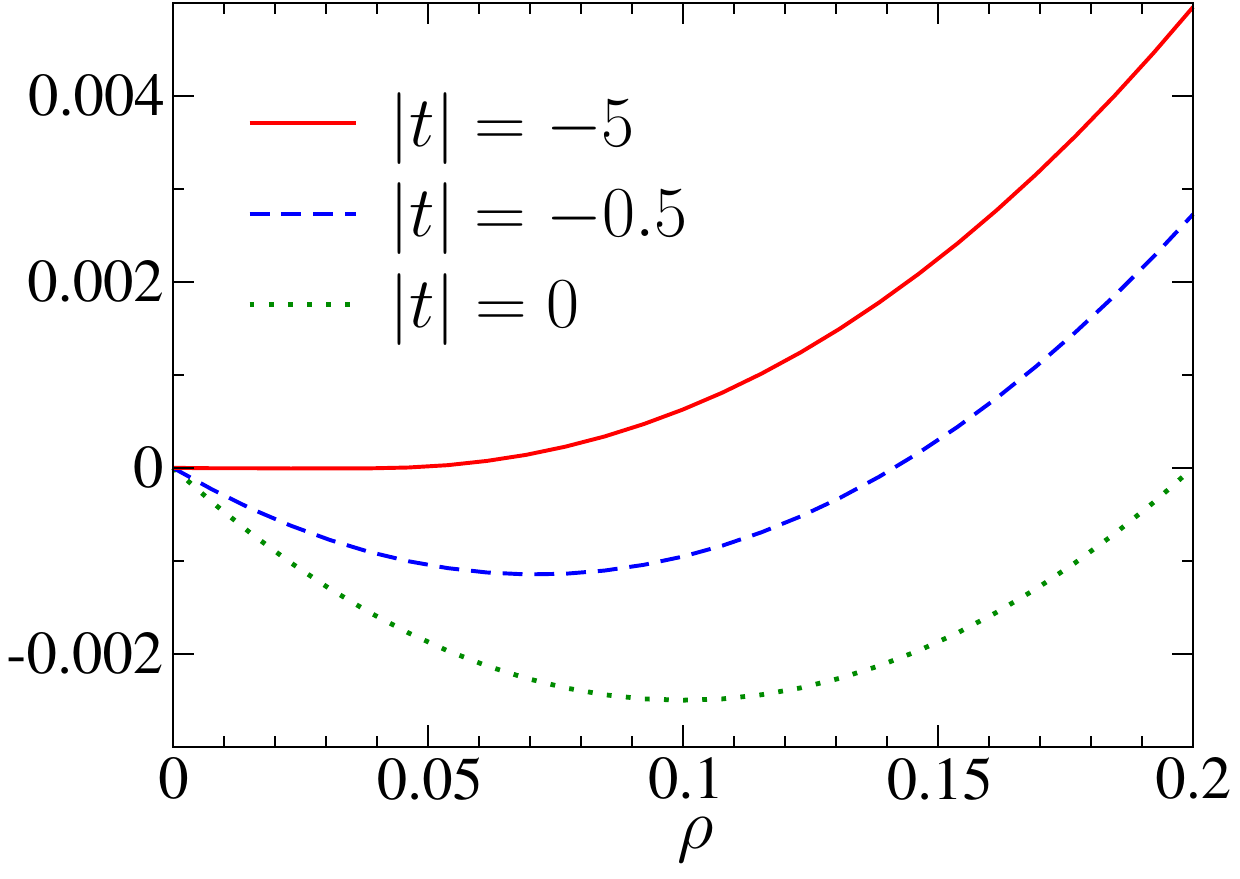}}
\caption{Effective potential $U_k(\rho)-U_k(0)$ 
	obtained from the LPA flow equation~(\ref{appDE:eq4}) 
	and the initial condition $U_\Lambda(\rho)=r_0\rho+(u_0/6)\rho^2$ for various values of the RG time $t=\ln(k/\Lambda)$ ($N=3$ and $d=3$). 
	Left panel: disordered phase, $\rho_{0,\Lambda}=0.05$ and $\rho_{0}=0$. 
	Right panel: ordered phase, $\rho_{0,\Lambda}=0.1$ and $\rho_{0}\simeq 0.05$. 
	Note that the potential becomes flat for $\rho\in [0,\rho_0]$ as $k\to 0$, showing the approach to the convex effective potential. Since we are using the variable $\rho$ (rather than $\phibf$), the usual picture of the Mexican-hat shape potential when $\rho_{0,k}>0$ is partially lost (in particular because $U_k'(0)\neq 0$).}
\label{appDE:fig_U} 
\end{figure}

Equation~(\ref{appDE:eq4}) can be integrated with the initial condition $U_\Lambda(\rho)=r_0\rho+(u_0/6)\rho^2$ at scale $k=\Lambda$ corresponding to the classical (or mean-field) effective potential of the O($N$) model~(\ref{sec_frg:Smicro}). $U_\Lambda(\rho)$ exhibits a minimum at $\rho_{0,\Lambda}$ defined by~(\ref{appEA:eq5a}). Since fluctuations tend to disorder the system, $\rho_{0,k}$ can only decrease during the flow. The equilibrium state depends on the value of $\rho_0=\lim_{k\to 0}\rho_{0,k}$. When $r_0$ is larger than a critical value $r_{0c}<0$, related to the critical temperature {\it via} $r_{0c}=\bar r_0(T_c-T_0)$ (see Sec.~\ref{sec_frg}), $\rho_0=0$ and the system is in the symmetric (high-temperature) phase. When $r_0<r_{0c}$, the system is in the ordered (low-temperature) phase and the O($N$) symmetry is spontaneously broken; the uniform field in the equilibrium state has a fixed length $|\phibf_{\rm eq}|=\sqrt{2\rho_0}$ but its direction is arbitrary. Note that when $r_0=r_{0c}$, $\rho_{0,k}$ remains nonzero for all $k>0$ and vanishes only when $k=0$. Typical flows of the effective potential are shown in Fig.~\ref{appDE:fig_U}. 

\subsubsection{Scaling form of the LPA flow equation} 

For $r_0=r_{0c}$ the system is scale invariant (the correlation length diverges) and the only length scale in the problem, $k^{-1}$, comes from the infrared regulator. Thus we expect that the effective potential $U_k(\rho)$ will exhibit scale invariance, i.e., reach a fixed point of the flow equation $\dt U_k$, if we measure all quantities in units of $k$. In practice, this is achieved by introducing dimensionless quantities,
\begin{equation}
\tilde \r= k \r, \quad \tilde\phibf(\r) =k^{-(d-2)/2} \phibf(\r), \quad \tilde U_k\bigl(\trho(\tilde \r)\bigr) = k^{-d} U_k\bigl(\rho(\r)\bigr) , 
\label{appDE:eq5}
\end{equation}
in terms of which the LPA effective action reads 
\begin{equation}
\Gamma_k[\phibf] = \int_{\tilde \r} \left\{ \half (\nablabf_{\tilde\r}\tilde\phibf)^2 + \tilde U_k(\tilde\rho) \right\} .
\end{equation} 
The flow equation 
\begin{equation}
\dt \tilde U_k = -d\tilde U_k + (d-2)\trho \tilde U_k' +2 v_d[ l_0^d(\tilde U_k'+2\trho \tilde U''_k)+(N-1)l_0^d(\tilde U_k') ]
\label{appDE:eq6}
\end{equation}
of the dimensionless effective potential $\tilde U_k(\trho)$ follows from~(\ref{appDE:eq4}) and (\ref{appDE:eq5}) once the derivative wrt $t$ at fixed $\rho$ in~(\ref{appDE:eq4}) has been changed for a derivative at fixed $\trho$: $\dt|_{\trho}=\dt|_{\rho}+\dt\rho|_{\trho}\partial_\rho$. To eliminate a field-independent, relevant, constant it is often more convenient to consider the RG equation  
\begin{equation}
\dt \tilde U'_k = -2\tilde U'_k + (d-2)\trho \tilde U_k'' - 2 v_d[ (3\tilde U_k''+2\trho \tilde U'''_k) l_1^d(\tilde U_k'+2\trho \tilde U''_k)+(N-1) \tilde U''_k l_1^d(\tilde U_k') ] 
\label{appDE:eq6a}
\end{equation}
satisfied by the derivative of the effective potential. In Eqs.~(\ref{appDE:eq6},\ref{appDE:eq6a}) we have introduced the ``threshold'' functions 
\begin{equation}
l_n^d(w) = - (n+\delta_{n,0}) \int_0^\infty dy\, y^{d/2+1} \frac{r'}{[y(1+r)+w]^{n+1}} .
\end{equation}

\subsubsection{Fixed-point solutions and critical exponents}

The fixed-point solutions $\tilde U^*{}'(\trho)$ of the flow equation are obtained from  
\begin{equation}
0 = \dt\tilde U^*{}'  = -2\tilde U^*{}' + (d-2)\trho \tilde U^*{}'' - 2 v_d[ (3\tilde U^*{}''+2\trho \tilde U^*{}''') l_1^d(\tilde U^*{}'+2\trho \tilde U^*{}'')+(N-1) \tilde U^*{}'' l_1^d(\tilde U^*{}') ]  .
\label{appDE:eq7}
\end{equation}
This equation admits the trivial solution $\tilde U^*{}'(\trho)=0$ corresponding to the (noninteracting) Gaussian fixed point with all $n$-point vertices $\Gamma^{(n)}$ ($n>3$) vanishing while $\Gamma^{(2)}(\p)=\p^2$. For simplicity we now consider the case $N=1$ and the theta regulator for which $l_1^d(w)=(2/d)(1+w)^{-2}$. To obtain the critical exponents associated with the Gaussian fixed point, we must linearize the flow equation about $\tilde U^*{}'(\trho)=0$. Taking $\phi$ as the variable,
\begin{equation}
\dt \tilde U'_k = - \left( \frac{d}{2} + 1 \right) \tilde U'_k + \left( \frac{d}{2} - 1 \right) \phi \tilde U''_k - 4 \frac{v_d}{d} \tilde U''_k . 
\label{appDE:eq7a}
\end{equation}
Setting $\tilde U_k'(\phi)=f(\sqrt{\alpha}\phi)$ with $\alpha=d/4v_d$, $f(x)=h(\beta x)e^{-\lambda t}$ and $\beta=\sqrt{d-2}/2$ (assuming $d>2$), we finally obtain\footnote{The same linearized equation is obtained from the Wilson-Polchinski equation~\cite{Polchinski84} in the LPA. This follows from an exact map between the Wilson-Polchinski and FRG equations in the LPA when the theta regulator is used~\cite{Morris05}.}  
\begin{equation}
h''(y) - 2y h'(y) + \frac{2}{d-2}(2+d-2\lambda) h(y) = 0 . 
\end{equation}
This equation is known to have polynomial solutions,\footnote{Non-polynomial solutions imply a continuum of eigenvalues and can be discarded on physical grounds~\cite{Morris94a,Morris95}.} given by the Hermite polynomials $h(y)=\hat H_{2n-1}(y)=2^{n-1/2}H_{2n-1}(y)$ of degree $2n-1$, only for a set of discrete values of $\lambda$ satisfying
\begin{equation}
2n-1 = \frac{d+2-2\lambda_n}{d-2} , \quad \mbox{i.e.} \quad \lambda_n = d-n(d-2) , 
\label{appDE:eq8}
\end{equation}
where $n=1,2,3$... If one considers symmetric perturbations, even degree Hermite polynomials are not allowed since in that case the functions $\tilde U_k'(\phi)$ and $f_k(x)$ are odd. Note that the $\lambda_n$'s coincide with the scaling dimension $[v_{2n}]$ of the vertex $v_{2n}\int_\r \varphi^{2n}$ at the Gaussian fixed point. 

When $d>4$ all eigenvalues $\lambda_n$ are negative except $\lambda_1$ which determines the correlation-length critical exponent $\nu=1/\lambda_1=1/2$. The associated eigenvector is given by $\hat H_1(y)\propto y$, which corresponds to a $\phi^2$ term in $U_k(\phi)$. The less negative eigenvalue $\lambda_2$ determines the correction-to-scaling exponent $\omega=-\lambda_2=d-4$, i.e. the speed at which $\tilde U'_k$ approaches the fixed-point solution $\tilde U^*{}'$ when the system is critical. When $d<4$,\footnote{To determine whether the field associated with the eigenvalue $\lambda_2=4-d$ is relevant or irrelevant in four dimensions, one must go beyond the linear approximation~(\ref{appDE:eq7a}). We do not discuss the case $d=4$ here.} $\lambda_2$ becomes positive and we expect the phase transition to be described by a nontrivial fixed point with a single relevant direction (i.e. the Wilson-Fisher fixed point for the O($N$) model~\cite{Wilson72}). The Gaussian fixed point then becomes a tricritical fixed point. 

Nontrivial fixed points cannot be found analytically and one must rely on a numerical solution. Since Eq.~(\ref{appDE:eq7}) is a second-order differential equation, a solution is {\it a priori} parameterized by two arbitrary constants. However, if we require $\tilde U^*{}'(\trho)$ to be regular at the origin, setting $\trho=0$ in~(\ref{appDE:eq7}) yields $\tilde U^*{}'(0)+ v_d (N+2) \tilde U^*{}''(0) l_1^d(\tilde U^*{}'(0))=0$. Moreover, if $\tilde U^*{}'(\trho)$ does not vanish for $\trho\to\infty$, it must behave as $\tilde U^*{}'(\trho)\sim \trho^{2/(d-2)}$.\footnote{The large-$\trho$ behavior of $\tilde U^*{}'(\trho)$ and $g(\trho)$ is obtained by noting that the threshold functions $l_n^d$ give a subleading contribution when $\trho\to\infty$.\label{appDE:footnote2}} Thus we now have a second-order differential equation with two boundary conditions. Most solutions are found to be singular at some $\trho_c$ and must be discarded, so that we end up with a discrete set of acceptable solutions. In practice, the fixed-point solution $\tilde U^*{}'$ can be determined by the shooting method, i.e., by fine tuning $\tilde U^*{}'(0)$ and $\tilde U^*{}''(0)$ until a regular solution satisfying the above-mentioned boundary conditions for $\trho\to 0$ and $\trho\to\infty$ is found~\cite{Morris94a,Morris94b,Morris95}. For $d\geq 4$ only the Gaussian fixed point is found. For $3\leq d<4$ a nontrivial fixed point (the Wilson-Fisher fixed point) appears. A new nontrivial fixed point emanates from the Gaussian fixed point each time that one of the eigenvalues $\lambda_n$ [Eq.~(\ref{appDE:eq8})] becomes negative, which occurs at the dimensional thresholds $d_n=2n/(n-1)$ ($n\geq 2$). 

Once a fixed point is identified, one can determine the critical exponents by linearizing the flow about $\tilde U^*{}'$. Setting 
\begin{equation}
\tilde U_k'(\trho) = \tilde U^*{}'(\trho) + e^{-\lambda t} g(\trho) , 
\label{appDE:eq9}
\end{equation}
we find 
\begin{align}
& (\lambda-2) g + (d-2) \trho g' - 2 v_d \{ (3g'+2\trho g'') l_1^d(\tilde U^*{}'+2\trho \tilde U^*{}'') \nonumber \\ &
- (g+2\trho g') (3\tilde U^*{}''+2\trho \tilde U^*{}''') l_2^d(\tilde U^*{}'+2\trho \tilde U^*{}'') 
+ (N-1) [ g' l_1^d(\tilde U^*{}') 
- g \tilde U^*{}'' l_2^d(\tilde U^*{}') ] \} 
 = 0 .
\end{align}
For simplicity we consider here perturbations respecting the O($N$) symmetry (non-symmetric perturbations can be considered in a similar way). Again the solutions of this second-order differential equation are labeled by two parameters.  However, one can choose $g(0)=1$ (arbitrary normalization) while $g(0)$ and $g'(0)$ are not independent if $g(\trho)$ is regular at the origin. The solution is then unique for a given $\lambda$. Regular solutions are obtained only for a countable set of $\lambda$'s and behave as $g(\trho)\sim \trho^{(2-\lambda)/(d-2)}$ for $\trho\to\infty$;$^{\ref{appDE:footnote2}}$ they can be determined by the shooting method. For $N=1$ and $d=3$, the LPA with the exponential regulator function $R_k(\q)=\q^2/(e^{\q^2/k^2}-1)$ gives $\nu=0.6589$ and $\omega=0.6440$~\cite{DePolsi20}, to be compared with the values shown in Table~\ref{sec_frg:table_critexp}. Since the anomalous dimension $\eta$ obviously vanishes in the LPA, all other critical exponents ($\alpha$, $\beta$, $\gamma$ and $\delta$) can be deduced from the value of $\nu$ and the usual scaling laws~\cite{Ma_book}. For a more detailed discussion of the shooting method, we refer to Refs.~\cite{Bagnuls01,Ball95,Comellas98,Morris94a,Morris94b,Morris95,Morris98,Zumbach94,Codello12}. 

\begin{figure}
\centerline{\includegraphics[width=6.5cm]{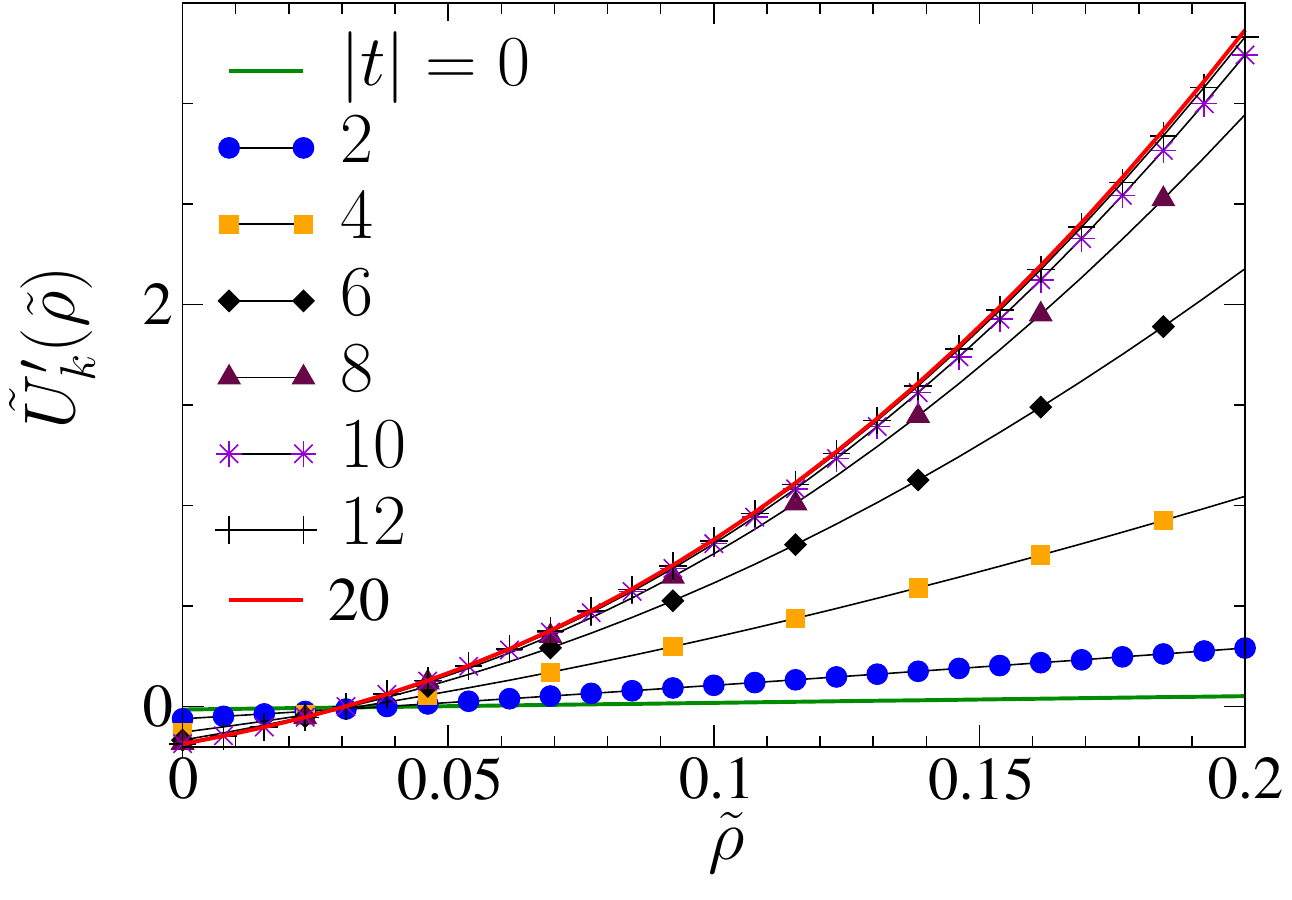}
\hspace{0.5cm}
\includegraphics[width=6.5cm]{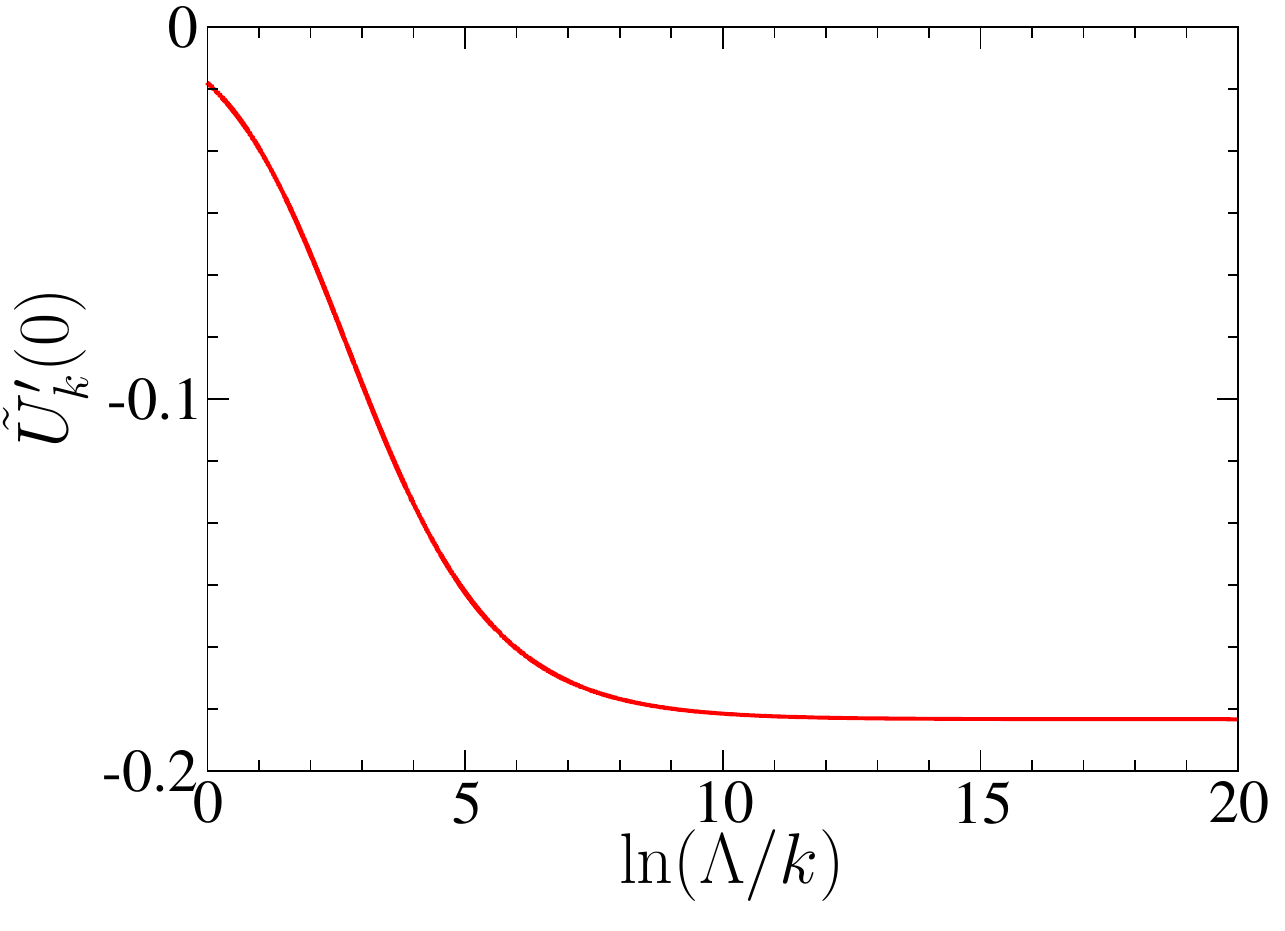}}
\caption{(Left) Derivative $\tilde U_k'(\trho)$ of the effective potential in the LPA for a system near criticality and an RG time $t=\ln(k/\Lambda)$ between 0 and $-20$ ($d=3$, $N=1$). (Right) $\tilde U'_k(0)$ vs $|t|$. For $|t|$ larger than the Ginzburg scale $|t_G|=\ln(\Lambda \xi_G)\simeq 10$ (see \ref{appDE:subsubsec_lpapcritical} for the definition of the Ginzburg length $\xi_G$), $\tilde U'_k(\tilde\rho)$ is nearly equal to the fixed-point value $\tilde U^*{}'(\tilde\rho)$.}
\label{appDE:fig_Utilde}
\end{figure}

The shooting method suffers from an important drawback. Beyond the LPA, there are other functions to be considered in addition to the effective potential, e.g. $Z_k(\rho)$ and $Y_k(\rho)$ in the derivative expansion to second order (\ref{appDE:subsec_DE2}), which may require to fine tune additional parameters. As an alternative to the shooting method, which is not restricted to the LPA, one can numerically solve the flow equation for a system near criticality. In the O($N$) model, the initial effective potential $\tilde U'_\Lambda(\trho)= \tilde r_0 +(\tilde u_0/3)\trho$ depends on two parameters; the critical point can be reached by fine tuning $\tilde r_0$ (with $\tilde u_0$ fixed). When $\tilde r_0$ is near the critical value $\tilde r_{0c}$, the solution of the flow equation takes the form 
\begin{equation}
\tilde U'_k(\trho) \simeq \tilde U^*{}'(\trho) + g_1(\trho) e^{\omega t} + g_2(\trho) e^{-t/\nu} 
\label{appDE:eq9a}
\end{equation}
for large $|t|$. For $\tilde r_0$ slightly detuned from the critical value $\tilde r_{0c}$, $\tilde U'_k$ first moves closer to $\tilde U^*{}'$ with a speed controlled by the correction-to-scaling exponent $\omega$, remains nearly equal to $\tilde U^*{}'$ for a long time, before eventually running away with a rate $1/\nu$ given by the inverse of the correlation-length exponent. This allows us to determine both $\tilde U^*{}'$ and the critical exponents $\nu$ and $\omega$.\footnote{To determine $\nu$ and $\omega$ it is sufficient to consider, e.g., the flow of $\tilde U'_k(0)$ and use~(\ref{appDE:eq9a}).} A typical flow for a system near criticality is shown in Fig.~\ref{appDE:fig_Utilde}. Note that the vanishing of $\tilde U^*{}'$ at a nonzero $\trho^*_0$, corresponding to the minimum of $\tilde U^*$, is not in contradiction with the system being critical; going back to dimensionful variables one finds that $U_k(\rho)$ exhibits a minimum at $\rho_{0,k}=k^{d-2}\tilde \rho^*_0$ which vanishes for $k\to 0$. 

A third possible procedure to determine the critical exponents is the following. The (approximate) fixed-point potential $\tilde U^*{}'$ found from the numerical solution of the flow equation can also be used as an initial guess of the solution of Eq.~(\ref{appDE:eq7}). Discretizing the $\trho$ variable, i.e. $\trho_i=i \Delta\trho$ ($i=0,\cdots,M-1$), one is then left with a system of $M$ equations for the $M$ variables $\tilde U^*{}'(\trho_i)$, which can be solved with standard numerical algorithms. The initial guess, being close to the exact solution, ensures convergence to the physical solution, the accuracy being limited by the finite number $M$ of $\trho_i$ variables and the maximum value $\trho_{M-1}$. To compute the correlation-length exponent $\nu$, one writes $\tilde U_k'(\trho_i)=\tilde U^*{}'(\trho_i)+e^{-\lambda t} g(\trho_i)$, see Eq.~(\ref{appDE:eq9}), and linearize the flow equation $\dt\tilde U'_k$ about $\tilde U^*{}'$. The possible values of $\lambda$ are then given by the eigenvalues of the stability matrix 
\begin{equation}
L_{ij} = - \frac{\delta \dt\tilde U'_k(\trho_i)}{\delta\tilde U'_k(\trho_j)} \biggl|_{\tilde U'_k=\tilde U^*{}'} . 
\end{equation} 
The largest eigenvalue (the only positive one for a standard bicritical point) determines $1/\nu$. The absolute value of the less negative eigenvalue gives the correction-to-scaling exponent $\omega$.

Finally, we note that it is possible to determine the critical exponents directly from physical quantities computed at $k=0$. For instance, in the disordered phase $G_{k=0,ii}(\p,\phibf=0)=(\p^2+U_{k=0}'(0))^{-1}$ and the correlation length is given by $\xi=1/\sqrt{U'_{k=0}(0)}$. This allows us to obtain $\nu$ using 
\begin{equation}
U'_{k=0}(0) \sim (r_0-r_{0c})^{2\nu} \quad \mbox{for} \quad r_0 \to r_{0c}^+ . 
\end{equation} 
Similarly, in the ordered phase one can compute the magnitude of the order parameter,
\begin{equation} 
|\phibf_{\rm eq}| = \lim_{k\to 0} \sqrt{2\rho_{0,k}} \sim (r_{0c}-r_0)^\beta \quad \mbox{for} \quad r_0 \to r_{0c}^- ,
\end{equation}
to obtain the exponent $\beta$.

\subsubsection[Upper and lower critical dimensions]{Upper and lower critical dimensions -- Consistency with the Mermin-Wagner theorem} 

The LPA recovers the well-known result that the upper critical dimension is $d_c^+=4$ for the O($N$) model: For $d\geq d_c^+$, the fixed point controlling the transition is the (trivial) Gaussian fixed point and the critical exponents take their mean-field values $\nu=1/2$, $\eta=0$, etc. For $d\leq 2$ and $N>1$ no fixed point is obtained, the RG flow always drives the system to the disordered phase. This identifies $d_c^-=2$ as the lower critical dimension (for $N>1$)~\cite{Codello13,Defenu15}. This can be understood from the flow equation~(\ref{appDE:eq4}). Let us assume that the effective potential exhibits a minimum at $\rho_{0,k}>0$. Differentiating $U_k'(\rho_{0,k})=0$ wrt $t$ we find $\dt U_k'(\rho_{0,k})+ U_k''(\rho_{0,k})\dt \rho_{0,k}=0$ (with $\dt U_k'(\rho_{0,k})\equiv \dt U_k'|_{\rho_{0,k}}$), i.e. 
\begin{equation}
\dt \rho_{0,k} = - \frac{\dt U_k'(\rho_{0,k})}{U_k''(\rho_{0,k})} .
\end{equation}
Considering the theta regulator for simplicity, 
\begin{equation}
\dt U'_k = - 4 \frac{v_d}{d} k^{d+2} \left[ \frac{3U_k'' + 2\rho U_k'''}{(k^{2}+U_k' + 2\rho U_k'')^2} + \frac{(N-1)U_k''}{(k^{2}+U_k')^2} \right]
\end{equation}
(see Eq.~(\ref{appDE:eq4})). Thus $\dt U'_k(\rho_{0,k})\sim -(N-1) k^{d-2}$ and $U'_k(\rho_{0,k})$ diverges for $k\to 0$ if $d\leq 2$ and $N>1$ (the divergence is logarithmic for $d=2$), which is incompatible with $U_k'(\rho_{0,k})=0$ and therefore $\rho_{0,k}>0$. It follows that $\rho_{0,k}$ must vanish at a nonzero momentum scale $k$ (the numerical solution of the LPA flow equation confirms this analysis) in agreement with the Mermin-Wagner theorem forbidding the spontaneous breaking of a continuous symmetry in two- and one-dimensional classical systems~\cite{Mermin66}. In two dimensions the LPA fails when $N=1$, since it does not allow for spontaneous symmetry breaking, and does not describe the Berezinskii-Kosterliz-Thouless transition when $N=2$.\footnote{The breakdown of the LPA for a two-dimensional critical system can be seen from the expected large-field behavior of the effective potential at the fixed point: $\tilde U^*(\trho)\sim\trho^{d/(d-2)}$. The DE to second order gives results in two dimensions for both $N=1$ and $N=2$~\cite{Jakubczyk14}.}

\subsubsection{Spontaneous symmetry breaking and approach to convexity}

The stability of the LPA flow equation~(\ref{sec_frg:eqLPA}) or (\ref{appDE:eq4}) requires 
\begin{equation}
\begin{split} 
	& \q^2 + R_k(\q^2) + U'_k(\rho) \geq 0 , \\ 
	& \q^2 + R_k(\q^2) + U'_k(\rho) + 2 \rho U''_k(\rho) \geq 0 . 
\end{split}
\label{appDE:eq10} 
\end{equation}
If conditions~(\ref{appDE:eq10}) are not fulfilled, a pole appears in the propagator $G_{k,\rm L}$ or $G_{k,\rm T}$ thus making the threshold functions $l^d_n(w)$ ill-defined when evaluated at $w=\tilde U_k'(\trho)$ or $\tilde U_k'(\trho)+2\trho\tilde U_k''(\trho)$. In the broken-symmetry phase, $U'_k(\rho)$ is negative for $0\leq \rho\leq\rho_{0,k}$ and there is no guarantee that Eqs.~(\ref{appDE:eq10}) are satisfied. In the LPA, and for some regulators, the singularity of the threshold function is approached but never reached. This is the case for the theta regulator or the exponential regulator (with a prefactor $\alpha>2$)~\cite{Tetradis92,Tetradis96,Berges:2000ew,Pelaez16}. A detailed analysis shows that for these regulators the potential behaves as $U'_k(\rho)\sim -k^2$ in the internal region $0\leq\rho\lesssim\rho_{0,k}$ but both stability conditions~(\ref{appDE:eq10}) remain fulfilled. This implies that the free energy $U_{k=0}(\rho)$ is convex, with a flat part $U_{k=0}(\rho)={\rm const}$ for $0\leq \rho\leq\rho_0=\lim_{k\to 0}\rho_{0,k}$, as shown in Fig.~\ref{appDE:fig_U}. It should be emphasized that, in contrast to most perturbative approaches, the convexity of the thermodynamic potential is not obtained here from the Maxwell construction but is a genuine property of the effective potential in the LPA.  

From a practical point of view, the approach to convexity of the potential makes its numerical determination difficult. Indeed, a tiny numerical error can lead to negative values of $\q^2 + R_k(\q^2) + U'_k(\rho)$, even in cases where we know that conditions~(\ref{appDE:eq10}) should remain fulfilled. This difficulty is more pronounced for $N=1$, mainly because $U''_k(\rho)$ becomes discontinuous at $\rho=\rho_{0,k}$ in the limit $k\to 0$: $U''_k(\rho)\to 0$ for $\rho<\rho_{0,k}$ due to the approach to convexity whereas $U''_k(\rho_{0,k}+0^+)$ remains nonzero.\footnote{In the O($N>1$) model, $U_k''(\rho_{0,k}+0^+)$ vanishes for $k\to 0$ (this is due to the divergence of the longitudinal susceptibility discussed in~\ref{appDE:subsubsec_lowT}) and $U'_k(\rho)$ remains a regular function around $\rho_{0,k}$.} Nevertheless, analytical solutions of the potential in the internal region can be exploited to set up efficient algorithms for the broken-symmetry phase~\cite{Caillol12a,Pelaez16}.\footnote{We refer to Ref.~\cite{Pelaez16} for a discussion of the approach to convexity in the derivative expansion beyond the LPA.}  
%commentaires sur transitions du 1er ordre? 

\subsection{Improving the LPA: the LPA$'$}

The LPA$'$, defined by the effective action~(\ref{sec_frg:gammalpap}), is the minimal improvement of the LPA that allows one to compute the anomalous dimension $\eta$. In addition to the effective potential $U_k(\rho)$, the effective action includes a field renormalization factor $Z_k$ which diverges as $Z_k\sim k^{-\eta}$ when the system is critical. 

The flow equation of the effective potential is given by~(\ref{sec_frg:eqLPA}) where the longitudinal and transverse propagators 
\begin{equation}
\begin{split}
G_{k,\rm L}({\bf q},\rho) &= [\Gamma^{(2)}_{k,\rm L}({\bf q},\rho) + R_k({\bf q})]^{-1} 
= [Z_k {\bf q}^2+U'_k(\rho)+2\rho U_k''(\rho)+R_k({\bf q})]^{-1} , \\  
G_{k,\rm T}({\bf q},\rho) &= [\Gamma^{(2)}_{k,\rm T}({\bf q},\rho) + R_k({\bf q})]^{-1} = [ Z_k {\bf q}^2+U'_k(\rho)+R_k({\bf q})]^{-1} 
\end{split}
\end{equation}
now include the field renormalization factor $Z_k$. Since 
\begin{equation} 
Z_k = \lim_{\p\to 0} \frac{\partial}{\partial\p^2} \Gamma^{(2)}_{k,\rm T}(\p,\rho_{0,k}) , 
\label{appDE:eq11}
\end{equation} 
the flow equation $\dt Z_k$ is deduced from that of the two-point vertex in a uniform field $\phibf(\r)=\phibf$, 
\begin{align} 
\dt \Gamma^{(2)}_{k,ij}(\p,\phibf) ={}& \half \tilde\partial_t \int_\q \Bigl\{ 
G_{k,i_1,i_2}(\q,\phibf) \Gamma^{(4)}_{k,iji_2i_1}(\p,-\p,\q,-\q,\phibf) \nonumber \\ & 
- G_{k,i_1,i_2}(\q,\phibf) \Gamma^{(3)}_{k,ii_2i_3}(\p,\q,-\p-\q,\phibf) 
G_{k,i_3,i_4}(\p+\q,\phibf) \Gamma^{(3)}_{k,ji_4i_1}(-\p,\p+\q,-\q,\phibf)   
\Bigr\} ,
\label{appDE:eq12}
\end{align}
where $\tilde\dt$ acts only on the $k$ dependence of $R_k$: $\tilde\dt G_{k,i_1i_2}(\q,\phibf)=- G_{k,i_1i_3}(\q,\phibf) \dt R_k(\q)G_{k,i_3i_2}(\q,\phibf)$. The reason for defining $Z_k$ from the transverse 2-point vertex is that in the full derivative expansion to order $\calO(\partial^2)$ the longitudinal vertex $\Gamma^{(2)}_{k,\rm L}$ includes a term $Y_k\p^2$ (see \ref{appDE:subsec_DE2}). Although this term is neglected in the LPA$'$, the equation $\dt\Gamma^{(2)}_{k,\rm L}$ yields contributions to both $\dt Z_k$ and $\dt Y_k$ whereas $\dt\Gamma^{(2)}_{k,\rm T}$ contributes only to $\dt Z_k$. Equations~(\ref{appDE:eq11}) and (\ref{appDE:eq12}) imply 
\begin{equation}
\dt Z_k = 16 \frac{v_d}{d} \rho_{0,k} U_k''(\rho_{0,k})^2 \tilde \dt \int_0^\infty d|\p| \, |\p|^{d+1} G'_{k,\rm L}(\p,\rho_{0,k}) G'_{k,\rm T}(\p,\rho_{0,k}) ,
\label{appDE:eq13}
\end{equation}
where $G'_{k}(\p,\rho_{0,k})=\partial_{\p^2} G_{k}(\p,\rho_{0,k})$.\footnote{For a derivation of~(\ref{appDE:eq13}) see, e.g., Ref.~\cite{Tetradis94}.} 

The LPA$'$ flow equations can be written in dimensionless form by making use of the dimensionless variables~(\ref{appDE:eq5}) provided that we define $\tilde\phibf(\tilde\r)=\sqrt{Z_k}k^{-(d-2)/2}\phibf(\r)$ with a factor $\sqrt{Z_k}$ and write the regulator function in the form $R_k(\q)=Z_k \q^2 r(\q^2/k^2)$: 
\begin{equation}
\begin{split} 
\dt \tilde U'_k &= (\eta_k-2)\tilde U'_k + (d-2+\eta_k)\trho \tilde U_k'' - 2 v_d[ (3\tilde U_k''+2\trho \tilde U'''_k) l_1^d(\tilde U_k'+2\trho \tilde U''_k,\eta_k)+(N-1) \tilde U''_k l_1^d(\tilde U_k',\eta_k) ] , \\  
\eta_k &= 16 \frac{v_d}{d} \trho_{0,k} \tilde U_k''(\trho_{0,k})^2 m^d_{22}(2\trho_{0,k}\tilde U_k''(\trho_{0,k}),\eta_k) ,
\end{split}
\label{appDE:eq14}
\end{equation}
where $\eta_k=-\dt \ln Z_k$ is the ``running'' anomalous dimension. The threshold functions are defined by 
\begin{equation}
\begin{split}
l_n^d(w,\eta) &= - \frac{n+\delta_{n,0}}{2} \int_0^\infty dy \, y^{d/2} \frac{\eta r+2yr'}{P(w)^{n+1}} , \\
m_{22}^d(w,\eta) &= - \int_0^\infty dy \, y^{d/2} \frac{1+r+yr'}{P(w)^2 P(0)^2} \biggl\lbrace y(\eta r+2yr') (1+r+yr')\biggl[\frac{1}{P(w)} + \frac{1}{P(0)} \biggr]
- \eta r - (\eta+4)yr' - 2y^2 r'' \biggr\rbrace ,
\end{split}
\end{equation}
where $P(w)=y(1+r)+w$. To find the fixed-point solution $\tilde U^*{}'$ and the critical exponents, we can use the same methods as in the LPA. When the fixed point is reached by fine tuning $r_0$, the anomalous dimension $\eta$ can be estimated from the value of $\eta_k$ on the plateau $\xi^{-1}\ll k\ll \xi_G^{-1}$ (see~\ref{appDE:subsubsec_lpapcritical} and Fig.~\ref{appDE:fig_lpapcrit}). 

The LPA$'$ can be simplified by expanding the effective potential about $\rho_{0,k}$ as in Eq.~(\ref{sec_frg:Utroncated}). The flow equations then reduce to coupled ordinary differential equations for the dimensionless variables $\trho_{0,k}=Z_k k^{2-d}\rho_{0,k}$, $\tdelta_k=(Z_kk^2)^{-1}\delta_k$, $\tlambda_k=Z_k^{-2}k^{d-4}\lambda_k$ and $\eta_k$. In spite of its simplicity  the truncated LPA$'$ turns out to be highly nontrivial and fundamentally different from a perturbative RG based on a loop expansion (the flow equations are still nonperturbative since they are of infinite order in the coupling constant $\tlambda_k$). Although it is not reliable for an accurate estimate of the critical exponents, it yields a rather complete picture of the long-distance physics at and near criticality, as well as in the low-temperature phase, regardless of the value of $N$ and for all dimensions $d\geq 2$.\footnote{Note however that a polynomial approximation of the effective potential, although stable to quadratic order in $\rho$, does not always converge to higher orders below three dimensions.} It also captures some features of the Berezinskii-Kosterliz-Thouless transition when $d=2$ and $N=2$~\cite{Graeter95} although a more complete analysis requires the DE to second order~\cite{,Gersdorff01,Jakubczyk14}.

\subsubsection{Critical behavior: the limits $d\to 4$, $d\to 2$ and $N\to\infty$} 
\label{appDE:subsubsec_lpapcritical} 

\begin{figure} 
	\centerline{\includegraphics[height=4.1cm]{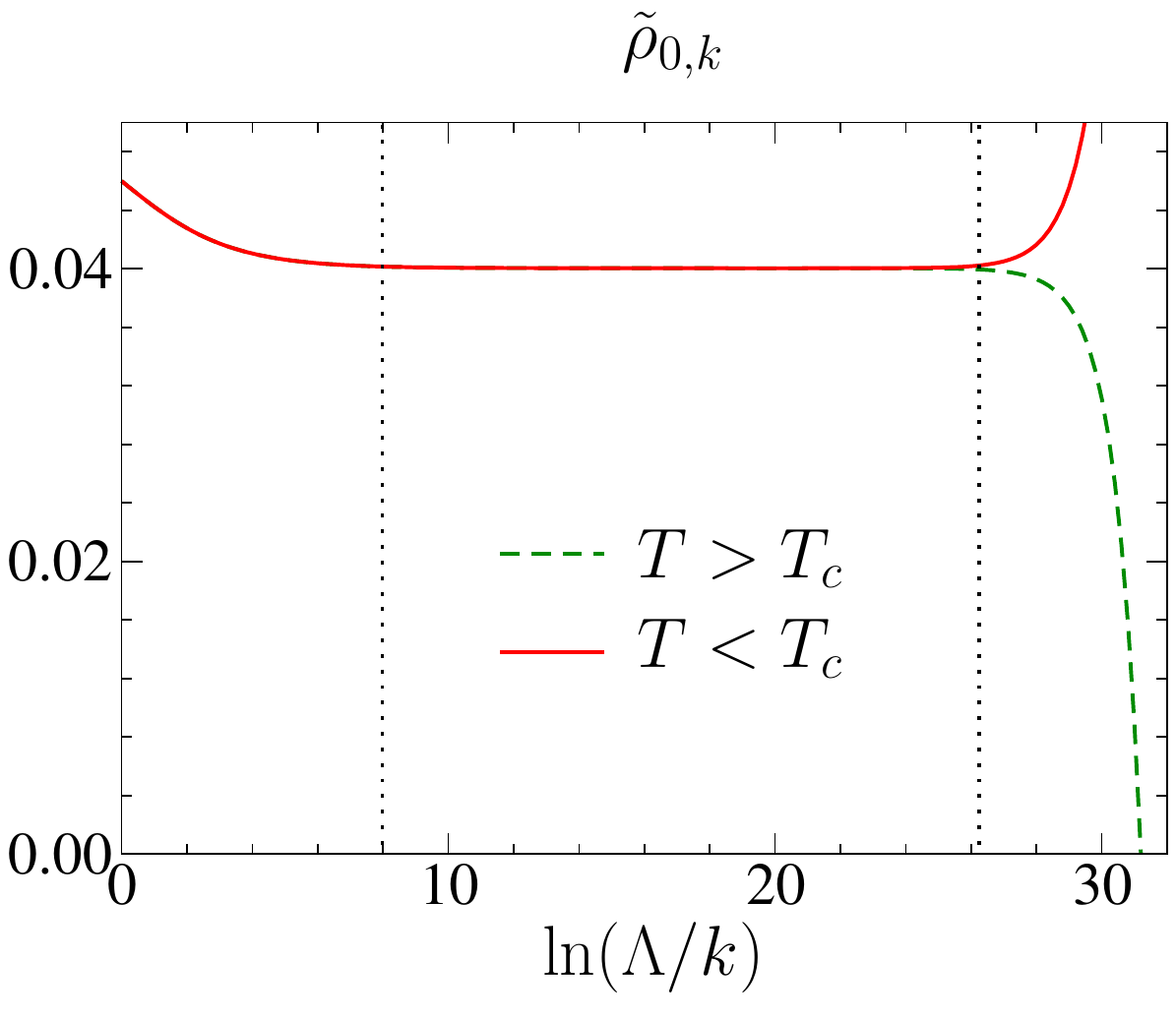}
	\hspace{0.45cm}\includegraphics[height=4.2cm]{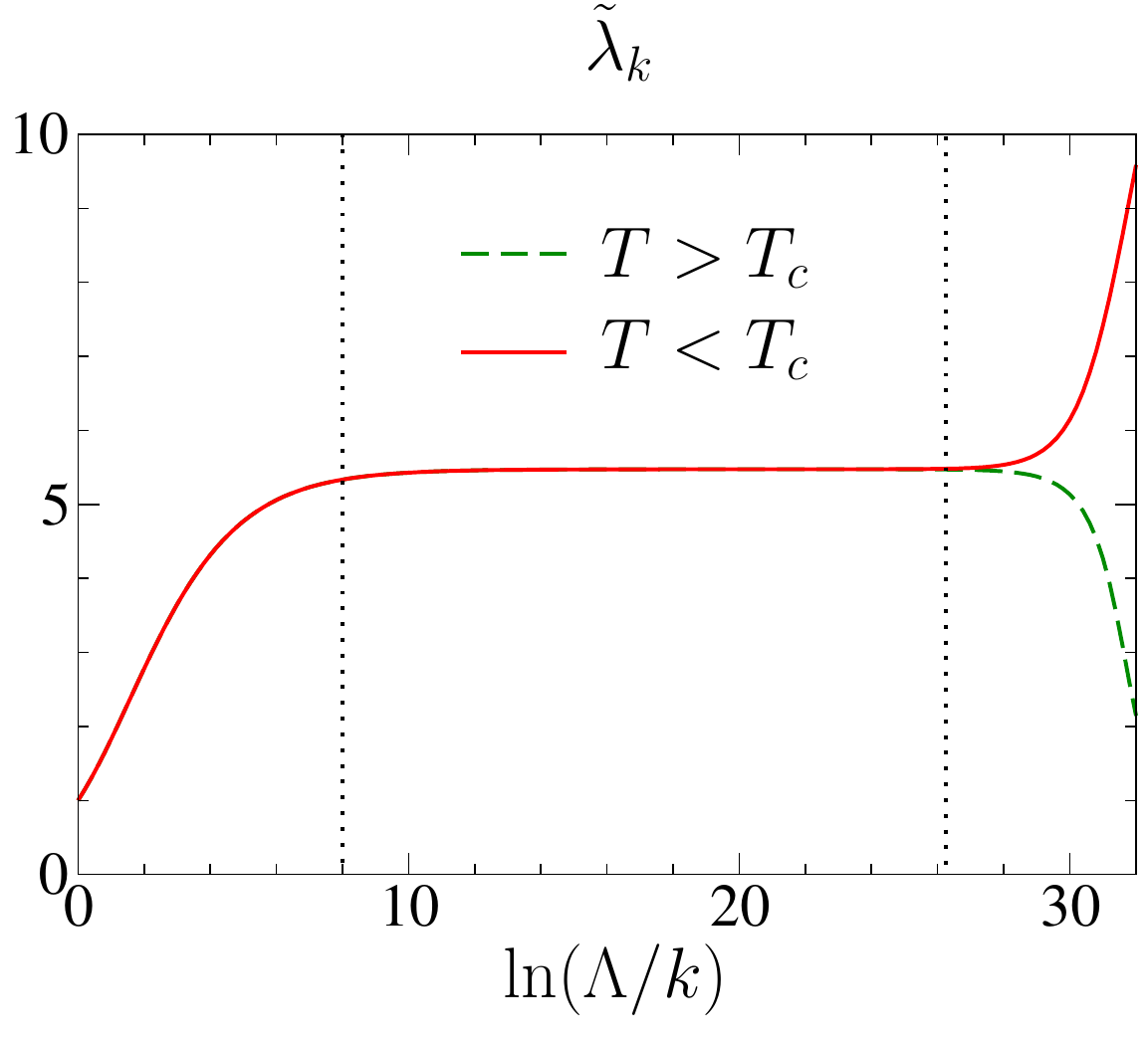}
	\hspace{0.45cm}\includegraphics[height=4.1cm]{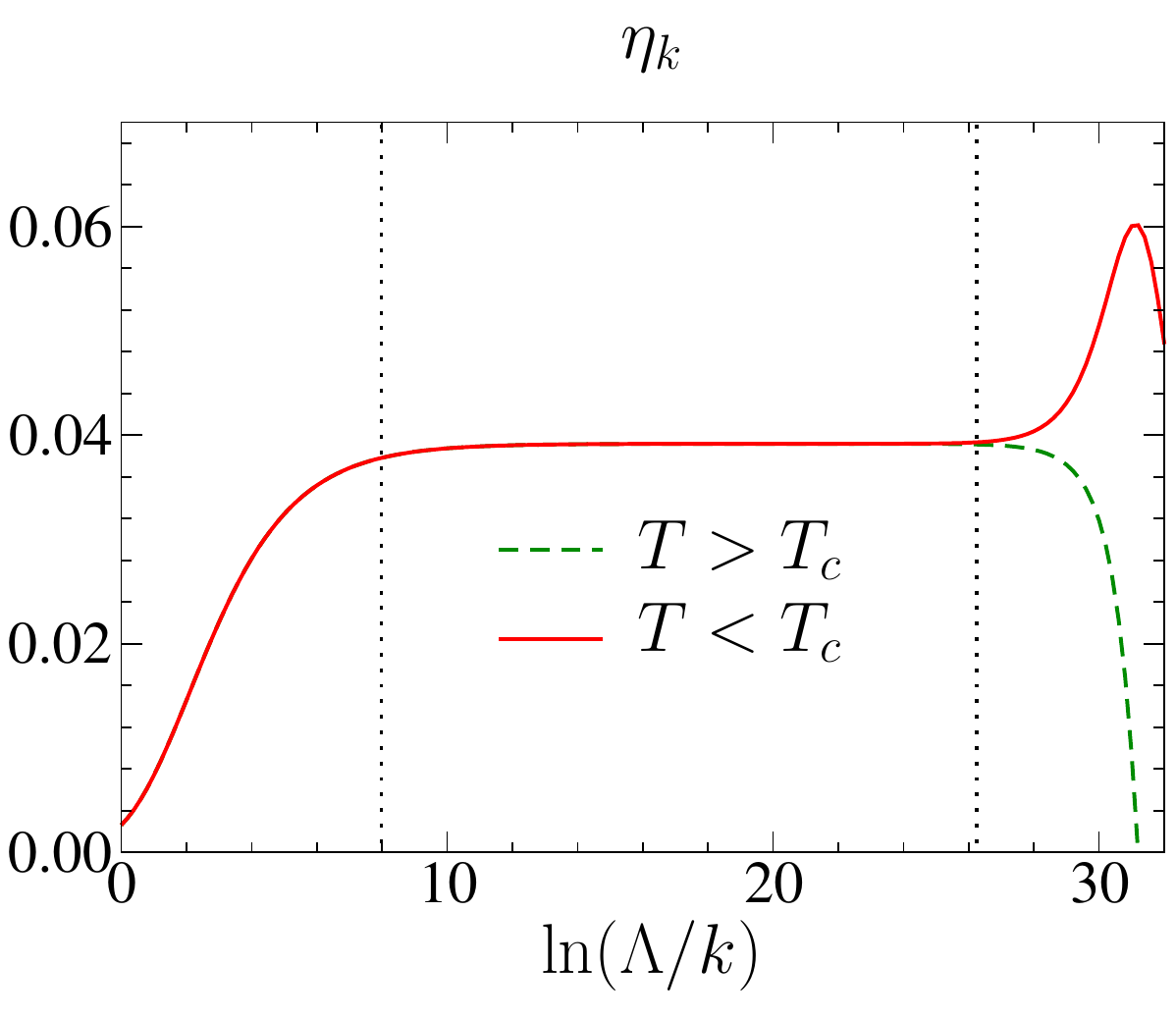}}
	\caption{$\trho_{0,k}$, $\tlambda_k$ and $\eta_k$ vs $\ln(\Lambda/k)=-t$ near criticality  in the LPA$'$ ($d=3$ and $N=3$). The vertical dotted lines show the Ginzburg scale $|t_G|=\ln(\Lambda\xi_G)$ and the correlation length scale $|t_\xi|=\ln(\Lambda\xi)$ ($T>T_c$ i.e. $r_0>r_{0c}$) or the Josephson scale $|t_J|=\ln(\Lambda\xi_J)$ ($T<T_c$ i.e. $r_0<r_{0c}$).}
\label{appDE:fig_lpapcrit}
\end{figure}

Figure~\ref{appDE:fig_lpapcrit} shows typical solutions of the truncated LPA$'$ flow equations for a nearly critical system. One clearly observes a critical regime where $\trho_{0,k}$, $\tlambda_k$ and $\eta_k$ are very close to their fixed-point values $\trho^*_{0}$, $\tlambda^*$ and $\eta$. The critical regime starts when $k\sim\xi_G^{-1}$ is of the order of the inverse Ginzburg length $\xi_G\sim u_0^{1/(d-4)}$ (the Ginzburg scale $k_G= \xi_G^{-1}$ is defined by $\tlambda_{k_G}\sim 1$ and signals the breakdown of perturbation theory). It ends when $k\sim\xi^{-1}$ (disordered phase) or $k\sim\xi_J^{-1}$ (ordered phase) where $\xi$ and $\xi_J$ are the correlation and Josephson lengths, respectively.

Near four dimensions, i.e. $d=4-\epsilon$, the truncated flow equations give $\nu=1/2+(\epsilon/4)(N+2)/(N+4)+\calO(\epsilon^2)$ and $\omega=\epsilon+\calO(\epsilon^2)$, in agreement with the one-loop perturbative RG~\cite{Ma_book,Zinn_book}. More surprisingly, one also recovers the critical exponents obtained from the nonlinear sigma model near two dimensions, i.e. for $d=2+\epsilon$: $\nu=1/\epsilon+\calO(\epsilon^0)$ and $\eta=\epsilon/(N-2)+\calO(\epsilon^2)$~\cite{Zinn_book}. Since $\trho^*_{0}=\calO(\epsilon^{-1})$ and $\tlambda^*=\calO(\epsilon^0)$, in the critical regime $\xi^{-1},\xi_J^{-1}\ll k\ll \xi_G^{-1}$ the longitudinal mode has a large (square) dimensionless mass $2\tlambda_k\trho_{0,k}\sim \epsilon^{-1}$  which suppresses the threshold functions. We thus recover the fact that near two dimensions the critical behavior is entirely determined by the Goldstone modes, which is the basic assumption when studying the critical behavior of the O($N$) model in the framework of the nonlinear sigma model. The relation to the nonlinear sigma model can be further understood~\cite{Delamotte04} by considering the Wilsonian action of the dimensionless field $\tvarphibf$, 
\begin{equation}
S[\tvarphibf] = \int_{\tilde \r} \biggl\lbrace \half (\nablabf \tvarphibf)^2 + \frac{\tlambda}{2}(\trho-\trho_0)^2 \biggr\rbrace , 
\end{equation}
where $\trho=\tvarphibf^2/2$. Rescaling the field, $\tvarphibf\to\sqrt{2\trho_0}\tvarphibf$, we obtain 
\begin{equation}
S[\tvarphibf] = \trho_0 \int_{\tilde \r} \biggl\lbrace (\nablabf \tvarphibf)^2 + \frac{\tlambda\trho_0}{2} (\tvarphibf^2-1)^2 \biggr\rbrace .
\label{appDE:eq15}
\end{equation}
In the limit $\tlambda\trho_0\to\infty$ (corresponding to an infinite dimensionless mass for the longitudinal mode), the last term in~(\ref{appDE:eq15}) imposes the constraint $\tvarphibf^2=1$, and we obtain a nonlinear sigma model 
\begin{equation}
S[\tvarphibf] = \frac{1}{2\tilde g} \int_{\tilde \r} (\nablabf \tvarphibf)^2, \qquad (\tvarphibf^2=1) 
\end{equation}
with (dimensionless) coupling constant $\tilde g=1/2\trho_0$. From the flow equation satisfied by $\dt\trho_{0,k}$, we deduce  
\begin{equation}
\dt \tilde g_k = \epsilon \tilde g_k - \frac{N-2}{2\pi} \tilde g_k^2 + \calO(\epsilon^3) , 
\label{appDE:eq16}
\end{equation}
which is identical, to order $\epsilon^2$, to the RG equation obtained from the nonlinear sigma model to one-loop order~\cite{Polyakov75,Nelson77,Zinn_book}. 
There is nevertheless an important difference between the usual perturbative RG approach to the nonlinear sigma model and the nonperturbative FRG approach to the linear O($N$) model near two dimensions. While in the former case the RG equation~(\ref{appDE:eq16}) is not valid anymore when the coupling constant $\tilde g_k$ becomes of order one, in the latter case there is no difficulty to continue the LPA$'$ flow in the strong-coupling regime and describe the disordered phase with a finite correlation length. 

Finally we note that the truncated LPA$'$ flow equations are also exact to leading order in the large-$N$ limit where $\nu=1/(d-2)+\calO(N^{-1})$ and $\eta=\calO(N^{-1})$.

\subsubsection{Low-temperature phase}
\label{appDE:subsubsec_lowT}

The physics of the O($N$) model remains nontrivial in the whole ordered phase when $N\geq 2$ because of the Goldstone modes associated with the spontaneously broken symmetry. Mean-field (or Gaussian theory) predicts $G_{\rm L}(\p)=1/(\p^2+2|r_0|)$ and $G_{\rm T}(\p)=1/\p^2$ when $r_0<0$~\cite{Ma_book}. The transverse propagator is gapless, in agreement with Goldstone's theorem while longitudinal fluctuations have a finite correlation length $\xi=(2|r_0|)^{-1/2}$. The last result is however an artifact of mean-field theory~\cite{Patasinskij73,Dupuis11}. The coupling between transverse and longitudinal fluctuations implies that $G_{\rm L}(\r)\sim (N-1) [G_{\rm T}(\r)]^2\sim 1/|\r|^{2d-4}$ when $d\leq 4$, i.e. $G_{\rm L}(\p)\sim 1/|\p|^{4-d}$ (the divergence for $\p\to 0$ is logarithmic when $d=4$). The presence of a singularity in the longitudinal channel, driven by transverse fluctuations, is a general phenomenon in systems with a continuous broken symmetry~\cite{Patasinskij73}.

\begin{figure}
	\centerline{\includegraphics[clip,height=4cm]{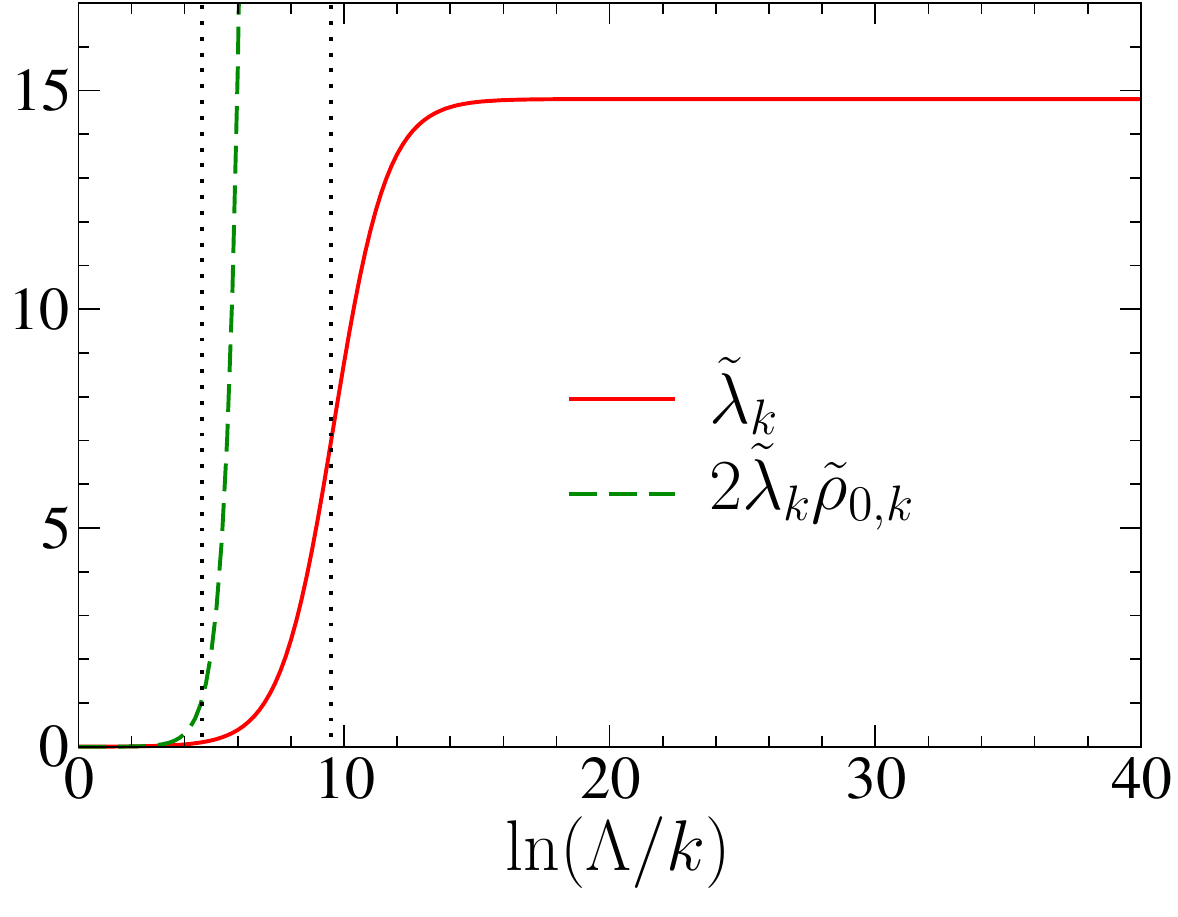}
	\hspace{0.5cm}
	\includegraphics[clip,height=4cm]{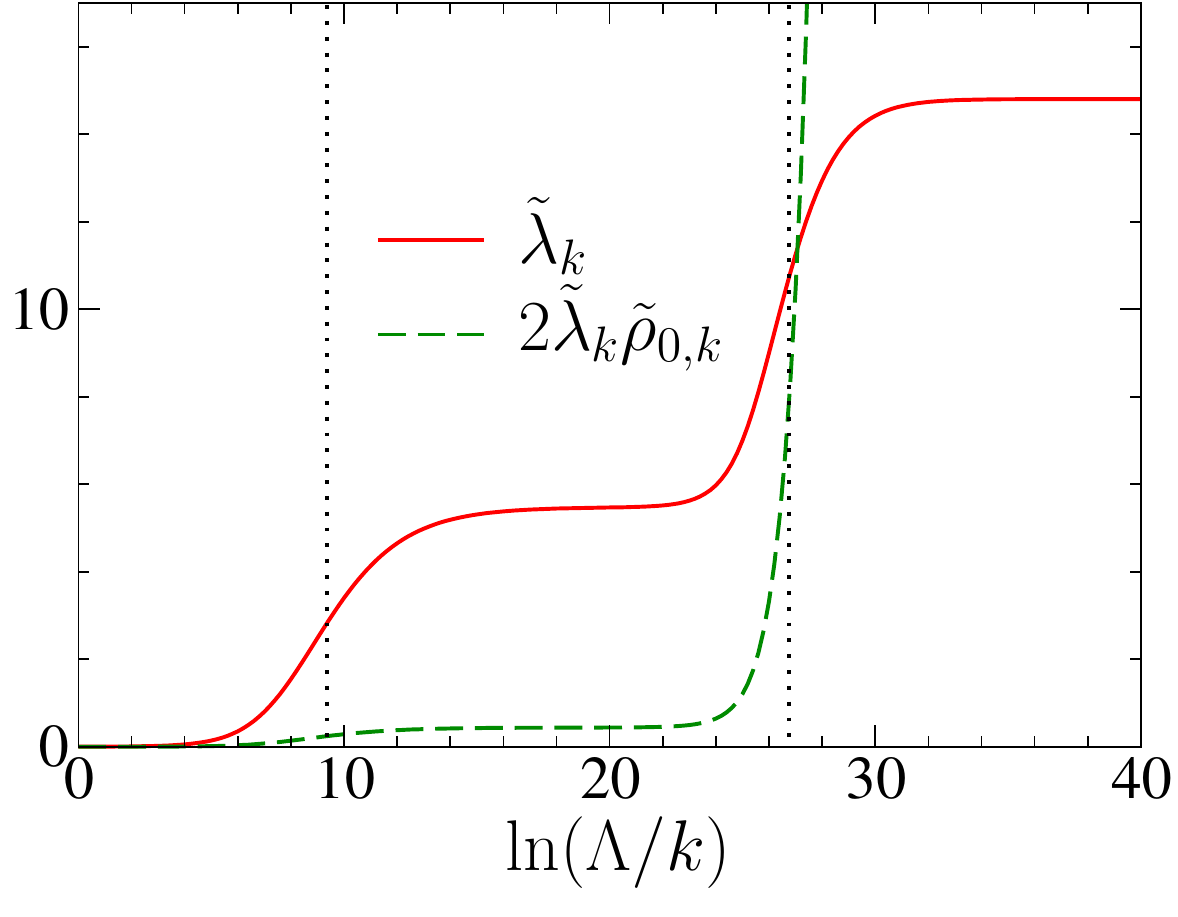}}
	\caption{$\tlambda_k$ and $2\tilde \lambda_k\trho_{0,k}$ vs $\ln(\Lambda/k)$ in the ordered phase for $d=3$ and $N=3$ in the LPA$'$. Left panel: deep in the ordered phase $T_c-T_G\ll T_c-T$ ($T_G$ is the Ginzburg temperature). The 
		dotted vertical lines show the hydrodynamic ($k_c$) and Ginzburg ($k_G$) scales. Right panel: critical regime $T_c-T\ll T_c-T_G$. The dotted vertical lines show the Ginzburg ($k_G$) and Josephson ($k_J$) scales. The first plateau, $\tlambda_k\sim\tlambda^*_{\rm crit}$, corresponds to the critical regime and the second one, $\tlambda_k\sim\tlambda^*$ (visible for $\ln(\Lambda/k)\gtrsim 30$), to the low-temperature fixed point.}
\label{appDE:fig_lpaplowT} 
\end{figure}

In the LPA$'$ the divergence of the longitudinal susceptibility $G_{\rm L}(\p=0)$ can be traced back to the infrared behavior of the coupling constant $\tlambda_k\to \tlambda^*$ for $d<4$ (here $\tlambda^*$ should not be confused with the critical value $\tlambda^*_{\rm crit}$ discussed in~\ref{appDE:subsubsec_lpapcritical} and shown in Fig.~\ref{appDE:fig_lpapcrit}); see Fig.~\ref{appDE:fig_lpaplowT}. Thus the mass of the longitudinal mode behaves as $2\rho_{k,0}\lambda_k\sim k^{4-d}$ since $Z_k\to Z$ and $\rho_{0,k}\to \rho_0$ when $k\to 0$. If we approximate $G_{k=0,\rm L}(\p,\rho_0)$ by $G_{k\sim|\p|,\rm L}(\p,\rho_0)$, arguing that $\p$ acts as an effective infrared cutoff when $|{\bf p}|\gg k$, and use the LPA$'$ result for $G_{k\sim|\p|,\rm L}(\p,\rho_0)$,  one finds $G_{k=0,\rm L}(\p,\rho_0)\sim 1/|\p|^{4-d}$. For $d=4$, $\tlambda_k$ vanishes logarithmically and $G_{k=0,\rm L}(\p,\rho_0)\sim \ln(1/|\p|)$. 

The criterion $2\tlambda_k\trho_{0,k}\sim 1$ defines a hydrodynamic momentum scale $k_c$ below which the contribution of the longitudinal mode to the flow equations is suppressed. Deep in the ordered phase, when $T_c-T_G\ll T_c-T$ (with $T_G$ the Ginzburg temperature), we observe a crossover for $k\sim k_G\ll k_c$ between a Gaussian regime, where mean-field theory is essentially correct, and a Goldstone regime characterized by the divergence of the longitudinal propagator. Sufficiently close to the critical point ($T_c-T\ll T_c-T_G$), where $k_c<k_G$, there is first a crossover for $k\sim k_G$ from the Gaussian to the critical regime, followed by a second crossover to the Goldstone regime at a characteristic scale $k_J\sim \xi_J^{-1}$ defined by $2\tlambda_k\trho_{0,k}\sim 1$ and associated with the Josephson length $\xi_J$ (see the left and right panels of Fig.~\ref{appDE:fig_lpaplowT})~\cite{Dupuis11}.

\subsection{Second-order of the derivative expansion} 
\label{appDE:subsec_DE2}

In the second order of the derivative expansion, the scale-dependent effective action is defined by
\begin{equation}
\Gamma_k^{{\rm DE}_2}[\phibf]=\int_{\bf r} \left\{\frac{Z_k(\rho)}{2}({\bf \nabla}\phibf)^2+
\frac{Y_k(\rho)}{4}({\bf \nabla}\rho)^2+U_k(\rho)\right\} ,
\label{appDE:eq17}
\end{equation}
which gives the two-point vertices 
\begin{equation} 
\begin{split}
\Gamma^{(2)}_{k,\rm L}({\bf q},\rho) &= [Z_k(\rho)+\rho Y_k(\rho)]{\bf q}^2+U'_k(\rho)+2\rho U_k''(\rho) , \\  
\Gamma^{(2)}_{k,\rm T}({\bf q},\rho) &= Z_k(\rho) {\bf q}^2+U'_k(\rho) 
\end{split}
\end{equation}
in a uniform field. The potential $U_k(\rho)$ still satisfies Eq.~(\ref{sec_frg:eqLPA}), and the flow equations of $Z_k(\rho)$ and $Y_k(\rho)$ are obtained from 
\begin{equation}
Z_k(\rho) = \lim_{\p\to 0} \frac{\partial}{\partial \p^2} \Gamma_{k,\rm T}^{(2)}(\p,\rho) , \qquad 
Z_k(\rho) + \rho Y_k(\rho) = \lim_{\p\to 0} \frac{\partial}{\partial \p^2} \Gamma_{k,\rm L}^{(2)}(\p,\rho) .
\end{equation} 
Inserting the vertices derived from~(\ref{appDE:eq17}) into the RG equations $\dt U_k(\rho)$, $\dt Z_k(\rho)$ and $\dt Y_k(\rho)$ yields the standard
(or ``complete'') flow equations at order DE$_2$.\footnote{The regulator function $R_k(\p)=Z_k \p^2 r(\p^2/k^2)$ has the same form as in the LPA$'$ where the field renormalization factor $Z_k$ is defined by~(\ref{sec_frg:adim2}). The expression of the ``complete'' flow equations can be found in Ref.~\cite{Gersdorff01}.}  
However, the DE being a {\it low-momentum expansion}, one can also truncate the rhs of the flow equations to order $\p^2$, which is sometimes referred to as the ``strict'' DE$_2$ expansion. At order LPA or LPA$'$ both approximation schemes coincide. At the level of the DE$_2$ they differ by terms at least of order $|\p|^4$ and, as expected, their relative difference at a given order is bounded by 1/4. The estimates of the standard and strict DE$_2$ for the critical exponents agree within error bars~\cite{DePolsi20}

\subsubsection{Choice of the regulator} 
\label{appDE:subsubsec_choiceRk}

Let us now discuss the shape of the regulator function~\cite{Balog19}. The DE, like any approximation scheme, introduces a dependence of the end results on the choice of $R_k$.  There exists some constraints and {\it a priori} guidelines to choose $R_k$ so that its influence stays minimal. First, $R_k$ must freeze the small momentum modes $|\q|\ll k$ in $\calZ_k[\J]$ [Eq.~(\ref{appDE:eq1a})]. It must also leave unchanged the large momentum modes $|\q|\gg k$. Second, because the DE is a Taylor expansion of the $\Gamma^{(n)}_k(\{\p_i\})$'s in powers of $\p_i\cdot\p_j$, it is valid provided $|\p_i\cdot\p_j|<\calR$ where $\calR$, typically between $4k^2$ and $9k^2$, is the radius of convergence of the expansion (see Sec.~\ref{sec_frg:validityDE}). This implies that whenever the $\Gamma^{(n)}_k$'s are replaced in a flow equation by their DE, the momentum region beyond $\calR$ must be efficiently cut off. This is achieved by the $\dt R_k(\q)$ term in~(\ref{sec_frg:eqwet}) provided that $R_k(\q)$ nearly vanishes for $|\q|\gtrsim k$. On the other hand, modes with momenta $|\q|\ll k$ are almost frozen if $R_k(\q)$ is of order $k^2$ in that momentum range. These characteristic features give the overall shape of the regulator function. Note also that if a nonanalytic regulator is chosen, one must make sure that the nonanalyticities introduced in the complex plane of $\q^2$ are at least at a distance $\calR$ from the origin. Finally, at order $s$ of the DE (\ref{appDE:subsec_DE4}), the flow equations involve $\dt R_k(\q)$ and $\partial^n_{\q^2}R_k(\q)$ with $n$ varying from 1 to $s/2$. Since the DE is performed around $\q=0$, it is important that these derivatives decrease monotonically; if not, a ``bump'' at a finite value $\q^2=q_0^2>0$ would magnify a region around $q_0$ which is less accurately described by the DE. 

\begin{figure}
\centerline{\includegraphics[width=8cm]{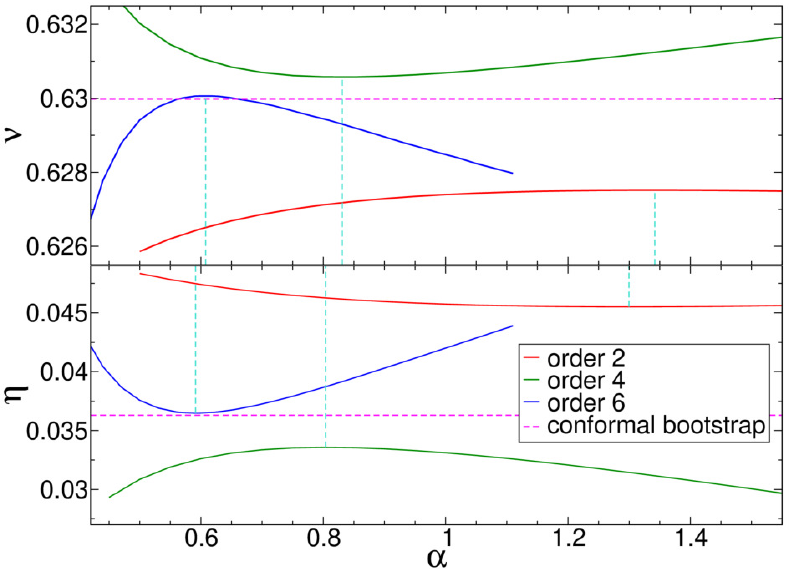}}
\caption{Exponent values $\nu(\alpha)$ and $\eta(\alpha)$ in the three-dimensional Ising universality class at different orders of the
DE with the regulator $E_k(\q)$ [Eq.~(\ref{appDE:eq19})]. Vertical lines indicate the value $\alpha_{\rm opt}$ obtained from the PMS and the horizontal dashed lines the conformal bootstrap results~\cite{Kos14}. LPA results do not appear within the narrow ranges of values chosen
here (see Table~\ref{sec_frg:table_critexp}). Reprinted from Ref.~\cite{Balog19}.}
\label{appDE:fig_PMS} 
\end{figure}

Possible choices of the regulator functions are~\cite{Balog19,DePolsi20}  
\begin{equation}
\begin{split}
W_k(\q) &= \alpha Z_k k^2 \frac{y}{e^y-1} , \\ 
\Theta^n_k(\q) &= \alpha Z_k k^2 (1-y)^n \theta(1-y) \quad n\in \mathbb{N} , \\ 
E_k(\q) &= \alpha Z_k k^2 e^{-y} ,
\end{split}
\label{appDE:eq19}
\end{equation}
where $y=\q^2/k^2$ and $\alpha$ is a variational parameter determined by the principle of minimum sensitivity (PMS), that is by demanding that locally critical exponents be independent of $\alpha$, e.g. $d\nu/d\alpha|_{\alpha_{\rm opt}}=0$ (see Sec.~\ref{sec_frg:subsubsec_DE2}). The determination of the critical exponents $\nu$ and $\eta$ from the PMS is illustrated in Fig.~\ref{appDE:fig_PMS}. 

For all studied values of $N$ the anomalous dimension $\eta$ alternates around the most precise estimates reported in the literature (whenever these estimates are more precise than the DE) when one considers higher and higher orders of the DE. This also occurs for the exponent $\nu$ within LPA and DE$_2$ and, for $1\leq N\leq 5$ (but not for any $N$), also at order DE$_4$ (see Fig.~\ref{appDE:fig_PMS}). The alternating behavior is not seen between DE$_4$ and DE$_6$ when $N=1$ (the only case that has been considered at order DE$_6$) and is also absent for the exponent $\omega$. When the alternating behavior is observed, the PMS choice of $\alpha$ shows the fastest apparent convergence (in the sense that $\alpha_{\rm opt}$ is also the value obtained by minimizing the difference between two successive estimates of the critical exponent). As discussed in the next section, this observation can be exploited in order to improve the estimates and error bars of the exponents.

\subsubsection{Estimate of the error}

The small expansion parameter discussed in Sec.~\ref{sec_frg:validityDE}~\cite{Balog19} can be exploited in order to give a well-founded estimate of error bars in the DE~\cite{DePolsi20}. Since the radius of convergence of the DE is expected to be of the order of $4-9$ for $|\p_i\cdot\p_j|/k^2$ and, at the same time, the derivative $\partial_t R_k(\q)$ of the regulator function restricts the internal
momentum to the range ${\bf q}^2\lesssim k^2$, one expects successive orders of the expansion in momentum to be suppressed by a factor $1/9-1/4$ for a given family of regulators. As a consequence, for a quantity $Q$ a typical (somewhat pessimistic\footnote{Equation~(\ref{appDE:eq18}) can be too optimistic if $Q^{(s)}$ and $Q^{(s-2)}$ accidentally cross when varying a certain parameter, e.g. $N$ or $d$. In Ref.~\cite{DePolsi20} this issue is taken into account by estimating the error bar using a value of the parameter $N$ away from the accidental crossing.}) estimate of the error  is
\begin{equation}
\Delta Q^{(s)}=\frac{|Q^{(s)}-Q^{(s-2)}|}{4} ,
\label{appDE:eq18}
\end{equation}
where $Q^{(s)}$ is the value obtained from the PMS at order $s$ of the DE. When one considers several families of regulators, a natural estimate is given by the middle value. This introduces another potential error, in addition to~(\ref{appDE:eq18}), which can be estimated by $Q^{(s)}_{max}-Q^{(s)}_{min}$ where the maximum and minimum values are derived from the PMS values obtained among some reasonable families of regulators~\cite{DePolsi20}.

In some cases, there are strong reasons to believe that successive orders of the DE give strict (upper or lower) bounds to a given critical exponent (at least up to a given order). This  significantly improves the estimates of central values and error bars, as discussed in Ref.~\cite{DePolsi20}.

These error estimates, whether or not one can use the improved ones based on strict bounds, have been tested numerically in Ref.~\cite{DePolsi20}. For the O($N$) model up to order DE$_4$, with $1\leq N\leq 5$, and up to order DE$_6$ for $N=1$, the obtained values of the critical exponents, when compared with the most precise results available in the literature (when the latter are more precise than the DE), are always within the estimated error bars.

\subsection{Fourth- and sixth-order of the derivative expansion} 
\label{appDE:subsec_DE4}

The DE has been pushed to fourth order for the O($N$) model with $0\leq N\leq 4$~\cite{DePolsi20} and to sixth order for the $\mathbb{Z}_2$ (Ising) universality class~\cite{Balog19}. The ``strict'' DE has been used since it keeps equations to a more manageable size. The corresponding ansatz for the effective action reads 
\begin{align}
\nonumber\Gamma_k^{\partial^4} [\phi]={}& \int_\r \bigg\{  
U_k(\rho)+\frac{1}{2}Z_k(\rho)\big(\partial_\mu\phi^a\big)^2
+\frac{1}{4}Y_k(\rho)\big(\partial_\mu\rho\big)^2
+\frac{W_{1,k}(\rho)}{2}\big(\partial_\mu\partial_\nu \phi^a\big)^2  \nonumber
+\frac{W_{2,k}(\rho)}{2}\big(\phi^a \partial_\mu\partial_\nu \phi^a\big)^2 \\
& \nonumber + W_{3,k}(\rho)\partial_\mu\rho\partial_\nu\phi^a\partial_\mu\partial_\nu \phi^a
+\frac{W_{4,k}(\rho)}{2}
\phi^b\partial_\mu\phi^a\partial_\nu\phi^a\partial_\mu\partial_\nu \phi^b 
+ \frac{W_{5,k}(\rho)}{2}
\varphi^a\partial_\mu\rho\partial_\nu\rho\partial_\mu\partial_\nu 
\phi^a
+\frac{W_{6,k}(\rho)}{4}\Big(\big(\partial_\mu\phi^a\big)^2\Big)^2 \\
& +\frac{W_{7,k}(\rho)}{4}\big(\partial_\mu\phi^a\partial_\nu\phi^a\big)^2
+\frac{W_{8,k}(\rho)}{2}
\partial_\mu\phi^a\partial_\nu\phi^a\partial_\mu\rho\partial_\nu\rho
+\frac{W_{9,k}(\rho)}{2}\big(\partial_\mu\phi^a\big)^2 \big(\partial_\nu\rho\big)^2 
+\frac{W_{10,k}(\rho)}{4} \Big(\big(\partial_\mu \rho\big)^2\Big)^2  \bigg\} 
\end{align}
to fourth order~\cite{DePolsi20}, and 
\begin{multline}
\Gamma_k^{\partial^6,\mathbb{Z}_2}[\phi]= \int_\r \bigg\{U_k(\phi) + \tfrac{1}{2} 
Z_k(\phi) (\partial_\mu\phi)^2 
+ \tfrac{1}{2} W^a_k(\phi)(\partial_\mu\partial_\nu\phi)^2 + 
\tfrac{1}{2} \phi W^b_k (\phi)(\partial^2\phi)(\partial_\mu\phi)^2 
+ \tfrac{1}{2}  W^c_k 
(\phi)\left((\partial_\mu\phi)^2\right)^2 \\ + \tfrac{1}{2} \tilde X^a_k 
(\phi)(\partial_\mu\partial_\nu\partial_\rho\phi)^2 
+ \tfrac{1}{2} \phi \tilde X^b_k 
(\phi)(\partial_\mu\partial_\nu\phi)(\partial_\nu\partial_\rho\phi)(\partial_\mu
\partial_\rho\phi) 
+ \tfrac{1}{2}  \phi\tilde 
X^c_k(\phi)\left(\partial^2\phi\right)^3 
+\tfrac{1}{2}  \tilde X^d_k (\phi)\left(\partial^2\phi\right)^2 
(\partial_\mu\phi)^2 \\
+ \tfrac{1}{2} \tilde X^e_k (\phi)(\partial_\nu\phi)^2 
(\partial_\mu\phi)(\partial^2\partial_\mu\phi)
+ \tfrac{1}{2} \tilde X^f_k (\phi)(\partial_\rho\phi)^2 
(\partial_\mu\partial_\nu\phi)^2 
+ \tfrac{1}{2}  \phi\tilde X^g_k 
(\phi)\left(\partial^2\phi\right)\left((\partial_\mu\phi)^2\right)^2
+ \tfrac{1}{96} \tilde{X}^h_k (\phi) \left((\partial_\mu\phi)^2\right)^3 \bigg\} 
\end{multline}
to sixth order~\cite{Balog19}.

%% file: frg_review.bbl
\begin{thebibliography}{1000}
\expandafter\ifx\csname url\endcsname\relax
\def\url#1{\texttt{#1}}\fi
\expandafter\ifx\csname urlprefix\endcsname\relax\def\urlprefix{URL }\fi
\expandafter\ifx\csname href\endcsname\relax
\def\href#1#2{#2} \def\path#1{#1}\fi

\bibitem{Petermann:1953wpa}
E.~C.~G. Stueckelberg~de Breidenbach, A.~Petermann, {Normalization of constants
in the quanta theory}, Helv. Phys. Acta 26 (1953) 499--520.
\newblock \href {https://doi.org/10.5169/seals-112426}
{\path{doi:10.5169/seals-112426}}.

\bibitem{GellMann:1954fq}
M.~Gell-Mann, F.~Low, {Quantum electrodynamics at small distances}, Phys. Rev.
95 (1954) 1300--1312.
\newblock \href {https://doi.org/10.1103/PhysRev.95.1300}
{\path{doi:10.1103/PhysRev.95.1300}}.

\bibitem{Bogolyubov:1983gp}
N.~Bogolyubov, D.~Shirkov, {QUANTUM FIELDS}, 1983.

\bibitem{Symanzik:1970rt}
K.~Symanzik, {Small distance behavior in field theory and power counting},
Commun. Math. Phys. 18 (1970) 227--246.
\newblock \href {https://doi.org/10.1007/BF01649434}
{\path{doi:10.1007/BF01649434}}.

\bibitem{Callan:1970yg}
J.~Callan, Curtis~G., {Broken scale invariance in scalar field theory}, Phys.
Rev. D 2 (1970) 1541--1547.
\newblock \href {https://doi.org/10.1103/PhysRevD.2.1541}
{\path{doi:10.1103/PhysRevD.2.1541}}.

\bibitem{Kadanoff66}
L.~P. Kadanoff, Scaling laws for ising models near ${T}_{c}$, Physics 2 (1966)
263--272.
\newblock \href {https://doi.org/10.1103/PhysicsPhysiqueFizika.2.263}
{\path{doi:10.1103/PhysicsPhysiqueFizika.2.263}}.

\bibitem{Wilson71}
K.~G. Wilson, {Renormalization Group and Critical Phenomena. I. Renormalization
Group and the Kadanoff Scaling Picture}, Phys. Rev. B 4 (1971) 3174--3183.
\newblock \href {https://doi.org/10.1103/PhysRevB.4.3174}
{\path{doi:10.1103/PhysRevB.4.3174}}.

\bibitem{Wilson71a}
K.~G. Wilson, {Renormalization Group and Critical Phenomena. II. Phase-Space
Cell Analysis of Critical Behavior}, Phys. Rev. B 4 (1971) 3184--3205.
\newblock \href {https://doi.org/10.1103/PhysRevB.4.3184}
{\path{doi:10.1103/PhysRevB.4.3184}}.

\bibitem{Wilson74}
K.~G. Wilson, J.~B. Kogut, {The renormalization group and the $\epsilon$
expansion}, Phys. Rep. 12 (1974) 75.
\newblock \href {https://doi.org/doi:10.1016/0370-1573(74)90023-4}
{\path{doi:doi:10.1016/0370-1573(74)90023-4}}.

\bibitem{Wegner73}
F.~J. Wegner, A.~Houghton, {Renormalization Group Equation for Critical
Phenomena}, Phys. Rev. A 8 (1973) 401--412.
\newblock \href {https://doi.org/10.1103/PhysRevA.8.401}
{\path{doi:10.1103/PhysRevA.8.401}}.

\bibitem{Polchinski84}
J.~Polchinski, {Renormalization and effective Lagrangians}, Nucl. Phys. B 231
(1984) 269.
\newblock \href {https://doi.org/doi:10.1016/0550-3213(84)90287-6}
{\path{doi:doi:10.1016/0550-3213(84)90287-6}}.

\bibitem{Fisher98}
M.~Fisher, {Renormalization group theory: Its basis and formulation in
statistical physics}, Rev. Mod. Phys. 70 (1998) 653--681.
\newblock \href {https://doi.org/10.1103/RevModPhys.70.653}
{\path{doi:10.1103/RevModPhys.70.653}}.

\bibitem{Bagnuls01}
C.~Bagnuls, C.~Bervillier, {Exact renormalization equations: an introductory
review}, Phys. Rep. 348 (2001) 91.
\newblock \href {https://doi.org/doi:10.1016/S0370-1573(00)00137-X}
{\path{doi:doi:10.1016/S0370-1573(00)00137-X}}.

\bibitem{Guida98}
R.~Guida, J.~Zinn-Justin, {Critical exponents of the $N$-vector model}, J.
Phys. A 31~(40) (1998) 8103.
\newblock \href {https://doi.org/doi:10.1088/0305-4470/31/40/006}
{\path{doi:doi:10.1088/0305-4470/31/40/006}}.

\bibitem{Pelissetto02}
A.~Pelissetto, E.~Vicari, {Critical phenomena and renormalization-group
theory}, Phys. Rep. 368 (2002) 549.
\newblock \href
{https://doi.org/http://dx.doi.org/10.1016/S0370-1573(02)00219-3}
{\path{doi:http://dx.doi.org/10.1016/S0370-1573(02)00219-3}}.

\bibitem{Gawedzki1985}
K.~Gawedzki, A.~Kupiainen, Massless lattice $\varphi^4_4$ theory: Rigorous
control of a renormalizable asymptotically free model, Communications in
Mathematical Physics 99~(2) (1985) 197--252.
\newblock \href {https://doi.org/10.1007/BF01212281}
{\path{doi:10.1007/BF01212281}}.

\bibitem{Balaban:1988rr}
T.~Balaban, {Convergent Renormalization Expansions for Lattice Gauge Theories},
Commun. Math. Phys. 119 (1988) 243--285.
\newblock \href {https://doi.org/10.1007/BF01217741}
{\path{doi:10.1007/BF01217741}}.

\bibitem{Brydges1987}
D.~C. Brydges, T.~Kennedy, Mayer expansions and the hamilton-jacobi equation,
Journal of Statistical Physics 48~(1) (1987) 19--49.
\newblock \href {https://doi.org/10.1007/BF01010398}
{\path{doi:10.1007/BF01010398}}.

\bibitem{Feldman:1987zq}
J.~Feldman, J.~Magnen, V.~Rivasseau, R.~Seneor, {Construction and Borel
Summability of Infrared $\phi^4$ in Four-dimensions by a Phase Space
Expansion}, Commun. Math. Phys. 109 (1987) 437--480.
\newblock \href {https://doi.org/10.1007/BF01206146}
{\path{doi:10.1007/BF01206146}}.

\bibitem{rivasseau_book}
V.~Rivasseau, From perturbative to constructive renormalization, 2nd Edition,
Princeton Series in Physics, Princeton Univ Pr, 1991.

\bibitem{Wilson:1974mb}
K.~G. Wilson, {The Renormalization Group: Critical Phenomena and the Kondo
Problem}, Rev. Mod. Phys. 47 (1975) 773.
\newblock \href {https://doi.org/10.1103/RevModPhys.47.773}
{\path{doi:10.1103/RevModPhys.47.773}}.

\bibitem{Schollwock:2005zz}
U.~Schollw{\"o}ck, {The density-matrix renormalization group}, Rev. Mod. Phys.
77 (2005) 259--315.
\newblock \href {http://arxiv.org/abs/cond-mat/0409292}
{\path{arXiv:cond-mat/0409292}}, \href
{https://doi.org/10.1103/RevModPhys.77.259}
{\path{doi:10.1103/RevModPhys.77.259}}.

\bibitem{Wilson72}
K.~G. Wilson, M.~E. Fisher, {Critical Exponents in 3.99 Dimensions}, Phys. Rev.
Lett. 28 (1972) 240--243.
\newblock \href {https://doi.org/10.1103/PhysRevLett.28.240}
{\path{doi:10.1103/PhysRevLett.28.240}}.

\bibitem{Migdal75}
A.~A. Migdal, {Phase transitions in gauge and spin-lattice systems}, JETP 42
(1975) 743.

\bibitem{Polyakov75}
A.~M. Polyakov, {Interaction of goldstone particles in two dimensions.
Applications to ferromagnets and massive Yang-Mills fields}, Phys. Lett. B 59
(1975) 79.
\newblock \href
{https://doi.org/http://dx.doi.org/10.1016/0370-2693(75)90161-6}
{\path{doi:http://dx.doi.org/10.1016/0370-2693(75)90161-6}}.

\bibitem{Brezin76a}
E.~Br\'ezin, J.~Zinn-Justin, {Renormalization of the Nonlinear $\sigma{{}}$
Model in $2+\epsilon{{}}$ Dimensions---Application to the Heisenberg
Ferromagnets}, Phys. Rev. Lett. 36 (1976) 691.
\newblock \href {https://doi.org/10.1103/PhysRevLett.36.691}
{\path{doi:10.1103/PhysRevLett.36.691}}.

\bibitem{Brezin76b}
E.~Br\'ezin, J.~Zinn-Justin, {Spontaneous breakdown of continuous symmetries
near two dimensions}, Phys. Rev. B 14~(7) (1976) 3110.
\newblock \href {https://doi.org/10.1103/PhysRevB.14.3110}
{\path{doi:10.1103/PhysRevB.14.3110}}.

\bibitem{Nelson77}
D.~R. Nelson, R.~A. Pelcovits, {Momentum-shell recursion relations, anisotropic
spins, and liquid crystals in $2+\epsilon$ dimensions}, Phys. Rev. B 16
(1977) 2191.
\newblock \href {https://doi.org/10.1103/PhysRevB.16.2191}
{\path{doi:10.1103/PhysRevB.16.2191}}.

\bibitem{Moshe03}
M.~Moshe, J.~Zinn-Justin, {Quantum field theory in the large $N$ limit: a
review}, Phys. Rep. 385 (2003) 69.
\newblock \href {https://doi.org/https://doi.org/10.1016/S0370-1573(03)00263-1}
{\path{doi:https://doi.org/10.1016/S0370-1573(03)00263-1}}.

\bibitem{Berezinskii71}
V.~L. Berezinskii,
\href{http://www.jetp.ac.ru/cgi-bin/e/index/e/32/3/p493?a=list}{Destruction
of long-range order in one-dimensional and two-dimensional systems having a
continuous symmetry group i. classical systems}, Sov. Phys. JETP 32 (1971)
493.
\newline\urlprefix\url{http://www.jetp.ac.ru/cgi-bin/e/index/e/32/3/p493?a=list}

\bibitem{Berezinskii72}
V.~L. Berezinskii,
\href{http://www.jetp.ac.ru/cgi-bin/e/index/e/34/3/p610?a=list}{Destruction
of long-range order in one-dimensional and two-dimensional systems possessing
a continuous symmetry group. ii. quantum systems}, Sov. Phys. JETP 34 (1972)
610.
\newline\urlprefix\url{http://www.jetp.ac.ru/cgi-bin/e/index/e/34/3/p610?a=list}

\bibitem{Kosterlitz73}
J.~M. Kosterlitz, D.~J. Thouless, {Ordering, metastability and phase
transitions in two-dimensional systems}, J. of Phys. C 6 (1973) 1181.
\newblock \href {https://doi.org/doi:10.1088/0022-3719/6/7/010}
{\path{doi:doi:10.1088/0022-3719/6/7/010}}.

\bibitem{Kosterlitz74}
J.~M. Kosterlitz, D.~J. Thouless, {The critical properties of the
two-dimensional XY model}, J. Phys. C 7 (1974) 1046.
\newblock \href {https://doi.org/doi:10.1088/0022-3719/7/6/005}
{\path{doi:doi:10.1088/0022-3719/7/6/005}}.

\bibitem{Chaikin_book}
P.~M. Chaikin, T.~C. Lubensky, {Principles of Condensed Matter Physics},
Cambridge University Press, 1995.

\bibitem{Hasenfratz86}
A.~Hasenfratz, P.~Hasenfratz, {Renormalization group study of scalar field
theories}, Nucl. Phys. B 270 (1986) 687--701.
\newblock \href {https://doi.org/https://doi.org/10.1016/0550-3213(86)90573-0}
{\path{doi:https://doi.org/10.1016/0550-3213(86)90573-0}}.

\bibitem{Chang92}
T.~Chang, D.~Vvedensky, J.~Nicoll,
\href{http://www.sciencedirect.com/science/article/pii/037015739290041W}{Differential
renormalization-group generators for static and dynamic critical phenomena},
Phys. Rep. 217 (1992) 279 -- 360.
\newblock \href {https://doi.org/https://doi.org/10.1016/0370-1573(92)90041-W}
{\path{doi:https://doi.org/10.1016/0370-1573(92)90041-W}}.
\newline\urlprefix\url{http://www.sciencedirect.com/science/article/pii/037015739290041W}

\bibitem{Parola95}
A.~Parola, L.~Reatto, {Liquid state theory and critical phenomena}, Adv. Phys.
44 (1995) 211.
\newblock \href {https://doi.org/10.1080/00018739500101536}
{\path{doi:10.1080/00018739500101536}}.

\bibitem{Nicoll74}
J.~F. Nicoll, T.~S. Chang, H.~E. Stanley, {Approximate Renormalization Group
Based on the Wegner-Houghton Differential Generator}, Phys. Rev. Lett. 33
(1974) 540.
\newblock \href {https://doi.org/10.1103/PhysRevLett.33.540}
{\path{doi:10.1103/PhysRevLett.33.540}}.

\bibitem{Nicoll77}
J.~Nicoll, T.~Chang, An exact one-particle-irreducible renormalization-group
generator for critical phenomena, Phys. Lett. A 62~(5) (1977) 287 -- 289.
\newblock \href {https://doi.org/https://doi.org/10.1016/0375-9601(77)90417-0}
{\path{doi:https://doi.org/10.1016/0375-9601(77)90417-0}}.

\bibitem{Parola84}
A.~Parola, L.~Reatto, {Liquid-State Theory for Critical Phenomena}, Phys. Rev.
Lett. 53 (1984) 2417.
\newblock \href {https://doi.org/10.1103/PhysRevLett.53.2417}
{\path{doi:10.1103/PhysRevLett.53.2417}}.

\bibitem{Nicoll76}
J.~F. Nicoll, T.~S. Chang, H.~E. Stanley, {Exact and approximate differential
renormalization-group generators}, Phys. Rev. A 13 (1976) 1251--1264.
\newblock \href {https://doi.org/10.1103/PhysRevA.13.1251}
{\path{doi:10.1103/PhysRevA.13.1251}}.

\bibitem{Newman84}
K.~E. Newman, E.~K. Riedel, Critical exponents by the scaling-field method: The
isotropic $n$-vector model in three dimensions, Phys. Rev. B 30 (1984)
6615--6638.
\newblock \href {https://doi.org/10.1103/PhysRevB.30.6615}
{\path{doi:10.1103/PhysRevB.30.6615}}.

\bibitem{Newman84a}
K.~E. Newman, E.~K. Riedel, S.~Muto, $q$-state potts model by wilson's exact
renormalization-group equation, Phys. Rev. B 29 (1984) 302--313.
\newblock \href {https://doi.org/10.1103/PhysRevB.29.302}
{\path{doi:10.1103/PhysRevB.29.302}}.

\bibitem{Golner86}
G.~R. Golner, Nonperturbative renormalization-group calculations for continuum
spin systems, Phys. Rev. B 33 (1986) 7863--7866.
\newblock \href {https://doi.org/10.1103/PhysRevB.33.7863}
{\path{doi:10.1103/PhysRevB.33.7863}}.

\bibitem{Hasenfratz88}
P.~Hasenfratz, J.~Nager, {The cut-off dependence of the Higgs meson mass and
the onset of new physics in the standard model}, Z. Phys. C 37 (1988)
477--487.
\newblock \href {https://doi.org/10.1007/BF01578143}
{\path{doi:10.1007/BF01578143}}.

\bibitem{Zumbach93}
G.~Zumbach, {Almost second order phase transitions}, Phys. Rev. Lett. 71 (1993)
2421--2424.
\newblock \href {https://doi.org/10.1103/PhysRevLett.71.2421}
{\path{doi:10.1103/PhysRevLett.71.2421}}.

\bibitem{Zumbach94a}
G.~Zumbach, {Phase transitions with O(n) symmetry broken down to O(n-p)}, Nucl.
Phys. B 413 (1994) 771--791.
\newblock \href
{https://doi.org/http://dx.doi.org/10.1016/0550-3213(94)90012-4}
{\path{doi:http://dx.doi.org/10.1016/0550-3213(94)90012-4}}.

\bibitem{Zumbach94b}
G.~Zumbach, {The local potential approximation of the renormalization group and
its applications}, Phys. Lett. A 190 (1994) 225--230.
\newblock \href
{https://doi.org/http://dx.doi.org/10.1016/0375-9601(94)90746-3}
{\path{doi:http://dx.doi.org/10.1016/0375-9601(94)90746-3}}.

\bibitem{Fisher85}
D.~S. Fisher, {Random fields, random anisotropies, nonlinear
\ensuremath{\sigma} models, and dimensional reduction}, Phys. Rev. B 31
(1985) 7233.
\newblock \href {https://doi.org/10.1103/PhysRevB.31.7233}
{\path{doi:10.1103/PhysRevB.31.7233}}.

\bibitem{Narayan92}
O.~Narayan, D.~S. Fisher, {Dynamics of sliding charge-density waves in
4-\ensuremath{\epsilon} dimensions}, Phys. Rev. Lett. 68 (1992) 3615.
\newblock \href {https://doi.org/10.1103/PhysRevLett.68.3615}
{\path{doi:10.1103/PhysRevLett.68.3615}}.

\bibitem{Nattermann92}
{T. Nattermann}, {S. Stepanow}, {L.-H. Tang}, {H. Leschhorn}, {Dynamics of
interface depinning in a disordered medium}, J. Phys. II France 2~(8) (1992)
1483.
\newblock \href {https://doi.org/10.1051/jp2:1992214}
{\path{doi:10.1051/jp2:1992214}}.

\bibitem{Chauve01}
P.~Chauve, P.~L. Doussal, K.~J. Wiese, {Renormalization of Pinned Elastic
Systems: How Does It Work Beyond One Loop?}, Phys. Rev. Lett. 86~(9) (2001)
1785.
\newblock \href {https://doi.org/10.1103/physrevlett.86.1785}
{\path{doi:10.1103/physrevlett.86.1785}}.

\bibitem{Ledoussal04}
P.~Le~Doussal, K.~J. Wiese, P.~Chauve, {Functional renormalization group and
the field theory of disordered elastic systems}, Phys. Rev. E 69 (2004)
026112.
\newblock \href {https://doi.org/10.1103/PhysRevE.69.026112}
{\path{doi:10.1103/PhysRevE.69.026112}}.

\bibitem{Tarjus04}
G.~Tarjus, M.~Tissier, {Nonperturbative Functional Renormalization Group for
Random-Field Models: The Way Out of Dimensional Reduction}, Phys. Rev. Lett.
93 (2004) 267008.
\newblock \href {https://doi.org/10.1103/PhysRevLett.93.267008}
{\path{doi:10.1103/PhysRevLett.93.267008}}.

\bibitem{Ringwald90}
A.~Ringwald, C.~Wetterich, {Average action for the $N$-component $\varphi^4$
theory}, Nucl. Phys. B 334 (1990) 506.
\newblock \href {https://doi.org/doi:10.1016/0550-3213(90)90489-Z}
{\path{doi:doi:10.1016/0550-3213(90)90489-Z}}.

\bibitem{Wetterich91}
C.~Wetterich, {Average action and the renormalization group equations}, Nucl.
Phys. B 352 (1991) 529.
\newblock \href {https://doi.org/doi:10.1016/0550-3213(91)90099-J}
{\path{doi:doi:10.1016/0550-3213(91)90099-J}}.

\bibitem{Wetterich93a}
C.~Wetterich, {The average action for scalar fields near phase transitions}, Z.
Phys. C 57 (1993) 451--469.
\newblock \href {https://doi.org/10.1007/BF01474340}
{\path{doi:10.1007/BF01474340}}.

\bibitem{Wetterich93b}
C.~Wetterich, {Improvement of the average action}, Z. Phys. C 60 (1993)
461–469.
\newblock \href {https://doi.org/10.1007/BF01560044}
{\path{doi:10.1007/BF01560044}}.

\bibitem{Wetterich93}
C.~Wetterich, {Exact evolution equation for the effective potential}, Phys.
Lett. B 301 (1993) 90.
\newblock \href {https://doi.org/doi:10.1016/0370-2693(93)90726-X}
{\path{doi:doi:10.1016/0370-2693(93)90726-X}}.

\bibitem{Berges:2000ew}
J.~Berges, N.~Tetradis, C.~Wetterich, {Non-perturbative renormalization flow in
quantum field theory and statistical physics}, Phys. Rep. 363 (2002)
223--386.
\newblock \href {http://arxiv.org/abs/hep-ph/0005122}
{\path{arXiv:hep-ph/0005122}}, \href
{https://doi.org/10.1016/S0370-1573(01)00098-9}
{\path{doi:10.1016/S0370-1573(01)00098-9}}.

\bibitem{Ellwanger94}
U.~Ellwanger, Flow equations for $n$ point functions and bound states, Z. Phys.
C 62 (1994) 503.
\newblock \href {https://doi.org/10.1007/BF01555911}
{\path{doi:10.1007/BF01555911}}.

\bibitem{Morris94}
T.~R. Morris, {The exact renormalization group and approximate solutions}, Int.
J. Mod. Phys. A 09 (1994) 2411.
\newblock \href {https://doi.org/10.1142/S0217751X94000972}
{\path{doi:10.1142/S0217751X94000972}}.

\bibitem{Bonini93}
M.~Bonini, M.~D'Attanasio, G.~Marchesini, {Perturbative renormalization and
infrared finiteness in the Wilson renormalization group: the massless scalar
case}, Nucl. Phys. B 409 (1993) 441 -- 464.
\newblock \href {https://doi.org/https://doi.org/10.1016/0550-3213(93)90588-G}
{\path{doi:https://doi.org/10.1016/0550-3213(93)90588-G}}.

\bibitem{Aoki00}
K.~I. Aoki, Introduction to the non-perturbative renormalization group and its
recent applications, Int. J. Mod. Phys. B 14 (2000) 1249.
\newblock \href {https://doi.org/10.1142/S0217979200000923}
{\path{doi:10.1142/S0217979200000923}}.

\bibitem{Polonyi:2001se}
J.~Polonyi, {Lectures on the functional renormalization group method}, Central
Eur.J.Phys. 1 (2003) 1--71.
\newblock \href {http://arxiv.org/abs/hep-th/0110026}
{\path{arXiv:hep-th/0110026}}, \href {https://doi.org/10.2478/BF02475552}
{\path{doi:10.2478/BF02475552}}.

\bibitem{Delamotte04}
B.~Delamotte, D.~Mouhanna, M.~Tissier, {Nonperturbative renormalization-group
approach to frustrated magnets}, Phys. Rev. B 69 (2004) 134413.
\newblock \href {https://doi.org/10.1103/PhysRevB.69.134413}
{\path{doi:10.1103/PhysRevB.69.134413}}.

\bibitem{Pawlowski:2005xe}
J.~M. Pawlowski, {Aspects of the functional renormalisation group}, Annals
Phys. 322 (2007) 2831--2915.
\newblock \href {http://arxiv.org/abs/hep-th/0512261}
{\path{arXiv:hep-th/0512261}}, \href
{https://doi.org/10.1016/j.aop.2007.01.007}
{\path{doi:10.1016/j.aop.2007.01.007}}.

\bibitem{Rosten:2010vm}
O.~J. Rosten, {Fundamentals of the Exact Renormalization Group}, Phys. Rept.
511 (2012) 177--272.
\newblock \href {http://arxiv.org/abs/1003.1366} {\path{arXiv:1003.1366}},
\href {https://doi.org/10.1016/j.physrep.2011.12.003}
{\path{doi:10.1016/j.physrep.2011.12.003}}.

\bibitem{Kopietz_book}
P.~Kopietz, L.~Bartosch, F.~Sch\"utz, {Introduction to the Functional
Renormalization Group}, Springer, Berlin, 2010.
\newblock \href {https://doi.org/10.1007/978-3-642-05094-7}
{\path{doi:10.1007/978-3-642-05094-7}}.

\bibitem{Braun:2011pp}
J.~Braun, {Fermion Interactions and Universal Behavior in Strongly Interacting
Theories}, J.Phys. G39 (2012) 033001.
\newblock \href {http://arxiv.org/abs/1108.4449} {\path{arXiv:1108.4449}},
\href {https://doi.org/10.1088/0954-3899/39/3/033001}
{\path{doi:10.1088/0954-3899/39/3/033001}}.

\bibitem{Delamotte12}
B.~Delamotte, {An Introduction to the Nonperturbative Renormalization Group},
in: A.~Schwenk, J.~Polonyi (Eds.), {Renormalization Group and Effective Field
Theory Approaches to Many-Body Systems}, Vol. 852 of Lecture Notes in
Physics, Springer Berlin Heidelberg, 2012, pp. 49--132.
\newblock \href {https://doi.org/10.1007/978-3-642-27320-9_2}
{\path{doi:10.1007/978-3-642-27320-9_2}}.

\bibitem{Gies12}
H.~Gies, \href{https://doi.org/10.1007/978-3-642-27320-9_6}{{Introduction to
the Functional RG and Applications to Gauge Theories}}, Springer Berlin
Heidelberg, Berlin, Heidelberg, 2012, pp. 287--348.
\newblock \href {https://doi.org/10.1007/978-3-642-27320-9_6}
{\path{doi:10.1007/978-3-642-27320-9_6}}.
\newline\urlprefix\url{https://doi.org/10.1007/978-3-642-27320-9_6}

\bibitem{Zinn_book}
J.~Zinn-Justin, {Quantum Field Theory and Critical Phenomena}, Fourth Edition,
Clarendon Press, Oxford, 2002.

\bibitem{Litim01}
D.~F. Litim, {Optimized renormalization group flows}, Phys. Rev. D 64 (2001)
105007.
\newblock \href {https://doi.org/10.1103/PhysRevD.64.105007}
{\path{doi:10.1103/PhysRevD.64.105007}}.

\bibitem{Morris05}
T.~R. Morris, {Equivalence of local potential approximations}, J. High Energy
Phys. 07 (2005) 027.
\newblock \href {https://doi.org/10.1088/1126-6708/2005/07/027}
{\path{doi:10.1088/1126-6708/2005/07/027}}.

\bibitem{Papenbrock95}
T.~Papenbrock, C.~Wetterich, Two-loop results from improved one loop
computations, Z. Phys. C 65 (1995) 519.
\newblock \href {https://doi.org/10.1007/BF01556140}
{\path{doi:10.1007/BF01556140}}.

\bibitem{Bonini97}
M.~Bonini, G.~Marchesini, M.~Simionato, Beta function and flowing couplings in
the exact wilson renormalization group in yang-mills theory, Nucl. Phys. B
483 (1997) 475 -- 494.
\newblock \href {https://doi.org/https://doi.org/10.1016/S0550-3213(96)00571-8}
{\path{doi:https://doi.org/10.1016/S0550-3213(96)00571-8}}.

\bibitem{Morris99}
T.~R. Morris, J.~F. Tighe, {Convergence of derivative expansions of the
renormalization group}, J. High Energy Phys. 08 (1999) 007.
\newblock \href {https://doi.org/10.1088/1126-6708/1999/08/007}
{\path{doi:10.1088/1126-6708/1999/08/007}}.

\bibitem{Kopietz:2000bh}
P.~Kopietz, {Two loop beta function from the exact renormalization group},
Nucl. Phys. B 595 (2001) 493--518.
\newblock \href {http://arxiv.org/abs/hep-th/0007128}
{\path{arXiv:hep-th/0007128}}, \href
{https://doi.org/10.1016/S0550-3213(00)00680-5}
{\path{doi:10.1016/S0550-3213(00)00680-5}}.

\bibitem{Latorre:2000qc}
J.~I. Latorre, T.~R. Morris, {Exact scheme independence}, JHEP 11 (2000) 004.
\newblock \href {http://arxiv.org/abs/hep-th/0008123}
{\path{arXiv:hep-th/0008123}}, \href
{https://doi.org/10.1088/1126-6708/2000/11/004}
{\path{doi:10.1088/1126-6708/2000/11/004}}.

\bibitem{Latorre01}
J.~I. Latorre, T.~R. Morris, {Scheme Independence as an Inherent Redundancy in
Quantum Field Theory}, Int. J. Mod. Phys. A16 (2001) 2071--2074, [,72(2002)].
\newblock \href {http://arxiv.org/abs/hep-th/0102037}
{\path{arXiv:hep-th/0102037}}, \href
{https://doi.org/10.1142/S0217751X01004724}
{\path{doi:10.1142/S0217751X01004724}}.

\bibitem{Arnone:2002yh}
S.~Arnone, A.~Gatti, T.~R. Morris, {Exact scheme independence at one loop},
JHEP 05 (2002) 059.
\newblock \href {http://arxiv.org/abs/hep-th/0201237}
{\path{arXiv:hep-th/0201237}}, \href
{https://doi.org/10.1088/1126-6708/2002/05/059}
{\path{doi:10.1088/1126-6708/2002/05/059}}.

\bibitem{Arnone:2003pa}
S.~Arnone, A.~Gatti, T.~R. Morris, O.~J. Rosten, {Exact scheme independence at
two loops}, Phys. Rev. D69 (2004) 065009.
\newblock \href {http://arxiv.org/abs/hep-th/0309242}
{\path{arXiv:hep-th/0309242}}, \href
{https://doi.org/10.1103/PhysRevD.69.065009}
{\path{doi:10.1103/PhysRevD.69.065009}}.

\bibitem{Litim:1996nw}
D.~F. Litim, {Scheme independence at first order phase transitions and the
renormalization group}, Phys.Lett. B393 (1997) 103--109.
\newblock \href {http://arxiv.org/abs/hep-th/9609040}
{\path{arXiv:hep-th/9609040}}, \href
{https://doi.org/10.1016/S0370-2693(96)01613-9}
{\path{doi:10.1016/S0370-2693(96)01613-9}}.

\bibitem{Pernici:1997ie}
M.~Pernici, M.~Raciti, F.~Riva, {Hard - soft renormalization and the exact
renormalization group}, Nucl. Phys. B520 (1998) 469--492.
\newblock \href {http://arxiv.org/abs/hep-th/9710145}
{\path{arXiv:hep-th/9710145}}, \href
{https://doi.org/10.1016/S0550-3213(98)00176-X}
{\path{doi:10.1016/S0550-3213(98)00176-X}}.

\bibitem{Ellwanger:1997tp}
U.~Ellwanger, {The Running gauge coupling in the exact renormalization group
approach}, Z. Phys. C76 (1997) 721--727.
\newblock \href {http://arxiv.org/abs/hep-ph/9702309}
{\path{arXiv:hep-ph/9702309}}, \href {https://doi.org/10.1007/s002880050593}
{\path{doi:10.1007/s002880050593}}.

\bibitem{Pernici:1998tp}
M.~Pernici, M.~Raciti, {Wilsonian flow and mass independent renormalization},
Nucl. Phys. B531 (1998) 560--592.
\newblock \href {http://arxiv.org/abs/hep-th/9803212}
{\path{arXiv:hep-th/9803212}}, \href
{https://doi.org/10.1016/S0550-3213(98)80007-2}
{\path{doi:10.1016/S0550-3213(98)80007-2}}.

\bibitem{Rosten:2006pd}
O.~J. Rosten, {Universality From Very General Nonperturbative Flow Equations in
QCD}, Phys. Lett. B645 (2007) 466--469.
\newblock \href {http://arxiv.org/abs/hep-th/0611323}
{\path{arXiv:hep-th/0611323}}, \href
{https://doi.org/10.1016/j.physletb.2006.12.057}
{\path{doi:10.1016/j.physletb.2006.12.057}}.

\bibitem{Codello:2013bra}
A.~Codello, M.~Demmel, O.~Zanusso, {Scheme dependence and universality in the
functional renormalization group}, Phys. Rev. D90~(2) (2014) 027701.
\newblock \href {http://arxiv.org/abs/1310.7625} {\path{arXiv:1310.7625}},
\href {https://doi.org/10.1103/PhysRevD.90.027701}
{\path{doi:10.1103/PhysRevD.90.027701}}.

\bibitem{Schnoerr13}
D.~Schnoerr, I.~Boettcher, J.~M. Pawlowski, C.~Wetterich, {Error estimates and
specification parameters for functional renormalization }, Ann. Phys. 334
(2013) 83--99.
\newblock \href {https://doi.org/http://dx.doi.org/10.1016/j.aop.2013.03.013}
{\path{doi:http://dx.doi.org/10.1016/j.aop.2013.03.013}}.

\bibitem{Reuter93}
M.~Reuter, N.~Tetradis, C.~Wetterich, {The large-N limit and the
high-temperature phase transition for the $\phi^4$ theory}, Nucl. Phys. B 401
(1993) 567--590.
\newblock \href {https://doi.org/10.1016/0550-3213(93)90314-F}
{\path{doi:10.1016/0550-3213(93)90314-F}}.

\bibitem{Morris94b}
T.~R. Morris, {On truncations of the exact renormalization group }, Phys. Lett.
B 334 (1994) 355.
\newblock \href {https://doi.org/10.1016/0370-2693(94)90700-5}
{\path{doi:10.1016/0370-2693(94)90700-5}}.

\bibitem{Tetradis92}
N.~Tetradis, C.~Wetterich, {Scale dependence of the average potential around
the maximum in $\phi^4$ theory.}, Nucl. Phys. B 383 (1992) 197.
\newblock \href {https://doi.org/doi:10.1016/0550-3213(92)90676-3}
{\path{doi:doi:10.1016/0550-3213(92)90676-3}}.

\bibitem{Tetradis96}
N.~Tetradis, D.~Litim, {Analytical solutions of exact renormalization group
equations }, Nucl. Phys. B 464 (1996) 492--511.
\newblock \href
{https://doi.org/http://dx.doi.org/10.1016/0550-3213(95)00642-7}
{\path{doi:http://dx.doi.org/10.1016/0550-3213(95)00642-7}}.

\bibitem{Pelaez16}
M.~Pel\'aez, N.~Wschebor, {Ordered phase of the $\mathrm{O}(N)$ model within
the nonperturbative renormalization group}, Phys. Rev. E 94 (2016) 042136.
\newblock \href {https://doi.org/10.1103/PhysRevE.94.042136}
{\path{doi:10.1103/PhysRevE.94.042136}}.

\bibitem{Litim:2006nn}
D.~F. Litim, J.~M. Pawlowski, L.~Vergara, {Convexity of the effective action
from functional flows} (2006).
\newblock \href {http://arxiv.org/abs/hep-th/0602140}
{\path{arXiv:hep-th/0602140}}.

\bibitem{Morris94a}
T.~R. Morris, {Derivative expansion of the exact renormalization group }, Phys.
Lett. B 329 (1994) 241.
\newblock \href
{https://doi.org/http://dx.doi.org/10.1016/0370-2693(94)90767-6}
{\path{doi:http://dx.doi.org/10.1016/0370-2693(94)90767-6}}.

\bibitem{Ball95}
R.~D. Ball, P.~E. Haagensen, J.~I. Latorre, E.~Moreno, {Scheme independence and
the exact renormalization group}, Phys. Lett. B 347 (1995) 80.
\newblock \href {https://doi.org/doi:10.1016/0370-2693(95)00025-G}
{\path{doi:doi:10.1016/0370-2693(95)00025-G}}.

\bibitem{Comellas98}
J.~Comellas, {Polchinski equation, reparameterization invariance and the
derivative e xpansion }, Nucl. Phys. B 509 (1998) 662--686.
\newblock \href
{https://doi.org/http://dx.doi.org/10.1016/S0550-3213(97)00692-5}
{\path{doi:http://dx.doi.org/10.1016/S0550-3213(97)00692-5}}.

\bibitem{Morris98}
T.~R. Morris, M.~D. Turner, {Derivative expansion of the renormalization group
in O(N) scalar field theory }, Nucl. Phys. B 509 (1998) 637--661.
\newblock \href
{https://doi.org/http://dx.doi.org/10.1016/S0550-3213(97)00640-8}
{\path{doi:http://dx.doi.org/10.1016/S0550-3213(97)00640-8}}.

\bibitem{Zumbach94}
G.~Zumbach, {The renormalization group in the local potential approximation and
its applications to the O (n) model}, Nucl. Phys. B 413 (1994) 754 -- 770.
\newblock \href
{https://doi.org/http://dx.doi.org/10.1016/0550-3213(94)90011-6}
{\path{doi:http://dx.doi.org/10.1016/0550-3213(94)90011-6}}.

\bibitem{Defenu15}
N.~Defenu, P.~Mati, I.~G. M{\'a}ri{\'a}n, I.~N{\'a}ndori, A.~Trombettoni,
Truncation effects in the functional renormalization group study of
spontaneous symmetry breaking, J. High Energy Phys. 05 (2015) 141.
\newblock \href {https://doi.org/10.1007/JHEP05(2015)141}
{\path{doi:10.1007/JHEP05(2015)141}}.

\bibitem{Mermin66}
N.~D. Mermin, H.~Wagner, {Absence of Ferromagnetism or Antiferromagnetism in
One- or Two-Dimensional Isotropic Heisenberg Models}, Phys. Rev. Lett. 17
(1966) 1133--1136.
\newblock \href {https://doi.org/10.1103/PhysRevLett.17.1133}
{\path{doi:10.1103/PhysRevLett.17.1133}}.

\bibitem{Hohenberg67}
P.~C. Hohenberg, Existence of long-range order in one and two dimensions, Phys.
Rev. 158 (1967) 383--386.
\newblock \href {https://doi.org/10.1103/PhysRev.158.383}
{\path{doi:10.1103/PhysRev.158.383}}.

\bibitem{Coleman73}
S.~Coleman, There are no goldstone bosons in two dimensions, Commun. Math.
Phys. 31 (1973) 259--264.
\newblock \href {https://doi.org/10.1007/BF01646487}
{\path{doi:10.1007/BF01646487}}.

\bibitem{Labus:2015ska}
P.~Labus, R.~Percacci, G.~P. Vacca, {Asymptotic safety in $O(N)$ scalar models
coupled to gravity}, Phys. Lett. B753 (2016) 274--281.
\newblock \href {http://arxiv.org/abs/1505.05393} {\path{arXiv:1505.05393}},
\href {https://doi.org/10.1016/j.physletb.2015.12.022}
{\path{doi:10.1016/j.physletb.2015.12.022}}.

\bibitem{Delamotte16a}
B.~Delamotte, M.~Tissier, N.~Wschebor, {Scale invariance implies conformal
invariance for the three-dimensional Ising model}, Phys. Rev. E 93 (2016)
012144.
\newblock \href {https://doi.org/10.1103/PhysRevE.93.012144}
{\path{doi:10.1103/PhysRevE.93.012144}}.

\bibitem{Blaizot06a}
J.-P. Blaizot, R.~M\'endez-Galain, N.~Wschebor, {Nonperturbative
renormalization group and momentum dependence of $n$-point functions. I},
Phys. Rev. E 74 (2006) 051116.
\newblock \href {https://doi.org/10.1103/PhysRevE.74.051116}
{\path{doi:10.1103/PhysRevE.74.051116}}.

\bibitem{Codello12}
A.~Codello, Scaling solutions in a continuous dimension, J. Phys. A: Math.
Theor. 45 (2012) 465006.
\newblock \href {https://doi.org/10.1088/1751-8113/45/46/465006}
{\path{doi:10.1088/1751-8113/45/46/465006}}.

\bibitem{Codello13}
A.~Codello, G.~D'Odorico, {$O(N)$-Universality Classes and the Mermin-Wagner
Theorem}, Phys. Rev. Lett. 110 (2013) 141601.
\newblock \href {https://doi.org/10.1103/PhysRevLett.110.141601}
{\path{doi:10.1103/PhysRevLett.110.141601}}.

\bibitem{Codello15}
A.~Codello, N.~Defenu, G.~D'Odorico, {Critical exponents of $O(N)$ models in
fractional dimensions}, Phys. Rev. D 91 (2015) 105003.
\newblock \href {https://doi.org/10.1103/PhysRevD.91.105003}
{\path{doi:10.1103/PhysRevD.91.105003}}.

\bibitem{Tetradis94}
N.~Tetradis, C.~Wetterich, {Critical exponents from the effective average
action}, Nucl. Phys. B 422 (1994) 541.
\newblock \href {https://doi.org/doi:10.1016/0550-3213(94)90446-4}
{\path{doi:doi:10.1016/0550-3213(94)90446-4}}.

\bibitem{Aoki98}
K.-I. Aoki, K.~Morikawa, W.~Souma, J.-I. Sumi, H.~Terao, {Rapidly Converging
Truncation Scheme of the Exact Renormalization Group}, Prog. Theor. Phys. 99
(1998) 451.
\newblock \href {https://doi.org/10.1143/PTP.99.451}
{\path{doi:10.1143/PTP.99.451}}.

\bibitem{Morris97}
T.~R. Morris, {Three-dimensional massive scalar field theory and the derivative
expansion of the renormalization group}, Nucl. Phys. B 495 (1997) 477.
\newblock \href {https://doi.org/https://doi.org/10.1016/S0550-3213(97)00233-2}
{\path{doi:https://doi.org/10.1016/S0550-3213(97)00233-2}}.

\bibitem{Seide99}
S.~Seide, C.~Wetterich, {Equation of state near the endpoint of the critical
line}, Nucl. Phys. B 562 (1999) 524--546.
\newblock \href
{https://doi.org/http://dx.doi.org/10.1016/S0550-3213(99)00545-3}
{\path{doi:http://dx.doi.org/10.1016/S0550-3213(99)00545-3}}.

\bibitem{Gersdorff01}
G.~v. Gersdorff, C.~Wetterich, {Nonperturbative renormalization flow and
essential scaling for the Kosterlitz-Thouless transition}, Phys. Rev. B 64
(2001) 054513.
\newblock \href {https://doi.org/10.1103/PhysRevB.64.054513}
{\path{doi:10.1103/PhysRevB.64.054513}}.

\bibitem{DePolsi20}
G.~De~Polsi, I.~Balog, M.~Tissier, N.~Wschebor, {Precision calculation of
critical exponents in the O($N$) universality classes with the
nonperturbative renormalization group}, Phys. Rev. E 101 (2020) 042113.
\newblock \href {https://doi.org/10.1103/PhysRevE.101.042113}
{\path{doi:10.1103/PhysRevE.101.042113}}.

\bibitem{Balog19}
I.~Balog, H.~Chat\'e, B.~Delamotte, M.~Marohni\'c, N.~Wschebor, Convergence of
nonperturbative approximations to the renormalization group, Phys. Rev. Lett.
123 (2019) 240604.
\newblock \href {https://doi.org/10.1103/PhysRevLett.123.240604}
{\path{doi:10.1103/PhysRevLett.123.240604}}.

\bibitem{Hasselmann12}
N.~Hasselmann, {Effective-average-action-based approach to correlation
functions at finite momenta}, Phys. Rev. E 86 (2012) 041118.
\newblock \href {https://doi.org/10.1103/PhysRevE.86.041118}
{\path{doi:10.1103/PhysRevE.86.041118}}.

\bibitem{Rose18}
F.~Rose, N.~Dupuis, {Nonperturbative renormalization-group approach preserving
the momentum dependence of correlation functions}, Phys. Rev. B 97 (2018)
174514.
\newblock \href {https://doi.org/10.1103/PhysRevB.97.174514}
{\path{doi:10.1103/PhysRevB.97.174514}}.

\bibitem{Benitez09}
F.~Benitez, J.-P. Blaizot, H.~Chat\'e, B.~Delamotte, R.~M\'endez-Galain,
N.~Wschebor, {Solutions of renormalization group flow equations with full
momentum dependence}, Phys. Rev. E 80 (2009) 030103(R).
\newblock \href {https://doi.org/10.1103/PhysRevE.80.030103}
{\path{doi:10.1103/PhysRevE.80.030103}}.

\bibitem{Benitez12}
F.~Benitez, J.-P. Blaizot, H.~Chat\'e, B.~Delamotte, R.~M\'endez-Galain,
N.~Wschebor, {Nonperturbative renormalization group preserving full-momentum
dependence: Implementation and quantitative evaluation}, Phys. Rev. E 85
(2012) 026707.
\newblock \href {https://doi.org/10.1103/PhysRevE.85.026707}
{\path{doi:10.1103/PhysRevE.85.026707}}.

\bibitem{Hasenbusch10}
M.~Hasenbusch, {Finite size scaling study of lattice models in the
three-dimensional Ising universality class}, Phys. Rev. B 82 (2010) 174433.
\newblock \href {https://doi.org/10.1103/PhysRevB.82.174433}
{\path{doi:10.1103/PhysRevB.82.174433}}.

\bibitem{Campostrini06}
M.~Campostrini, M.~Hasenbusch, A.~Pelissetto, E.~Vicari, {Theoretical estimates
of the critical exponents of the superfluid transition in
$^{{4}}\mathrm{{He}}$ by lattice methods}, Phys. Rev. B 74 (2006) 144506.
\newblock \href {https://doi.org/10.1103/PhysRevB.74.144506}
{\path{doi:10.1103/PhysRevB.74.144506}}.

\bibitem{Campostrini02}
M.~Campostrini, M.~Hasenbusch, A.~Pelissetto, P.~Rossi, E.~Vicari, {Critical
exponents and equation of state of the three-dimensional Heisenberg
universality class}, Phys. Rev. B 65 (2002) 144520.
\newblock \href {https://doi.org/10.1103/PhysRevB.65.144520}
{\path{doi:10.1103/PhysRevB.65.144520}}.

\bibitem{Hasenbusch:2019jkj}
M.~Hasenbusch, {Monte Carlo study of an improved clock model in three
dimensions}, Phys. Rev. B100~(22) (2019) 224517.
\newblock \href {http://arxiv.org/abs/1910.05916} {\path{arXiv:1910.05916}},
\href {https://doi.org/10.1103/PhysRevB.100.224517}
{\path{doi:10.1103/PhysRevB.100.224517}}.

\bibitem{Clisby16}
N.~Clisby, B.~D\"unweg,
\href{https://link.aps.org/doi/10.1103/PhysRevE.94.052102}{High-precision
estimate of the hydrodynamic radius for self-avoiding walks}, Phys. Rev. E 94
(2016) 052102.
\newblock \href {https://doi.org/10.1103/PhysRevE.94.052102}
{\path{doi:10.1103/PhysRevE.94.052102}}.
\newline\urlprefix\url{https://link.aps.org/doi/10.1103/PhysRevE.94.052102}

\bibitem{Clisby_2017}
N.~Clisby, \href{https://doi.org/10.1088%2F1751-8121%2Faa7231}{Scale-free monte
carlo method for calculating the critical exponent $\gamma$ of self-avoiding
walks}, Journal of Physics A: Mathematical and Theoretical 50~(26) (2017)
264003.
\newblock \href {https://doi.org/10.1088/1751-8121/aa7231}
{\path{doi:10.1088/1751-8121/aa7231}}.
\newline\urlprefix\url{https://doi.org/10.1088%2F1751-8121%2Faa7231}

\bibitem{Kompaniets:2017yct}
M.~V. Kompaniets, E.~Panzer, {Minimally subtracted six loop renormalization of
$O(n)$-symmetric $\phi^4$ theory and critical exponents}, Phys. Rev. D96~(3)
(2017) 036016.
\newblock \href {http://arxiv.org/abs/1705.06483} {\path{arXiv:1705.06483}},
\href {https://doi.org/10.1103/PhysRevD.96.036016}
{\path{doi:10.1103/PhysRevD.96.036016}}.

\bibitem{Shimada:2015gda}
H.~Shimada, S.~Hikami, {Fractal dimensions of self-avoiding walks and Ising
high-temperature graphs in 3D conformal bootstrap}, J. Statist. Phys. 165
(2016) 1006.
\newblock \href {http://arxiv.org/abs/1509.04039} {\path{arXiv:1509.04039}},
\href {https://doi.org/10.1007/s10955-016-1658-x}
{\path{doi:10.1007/s10955-016-1658-x}}.

\bibitem{Kos16}
F.~Kos, D.~Poland, D.~Simmons-Duffin, A.~Vichi, {Precision islands in the Ising
and O($N$) models}, J. High Energy Phys. 08 (2016) 036.
\newblock \href {https://doi.org/10.1007/JHEP08(2016)036}
{\path{doi:10.1007/JHEP08(2016)036}}.

\bibitem{Simmons-Duffin:2016wlq}
D.~Simmons-Duffin, {The Lightcone Bootstrap and the Spectrum of the 3d Ising
CFT}, JHEP 03 (2017) 086.
\newblock \href {http://arxiv.org/abs/1612.08471} {\path{arXiv:1612.08471}},
\href {https://doi.org/10.1007/JHEP03(2017)086}
{\path{doi:10.1007/JHEP03(2017)086}}.

\bibitem{Echeverri2016}
A.~C. Echeverri, B.~von Harling, M.~Serone,
\href{https://doi.org/10.1007/JHEP09(2016)097}{The effective bootstrap},
Journal of High Energy Physics 2016~(9) (2016) 97.
\newblock \href {https://doi.org/10.1007/JHEP09(2016)097}
{\path{doi:10.1007/JHEP09(2016)097}}.
\newline\urlprefix\url{https://doi.org/10.1007/JHEP09(2016)097}

\bibitem{chester2019carving}
S.~M. Chester, W.~Landry, J.~Liu, D.~Poland, D.~Simmons-Duffin, N.~Su,
A.~Vichi, Carving out ope space and precise $o(2)$ model critical exponents
(2019).
\newblock \href {http://arxiv.org/abs/1912.03324} {\path{arXiv:1912.03324}}.

\bibitem{Litim00}
D.~Litim, {Optimisation of the exact renormalization group}, Phys. Lett. B 486
(2000) 92.
\newblock \href
{https://doi.org/http://dx.doi.org/10.1016/S0370-2693(00)00748-6}
{\path{doi:http://dx.doi.org/10.1016/S0370-2693(00)00748-6}}.

\bibitem{Litim01b}
D.~F. Litim, Mind the gap, Int. J. Mod. Phys. A 16 (2001) 2081--2087.
\newblock \href {https://doi.org/10.1142/S0217751X01004748}
{\path{doi:10.1142/S0217751X01004748}}.

\bibitem{Litim02}
D.~F. Litim, Critical exponents from optimised renormalisation group flows,
Nucl. Phys. B 631 (2002) 128--158.
\newblock \href {https://doi.org/https://doi.org/10.1016/S0550-3213(02)00186-4}
{\path{doi:https://doi.org/10.1016/S0550-3213(02)00186-4}}.

\bibitem{Litim05}
D.~F. Litim,
\href{http://stacks.iop.org/1126-6708/2005/i=07/a=005}{Universality and the
renormalisation group}, J. High Energy Phys. 07 (2005) 005.
\newline\urlprefix\url{http://stacks.iop.org/1126-6708/2005/i=07/a=005}

\bibitem{Liao00}
S.-B. Liao, J.~Polonyi, M.~Strickland, Optimization of renormalization group
flow, Nucl. Phys. B 567 (2000) 493--514.
\newblock \href {https://doi.org/https://doi.org/10.1016/S0550-3213(99)00496-4}
{\path{doi:https://doi.org/10.1016/S0550-3213(99)00496-4}}.

\bibitem{Canet03a}
L.~Canet, B.~Delamotte, D.~Mouhanna, J.~Vidal, {Optimization of the derivative
expansion in the nonperturbative renormalization group}, Phys. Rev. D 67
(2003) 065004.
\newblock \href {https://doi.org/10.1103/PhysRevD.67.065004}
{\path{doi:10.1103/PhysRevD.67.065004}}.

\bibitem{Canet03b}
L.~Canet, B.~Delamotte, D.~Mouhanna, J.~Vidal, {Nonperturbative renormalization
group approach to the Ising model: A derivative expansion at order
$\partial{{}}^{{4}}$ }, Phys. Rev. B 68 (2003) 064421.
\newblock \href {https://doi.org/10.1103/PhysRevB.68.064421}
{\path{doi:10.1103/PhysRevB.68.064421}}.

\bibitem{Canet05}
L.~Canet, {Optimization of field-dependent nonperturbative renormalization
group flows}, Phys. Rev. B71 (2005) 012418.
\newblock \href {http://arxiv.org/abs/hep-th/0409300}
{\path{arXiv:hep-th/0409300}}, \href
{https://doi.org/10.1103/PhysRevB.71.012418}
{\path{doi:10.1103/PhysRevB.71.012418}}.

\bibitem{Nandori14}
I.~Nandori, I.~G. Marian, V.~Bacso, {Spontaneous symmetry breaking and
optimization of functional renormalization group}, Phys. Rev. D89~(4) (2014)
047701.
\newblock \href {http://arxiv.org/abs/1303.4508} {\path{arXiv:1303.4508}},
\href {https://doi.org/10.1103/PhysRevD.89.047701}
{\path{doi:10.1103/PhysRevD.89.047701}}.

\bibitem{Pawlowski15b}
J.~M. {Pawlowski}, M.~M. {Scherer}, R.~{Schmidt}, S.~J. {Wetzel}, {Physics and
the choice of regulators in functional renormalisation group flows} (Dec.
2015).
\newblock \href {http://arxiv.org/abs/1512.03598} {\path{arXiv:1512.03598}}.

\bibitem{Litim11}
D.~F. Litim, D.~Zappal\`a, Ising exponents from the functional renormalization
group, Phys. Rev. D 83 (2011) 085009.
\newblock \href {https://doi.org/10.1103/PhysRevD.83.085009}
{\path{doi:10.1103/PhysRevD.83.085009}}.

\bibitem{Morris95}
T.~R. Morris, {The renormalization group and two-dimensional multicritical
effective scalar field theory}, Phys. Lett. B 345 (1995) 139--148.
\newblock \href {https://doi.org/https://doi.org/10.1016/0370-2693(94)01603-A}
{\path{doi:https://doi.org/10.1016/0370-2693(94)01603-A}}.

\bibitem{Ballhausen04}
H.~Ballhausen, J.~Berges, C.~Wetterich, Critical phenomena in continuous
dimension, Phys. Lett. B 582 (2004) 144--150.
\newblock \href
{https://doi.org/https://doi.org/10.1016/j.physletb.2003.12.033}
{\path{doi:https://doi.org/10.1016/j.physletb.2003.12.033}}.

\bibitem{Defenu18}
N.~Defenu, A.~Codello, Scaling solutions in the derivative expansion, Phys.
Rev. D 98 (2018) 016013.
\newblock \href {https://doi.org/10.1103/PhysRevD.98.016013}
{\path{doi:10.1103/PhysRevD.98.016013}}.

\bibitem{DAttanasio97}
M.~D'Attanasio, T.~R. Morris, {Large N and the renormalization group}, Phys.
Lett. B 409 (1997) 363.
\newblock \href
{https://doi.org/http://dx.doi.org/10.1016/S0370-2693(97)00866-6}
{\path{doi:http://dx.doi.org/10.1016/S0370-2693(97)00866-6}}.

\bibitem{Yabunaka17}
S.~Yabunaka, B.~Delamotte, Surprises in $o(n)$ models: Nonperturbative fixed
points, large $n$ limits, and multicriticality, Phys. Rev. Lett. 119 (2017)
191602.
\newblock \href {https://doi.org/10.1103/PhysRevLett.119.191602}
{\path{doi:10.1103/PhysRevLett.119.191602}}.

\bibitem{Yabunaka18}
S.~Yabunaka, B.~Delamotte, Why might the standard large $n$ analysis fail in
the $\mathrm{O}(n)$ model: The role of cusps in fixed point potentials, Phys.
Rev. Lett. 121 (2018) 231601.
\newblock \href {https://doi.org/10.1103/PhysRevLett.121.231601}
{\path{doi:10.1103/PhysRevLett.121.231601}}.

\bibitem{Katsis:2018bvc}
A.~Katsis, N.~Tetradis, {Multicritical points of the O($N$) scalar theory in $2
< d < 4$ for large $N$}, Phys. Lett. B780 (2018) 491--494.
\newblock \href {http://arxiv.org/abs/1801.07659} {\path{arXiv:1801.07659}},
\href {https://doi.org/10.1016/j.physletb.2018.03.038}
{\path{doi:10.1016/j.physletb.2018.03.038}}.

\bibitem{Graeter95}
M.~Gr\"ater, C.~Wetterich, {Kosterlitz-Thouless Phase Transition in the Two
Dimensional Linear $\sigma$ Model}, Phys. Rev. Lett. 75 (1995) 378--381.
\newblock \href {https://doi.org/10.1103/PhysRevLett.75.378}
{\path{doi:10.1103/PhysRevLett.75.378}}.

\bibitem{Jakubczyk14}
P.~Jakubczyk, N.~Dupuis, B.~Delamotte, {Reexamination of the nonperturbative
renormalization-group approach to the Kosterlitz-Thouless transition}, Phys.
Rev. E 90 (2014) 062105.
\newblock \href {https://doi.org/10.1103/PhysRevE.90.062105}
{\path{doi:10.1103/PhysRevE.90.062105}}.

\bibitem{Jakubczyk16}
P.~Jakubczyk, A.~Eberlein, {Thermodynamics of the two-dimensional $\mathit{XY}$
model from functional renormalization}, Phys. Rev. E 93 (2016) 062145.
\newblock \href {https://doi.org/10.1103/PhysRevE.93.062145}
{\path{doi:10.1103/PhysRevE.93.062145}}.

\bibitem{Rancon17}
A.~Ran\c{c}on, N.~Dupuis, {Kosterlitz-Thouless signatures in the
low-temperature phase of layered three-dimensional systems}, Phys. Rev. B 96
(2017) 214512.
\newblock \href {https://doi.org/10.1103/PhysRevB.96.214512}
{\path{doi:10.1103/PhysRevB.96.214512}}.

\bibitem{Jakubczyk17}
P.~Jakubczyk, W.~Metzner, {Longitudinal fluctuations in the
Berezinskii-Kosterlitz-Thouless phase}, Phys. Rev. B 95 (2017) 085113.
\newblock \href {https://doi.org/10.1103/PhysRevB.95.085113}
{\path{doi:10.1103/PhysRevB.95.085113}}.

\bibitem{Jakubczyk17a}
P.~Jakubczyk, {Renormalization theory for the Fulde-Ferrell-Larkin-Ovchinnikov
states at $T>0$}, Phys. Rev. A95 (2017) 063626.
\newblock \href {http://arxiv.org/abs/1706.01524} {\path{arXiv:1706.01524}},
\href {https://doi.org/10.1103/PhysRevA.95.063626}
{\path{doi:10.1103/PhysRevA.95.063626}}.

\bibitem{Defenu17}
N.~Defenu, A.~Trombettoni, I.~N\'andori, T.~Enss, {Nonperturbative
renormalization group treatment of amplitude fluctuations for $\varphi^4$
topological phase transitions}, Phys. Rev. B 96 (2017) 174505.
\newblock \href {https://doi.org/10.1103/PhysRevB.96.174505}
{\path{doi:10.1103/PhysRevB.96.174505}}.

\bibitem{Krieg17a}
J.~Krieg, P.~Kopietz, {Dual lattice functional renormalization group for the
Berezinskii-Kosterlitz-Thouless transition: Irrelevance of amplitude and
out-of-plane fluctuations}, Phys. Rev. E 96 (2017) 042107.
\newblock \href {https://doi.org/10.1103/PhysRevE.96.042107}
{\path{doi:10.1103/PhysRevE.96.042107}}.

\bibitem{Villain75}
{Villain, J.}, Theory of one- and two-dimensional magnets with an easy
magnetization plane. ii. the planar, classical, two-dimensional magnet, J.
Phys. France 36 (1975) 581--590.
\newblock \href {https://doi.org/10.1051/jphys:01975003606058100}
{\path{doi:10.1051/jphys:01975003606058100}}.

\bibitem{Fischer:2004uk}
C.~S. Fischer, H.~Gies, {Renormalization flow of Yang-Mills propagators}, JHEP
10 (2004) 048.
\newblock \href {http://arxiv.org/abs/hep-ph/0408089}
{\path{arXiv:hep-ph/0408089}}, \href
{https://doi.org/10.1088/1126-6708/2004/10/048}
{\path{doi:10.1088/1126-6708/2004/10/048}}.

\bibitem{Borchardt16}
J.~Borchardt, B.~Knorr, Solving functional flow equations with pseudospectral
methods, Phys. Rev. D 94 (2016) 025027.
\newblock \href {https://doi.org/10.1103/PhysRevD.94.025027}
{\path{doi:10.1103/PhysRevD.94.025027}}.

\bibitem{Borchardt15}
J.~Borchardt, B.~Knorr, Global solutions of functional fixed point equations
via pseudospectral methods, Phys. Rev. D 91 (2015) 105011.
\newblock \href {https://doi.org/10.1103/PhysRevD.91.105011}
{\path{doi:10.1103/PhysRevD.91.105011}}.

\bibitem{Rose16a}
F.~Rose, F.~Benitez, F.~L\'eonard, B.~Delamotte, {Bound states of the
${{\ensuremath{{\phi}}}}^{{4}}$ model via the nonperturbative renormalization
group}, Phys. Rev. D 93 (2016) 125018.
\newblock \href {https://doi.org/10.1103/PhysRevD.93.125018}
{\path{doi:10.1103/PhysRevD.93.125018}}.

\bibitem{Grossi:2019urj}
E.~Grossi, N.~Wink, {Resolving phase transitions with Discontinuous Galerkin
methods} (3 2019).
\newblock \href {http://arxiv.org/abs/1903.09503} {\path{arXiv:1903.09503}}.

\bibitem{Litim01a}
D.~F. Litim, Derivative expansion and renormalisation group flows, J. High
Energy Phys. 11 (2001) 059.
\newblock \href {https://doi.org/10.1088/1126-6708/2001/11/059}
{\path{doi:10.1088/1126-6708/2001/11/059}}.

\bibitem{Kubyshin02}
{\relax Yu}.~A. Kubyshin, R.~Neves, R.~Potting, {Solutions of the polchinski
erg equation in the O(n) scalar model}, Int. J. Mod. Phys. A17 (2002) 4871.
\newblock \href {http://arxiv.org/abs/hep-th/0202199}
{\path{arXiv:hep-th/0202199}}, \href
{https://doi.org/10.1142/S0217751X02011400}
{\path{doi:10.1142/S0217751X02011400}}.

\bibitem{Berges96}
J.~Berges, N.~Tetradis, C.~Wetterich, {Critical Equation of State from the
Average Action}, Phys. Rev. Lett. 77 (1996) 873--876.
\newblock \href {https://doi.org/10.1103/PhysRevLett.77.873}
{\path{doi:10.1103/PhysRevLett.77.873}}.

\bibitem{Rancon13a}
A.~Ran\c{c}on, O.~Kodio, N.~Dupuis, P.~Lecheminant, {Thermodynamics in the
vicinity of a relativistic quantum critical point in $2+1$ dimensions}, Phys.
Rev. E 88 (2013) 012113.
\newblock \href {https://doi.org/10.1103/PhysRevE.88.012113}
{\path{doi:10.1103/PhysRevE.88.012113}}.

\bibitem{Rancon13b}
A.~Ran\c{c}on, N.~Dupuis, {Quantum XY criticality in a two-dimensional Bose gas
near the Mott transition}, Europhys. Lett. 104 (2013) 16002.
\newblock \href {https://doi.org/10.1209/0295-5075/104/16002}
{\path{doi:10.1209/0295-5075/104/16002}}.

\bibitem{Rancon16}
A.~Ran\c{c}on, L.-P. Henry, F.~Rose, D.~L. Cardozo, N.~Dupuis, P.~C.~W.
Holdsworth, T.~Roscilde, {Critical Casimir forces from the equation of state
of quantum critical systems}, Phys. Rev. B 94 (2016) 140506(R).
\newblock \href {https://doi.org/10.1103/PhysRevB.94.140506}
{\path{doi:10.1103/PhysRevB.94.140506}}.

\bibitem{Dupuis11}
N.~Dupuis, {Infrared behavior in systems with a broken continuous symmetry:
Classical O($N$) model versus interacting bosons}, Phys. Rev. E 83 (2011)
031120.
\newblock \href {https://doi.org/10.1103/PhysRevE.83.031120}
{\path{doi:10.1103/PhysRevE.83.031120}}.

\bibitem{Caillol12a}
J.-M. Caillol, {The non-perturbative renormalization group in the ordered
phase}, Nucl. Phys. B 855 (2012) 854.
\newblock \href {https://doi.org/10.1016/j.nuclphysb.2011.10.026}
{\path{doi:10.1016/j.nuclphysb.2011.10.026}}.

\bibitem{Patasinskij73}
A.~Z. Patasinskij, V.~L. Pokrovskij, Longitudinal susceptibility and
correlations in degenerate systems, Sov. Phys. JETP 37 (1973) 733, zh. Eksp.
Teor. Fiz. {\bf 64}, 1445 (1973).

\bibitem{Zwerger04}
W.~Zwerger, {Anomalous Fluctuations in Phases with a Broken Continuous
Symmetry}, Phys. Rev. Lett. 92 (2004) 027203.
\newblock \href {https://doi.org/10.1103/PhysRevLett.92.027203}
{\path{doi:10.1103/PhysRevLett.92.027203}}.

\bibitem{Hellwig:2015woa}
T.~Hellwig, A.~Wipf, O.~Zanusso, {Scaling and superscaling solutions from the
functional renormalization group}, Phys. Rev. D92~(8) (2015) 085027.
\newblock \href {http://arxiv.org/abs/1508.02547} {\path{arXiv:1508.02547}},
\href {https://doi.org/10.1103/PhysRevD.92.085027}
{\path{doi:10.1103/PhysRevD.92.085027}}.

\bibitem{Eichhorn:2013zza}
A.~Eichhorn, D.~Mesterh\'azy, M.~M. Scherer, {Multicritical behavior in models
with two competing order parameters}, Phys.\ Rev.\ E 88 (2013) 042141.
\newblock \href {http://arxiv.org/abs/1306.2952} {\path{arXiv:1306.2952}},
\href {https://doi.org/10.1103/PhysRevE.88.042141}
{\path{doi:10.1103/PhysRevE.88.042141}}.

\bibitem{Codello:2017hhh}
A.~Codello, M.~Safari, G.~P. Vacca, O.~Zanusso, {Functional perturbative RG and
CFT data in the $\epsilon$-expansion}, Eur. Phys. J. C78~(1) (2018) 30.
\newblock \href {http://arxiv.org/abs/1705.05558} {\path{arXiv:1705.05558}},
\href {https://doi.org/10.1140/epjc/s10052-017-5505-2}
{\path{doi:10.1140/epjc/s10052-017-5505-2}}.

\bibitem{Marchais:2017jqc}
D.~F. Litim, E.~Marchais, P.~Mati, {Fixed points and the spontaneous breaking
of scale invariance}, Phys. Rev. D95~(12) (2017) 125006.
\newblock \href {http://arxiv.org/abs/1702.05749} {\path{arXiv:1702.05749}},
\href {https://doi.org/10.1103/PhysRevD.95.125006}
{\path{doi:10.1103/PhysRevD.95.125006}}.

\bibitem{Eichhorn16}
A.~Eichhorn, L.~Janssen, M.~M. Scherer, Critical $o(n)$ models above four
dimensions: Small-$n$ solutions and stability, Phys. Rev. D 93 (2016) 125021.
\newblock \href {https://doi.org/10.1103/PhysRevD.93.125021}
{\path{doi:10.1103/PhysRevD.93.125021}}.

\bibitem{Percacci:2014tfa}
R.~Percacci, G.~P. Vacca, {Are there scaling solutions in the $O(N)$-models for
large $N$ in $d>4$ ?}, Phys. Rev. D90 (2014) 107702.
\newblock \href {http://arxiv.org/abs/1405.6622} {\path{arXiv:1405.6622}},
\href {https://doi.org/10.1103/PhysRevD.90.107702}
{\path{doi:10.1103/PhysRevD.90.107702}}.

\bibitem{Mati:2014xma}
P.~Mati, {Vanishing beta function curves from the functional renormalization
group}, Phys. Rev. D91~(12) (2015) 125038.
\newblock \href {http://arxiv.org/abs/1501.00211} {\path{arXiv:1501.00211}},
\href {https://doi.org/10.1103/PhysRevD.91.125038}
{\path{doi:10.1103/PhysRevD.91.125038}}.

\bibitem{Defenu:2014bea}
N.~Defenu, A.~Trombettoni, A.~Codello, {Fixed-point structure and effective
fractional dimensionality for O$(N)$ models with long-range interactions},
Phys. Rev. E92~(5) (2015) 052113.
\newblock \href {http://arxiv.org/abs/1409.8322} {\path{arXiv:1409.8322}},
\href {https://doi.org/10.1103/PhysRevE.92.052113}
{\path{doi:10.1103/PhysRevE.92.052113}}.

\bibitem{Defenu:2017utb}
N.~Defenu, A.~Trombettoni, S.~Ruffo, {Criticality and phase diagram of quantum
long-range O( N ) models}, Phys. Rev. B96~(10) (2017) 104432.
\newblock \href {http://arxiv.org/abs/1704.00528} {\path{arXiv:1704.00528}},
\href {https://doi.org/10.1103/PhysRevB.96.104432}
{\path{doi:10.1103/PhysRevB.96.104432}}.

\bibitem{Goll:2018vdj}
R.~Goll, P.~Kopietz, {Renormalization group for $\varphi^4$-theory with
long-range interaction and the critical exponent $\eta$ of the Ising model},
Phys. Rev. E98~(2) (2018) 022135.
\newblock \href {http://arxiv.org/abs/1804.04150} {\path{arXiv:1804.04150}},
\href {https://doi.org/10.1103/PhysRevE.98.022135}
{\path{doi:10.1103/PhysRevE.98.022135}}.

\bibitem{Defenu:2020umv}
N.~Defenu, A.~Codello, S.~Ruffo, A.~Trombettoni, {Criticality of spin systems
with weak long-range interactions}, J. Phys. A 53~(14) (2020) 143001.
\newblock \href {http://arxiv.org/abs/1908.05158} {\path{arXiv:1908.05158}},
\href {https://doi.org/10.1088/1751-8121/ab6a6c}
{\path{doi:10.1088/1751-8121/ab6a6c}}.

\bibitem{Litim:2016hlb}
D.~F. Litim, E.~Marchais, {Critical $O(N)$ models in the complex field plane},
Phys. Rev. D95~(2) (2017) 025026.
\newblock \href {http://arxiv.org/abs/1607.02030} {\path{arXiv:1607.02030}},
\href {https://doi.org/10.1103/PhysRevD.95.025026}
{\path{doi:10.1103/PhysRevD.95.025026}}.

\bibitem{Juttner:2017cpr}
A.~J\"uttner, D.~F. Litim, E.~Marchais, {Global Wilson-Fisher fixed points},
Nucl. Phys. B921 (2017) 769–795.
\newblock \href {http://arxiv.org/abs/1701.05168} {\path{arXiv:1701.05168}},
\href {https://doi.org/10.1016/j.nuclphysb.2017.06.010}
{\path{doi:10.1016/j.nuclphysb.2017.06.010}}.

\bibitem{Halpern95}
K.~Halpern, K.~Huang,
\href{https://link.aps.org/doi/10.1103/PhysRevLett.74.3526}{{Fixed-Point
Structure of Scalar Fields}}, Phys. Rev. Lett. 74 (1995) 3526–3529.
\newblock \href {https://doi.org/10.1103/PhysRevLett.74.3526}
{\path{doi:10.1103/PhysRevLett.74.3526}}.
\newline\urlprefix\url{https://link.aps.org/doi/10.1103/PhysRevLett.74.3526}

\bibitem{Halpern96}
K.~Halpern, K.~Huang,
\href{https://link.aps.org/doi/10.1103/PhysRevLett.77.1659}{{Halpern and
Huang Reply:}}, Phys. Rev. Lett. 77 (1996) 1659–1659.
\newblock \href {https://doi.org/10.1103/PhysRevLett.77.1659}
{\path{doi:10.1103/PhysRevLett.77.1659}}.
\newline\urlprefix\url{https://link.aps.org/doi/10.1103/PhysRevLett.77.1659}

\bibitem{Morris:1996nx}
T.~R. Morris, {On the fixed point structure of scalar fields}, Phys. Rev. Lett.
77 (1996) 1658.
\newblock \href {http://arxiv.org/abs/hep-th/9601128}
{\path{arXiv:hep-th/9601128}}, \href
{https://doi.org/10.1103/PhysRevLett.77.1658}
{\path{doi:10.1103/PhysRevLett.77.1658}}.

\bibitem{Gies:2000xr}
H.~Gies, {Flow equation for Halpern-Huang directions of scalar O(N) models},
Phys. Rev. D63 (2001) 065011.
\newblock \href {http://arxiv.org/abs/hep-th/0009041}
{\path{arXiv:hep-th/0009041}}, \href
{https://doi.org/10.1103/PhysRevD.63.065011}
{\path{doi:10.1103/PhysRevD.63.065011}}.

\bibitem{Bridle:2016nsu}
I.~{Hamzaan Bridle}, T.~R. Morris, {Fate of nonpolynomial interactions in
scalar field theory}, Phys. Rev. D94 (2016) 065040.
\newblock \href {http://arxiv.org/abs/1605.06075} {\path{arXiv:1605.06075}},
\href {https://doi.org/10.1103/PhysRevD.94.065040}
{\path{doi:10.1103/PhysRevD.94.065040}}.

\bibitem{Jakubczyk:2012iza}
P.~Jakubczyk, M.~Napiorkowski, {Critical Casimir forces for O(N) models from
functional renormalization}, Phys.Rev. B87 (2013) 165439.
\newblock \href {http://arxiv.org/abs/1212.2647} {\path{arXiv:1212.2647}},
\href {https://doi.org/10.1103/PhysRevB.87.165439}
{\path{doi:10.1103/PhysRevB.87.165439}}.

\bibitem{Codello:2008qq}
A.~Codello, R.~Percacci, {Fixed Points of Nonlinear Sigma Models in d$>$2},
Phys. Lett. B672 (2009) 280--283.
\newblock \href {http://arxiv.org/abs/0810.0715} {\path{arXiv:0810.0715}},
\href {https://doi.org/10.1016/j.physletb.2009.01.032}
{\path{doi:10.1016/j.physletb.2009.01.032}}.

\bibitem{Flore:2012ma}
R.~Flore, A.~Wipf, O.~Zanusso, {Functional renormalization group of the
non-linear sigma model and the $O(N)$ universality class}, Phys. Rev. D87~(6)
(2013) 065019.
\newblock \href {http://arxiv.org/abs/1207.4499} {\path{arXiv:1207.4499}},
\href {https://doi.org/10.1103/PhysRevD.87.065019}
{\path{doi:10.1103/PhysRevD.87.065019}}.

\bibitem{Percacci:2013jpa}
R.~Percacci, M.~Safari, {Functional renormalization of N scalars with O(N)
invariance}, Phys. Rev. D88 (2013) 085007.
\newblock \href {http://arxiv.org/abs/1306.3918} {\path{arXiv:1306.3918}},
\href {https://doi.org/10.1103/PhysRevD.88.085007}
{\path{doi:10.1103/PhysRevD.88.085007}}.

\bibitem{Zinati18}
R.~B.~A. {Zinati}, A.~{Codello}, Functional rg approach to the potts model, J.
Stat. Mech: Theory Exp. 2018 (2018) 013206.
\newblock \href {https://doi.org/10.1088/1742-5468/aa9dcc}
{\path{doi:10.1088/1742-5468/aa9dcc}}.

\bibitem{Nagy09}
S.~Nagy, I.~N\'andori, J.~Polonyi, K.~Sailer, {Functional Renormalization Group
Approach to the Sine-Gordon Model}, Phys. Rev. Lett. 102 (2009) 241603.
\newblock \href {https://doi.org/10.1103/PhysRevLett.102.241603}
{\path{doi:10.1103/PhysRevLett.102.241603}}.

\bibitem{Pangon12}
V.~Pangon, {Structure of the broken phase of the sine-Gordon model using
functional renormalisation}, Int. J. Mod. Phys. A 27 (2012) 1250014.
\newblock \href {https://doi.org/10.1142/S0217751X12500145}
{\path{doi:10.1142/S0217751X12500145}}.

\bibitem{Pangon11}
V.~{Pangon}, {Generating the mass gap of the sine-Gordon model} (Nov. 2011).
\newblock \href {http://arxiv.org/abs/1111.6425} {\path{arXiv:1111.6425}}.

\bibitem{Bacso15}
V.~Bacs\'o, N.~Defenu, A.~Trombettoni, I.~N\'andori, {c-function and central
charge of the sine-Gordon model from the non-perturbative renormalization
group flow }, Nucl. Phys. B 901 (2015) 444 -- 460.
\newblock \href
{https://doi.org/http://dx.doi.org/10.1016/j.nuclphysb.2015.11.001}
{\path{doi:http://dx.doi.org/10.1016/j.nuclphysb.2015.11.001}}.

\bibitem{Oak17}
P.~Oak, B.~Sathiapalan, Exact renormalization group and sine gordon theory, J.
High Energy Phys. 2017~(7) (2017) 103.
\newblock \href {https://doi.org/10.1007/JHEP07(2017)103}
{\path{doi:10.1007/JHEP07(2017)103}}.

\bibitem{Daviet19}
R.~Daviet, N.~Dupuis, {Nonperturbative functional renormalization-group
approach to the sine-Gordon model and the Lukyanov-Zamolodchikov conjecture},
Phys. Rev. Lett. 122 (2019) 155301.
\newblock \href {https://doi.org/10.1103/PhysRevLett.122.155301}
{\path{doi:10.1103/PhysRevLett.122.155301}}.

\bibitem{Lukyanov97}
S.~Lukyanov, A.~B. Zamolodchikov, {Exact expectation values of local fields in
the quantum sine-Gordon model}, Nucl. Phys. B 493~(3) (1997) 571.
\newblock \href {https://doi.org/https://doi.org/10.1016/S0550-3213(97)00123-5}
{\path{doi:https://doi.org/10.1016/S0550-3213(97)00123-5}}.

\bibitem{Blaizot:2004qa}
J.-P. Blaizot, R.~Mend{\'e}z-Galain, N.~Wschebor, {Non perturbative
renormalization group, momentum dependence of n-point functions and the
transition temperature of the weakly interacting Bose gas}, Europhys.Lett. 72
(2005) 705--711.
\newblock \href {http://arxiv.org/abs/cond-mat/0412481}
{\path{arXiv:cond-mat/0412481}}, \href
{https://doi.org/10.1209/epl/i2005-10318-5}
{\path{doi:10.1209/epl/i2005-10318-5}}.

\bibitem{Blaizot06b}
J.-P. Blaizot, R.~M\'endez-Galain, N.~Wschebor, {Nonperturbative
renormalization group and momentum dependence of $n$-point functions. II},
Phys. Rev. E 74 (2006) 051117.
\newblock \href {https://doi.org/10.1103/PhysRevE.74.051117}
{\path{doi:10.1103/PhysRevE.74.051117}}.

\bibitem{Blaizot07}
J.-P. Blaizot, R.~M\'endez-Galain, N.~Wschebor, {Non-perturbative
renormalization group calculation of the scalar self-energy}, Eur. Phys. J. B
58 (2007) 297.
\newblock \href {https://doi.org/10.1140/epjb/e2007-00223-3}
{\path{doi:10.1140/epjb/e2007-00223-3}}.

\bibitem{Ledowski04}
S.~Ledowski, N.~Hasselmann, P.~Kopietz, {Self-energy and critical temperature
of weakly interacting bosons}, Phys. Rev. A 69 (2004) 061601.
\newblock \href {https://doi.org/10.1103/PhysRevA.69.061601}
{\path{doi:10.1103/PhysRevA.69.061601}}.

\bibitem{Hasselmann07}
N.~Hasselmann, A.~Sinner, P.~Kopietz, Two-parameter scaling of correlation
functions near continuous phase transitions, Phys. Rev. E 76 (2007) 040101.
\newblock \href {https://doi.org/10.1103/PhysRevE.76.040101}
{\path{doi:10.1103/PhysRevE.76.040101}}.

\bibitem{Sinner08}
A.~Sinner, N.~Hasselmann, P.~Kopietz, {Functional renormalization group in the
broken symmetry phase: momentum dependence and two-parameter scaling of the
self-energy}, J. Phys.: Condens. Matter 20 (2008) 075208.
\newblock \href {https://doi.org/10.1088/0953-8984/20/7/075208}
{\path{doi:10.1088/0953-8984/20/7/075208}}.

\bibitem{Guerra07}
D.~Guerra, R.~M\'endez-Galain, N.~Wschebor, {Correlation functions in the non
perturbative renormalization group and field expansion}, Eur. Phys. J. B 59
(2007) 357.
\newblock \href {https://doi.org/DOI: 10.1140/epjb/e2007-00296-x}
{\path{doi:DOI: 10.1140/epjb/e2007-00296-x}}.

\bibitem{Hasselmann11}
N.~Hasselmann, F.~L. Braghin, Nonlocal effective-average-action approach to
crystalline phantom membranes, Phys. Rev. E 83 (2011) 031137.
\newblock \href {https://doi.org/10.1103/PhysRevE.83.031137}
{\path{doi:10.1103/PhysRevE.83.031137}}.

\bibitem{Mathey14}
S.~Mathey, T.~Gasenzer, J.~M. Pawlowski,
\href{http://link.aps.org/doi/10.1103/PhysRevA.92.023635}{Anomalous scaling
at nonthermal fixed points of {B}urgers' and gross-pitaevskii turbulence},
Phys. Rev. A 92 (2015) 023635.
\newblock \href {https://doi.org/10.1103/PhysRevA.92.023635}
{\path{doi:10.1103/PhysRevA.92.023635}}.
\newline\urlprefix\url{http://link.aps.org/doi/10.1103/PhysRevA.92.023635}

\bibitem{Canet11b}
L.~Canet, H.~Chat{\'{e}}, B.~Delamotte,
\href{https://doi.org/10.1088%2F1751-8113%2F44%2F49%2F495001}{General
framework of the non-perturbative renormalization group for non-equilibrium
steady states}, Journal of Physics A: Mathematical and Theoretical 44~(49)
(2011) 495001.
\newblock \href {https://doi.org/10.1088/1751-8113/44/49/495001}
{\path{doi:10.1088/1751-8113/44/49/495001}}.
\newline\urlprefix\url{https://doi.org/10.1088%2F1751-8113%2F44%2F49%2F495001}

\bibitem{Canet16sm}
L.~Canet, B.~Delamotte, N.~Wschebor,
\href{http://link.aps.org/doi/10.1103/PhysRevE.93.063101}{Fully developed
isotropic turbulence: Nonperturbative renormalization group formalism and
fixed-point solution}, Phys. Rev. E 93 (2016) 063101.
\newblock \href {https://doi.org/10.1103/PhysRevE.93.063101}
{\path{doi:10.1103/PhysRevE.93.063101}}.
\newline\urlprefix\url{http://link.aps.org/doi/10.1103/PhysRevE.93.063101}

\bibitem{Feldmann:2017ooy}
P.~Feldmann, A.~Wipf, L.~Zambelli, {Critical Wess-Zumino models with four
supercharges in the functional renormalization group approach}, Phys. Rev.
D98~(9) (2018) 096005.
\newblock \href {http://arxiv.org/abs/1712.03910} {\path{arXiv:1712.03910}},
\href {https://doi.org/10.1103/PhysRevD.98.096005}
{\path{doi:10.1103/PhysRevD.98.096005}}.

\bibitem{Blaizot06}
J.-P. Blaizot, R.~M\'endez-Galain, N.~Wschebor, {A new method to solve the
non-perturbative renormalization group equations}, Phys. Lett. B 632 (2006)
571.
\newblock \href {https://doi.org/10.1016/j.physletb.2005.10.086}
{\path{doi:10.1016/j.physletb.2005.10.086}}.

\bibitem{Benitez08}
F.~Benitez, R.~M\'{e}ndez-Galain, N.~Wschebor, {Calculations on the two-point
function of the O(N) model}, Phys. Rev. B 77 (2008) 024431.
\newblock \href {https://doi.org/10.1103/PhysRevB.77.024431}
{\path{doi:10.1103/PhysRevB.77.024431}}.

\bibitem{Rose15}
F.~Rose, F.~L\'eonard, N.~Dupuis, {Higgs amplitude mode in the vicinity of a
$(2+1)$-dimensional quantum critical point: A nonperturbative
renormalization-group approach}, Phys. Rev. B 91 (2015) 224501.
\newblock \href {https://doi.org/10.1103/PhysRevB.91.224501}
{\path{doi:10.1103/PhysRevB.91.224501}}.

\bibitem{Pogorelov07}
A.~A. Pogorelov, I.~M. Suslov, Renormalization group functions for
two-dimensional phase transitions: To the problem of singular contributions,
J. Exp. Theor. Phys. 105 (2007) 360--370.
\newblock \href {https://doi.org/10.1134/S1063776107080080}
{\path{doi:10.1134/S1063776107080080}}.

\bibitem{Dupuis08}
N.~Dupuis, K.~Sengupta, {Non-perturbative renormalization-group approach to
lattice models}, Eur. Phys. J. B 66 (2008) 271.
\newblock \href {https://doi.org/10.1140/epjb/e2008-00417-1}
{\path{doi:10.1140/epjb/e2008-00417-1}}.

\bibitem{Machado10}
T.~Machado, N.~Dupuis, {From local to critical fluctuations in lattice models:
A nonperturbative renormalization-group approach}, Phys. Rev. E 82 (2010)
041128.
\newblock \href {https://doi.org/10.1103/PhysRevE.82.041128}
{\path{doi:10.1103/PhysRevE.82.041128}}.

\bibitem{Caillol12b}
J.-M. Caillol, {Critical line of the theory on a simple cubic lattice in the
local potential approximation }, Nucl. Phys. B 865 (2012) 291.
\newblock \href
{https://doi.org/http://dx.doi.org/10.1016/j.nuclphysb.2012.07.032}
{\path{doi:http://dx.doi.org/10.1016/j.nuclphysb.2012.07.032}}.

\bibitem{Caillol13}
J.-M. Caillol, {Critical line of the $\Phi^4$ scalar field theory on a 4D cubic
lattice in the local potential approximation}, Condens. Matter Phys. 16
(2013) 43005.
\newblock \href {https://doi.org/10.5488/CMP.16.43005}
{\path{doi:10.5488/CMP.16.43005}}.

\bibitem{Banerjee:2018hqr}
R.~Banerjee, M.~Niedermaier, {Graph rules for the linked cluster expansion of
the Legendre effective action}, J. Math. Phys. 60~(1) (2019) 013504.
\newblock \href {http://arxiv.org/abs/1812.06602} {\path{arXiv:1812.06602}},
\href {https://doi.org/10.1063/1.5031429} {\path{doi:10.1063/1.5031429}}.

\bibitem{Banerjee:2018pkt}
R.~Banerjee, {Critical behavior of the hopping expansion from the Functional
Renormalization Group}, PoS LATTICE2018 (2018) 249.
\newblock \href {http://arxiv.org/abs/1812.02251} {\path{arXiv:1812.02251}},
\href {https://doi.org/10.22323/1.334.0249} {\path{doi:10.22323/1.334.0249}}.

\bibitem{Rancon14a}
A.~Ran\c{c}on, {Nonperturbative renormalization group approach to quantum $XY$
spin models}, Phys. Rev. B 89 (2014) 214418.
\newblock \href {https://doi.org/10.1103/PhysRevB.89.214418}
{\path{doi:10.1103/PhysRevB.89.214418}}.

\bibitem{Krieg19}
J.~Krieg, P.~Kopietz,
\href{https://link.aps.org/doi/10.1103/PhysRevB.99.060403}{Exact
renormalization group for quantum spin systems}, Phys. Rev. B 99 (2019)
060403.
\newblock \href {https://doi.org/10.1103/PhysRevB.99.060403}
{\path{doi:10.1103/PhysRevB.99.060403}}.
\newline\urlprefix\url{https://link.aps.org/doi/10.1103/PhysRevB.99.060403}

\bibitem{Rancon11a}
A.~Ran\c{c}on, N.~Dupuis, {Nonperturbative renormalization group approach to
the Bose-Hubbard model}, Phys. Rev. B 83 (2011) 172501.
\newblock \href {https://doi.org/10.1103/PhysRevB.83.172501}
{\path{doi:10.1103/PhysRevB.83.172501}}.

\bibitem{Rancon11b}
A.~Ran\c{c}on, N.~Dupuis, {Nonperturbative renormalization group approach to
strongly correlated lattice bosons}, Phys. Rev. B 84 (2011) 174513.
\newblock \href {https://doi.org/10.1103/PhysRevB.84.174513}
{\path{doi:10.1103/PhysRevB.84.174513}}.

\bibitem{Rancon12a}
A.~Ran\c{c}on, N.~Dupuis, {Quantum criticality of a Bose gas in an optical
lattice near the Mott transition}, Phys. Rev. A 85 (2012) 011602(R).
\newblock \href {https://doi.org/10.1103/PhysRevA.85.011602}
{\path{doi:10.1103/PhysRevA.85.011602}}.

\bibitem{Rancon12d}
A.~Ran\c{c}on, N.~Dupuis, {Thermodynamics of a Bose gas near the
superfluid--Mott-insulator transition}, Phys. Rev. A 86 (2012) 043624.
\newblock \href {https://doi.org/10.1103/PhysRevA.86.043624}
{\path{doi:10.1103/PhysRevA.86.043624}}.

\bibitem{Reuther14}
J.~Reuther, R.~Thomale, {Cluster functional renormalization group}, Phys. Rev.
B 89 (2014) 024412.
\newblock \href {https://doi.org/10.1103/PhysRevB.89.024412}
{\path{doi:10.1103/PhysRevB.89.024412}}.

\bibitem{Wentzell15}
N.~Wentzell, C.~Taranto, A.~Katanin, A.~Toschi, S.~Andergassen, {Correlated
starting points for the functional renormalization group}, Phys. Rev. B 91
(2015) 045120.
\newblock \href {https://doi.org/10.1103/PhysRevB.91.045120}
{\path{doi:10.1103/PhysRevB.91.045120}}.

\bibitem{Taranto14}
C.~Taranto, S.~Andergassen, J.~Bauer, K.~Held, A.~Katanin, W.~Metzner,
G.~Rohringer, A.~Toschi, {From Infinite to Two Dimensions through the
Functional Renormalization Group}, Phys. Rev. Lett. 112 (2014) 196402.
\newblock \href {https://doi.org/10.1103/PhysRevLett.112.196402}
{\path{doi:10.1103/PhysRevLett.112.196402}}.

\bibitem{Sachdev_book}
S.~Sachdev, {Quantum Phase Transitions}, 2nd Edition, Cambridge University
Press, Cambridge, England, 2011.

\bibitem{Tetradis93}
N.~Tetradis, C.~Wetterich, {The high temperature phase transition for $\phi^4$
theories}, Nucl. Phys. B 398 (1993) 659--696.
\newblock \href {https://doi.org/10.1016/0550-3213(93)90608-R}
{\path{doi:10.1016/0550-3213(93)90608-R}}.

\bibitem{Litim06a}
D.~F. Litim, J.~M. Pawlowski, Non-perturbative thermal flows and resummations,
J. High Energy Phys. 11 (2006) 026.
\newblock \href {https://doi.org/10.1088/1126-6708/2006/11/026}
{\path{doi:10.1088/1126-6708/2006/11/026}}.

\bibitem{Schlessinger68}
L.~Schlessinger, Use of analyticity in the calculation of nonrelativistic
scattering amplitudes, Phys. Rev. 167 (1968) 1411.
\newblock \href {https://doi.org/10.1103/PhysRev.167.1411}
{\path{doi:10.1103/PhysRev.167.1411}}.

\bibitem{Vidberg77}
H.~J. Vidberg, J.~W. Serene, {Solving the Eliashberg equations by means of
$N$-point Pad\'e approximants}, J. Low Temp. Phys. 29 (1977) 179.
\newblock \href {https://doi.org/10.1007/BF00655090}
{\path{doi:10.1007/BF00655090}}.

\bibitem{Tripolt19}
R.-A. Tripolt, P.~Gubler, M.~Ulybyshev, L.~von Smekal, {Numerical analytic
continuation of Euclidean data}, Comput. Phys. Commun. 237 (2019) 129.
\newblock \href {https://doi.org/https://doi.org/10.1016/j.cpc.2018.11.012}
{\path{doi:https://doi.org/10.1016/j.cpc.2018.11.012}}.

\bibitem{Dupuis09b}
N.~Dupuis, {Infrared behavior and spectral function of a Bose superfluid at
zero temperature}, Phys. Rev. A 80 (2009) 043627.
\newblock \href {https://doi.org/10.1103/PhysRevA.80.043627}
{\path{doi:10.1103/PhysRevA.80.043627}}.

\bibitem{Sinner10}
A.~Sinner, N.~Hasselmann, P.~Kopietz, {Functional renormalization-group
approach to interacting bosons at zero temperature}, Phys. Rev. A 82 (2010)
063632.
\newblock \href {https://doi.org/10.1103/PhysRevA.82.063632}
{\path{doi:10.1103/PhysRevA.82.063632}}.

\bibitem{Schmidt11}
R.~Schmidt, T.~Enss, {Excitation spectra and rf response near the
polaron-to-molecule transition from the functional renormalization group},
Phys. Rev. A 83 (2011) 063620.
\newblock \href {https://doi.org/10.1103/PhysRevA.83.063620}
{\path{doi:10.1103/PhysRevA.83.063620}}.

\bibitem{Rose17a}
F.~Rose, N.~Dupuis, {Superuniversal transport near a $(2+1)$-dimensional
quantum critical point}, Phys. Rev. B 96 (2017) 100501(R).
\newblock \href {https://doi.org/10.1103/PhysRevB.96.100501}
{\path{doi:10.1103/PhysRevB.96.100501}}.

\bibitem{Tripolt17}
R.-A. Tripolt, I.~Haritan, J.~Wambach, N.~Moiseyev, {Threshold energies and
poles for hadron physical problems by a model-independent universal
algorithm}, Phys. Lett. B 774 (2017) 411.
\newblock \href
{https://doi.org/https://doi.org/10.1016/j.physletb.2017.10.001}
{\path{doi:https://doi.org/10.1016/j.physletb.2017.10.001}}.

\bibitem{Rohe05}
D.~Rohe, W.~Metzner, {Pseudogap at hot spots in the two-dimensional Hubbard
model at weak coupling}, Phys. Rev. B 71 (2005) 115116.
\newblock \href {https://doi.org/10.1103/PhysRevB.71.115116}
{\path{doi:10.1103/PhysRevB.71.115116}}.

\bibitem{Jakobs10a}
S.~G. Jakobs, M.~Pletyukhov, H.~Schoeller, {Nonequilibrium functional
renormalization group with frequency-dependent vertex function: A study of
the single-impurity Anderson model}, Phys. Rev. B 81 (2010) 195109.
\newblock \href {https://doi.org/10.1103/PhysRevB.81.195109}
{\path{doi:10.1103/PhysRevB.81.195109}}.

\bibitem{Floerchinger:2011sc}
S.~Floerchinger, {Analytic Continuation of Functional Renormalization Group
Equations}, JHEP 05 (2012) 021.
\newblock \href {http://arxiv.org/abs/1112.4374} {\path{arXiv:1112.4374}},
\href {https://doi.org/10.1007/JHEP05(2012)021}
{\path{doi:10.1007/JHEP05(2012)021}}.

\bibitem{Tripolt14}
R.-A. Tripolt, N.~Strodthoff, L.~von Smekal, J.~Wambach, {Spectral functions
from the functional renormalization group }, Nucl. Phys. A 931 (2014)
790--795.
\newblock \href
{https://doi.org/http://dx.doi.org/10.1016/j.nuclphysa.2014.09.061}
{\path{doi:http://dx.doi.org/10.1016/j.nuclphysa.2014.09.061}}.

\bibitem{Tripolt:2013jra}
R.-A. Tripolt, N.~Strodthoff, L.~von Smekal, J.~Wambach, {Spectral Functions
for the Quark-Meson Model Phase Diagram from the Functional Renormalization
Group}, Phys.Rev. D89 (2014) 034010.
\newblock \href {http://arxiv.org/abs/1311.0630} {\path{arXiv:1311.0630}},
\href {https://doi.org/10.1103/PhysRevD.89.034010}
{\path{doi:10.1103/PhysRevD.89.034010}}.

\bibitem{Kamikado:2013sia}
K.~Kamikado, N.~Strodthoff, L.~von Smekal, J.~Wambach, {Real-Time Correlation
Functions in the O(N) Model from the Functional Renormalization Group},
Eur.Phys.J. C74 (2014) 2806.
\newblock \href {http://arxiv.org/abs/1302.6199} {\path{arXiv:1302.6199}},
\href {https://doi.org/10.1140/epjc/s10052-014-2806-6}
{\path{doi:10.1140/epjc/s10052-014-2806-6}}.

\bibitem{Haas:2013hpa}
M.~Haas, L.~Fister, J.~M. Pawlowski, {Gluon spectral functions and transport
coefficients in Yang--Mills theory}, Phys.Rev. D90~(9) (2014) 091501.
\newblock \href {http://arxiv.org/abs/1308.4960} {\path{arXiv:1308.4960}},
\href {https://doi.org/10.1103/PhysRevD.90.091501}
{\path{doi:10.1103/PhysRevD.90.091501}}.

\bibitem{Christiansen:2014ypa}
N.~Christiansen, M.~Haas, J.~M. Pawlowski, N.~Strodthoff, {Transport
Coefficients in Yang--Mills Theory and QCD}, Phys. Rev. Lett. 115~(11) (2015)
112002.
\newblock \href {http://arxiv.org/abs/1411.7986} {\path{arXiv:1411.7986}},
\href {https://doi.org/10.1103/PhysRevLett.115.112002}
{\path{doi:10.1103/PhysRevLett.115.112002}}.

\bibitem{Wambach14}
J.~Wambach, R.-A. Tripolt, N.~Strodthoff, L.~von Smekal, {Spectral functions
from the functional renormalization group }, Nucl. Phys. A 928 (2014)
156--167.
\newblock \href
{https://doi.org/http://dx.doi.org/10.1016/j.nuclphysa.2014.04.027}
{\path{doi:http://dx.doi.org/10.1016/j.nuclphysa.2014.04.027}}.

\bibitem{Pawlowski:2015mia}
J.~M. Pawlowski, N.~Strodthoff, {Real time correlation functions and the
functional renormalization group}, Phys. Rev. D92~(9) (2015) 094009.
\newblock \href {http://arxiv.org/abs/1508.01160} {\path{arXiv:1508.01160}},
\href {https://doi.org/10.1103/PhysRevD.92.094009}
{\path{doi:10.1103/PhysRevD.92.094009}}.

\bibitem{Pawlowski18}
J.~M. Pawlowski, N.~Strodthoff, N.~Wink, {Finite temperature spectral functions
in the $O(N)$ model}, Phys. Rev. D 98 (2018) 074008.
\newblock \href {https://doi.org/10.1103/PhysRevD.98.074008}
{\path{doi:10.1103/PhysRevD.98.074008}}.

\bibitem{Khedri18}
A.~Kedri, T.~A. Costi, V.~Meden, {Nonequilibrium thermoelectric transport
through vibrating molecular quantum dots}, Phys. Rev. B 98 (2018) 195138.
\newblock \href {https://doi.org/10.1103/PhysRevB.98.195138}
{\path{doi:10.1103/PhysRevB.98.195138}}.

\bibitem{Cyrol:2018xeq}
A.~K. Cyrol, J.~M. Pawlowski, A.~Rothkopf, N.~Wink, {Reconstructing the gluon},
SciPost Phys. 5 (2018) 065.
\newblock \href {http://arxiv.org/abs/1804.00945} {\path{arXiv:1804.00945}},
\href {https://doi.org/10.21468/SciPostPhys.5.6.065}
{\path{doi:10.21468/SciPostPhys.5.6.065}}.

\bibitem{Debelhoir16a}
T.~Debelhoir, N.~Dupuis, {Simulating frustrated magnetism with spinor Bose
gases}, Phys. Rev. A 93 (2016) 051603.
\newblock \href {https://doi.org/10.1103/PhysRevA.93.051603}
{\path{doi:10.1103/PhysRevA.93.051603}}.

\bibitem{Debelhoir16b}
T.~Debelhoir, N.~Dupuis, {First-order phase transitions in spinor Bose gases
and frustrated magnets}, Phys. Rev. A 94 (2016) 053623.
\newblock \href {https://doi.org/10.1103/PhysRevA.94.053623}
{\path{doi:10.1103/PhysRevA.94.053623}}.

\bibitem{Antonenko95a}
S.~Antonenko, A.~Sokolov, K.~Varnashev, {Chiral transitions in
three-dimensional magnets and higher order $\epsilon$ expansion}, Phys. Lett.
A 208~(1–2) (1995) 161 -- 164.
\newblock \href
{https://doi.org/http://dx.doi.org/10.1016/0375-9601(95)00736-M}
{\path{doi:http://dx.doi.org/10.1016/0375-9601(95)00736-M}}.

\bibitem{Calabrese03sm}
P.~{Calabrese}, P.~{Parruccini}, {Five-loop $\epsilon$ expansion for
O(n)$\times$O(m) spin models}, Nuclear Physics B 679 (2004) 568–596.
\newblock \href {http://arxiv.org/abs/cond-mat/0308037}
{\path{arXiv:cond-mat/0308037}}, \href
{https://doi.org/10.1016/j.nuclphysb.2003.12.002}
{\path{doi:10.1016/j.nuclphysb.2003.12.002}}.

\bibitem{Pelissetto:2000ne}
A.~Pelissetto, P.~Rossi, E.~Vicari, {The Critical behavior of frustrated spin
models with noncollinear order}, Phys. Rev. B63 (2001) 140414.
\newblock \href {http://arxiv.org/abs/cond-mat/0007389}
{\path{arXiv:cond-mat/0007389}}, \href
{https://doi.org/10.1103/PhysRevB.63.140414}
{\path{doi:10.1103/PhysRevB.63.140414}}.

\bibitem{Calabrese02}
P.~Calabrese, P.~Parruccini, A.~I. Sokolov, {Chiral phase transitions: Focus
driven critical behavior in systems with planar and vector ordering}, Phys.
Rev. B 66 (2002) 180403.
\newblock \href {https://doi.org/10.1103/PhysRevB.66.180403}
{\path{doi:10.1103/PhysRevB.66.180403}}.

\bibitem{Nakayama:2014sba}
Y.~Nakayama, T.~Ohtsuki, {Bootstrapping phase transitions in QCD and frustrated
spin systems}, Phys. Rev. D91~(2) (2015) 021901.
\newblock \href {http://arxiv.org/abs/1407.6195} {\path{arXiv:1407.6195}},
\href {https://doi.org/10.1103/PhysRevD.91.021901}
{\path{doi:10.1103/PhysRevD.91.021901}}.

\bibitem{Stergiou:2019dcv}
A.~Stergiou, {Bootstrapping MN and Tetragonal CFTs in Three Dimensions},
SciPost Phys. 7 (2019) 010.
\newblock \href {http://arxiv.org/abs/1904.00017} {\path{arXiv:1904.00017}},
\href {https://doi.org/10.21468/SciPostPhys.7.1.010}
{\path{doi:10.21468/SciPostPhys.7.1.010}}.

\bibitem{Henriksson:2020fqi}
J.~Henriksson, S.~R. Kousvos, A.~Stergiou, {Analytic and Numerical Bootstrap of
CFTs with $O(m)\times O(n)$ Global Symmetry in 3D} (4 2020).
\newblock \href {http://arxiv.org/abs/2004.14388} {\path{arXiv:2004.14388}}.

\bibitem{Poland19}
D.~Poland, S.~Rychkov, A.~Vichi,
\href{https://link.aps.org/doi/10.1103/RevModPhys.91.015002}{The conformal
bootstrap: Theory, numerical techniques, and applications}, Rev. Mod. Phys.
91 (2019) 015002.
\newblock \href {https://doi.org/10.1103/RevModPhys.91.015002}
{\path{doi:10.1103/RevModPhys.91.015002}}.
\newline\urlprefix\url{https://link.aps.org/doi/10.1103/RevModPhys.91.015002}

\bibitem{Itakura01}
M.~Itakura, {Monte Carlo renormalization group study of the Heisenberg and XY
antiferromagnet on the stacked triangular lattice and the chiral phi**4
model}, J. Phys. Soc. Jap. 72 (2003) 74–82.
\newblock \href {http://arxiv.org/abs/cond-mat/0110306}
{\path{arXiv:cond-mat/0110306}}, \href {https://doi.org/10.1143/JPSJ.72.74}
{\path{doi:10.1143/JPSJ.72.74}}.

\bibitem{Ngo08}
V.~T. Ngo, H.~T. Diep,
\href{https://link.aps.org/doi/10.1103/PhysRevE.78.031119}{Phase transition
in heisenberg stacked triangular antiferromagnets: End of a controversy},
Phys. Rev. E 78 (2008) 031119.
\newblock \href {https://doi.org/10.1103/PhysRevE.78.031119}
{\path{doi:10.1103/PhysRevE.78.031119}}.
\newline\urlprefix\url{https://link.aps.org/doi/10.1103/PhysRevE.78.031119}

\bibitem{Delamotte:1998ay}
B.~Delamotte, D.~Mouhanna, P.~Lecheminant, {The Wilson renormalization group
approach of the principal chiral model around two-dimensions}, Phys. Rev. B59
(1999) 6006–6009.
\newblock \href {http://arxiv.org/abs/hep-th/9805210}
{\path{arXiv:hep-th/9805210}}, \href
{https://doi.org/10.1103/PhysRevB.59.6006}
{\path{doi:10.1103/PhysRevB.59.6006}}.

\bibitem{Tissier:1999hv}
M.~Tissier, D.~Mouhanna, B.~Delamotte, {A Nonperturbative approach of the
principal chiral model between two-dimensions and four-dimensions}, Phys.
Rev. B61 (2000) 15327--15330.
\newblock \href {http://arxiv.org/abs/cond-mat/9908352}
{\path{arXiv:cond-mat/9908352}}, \href
{https://doi.org/10.1103/PhysRevB.61.15327}
{\path{doi:10.1103/PhysRevB.61.15327}}.

\bibitem{Tissier:2000tz}
M.~Tissier, B.~Delamotte, D.~Mouhanna, {Heisenberg frustrated magnets: A
Nonperturbative approach}, Phys. Rev. Lett. 84 (2000) 5208--5211.
\newblock \href {http://arxiv.org/abs/cond-mat/0001350}
{\path{arXiv:cond-mat/0001350}}, \href
{https://doi.org/10.1103/PhysRevLett.84.5208}
{\path{doi:10.1103/PhysRevLett.84.5208}}.

\bibitem{Tissier:2001uk}
M.~Tissier, B.~Delamotte, D.~Mouhanna, {XY frustrated systems: Continuous
exponents in discontinuous phase transitions}, Phys. Rev. B67 (2003) 134422.
\newblock \href {http://arxiv.org/abs/cond-mat/0107183}
{\path{arXiv:cond-mat/0107183}}, \href
{https://doi.org/10.1103/PhysRevB.67.134422}
{\path{doi:10.1103/PhysRevB.67.134422}}.

\bibitem{Delamotte:2016acs}
B.~Delamotte, M.~Dudka, D.~Mouhanna, S.~Yabunaka, {Functional renormalization
group approach to noncollinear magnets}, Phys. Rev. B93~(6) (2016) 064405.
\newblock \href {http://arxiv.org/abs/1510.00169} {\path{arXiv:1510.00169}},
\href {https://doi.org/10.1103/PhysRevB.93.064405}
{\path{doi:10.1103/PhysRevB.93.064405}}.

\bibitem{Pelissetto:2001fi}
A.~Pelissetto, P.~Rossi, E.~Vicari, {Large n critical behavior of $O(n) \times
O(m)$ spin models}, Nucl. Phys. B607 (2001) 605–634.
\newblock \href {http://arxiv.org/abs/hep-th/0104024}
{\path{arXiv:hep-th/0104024}}, \href
{https://doi.org/10.1016/S0550-3213(01)00223-1}
{\path{doi:10.1016/S0550-3213(01)00223-1}}.

\bibitem{Calabrese04a}
P.~Calabrese, P.~Parruccini, A.~Pelissetto, E.~Vicari, {Critical behavior of
$\mathrm{O}(2)\ensuremath{\otimes}\mathrm{O}(N)$ symmetric models}, Phys.
Rev. B 70 (2004) 174439.
\newblock \href {https://doi.org/10.1103/PhysRevB.70.174439}
{\path{doi:10.1103/PhysRevB.70.174439}}.

\bibitem{Berges:1996ja}
J.~Berges, C.~Wetterich, {Equation of state and coarse grained free energy for
matrix models}, Nucl. Phys. B487 (1997) 675.
\newblock \href {http://arxiv.org/abs/hep-th/9609019}
{\path{arXiv:hep-th/9609019}}, \href
{https://doi.org/10.1016/S0550-3213(96)00670-0}
{\path{doi:10.1016/S0550-3213(96)00670-0}}.

\bibitem{Kindermann:2000vj}
M.~Kindermann, C.~Wetterich, {Phase transitions in liquid helium 3}, Phys. Rev.
Lett. 86 (2001) 1034–1037.
\newblock \href {http://arxiv.org/abs/cond-mat/0008332}
{\path{arXiv:cond-mat/0008332}}, \href
{https://doi.org/10.1103/PhysRevLett.86.1034}
{\path{doi:10.1103/PhysRevLett.86.1034}}.

\bibitem{Bornholdt:1994rf}
S.~Bornholdt, N.~Tetradis, C.~Wetterich, {Coleman-Weinberg phase transition in
two scalar models}, Phys. Lett. B348 (1995) 89–99.
\newblock \href {http://arxiv.org/abs/hep-th/9408132}
{\path{arXiv:hep-th/9408132}}, \href
{https://doi.org/10.1016/0370-2693(95)00045-M}
{\path{doi:10.1016/0370-2693(95)00045-M}}.

\bibitem{Bornholdt:1995rn}
S.~Bornholdt, N.~Tetradis, C.~Wetterich, {High temperature phase transition in
two scalar theories}, Phys. Rev. D53 (1996) 4552–4569.
\newblock \href {http://arxiv.org/abs/hep-ph/9503282}
{\path{arXiv:hep-ph/9503282}}, \href
{https://doi.org/10.1103/PhysRevD.53.4552}
{\path{doi:10.1103/PhysRevD.53.4552}}.

\bibitem{Bornholdt:1996ir}
S.~Bornholdt, P.~Buttner, N.~Tetradis, C.~Wetterich, {Flow of the coarse
grained free energy for crossover phenomena}, Int. J. Mod. Phys. A14 (1999)
899–918.
\newblock \href {http://arxiv.org/abs/cond-mat/9603129}
{\path{arXiv:cond-mat/9603129}}, \href
{https://doi.org/10.1142/S0217751X99000440}
{\path{doi:10.1142/S0217751X99000440}}.

\bibitem{Tissier:2002zz}
M.~Tissier, D.~Mouhanna, J.~Vidal, B.~Delamotte, {Randomly dilute Ising model:
A nonperturbative approach}, Phys. Rev. B65 (2002) 140402.
\newblock \href {https://doi.org/10.1103/PhysRevB.65.140402}
{\path{doi:10.1103/PhysRevB.65.140402}}.

\bibitem{Chlebicki:2019yks}
A.~Chlebicki, P.~Jakubczyk, {Criticality of the $O(2)$ model with cubic
anisotropies from nonperturbative renormalization}, Phys. Rev. E100~(5)
(2019) 052106.
\newblock \href {http://arxiv.org/abs/1909.10600} {\path{arXiv:1909.10600}},
\href {https://doi.org/10.1103/PhysRevE.100.052106}
{\path{doi:10.1103/PhysRevE.100.052106}}.

\bibitem{An:2016lni}
X.~An, D.~Mesterh\'azy, M.~A. Stephanov, {Functional renormalization group
approach to the Yang-Lee edge singularity}, JHEP 07 (2016) 041.
\newblock \href {http://arxiv.org/abs/1605.06039} {\path{arXiv:1605.06039}},
\href {https://doi.org/10.1007/JHEP07(2016)041}
{\path{doi:10.1007/JHEP07(2016)041}}.

\bibitem{Zambelli:2016cbw}
L.~Zambelli, O.~Zanusso, {Lee-Yang model from the functional renormalization
group}, Phys. Rev. D95~(8) (2017) 085001.
\newblock \href {http://arxiv.org/abs/1612.08739} {\path{arXiv:1612.08739}},
\href {https://doi.org/10.1103/PhysRevD.95.085001}
{\path{doi:10.1103/PhysRevD.95.085001}}.

\bibitem{Lauscher:2000ux}
O.~Lauscher, M.~Reuter, C.~Wetterich, {Rotation symmetry breaking condensate in
a scalar theory}, Phys. Rev. D62 (2000) 125021.
\newblock \href {http://arxiv.org/abs/hep-th/0006099}
{\path{arXiv:hep-th/0006099}}, \href
{https://doi.org/10.1103/PhysRevD.62.125021}
{\path{doi:10.1103/PhysRevD.62.125021}}.

\bibitem{Leonard15}
F.~L\'eonard, B.~Delamotte, {Critical Exponents Can Be Different on the Two
Sides of a Transition: A Generic Mechanism}, Phys. Rev. Lett. 115 (2015)
200601.
\newblock \href {https://doi.org/10.1103/PhysRevLett.115.200601}
{\path{doi:10.1103/PhysRevLett.115.200601}}.

\bibitem{Meurice:2007zg}
Y.~Meurice, {Nonlinear Aspects of the Renormalization Group Flows of Dyson's
Hierarchical Model}, J. Phys. A40 (2007) R39.
\newblock \href {http://arxiv.org/abs/hep-th/0701191}
{\path{arXiv:hep-th/0701191}}, \href
{https://doi.org/10.1088/1751-8113/40/23/R01}
{\path{doi:10.1088/1751-8113/40/23/R01}}.

\bibitem{Litim:2007jb}
D.~F. Litim, {Towards functional flows for hierarchical models}, Phys.Rev. D76
(2007) 105001.
\newblock \href {http://arxiv.org/abs/0704.1514} {\path{arXiv:0704.1514}},
\href {https://doi.org/10.1103/PhysRevD.76.105001}
{\path{doi:10.1103/PhysRevD.76.105001}}.

\bibitem{Eichhorn:2014asa}
A.~Eichhorn, D.~Mesterh\'azy, M.~M. Scherer, {Stability of fixed points and
generalized critical behavior in multifield models}, Phys. Rev. E90~(5)
(2014) 052129.
\newblock \href {http://arxiv.org/abs/1407.7442} {\path{arXiv:1407.7442}},
\href {https://doi.org/10.1103/PhysRevE.90.052129}
{\path{doi:10.1103/PhysRevE.90.052129}}.

\bibitem{Eichhorn:2015woa}
A.~Eichhorn, T.~Helfer, D.~Mesterh\'azy, M.~M. Scherer, {Discovering and
quantifying nontrivial fixed points in multi-field models}, Eur. Phys. J.
C76~(2) (2016) 88.
\newblock \href {http://arxiv.org/abs/1510.04807} {\path{arXiv:1510.04807}},
\href {https://doi.org/10.1140/epjc/s10052-016-3921-3}
{\path{doi:10.1140/epjc/s10052-016-3921-3}}.

\bibitem{Boettcher:2015pja}
I.~Boettcher, {Scaling relations and multicritical phenomena from Functional
Renormalization}, Phys. Rev. E91~(6) (2015) 062112.
\newblock \href {http://arxiv.org/abs/1503.07817} {\path{arXiv:1503.07817}},
\href {https://doi.org/10.1103/PhysRevE.91.062112}
{\path{doi:10.1103/PhysRevE.91.062112}}.

\bibitem{Borchardt:2016kco}
J.~Borchardt, A.~Eichhorn, {Universal behavior of coupled order parameters
below three dimensions}, Phys. Rev. E94~(4) (2016) 042105.
\newblock \href {http://arxiv.org/abs/1606.07449} {\path{arXiv:1606.07449}},
\href {https://doi.org/10.1103/PhysRevE.94.042105}
{\path{doi:10.1103/PhysRevE.94.042105}}.

\bibitem{Kownacki09}
J.-P. Kownacki, D.~Mouhanna, Crumpling transition and flat phase of polymerized
phantom membranes, Phys. Rev. E 79 (2009) 040101(R).
\newblock \href {https://doi.org/10.1103/PhysRevE.79.040101}
{\path{doi:10.1103/PhysRevE.79.040101}}.

\bibitem{Essafi11}
K.~{Essafi}, J.~P. {Kownacki}, D.~{Mouhanna}, {Crumpled-to-Tubule Transition in
Anisotropic Polymerized Membranes: Beyond the {\ensuremath{\gamma}}
Expansion}, \prl 106 (2011) 128102.
\newblock \href {http://arxiv.org/abs/1011.6173} {\path{arXiv:1011.6173}},
\href {https://doi.org/10.1103/PhysRevLett.106.128102}
{\path{doi:10.1103/PhysRevLett.106.128102}}.

\bibitem{Essafi14a}
K.~{Essafi}, J.~P. {Kownacki}, D.~{Mouhanna}, {First-order phase transitions in
polymerized phantom membranes}, \pre 89 (2014) 042101.
\newblock \href {http://arxiv.org/abs/1402.0426} {\path{arXiv:1402.0426}},
\href {https://doi.org/10.1103/PhysRevE.89.042101}
{\path{doi:10.1103/PhysRevE.89.042101}}.

\bibitem{Essafi14b}
K.~{Essafi}, J.-P. {Kownacki}, D.~{Mouhanna}, {Nonperturbative Renormalization
Group Approach to Polymerized Membranes}, in: APS March Meeting Abstracts,
Vol. 2014, 2014, p. L18.014.

\bibitem{Coquand2016}
O.~{Coquand}, D.~{Mouhanna}, {Flat phase of quantum polymerized membranes},
\pre 94 (2016) 032125.
\newblock \href {http://arxiv.org/abs/1607.03335} {\path{arXiv:1607.03335}},
\href {https://doi.org/10.1103/PhysRevE.94.032125}
{\path{doi:10.1103/PhysRevE.94.032125}}.

\bibitem{Coquand2017}
O.~Coquand, K.~Essafi, J.~P. Kownacki, D.~Mouhanna, {Glassy phase in quenched
disordered crystalline membranes}, Phys. Rev. E97~(3) (2018) 030102.
\newblock \href {http://arxiv.org/abs/1708.08364} {\path{arXiv:1708.08364}},
\href {https://doi.org/10.1103/PhysRevE.97.030102}
{\path{doi:10.1103/PhysRevE.97.030102}}.

\bibitem{PhysRevE.84.021124}
P.~Jakubczyk, Capillary-wave models and the effective-average-action scheme of
functional renormalization group, Phys. Rev. E 84 (2011) 021124.
\newblock \href {https://doi.org/10.1103/PhysRevE.84.021124}
{\path{doi:10.1103/PhysRevE.84.021124}}.

\bibitem{PhysRevB.86.075142}
P.~Jakubczyk, Quantum interface unbinding transitions, Phys. Rev. B 86 (2012)
075142.
\newblock \href {https://doi.org/10.1103/PhysRevB.86.075142}
{\path{doi:10.1103/PhysRevB.86.075142}}.

\bibitem{Jakubczyk_2015}
P.~Jakubczyk, M.~Napi{\'{o}}rkowski, F.~Benitez, Quantum wetting transitions in
two dimensions: An alternative path to non-universal interfacial
singularities, {EPL} (Europhysics Letters) 110~(1) (2015) 16002.
\newblock \href {https://doi.org/10.1209/0295-5075/110/16002}
{\path{doi:10.1209/0295-5075/110/16002}}.

\bibitem{Hofling:2002hj}
F.~Hofling, C.~Nowak, C.~Wetterich, {Phase transition and critical behavior of
the D = 3 Gross-Neveu model}, Phys. Rev. B66 (2002) 205111.
\newblock \href {http://arxiv.org/abs/cond-mat/0203588}
{\path{arXiv:cond-mat/0203588}}, \href
{https://doi.org/10.1103/PhysRevB.66.205111}
{\path{doi:10.1103/PhysRevB.66.205111}}.

\bibitem{Jaeckel:2002rm}
J.~Jaeckel, C.~Wetterich, {Flow equations without mean field ambiguity},
Phys.Rev. D68 (2003) 025020.
\newblock \href {http://arxiv.org/abs/hep-ph/0207094}
{\path{arXiv:hep-ph/0207094}}, \href
{https://doi.org/10.1103/PhysRevD.68.025020}
{\path{doi:10.1103/PhysRevD.68.025020}}.

\bibitem{Wetterich:2002ky}
C.~Wetterich, {Bosonic effective action for interacting fermions}, Phys. Rev.
B75 (2007) 085102.
\newblock \href {http://arxiv.org/abs/cond-mat/0208361}
{\path{arXiv:cond-mat/0208361}}, \href
{https://doi.org/10.1103/PhysRevB.75.085102}
{\path{doi:10.1103/PhysRevB.75.085102}}.

\bibitem{Gies:2009da}
H.~Gies, L.~Janssen, S.~Rechenberger, M.~M. Scherer, {Phase transition and
critical behavior of d=3 chiral fermion models with left/right asymmetry},
Phys. Rev. D81 (2010) 025009.
\newblock \href {http://arxiv.org/abs/0910.0764} {\path{arXiv:0910.0764}},
\href {https://doi.org/10.1103/PhysRevD.81.025009}
{\path{doi:10.1103/PhysRevD.81.025009}}.

\bibitem{Gies:2010st}
H.~Gies, L.~Janssen, {UV fixed-point structure of the three-dimensional
Thirring model}, Phys. Rev. D 82 (2010) 085018.
\newblock \href {http://arxiv.org/abs/1006.3747} {\path{arXiv:1006.3747}},
\href {https://doi.org/10.1103/PhysRevD.82.085018}
{\path{doi:10.1103/PhysRevD.82.085018}}.

\bibitem{Braun:2010tt}
J.~Braun, H.~Gies, D.~D. Scherer, {Asymptotic safety: a simple example}, Phys.
Rev. D 83 (2011) 085012.
\newblock \href {http://arxiv.org/abs/1011.1456} {\path{arXiv:1011.1456}},
\href {https://doi.org/10.1103/PhysRevD.83.085012}
{\path{doi:10.1103/PhysRevD.83.085012}}.

\bibitem{Scherer:2012nn}
D.~D. Scherer, H.~Gies, {Renormalization Group Study of Magnetic Catalysis in
the 3d Gross-Neveu Model}, Phys. Rev. B85 (2012) 195417.
\newblock \href {http://arxiv.org/abs/1201.3746} {\path{arXiv:1201.3746}},
\href {https://doi.org/10.1103/PhysRevB.85.195417}
{\path{doi:10.1103/PhysRevB.85.195417}}.

\bibitem{Janssen:2012pq}
L.~Janssen, H.~Gies, {Critical behavior of the (2+1)-dimensional Thirring
model}, Phys. Rev. D86 (2012) 105007.
\newblock \href {http://arxiv.org/abs/1208.3327} {\path{arXiv:1208.3327}},
\href {https://doi.org/10.1103/PhysRevD.86.105007}
{\path{doi:10.1103/PhysRevD.86.105007}}.

\bibitem{Scherer:2013pda}
D.~D. Scherer, J.~Braun, H.~Gies, {Many-flavor Phase Diagram of the (2+1)d
Gross-Neveu Model at Finite Temperature}, J. Phys. A46 (2013) 285002.
\newblock \href {http://arxiv.org/abs/1212.4624} {\path{arXiv:1212.4624}},
\href {https://doi.org/10.1088/1751-8113/46/28/285002}
{\path{doi:10.1088/1751-8113/46/28/285002}}.

\bibitem{Strack:2013jta}
P.~Strack, P.~Jakubczyk, {Fluctuations of imbalanced fermionic superfluids in
two dimensions induce continuous quantum phase transitions and non-Fermi
liquid behavior}, Phys. Rev. X4~(2) (2014) 021012.
\newblock \href {http://arxiv.org/abs/1311.4885} {\path{arXiv:1311.4885}},
\href {https://doi.org/10.1103/PhysRevX.4.021012}
{\path{doi:10.1103/PhysRevX.4.021012}}.

\bibitem{Janssen:2014gea}
L.~Janssen, I.~F. Herbut, {Antiferromagnetic critical point on graphene's
honeycomb lattice: A functional renormalization group approach}, Phys. Rev.
B89 (2014) 205403.
\newblock \href {http://arxiv.org/abs/1402.6277} {\path{arXiv:1402.6277}},
\href {https://doi.org/10.1103/PhysRevB.89.205403}
{\path{doi:10.1103/PhysRevB.89.205403}}.

\bibitem{Bauer:2015qwa}
C.~Bauer, A.~Rückriegel, A.~Sharma, P.~Kopietz, {Nonperturbative
renormalization group calculation of quasiparticle velocity and dielectric
function of graphene}, Phys. Rev. B92~(12) (2015) 121409.
\newblock \href {http://arxiv.org/abs/1506.08825} {\path{arXiv:1506.08825}},
\href {https://doi.org/10.1103/PhysRevB.92.121409}
{\path{doi:10.1103/PhysRevB.92.121409}}.

\bibitem{Gehring:2015vja}
F.~Gehring, H.~Gies, L.~Janssen, {Fixed-point structure of low-dimensional
relativistic fermion field theories: Universality classes and emergent
symmetry}, Phys. Rev. D 92~(8) (2015) 085046.
\newblock \href {http://arxiv.org/abs/1506.07570} {\path{arXiv:1506.07570}},
\href {https://doi.org/10.1103/PhysRevD.92.085046}
{\path{doi:10.1103/PhysRevD.92.085046}}.

\bibitem{Knorr:2016sfs}
B.~Knorr, {Ising and Gross-Neveu model in next-to-leading order}, Phys. Rev. B
94~(24) (2016) 245102.
\newblock \href {http://arxiv.org/abs/1609.03824} {\path{arXiv:1609.03824}},
\href {https://doi.org/10.1103/PhysRevB.94.245102}
{\path{doi:10.1103/PhysRevB.94.245102}}.

\bibitem{Janssen:2016xvc}
L.~Janssen, I.~F. Herbut, {Phase diagram of electronic systems with quadratic
Fermi nodes in $2 < d < 4$: $2+\epsilon$ expansion, $4-\epsilon$ expansion,
and functional renormalization group}, Phys. Rev. B 95~(7) (2017) 075101.
\newblock \href {http://arxiv.org/abs/1611.04594} {\path{arXiv:1611.04594}},
\href {https://doi.org/10.1103/PhysRevB.95.075101}
{\path{doi:10.1103/PhysRevB.95.075101}}.

\bibitem{Ihrig:2018hho}
B.~Ihrig, L.~N. Mihaila, M.~M. Scherer, {Critical behavior of Dirac fermions
from perturbative renormalization}, Phys. Rev. B98~(12) (2018) 125109.
\newblock \href {http://arxiv.org/abs/1806.04977} {\path{arXiv:1806.04977}},
\href {https://doi.org/10.1103/PhysRevB.98.125109}
{\path{doi:10.1103/PhysRevB.98.125109}}.

\bibitem{Berges97}
J.~Berges, N.~Tetradis, C.~Wetterich, {Coarse graining and first order phase
transitions}, Phys. Lett. B 393 (1997) 387--394.
\newblock \href
{https://doi.org/http://dx.doi.org/10.1016/S0370-2693(96)01654-1}
{\path{doi:http://dx.doi.org/10.1016/S0370-2693(96)01654-1}}.

\bibitem{Strumia99}
A.~Strumia, N.~Tetradis, {A consistent calculation of bubble-nucleation rates},
Nucl. Phys. B 542 (1999) 719--741.
\newblock \href {https://doi.org/https://doi.org/10.1016/S0550-3213(98)00804-9}
{\path{doi:https://doi.org/10.1016/S0550-3213(98)00804-9}}.

\bibitem{Strumia99a}
A.~Strumia, N.~Tetradis, C.~Wetterich, The region of validity of homogeneous
nucleation theory, Phys. Lett. B 467 (1999) 279--288.
\newblock \href {https://doi.org/https://doi.org/10.1016/S0370-2693(99)01158-2}
{\path{doi:https://doi.org/10.1016/S0370-2693(99)01158-2}}.

\bibitem{Strumia99b}
A.~Strumia, N.~Tetradis, Bubble-nucleation rates for radiatively induced
first-order phase transitions, Nucl. Phys. B 554 (1999) 697--718.
\newblock \href {https://doi.org/https://doi.org/10.1016/S0550-3213(99)00285-0}
{\path{doi:https://doi.org/10.1016/S0550-3213(99)00285-0}}.

\bibitem{Strumia99c}
A.~Strumia, N.~Tetradis, Testing nucleation theory in two dimensions, Nucl.
Phys. B 560 (1999) 482 -- 496.
\newblock \href {https://doi.org/https://doi.org/10.1016/S0550-3213(99)00455-1}
{\path{doi:https://doi.org/10.1016/S0550-3213(99)00455-1}}.

\bibitem{Tetradis98}
N.~Tetradis, {Renormalization-group study of weakly first-order phase
transitions }, Phys. Lett. B 431 (1998) 380--386.
\newblock \href
{https://doi.org/http://dx.doi.org/10.1016/S0370-2693(98)00575-9}
{\path{doi:http://dx.doi.org/10.1016/S0370-2693(98)00575-9}}.

\bibitem{Munster:2000kk}
G.~Munster, A.~Strumia, N.~Tetradis, {Comparison of two methods for calculating
nucleation rates}, Phys. Lett. A271 (2000) 80–86.
\newblock \href {http://arxiv.org/abs/cond-mat/0002278}
{\path{arXiv:cond-mat/0002278}}, \href
{https://doi.org/10.1016/S0375-9601(00)00349-2}
{\path{doi:10.1016/S0375-9601(00)00349-2}}.

\bibitem{Tissier00a}
M.~Tissier, B.~Delamotte, D.~Mouhanna, {Frustrated Heisenberg Magnets: A
Nonperturbative Approach}, Phys. Rev. Lett. 84 (2000) 5208--5211.
\newblock \href {https://doi.org/10.1103/PhysRevLett.84.5208}
{\path{doi:10.1103/PhysRevLett.84.5208}}.

\bibitem{Qin:2018xox}
B.~Qin, D.~Hou, M.~Huang, D.~Li, H.~Zhang, {Nonperturbative renormalization
group for the Landau--de Gennes model}, Phys. Rev. B 98~(1) (2018) 014102.
\newblock \href {http://arxiv.org/abs/1803.03683} {\path{arXiv:1803.03683}},
\href {https://doi.org/10.1103/PhysRevB.98.014102}
{\path{doi:10.1103/PhysRevB.98.014102}}.

\bibitem{Kapoyannis:2000sp}
A.~S. Kapoyannis, N.~Tetradis, {Quantum mechanical tunneling and the
renormalization group}, Phys. Lett. A276 (2000) 225–232.
\newblock \href {http://arxiv.org/abs/hep-th/0010180}
{\path{arXiv:hep-th/0010180}}, \href
{https://doi.org/10.1016/S0375-9601(00)00671-X}
{\path{doi:10.1016/S0375-9601(00)00671-X}}.

\bibitem{Zappala:2001nv}
D.~Zappala, {Improving the renormalization group approach to the quantum
mechanical double well potential}, Phys.Lett. A290 (2001) 35--40.
\newblock \href {http://arxiv.org/abs/quant-ph/0108019}
{\path{arXiv:quant-ph/0108019}}, \href
{https://doi.org/10.1016/S0375-9601(01)00642-9}
{\path{doi:10.1016/S0375-9601(01)00642-9}}.

\bibitem{Rulquin15}
C.~{Rulquin}, P.~{Urbani}, G.~{Biroli}, G.~{Tarjus}, M.~{Tarzia},
{Nonperturbative fluctuations and metastability in a simple model: from
observables to microscopic theory and back}, arXiv e-prints (2015)
arXiv:1507.02498\href {http://arxiv.org/abs/1507.02498}
{\path{arXiv:1507.02498}}.

\bibitem{Bergerhoff:1995zq}
B.~Bergerhoff, F.~Freire, D.~Litim, S.~Lola, C.~Wetterich, {Phase diagram of
superconductors}, Phys.Rev. B53 (1996) 5734--5757.
\newblock \href {http://arxiv.org/abs/hep-ph/9503334}
{\path{arXiv:hep-ph/9503334}}, \href
{https://doi.org/10.1103/PhysRevB.53.5734}
{\path{doi:10.1103/PhysRevB.53.5734}}.

\bibitem{Bergerhoff:1995zm}
B.~Bergerhoff, D.~Litim, S.~Lola, C.~Wetterich, {Phase transition of N
component superconductors}, Int.J.Mod.Phys. A11 (1996) 4273--4306.
\newblock \href {http://arxiv.org/abs/cond-mat/9502039}
{\path{arXiv:cond-mat/9502039}}, \href
{https://doi.org/10.1142/S0217751X96001991}
{\path{doi:10.1142/S0217751X96001991}}.

\bibitem{Caillol06}
J.~M. Caillol, {Non-perturbative renormalization group for simple fluids}, Mol.
Phys. 104 (2006) 1931.
\newblock \href {https://doi.org/10.1080/00268970600740774}
{\path{doi:10.1080/00268970600740774}}.

\bibitem{Tarjus11}
G.~{Tarjus}, M.-L. {Rosinberg}, E.~{Kierlik}, M.~{Tissier}, {Hierarchical
reference theory of critical fluids in disordered porous media}, Molecular
Physics 109 (2011) 2863–2887.
\newblock \href {http://arxiv.org/abs/1107.1605} {\path{arXiv:1107.1605}},
\href {https://doi.org/10.1080/00268976.2011.620024}
{\path{doi:10.1080/00268976.2011.620024}}.

\bibitem{Boettcher:2013kia}
I.~Boettcher, J.~M. Pawlowski, C.~Wetterich, {Critical temperature and
superfluid gap of the Unitary Fermi Gas from Functional Renormalization},
Phys.Rev. A89 (2014) 053630.
\newblock \href {http://arxiv.org/abs/1312.0505} {\path{arXiv:1312.0505}},
\href {https://doi.org/10.1103/PhysRevA.89.053630}
{\path{doi:10.1103/PhysRevA.89.053630}}.

\bibitem{Rancon12b}
A.~Ran\c{c}on, N.~Dupuis, {Universal thermodynamics of a two-dimensional Bose
gas}, Phys. Rev. A 85 (2012) 063607.
\newblock \href {https://doi.org/10.1103/PhysRevA.85.063607}
{\path{doi:10.1103/PhysRevA.85.063607}}.

\bibitem{Ellwanger:1994iz}
U.~Ellwanger, {Flow equations and BRS invariance for Yang-Mills theories},
Phys. Lett. B335 (1994) 364--370.
\newblock \href {http://arxiv.org/abs/hep-th/9402077}
{\path{arXiv:hep-th/9402077}}, \href
{https://doi.org/10.1016/0370-2693(94)90365-4}
{\path{doi:10.1016/0370-2693(94)90365-4}}.

\bibitem{Codello:2012sn}
A.~Codello, G.~D'Odorico, C.~Pagani, R.~Percacci, {The Renormalization Group
and Weyl-invariance}, Class. Quant. Grav. 30 (2013) 115015.
\newblock \href {http://arxiv.org/abs/1210.3284} {\path{arXiv:1210.3284}},
\href {https://doi.org/10.1088/0264-9381/30/11/115015}
{\path{doi:10.1088/0264-9381/30/11/115015}}.

\bibitem{Codello:2015ana}
A.~Codello, G.~D'Odorico, C.~Pagani, {Functional and Local Renormalization
Groups}, Phys. Rev. D91~(12) (2015) 125016.
\newblock \href {http://arxiv.org/abs/1502.02439} {\path{arXiv:1502.02439}},
\href {https://doi.org/10.1103/PhysRevD.91.125016}
{\path{doi:10.1103/PhysRevD.91.125016}}.

\bibitem{Rosten:2016zap}
O.~J. Rosten, {A Conformal Fixed-Point Equation for the Effective Average
Action}, Int. J. Mod. Phys. A34~(05) (2019) 1950027.
\newblock \href {http://arxiv.org/abs/1605.01729} {\path{arXiv:1605.01729}},
\href {https://doi.org/10.1142/S0217751X19500271}
{\path{doi:10.1142/S0217751X19500271}}.

\bibitem{Pagani:2017tdr}
C.~Pagani, H.~Sonoda, {Products of composite operators in the exact
renormalization group formalism}, PTEP 2018~(2) (2018) 023B02.
\newblock \href {http://arxiv.org/abs/1707.09138} {\path{arXiv:1707.09138}},
\href {https://doi.org/10.1093/ptep/ptx189} {\path{doi:10.1093/ptep/ptx189}}.

\bibitem{DePolsi:2018vxc}
G.~{De Polsi}, M.~Tissier, N.~Wschebor, {Exact critical exponents for vector
operators in the 3d Ising model and conformal invariance} (2018).
\newblock \href {http://arxiv.org/abs/1804.08374} {\path{arXiv:1804.08374}}.

\bibitem{Morris:2018zgy}
T.~R. Morris, R.~Percacci, {Trace anomaly and infrared cutoffs}, Phys. Rev.
D99~(10) (2019) 105007.
\newblock \href {http://arxiv.org/abs/1810.09824} {\path{arXiv:1810.09824}},
\href {https://doi.org/10.1103/PhysRevD.99.105007}
{\path{doi:10.1103/PhysRevD.99.105007}}.

\bibitem{Rosten:2014oja}
O.~J. Rosten, {On Functional Representations of the Conformal Algebra}, Eur.
Phys. J. C77~(7) (2017) 477.
\newblock \href {http://arxiv.org/abs/1411.2603} {\path{arXiv:1411.2603}},
\href {https://doi.org/10.1140/epjc/s10052-017-5049-5}
{\path{doi:10.1140/epjc/s10052-017-5049-5}}.

\bibitem{Rosten:2016nmc}
O.~J. Rosten, {A Wilsonian Energy-Momentum Tensor}, Eur. Phys. J. C78~(4)
(2018) 312.
\newblock \href {http://arxiv.org/abs/1605.01055} {\path{arXiv:1605.01055}},
\href {https://doi.org/10.1140/epjc/s10052-018-5783-3}
{\path{doi:10.1140/epjc/s10052-018-5783-3}}.

\bibitem{Sonoda:2017zgl}
H.~Sonoda, {Conformal invariance for Wilson actions}, PTEP 2017~(8) (2017)
083B05.
\newblock \href {http://arxiv.org/abs/1705.01239} {\path{arXiv:1705.01239}},
\href {https://doi.org/10.1093/ptep/ptx114} {\path{doi:10.1093/ptep/ptx114}}.

\bibitem{Sonoda:2015pva}
H.~Sonoda, {Construction of the Energy-Momentum Tensor for Wilson Actions},
Phys. Rev. D92~(6) (2015) 065016.
\newblock \href {http://arxiv.org/abs/1504.02831} {\path{arXiv:1504.02831}},
\href {https://doi.org/10.1103/PhysRevD.92.065016}
{\path{doi:10.1103/PhysRevD.92.065016}}.

\bibitem{DePolsi:2019owi}
G.~De~Polsi, M.~Tissier, N.~Wschebor, {Conformal invariance and vector
operators in the $O(N)$ model}, J. Statist. Phys. 177 (2019) 1089.
\newblock \href {http://arxiv.org/abs/1907.09981} {\path{arXiv:1907.09981}},
\href {https://doi.org/10.1007/s10955-019-02411-3}
{\path{doi:10.1007/s10955-019-02411-3}}.

\bibitem{Pagani:2020ejb}
C.~Pagani, H.~Sonoda, {Operator product expansion coefficients in the exact
renormalization group formalism} (2020).
\newblock \href {http://arxiv.org/abs/2001.07015} {\path{arXiv:2001.07015}}.

\bibitem{Zomolodchikov66}
A.~B. Zomolodchikov, {"Irreversibility" of the flux of the renormalization
group in a 2D field theory}, JETP Lett. 43 (1966) 730.

\bibitem{Generowicz:1997he}
J.~Generowicz, C.~Harvey-Fros, T.~R. Morris, {C function representation of the
local potential approximation}, Phys. Lett. B407 (1997) 27–32.
\newblock \href {http://arxiv.org/abs/hep-th/9705088}
{\path{arXiv:hep-th/9705088}}, \href
{https://doi.org/10.1016/S0370-2693(97)00729-6}
{\path{doi:10.1016/S0370-2693(97)00729-6}}.

\bibitem{Codello14}
A.~Codello, G.~D’Odorico, C.~Pagani, {A functional RG equation for the
c-function}, J. High Energy Phys. 2014~(7) (2014).
\newblock \href {https://doi.org/10.1007/JHEP07(2014)040}
{\path{doi:10.1007/JHEP07(2014)040}}.

\bibitem{Aharony96}
A.~Aharony, A.~B. Harris, {Absence of Self-Averaging and Universal Fluctuations
in Random Systems near Critical Points}, Phys. Rev. Lett. 77 (1996) 3700.
\newblock \href {https://doi.org/10.1103/PhysRevLett.77.3700}
{\path{doi:10.1103/PhysRevLett.77.3700}}.

\bibitem{tissier12-2}
M.~Tissier, G.~Tarjus, Nonperturbative functional renormalization group for
random field models and related disordered systems. iii. superfield formalism
and ground-state dominance, Phys. Rev. B 85 (2012) 104202.
\newblock \href {https://doi.org/10.1103/PhysRevB.85.104202}
{\path{doi:10.1103/PhysRevB.85.104202}}.

\bibitem{Balents96}
L.~Balents, J.-P. Bouchaud, M.~M{\'{e}}zard, {The Large Scale Energy Landscape
of Randomly Pinned Objects}, J. Phys. I 6~(8) (1996) 1007.
\newblock \href {https://doi.org/10.1051/jp1:1996112}
{\path{doi:10.1051/jp1:1996112}}.

\bibitem{Balents_1993}
L.~Balents, Localization of elastic layers by correlated disorder, Europhysics
Letters ({EPL}) 24~(6) (1993) 489–494.
\newblock \href {https://doi.org/10.1209/0295-5075/24/6/011}
{\path{doi:10.1209/0295-5075/24/6/011}}.

\bibitem{PhysRevLett.56.1964}
D.~S. Fisher, Interface fluctuations in disordered systems:
$5\ensuremath{-}\ensuremath{\epsilon}$ expansion and failure of dimensional
reduction, Phys. Rev. Lett. 56 (1986) 1964–1967.
\newblock \href {https://doi.org/10.1103/PhysRevLett.56.1964}
{\path{doi:10.1103/PhysRevLett.56.1964}}.

\bibitem{PhysRevB.48.5949}
L.~Balents, D.~S. Fisher,
\href{https://link.aps.org/doi/10.1103/PhysRevB.48.5949}{Large-n expansion of
(4-\ensuremath{\epsilon})-dimensional oriented manifolds in random media},
Phys. Rev. B 48 (1993) 5949–5963.
\newblock \href {https://doi.org/10.1103/PhysRevB.48.5949}
{\path{doi:10.1103/PhysRevB.48.5949}}.
\newline\urlprefix\url{https://link.aps.org/doi/10.1103/PhysRevB.48.5949}

\bibitem{PhysRevB.46.11520}
O.~Narayan, D.~S. Fisher, Critical behavior of sliding charge-density waves in
4-\ensuremath{\epsilon} dimensions, Phys. Rev. B 46 (1992) 11520–11549.
\newblock \href {https://doi.org/10.1103/PhysRevB.46.11520}
{\path{doi:10.1103/PhysRevB.46.11520}}.

\bibitem{PhysRevB.62.6241}
P.~Chauve, T.~Giamarchi, P.~{Le Doussal}, Creep and depinning in disordered
media, Phys. Rev. B 62 (2000) 6241–6267.
\newblock \href {https://doi.org/10.1103/PhysRevB.62.6241}
{\path{doi:10.1103/PhysRevB.62.6241}}.

\bibitem{PhysRevLett.88.177202}
D.~E. Feldman, Critical exponents of the random-field $o(\mathit{N})$ model,
Phys. Rev. Lett. 88 (2002) 177202.
\newblock \href {https://doi.org/10.1103/PhysRevLett.88.177202}
{\path{doi:10.1103/PhysRevLett.88.177202}}.

\bibitem{Ledoussal10}
P.~Le~Doussal, {Exact results and open questions in first principle functional
{RG}}, Ann. Phys. 325~(1) (2010) 49.
\newblock \href {https://doi.org/10.1016/j.aop.2009.10.010}
{\path{doi:10.1016/j.aop.2009.10.010}}.

\bibitem{Tarjus19}
G.~Tarjus, M.~Tissier, {Random-field Ising and O($N$) models: theoretical
description through the functional ren ormalization group}, Eur. Phys. J. B
93~(3) (2020) 50.
\newblock \href {https://doi.org/10.1140/epjb/e2020-100489-1}
{\path{doi:10.1140/epjb/e2020-100489-1}}.

\bibitem{imry75}
Y.~Imry, S.-k. Ma, Random-field instability of the ordered state of continuous
symmetry, Phys. Rev. Lett. 35 (1975) 1399--1401.
\newblock \href {https://doi.org/10.1103/PhysRevLett.35.1399}
{\path{doi:10.1103/PhysRevLett.35.1399}}.

\bibitem{1984PhRvB..29..505C}
J.~L. {Cardy}, {Random-field effects in site-disordered Ising
antiferromagnets}, \prb 29 (1984) 505–507.
\newblock \href {https://doi.org/10.1103/PhysRevB.29.505}
{\path{doi:10.1103/PhysRevB.29.505}}.

\bibitem{brochard83}
F.~Brochard, P.~G. de~Gennes, { Phase transitions of binary mixtures in random
media}, J. Phys. (France) Lett. 44 (1983) L44.
\newblock \href {https://doi.org/10.1051/jphyslet:019830044018078500}
{\path{doi:10.1051/jphyslet:019830044018078500}}.

\bibitem{grinstein76}
G.~Grinstein,
\href{https://link.aps.org/doi/10.1103/PhysRevLett.37.944}{Ferromagnetic
phase transitions in random fields: The breakdown of scaling laws}, Phys.
Rev. Lett. 37 (1976) 944--947.
\newblock \href {https://doi.org/10.1103/PhysRevLett.37.944}
{\path{doi:10.1103/PhysRevLett.37.944}}.
\newline\urlprefix\url{https://link.aps.org/doi/10.1103/PhysRevLett.37.944}

\bibitem{aharony76}
A.~Aharony, Y.~Imry, S.-K. Ma, {Comments on the critical behavior of random
systems}, Phys. Rev. B 13 (1976) 466.
\newblock \href {https://doi.org/10.1103/PhysRevB.13.466}
{\path{doi:10.1103/PhysRevB.13.466}}.

\bibitem{young77}
A.~P. Young, \href{https://doi.org/10.1088%2F0022-3719%2F10%2F9%2F007}{On the
lowering of dimensionality in phase transitions with random fields}, Journal
of Physics C: Solid State Physics 10~(9) (1977) L257--L256.
\newblock \href {https://doi.org/10.1088/0022-3719/10/9/007}
{\path{doi:10.1088/0022-3719/10/9/007}}.
\newline\urlprefix\url{https://doi.org/10.1088%2F0022-3719%2F10%2F9%2F007}

\bibitem{parisi79}
G.~Parisi, N.~Sourlas,
\href{https://link.aps.org/doi/10.1103/PhysRevLett.43.744}{Random magnetic
fields, supersymmetry, and negative dimensions}, Phys. Rev. Lett. 43 (1979)
744--745.
\newblock \href {https://doi.org/10.1103/PhysRevLett.43.744}
{\path{doi:10.1103/PhysRevLett.43.744}}.
\newline\urlprefix\url{https://link.aps.org/doi/10.1103/PhysRevLett.43.744}

\bibitem{imbrie84}
J.~Z. Imbrie, \href{https://link.aps.org/doi/10.1103/PhysRevLett.53.1747}{Lower
critical dimension of the random-field ising model}, Phys. Rev. Lett. 53
(1984) 1747--1750.
\newblock \href {https://doi.org/10.1103/PhysRevLett.53.1747}
{\path{doi:10.1103/PhysRevLett.53.1747}}.
\newline\urlprefix\url{https://link.aps.org/doi/10.1103/PhysRevLett.53.1747}

\bibitem{bricmont87}
J.~Bricmont, A.~Kupiainen,
\href{https://link.aps.org/doi/10.1103/PhysRevLett.59.1829}{Lower critical
dimension for the random-field ising model}, Phys. Rev. Lett. 59 (1987)
1829--1832.
\newblock \href {https://doi.org/10.1103/PhysRevLett.59.1829}
{\path{doi:10.1103/PhysRevLett.59.1829}}.
\newline\urlprefix\url{https://link.aps.org/doi/10.1103/PhysRevLett.59.1829}

\bibitem{bricmont88}
J.~Bricmont, A.~Kupiainen,
\href{https://projecteuclid.org:443/euclid.cmp/1104161515}{Phase transition
in the $3$d random field ising model}, Comm. Math. Phys. 116~(4) (1988)
539--572.
\newline\urlprefix\url{https://projecteuclid.org:443/euclid.cmp/1104161515}

\bibitem{aharony76sm}
A.~Aharony, Y.~Imry, S.-k. Ma,
\href{https://link.aps.org/doi/10.1103/PhysRevLett.37.1364}{Lowering of
dimensionality in phase transitions with random fields}, Phys. Rev. Lett. 37
(1976) 1364--1367.
\newblock \href {https://doi.org/10.1103/PhysRevLett.37.1364}
{\path{doi:10.1103/PhysRevLett.37.1364}}.
\newline\urlprefix\url{https://link.aps.org/doi/10.1103/PhysRevLett.37.1364}

\bibitem{tissier06}
M.~Tissier, G.~Tarjus, {Two-loop functional renormalization group of the random
field and random anisotropy $O(N)$ models}, Phys. Rev. B 74 (2006) 214419.
\newblock \href {https://doi.org/10.1103/PhysRevB.74.214419}
{\path{doi:10.1103/PhysRevB.74.214419}}.

\bibitem{tissier06b}
M.~Tissier, G.~Tarjus,
\href{https://link.aps.org/doi/10.1103/PhysRevLett.96.087202}{Unified picture
of ferromagnetism, quasi-long-range order, and criticality in random-field
models}, Phys. Rev. Lett. 96 (2006) 087202.
\newblock \href {https://doi.org/10.1103/PhysRevLett.96.087202}
{\path{doi:10.1103/PhysRevLett.96.087202}}.
\newline\urlprefix\url{https://link.aps.org/doi/10.1103/PhysRevLett.96.087202}

\bibitem{tissier08a}
G.~Tarjus, M.~Tissier,
\href{https://link.aps.org/doi/10.1103/PhysRevB.78.024203}{Nonperturbative
functional renormalization group for random field models and related
disordered systems. i. effective average action formalism}, Phys. Rev. B 78
(2008) 024203.
\newblock \href {https://doi.org/10.1103/PhysRevB.78.024203}
{\path{doi:10.1103/PhysRevB.78.024203}}.
\newline\urlprefix\url{https://link.aps.org/doi/10.1103/PhysRevB.78.024203}

\bibitem{tissier08}
M.~Tissier, G.~Tarjus, {Nonperturbative functional renormalization group for
random field models and related disordered systems. II. Results for the
random field $O(N)$ model}, Phys. Rev. B 78 (2008) 024204.
\newblock \href {https://doi.org/10.1103/PhysRevB.78.024204}
{\path{doi:10.1103/PhysRevB.78.024204}}.

\bibitem{tissier11b-2}
M.~Tissier, G.~Tarjus, Supersymmetry and its spontaneous breaking in the random
field ising model, Phys. Rev. Lett. 107 (2011) 041601.
\newblock \href {https://doi.org/10.1103/PhysRevLett.107.041601}
{\path{doi:10.1103/PhysRevLett.107.041601}}.

\bibitem{tissier12b-2}
M.~Tissier, G.~Tarjus, Nonperturbative functional renormalization group for
random field models and related disordered systems. iv. supersymmetry and its
spontaneous breaking, Phys. Rev. B 85 (2012) 104203.
\newblock \href {https://doi.org/10.1103/PhysRevB.85.104203}
{\path{doi:10.1103/PhysRevB.85.104203}}.

\bibitem{Tarjus_2013}
G.~Tarjus, I.~Balog, M.~Tissier, Critical scaling in random-field systems: 2 or
3 independent exponents?, {EPL} (Europhysics Letters) 103~(6) (2013) 61001.
\newblock \href {https://doi.org/10.1209/0295-5075/103/61001}
{\path{doi:10.1209/0295-5075/103/61001}}.

\bibitem{PhysRevB.64.214419}
A.~K. Hartmann, A.~P. Young, Specific-heat exponent of random-field systems via
ground-state calculations, Phys. Rev. B 64 (2001) 214419.
\newblock \href {https://doi.org/10.1103/PhysRevB.64.214419}
{\path{doi:10.1103/PhysRevB.64.214419}}.

\bibitem{PhysRevB.65.134411}
A.~A. Middleton, D.~S. Fisher, Three-dimensional random-field ising magnet:
Interfaces, scaling, and the nature of states, Phys. Rev. B 65 (2002) 134411.
\newblock \href {https://doi.org/10.1103/PhysRevB.65.134411}
{\path{doi:10.1103/PhysRevB.65.134411}}.

\bibitem{PhysRevLett.110.227201}
N.~G. Fytas, V.~Mart{\'i}n-Mayor, Universality in the three-dimensional
random-field ising model, Phys. Rev. Lett. 110 (2013) 227201.
\newblock \href {https://doi.org/10.1103/PhysRevLett.110.227201}
{\path{doi:10.1103/PhysRevLett.110.227201}}.

\bibitem{PhysRevE.95.042117}
N.~G. Fytas, V.~Mart{\'i}n-Mayor, M.~Picco, N.~Sourlas, Restoration of
dimensional reduction in the random-field ising model at five dimensions,
Phys. Rev. E 95 (2017) 042117.
\newblock \href {https://doi.org/10.1103/PhysRevE.95.042117}
{\path{doi:10.1103/PhysRevE.95.042117}}.

\bibitem{PhysRevLett.110.135703}
G.~Tarjus, M.~Baczyk, M.~Tissier, Avalanches and dimensional reduction
breakdown in the critical behavior of disordered systems, Phys. Rev. Lett.
110 (2013) 135703.
\newblock \href {https://doi.org/10.1103/PhysRevLett.110.135703}
{\path{doi:10.1103/PhysRevLett.110.135703}}.

\bibitem{Balog_2014}
I.~Balog, G.~Tarjus, M.~Tissier, Critical behaviour of the random-field ising
model with long-range interactions in one dimension, Journal of Statistical
Mechanics: Theory and Experiment 2014~(10) (2014) P10017.
\newblock \href {https://doi.org/10.1088/1742-5468/2014/10/p10017}
{\path{doi:10.1088/1742-5468/2014/10/p10017}}.

\bibitem{PhysRevB.88.014204}
M.~Baczyk, M.~Tissier, G.~Tarjus, Y.~Sakamoto, Dimensional reduction and its
breakdown in the three-dimensional long-range random-field ising model, Phys.
Rev. B 88 (2013) 014204.
\newblock \href {https://doi.org/10.1103/PhysRevB.88.014204}
{\path{doi:10.1103/PhysRevB.88.014204}}.

\bibitem{PhysRevB.91.214201}
I.~Balog, G.~Tarjus, Activated dynamic scaling in the random-field ising model:
A nonperturbative functional renormalization group approach, Phys. Rev. B 91
(2015) 214201.
\newblock \href {https://doi.org/10.1103/PhysRevB.91.214201}
{\path{doi:10.1103/PhysRevB.91.214201}}.

\bibitem{PhysRevB.89.104201}
I.~Balog, M.~Tissier, G.~Tarjus, Same universality class for the critical
behavior in and out of equilibrium in a quenched random field, Phys. Rev. B
89 (2014) 104201.
\newblock \href {https://doi.org/10.1103/PhysRevB.89.104201}
{\path{doi:10.1103/PhysRevB.89.104201}}.

\bibitem{PhysRevB.97.094204}
I.~Balog, G.~Tarjus, M.~Tissier, Criticality of the random field ising model in
and out of equilibrium: A nonperturbative functional renormalization group
description, Phys. Rev. B 97 (2018) 094204.
\newblock \href {https://doi.org/10.1103/PhysRevB.97.094204}
{\path{doi:10.1103/PhysRevB.97.094204}}.

\bibitem{Balog19a}
I.~Balog, G.~Tarjus, M.~Tissier, Benchmarking the nonperturbative functional
renormalization group approach on the random elastic manifold model in and
out of equilibrium, J. Stat. Mech: Theory Exp. 2019~(10) (2019) 103301.
\newblock \href {https://doi.org/10.1088/1742-5468/ab3da5}
{\path{doi:10.1088/1742-5468/ab3da5}}.

\bibitem{SciPostPhys.1.1.007}
G.~Biroli, C.~Rulquin, G.~Tarjus, M.~Tarzia, {Role of fluctuations in the phase
transitions of coupled plaquette spin models of glasses}, SciPost Phys. 1
(2016) 007.
\newblock \href {https://doi.org/10.21468/SciPostPhys.1.1.007}
{\path{doi:10.21468/SciPostPhys.1.1.007}}.

\bibitem{Dupuis19}
N.~Dupuis, {Glassy properties of the Bose-glass phase of a one-dimensional
disordered Bose fluid}, Phys. Rev. E 100 (2019) 030102(R).
\newblock \href {https://doi.org/10.1103/PhysRevE.100.030102}
{\path{doi:10.1103/PhysRevE.100.030102}}.

\bibitem{Dupuis20}
N.~Dupuis, R.~Daviet, Bose-glass phase of a one-dimensional disordered bose
fluid: Metastable states, quantum tunneling, and droplets, Phys. Rev. E 101
(2020) 042139.
\newblock \href {https://doi.org/10.1103/PhysRevE.101.042139}
{\path{doi:10.1103/PhysRevE.101.042139}}.

\bibitem{Dupuis20a}
N.~Dupuis, Is there a mott-glass phase in a one-dimensional disordered quantum
fluid with linearly confining interactions?, Europhys. Lett. 130~(5) (2020)
56002.
\newblock \href {https://doi.org/10.1209/0295-5075/130/56002}
{\path{doi:10.1209/0295-5075/130/56002}}.

\bibitem{Koenig99}
H.~Schoeller, J.~K\"onig,
\href{https://link.aps.org/doi/10.1103/PhysRevLett.84.3686}{Real-time
renormalization group and charge fluctuations in quantum dots}, Phys. Rev.
Lett. 84 (2000) 3686--3689.
\newblock \href {https://doi.org/10.1103/PhysRevLett.84.3686}
{\path{doi:10.1103/PhysRevLett.84.3686}}.
\newline\urlprefix\url{https://link.aps.org/doi/10.1103/PhysRevLett.84.3686}

\bibitem{Jakobs07}
S.~G. Jakobs, V.~Meden, H.~Schoeller, {Nonequilibrium Functional
Renormalization Group for Interacting Quantum Systems}, Phys. Rev. Lett. 99
(2007) 150603.
\newblock \href {https://doi.org/10.1103/PhysRevLett.99.150603}
{\path{doi:10.1103/PhysRevLett.99.150603}}.

\bibitem{Gasenzer:2007za}
T.~Gasenzer, J.~M. Pawlowski, {Towards far-from-equilibrium quantum field
dynamics: A functional renormalisation-group approach}, Phys. Lett. B670
(2008) 135--140.
\newblock \href {http://arxiv.org/abs/0710.4627} {\path{arXiv:0710.4627}},
\href {https://doi.org/10.1016/j.physletb.2008.10.049}
{\path{doi:10.1016/j.physletb.2008.10.049}}.

\bibitem{Pietroni:2008jx}
M.~Pietroni, {Flowing with Time: a New Approach to Nonlinear Cosmological
Perturbations}, JCAP 0810 (2008) 036.
\newblock \href {http://arxiv.org/abs/0806.0971} {\path{arXiv:0806.0971}},
\href {https://doi.org/10.1088/1475-7516/2008/10/036}
{\path{doi:10.1088/1475-7516/2008/10/036}}.

\bibitem{Berges:2008sr}
J.~Berges, G.~Hoffmeister, {Nonthermal fixed points and the functional
renormalization group}, Nucl.Phys. B813 (2009) 383--407.
\newblock \href {http://arxiv.org/abs/0809.5208} {\path{arXiv:0809.5208}},
\href {https://doi.org/10.1016/j.nuclphysb.2008.12.017}
{\path{doi:10.1016/j.nuclphysb.2008.12.017}}.

\bibitem{Berges12}
J.~Berges, D.~Mesterh\'azy,
\href{http://www.sciencedirect.com/science/article/pii/S0920563212001600}{Introduction
to the nonequilibrium functional {R}enormalization {G}roup}, Nucl. Phys. B -
Proceedings Supplements 228~(0) (2012) 37 -- 60, “Physics at all scales:
The Renormalization Group” Proceedings of the 49th Internationale
Universitätswochen fur Theoretische Physik.
\newblock \href {https://doi.org/10.1016/j.nuclphysbps.2012.06.003}
{\path{doi:10.1016/j.nuclphysbps.2012.06.003}}.
\newline\urlprefix\url{http://www.sciencedirect.com/science/article/pii/S0920563212001600}

\bibitem{Sieberer13}
L.~M. Sieberer, S.~D. Huber, E.~Altman, S.~Diehl, {Dynamical Critical Phenomena
in Driven-Dissipative Systems}, Phys. Rev. Lett. 110 (2013) 195301.
\newblock \href {https://doi.org/10.1103/PhysRevLett.110.195301}
{\path{doi:10.1103/PhysRevLett.110.195301}}.

\bibitem{Chiocchetta17}
A.~Chiocchetta, A.~Gambassi, S.~Diehl, J.~Marino,
\href{https://link.aps.org/doi/10.1103/PhysRevLett.118.135701}{Dynamical
crossovers in prethermal critical states}, Phys. Rev. Lett. 118 (2017)
135701.
\newblock \href {https://doi.org/10.1103/PhysRevLett.118.135701}
{\path{doi:10.1103/PhysRevLett.118.135701}}.
\newline\urlprefix\url{https://link.aps.org/doi/10.1103/PhysRevLett.118.135701}

\bibitem{Chiocchetta16}
A.~Chiocchetta, A.~Gambassi, S.~Diehl, J.~Marino,
\href{https://link.aps.org/doi/10.1103/PhysRevB.94.174301}{Universal
short-time dynamics: Boundary functional renormalization group for a
temperature quench}, Phys. Rev. B 94 (2016) 174301.
\newblock \href {https://doi.org/10.1103/PhysRevB.94.174301}
{\path{doi:10.1103/PhysRevB.94.174301}}.
\newline\urlprefix\url{https://link.aps.org/doi/10.1103/PhysRevB.94.174301}

\bibitem{Martin73}
P.~C. Martin, E.~D. Siggia, H.~A. Rose,
\href{http://link.aps.org/doi/10.1103/PhysRevA.8.423}{Statistical dynamics of
classical systems}, Phys. Rev. A 8 (1973) 423--437.
\newblock \href {https://doi.org/10.1103/PhysRevA.8.423}
{\path{doi:10.1103/PhysRevA.8.423}}.
\newline\urlprefix\url{http://link.aps.org/doi/10.1103/PhysRevA.8.423}

\bibitem{Janssen76}
H.-K. Janssen, \href{http://dx.doi.org/10.1007/BF01316547}{On a lagrangean for
classical field dynamics and renormalization group calculations of dynamical
critical properties}, Z. Phys. B 23 (1976) 377--380.
\newblock \href {https://doi.org/10.1007/BF01316547}
{\path{doi:10.1007/BF01316547}}.
\newline\urlprefix\url{http://dx.doi.org/10.1007/BF01316547}

\bibitem{Dominicis76}
{de Dominicis, C.},
\href{http://dx.doi.org/10.1051/jphyscol:1976138}{Techniques de
renormalisation de la th\'eorie des champs et dynamique des ph\'enom\`enes
critiques}, J. Phys. (Paris) Colloq. 37~(C1) (1976) 247--253.
\newblock \href {https://doi.org/10.1051/jphyscol:1976138}
{\path{doi:10.1051/jphyscol:1976138}}.
\newline\urlprefix\url{http://dx.doi.org/10.1051/jphyscol:1976138}

\bibitem{Canet04a}
L.~Canet, B.~Delamotte, O.~Deloubri\`ere, N.~Wschebor, Nonperturbative
renormalization-group study of reaction-diffusion processes, Phys. Rev. Lett.
92~(19) (2004) 195703.
\newblock \href {https://doi.org/10.1103/PhysRevLett.92.195703}
{\path{doi:10.1103/PhysRevLett.92.195703}}.

\bibitem{Benitez13}
F.~Benitez, N.~Wschebor,
\href{http://link.aps.org/doi/10.1103/PhysRevE.87.052132}{Branching and
annihilating random walks: Exact results at low branching rate}, Phys. Rev. E
87 (2013) 052132.
\newblock \href {https://doi.org/10.1103/PhysRevE.87.052132}
{\path{doi:10.1103/PhysRevE.87.052132}}.
\newline\urlprefix\url{http://link.aps.org/doi/10.1103/PhysRevE.87.052132}

\bibitem{Duclut17}
C.~Duclut, B.~Delamotte,
\href{https://link.aps.org/doi/10.1103/PhysRevE.95.012107}{Frequency
regulators for the nonperturbative renormalization group: A general study and
the model a as a benchmark}, Phys. Rev. E 95 (2017) 012107.
\newblock \href {https://doi.org/10.1103/PhysRevE.95.012107}
{\path{doi:10.1103/PhysRevE.95.012107}}.
\newline\urlprefix\url{https://link.aps.org/doi/10.1103/PhysRevE.95.012107}

\bibitem{Mesterhazy13}
D.~Mesterh\'azy, J.~H. Stockemer, L.~F. Palhares, J.~Berges,
\href{http://link.aps.org/doi/10.1103/PhysRevB.88.174301}{Dynamic
universality class of model c from the functional renormalization group},
Phys. Rev. B 88 (2013) 174301.
\newblock \href {https://doi.org/10.1103/PhysRevB.88.174301}
{\path{doi:10.1103/PhysRevB.88.174301}}.
\newline\urlprefix\url{http://link.aps.org/doi/10.1103/PhysRevB.88.174301}

\bibitem{Hohenberg77}
P.~C. Hohenberg, B.~I. Halperin, {Theory of dynamic critical phenomena}, Rev.
Mod. Phys. 49 (1977) 435--479.
\newblock \href {https://doi.org/10.1103/RevModPhys.49.435}
{\path{doi:10.1103/RevModPhys.49.435}}.

\bibitem{Andreanov06}
A.~Andreanov, G.~Biroli, A.~Lef\`evre,
\href{http://stacks.iop.org/1742-5468/2006/i=07/a=P07008}{Dynamical field
theory for glass-forming liquids, self-consistent resummations and
time-reversal symmetry}, J. Stat. Mech.: Theor. Exp. 2006~(07) (2006) P07008.
\newblock \href {https://doi.org/10.1088/1742-5468/2006/07/P07008}
{\path{doi:10.1088/1742-5468/2006/07/P07008}}.
\newline\urlprefix\url{http://stacks.iop.org/1742-5468/2006/i=07/a=P07008}

\bibitem{Canet07b}
L.~Canet, H.~Chat\'e, \href{http://stacks.iop.org/1751-8121/40/i=9/a=002}{A
non-perturbative approach to critical dynamics}, Journal of Physics A:
Mathematical and Theoretical 40~(9) (2007) 1937--1949.
\newblock \href {https://doi.org/10.1088/1751-8113/40/9/002}
{\path{doi:10.1088/1751-8113/40/9/002}}.
\newline\urlprefix\url{http://stacks.iop.org/1751-8121/40/i=9/a=002}

\bibitem{Prudnikov06}
A.~S. Krinitsyn, V.~V. Prudnikov, P.~V. Prudnikov, Calculations of the
dynamical critical exponent using the asymptotic series summation method,
Theor. Math. Phys. 147 (2006) 561--575.
\newblock \href {https://doi.org/10.1007/s11232-006-0063-z}
{\path{doi:10.1007/s11232-006-0063-z}}.

\bibitem{Ito00}
N.~Ito, K.~Hukushima, K.~Ogawa, Y.~Ozeki,
\href{https://doi.org/10.1143/JPSJ.69.1931}{Nonequilibrium relaxation of
fluctuations of physical quantities}, Journal of the Physical Society of
Japan 69~(7) (2000) 1931--1934.
\newblock \href {https://doi.org/10.1143/JPSJ.69.1931}
{\path{doi:10.1143/JPSJ.69.1931}}.
\newline\urlprefix\url{https://doi.org/10.1143/JPSJ.69.1931}

\bibitem{Grassberger95}
P.~Grassberger,
\href{http://www.sciencedirect.com/science/article/pii/0378437194002852}{Damage
spreading and critical exponents for “model a” ising dynamics}, Physica
A: Statistical Mechanics and its Applications 214~(4) (1995) 547 -- 559.
\newblock \href {https://doi.org/https://doi.org/10.1016/0378-4371(94)00285-2}
{\path{doi:https://doi.org/10.1016/0378-4371(94)00285-2}}.
\newline\urlprefix\url{http://www.sciencedirect.com/science/article/pii/0378437194002852}

\bibitem{kardar86}
M.~Kardar, G.~Parisi, Y.-C. Zhang, Dynamic scaling of growing interfaces, Phys.
Rev. Lett. 56~(9) (1986) 889--892.
\newblock \href {https://doi.org/10.1103/PhysRevLett.56.889}
{\path{doi:10.1103/PhysRevLett.56.889}}.

\bibitem{Forster77}
D.~Forster, D.~R. Nelson, M.~J. Stephen, Large-distance and long-time
properties of a randomly stirred fluid, Phys. Rev. A 16~(2) (1977) 732--749.
\newblock \href {https://doi.org/10.1103/PhysRevA.16.732}
{\path{doi:10.1103/PhysRevA.16.732}}.

\bibitem{Kardar87}
M.~Kardar, Replica {B}ethe ansatz studies of two-dimensional interfaces with
quenched random impurities, Nucl. Phys. B 290~(0) (1987) 582 -- 602.
\newblock \href {https://doi.org/10.1016/0550-3213(87)90203-3}
{\path{doi:10.1016/0550-3213(87)90203-3}}.

\bibitem{Squizzato18}
D.~Squizzato, L.~Canet, A.~Minguzzi,
\href{https://link.aps.org/doi/10.1103/PhysRevB.97.195453}{{K}ardar-{P}arisi-{Z}hang
universality in the phase distributions of one-dimensional
exciton-polaritons}, Phys. Rev. B 97 (2018) 195453.
\newblock \href {https://doi.org/10.1103/PhysRevB.97.195453}
{\path{doi:10.1103/PhysRevB.97.195453}}.
\newline\urlprefix\url{https://link.aps.org/doi/10.1103/PhysRevB.97.195453}

\bibitem{Halpin-Healy95}
T.~Halpin-Healy, Y.-C. Zhang,
\href{http://www.sciencedirect.com/science/article/B6TVP-3YGTS2M-5/2/6497844a9ae9cef6fba31e542e177217}{Kinetic
roughening phenomena, stochastic growth, directed polymers and all that.
aspects of multidisciplinary statistical mechanics}, Phys. Rep. 254~(4-6)
(1995) 215 -- 414.
\newblock \href {https://doi.org/10.1016/0370-1573(94)00087-J}
{\path{doi:10.1016/0370-1573(94)00087-J}}.
\newline\urlprefix\url{http://www.sciencedirect.com/science/article/B6TVP-3YGTS2M-5/2/6497844a9ae9cef6fba31e542e177217}

\bibitem{Corwin12}
I.~Corwin,
\href{http://www.worldscientific.com/doi/abs/10.1142/S2010326311300014}{The
{K}ardar{-P}arisi{-Z}hang equation and universality classes}, Random Matrices
01~(01) (2012) 1130001.
\newblock \href {https://doi.org/10.1142/S2010326311300014}
{\path{doi:10.1142/S2010326311300014}}.
\newline\urlprefix\url{http://www.worldscientific.com/doi/abs/10.1142/S2010326311300014}

\bibitem{Takeuchi10sm}
K.~A. Takeuchi, M.~Sano, Universal fluctuations of growing interfaces: Evidence
in turbulent liquid crystals, Phys. Rev. Lett. 104~(23) (2010) 230601.
\newblock \href {https://doi.org/10.1103/PhysRevLett.104.230601}
{\path{doi:10.1103/PhysRevLett.104.230601}}.

\bibitem{Takeuchi12}
K.~Takeuchi, M.~Sano,
\href{http://dx.doi.org/10.1007/s10955-012-0503-0}{Evidence for
geometry-dependent universal fluctuations of the {K}ardar-{P}arisi-{Z}hang
interfaces in liquid-crystal turbulence}, J. Stat. Phys. 147 (2012) 853--890.
\newblock \href {https://doi.org/10.1007/s10955-012-0503-0}
{\path{doi:10.1007/s10955-012-0503-0}}.
\newline\urlprefix\url{http://dx.doi.org/10.1007/s10955-012-0503-0}

\bibitem{Canet10sm}
L.~Canet, H.~Chat\'e, B.~Delamotte, N.~Wschebor, Nonperturbative
renormalization group for the {K}ardar-{P}arisi-{Z}hang equation, Phys. Rev.
Lett. 104~(15) (2010) 150601.
\newblock \href {https://doi.org/10.1103/PhysRevLett.104.150601}
{\path{doi:10.1103/PhysRevLett.104.150601}}.

\bibitem{Lebedev94}
V.~V. Lebedev, V.~S. L'vov, Hidden symmetry, exact relations, and a small
parameter in the {K}ardar-{P}arisi-{Z}hang problem with strong coupling,
Phys. Rev. E 49~(2) (1994) R959--R962.
\newblock \href {https://doi.org/10.1103/PhysRevE.49.R959}
{\path{doi:10.1103/PhysRevE.49.R959}}.

\bibitem{Canet11asm}
L.~Canet, H.~Chat\'e, B.~Delamotte, N.~Wschebor,
\href{http://link.aps.org/doi/10.1103/PhysRevE.84.061128}{Nonperturbative
{R}enormalization {G}roup for the {K}ardar-{P}arisi-{Z}hang equation: General
framework and first applications}, Phys. Rev. E 84 (2011) 061128.
\newblock \href {https://doi.org/10.1103/PhysRevE.84.061128}
{\path{doi:10.1103/PhysRevE.84.061128}}.
\newline\urlprefix\url{http://link.aps.org/doi/10.1103/PhysRevE.84.061128}

\bibitem{Canet05b}
L.~Canet, \href{http://arxiv.org/abs/cond-mat/0509541}{Strong-coupling fixed
point of the {K}ardar-{P}arisi-{Z}hang equation}, arXiv:cond-mat/0509541
(2005).
\newblock \href {http://arxiv.org/abs/arXiv:cond-mat/0509541}
{\path{arXiv:arXiv:cond-mat/0509541}}.
\newline\urlprefix\url{http://arxiv.org/abs/cond-mat/0509541}

\bibitem{Wiese97}
K.~J. Wiese, Critical discussion of the two-loop calculations for the
{K}ardar-{P}arisi-{Z}hang equation, Phys. Rev. E 56~(5) (1997) 5013--5017.
\newblock \href {https://doi.org/10.1103/PhysRevE.56.5013}
{\path{doi:10.1103/PhysRevE.56.5013}}.

\bibitem{Praehofer04}
M.~Pr\"ahofer, H.~Spohn,
\href{http://www.springerlink.com/content/x6x002532vl4q877/?p=81653f9d9d8e4cd2bea4376198ef8da2&pi=2&hl=u}{Exact
scaling functions for one-dimensional stationary {KPZ} growth}, J. Stat.
Phys. 115~(1-2) (2004) 255--279.
\newblock \href {https://doi.org/10.1023/B:JOSS.0000019810.21828.fc}
{\path{doi:10.1023/B:JOSS.0000019810.21828.fc}}.
\newline\urlprefix\url{http://www.springerlink.com/content/x6x002532vl4q877/?p=81653f9d9d8e4cd2bea4376198ef8da2&pi=2&hl=u}

\bibitem{Canet12Err}
L.~Canet, H.~Chat\'e, B.~Delamotte, N.~Wschebor,
\href{http://link.aps.org/doi/10.1103/PhysRevE.86.019904}{Erratum:
Nonperturbative {R}enormalization {G}roup for the {K}ardar-{P}arisi-{Z}hang
equation: General framework and first applications [phys. rev. e 84, 061128
(2011)]}, Phys. Rev. E 86 (2012) 019904.
\newblock \href {https://doi.org/10.1103/PhysRevE.86.019904}
{\path{doi:10.1103/PhysRevE.86.019904}}.
\newline\urlprefix\url{http://link.aps.org/doi/10.1103/PhysRevE.86.019904}

\bibitem{Kloss12}
T.~Kloss, L.~Canet, N.~Wschebor,
\href{http://link.aps.org/doi/10.1103/PhysRevE.86.051124}{Nonperturbative
{R}enormalization {G}roup for the stationary {K}ardar-{P}arisi-{Z}hang
equation: Scaling functions and amplitude ratios in 1+1, 2+1, and 3+1
dimensions}, Phys. Rev. E 86 (2012) 051124.
\newblock \href {https://doi.org/10.1103/PhysRevE.86.051124}
{\path{doi:10.1103/PhysRevE.86.051124}}.
\newline\urlprefix\url{http://link.aps.org/doi/10.1103/PhysRevE.86.051124}

\bibitem{Halpin-Healy13}
T.~Halpin-Healy,
\href{http://link.aps.org/doi/10.1103/PhysRevE.88.042118}{Extremal paths, the
stochastic heat equation, and the three-dimensional {K}ardar-{P}arisi-{Z}hang
universality class}, Phys. Rev. E 88 (2013) 042118.
\newblock \href {https://doi.org/10.1103/PhysRevE.88.042118}
{\path{doi:10.1103/PhysRevE.88.042118}}.
\newline\urlprefix\url{http://link.aps.org/doi/10.1103/PhysRevE.88.042118}

\bibitem{Halpin-Healy13Err}
T.~Halpin-Healy,
\href{http://link.aps.org/doi/10.1103/PhysRevE.88.069903}{Erratum: Extremal
paths, the stochastic heat equation, and the three-dimensional
{K}ardar-{P}arisi-{Z}hang universality class [phys. rev. e \textbf{88} ,
042118 (2013)]}, Phys. Rev. E 88 (2013) 069903.
\newblock \href {https://doi.org/10.1103/PhysRevE.88.069903}
{\path{doi:10.1103/PhysRevE.88.069903}}.
\newline\urlprefix\url{http://link.aps.org/doi/10.1103/PhysRevE.88.069903}

\bibitem{Kloss14b}
T.~Kloss, L.~Canet, N.~Wschebor,
\href{http://link.aps.org/doi/10.1103/PhysRevE.90.062133}{Strong-coupling
phases of the anisotropic {K}ardar-{P}arisi-{Z}hang equation}, Phys. Rev. E
90 (2014) 062133.
\newblock \href {https://doi.org/10.1103/PhysRevE.90.062133}
{\path{doi:10.1103/PhysRevE.90.062133}}.
\newline\urlprefix\url{http://link.aps.org/doi/10.1103/PhysRevE.90.062133}

\bibitem{Kloss14a}
T.~Kloss, L.~Canet, B.~Delamotte, N.~Wschebor,
\href{http://link.aps.org/doi/10.1103/PhysRevE.89.022108}{{K}ardar-{P}arisi-{Z}hang
equation with spatially correlated noise: A unified picture from
nonperturbative {R}enormalization {G}roup}, Phys. Rev. E 89 (2014) 022108.
\newblock \href {https://doi.org/10.1103/PhysRevE.89.022108}
{\path{doi:10.1103/PhysRevE.89.022108}}.
\newline\urlprefix\url{http://link.aps.org/doi/10.1103/PhysRevE.89.022108}

\bibitem{Mathey17sm}
S.~Mathey, E.~Agoritsas, T.~Kloss, V.~Lecomte, L.~Canet,
\href{https://link.aps.org/doi/10.1103/PhysRevE.95.032117}{{K}ardar-{P}arisi-{Z}hang
equation with short-range correlated noise: Emergent symmetries and
nonuniversal observables}, Phys. Rev. E 95 (2017) 032117.
\newblock \href {https://doi.org/10.1103/PhysRevE.95.032117}
{\path{doi:10.1103/PhysRevE.95.032117}}.
\newline\urlprefix\url{https://link.aps.org/doi/10.1103/PhysRevE.95.032117}

\bibitem{Strack15}
P.~Strack, \href{https://link.aps.org/doi/10.1103/PhysRevE.91.032131}{Dynamic
criticality far from equilibrium: One-loop flow of
{B}urgers-{K}ardar-{P}arisi-{Z}hang systems with broken galilean invariance},
Phys. Rev. E 91 (2015) 032131.
\newblock \href {https://doi.org/10.1103/PhysRevE.91.032131}
{\path{doi:10.1103/PhysRevE.91.032131}}.
\newline\urlprefix\url{https://link.aps.org/doi/10.1103/PhysRevE.91.032131}

\bibitem{Squizzato19}
D.~Squizzato, L.~Canet,
\href{https://link.aps.org/doi/10.1103/PhysRevE.100.062143}{Kardar-parisi-zhang
equation with temporally correlated noise: A nonperturbative renormalization
group approach}, Phys. Rev. E 100 (2019) 062143.
\newblock \href {https://doi.org/10.1103/PhysRevE.100.062143}
{\path{doi:10.1103/PhysRevE.100.062143}}.
\newline\urlprefix\url{https://link.aps.org/doi/10.1103/PhysRevE.100.062143}

\bibitem{Frisch95}
U.~Frisch, Turbulence: the legacy of {A. N. Kolmogorov}, Cambridge University
Press, Cambridge, 1995.
\newblock \href {https://doi.org/10.1017/CBO9781139170666}
{\path{doi:10.1017/CBO9781139170666}}.

\bibitem{Kolmogorov41a}
A.~N. Kolmogorov,
\href{http://rspa.royalsocietypublishing.org/content/434/1890/9}{The local
structure of turbulence in incompressible viscous fluid for very large
{R}eynolds numbers}, Proceedings of the Royal Society of London A:
Mathematical, Physical and Engineering Sciences 434~(1890) (1991) 9--13.
\newblock \href {https://doi.org/10.1098/rspa.1991.0075}
{\path{doi:10.1098/rspa.1991.0075}}.
\newline\urlprefix\url{http://rspa.royalsocietypublishing.org/content/434/1890/9}

\bibitem{Kolmogorov41c}
A.~N. Kolmogorov,
\href{http://rspa.royalsocietypublishing.org/content/434/1890/15}{Dissipation
of energy in the locally isotropic turbulence}, Proceedings of the Royal
Society of London A: Mathematical, Physical and Engineering Sciences
434~(1890) (1991) 15--17.
\newblock \href {https://doi.org/10.1098/rspa.1991.0076}
{\path{doi:10.1098/rspa.1991.0076}}.
\newline\urlprefix\url{http://rspa.royalsocietypublishing.org/content/434/1890/15}

\bibitem{Adzhemyan99}
L.~T. Adzhemyan, N.~V. Antonov, A.~N. Vasil'ev, The Field Theoretic
{R}enormalization {G}roup in Fully Developed Turbulence, Gordon and Breach,
London, 1999.

\bibitem{Zhou10sm}
Y.~Zhou,
\href{http://www.sciencedirect.com/science/article/pii/S0370157309002129}{Renormalization
{G}roup theory for fluid and plasma turbulence}, Phys. Rep. 488~(1) (2010) 1
-- 49.
\newblock \href {https://doi.org/10.1016/j.physrep.2009.04.004}
{\path{doi:10.1016/j.physrep.2009.04.004}}.
\newline\urlprefix\url{http://www.sciencedirect.com/science/article/pii/S0370157309002129}

\bibitem{Tomassini97}
P.~Tomassini,
\href{http://www.sciencedirect.com/science/article/pii/S0370269397009805}{An
exact renormalization group analysis of 3{D} well developed turbulence},
Physics Letters B 411~(1) (1997) 117 -- 126.
\newblock \href {https://doi.org/10.1016/S0370-2693(97)00980-5}
{\path{doi:10.1016/S0370-2693(97)00980-5}}.
\newline\urlprefix\url{http://www.sciencedirect.com/science/article/pii/S0370269397009805}

\bibitem{Monasterio12}
C.~Mej\'ia-Monasterio, P.~Muratore-Ginanneschi,
\href{http://link.aps.org/doi/10.1103/PhysRevE.86.016315}{Nonperturbative
renormalization group study of the stochastic {N}avier-{S}tokes equation},
Phys. Rev. E 86 (2012) 016315.
\newblock \href {https://doi.org/10.1103/PhysRevE.86.016315}
{\path{doi:10.1103/PhysRevE.86.016315}}.
\newline\urlprefix\url{http://link.aps.org/doi/10.1103/PhysRevE.86.016315}

\bibitem{Canet17}
L.~Canet, V.~Rossetto, N.~Wschebor, G.~Balarac,
\href{https://link.aps.org/doi/10.1103/PhysRevE.95.023107}{Spatiotemporal
velocity-velocity correlation function in fully developed turbulence}, Phys.
Rev. E 95 (2017) 023107.
\newblock \href {https://doi.org/10.1103/PhysRevE.95.023107}
{\path{doi:10.1103/PhysRevE.95.023107}}.
\newline\urlprefix\url{https://link.aps.org/doi/10.1103/PhysRevE.95.023107}

\bibitem{Tarpin18}
M.~Tarpin, L.~Canet, N.~Wschebor,
\href{https://doi.org/10.1063/1.5020022}{Breaking of scale invariance in the
time dependence of correlation functions in isotropic and homogeneous
turbulence}, Physics of Fluids 30~(5) (2018) 055102.
\newblock \href {https://doi.org/10.1063/1.5020022}
{\path{doi:10.1063/1.5020022}}.
\newline\urlprefix\url{https://doi.org/10.1063/1.5020022}

\bibitem{Tarpin19}
M.~Tarpin, L.~Canet, C.~Pagani, N.~Wschebor,
\href{https://doi.org/10.1088%2F1751-8121%2Faaf3f0}{Stationary, isotropic and
homogeneous two-dimensional turbulence: a first non-perturbative
renormalization group approach}, Journal of Physics A: Mathematical and
Theoretical 52~(8) (2019) 085501.
\newblock \href {https://doi.org/10.1088/1751-8121/aaf3f0}
{\path{doi:10.1088/1751-8121/aaf3f0}}.
\newline\urlprefix\url{https://doi.org/10.1088%2F1751-8121%2Faaf3f0}

\bibitem{Canet15sm}
L.~Canet, B.~Delamotte, N.~Wschebor,
\href{http://link.aps.org/doi/10.1103/PhysRevE.91.053004}{Fully developed
isotropic turbulence: Symmetries and exact identities}, Phys. Rev. E 91
(2015) 053004.
\newblock \href {https://doi.org/10.1103/PhysRevE.91.053004}
{\path{doi:10.1103/PhysRevE.91.053004}}.
\newline\urlprefix\url{http://link.aps.org/doi/10.1103/PhysRevE.91.053004}

\bibitem{Debue18}
P.~Debue, D.~Kuzzay, E.-W. Saw, F.~m.~c. Daviaud, B.~Dubrulle, L.~Canet,
V.~Rossetto, N.~Wschebor,
\href{https://link.aps.org/doi/10.1103/PhysRevFluids.3.024602}{Experimental
test of the crossover between the inertial and the dissipative range in a
turbulent swirling flow}, Phys. Rev. Fluids 3 (2018) 024602.
\newblock \href {https://doi.org/10.1103/PhysRevFluids.3.024602}
{\path{doi:10.1103/PhysRevFluids.3.024602}}.
\newline\urlprefix\url{https://link.aps.org/doi/10.1103/PhysRevFluids.3.024602}

\bibitem{Gorbunova19}
A.~Gorbunova, G.~Balarac, M.~Bourgoin, L.~Canet, N.~Mordant, V.~Rossetto,
Analysis of the dissipative range of the energy spectrum in grid turbulence
and in direct numerical simulations, Phys. Rev. Fluids 5 (2020) 044604.
\newblock \href {https://doi.org/10.1103/PhysRevFluids.5.044604}
{\path{doi:10.1103/PhysRevFluids.5.044604}}.

\bibitem{Hinrichsen00}
H.~Hinrichsen, \href{https://doi.org/10.1080/00018730050198152}{Non-equilibrium
critical phenomena and phase transitions into absorbing states}, Advances in
Physics 49~(7) (2000) 815--958.
\newblock \href {https://doi.org/10.1080/00018730050198152}
{\path{doi:10.1080/00018730050198152}}.
\newline\urlprefix\url{https://doi.org/10.1080/00018730050198152}

\bibitem{Odor04}
G.~\'Odor,
\href{https://link.aps.org/doi/10.1103/RevModPhys.76.663}{Universality
classes in nonequilibrium lattice systems}, Rev. Mod. Phys. 76 (2004)
663--724.
\newblock \href {https://doi.org/10.1103/RevModPhys.76.663}
{\path{doi:10.1103/RevModPhys.76.663}}.
\newline\urlprefix\url{https://link.aps.org/doi/10.1103/RevModPhys.76.663}

\bibitem{Henkel08}
M.~Henkel, H.~Hinrichsen, S.~L\"ubeck, Non-equilibrium Phase Transitions,
Volume I: Absorbing Phase Transitions, Mathematical and Theoretical Physics,
Springer, The Netherlands, 2008.
\newblock \href {https://doi.org/10.1007/978-1-4020-8765-3}
{\path{doi:10.1007/978-1-4020-8765-3}}.

\bibitem{Bramson85}
M.~Bramson, L.~Gray, \href{https://doi.org/10.1007/BF00535338}{The survival of
branching annihilating random walk}, Zeitschrift f{\"u}r
Wahrscheinlichkeitstheorie und Verwandte Gebiete 68~(4) (1985) 447--460.
\newblock \href {https://doi.org/10.1007/BF00535338}
{\path{doi:10.1007/BF00535338}}.
\newline\urlprefix\url{https://doi.org/10.1007/BF00535338}

\bibitem{Cardy96}
J.~Cardy, U.~C. T\"auber,
\href{https://link.aps.org/doi/10.1103/PhysRevLett.77.4780}{Theory of
branching and annihilating random walks}, Phys. Rev. Lett. 77 (1996)
4780--4783.
\newblock \href {https://doi.org/10.1103/PhysRevLett.77.4780}
{\path{doi:10.1103/PhysRevLett.77.4780}}.
\newline\urlprefix\url{https://link.aps.org/doi/10.1103/PhysRevLett.77.4780}

\bibitem{Cardy98}
J.~L. Cardy, U.~C. T{\"a}uber,
\href{https://doi.org/10.1023/A:1023233431588}{Field theory of branching and
annihilating random walks}, Journal of Statistical Physics 90~(1) (1998)
1--56.
\newblock \href {https://doi.org/10.1023/A:1023233431588}
{\path{doi:10.1023/A:1023233431588}}.
\newline\urlprefix\url{https://doi.org/10.1023/A:1023233431588}

\bibitem{Tauber05}
U.~C. T\"auber, M.~Howard, B.~P. Vollmayr-Lee, Applications of field-theoretic
renormalization group methods to reaction{\textendash}diffusion problems,
Journal of Physics A: Mathematical and General 38~(17) (2005) R79--R131.
\newblock \href {https://doi.org/10.1088/0305-4470/38/17/r01}
{\path{doi:10.1088/0305-4470/38/17/r01}}.

\bibitem{doi76}
M.~Doi, \href{https://doi.org/10.1088%2F0305-4470%2F9%2F9%2F009}{Stochastic
theory of diffusion-controlled reaction}, Journal of Physics A: Mathematical
and General 9~(9) (1976) 1479--1495.
\newblock \href {https://doi.org/10.1088/0305-4470/9/9/009}
{\path{doi:10.1088/0305-4470/9/9/009}}.
\newline\urlprefix\url{https://doi.org/10.1088%2F0305-4470%2F9%2F9%2F009}

\bibitem{peliti84}
{Peliti, L.}, \href{https://doi.org/10.1051/jphys:019850046090146900}{Path
integral approach to birth-death processes on a lattice}, J. Phys. France
46~(9) (1985) 1469--1483.
\newblock \href {https://doi.org/10.1051/jphys:019850046090146900}
{\path{doi:10.1051/jphys:019850046090146900}}.
\newline\urlprefix\url{https://doi.org/10.1051/jphys:019850046090146900}

\bibitem{Tauber14}
U.~C. T\"auber, Critical Dynamics: A Field Theory Approach to Equilibrium and
Non-Equilibrium Scaling Behavior, Cambridge University Press, 2014.
\newblock \href {https://doi.org/10.1017/CBO9781139046213}
{\path{doi:10.1017/CBO9781139046213}}.

\bibitem{Wijland98}
F.~{van Wijland}, K.~Oerding, H.~Hilhorst,
\href{http://www.sciencedirect.com/science/article/pii/S0378437197006031}{Wilson
renormalization of a reaction{-}diffusion process}, Physica A: Statistical
Mechanics and its Applications 251~(1) (1998) 179 -- 201.
\newblock \href {https://doi.org/https://doi.org/10.1016/S0378-4371(97)00603-1}
{\path{doi:https://doi.org/10.1016/S0378-4371(97)00603-1}}.
\newline\urlprefix\url{http://www.sciencedirect.com/science/article/pii/S0378437197006031}

\bibitem{Buchhold16}
M.~Buchhold, S.~Diehl,
\href{https://link.aps.org/doi/10.1103/PhysRevE.94.012138}{Background field
functional renormalization group for absorbing state phase transitions},
Phys. Rev. E 94 (2016) 012138.
\newblock \href {https://doi.org/10.1103/PhysRevE.94.012138}
{\path{doi:10.1103/PhysRevE.94.012138}}.
\newline\urlprefix\url{https://link.aps.org/doi/10.1103/PhysRevE.94.012138}

\bibitem{Bartels16}
J.~Bartels, C.~Contreras, G.~P. Vacca,
\href{https://doi.org/10.1007/JHEP03(2016)201}{Could reggeon field theory be
an effective theory for qcd in the regge limit?}, Journal of High Energy
Physics 2016~(3) (2016) 201.
\newblock \href {https://doi.org/10.1007/JHEP03(2016)201}
{\path{doi:10.1007/JHEP03(2016)201}}.
\newline\urlprefix\url{https://doi.org/10.1007/JHEP03(2016)201}

\bibitem{Canet04b}
L.~Canet, H.~Chat\'e, B.~Delamotte, Quantitative phase diagrams of branching
and annihilating random walks, Phys. Rev. Lett. 92~(25) (2004) 255703.
\newblock \href {https://doi.org/10.1103/PhysRevLett.92.255703}
{\path{doi:10.1103/PhysRevLett.92.255703}}.

\bibitem{Odor04b}
G.~\'Odor, \href{https://link.aps.org/doi/10.1103/PhysRevE.70.066122}{Role of
diffusion in branching and annihilation random walk models}, Phys. Rev. E 70
(2004) 066122.
\newblock \href {https://doi.org/10.1103/PhysRevE.70.066122}
{\path{doi:10.1103/PhysRevE.70.066122}}.
\newline\urlprefix\url{https://link.aps.org/doi/10.1103/PhysRevE.70.066122}

\bibitem{Canet06b}
L.~Canet, H.~J. Hilhorst, Single-site approximation for reaction-diffusion
processes, J. Stat. Phys. 125~(3) (2006) 517--531.
\newblock \href {https://doi.org/10.1007/s10955-006-9206-8}
{\path{doi:10.1007/s10955-006-9206-8}}.

\bibitem{Benitez12b}
F.~Benitez, N.~Wschebor,
\href{http://link.aps.org/doi/10.1103/PhysRevE.86.010104}{Branching-rate
expansion around annihilating random walks}, Phys. Rev. E 86 (2012) 010104.
\newblock \href {https://doi.org/10.1103/PhysRevE.86.010104}
{\path{doi:10.1103/PhysRevE.86.010104}}.
\newline\urlprefix\url{http://link.aps.org/doi/10.1103/PhysRevE.86.010104}

\bibitem{Canet05sm}
L.~Canet, H.~Chat\'e, B.~Delamotte, I.~Dornic, M.~A. Mu\~noz, Nonperturbative
fixed point in a nonequilibrium phase transition, Phys. Rev. Lett. 95~(10)
(2005) 100601.
\newblock \href {https://doi.org/10.1103/PhysRevLett.95.100601}
{\path{doi:10.1103/PhysRevLett.95.100601}}.

\bibitem{Henkel04}
M.~Henkel, H.~Hinrichsen,
\href{https://doi.org/10.1088%2F0305-4470%2F37%2F28%2Fr01}{The
non-equilibrium phase transition of the pair-contact process with diffusion},
Journal of Physics A: Mathematical and General 37~(28) (2004) R117--R159.
\newblock \href {https://doi.org/10.1088/0305-4470/37/28/r01}
{\path{doi:10.1088/0305-4470/37/28/r01}}.
\newline\urlprefix\url{https://doi.org/10.1088%2F0305-4470%2F37%2F28%2Fr01}

\bibitem{Tarpin17}
M.~Tarpin, F.~Benitez, L.~Canet, N.~Wschebor,
\href{https://link.aps.org/doi/10.1103/PhysRevE.96.022137}{Nonperturbative
renormalization group for the diffusive epidemic process}, Phys. Rev. E 96
(2017) 022137.
\newblock \href {https://doi.org/10.1103/PhysRevE.96.022137}
{\path{doi:10.1103/PhysRevE.96.022137}}.
\newline\urlprefix\url{https://link.aps.org/doi/10.1103/PhysRevE.96.022137}

\bibitem{Berges:2015kfa}
J.~Berges, {Nonequilibrium Quantum Fields: From Cold Atoms to Cosmology} (3
2015).
\newblock \href {http://arxiv.org/abs/1503.02907} {\path{arXiv:1503.02907}}.

\bibitem{Schmied:2018mte}
C.-M. Schmied, A.~N. Mikheev, T.~Gasenzer, {Non-thermal fixed points: Universal
dynamics far from equilibrium}, Int. J. Mod. Phys. A 34~(29) (2019) 1941006.
\newblock \href {http://arxiv.org/abs/1810.08143} {\path{arXiv:1810.08143}},
\href {https://doi.org/10.1142/S0217751X19410069}
{\path{doi:10.1142/S0217751X19410069}}.

\bibitem{Gasenzer:2010rq}
T.~Gasenzer, S.~Kessler, J.~M. Pawlowski, {Far-from-equilibrium quantum
many-body dynamics}, Eur.Phys.J. C70 (2010) 423--443.
\newblock \href {http://arxiv.org/abs/1003.4163} {\path{arXiv:1003.4163}},
\href {https://doi.org/10.1140/epjc/s10052-010-1430-3}
{\path{doi:10.1140/epjc/s10052-010-1430-3}}.

\bibitem{Corell:2019jxh}
L.~Corell, A.~K. Cyrol, M.~Heller, J.~M. Pawlowski, {Flowing with the Temporal
Renormalisation Group} (2019).
\newblock \href {http://arxiv.org/abs/1910.09369} {\path{arXiv:1910.09369}}.

\bibitem{Berges09}
J.~Berges, G.~Hoffmeister,
\href{http://www.sciencedirect.com/science/article/pii/S0550321308007219}{Nonthermal
fixed points and the functional {R}enormalization {G}roup}, Nucl. Phys. B
813~(3) (2009) 383 -- 407.
\newblock \href {https://doi.org/10.1016/j.nuclphysb.2008.12.017}
{\path{doi:10.1016/j.nuclphysb.2008.12.017}}.
\newline\urlprefix\url{http://www.sciencedirect.com/science/article/pii/S0550321308007219}

\bibitem{Sieberer14}
L.~M. Sieberer, S.~D. Huber, E.~Altman, S.~Diehl,
\href{http://link.aps.org/doi/10.1103/PhysRevB.89.134310}{Nonequilibrium
functional renormalization for driven-dissipative {B}ose-{E}instein
condensation}, Phys. Rev. B 89 (2014) 134310.
\newblock \href {https://doi.org/10.1103/PhysRevB.89.134310}
{\path{doi:10.1103/PhysRevB.89.134310}}.
\newline\urlprefix\url{http://link.aps.org/doi/10.1103/PhysRevB.89.134310}

\bibitem{Mesterhazy:2015uja}
D.~Mesterh\'{a}zy, J.~H. Stockemer, Y.~Tanizaki, {From quantum to classical
dynamics: The relativistic $O(N)$ model in the framework of the real-time
functional renormalization group}, Phys. Rev. D92~(7) (2015) 076001.
\newblock \href {http://arxiv.org/abs/1504.07268} {\path{arXiv:1504.07268}},
\href {https://doi.org/10.1103/PhysRevD.92.076001}
{\path{doi:10.1103/PhysRevD.92.076001}}.

\bibitem{Lesgourgues:2009am}
J.~Lesgourgues, S.~Matarrese, M.~Pietroni, A.~Riotto, {Non-linear Power
Spectrum including Massive Neutrinos: the Time-RG Flow Approach}, JCAP 0906
(2009) 017.
\newblock \href {http://arxiv.org/abs/0901.4550} {\path{arXiv:0901.4550}},
\href {https://doi.org/10.1088/1475-7516/2009/06/017}
{\path{doi:10.1088/1475-7516/2009/06/017}}.

\bibitem{Bartolo:2009rb}
N.~Bartolo, J.~P.~B. Almeida, S.~Matarrese, M.~Pietroni, A.~Riotto, {Signatures
of Primordial non-Gaussianities in the Matter Power-Spectrum and Bispectrum:
the Time-RG Approach}, JCAP 1003 (2010) 011.
\newblock \href {http://arxiv.org/abs/0912.4276} {\path{arXiv:0912.4276}},
\href {https://doi.org/10.1088/1475-7516/2010/03/011}
{\path{doi:10.1088/1475-7516/2010/03/011}}.

\bibitem{Audren:2011ne}
B.~Audren, J.~Lesgourgues, {Non-linear matter power spectrum from Time
Renormalisation Group: efficient computation and comparison with one-loop},
JCAP 10 (2011) 037.
\newblock \href {http://arxiv.org/abs/1106.2607} {\path{arXiv:1106.2607}},
\href {https://doi.org/10.1088/1475-7516/2011/10/037}
{\path{doi:10.1088/1475-7516/2011/10/037}}.

\bibitem{Juergens12}
G.~{J{\"u}rgens}, M.~{Bartelmann}, {Perturbation Theory Trispectrum in the Time
Renormalisation Approach}, arXiv e-prints (2012) arXiv:1204.6524\href
{http://arxiv.org/abs/1204.6524} {\path{arXiv:1204.6524}}.

\bibitem{Vollmer:2014pma}
A.~Vollmer, L.~Amendola, R.~Catena, {Efficient implementation of the Time
Renormalization Group}, Phys. Rev. D 93~(4) (2016) 043526.
\newblock \href {http://arxiv.org/abs/1412.1650} {\path{arXiv:1412.1650}},
\href {https://doi.org/10.1103/PhysRevD.93.043526}
{\path{doi:10.1103/PhysRevD.93.043526}}.

\bibitem{Floerchinger:2016hja}
S.~Floerchinger, M.~Garny, N.~Tetradis, U.~A. Wiedemann, {Renormalization-group
flow of the effective action of cosmological large-scale structures}, JCAP 01
(2017) 048.
\newblock \href {http://arxiv.org/abs/1607.03453} {\path{arXiv:1607.03453}},
\href {https://doi.org/10.1088/1475-7516/2017/01/048}
{\path{doi:10.1088/1475-7516/2017/01/048}}.

\bibitem{Boettcher:2012cm}
I.~Boettcher, J.~M. Pawlowski, S.~Diehl, {Ultracold atoms and the Functional
Renormalization Group}, Nucl. Phys. B Proc. Suppl. 228 (2012) 63--135.
\newblock \href {http://arxiv.org/abs/1204.4394} {\path{arXiv:1204.4394}},
\href {https://doi.org/10.1016/j.nuclphysbps.2012.06.004}
{\path{doi:10.1016/j.nuclphysbps.2012.06.004}}.

\bibitem{Metzner12}
W.~Metzner, M.~Salmhofer, C.~Honerkamp, V.~Meden, K.~Sch\"onhammer, {Functional
renormalization group approach to correlated fermion systems}, Rev. Mod.
Phys. 84 (2012) 299--352.
\newblock \href {https://doi.org/10.1103/RevModPhys.84.299}
{\path{doi:10.1103/RevModPhys.84.299}}.

\bibitem{Floerchinger:2011yv}
S.~Floerchinger, S.~Moroz, R.~Schmidt, {Efimov physics from the functional
renormalization group}, Few Body Syst. 51 (2011) 153--180.
\newblock \href {http://arxiv.org/abs/1102.0896} {\path{arXiv:1102.0896}},
\href {https://doi.org/10.1007/s00601-011-0231-z}
{\path{doi:10.1007/s00601-011-0231-z}}.

\bibitem{Bogoliubov47}
N.~N. Bogoliubov, {On the theory of superfluidity}, J. Phys. (USSR) 11 (1947)
23--32.

\bibitem{Beliaev58a}
S.~T. Beliaev, {Application of the methods of quantum field theory to a system
of bosons}, Sov. Phys. JETP 7 (1958) 289, zh. Eksp. Teor. Fiz. 34, 417
(1958).

\bibitem{Beliaev58b}
S.~T. Beliaev, {Energy spectrum of a non-ideal Bose gas}, Sov. Phys. JETP 7
(1958) 299, zh. Eksp. Teor. Fiz. 34, 433 (1958).

\bibitem{Hugenholtz59}
N.~M. Hugenholtz, D.~Pines, {Ground-State Energy and Excitation Spectrum of a
System of Interacting Bosons}, Phys. Rev. 116 (1959) 489.
\newblock \href {https://doi.org/10.1103/PhysRev.116.489}
{\path{doi:10.1103/PhysRev.116.489}}.

\bibitem{Gavoret64}
J.~Gavoret, P.~Nozi\`eres, {Structure of the perturbation expansion for the
Bose liquid at zero temperature}, Ann. Phys. (N.Y.) 28 (1964) 349.
\newblock \href {https://doi.org/10.1016/0003-4916(64)90200-3}
{\path{doi:10.1016/0003-4916(64)90200-3}}.

\bibitem{Popov79}
V.~N. Popov, A.~V. Seredniakov, {Low-frequency asymptotic form of the
self-energy parts of a superfluid Bose system at $T=0$}, Sov. Phys. JETP 50
(1979) 193, zh. Eksp. Teor. Fiz. 77, 377 (1979).

\bibitem{Popov_book2}
V.~Popov, Functional Integrals in Quantum Field Theory and Statistical Physics
(Mathematical Physics and Applied Mathematics), 1st Edition, Kluwer, 1983.

\bibitem{Nepomnyashchii75}
A.~A. Nepomnyashchii, Y.~A. Nepomnyashchii, {Contribution to the theory of the
spectrum of a Bose system with condensate at small momenta}, JETP Lett. 21
(1975) 1, pis'ma Zh. Eksp. Teor. Fiz. 21, 3 (1975).

\bibitem{Nepomnyashchii78}
Y.~A. Nepomnyashchii, A.~A. Nepomnyashchii, {Infrared divergence in field
theory of a Bose system with a condensate}, Sov. Phys. JETP 48 (1978) 493,
zh. Eksp. Teor. Fiz. 75, 976 (1978).

\bibitem{Nepomnyashchii83}
Y.~A. Nepomnyashchii, {Concerning the nature of the $\lambda$-transition order
parameter}, Sov. Phys. JETP 58 (1983) 722, zh. Eksp. Teor. Fiz. 85, 1244
(1983).

\bibitem{Castellani97}
C.~Castellani, C.~Di~Castro, F.~Pistolesi, G.~C. Strinati, {Infrared Behavior
of Interacting Bosons at Zero Temperature}, Phys. Rev. Lett. 78 (1997) 1612.
\newblock \href {https://doi.org/10.1103/PhysRevLett.78.1612}
{\path{doi:10.1103/PhysRevLett.78.1612}}.

\bibitem{Pistolesi04}
F.~Pistolesi, C.~Castellani, C.~Di~Castro, G.~C. Strinati,
{Renormalization-group approach to the infrared behavior of a
zero-temperature Bose system}, Phys. Rev. B 69 (2004) 024513.
\newblock \href {https://doi.org/10.1103/PhysRevB.69.024513}
{\path{doi:10.1103/PhysRevB.69.024513}}.

\bibitem{Dupuis09a}
N.~Dupuis, {Unified Picture of Superfluidity: From Bogoliubov's Approximation
to Popov's Hydrodynamic Theory}, Phys. Rev. Lett. 102 (2009) 190401.
\newblock \href {https://doi.org/10.1103/PhysRevLett.102.190401}
{\path{doi:10.1103/PhysRevLett.102.190401}}.

\bibitem{Wetterich08}
C.~Wetterich, {Functional renormalization for quantum phase transitions with
nonrelativistic bosons}, Phys. Rev. B 77 (2008) 064504.
\newblock \href {https://doi.org/10.1103/PhysRevB.77.064504}
{\path{doi:10.1103/PhysRevB.77.064504}}.

\bibitem{Dupuis07}
N.~Dupuis, K.~Sengupta, {Non-perturbative renormalization group approach to
zero-temperature Bose systems}, Europhys. Lett. 80 (2007) 50007.
\newblock \href {https://doi.org/10.1209/0295-5075/80/50007}
{\path{doi:10.1209/0295-5075/80/50007}}.

\bibitem{Sinner09}
A.~Sinner, N.~Hasselmann, P.~Kopietz, {Spectral Function and Quasiparticle
Damping of Interacting Bosons in Two Dimensions}, Phys. Rev. Lett. 102 (2009)
120601.
\newblock \href {https://doi.org/10.1103/PhysRevLett.102.120601}
{\path{doi:10.1103/PhysRevLett.102.120601}}.

\bibitem{Floerchinger08}
S.~Floerchinger, C.~Wetterich, {Functional renormalization for Bose-Einstein
condensation}, Phys. Rev. A 77 (2008) 053603.
\newblock \href {https://doi.org/10.1103/PhysRevA.77.053603}
{\path{doi:10.1103/PhysRevA.77.053603}}.

\bibitem{Floerchinger09a}
S.~Floerchinger, C.~Wetterich, {Superfluid Bose gas in two dimensions}, Phys.
Rev. A 79 (2009) 013601.
\newblock \href {https://doi.org/10.1103/PhysRevA.79.013601}
{\path{doi:10.1103/PhysRevA.79.013601}}.

\bibitem{Floerchinger09b}
S.~Floerchinger, C.~Wetterich, {Nonperturbative thermodynamics of an
interacting Bose gas}, Phys. Rev. A 79 (2009) 063602.
\newblock \href {https://doi.org/10.1103/PhysRevA.79.063602}
{\path{doi:10.1103/PhysRevA.79.063602}}.

\bibitem{Eichler09}
C.~Eichler, N.~Hasselmann, P.~Kopietz, {Condensate density of interacting
bosons: A functional renormalization group approach}, Phys. Rev. E 80 (2009)
051129.
\newblock \href {https://doi.org/10.1103/PhysRevE.80.051129}
{\path{doi:10.1103/PhysRevE.80.051129}}.

\bibitem{Krieg17}
J.~Krieg, D.~Strassel, S.~Streib, S.~Eggert, P.~Kopietz, {Thermodynamics and
renormalized quasiparticles in the vicinity of the dilute Bose gas quantum
critical point in two dimensions}, Phys. Rev. B 95 (2017) 024414.
\newblock \href {https://doi.org/10.1103/PhysRevB.95.024414}
{\path{doi:10.1103/PhysRevB.95.024414}}.

\bibitem{Isaule18}
F.~Isaule, M.~C. Birse, N.~R. Walet, Application of the functional
renormalization group to bose gases: From linear to hydrodynamic
fluctuations, Phys. Rev. B 98 (2018) 144502.
\newblock \href {https://doi.org/10.1103/PhysRevB.98.144502}
{\path{doi:10.1103/PhysRevB.98.144502}}.

\bibitem{Isaule20}
F.~Isaule, M.~C. Birse, N.~R. Walet, Thermodynamics of bose gases from
functional renormalization with a hydrodynamic low-energy effective action,
Ann. Phys. 412 (2020) 168006.
\newblock \href {https://doi.org/https://doi.org/10.1016/j.aop.2019.168006}
{\path{doi:https://doi.org/10.1016/j.aop.2019.168006}}.

\bibitem{Lee58}
T.~D. Lee, C.~N. Yang, {Low-Temperature Behavior of a Dilute Bose System of
Hard Spheres. I. Equilibrium Properties}, Phys. Rev. 112 (1958) 1419--1429.
\newblock \href {https://doi.org/10.1103/PhysRev.112.1419}
{\path{doi:10.1103/PhysRev.112.1419}}.

\bibitem{Toyoda82}
T.~Toyoda, {A microscopic theory of the lambda transition}, Ann. Phys. 141~(1)
(1982) 154.
\newblock \href
{https://doi.org/http://dx.doi.org/10.1016/0003-4916(82)90277-9}
{\path{doi:http://dx.doi.org/10.1016/0003-4916(82)90277-9}}.

\bibitem{Huang99}
K.~Huang, {Transition Temperature of a Uniform Imperfect Bose Gas}, Phys. Rev.
Lett. 83 (1999) 3770--3771.
\newblock \href {https://doi.org/10.1103/PhysRevLett.83.3770}
{\path{doi:10.1103/PhysRevLett.83.3770}}.

\bibitem{Gruter97}
P.~Gr\"uter, D.~Ceperley, F.~Lalo\"e, {Critical Temperature of Bose-Einstein
Condensation of Hard-Sphere Gases}, Phys. Rev. Lett. 79 (1997) 3549--3552.
\newblock \href {https://doi.org/10.1103/PhysRevLett.79.3549}
{\path{doi:10.1103/PhysRevLett.79.3549}}.

\bibitem{Holzmann99}
M.~Holzmann, W.~Krauth, {Transition Temperature of the Homogeneous, Weakly
Interacting Bose Gas}, Phys. Rev. Lett. 83 (1999) 2687.
\newblock \href {https://doi.org/10.1103/PhysRevLett.83.2687}
{\path{doi:10.1103/PhysRevLett.83.2687}}.

\bibitem{Holzmann99a}
{Holzmann, M.}, {Gr\"uter, P.}, {Lalo\"e, F.}, {Bose-Einstein condensation in
interacting gases}, Eur. Phys. J. B 10 (1999) 739.
\newblock \href {https://doi.org/10.1007/s100510050905}
{\path{doi:10.1007/s100510050905}}.

\bibitem{Baym99}
G.~Baym, J.-P. Blaizot, M.~Holzmann, F.~Lalo\"e, D.~Vautherin, {The Transition
Temperature of the Dilute Interacting Bose Gas}, Phys. Rev. Lett. 83 (1999)
1703--1706.
\newblock \href {https://doi.org/10.1103/PhysRevLett.83.1703}
{\path{doi:10.1103/PhysRevLett.83.1703}}.

\bibitem{Baym01}
{Baym, G.}, {Blaizot, J.-P.}, {Holzmann, M.}, {Lalo\"e, F.}, {Vautherin, D.},
{Bose-Einstein transition in a dilute interacting gas}, Eur. Phys. J. B 24
(2001) 107.
\newblock \href {https://doi.org/10.1007/s100510170028}
{\path{doi:10.1007/s100510170028}}.

\bibitem{Blaizot05}
J.-P. Blaizot, R.~M. Galain, N.~Wschebor, {Non-Perturbative Renormalization
Group calculation of the transition temperature of the weakly interacting
Bose gas}, Europhys. Lett. 72 (2005) 705.
\newblock \href {https://doi.org/10.1209/epl/i2005-10318-5}
{\path{doi:10.1209/epl/i2005-10318-5}}.

\bibitem{Arnold01}
P.~Arnold, G.~Moore, {BEC Transition Temperature of a Dilute Homogeneous
Imperfect Bose Gas}, Phys. Rev. Lett. 87 (2001) 120401.
\newblock \href {https://doi.org/10.1103/PhysRevLett.87.120401}
{\path{doi:10.1103/PhysRevLett.87.120401}}.

\bibitem{Kashurnikov01}
V.~A. Kashurnikov, N.~V. Prokof'ev, B.~V. Svistunov, {Critical Temperature
Shift in Weakly Interacting Bose Gas}, Phys. Rev. Lett. 87 (2001) 120402.
\newblock \href {https://doi.org/10.1103/PhysRevLett.87.120402}
{\path{doi:10.1103/PhysRevLett.87.120402}}.

\bibitem{Kastening04}
B.~Kastening, {Bose-Einstein condensation temperature of a homogenous weakly
interacting Bose gas in variational perturbation theory through seven loops},
Phys. Rev. A 69 (2004) 043613.
\newblock \href {https://doi.org/10.1103/PhysRevA.69.043613}
{\path{doi:10.1103/PhysRevA.69.043613}}.

\bibitem{Hasselmann04}
N.~Hasselmann, S.~Ledowski, P.~Kopietz, {Critical behavior of weakly
interacting bosons: A functional renormalization-group approach}, Phys. Rev.
A 70 (2004) 063621.
\newblock \href {https://doi.org/10.1103/PhysRevA.70.063621}
{\path{doi:10.1103/PhysRevA.70.063621}}.

\bibitem{Capogrosso07}
B.~Capogrosso-Sansone, N.~V. Prokof'ev, B.~V. Svistunov, {Phase diagram and
thermodynamics of the three-dimensional Bose-Hubbard model}, Phys. Rev. B 75
(2007) 134302.
\newblock \href {https://doi.org/10.1103/PhysRevB.75.134302}
{\path{doi:10.1103/PhysRevB.75.134302}}.

\bibitem{Anders11}
P.~Anders, E.~Gull, L.~Pollet, M.~Troyer, P.~Werner, {Dynamical mean-field
theory for bosons}, New J. Phys. 13 (2011) 075013.
\newblock \href {https://doi.org/10.1088/1367-2630/13/7/075013}
{\path{doi:10.1088/1367-2630/13/7/075013}}.

\bibitem{Capogrosso08}
B.~Capogrosso-Sansone, S.~G. S\"oyler, N.~Prokof'ev, B.~Svistunov, {Monte Carlo
study of the two-dimensional Bose-Hubbard model}, Phys. Rev. A 77 (2008)
015602.
\newblock \href {https://doi.org/10.1103/PhysRevA.77.015602}
{\path{doi:10.1103/PhysRevA.77.015602}}.

\bibitem{Fisher89}
M.~P.~A. Fisher, P.~B. Weichman, G.~Grinstein, D.~S. Fisher, {Boson
localization and the superfluid-insulator transition}, Phys. Rev. B 40 (1989)
546.
\newblock \href {https://doi.org/10.1103/PhysRevB.40.546}
{\path{doi:10.1103/PhysRevB.40.546}}.

\bibitem{Panas15}
J.~Panas, A.~Kauch, J.~Kune\ifmmode~\check{s}\else \v{s}\fi{}, D.~Vollhardt,
K.~Byczuk, Numerical calculation of spectral functions of the bose-hubbard
model using bosonic dynamical mean-field theory, Phys. Rev. B 92 (2015)
045102.
\newblock \href {https://doi.org/10.1103/PhysRevB.92.045102}
{\path{doi:10.1103/PhysRevB.92.045102}}.

\bibitem{Vasilyev09}
O.~Vasilyev, A.~Gambassi, A.~Macio\l{}ek, S.~Dietrich, {Universal scaling
functions of critical Casimir forces obtained by Monte Carlo simulations},
Phys. Rev. E 79 (2009) 041142.
\newblock \href {https://doi.org/10.1103/PhysRevE.79.041142}
{\path{doi:10.1103/PhysRevE.79.041142}}.

\bibitem{Hucht11}
A.~Hucht, D.~Gr\"uneberg, F.~M. Schmidt, {Aspect-ratio dependence of
thermodynamic Casimir forces}, Phys. Rev. E 83 (2011) 051101.
\newblock \href {https://doi.org/10.1103/PhysRevE.83.051101}
{\path{doi:10.1103/PhysRevE.83.051101}}.

\bibitem{LopesCardozo15}
D.~Lopes~Cardozo, {Finite size scaling and the critical Casimir force : Ising
magnets and binary fluids}, Ph.D. thesis (2015).

\bibitem{Dantchev04}
D.~Dantchev, M.~Krech, {Critical Casimir force and its fluctuations in lattice
spin models: Exact and Monte Carlo results}, Phys. Rev. E 69 (2004) 046119.
\newblock \href {https://doi.org/10.1103/PhysRevE.69.046119}
{\path{doi:10.1103/PhysRevE.69.046119}}.

\bibitem{Blaizot07a}
J.-P. Blaizot, A.~Ipp, R.~M\'endez-Galain, N.~Wschebor, {Perturbation theory
and non-perturbative renormalization flow in scalar field theory at finite
temperature}, Nucl. Phys. A 784 (2007) 376.
\newblock \href {https://doi.org/10.1016/j.nuclphysa.2006.11.139}
{\path{doi:10.1016/j.nuclphysa.2006.11.139}}.

\bibitem{Blaizot11}
J.-P. Blaizot, A.~Ipp, N.~Wschebor, {Calculation of the pressure of a hot
scalar theory within the Non-Perturbative Renormalization Group}, Nucl. Phys.
A 849 (2011) 165.
\newblock \href {https://doi.org/10.1016/j.nuclphysa.2010.10.007}
{\path{doi:10.1016/j.nuclphysa.2010.10.007}}.

\bibitem{Agostini97}
V.~Agostini, G.~Carlino, M.~Caselle, M.~Hasenbusch, The spectrum of the 2 +
1-dimensional gauge ising model, Nucl. Phys. B 484 (1997) 331.
\newblock \href {https://doi.org/https://doi.org/10.1016/S0550-3213(96)00539-1}
{\path{doi:https://doi.org/10.1016/S0550-3213(96)00539-1}}.

\bibitem{Caselle99}
M.~Caselle, M.~Hasenbusch, P.~Provero, {Non-perturbative states in the 3D
$\varphi^4$ theory}, Nucl. Phys. B 556 (1999) 575 -- 600.
\newblock \href {https://doi.org/https://doi.org/10.1016/S0550-3213(99)00333-8}
{\path{doi:https://doi.org/10.1016/S0550-3213(99)00333-8}}.

\bibitem{Nishiyama14}
Y.~Nishiyama, {Universal critical behavior of the two-magnon-bound-state mass
gap for the (2+1)-dimensional Ising model}, Physica A 413 (2014) 577 -- 582.
\newblock \href {https://doi.org/https://doi.org/10.1016/j.physa.2014.07.025}
{\path{doi:https://doi.org/10.1016/j.physa.2014.07.025}}.

\bibitem{Podolsky11}
D.~Podolsky, A.~Auerbach, D.~P. Arovas, {Visibility of the amplitude (Higgs)
mode in condensed matter}, Phys. Rev. B 84 (2011) 174522.
\newblock \href {https://doi.org/10.1103/PhysRevB.84.174522}
{\path{doi:10.1103/PhysRevB.84.174522}}.

\bibitem{Rancon14}
A.~Ran\c{c}on, N.~Dupuis, {Higgs amplitude mode in the vicinity of a
$(2+1)$-dimensional quantum critical point}, Phys. Rev. B 89 (2014)
180501(R).
\newblock \href {https://doi.org/10.1103/PhysRevB.89.180501}
{\path{doi:10.1103/PhysRevB.89.180501}}.

\bibitem{Gazit13}
S.~Gazit, D.~Podolsky, A.~Auerbach, {Fate of the Higgs Mode Near Quantum
Criticality}, Phys. Rev. Lett. 110 (2013) 140401.
\newblock \href {https://doi.org/10.1103/PhysRevLett.110.140401}
{\path{doi:10.1103/PhysRevLett.110.140401}}.

\bibitem{Nishiyama15}
Y.~Nishiyama, {Critical behavior of the Higgs- and Goldstone-mass gaps for the
two-dimensional \{XY\} model}, Nucl. Phys. B 897 (2015) 555.
\newblock \href {https://doi.org/10.1016/j.nuclphysb.2015.06.006}
{\path{doi:10.1016/j.nuclphysb.2015.06.006}}.

\bibitem{Nishiyama16}
Y.~Nishiyama, {Universal scaled Higgs-mass gap for the bilayer Heisenberg model
in the ordered phase}, Eur. Phys. J. B 89 (2016) 1.
\newblock \href {https://doi.org/10.1140/epjb/e2016-60885-0}
{\path{doi:10.1140/epjb/e2016-60885-0}}.

\bibitem{Lohofer15}
M.~Loh\"ofer, T.~Coletta, D.~G. Joshi, F.~F. Assaad, M.~Vojta, S.~Wessel,
F.~Mila, {Dynamical structure factors and excitation modes of the bilayer
Heisenberg model}, Phys. Rev. B 92 (2015) 245137.
\newblock \href {https://doi.org/10.1103/PhysRevB.92.245137}
{\path{doi:10.1103/PhysRevB.92.245137}}.

\bibitem{Rose17}
F.~Rose, N.~Dupuis, {Nonperturbative functional renormalization-group approach
to transport in the vicinity of a $(2+1)$-dimensional O($N$)-symmetric
quantum critical point}, Phys. Rev. B 95 (2017) 014513.
\newblock \href {https://doi.org/10.1103/PhysRevB.95.014513}
{\path{doi:10.1103/PhysRevB.95.014513}}.

\bibitem{Negele_book}
J.~W. Negele, H.~Orland, {Quantum Many-particle Systems}, Westview Press, 1998.

\bibitem{Salmhofer19}
M.~Salmhofer,
\href{http://dx.doi.org/10.1016/j.nuclphysb.2018.07.004}{Renormalization in
condensed matter: Fermionic systems – from mathematics to materials},
Nuclear Physics B 941 (2019) 868–899.
\newblock \href {https://doi.org/10.1016/j.nuclphysb.2018.07.004}
{\path{doi:10.1016/j.nuclphysb.2018.07.004}}.
\newline\urlprefix\url{http://dx.doi.org/10.1016/j.nuclphysb.2018.07.004}

\bibitem{Metzner98}
W.~Metzner, C.~Castellani, C.~D. Castro, Fermi systems with strong forward
scattering, Adv. Phys. 47~(3) (1998) 317--445.
\newblock \href {https://doi.org/10.1080/000187398243528}
{\path{doi:10.1080/000187398243528}}.

\bibitem{Husemann09}
C.~Husemann, M.~Salmhofer, Efficient parametrization of the vertex function,
$\omega$ scheme, and the $t,t'$ hubbard model at van hove filling, Phys. Rev.
B 79 (2009) 195125.
\newblock \href {https://doi.org/10.1103/PhysRevB.79.195125}
{\path{doi:10.1103/PhysRevB.79.195125}}.

\bibitem{Andergassen04}
S.~Andergassen, T.~Enss, V.~Meden, W.~Metzner, U.~Scholl\-w{\"o}ck,
K.~Sch{\"o}nhammer, {Functional renormalization group for Luttinger liquids
with impurities}, Phys. Rev. B 70 (2004) 075102.
\newblock \href {https://doi.org/10.1103/PhysRevB.70.075102}
{\path{doi:10.1103/PhysRevB.70.075102}}.

\bibitem{Honerkamp04}
C.~Honerkamp, D.~Rohe, S.~Andergassen, T.~Enss, {Interaction flow method for
many-fermion systems}, Phys. Rev. B 70 (2004) 235115.
\newblock \href {https://doi.org/10.1103/PhysRevB.70.235115}
{\path{doi:10.1103/PhysRevB.70.235115}}.

\bibitem{Honerkamp01b}
C.~Honerkamp, M.~Salmhofer, {Temperature-flow renormalization group and the
competition between superconductivity and ferromagnetism}, Phys. Rev. B 64
(2001) 184516.
\newblock \href {https://doi.org/10.1103/PhysRevB.64.184516}
{\path{doi:10.1103/PhysRevB.64.184516}}.

\bibitem{Shankar94}
R.~Shankar, {Renormalization-group approach to interacting fermions}, Rev. Mod.
Phys. 66~(1) (1994) 129.
\newblock \href {https://doi.org/10.1103/RevModPhys.66.129}
{\path{doi:10.1103/RevModPhys.66.129}}.

\bibitem{Anderson87}
P.~W. Anderson, {The Resonating Valence Bond State in La$_2$CuO$_4$ and
Superconductivity}, Science 235 (1987) 1196.
\newblock \href {https://doi.org/10.1126/science.235.4793.1196}
{\path{doi:10.1126/science.235.4793.1196}}.

\bibitem{Scalapino95}
D.~J. Scalapino, {The case for $d_{{x^2-y^2}}$ pairing in the cuprate
superconductors}, Phys. Rep. 250 (1995) 329.
\newblock \href {https://doi.org/10.1016/0370-1573(94)00086-I}
{\path{doi:10.1016/0370-1573(94)00086-I}}.

\bibitem{Zanchi98}
D.~Zanchi, H.~Schulz, {Weakly correlated electrons on a square lattice: A
renormalization group theory}, Europhys. Lett. 44~(2) (1998) 235--241.
\newblock \href {https://doi.org/10.1209/epl/i1998-00462-x}
{\path{doi:10.1209/epl/i1998-00462-x}}.

\bibitem{Zanchi00}
D.~Zanchi, H.~J. Schulz, {Weakly correlated electrons on a square lattice:
Renormalization-group theory}, Phys. Rev. B 61 (2000) 13609--13632.
\newblock \href {https://doi.org/10.1103/PhysRevB.61.13609}
{\path{doi:10.1103/PhysRevB.61.13609}}.

\bibitem{Halboth00}
C.~J. Halboth, W.~Metzner, {Renormalization-group analysis of the
two-dimensional Hubbard model}, Phys. Rev. B 61 (2000) 7364--7377.
\newblock \href {https://doi.org/10.1103/PhysRevB.61.7364}
{\path{doi:10.1103/PhysRevB.61.7364}}.

\bibitem{Halboth00a}
C.~J. Halboth, W.~Metzner, {d-Wave Superconductivity and Pomeranchuk
Instability in the Two-Dimensional Hubbard Model}, Phys. Rev. Lett. 85 (2000)
5162.
\newblock \href {https://doi.org/10.1103/PhysRevLett.85.5162}
{\path{doi:10.1103/PhysRevLett.85.5162}}.

\bibitem{Honerkamp01}
C.~Honerkamp, M.~Salmhofer, N.~Furukawa, T.~M. Rice, {Breakdown of the
Landau-Fermi liquid in two dimensions due to umklapp scattering}, Phys. Rev.
B 63 (2001) 035109.
\newblock \href {https://doi.org/10.1103/PhysRevB.63.035109}
{\path{doi:10.1103/PhysRevB.63.035109}}.

\bibitem{Honerkamp01a}
C.~Honerkamp, M.~Salmhofer, {Magnetic and Superconducting Instabilities of the
Hubbard Model at the Van Hove Filling}, Phys. Rev. Lett. 87 (2001) 187004.
\newblock \href {https://doi.org/10.1103/PhysRevLett.87.187004}
{\path{doi:10.1103/PhysRevLett.87.187004}}.

\bibitem{Giering12}
K.-U. Giering, M.~Salmhofer, {Self-energy flows in the two-dimensional
repulsive Hubbard model}, Phys. Rev. B 86 (2012) 245122.
\newblock \href {https://doi.org/10.1103/PhysRevB.86.245122}
{\path{doi:10.1103/PhysRevB.86.245122}}.

\bibitem{Karrasch08}
C.~Karrasch, R.~Hedden, R.~Peters, T.~Pruschke, K.~Sch{\"o}n\-hammer, V.~Meden,
J. Phys.: Condensed Matter 20 (2008) 345205.
\newblock \href {https://doi.org/10.1088/0953-8984/20/34/345205}
{\path{doi:10.1088/0953-8984/20/34/345205}}.

\bibitem{Husemann12}
C.~Husemann, K.-U. Giering, M.~Salmhofer, {Frequency-dependent vertex functions
of the ($t,t'$) Hubbard model at weak coupling}, Phys. Rev. B 85 (2012)
075121.
\newblock \href {https://doi.org/10.1103/PhysRevB.85.075121}
{\path{doi:10.1103/PhysRevB.85.075121}}.

\bibitem{Vilardi17}
D.~Vilardi, C.~Taranto, W.~Metzner, {Nonseparable frequency dependence of the
two-particle vertex in interacting fermion systems}, Phys. Rev. B 96 (2017)
235110.
\newblock \href {https://doi.org/10.1103/PhysRevB.96.235110}
{\path{doi:10.1103/PhysRevB.96.235110}}.

\bibitem{Kugler18}
F.~B. Kugler, J.~von Delft, {Multiloop Functional Renormalization Group That
Sums Up All Parquet Diagrams}, Phys. Rev. Lett. 120 (2018) 057403.
\newblock \href {https://doi.org/10.1103/PhysRevLett.120.057403}
{\path{doi:10.1103/PhysRevLett.120.057403}}.

\bibitem{Kugler18a}
F.~Kugler, J.~von Delft, {Multiloop functional renormalization group for
general models}, Phys. Rev. B 97 (2018) 035162.
\newblock \href {https://doi.org/10.1103/PhysRevB.97.035162}
{\path{doi:10.1103/PhysRevB.97.035162}}.

\bibitem{Tagliavini19}
A.~Tagliavini, C.~Hille, F.~B. Kugler, S.~Andergassen, A.~Toschi, C.~Honerkamp,
{Multiloop functional renormalization group for the two-dimensional Hubbard
model: Loop convergence of the response functions}, SciPost Phys. 6 (2019)
009.
\newblock \href {https://doi.org/10.21468/SciPostPhys.6.1.009}
{\path{doi:10.21468/SciPostPhys.6.1.009}}.

\bibitem{Hille20}
C.~Hille, F.~B. Kugler, C.~J. Eckhardt, Y.-Y. He, A.~Kauch, C.~Honerkamp,
A.~Toschi, S.~Andergassen, {Quantitative functional renormalisation-group
description of the two-dimensional Hubbard model} (2020).
\newblock \href {http://arxiv.org/abs/2002.02733} {\path{arXiv:2002.02733}}.

\bibitem{Graser09}
S.~Graser, T.~A. Maier, P.~J. Hirschfeld, D.~J. Scalapino, {Near-degeneracy of
several pairing channels in multiorbital models for the Fe pnictides}, New J.
Phys. 11 (2009) 025016.
\newblock \href {https://doi.org/10.1088/1367-2630/11/2/025016}
{\path{doi:10.1088/1367-2630/11/2/025016}}.

\bibitem{Maier09}
T.~A. Maier, S.~Graser, D.~J. Scalapino, P.~J. Hirschfeld, {Origin of gap
anisotropy in spin fluctuation models of the iron pnictides}, Phys. Rev. B 79
(2009) 224510.
\newblock \href {https://doi.org/10.1103/PhysRevB.79.224510}
{\path{doi:10.1103/PhysRevB.79.224510}}.

\bibitem{Platt13}
C.~{Platt}, W.~{Hanke}, R.~{Thomale}, {Functional renormalization group for
multi-orbital Fermi surface instabilities}, Adv. in Phys. 62~(4-6) (2013)
453--562.
\newblock \href {http://arxiv.org/abs/1310.6191} {\path{arXiv:1310.6191}},
\href {https://doi.org/10.1080/00018732.2013.862020}
{\path{doi:10.1080/00018732.2013.862020}}.

\bibitem{Schober16}
G.~A.~H. Schober, K.-U. Giering, M.~M. Scherer, C.~Honerkamp, M.~Salmhofer,
{Functional renormalization and mean-field approach to multiband systems with
spin-orbit coupling: Application to the Rashba model with attractive
interaction}, Phys. Rev. B 93 (2016) 115111.
\newblock \href {https://doi.org/10.1103/PhysRevB.93.115111}
{\path{doi:10.1103/PhysRevB.93.115111}}.

\bibitem{Scherer18}
M.~M. Scherer, C.~Honerkamp, A.~N. Rudenko, E.~A. Stepanov, A.~I. Lichtenstein,
M.~I. Katsnelson, {Excitonic instability and unconventional pairing in the
nodal-line materials ZrSiS and ZrSiSe}, Phys. Rev. B 98 (2018) 241112(R).
\newblock \href {https://doi.org/10.1103/PhysRevB.98.241112}
{\path{doi:10.1103/PhysRevB.98.241112}}.

\bibitem{Classen19}
L.~Classen, C.~Honerkamp, M.~M. Scherer, {Competing phases of interacting
electrons on triangular lattices in moir\'e heterostructures}, Phys. Rev. B
99 (2019) 195120.
\newblock \href {https://doi.org/10.1103/PhysRevB.99.195120}
{\path{doi:10.1103/PhysRevB.99.195120}}.

\bibitem{Salmhofer04}
M.~Salmhofer, C.~Honerkamp, W.~Metzner, O.~Lauscher, {Renormalization Group
Flows into Phases with Broken Symmetry}, Prog. Theor. Phys. 112 (2004)
943--970.
\newblock \href {https://doi.org/10.1143/PTP.112.943}
{\path{doi:10.1143/PTP.112.943}}.

\bibitem{Eberlein13a}
A.~Eberlein, W.~Metzner, {Effective interactions and fluctuation effects in
spin-singlet superfluids}, Phys. Rev. B 87 (2013) 174523.
\newblock \href {https://doi.org/10.1103/PhysRevB.87.174523}
{\path{doi:10.1103/PhysRevB.87.174523}}.

\bibitem{Katanin04}
A.~A. Katanin, {Fulfillment of Ward identities in the functional
renormalization group approach}, Phys. Rev. B 70 (2004) 115109.
\newblock \href {https://doi.org/10.1103/PhysRevB.70.115109}
{\path{doi:10.1103/PhysRevB.70.115109}}.

\bibitem{Gersch05}
R.~Gersch, C.~Honerkamp, D.~Rohe, W.~Metzner, {Fermionic renormalization group
flow into phases with broken discrete symmetry: charge-density wave
mean-field model}, Eur. Phys. J. B 48~(3) (2005) 349--358.
\newblock \href {https://doi.org/10.1140/epjb/e2005-00416-8}
{\path{doi:10.1140/epjb/e2005-00416-8}}.

\bibitem{Gersch08}
R.~Gersch, C.~Honerkamp, W.~Metzner, {Superconductivity in the attractive
Hubbard model: functional renormalization group analysis}, New J. Phys. 10
(2008) 045003.
\newblock \href {https://doi.org/10.1088/1367-2630/10/4/045003}
{\path{doi:10.1088/1367-2630/10/4/045003}}.

\bibitem{Eberlein14fb}
A.~Eberlein, W.~Metzner, {Superconductivity in the two-dimensional
$t$-$t'$-Hubbard model}, Phys. Rev. B 89 (2014) 035126.
\newblock \href {https://doi.org/10.1103/PhysRevB.89.035126}
{\path{doi:10.1103/PhysRevB.89.035126}}.

\bibitem{Popov87}
V.~N. Popov, Functional integrals and collective excitations, Cambridge
University Press, Cambridge, 1987.

\bibitem{Baier04}
T.~Baier, E.~Bick, C.~Wetterich, {Temperature dependence of antiferromagnetic
order in the Hubbard model}, Phys. Rev. B 70 (2004) 125111.
\newblock \href {https://doi.org/10.1103/PhysRevB.70.125111}
{\path{doi:10.1103/PhysRevB.70.125111}}.

\bibitem{Birse05fb}
M.~C. Birse, B.~Krippa, J.~A. McGovern, N.~R. Walet, {Pairing in many-fermion
systems: an exact renormalization group treatment}, Phys. Lett. B 605 (2005)
287.
\newblock \href {https://doi.org/10.1016/j.physletb.2004.11.044}
{\path{doi:10.1016/j.physletb.2004.11.044}}.

\bibitem{Diehl07fb}
S.~Diehl, H.~Gies, J.~Pawlowski, C.~Wetterich, {Flow equations for the BCS-BEC
crossover}, Phys. Rev. A 76 (2007) 021602(R).
\newblock \href {https://doi.org/10.1103/PhysRevA.76.021602}
{\path{doi:10.1103/PhysRevA.76.021602}}.

\bibitem{Diehl:2007ri}
S.~Diehl, H.~Gies, J.~Pawlowski, C.~Wetterich, {Renormalisation flow and
universality for ultracold fermionic atoms}, Phys.Rev. A76 (2007) 053627.
\newblock \href {http://arxiv.org/abs/cond-mat/0703366}
{\path{arXiv:cond-mat/0703366}}, \href
{https://doi.org/10.1103/PhysRevA.76.053627}
{\path{doi:10.1103/PhysRevA.76.053627}}.

\bibitem{Krippa07}
B.~Krippa, {Superfluidity in many fermion systems: Exact renormalisation group
treatment}, Eur. Phys. J. A 31 (2007) 734.
\newblock \href {https://doi.org/10.1140/epja/i2006-10286-2}
{\path{doi:10.1140/epja/i2006-10286-2}}.

\bibitem{Gies:2001nw}
H.~Gies, C.~Wetterich, {Renormalization flow of bound states}, Phys.Rev. D65
(2002) 065001.
\newblock \href {http://arxiv.org/abs/hep-th/0107221}
{\path{arXiv:hep-th/0107221}}, \href
{https://doi.org/10.1103/PhysRevD.65.065001}
{\path{doi:10.1103/PhysRevD.65.065001}}.

\bibitem{Gies:2002hq}
H.~Gies, C.~Wetterich, {Universality of spontaneous chiral symmetry breaking in
gauge theories}, Phys.Rev. D69 (2004) 025001.
\newblock \href {http://arxiv.org/abs/hep-th/0209183}
{\path{arXiv:hep-th/0209183}}, \href
{https://doi.org/10.1103/PhysRevD.69.025001}
{\path{doi:10.1103/PhysRevD.69.025001}}.

\bibitem{Floerchinger:2008qc}
S.~Floerchinger, M.~Scherer, S.~Diehl, C.~Wetterich, {Particle-hole
fluctuations in the BCS-BEC Crossover}, Phys. Rev. B78 (2008) 174528.
\newblock \href {http://arxiv.org/abs/0808.0150} {\path{arXiv:0808.0150}},
\href {https://doi.org/10.1103/PhysRevB.78.174528}
{\path{doi:10.1103/PhysRevB.78.174528}}.

\bibitem{Bartosch09fb}
L.~Bartosch, P.~Kopietz, A.~Ferraz, {Renormalization of the BCS-BEC crossover
by order-parameter fluctuations}, Phys. Rev. B 80 (2009) 104514.
\newblock \href {https://doi.org/10.1103/PhysRevB.80.104514}
{\path{doi:10.1103/PhysRevB.80.104514}}.

\bibitem{Obert13}
B.~Obert, C.~Husemann, W.~Metzner, {Low-energy singularities in the ground
state of fermionic superfluids}, Phys. Rev. B 88 (2013) 144508.
\newblock \href {https://doi.org/10.1103/PhysRevB.88.144508}
{\path{doi:10.1103/PhysRevB.88.144508}}.

\bibitem{Krahl09a}
H.~C. Krahl, J.~A. M\"uller, C.~Wetterich, {Generation of d-wave coupling in
the two-dimensional Hubbard model from functional renormalization}, Phys.
Rev. B 79 (2009) 094526.
\newblock \href {https://doi.org/10.1103/PhysRevB.79.094526}
{\path{doi:10.1103/PhysRevB.79.094526}}.

\bibitem{Friederich10}
S.~Friederich, H.~C. Krahl, C.~Wetterich, {Four-point vertex in the Hubbard
model and partial bosonization}, Phys. Rev. B 81 (2010) 235108.
\newblock \href {https://doi.org/10.1103/PhysRevB.81.235108}
{\path{doi:10.1103/PhysRevB.81.235108}}.

\bibitem{Friederich11}
S.~Friederich, H.~C. Krahl, C.~Wetterich, {Functional renormalization for
spontaneous symmetry breaking in the Hubbard model}, Phys. Rev. B 83 (2011)
155125.
\newblock \href {https://doi.org/10.1103/PhysRevB.83.155125}
{\path{doi:10.1103/PhysRevB.83.155125}}.

\bibitem{Reiss07}
J.~Reiss, D.~Rohe, W.~Metzner, {Renormalized mean-field analysis of
antiferromagnetism and d-wave superconductivity in the two-dimensional
Hubbard model}, Phys. Rev. B 75 (2007) 075110.
\newblock \href {https://doi.org/10.1103/PhysRevB.75.075110}
{\path{doi:10.1103/PhysRevB.75.075110}}.

\bibitem{Wang14}
J.~Wang, A.~Eberlein, W.~Metzner, {Competing order in correlated electron
systems made simple: Consistent fusion of functional renormalization and
mean-field theory}, Phys. Rev. B 89 (2014) 121116(R).
\newblock \href {https://doi.org/10.1103/PhysRevB.89.121116}
{\path{doi:10.1103/PhysRevB.89.121116}}.

\bibitem{Yamase16}
H.~Yamase, A.~Eberlein, W.~Metzner, {Coexistence of Incommensurate Magnetism
and Superconductivity in the Two-Dimensional Hubbard Model}, Phys. Rev. Lett.
116 (2016) 096402.
\newblock \href {https://doi.org/10.1103/PhysRevLett.116.096402}
{\path{doi:10.1103/PhysRevLett.116.096402}}.

\bibitem{Nagy11}
S.~Nagy, K.~Sailer, {Functional renormalization group for quantized anharmonic
oscillator }, Ann. Phys. 326~(8) (2011) 1839 -- 1876.
\newblock \href {https://doi.org/http://dx.doi.org/10.1016/j.aop.2011.04.011}
{\path{doi:http://dx.doi.org/10.1016/j.aop.2011.04.011}}.

\bibitem{Polonyi05}
J.~Polonyi, K.~Sailer, {Renormalization group in internal space}, Phys. Rev. D
71 (2005) 025010.
\newblock \href {https://doi.org/10.1103/PhysRevD.71.025010}
{\path{doi:10.1103/PhysRevD.71.025010}}.

\bibitem{Blaizot11b}
J.-P. Blaizot, J.~M. Pawlowski, U.~Reinosa, {Exact renormalization group and
$\Phi$-derivable approximations }, Phys. Lett. B 696 (2011) 523 -- 528.
\newblock \href
{https://doi.org/http://dx.doi.org/10.1016/j.physletb.2010.12.058}
{\path{doi:http://dx.doi.org/10.1016/j.physletb.2010.12.058}}.

\bibitem{Luttinger60b}
J.~M. Luttinger, J.~C. Ward, {Ground-State Energy of a Many-Fermion System.
II}, Phys. Rev. 118 (1960) 1417--1427.
\newblock \href {https://doi.org/10.1103/PhysRev.118.1417}
{\path{doi:10.1103/PhysRev.118.1417}}.

\bibitem{Baym61}
G.~Baym, L.~P. Kadanoff, {Conservation Laws and Correlation Functions}, Phys.
Rev. 124 (1961) 287--299.
\newblock \href {https://doi.org/10.1103/PhysRev.124.287}
{\path{doi:10.1103/PhysRev.124.287}}.

\bibitem{Baym62}
G.~Baym, {Self-Consistent Approximations in Many-Body Systems}, Phys. Rev. 127
(1962) 1391--1401.
\newblock \href {https://doi.org/10.1103/PhysRev.127.1391}
{\path{doi:10.1103/PhysRev.127.1391}}.

\bibitem{Dedominicis64a}
C.~D. Dominicis, P.~C. Martin, {Stationary Entropy Principle and
Renormalization in Normal and Superfluid Systems. I. Algebraic Formulation},
J. Math. Phys. 5~(1) (1964) 14--30.
\newblock \href {https://doi.org/10.1063/1.1704062}
{\path{doi:10.1063/1.1704062}}.

\bibitem{Dedominicis64b}
C.~D. Dominicis, P.~C. Martin, {Stationary Entropy Principle and
Renormalization in Normal and Superfluid Systems. II. Diagrammatic
Formulation}, J. Math. Phys. 5~(1) (1964) 31--59.
\newblock \href {https://doi.org/10.1063/1.1704064}
{\path{doi:10.1063/1.1704064}}.

\bibitem{Cornwall74}
J.~M. Cornwall, R.~Jackiw, E.~Tomboulis, {Effective action for composite
operators}, Phys. Rev. D 10 (1974) 2428--2445.
\newblock \href {https://doi.org/10.1103/PhysRevD.10.2428}
{\path{doi:10.1103/PhysRevD.10.2428}}.

\bibitem{Dupuis:2005ij}
N.~Dupuis, {Renormalization group approach to interacting fermion systems in
the two-particle-irreducible formalism}, Eur. Phys. J. B48 (2005) 319.
\newblock \href {http://arxiv.org/abs/cond-mat/0506542}
{\path{arXiv:cond-mat/0506542}}, \href
{https://doi.org/10.1140/epjb/e2005-00409-7}
{\path{doi:10.1140/epjb/e2005-00409-7}}.

\bibitem{Dupuis14}
N.~Dupuis, {Nonperturbative renormalization-group approach to fermion systems
in the two-particle-irreducible effective action formalism}, Phys. Rev. B 89
(2014) 035113.
\newblock \href {https://doi.org/10.1103/PhysRevB.89.035113}
{\path{doi:10.1103/PhysRevB.89.035113}}.

\bibitem{Rentrop15}
J.~F. Rentrop, S.~G. Jakobs, V.~Meden, {Two-particle irreducible functional
renormalization group schemes—a comparative study}, J. Phys. A 48~(14)
(2015) 145002.
\newblock \href {https://doi.org/doi:10.1088/1751-8113/48/14/145002}
{\path{doi:doi:10.1088/1751-8113/48/14/145002}}.

\bibitem{Katanin19}
A.~A. Katanin, Extended dynamical mean field theory combined with the
two-particle irreducible functional renormalization-group approach as a tool
to study strongly correlated systems, Phys. Rev. B 99 (2019) 115112.
\newblock \href {https://doi.org/10.1103/PhysRevB.99.115112}
{\path{doi:10.1103/PhysRevB.99.115112}}.

\bibitem{Rentrop16}
J.~F. Rentrop, V.~Meden, S.~G. Jakobs, {Renormalization group flow of the
Luttinger-Ward functional: Conserving approximations and application to the
Anderson impurity model}, Phys. Rev. B 93 (2016) 195160.
\newblock \href {https://doi.org/10.1103/PhysRevB.93.195160}
{\path{doi:10.1103/PhysRevB.93.195160}}.

\bibitem{Polonyi02fb}
J.~Polonyi, K.~Sailer, {Effective action and density-functional theory}, Phys.
Rev. B 66 (2002) 155113.
\newblock \href {https://doi.org/10.1103/PhysRevB.66.155113}
{\path{doi:10.1103/PhysRevB.66.155113}}.

\bibitem{Kemler13}
S.~Kemler, J.~Braun, {Towards a renormalization group approach to density
functional theory—general formalism and case studies}, J. Phys. G: Nucl.
Part. Phys. 40~(8) (2013) 085105.
\newblock \href {https://doi.org/10.1088/0954-3899/40/8/085105}
{\path{doi:10.1088/0954-3899/40/8/085105}}.

\bibitem{Yokota19}
T.~Yokota, K.~Yoshida, T.~Kunihiro, {Functional renormalization-group
calculation of the equation of state of one-dimensional uniform matter
inspired by the Hohenberg-Kohn theorem}, Phys. Rev. C 99 (2019) 024302.
\newblock \href {https://doi.org/10.1103/PhysRevC.99.024302}
{\path{doi:10.1103/PhysRevC.99.024302}}.

\bibitem{Yokota19a}
T.~Yokota, K.~Yoshida, T.~Kunihiro, {Ab initio description of excited states of
1D uniform matter with the Hohenberg–Kohn-theorem-inspired
functional-renormalization-group method}, Progress of Theoretical and
Experimental Physics 2019~(1), 011D01 (01 2019).
\newblock \href {https://doi.org/10.1093/ptep/pty139}
{\path{doi:10.1093/ptep/pty139}}.

\bibitem{Yokota19b}
T.~Yokota, T.~Naito, {Functional-renormalization-group aided density functional
analysis for the correlation energy of the two-dimensional homogeneous
electron gas}, Phys. Rev. B 99 (2019) 115106.
\newblock \href {https://doi.org/10.1103/PhysRevB.99.115106}
{\path{doi:10.1103/PhysRevB.99.115106}}.

\bibitem{Hanson07}
R.~Hanson, L.~P. Kouwenhoven, J.~R. Petta, S.~Tarucha, L.~M.~K. Vandersypen,
{Spins in few-electron quantum dots}, Rev. Mod. Phys. 79 (2007) 1217.
\newblock \href {https://doi.org/10.1103/RevModPhys.79.1217}
{\path{doi:10.1103/RevModPhys.79.1217}}.

\bibitem{Mahan_book}
G.~D. Mahan, {Many-Particle Physics}, 3rd Edition, Kluwer Academic/Plenum
Publishers, New York, 2000.

\bibitem{Landauer57}
R.~Landauer, {Spatial Variation of Currents and Fields Due to Localized
Scatterers in Metallic Conduction}, IBM J. Res. Dev. 1 (1957) 223.
\newblock \href {https://doi.org/10.1147/rd.13.0223}
{\path{doi:10.1147/rd.13.0223}}.

\bibitem{Buettiker86}
M.~B{\"u}ttiker, {Four-Terminal Phase-Coherent Conductance}, Phys. Rev. Lett.
57 (1986) 1761.
\newblock \href {https://doi.org/10.1103/PhysRevLett.57.1761}
{\path{doi:10.1103/PhysRevLett.57.1761}}.

\bibitem{Oguri01}
A.~Oguri, {Transmission Probability for Interacting Electrons Connected to
Reservoirs}, J. Phys. Soc. Jpn. 70 (2001) 2666.
\newblock \href {https://doi.org/10.1143/JPSJ.70.2666}
{\path{doi:10.1143/JPSJ.70.2666}}.

\bibitem{Kane92}
C.~L. Kane, M.~P.~A. Fisher, Transmission through barriers and resonant
tunneling in an interacting one-dimensional electron gas, Phys. Rev. B 46
(1992) 15233--15262.
\newblock \href {https://doi.org/10.1103/PhysRevB.46.15233}
{\path{doi:10.1103/PhysRevB.46.15233}}.

\bibitem{Matveev93}
K.~A. Matveev, D.~Yue, L.~I. Glazman, {Tunneling in one-dimensional
non-Luttinger electron liquid}, Phys. Rev. Lett. 71 (1993) 3351.
\newblock \href {https://doi.org/10.1103/PhysRevLett.71.3351}
{\path{doi:10.1103/PhysRevLett.71.3351}}.

\bibitem{Giamarchi_book}
T.~Giamarchi, {Quantum physics in one dimension}, Oxford University Press,
Oxford, 2004.

\bibitem{Meden02}
V.~Meden, W.~Metzner, U.~Schollw{\"o}ck, K.~Sch{\"o}n\-hammer, {Scaling
behavior of impurities in mesoscopic Luttinger liquids}, Phys. Rev. B 65
(2002) 045318.
\newblock \href {https://doi.org/10.1103/PhysRevB.65.045318}
{\path{doi:10.1103/PhysRevB.65.045318}}.

\bibitem{Enss05}
T.~Enss, V.~Meden, S.~Andergassen, X.~Barnab\'e-Th\'eriault, W.~Metzner,
K.~Sch{\"o}nhammer, {Impurity and correlation effects on transport in
one-dimensional quantum wires}, Phys. Rev. B 71 (2005) 155401.
\newblock \href {https://doi.org/10.1103/PhysRevB.71.155401}
{\path{doi:10.1103/PhysRevB.71.155401}}.

\bibitem{Meden05}
V.~Meden, T.~Enss, S.~Andergassen, W.~Metzner, K.~Sch{\"o}nhammer, {Correlation
effects on resonant tunneling in one-dimensional quantum wires}, Phys. Rev. B
71 (2005) 041302.
\newblock \href {https://doi.org/10.1103/PhysRevB.71.041302}
{\path{doi:10.1103/PhysRevB.71.041302}}.

\bibitem{Meden03}
V.~Meden, U.~Schollw{\"o}ck, {Persistent currents in mesoscopic rings: A
numerical and renormalization group study}, Phys. Rev. B 67 (2003) 035106.
\newblock \href {https://doi.org/10.1103/PhysRevB.67.035106}
{\path{doi:10.1103/PhysRevB.67.035106}}.

\bibitem{Meden03a}
V.~Meden, U.~Schollw{\"o}ck, {Conductance of interacting nanowires}, Phys. Rev.
B 67 (2003) 193303.
\newblock \href {https://doi.org/10.1103/PhysRevB.67.193303}
{\path{doi:10.1103/PhysRevB.67.193303}}.

\bibitem{Barnabe05}
X.~Barnab\'e-Th\'eriault, A.~Sedeki, V.~Meden, K.~Sch{\"o}n\-hammer, {Junction
of Three Quantum Wires: Restoring Time-Reversal Symmetry by Interaction},
Phys. Rev. Lett. 94 (2005) 136405.
\newblock \href {https://doi.org/10.1103/PhysRevLett.94.136405}
{\path{doi:10.1103/PhysRevLett.94.136405}}.

\bibitem{Karrasch10}
C.~Karrasch, M.~Pletyukhov, L.~Borda, V.~Meden, {Functional renormalization
group study of the interacting resonant level model in and out of
equilibrium}, Phys. Rev. B 81 (2010) 125122.
\newblock \href {https://doi.org/10.1103/PhysRevB.81.125122}
{\path{doi:10.1103/PhysRevB.81.125122}}.

\bibitem{Gezzi07}
R.~Gezzi, T.~Pruschke, V.~Meden, {Functional renormalization group for
nonequilibrium quantum many-body problems}, Phys. Rev. B 75 (2007) 045324.
\newblock \href {https://doi.org/10.1103/PhysRevB.75.045324}
{\path{doi:10.1103/PhysRevB.75.045324}}.

\bibitem{Rammer86}
J.~Rammer, H.~Smith, {Quantum field-theoretical methods in transport theory of
metals}, Rev. Mod. Phys. 58 (1986) 323.
\newblock \href {https://doi.org/10.1103/RevModPhys.58.323}
{\path{doi:10.1103/RevModPhys.58.323}}.

\bibitem{Jakobs10}
S.~G. Jakobs, M.~Pletyukhov, H.~Schoeller, {Properties of multi-particle
Green's and vertex functions within Keldysh formalism}, J. Phys. A: Math.
Theor. 43 (2010) 103001.
\newblock \href {https://doi.org/10.1088/1751-8113/43/10/103001}
{\path{doi:10.1088/1751-8113/43/10/103001}}.

\bibitem{Kennes12}
D.~M. Kennes, S.~G. Jakobs, C.~Karrasch, V.~Meden, {Renormalization group
approach to time-dependent transport through correlated quantum dots}, Phys.
Rev. B 85 (2012) 085113.
\newblock \href {https://doi.org/10.1103/PhysRevB.85.085113}
{\path{doi:10.1103/PhysRevB.85.085113}}.

\bibitem{Kennes13}
D.~M. Kennes, V.~Meden, {Luttinger liquid properties of the steady state after
a quantum quench}, Phys. Rev. B 88 (2013) 165131.
\newblock \href {https://doi.org/10.1103/PhysRevB.88.165131}
{\path{doi:10.1103/PhysRevB.88.165131}}.

\bibitem{Eissing16}
K.~Eissing, V.~Meden, D.~M. Kennes, {Renormalization in Periodically Driven
Quantum Dots}, Phys. Rev. Lett. 116 (2016) 026801.
\newblock \href {https://doi.org/10.1103/PhysRevLett.116.026801}
{\path{doi:10.1103/PhysRevLett.116.026801}}.

\bibitem{Salmhofer01}
M.~Salmhofer, C.~Honerkamp, {Fermionic Renormalization Group Flows}, Prog.
Theor. Phys. 105 (2001) 1--35.
\newblock \href {https://doi.org/10.1143/PTP.105.1}
{\path{doi:10.1143/PTP.105.1}}.

\bibitem{Metzner89}
W.~Metzner, D.~Vollhardt, {Correlated Lattice Fermions in $d=\infty$
Dimensions}, Phys. Rev. Lett. 62 (1989) 324.
\newblock \href {https://doi.org/10.1103/PhysRevLett.62.324}
{\path{doi:10.1103/PhysRevLett.62.324}}.

\bibitem{Georges92}
A.~Georges, G.~Kotliar, {Hubbard model in infinite dimensions}, Phys. Rev. B 45
(1992) 6479--6483.
\newblock \href {https://doi.org/10.1103/PhysRevB.45.6479}
{\path{doi:10.1103/PhysRevB.45.6479}}.

\bibitem{Georges96}
A.~Georges, G.~Kotliar, W.~Krauth, M.~J. Rozenberg, {Dynamical mean-field
theory of strongly correlated fermion systems and the limit of infinite
dimensions}, Rev. Mod. Phys. 68 (1996) 13--125.
\newblock \href {https://doi.org/10.1103/RevModPhys.68.13}
{\path{doi:10.1103/RevModPhys.68.13}}.

\bibitem{Vilardi19}
D.~Vilardi, C.~Taranto, W.~Metzner, {Antiferromagnetic and d-wave pairing
correlations in the strongly interacting two-dimensional Hubbard model from
the functional renormalization group}, Phys. Rev. B 99 (2019) 104501.
\newblock \href {https://doi.org/10.1103/PhysRevB.99.104501}
{\path{doi:10.1103/PhysRevB.99.104501}}.

\bibitem{Reuther10}
J.~Reuther, P.~W{\"o}lfle, {$J_1$-$J_2$ frustrated two-dimensional Heisenberg
model: Random phase approximation and functional renormalization group},
Phys. Rev. B 81 (2010) 144410.
\newblock \href {https://doi.org/10.1103/PhysRevB.81.144410}
{\path{doi:10.1103/PhysRevB.81.144410}}.

\bibitem{Reuther11}
J.~Reuther, R.~Thomale, {Functional renormalization group for the anisotropic
triangular antiferromagnet}, Phys. Rev. B 83 (2011) 024402.
\newblock \href {https://doi.org/10.1103/PhysRevB.83.024402}
{\path{doi:10.1103/PhysRevB.83.024402}}.

\bibitem{Sachdev99}
S.~Sachdev, {Universal relaxational dynamics near two-dimensional quantum
critical points}, Phys. Rev. B 59~(21) (1999) 14054--14073.
\newblock \href {https://doi.org/10.1103/PhysRevB.59.14054}
{\path{doi:10.1103/PhysRevB.59.14054}}.

\bibitem{Loehneysen07}
H.~v.~L{\"o}hneysen, A.~Rosch, M.~Vojta, P.~W{\"o}lfle, {Fermi-liquid
instabilities at magnetic quantum phase transitions}, Rev. Mod. Phys. 79
(2007) 1015.
\newblock \href {https://doi.org/10.1103/RevModPhys.79.1015}
{\path{doi:10.1103/RevModPhys.79.1015}}.

\bibitem{Hertz76}
J.~A. Hertz, {Quantum critical phenomena}, Phys. Rev. B 14~(3) (1976)
1165--1184.
\newblock \href {https://doi.org/10.1103/PhysRevB.14.1165}
{\path{doi:10.1103/PhysRevB.14.1165}}.

\bibitem{Millis93}
A.~J. Millis, {Effect of a nonzero temperature on quantum critical points in
itinerant fermion systems}, Phys. Rev. B 48~(10) (1993) 7183--7196.
\newblock \href {https://doi.org/10.1103/PhysRevB.48.7183}
{\path{doi:10.1103/PhysRevB.48.7183}}.

\bibitem{Maier16}
S.~A. Maier, P.~Strack, {Universality of antiferromagnetic strange metals},
Phys. Rev. B 93 (2016) 165114.
\newblock \href {https://doi.org/10.1103/PhysRevB.93.165114}
{\path{doi:10.1103/PhysRevB.93.165114}}.

\bibitem{Jakubczyk09}
P.~Jakubczyk, W.~Metzner, H.~Yamase, {Turning a First Order Quantum Phase
Transition Continuous by Fluctuations: General Flow Equations and Application
to d-Wave Pomeranchuk Instability}, Phys. Rev. Lett. 103 (2009) 220602.
\newblock \href {https://doi.org/10.1103/PhysRevLett.103.220602}
{\path{doi:10.1103/PhysRevLett.103.220602}}.

\bibitem{Classen17}
L.~Classen, I.~F. Herbut, M.~M. Scherer, {Fluctuation-induced continuous
transition and quantum criticality in Dirac semimetals}, Phys. Rev. B 96
(2017) 115132.
\newblock \href {https://doi.org/10.1103/PhysRevB.96.115132}
{\path{doi:10.1103/PhysRevB.96.115132}}.

\bibitem{Maas:2011se}
A.~Maas, {Describing gauge bosons at zero and finite temperature}, Phys.Rept.
524 (2013) 203--300.
\newblock \href {http://arxiv.org/abs/1106.3942} {\path{arXiv:1106.3942}},
\href {https://doi.org/10.1016/j.physrep.2012.11.002}
{\path{doi:10.1016/j.physrep.2012.11.002}}.

\bibitem{Vandersickel:2012tz}
N.~Vandersickel, D.~Zwanziger, {The Gribov problem and QCD dynamics}, Phys.
Rept. 520 (2012) 175--251.
\newblock \href {http://arxiv.org/abs/1202.1491} {\path{arXiv:1202.1491}},
\href {https://doi.org/10.1016/j.physrep.2012.07.003}
{\path{doi:10.1016/j.physrep.2012.07.003}}.

\bibitem{Capri:2017bfd}
M.~Capri, D.~Fiorentini, A.~Pereira, S.~Sorella, {Renormalizability of the
refined Gribov-Zwanziger action in linear covariant gauges}, Phys. Rev. D
96~(5) (2017) 054022.
\newblock \href {http://arxiv.org/abs/1708.01543} {\path{arXiv:1708.01543}},
\href {https://doi.org/10.1103/PhysRevD.96.054022}
{\path{doi:10.1103/PhysRevD.96.054022}}.

\bibitem{Abbott:1983zw}
L.~F. Abbott, M.~T. Grisaru, R.~K. Schaefer, {The Background Field Method and
the S Matrix}, Nucl. Phys. B229 (1983) 372--380.
\newblock \href {https://doi.org/10.1016/0550-3213(83)90337-1}
{\path{doi:10.1016/0550-3213(83)90337-1}}.

\bibitem{Reuter:1993kw}
M.~Reuter, C.~Wetterich, {Effective average action for gauge theories and exact
evolution equations}, Nucl. Phys. B417 (1994) 181--214.
\newblock \href {https://doi.org/10.1016/0550-3213(94)90543-6}
{\path{doi:10.1016/0550-3213(94)90543-6}}.

\bibitem{Reuter:1994zn}
M.~Reuter, C.~Wetterich, {Indications for gluon condensation for
nonperturbative flow equations} (1994).
\newblock \href {http://arxiv.org/abs/hep-th/9411227}
{\path{arXiv:hep-th/9411227}}.

\bibitem{Wetterich:1996kf}
C.~Wetterich, {Integrating out gluons in flow equations}, Z. Phys. C72 (1996)
139--162.
\newblock \href {http://arxiv.org/abs/hep-ph/9604227}
{\path{arXiv:hep-ph/9604227}}, \href {https://doi.org/10.1007/s002880050232}
{\path{doi:10.1007/s002880050232}}.

\bibitem{Reuter:1997gx}
M.~Reuter, C.~Wetterich, {Gluon condensation in nonperturbative flow
equations}, Phys.Rev. D56 (1997) 7893--7916.
\newblock \href {http://arxiv.org/abs/hep-th/9708051}
{\path{arXiv:hep-th/9708051}}, \href
{https://doi.org/10.1103/PhysRevD.56.7893}
{\path{doi:10.1103/PhysRevD.56.7893}}.

\bibitem{Gies:2002af}
H.~Gies, {Running coupling in Yang-Mills theory: A flow equation study},
Phys.Rev. D66 (2002) 025006.
\newblock \href {http://arxiv.org/abs/hep-th/0202207}
{\path{arXiv:hep-th/0202207}}, \href
{https://doi.org/10.1103/PhysRevD.66.025006}
{\path{doi:10.1103/PhysRevD.66.025006}}.

\bibitem{Gies:2003ic}
H.~Gies, {Renormalizability of gauge theories in extra dimensions}, Phys. Rev.
D68 (2003) 085015.
\newblock \href {http://arxiv.org/abs/hep-th/0305208}
{\path{arXiv:hep-th/0305208}}, \href
{https://doi.org/10.1103/PhysRevD.68.085015}
{\path{doi:10.1103/PhysRevD.68.085015}}.

\bibitem{Codello:2013wxa}
A.~Codello, {Renormalization group flow equations for the proper vertices of
the background effective average action}, Phys. Rev. D91~(6) (2015) 065032.
\newblock \href {http://arxiv.org/abs/1304.2059} {\path{arXiv:1304.2059}},
\href {https://doi.org/10.1103/PhysRevD.91.065032}
{\path{doi:10.1103/PhysRevD.91.065032}}.

\bibitem{Gies:2004hy}
H.~Gies, J.~Jaeckel, {Renormalization flow of QED}, Phys. Rev. Lett. 93 (2004)
110405.
\newblock \href {http://arxiv.org/abs/hep-ph/0405183}
{\path{arXiv:hep-ph/0405183}}, \href
{https://doi.org/10.1103/PhysRevLett.93.110405}
{\path{doi:10.1103/PhysRevLett.93.110405}}.

\bibitem{Reuter:1992uk}
M.~Reuter, C.~Wetterich, {Average action for the Higgs model with Abelian gauge
symmetry}, Nucl. Phys. B391 (1993) 147--175.
\newblock \href {https://doi.org/10.1016/0550-3213(93)90145-F}
{\path{doi:10.1016/0550-3213(93)90145-F}}.

\bibitem{Reuter:1994sg}
M.~Reuter, C.~Wetterich, {Exact evolution equation for scalar electrodynamics},
Nucl. Phys. B427 (1994) 291--324.
\newblock \href {https://doi.org/10.1016/0550-3213(94)90278-X}
{\path{doi:10.1016/0550-3213(94)90278-X}}.

\bibitem{Freire:2000sx}
F.~Freire, D.~F. Litim, {Charge crossover at the U(1) Higgs phase transition},
Phys.Rev. D64 (2001) 045014.
\newblock \href {http://arxiv.org/abs/hep-ph/0002153}
{\path{arXiv:hep-ph/0002153}}, \href
{https://doi.org/10.1103/PhysRevD.64.045014}
{\path{doi:10.1103/PhysRevD.64.045014}}.

\bibitem{Reuter:1993nn}
M.~Reuter, C.~Wetterich, {Running gauge coupling in three-dimensions and the
electroweak phase transition}, Nucl. Phys. B408 (1993) 91--132.
\newblock \href {https://doi.org/10.1016/0550-3213(93)90134-B}
{\path{doi:10.1016/0550-3213(93)90134-B}}.

\bibitem{Pawlowski:2020qer}
J.~M. Pawlowski, M.~Reichert, {Quantum gravity: a fluctuating point of view} (7
2020).
\newblock \href {http://arxiv.org/abs/2007.10353} {\path{arXiv:2007.10353}}.

\bibitem{Bonini:1993sj}
M.~Bonini, M.~D'Attanasio, G.~Marchesini, {Renormalization group flow for SU(2)
Yang-Mills theory and gauge invariance}, Nucl. Phys. B421 (1994) 429--455.
\newblock \href {http://arxiv.org/abs/hep-th/9312114}
{\path{arXiv:hep-th/9312114}}, \href
{https://doi.org/10.1016/0550-3213(94)90335-2}
{\path{doi:10.1016/0550-3213(94)90335-2}}.

\bibitem{Bonini:1994kp}
M.~Bonini, M.~D'Attanasio, G.~Marchesini, {BRS symmetry for Yang-Mills theory
with exact renormalization group}, Nucl. Phys. B437 (1995) 163--186.
\newblock \href {http://arxiv.org/abs/hep-th/9410138}
{\path{arXiv:hep-th/9410138}}, \href
{https://doi.org/10.1016/0550-3213(94)00569-Z}
{\path{doi:10.1016/0550-3213(94)00569-Z}}.

\bibitem{Bonini:1994dz}
M.~Bonini, M.~D'Attanasio, G.~Marchesini, {BRS symmetry from renormalization
group flow}, Phys. Lett. B346 (1995) 87--93.
\newblock \href {http://arxiv.org/abs/hep-th/9412195}
{\path{arXiv:hep-th/9412195}}, \href
{https://doi.org/10.1016/0370-2693(94)01676-4}
{\path{doi:10.1016/0370-2693(94)01676-4}}.

\bibitem{Bonini:1995tx}
M.~Bonini, M.~D'Attanasio, G.~Marchesini, {Perturbative infrared finiteness of
Yang-Mills theory from renormalization group flow}, Nucl. Phys. B444 (1995)
602--616.
\newblock \href {https://doi.org/10.1016/0550-3213(95)00166-P}
{\path{doi:10.1016/0550-3213(95)00166-P}}.

\bibitem{Becchi:1996an}
C.~Becchi, {On the construction of renormalized gauge theories using
renormalization group techniques} (7 1996).
\newblock \href {http://arxiv.org/abs/hep-th/9607188}
{\path{arXiv:hep-th/9607188}}.

\bibitem{DAttanasio:1996tzp}
M.~D'Attanasio, T.~R. Morris, {Gauge invariance, the quantum action principle,
and the renormalization group}, Phys. Lett. B378 (1996) 213--221.
\newblock \href {http://arxiv.org/abs/hep-th/9602156}
{\path{arXiv:hep-th/9602156}}, \href
{https://doi.org/10.1016/0370-2693(96)00411-X}
{\path{doi:10.1016/0370-2693(96)00411-X}}.

\bibitem{Freire:2000bq}
F.~Freire, D.~F. Litim, J.~M. Pawlowski, {Gauge invariance and background field
formalism in the exact renormalization group}, Phys.Lett. B495 (2000)
256--262.
\newblock \href {http://arxiv.org/abs/hep-th/0009110}
{\path{arXiv:hep-th/0009110}}, \href
{https://doi.org/10.1016/S0370-2693(00)01231-4}
{\path{doi:10.1016/S0370-2693(00)01231-4}}.

\bibitem{Igarashi:1999rm}
Y.~Igarashi, K.~Itoh, H.~So, {Exact symmetries realized on the renormalization
group flow}, Phys. Lett. B479 (2000) 336--342.
\newblock \href {http://arxiv.org/abs/hep-th/9912262}
{\path{arXiv:hep-th/9912262}}, \href
{https://doi.org/10.1016/S0370-2693(00)00305-1}
{\path{doi:10.1016/S0370-2693(00)00305-1}}.

\bibitem{Igarashi:2000vf}
Y.~Igarashi, K.~Itoh, H.~So, {Exact BRS symmetry realized on the
renormalization group flow}, Prog. Theor. Phys. 104 (2000) 1053--1066.
\newblock \href {http://arxiv.org/abs/hep-th/0006180}
{\path{arXiv:hep-th/0006180}}, \href {https://doi.org/10.1143/PTP.104.1053}
{\path{doi:10.1143/PTP.104.1053}}.

\bibitem{Igarashi:2001mf}
Y.~Igarashi, K.~Itoh, H.~So, {BRS symmetry, the quantum master equation, and
the Wilsonian renormalization group}, Prog.Theor.Phys. 106 (2001) 149--166.
\newblock \href {http://arxiv.org/abs/hep-th/0101101}
{\path{arXiv:hep-th/0101101}}, \href {https://doi.org/10.1143/PTP.106.149}
{\path{doi:10.1143/PTP.106.149}}.

\bibitem{Igarashi:2001ey}
Y.~Igarashi, K.~Itoh, H.~So, {Regularized quantum master equation in the
Wilsonian renormalization group}, JHEP 10 (2001) 032.
\newblock \href {http://arxiv.org/abs/hep-th/0109202}
{\path{arXiv:hep-th/0109202}}, \href
{https://doi.org/10.1088/1126-6708/2001/10/032}
{\path{doi:10.1088/1126-6708/2001/10/032}}.

\bibitem{Igarashi:2001cv}
Y.~Igarashi, K.~Itoh, H.~So, {Realization of global symmetries in the Wilsonian
renormalization group}, Phys. Lett. B526 (2002) 164--172.
\newblock \href {http://arxiv.org/abs/hep-th/0111112}
{\path{arXiv:hep-th/0111112}}, \href
{https://doi.org/10.1016/S0370-2693(01)01461-7}
{\path{doi:10.1016/S0370-2693(01)01461-7}}.

\bibitem{Pawlowski:2003sk}
J.~M. Pawlowski, {Geometrical effective action and Wilsonian flows} (2003).
\newblock \href {http://arxiv.org/abs/hep-th/0310018}
{\path{arXiv:hep-th/0310018}}.

\bibitem{Sonoda:2007dj}
H.~Sonoda, {On the construction of QED using ERG}, J. Phys. A40 (2007)
9675--9690.
\newblock \href {http://arxiv.org/abs/hep-th/0703167}
{\path{arXiv:hep-th/0703167}}, \href
{https://doi.org/10.1088/1751-8113/40/31/034}
{\path{doi:10.1088/1751-8113/40/31/034}}.

\bibitem{Igarashi:2007fw}
Y.~Igarashi, K.~Itoh, H.~Sonoda, {Quantum master equation for QED in exact
renormalization group}, Prog. Theor. Phys. 118 (2007) 121--134.
\newblock \href {http://arxiv.org/abs/0704.2349} {\path{arXiv:0704.2349}},
\href {https://doi.org/10.1143/PTP.118.121} {\path{doi:10.1143/PTP.118.121}}.

\bibitem{Igarashi:2008bb}
Y.~Igarashi, K.~Itoh, H.~Sonoda, {Ward-Takahashi identity for Yang-Mills theory
in the Exact Renormalization Group}, Prog. Theor. Phys. 120 (2008)
1017--1028.
\newblock \href {http://arxiv.org/abs/0808.3430} {\path{arXiv:0808.3430}},
\href {https://doi.org/10.1143/PTP.120.1017}
{\path{doi:10.1143/PTP.120.1017}}.

\bibitem{Igarashi:2009tj}
Y.~Igarashi, K.~Itoh, H.~Sonoda, {Realization of Symmetry in the ERG Approach
to Quantum Field Theory}, Prog.Theor.Phys.Suppl. 181 (2010) 1--166.
\newblock \href {http://arxiv.org/abs/0909.0327} {\path{arXiv:0909.0327}},
\href {https://doi.org/10.1143/PTPS.181.1} {\path{doi:10.1143/PTPS.181.1}}.

\bibitem{Donkin:2012ud}
I.~Donkin, J.~M. Pawlowski, {The phase diagram of quantum gravity from
diffeomorphism-invariant RG-flows} (2012).
\newblock \href {http://arxiv.org/abs/1203.4207} {\path{arXiv:1203.4207}}.

\bibitem{Lavrov:2012xz}
P.~M. Lavrov, I.~L. Shapiro, {On the Functional Renormalization Group approach
for Yang-Mills fields}, JHEP 06 (2013) 086.
\newblock \href {http://arxiv.org/abs/1212.2577} {\path{arXiv:1212.2577}},
\href {https://doi.org/10.1007/JHEP06(2013)086}
{\path{doi:10.1007/JHEP06(2013)086}}.

\bibitem{Sonoda:2013dwa}
H.~Sonoda, {Gauge invariant composite operators of QED in the exact
renormalization group formalism}, J. Phys. A47 (2013) 015401.
\newblock \href {http://arxiv.org/abs/1309.3024} {\path{arXiv:1309.3024}},
\href {https://doi.org/10.1088/1751-8113/47/1/015401}
{\path{doi:10.1088/1751-8113/47/1/015401}}.

\bibitem{Safari:2015dva}
M.~Safari, {Splitting Ward identity}, Eur. Phys. J. C76~(4) (2016) 201.
\newblock \href {http://arxiv.org/abs/1508.06244} {\path{arXiv:1508.06244}},
\href {https://doi.org/10.1140/epjc/s10052-016-4036-6}
{\path{doi:10.1140/epjc/s10052-016-4036-6}}.

\bibitem{Safari:2016dwj}
M.~Safari, G.~P. Vacca, {Covariant and single-field effective action with the
background-field formalism}, Phys. Rev. D96~(8) (2017) 085001.
\newblock \href {http://arxiv.org/abs/1607.03053} {\path{arXiv:1607.03053}},
\href {https://doi.org/10.1103/PhysRevD.96.085001}
{\path{doi:10.1103/PhysRevD.96.085001}}.

\bibitem{Safari:2016gtj}
M.~Safari, G.~P. Vacca, {Covariant and background independent functional RG
flow for the effective average action}, JHEP 11 (2016) 139.
\newblock \href {http://arxiv.org/abs/1607.07074} {\path{arXiv:1607.07074}},
\href {https://doi.org/10.1007/JHEP11(2016)139}
{\path{doi:10.1007/JHEP11(2016)139}}.

\bibitem{Igarashi:2016gcf}
Y.~Igarashi, K.~Itoh, J.~M. Pawlowski, {Functional flows in QED and the
modified Ward–Takahashi identity}, J. Phys. A49~(40) (2016) 405401.
\newblock \href {http://arxiv.org/abs/1604.08327} {\path{arXiv:1604.08327}},
\href {https://doi.org/10.1088/1751-8113/49/40/405401}
{\path{doi:10.1088/1751-8113/49/40/405401}}.

\bibitem{Asnafi:2018pre}
S.~Asnafi, H.~Gies, L.~Zambelli, {BRST invariant RG flows}, Phys. Rev. D99~(8)
(2019) 085009.
\newblock \href {http://arxiv.org/abs/1811.03615} {\path{arXiv:1811.03615}},
\href {https://doi.org/10.1103/PhysRevD.99.085009}
{\path{doi:10.1103/PhysRevD.99.085009}}.

\bibitem{Igarashi:2019gkm}
Y.~Igarashi, K.~Itoh, T.~R. Morris, {BRST in the exact renormalization group},
PTEP 2019~(10) (2019) 103B01.
\newblock \href {http://arxiv.org/abs/1904.08231} {\path{arXiv:1904.08231}},
\href {https://doi.org/10.1093/ptep/ptz099} {\path{doi:10.1093/ptep/ptz099}}.

\bibitem{Barra:2019rhz}
V.~F. Barra, P.~M. Lavrov, E.~A. Dos~Reis, T.~de~Paula~Netto, I.~L. Shapiro,
{Functional renormalization group approach and gauge dependence in gravity
theories}, Phys. Rev. D 101~(6) (2020) 065001.
\newblock \href {http://arxiv.org/abs/1910.06068} {\path{arXiv:1910.06068}},
\href {https://doi.org/10.1103/PhysRevD.101.065001}
{\path{doi:10.1103/PhysRevD.101.065001}}.

\bibitem{Lavrov:2019agp}
P.~M. Lavrov, E.~A. dos Reis, T.~da~Paula~Netto, I.~L. Shapiro, {Gauge
invariance of the background average effective action}, Eur. Phys. J. C
79~(8) (2019) 661.
\newblock \href {http://arxiv.org/abs/1905.08296} {\path{arXiv:1905.08296}},
\href {https://doi.org/10.1140/epjc/s10052-019-7153-1}
{\path{doi:10.1140/epjc/s10052-019-7153-1}}.

\bibitem{Lavrov:2020exa}
P.~M. Lavrov, {BRST, Ward identities, gauge dependence and FRG} (2 2020).
\newblock \href {http://arxiv.org/abs/2002.05997} {\path{arXiv:2002.05997}}.

\bibitem{Bonini:1994xj}
M.~Bonini, M.~D'Attanasio, G.~Marchesini, {Axial anomalies in gauge theory by
exact renormalization group method}, Phys. Lett. B329 (1994) 249--258.
\newblock \href {http://arxiv.org/abs/hep-th/9403074}
{\path{arXiv:hep-th/9403074}}, \href
{https://doi.org/10.1016/0370-2693(94)90768-4}
{\path{doi:10.1016/0370-2693(94)90768-4}}.

\bibitem{Reuter:1996be}
M.~Reuter, {Renormalization of the topological charge in Yang-Mills theory},
Mod. Phys. Lett. A12 (1997) 2777--2802.
\newblock \href {http://arxiv.org/abs/hep-th/9604124}
{\path{arXiv:hep-th/9604124}}, \href
{https://doi.org/10.1142/S0217732397002922}
{\path{doi:10.1142/S0217732397002922}}.

\bibitem{Pawlowski:1996ch}
J.~Pawlowski, {Exact flow equations and the U(1) problem}, Phys.Rev. D58 (1998)
045011.
\newblock \href {http://arxiv.org/abs/hep-th/9605037}
{\path{arXiv:hep-th/9605037}}, \href
{https://doi.org/10.1103/PhysRevD.58.045011}
{\path{doi:10.1103/PhysRevD.58.045011}}.

\bibitem{Bergner:2012nu}
G.~Bergner, F.~Bruckmann, Y.~Echigo, Y.~Igarashi, J.~M. Pawlowski,
S.~Schierenberg, {Blocking-inspired supersymmetric actions: a status report},
Phys. Rev. D 87~(9) (2013) 094516.
\newblock \href {http://arxiv.org/abs/1212.0219} {\path{arXiv:1212.0219}},
\href {https://doi.org/10.1103/PhysRevD.87.094516}
{\path{doi:10.1103/PhysRevD.87.094516}}.

\bibitem{Ellwanger:1997wv}
U.~Ellwanger, {Confinement, monopoles and Wilsonian effective action}, Nucl.
Phys. B531 (1998) 593--612.
\newblock \href {http://arxiv.org/abs/hep-ph/9710326}
{\path{arXiv:hep-ph/9710326}}, \href
{https://doi.org/10.1016/S0550-3213(98)00542-2}
{\path{doi:10.1016/S0550-3213(98)00542-2}}.

\bibitem{Litim:1998qi}
D.~F. Litim, J.~M. Pawlowski, {Flow equations for Yang-Mills theories in
general axial gauges}, Phys.Lett. B435 (1998) 181--188.
\newblock \href {http://arxiv.org/abs/hep-th/9802064}
{\path{arXiv:hep-th/9802064}}, \href
{https://doi.org/10.1016/S0370-2693(98)00761-8}
{\path{doi:10.1016/S0370-2693(98)00761-8}}.

\bibitem{Simionato:1998iz}
M.~Simionato, {Gauge consistent Wilson renormalization group.2. NonAbelian
case}, Int. J. Mod. Phys. A15 (2000) 2153--2179.
\newblock \href {http://arxiv.org/abs/hep-th/9810117}
{\path{arXiv:hep-th/9810117}}, \href
{https://doi.org/10.1142/S0217751X00000896}
{\path{doi:10.1142/S0217751X00000896}}.

\bibitem{Simionato:1998te}
M.~Simionato, {Gauge consistent Wilson renormalization group: Abelian case},
Int. J. Mod. Phys. A15 (2000) 2121--2152.
\newblock \href {http://arxiv.org/abs/hep-th/9809004}
{\path{arXiv:hep-th/9809004}}, \href
{https://doi.org/10.1142/S0217751X00000884}
{\path{doi:10.1142/S0217751X00000884}}.

\bibitem{Litim:2002ce}
D.~F. Litim, J.~M. Pawlowski, {Renormalization group flows for gauge theories
in axial gauges}, JHEP 0209 (2002) 049.
\newblock \href {http://arxiv.org/abs/hep-th/0203005}
{\path{arXiv:hep-th/0203005}}.

\bibitem{Marhauser:2008fz}
F.~Marhauser, J.~M. Pawlowski, {Confinement in Polyakov Gauge} (2008).
\newblock \href {http://arxiv.org/abs/0812.1144} {\path{arXiv:0812.1144}}.

\bibitem{Kondo:2010ts}
K.-I. Kondo, {Toward a first-principle derivation of confinement and
chiral-symmetry-breaking crossover transitions in QCD}, Phys. Rev. D82 (2010)
065024.
\newblock \href {http://arxiv.org/abs/1005.0314} {\path{arXiv:1005.0314}},
\href {https://doi.org/10.1103/PhysRevD.82.065024}
{\path{doi:10.1103/PhysRevD.82.065024}}.

\bibitem{Gies:2001hk}
H.~Gies, {Wilsonian effective action for SU(2) Yang-Mills theory with
Cho-Faddeev-Niemi-Shabanov decomposition}, Phys. Rev. D63 (2001) 125023.
\newblock \href {http://arxiv.org/abs/hep-th/0102026}
{\path{arXiv:hep-th/0102026}}, \href
{https://doi.org/10.1103/PhysRevD.63.125023}
{\path{doi:10.1103/PhysRevD.63.125023}}.

\bibitem{Ellwanger:1998th}
U.~Ellwanger, {Field strength correlator and an infrared fixed point of the
Wilsonian exact renormalization group equations}, Eur. Phys. J. C7 (1999)
673--683.
\newblock \href {http://arxiv.org/abs/hep-ph/9807380}
{\path{arXiv:hep-ph/9807380}}, \href {https://doi.org/10.1007/s100529801033}
{\path{doi:10.1007/s100529801033}}.

\bibitem{Ellwanger:1999vc}
U.~Ellwanger, {Monopole condensation and antisymmetric tensor fields: compact
QED and the Wilsonian RG flow in Yang-Mills theories}, Nucl. Phys. B560
(1999) 587--600.
\newblock \href {http://arxiv.org/abs/hep-th/9906061}
{\path{arXiv:hep-th/9906061}}, \href
{https://doi.org/10.1016/S0550-3213(99)00460-5}
{\path{doi:10.1016/S0550-3213(99)00460-5}}.

\bibitem{Ellwanger:2002sj}
U.~Ellwanger, N.~Wschebor, {Massive Yang-Mills theory in Abelian gauges}, Int.
J. Mod. Phys. A18 (2003) 1595--1612.
\newblock \href {http://arxiv.org/abs/hep-th/0205057}
{\path{arXiv:hep-th/0205057}}, \href
{https://doi.org/10.1142/S0217751X03014198}
{\path{doi:10.1142/S0217751X03014198}}.

\bibitem{Ellwanger:2002xa}
U.~Ellwanger, N.~Wschebor, {Confinement and mass gap in Abelian gauge}, Eur.
Phys. J. C28 (2003) 415--424.
\newblock \href {http://arxiv.org/abs/hep-th/0211014}
{\path{arXiv:hep-th/0211014}}, \href
{https://doi.org/10.1140/epjc/s2003-01170-0}
{\path{doi:10.1140/epjc/s2003-01170-0}}.

\bibitem{Leder:2010ji}
M.~Leder, J.~M. Pawlowski, H.~Reinhardt, A.~Weber, {Hamiltonian Flow in Coulomb
Gauge Yang-Mills Theory}, Phys.Rev. D83 (2011) 025010.
\newblock \href {http://arxiv.org/abs/1006.5710} {\path{arXiv:1006.5710}},
\href {https://doi.org/10.1103/PhysRevD.83.025010}
{\path{doi:10.1103/PhysRevD.83.025010}}.

\bibitem{Leder:2011yc}
M.~Leder, H.~Reinhardt, A.~Weber, J.~M. Pawlowski, {Color Coulomb Potential in
Yang-Mills Theory from Hamiltonian Flows} (2011).
\newblock \href {http://arxiv.org/abs/1105.0800} {\path{arXiv:1105.0800}}.

\bibitem{Cyrol:2016tym}
A.~K. Cyrol, L.~Fister, M.~Mitter, J.~M. Pawlowski, N.~Strodthoff, {Landau
gauge Yang-Mills correlation functions}, Phys. Rev. D94~(5) (2016) 054005.
\newblock \href {http://arxiv.org/abs/1605.01856} {\path{arXiv:1605.01856}},
\href {https://doi.org/10.1103/PhysRevD.94.054005}
{\path{doi:10.1103/PhysRevD.94.054005}}.

\bibitem{Sternbeck:2006cg}
A.~Sternbeck, E.~M. Ilgenfritz, M.~Muller-Preussker, A.~Schiller, I.~L.
Bogolubsky, {Lattice study of the infrared behavior of QCD Green's functions
in Landau gauge}, PoS LAT2006 (2006) 076.
\newblock \href {http://arxiv.org/abs/hep-lat/0610053}
{\path{arXiv:hep-lat/0610053}}.

\bibitem{Cyrol:2017ewj}
A.~K. Cyrol, M.~Mitter, J.~M. Pawlowski, N.~Strodthoff, {Nonperturbative quark,
gluon, and meson correlators of unquenched QCD}, Phys. Rev. D97~(5) (2018)
054006.
\newblock \href {http://arxiv.org/abs/1706.06326} {\path{arXiv:1706.06326}},
\href {https://doi.org/10.1103/PhysRevD.97.054006}
{\path{doi:10.1103/PhysRevD.97.054006}}.

\bibitem{Fu:2019hdw}
W.-j. Fu, J.~M. Pawlowski, F.~Rennecke, {QCD phase structure at finite
temperature and density}, Phys. Rev. D 101~(5) (2020) 054032.
\newblock \href {http://arxiv.org/abs/1909.02991} {\path{arXiv:1909.02991}},
\href {https://doi.org/10.1103/PhysRevD.101.054032}
{\path{doi:10.1103/PhysRevD.101.054032}}.

\bibitem{Sternbeck:2012qs}
A.~Sternbeck, K.~Maltman, M.~Muller-Preussker, L.~von Smekal, {Determination of
LambdaMS from the gluon and ghost propagators in Landau gauge}, PoS
LATTICE2012 (2012) 243.
\newblock \href {http://arxiv.org/abs/1212.2039} {\path{arXiv:1212.2039}}.

\bibitem{Zafeiropoulos:2019flq}
S.~Zafeiropoulos, P.~Boucaud, F.~De~Soto, J.~Rodríguez-Quintero, J.~Segovia,
{Strong Running Coupling from the Gauge Sector of Domain Wall Lattice QCD
with Physical Quark Masses}, Phys. Rev. Lett. 122~(16) (2019) 162002.
\newblock \href {http://arxiv.org/abs/1902.08148} {\path{arXiv:1902.08148}},
\href {https://doi.org/10.1103/PhysRevLett.122.162002}
{\path{doi:10.1103/PhysRevLett.122.162002}}.

\bibitem{Boucaud:2018xup}
P.~Boucaud, F.~De~Soto, K.~Raya, J.~Rodríguez-Quintero, S.~Zafeiropoulos,
{Discretization effects on renormalized gauge-field Green's functions, scale
setting, and the gluon mass}, Phys. Rev. D 98~(11) (2018) 114515.
\newblock \href {http://arxiv.org/abs/1809.05776} {\path{arXiv:1809.05776}},
\href {https://doi.org/10.1103/PhysRevD.98.114515}
{\path{doi:10.1103/PhysRevD.98.114515}}.

\bibitem{Mitter:2014wpa}
M.~Mitter, J.~M. Pawlowski, N.~Strodthoff, {Chiral symmetry breaking in
continuum QCD}, Phys.Rev. D91 (2015) 054035.
\newblock \href {http://arxiv.org/abs/1411.7978} {\path{arXiv:1411.7978}},
\href {https://doi.org/10.1103/PhysRevD.91.054035}
{\path{doi:10.1103/PhysRevD.91.054035}}.

\bibitem{Cyrol:2017qkl}
A.~K. Cyrol, M.~Mitter, J.~M. Pawlowski, N.~Strodthoff, {Nonperturbative
finite-temperature Yang-Mills theory}, Phys. Rev. D97~(5) (2018) 054015.
\newblock \href {http://arxiv.org/abs/1708.03482} {\path{arXiv:1708.03482}},
\href {https://doi.org/10.1103/PhysRevD.97.054015}
{\path{doi:10.1103/PhysRevD.97.054015}}.

\bibitem{Hajizadeh:2019qrj}
O.~Hajizadeh, M.~Q. Huber, A.~Maas, J.~M. Pawlowski, {Exploring the Tan contact
term in Yang-Mills theory} (9 2019).
\newblock \href {http://arxiv.org/abs/1909.12727} {\path{arXiv:1909.12727}}.

\bibitem{Corell:2018yil}
L.~Corell, A.~K. Cyrol, M.~Mitter, J.~M. Pawlowski, N.~Strodthoff, {Correlation
functions of three-dimensional Yang-Mills theory from the FRG}, SciPost Phys.
5 (2018) 066.
\newblock \href {http://arxiv.org/abs/1803.10092} {\path{arXiv:1803.10092}},
\href {https://doi.org/10.21468/SciPostPhys.5.6.066}
{\path{doi:10.21468/SciPostPhys.5.6.066}}.

\bibitem{Ellwanger:1995qf}
U.~Ellwanger, M.~Hirsch, A.~Weber, {Flow equations for the relevant part of the
pure Yang-Mills action}, Z.Phys. C69 (1996) 687--698.
\newblock \href {http://arxiv.org/abs/hep-th/9506019}
{\path{arXiv:hep-th/9506019}}, \href {https://doi.org/10.1007/s002880050073}
{\path{doi:10.1007/s002880050073}}.

\bibitem{Ellwanger:1996wy}
U.~Ellwanger, M.~Hirsch, A.~Weber, {The Heavy quark potential from Wilson's
exact renormalization group}, Eur.Phys.J. C1 (1998) 563--578.
\newblock \href {http://arxiv.org/abs/hep-ph/9606468}
{\path{arXiv:hep-ph/9606468}}, \href {https://doi.org/10.1007/s100520050105}
{\path{doi:10.1007/s100520050105}}.

\bibitem{Bergerhoff:1997cv}
B.~Bergerhoff, C.~Wetterich, {Effective quark interactions and QCD
propagators}, Phys.Rev. D57 (1998) 1591--1604.
\newblock \href {http://arxiv.org/abs/hep-ph/9708425}
{\path{arXiv:hep-ph/9708425}}, \href
{https://doi.org/10.1103/PhysRevD.57.1591}
{\path{doi:10.1103/PhysRevD.57.1591}}.

\bibitem{Alkofer:2000wg}
R.~Alkofer, L.~von Smekal, {The infrared behavior of QCD Green's functions:
Confinement, dynamical symmetry breaking, and hadrons as relativistic bound
states}, Phys. Rept. 353 (2001) 281.
\newblock \href {http://arxiv.org/abs/hep-ph/0007355}
{\path{arXiv:hep-ph/0007355}}, \href
{https://doi.org/10.1016/S0370-1573(01)00010-2}
{\path{doi:10.1016/S0370-1573(01)00010-2}}.

\bibitem{Fischer:2006ub}
C.~S. Fischer, {Infrared properties of QCD from Dyson-Schwinger equations},
J.Phys.G G32 (2006) R253--R291.
\newblock \href {http://arxiv.org/abs/hep-ph/0605173}
{\path{arXiv:hep-ph/0605173}}, \href
{https://doi.org/10.1088/0954-3899/32/8/R02}
{\path{doi:10.1088/0954-3899/32/8/R02}}.

\bibitem{Fischer:2008uz}
C.~S. Fischer, A.~Maas, J.~M. Pawlowski, {On the infrared behavior of Landau
gauge Yang-Mills theory}, Annals Phys. 324 (2009) 2408--2437.
\newblock \href {http://arxiv.org/abs/0810.1987} {\path{arXiv:0810.1987}},
\href {https://doi.org/10.1016/j.aop.2009.07.009}
{\path{doi:10.1016/j.aop.2009.07.009}}.

\bibitem{Aguilar:2008xm}
A.~C. Aguilar, D.~Binosi, J.~Papavassiliou, {Gluon and ghost propagators in the
Landau gauge: Deriving lattice results from Schwinger-Dyson equations}, Phys.
Rev. D78 (2008) 025010.
\newblock \href {http://arxiv.org/abs/0802.1870} {\path{arXiv:0802.1870}},
\href {https://doi.org/10.1103/PhysRevD.78.025010}
{\path{doi:10.1103/PhysRevD.78.025010}}.

\bibitem{Binosi:2009qm}
D.~Binosi, J.~Papavassiliou, {Pinch Technique: Theory and Applications},
Phys.Rept. 479 (2009) 1--152.
\newblock \href {http://arxiv.org/abs/0909.2536} {\path{arXiv:0909.2536}},
\href {https://doi.org/10.1016/j.physrep.2009.05.001}
{\path{doi:10.1016/j.physrep.2009.05.001}}.

\bibitem{Boucaud:2011ug}
P.~Boucaud, J.~Leroy, A.~L. Yaouanc, J.~Micheli, O.~Pene, et~al., {The Infrared
Behaviour of the Pure Yang-Mills Green Functions}, Few Body Syst. 53 (2012)
387--436.
\newblock \href {http://arxiv.org/abs/1109.1936} {\path{arXiv:1109.1936}},
\href {https://doi.org/10.1007/s00601-011-0301-2}
{\path{doi:10.1007/s00601-011-0301-2}}.

\bibitem{Cucchieri:2011ig}
A.~Cucchieri, D.~Dudal, T.~Mendes, N.~Vandersickel, {Modeling the Gluon
Propagator in Landau Gauge: Lattice Estimates of Pole Masses and
Dimension-Two Condensates}, Phys. Rev. D85 (2012) 094513.
\newblock \href {http://arxiv.org/abs/1111.2327} {\path{arXiv:1111.2327}},
\href {https://doi.org/10.1103/PhysRevD.85.094513}
{\path{doi:10.1103/PhysRevD.85.094513}}.

\bibitem{Pelaez:2013cpa}
M.~Pel\'aez, M.~Tissier, N.~Wschebor, {Three-point correlation functions in
Yang-Mills theory}, Phys.Rev. D88 (2013) 125003.
\newblock \href {http://arxiv.org/abs/1310.2594} {\path{arXiv:1310.2594}},
\href {https://doi.org/10.1103/PhysRevD.88.125003}
{\path{doi:10.1103/PhysRevD.88.125003}}.

\bibitem{Aguilar:2013vaa}
A.~Aguilar, D.~Binosi, D.~Ib{\'a}{\~n}ez, J.~Papavassiliou, {Effects of
divergent ghost loops on the Green's functions of QCD}, Phys.Rev. D89 (2014)
085008.
\newblock \href {http://arxiv.org/abs/1312.1212} {\path{arXiv:1312.1212}},
\href {https://doi.org/10.1103/PhysRevD.89.085008}
{\path{doi:10.1103/PhysRevD.89.085008}}.

\bibitem{Reinosa:2017qtf}
U.~Reinosa, J.~Serreau, M.~Tissier, N.~Wschebor, {How nonperturbative is the
infrared regime of Landau gauge Yang-Mills correlators?}, Phys. Rev. D96~(1)
(2017) 014005.
\newblock \href {http://arxiv.org/abs/1703.04041} {\path{arXiv:1703.04041}},
\href {https://doi.org/10.1103/PhysRevD.96.014005}
{\path{doi:10.1103/PhysRevD.96.014005}}.

\bibitem{Huber:2018ned}
M.~Q. Huber, {Nonperturbative properties of Yang-Mills theories} (5 2018).
\newblock \href {http://arxiv.org/abs/1808.05227} {\path{arXiv:1808.05227}}.

\bibitem{Oliveira:2018lln}
O.~Oliveira, P.~J. Silva, J.-I. Skullerud, A.~Sternbeck, {Quark propagator with
two flavors of O(a)-improved Wilson fermions}, Phys. Rev. D 99~(9) (2019)
094506.
\newblock \href {http://arxiv.org/abs/1809.02541} {\path{arXiv:1809.02541}},
\href {https://doi.org/10.1103/PhysRevD.99.094506}
{\path{doi:10.1103/PhysRevD.99.094506}}.

\bibitem{Gracey:2019xom}
J.~A. Gracey, M.~Peláez, U.~Reinosa, M.~Tissier, {Two loop calculation of
Yang-Mills propagators in the Curci-Ferrari model}, Phys. Rev. D 100~(3)
(2019) 034023.
\newblock \href {http://arxiv.org/abs/1905.07262} {\path{arXiv:1905.07262}},
\href {https://doi.org/10.1103/PhysRevD.100.034023}
{\path{doi:10.1103/PhysRevD.100.034023}}.

\bibitem{Li:2019hyv}
S.~W. Li, P.~Lowdon, O.~Oliveira, P.~J. Silva, {The generalised infrared
structure of the gluon propagator}, Phys. Lett. B 803 (2020) 135329.
\newblock \href {http://arxiv.org/abs/1907.10073} {\path{arXiv:1907.10073}},
\href {https://doi.org/10.1016/j.physletb.2020.135329}
{\path{doi:10.1016/j.physletb.2020.135329}}.

\bibitem{Aguilar:2019uob}
A.~Aguilar, F.~De~Soto, M.~Ferreira, J.~Papavassiliou, J.~Rodríguez-Quintero,
S.~Zafeiropoulos, {Gluon propagator and three-gluon vertex with dynamical
quarks}, Eur. Phys. J. C 80~(2) (2020) 154.
\newblock \href {http://arxiv.org/abs/1912.12086} {\path{arXiv:1912.12086}},
\href {https://doi.org/10.1140/epjc/s10052-020-7741-0}
{\path{doi:10.1140/epjc/s10052-020-7741-0}}.

\bibitem{Huber:2020keu}
M.~Q. Huber, {Correlation functions of Landau gauge Yang-Mills theory} (3
2020).
\newblock \href {http://arxiv.org/abs/2003.13703} {\path{arXiv:2003.13703}}.

\bibitem{Kugo:1979gm}
T.~Kugo, I.~Ojima, {Local Covariant Operator Formalism of Nonabelian Gauge
Theories and Quark Confinement Problem}, Prog. Theor. Phys. Suppl. 66 (1979)
1.

\bibitem{Kugo:1995km}
T.~Kugo, {The universal renormalization factors Z(1) / Z(3) and color
confinement condition in non-Abelian gauge theory} (1995).
\newblock \href {http://arxiv.org/abs/hep-th/9511033}
{\path{arXiv:hep-th/9511033}}.

\bibitem{Fischer:2006vf}
C.~S. Fischer, J.~M. Pawlowski, {Uniqueness of infrared asymptotics in Landau
gauge Yang- Mills theory}, Phys. Rev. D75 (2007) 025012.
\newblock \href {http://arxiv.org/abs/hep-th/0609009}
{\path{arXiv:hep-th/0609009}}, \href
{https://doi.org/10.1103/PhysRevD.75.025012}
{\path{doi:10.1103/PhysRevD.75.025012}}.

\bibitem{Fischer:2009tn}
C.~S. Fischer, J.~M. Pawlowski, {Uniqueness of infrared asymptotics in Landau
gauge Yang- Mills theory II}, Phys. Rev. D80 (2009) 025023.
\newblock \href {http://arxiv.org/abs/0903.2193} {\path{arXiv:0903.2193}},
\href {https://doi.org/10.1103/PhysRevD.80.025023}
{\path{doi:10.1103/PhysRevD.80.025023}}.

\bibitem{Alkofer:2008jy}
R.~Alkofer, M.~Q. Huber, K.~Schwenzer, {Infrared singularities in Landau gauge
Yang-Mills theory}, Phys. Rev. D81 (2010) 105010.
\newblock \href {http://arxiv.org/abs/0801.2762} {\path{arXiv:0801.2762}},
\href {https://doi.org/10.1103/PhysRevD.81.105010}
{\path{doi:10.1103/PhysRevD.81.105010}}.

\bibitem{Denz:2016qks}
T.~Denz, J.~M. Pawlowski, M.~Reichert, {Towards apparent convergence in
asymptotically safe quantum gravity}, Eur. Phys. J. C78~(4) (2018) 336.
\newblock \href {http://arxiv.org/abs/1612.07315} {\path{arXiv:1612.07315}},
\href {https://doi.org/10.1140/epjc/s10052-018-5806-0}
{\path{doi:10.1140/epjc/s10052-018-5806-0}}.

\bibitem{Christiansen:2017bsy}
N.~Christiansen, K.~Falls, J.~M. Pawlowski, M.~Reichert, {Curvature dependence
of quantum gravity}, Phys. Rev. D97~(4) (2018) 046007.
\newblock \href {http://arxiv.org/abs/1711.09259} {\path{arXiv:1711.09259}},
\href {https://doi.org/10.1103/PhysRevD.97.046007}
{\path{doi:10.1103/PhysRevD.97.046007}}.

\bibitem{Carrington:2012ea}
M.~E. Carrington, {Renormalization group flow equations connected to the
$n$-particle-irreducible effective action}, Phys. Rev. D87~(4) (2013) 045011.
\newblock \href {http://arxiv.org/abs/1211.4127} {\path{arXiv:1211.4127}},
\href {https://doi.org/10.1103/PhysRevD.87.045011}
{\path{doi:10.1103/PhysRevD.87.045011}}.

\bibitem{Carrington:2014lba}
M.~E. Carrington, W.-J. Fu, D.~Pickering, J.~W. Pulver, {Renormalization group
methods and the 2PI effective action}, Phys. Rev. D91~(2) (2015) 025003.
\newblock \href {http://arxiv.org/abs/1404.0710} {\path{arXiv:1404.0710}},
\href {https://doi.org/10.1103/PhysRevD.91.025003}
{\path{doi:10.1103/PhysRevD.91.025003}}.

\bibitem{Carrington:2016zlc}
M.~E. Carrington, B.~A. Meggison, D.~Pickering, {The 2PI effective action at
four loop order in $\varphi^4$ theory}, Phys. Rev. D94~(2) (2016) 025018.
\newblock \href {http://arxiv.org/abs/1603.02085} {\path{arXiv:1603.02085}},
\href {https://doi.org/10.1103/PhysRevD.94.025018}
{\path{doi:10.1103/PhysRevD.94.025018}}.

\bibitem{Carrington:2017lry}
M.~E. Carrington, S.~A. Friesen, B.~A. Meggison, C.~D. Phillips, D.~Pickering,
K.~Sohrabi, {2PI effective theory at next-to-leading order using the
functional renormalization group}, Phys. Rev. D97~(3) (2018) 036005.
\newblock \href {http://arxiv.org/abs/1711.09135} {\path{arXiv:1711.09135}},
\href {https://doi.org/10.1103/PhysRevD.97.036005}
{\path{doi:10.1103/PhysRevD.97.036005}}.

\bibitem{Carrington:2019fwp}
M.~E. Carrington, S.~A. Friesen, C.~D. Phillips, D.~Pickering, {Renormalization
of the 4PI effective action using the functional renormalization group},
Phys. Rev. D99~(7) (2019) 074002.
\newblock \href {http://arxiv.org/abs/1901.00840} {\path{arXiv:1901.00840}},
\href {https://doi.org/10.1103/PhysRevD.99.074002}
{\path{doi:10.1103/PhysRevD.99.074002}}.

\bibitem{Alexander:2019cgw}
E.~Alexander, P.~Millington, J.~Nursey, P.~M. Saffin, {Alternative flow
equation for the functional renormalization group}, Phys. Rev. D100~(10)
(2019) 101702.
\newblock \href {http://arxiv.org/abs/1907.06503} {\path{arXiv:1907.06503}},
\href {https://doi.org/10.1103/PhysRevD.100.101702}
{\path{doi:10.1103/PhysRevD.100.101702}}.

\bibitem{Alexander:2019quf}
E.~Alexander, P.~Millington, J.~Nursey, P.~M. Saffin, {A new functional RG
flow: regulator-sourced 2PI versus average 1PI} (8 2019).
\newblock \href {http://arxiv.org/abs/1908.02214} {\path{arXiv:1908.02214}}.

\bibitem{Braun:2007bx}
J.~Braun, H.~Gies, J.~M. Pawlowski, {Quark Confinement from Color Confinement},
Phys.Lett. B684 (2010) 262--267.
\newblock \href {http://arxiv.org/abs/0708.2413} {\path{arXiv:0708.2413}},
\href {https://doi.org/10.1016/j.physletb.2010.01.009}
{\path{doi:10.1016/j.physletb.2010.01.009}}.

\bibitem{Fister:2013bh}
L.~Fister, J.~M. Pawlowski, {Confinement from Correlation Functions}, Phys.Rev.
D88 (2013) 045010.
\newblock \href {http://arxiv.org/abs/1301.4163} {\path{arXiv:1301.4163}},
\href {https://doi.org/10.1103/PhysRevD.88.045010}
{\path{doi:10.1103/PhysRevD.88.045010}}.

\bibitem{Cornwall:1981zr}
J.~M. Cornwall, {Dynamical Mass Generation in Continuum QCD}, Phys. Rev. D26
(1982) 1453.
\newblock \href {https://doi.org/10.1103/PhysRevD.26.1453}
{\path{doi:10.1103/PhysRevD.26.1453}}.

\bibitem{Eichhorn:2010zc}
A.~Eichhorn, H.~Gies, J.~M. Pawlowski, {Gluon condensation and scaling
exponents for the propagators in Yang-Mills theory}, Phys.Rev. D83 (2011)
045014.
\newblock \href {http://arxiv.org/abs/1010.2153} {\path{arXiv:1010.2153}},
\href {https://doi.org/10.1103/PhysRevD.83.045014,
10.1103/PhysRevD.83.069903} {\path{doi:10.1103/PhysRevD.83.045014,
10.1103/PhysRevD.83.069903}}.

\bibitem{Branchina:2003ek}
V.~Branchina, K.~A. Meissner, G.~Veneziano, {The Price of an exact, gauge
invariant RG flow equation}, Phys.Lett. B574 (2003) 319--324.
\newblock \href {http://arxiv.org/abs/hep-th/0309234}
{\path{arXiv:hep-th/0309234}}, \href
{https://doi.org/10.1016/j.physletb.2003.09.020}
{\path{doi:10.1016/j.physletb.2003.09.020}}.

\bibitem{Demmel:2014hla}
M.~Demmel, F.~Saueressig, O.~Zanusso, {RG flows of Quantum Einstein Gravity in
the linear-geometric approximation}, Annals Phys. 359 (2015) 141--165.
\newblock \href {http://arxiv.org/abs/1412.7207} {\path{arXiv:1412.7207}},
\href {https://doi.org/10.1016/j.aop.2015.04.018}
{\path{doi:10.1016/j.aop.2015.04.018}}.

\bibitem{Litim:2002hj}
D.~F. Litim, J.~M. Pawlowski, {Wilsonian flows and background fields},
Phys.Lett. B546 (2002) 279--286.
\newblock \href {http://arxiv.org/abs/hep-th/0208216}
{\path{arXiv:hep-th/0208216}}, \href
{https://doi.org/10.1016/S0370-2693(02)02693-X}
{\path{doi:10.1016/S0370-2693(02)02693-X}}.

\bibitem{Litim:2002xm}
D.~F. Litim, J.~M. Pawlowski, {Completeness and consistency of renormalisation
group flows}, Phys.Rev. D66 (2002) 025030.
\newblock \href {http://arxiv.org/abs/hep-th/0202188}
{\path{arXiv:hep-th/0202188}}, \href
{https://doi.org/10.1103/PhysRevD.66.025030}
{\path{doi:10.1103/PhysRevD.66.025030}}.

\bibitem{Folkerts:2011jz}
S.~Folkerts, D.~F. Litim, J.~M. Pawlowski, {Asymptotic freedom of Yang-Mills
theory with gravity}, Phys.Lett. B709 (2012) 234--241.
\newblock \href {http://arxiv.org/abs/1101.5552} {\path{arXiv:1101.5552}},
\href {https://doi.org/10.1016/j.physletb.2012.02.002}
{\path{doi:10.1016/j.physletb.2012.02.002}}.

\bibitem{Bridle:2013sra}
I.~H. Bridle, J.~A. Dietz, T.~R. Morris, {The local potential approximation in
the background field formalism}, JHEP 03 (2014) 093.
\newblock \href {http://arxiv.org/abs/1312.2846} {\path{arXiv:1312.2846}},
\href {https://doi.org/10.1007/JHEP03(2014)093}
{\path{doi:10.1007/JHEP03(2014)093}}.

\bibitem{Dietz:2015owa}
J.~A. Dietz, T.~R. Morris, {Background independent exact renormalization group
for conformally reduced gravity}, JHEP 04 (2015) 118.
\newblock \href {http://arxiv.org/abs/1502.07396} {\path{arXiv:1502.07396}},
\href {https://doi.org/10.1007/JHEP04(2015)118}
{\path{doi:10.1007/JHEP04(2015)118}}.

\bibitem{Labus:2016lkh}
P.~Labus, T.~R. Morris, Z.~H. Slade, {Background independence in a background
dependent renormalization group}, Phys. Rev. D94~(2) (2016) 024007.
\newblock \href {http://arxiv.org/abs/1603.04772} {\path{arXiv:1603.04772}},
\href {https://doi.org/10.1103/PhysRevD.94.024007}
{\path{doi:10.1103/PhysRevD.94.024007}}.

\bibitem{Wetterich:2016ewc}
C.~Wetterich, {Gauge invariant flow equation}, Nucl. Phys. B931 (2018)
262--282.
\newblock \href {http://arxiv.org/abs/1607.02989} {\path{arXiv:1607.02989}},
\href {https://doi.org/10.1016/j.nuclphysb.2018.04.020}
{\path{doi:10.1016/j.nuclphysb.2018.04.020}}.

\bibitem{Wetterich:2017aoy}
C.~Wetterich, {Gauge-invariant fields and flow equations for Yang–Mills
theories}, Nucl. Phys. B934 (2018) 265--316.
\newblock \href {http://arxiv.org/abs/1710.02494} {\path{arXiv:1710.02494}},
\href {https://doi.org/10.1016/j.nuclphysb.2018.07.002}
{\path{doi:10.1016/j.nuclphysb.2018.07.002}}.

\bibitem{Pawlowski:2018ixd}
J.~M. Pawlowski, M.~Reichert, C.~Wetterich, M.~Yamada, {Higgs scalar potential
in asymptotically safe quantum gravity}, Phys. Rev. D 99~(8) (2019) 086010.
\newblock \href {http://arxiv.org/abs/1811.11706} {\path{arXiv:1811.11706}},
\href {https://doi.org/10.1103/PhysRevD.99.086010}
{\path{doi:10.1103/PhysRevD.99.086010}}.

\bibitem{Wetterich:2019zdo}
C.~Wetterich, M.~Yamada, {Variable Planck mass from the gauge invariant flow
equation}, Phys. Rev. D100~(6) (2019) 066017.
\newblock \href {http://arxiv.org/abs/1906.01721} {\path{arXiv:1906.01721}},
\href {https://doi.org/10.1103/PhysRevD.100.066017}
{\path{doi:10.1103/PhysRevD.100.066017}}.

\bibitem{Morris:1999px}
T.~R. Morris, {A Gauge invariant exact renormalization group. 1.}, Nucl. Phys.
B573 (2000) 97--126.
\newblock \href {http://arxiv.org/abs/hep-th/9910058}
{\path{arXiv:hep-th/9910058}}, \href
{https://doi.org/10.1016/S0550-3213(99)00821-4}
{\path{doi:10.1016/S0550-3213(99)00821-4}}.

\bibitem{Morris:2000fs}
T.~R. Morris, {A Gauge invariant exact renormalization group. 2.}, JHEP 12
(2000) 012.
\newblock \href {http://arxiv.org/abs/hep-th/0006064}
{\path{arXiv:hep-th/0006064}}, \href
{https://doi.org/10.1088/1126-6708/2000/12/012}
{\path{doi:10.1088/1126-6708/2000/12/012}}.

\bibitem{Arnone:2001iy}
S.~Arnone, Y.~A. Kubyshin, T.~R. Morris, J.~F. Tighe, {Gauge invariant
regularization via SU(N|N)}, Int. J. Mod. Phys. A17 (2002) 2283--2330.
\newblock \href {http://arxiv.org/abs/hep-th/0106258}
{\path{arXiv:hep-th/0106258}}, \href
{https://doi.org/10.1142/S0217751X02009722}
{\path{doi:10.1142/S0217751X02009722}}.

\bibitem{Arnone:2002cs}
S.~Arnone, A.~Gatti, T.~R. Morris, {A Proposal for a manifestly gauge invariant
and universal calculus in Yang-Mills theory}, Phys. Rev. D67 (2003) 085003.
\newblock \href {http://arxiv.org/abs/hep-th/0209162}
{\path{arXiv:hep-th/0209162}}, \href
{https://doi.org/10.1103/PhysRevD.67.085003}
{\path{doi:10.1103/PhysRevD.67.085003}}.

\bibitem{Arnone:2005fb}
S.~Arnone, T.~R. Morris, O.~J. Rosten, {A Generalised Manifestly Gauge
Invariant Exact Renormalisation Group for SU(N) Yang-Mills}, Eur. Phys. J.
C50 (2007) 467--504.
\newblock \href {http://arxiv.org/abs/hep-th/0507154}
{\path{arXiv:hep-th/0507154}}, \href
{https://doi.org/10.1140/epjc/s10052-007-0258-y}
{\path{doi:10.1140/epjc/s10052-007-0258-y}}.

\bibitem{Arnone:2005vd}
S.~Arnone, T.~R. Morris, O.~J. Rosten, {Manifestly gauge invariant QED}, JHEP
10 (2005) 115.
\newblock \href {http://arxiv.org/abs/hep-th/0505169}
{\path{arXiv:hep-th/0505169}}, \href
{https://doi.org/10.1088/1126-6708/2005/10/115}
{\path{doi:10.1088/1126-6708/2005/10/115}}.

\bibitem{Morris:2005tv}
T.~R. Morris, O.~J. Rosten, {A Manifestly gauge invariant, continuum
calculation of the SU(N) Yang-Mills two-loop beta function}, Phys. Rev. D73
(2006) 065003.
\newblock \href {http://arxiv.org/abs/hep-th/0508026}
{\path{arXiv:hep-th/0508026}}, \href
{https://doi.org/10.1103/PhysRevD.73.065003}
{\path{doi:10.1103/PhysRevD.73.065003}}.

\bibitem{Morris:2006in}
T.~R. Morris, O.~J. Rosten, {Manifestly gauge invariant QCD}, J. Phys. A39
(2006) 11657--11681.
\newblock \href {http://arxiv.org/abs/hep-th/0606189}
{\path{arXiv:hep-th/0606189}}, \href
{https://doi.org/10.1088/0305-4470/39/37/020}
{\path{doi:10.1088/0305-4470/39/37/020}}.

\bibitem{Rosten:2006qx}
O.~J. Rosten, {General Computations Without Fixing the Gauge}, Phys. Rev. D74
(2006) 125006.
\newblock \href {http://arxiv.org/abs/hep-th/0604183}
{\path{arXiv:hep-th/0604183}}, \href
{https://doi.org/10.1103/PhysRevD.74.125006}
{\path{doi:10.1103/PhysRevD.74.125006}}.

\bibitem{Rosten:2006tk}
O.~J. Rosten, {A Manifestly gauge invariant and universal calculus for SU(N)
Yang-Mills}, Int. J. Mod. Phys. A21 (2006) 4627--4762.
\newblock \href {http://arxiv.org/abs/hep-th/0602229}
{\path{arXiv:hep-th/0602229}}, \href
{https://doi.org/10.1142/S0217751X06033040}
{\path{doi:10.1142/S0217751X06033040}}.

\bibitem{Rosten:2008zp}
O.~J. Rosten, {A Resummable beta-Function for Massless QED}, Phys. Lett. B662
(2008) 237--243.
\newblock \href {http://arxiv.org/abs/0801.2462} {\path{arXiv:0801.2462}},
\href {https://doi.org/10.1016/j.physletb.2008.03.006}
{\path{doi:10.1016/j.physletb.2008.03.006}}.

\bibitem{deAlwis:2017ysy}
S.~P. de~Alwis, {Exact RG Flow Equations and Quantum Gravity}, JHEP 03 (2018)
118.
\newblock \href {http://arxiv.org/abs/1707.09298} {\path{arXiv:1707.09298}},
\href {https://doi.org/10.1007/JHEP03(2018)118}
{\path{doi:10.1007/JHEP03(2018)118}}.

\bibitem{Bonanno:2019ukb}
A.~Bonanno, S.~Lippoldt, R.~Percacci, G.~P. Vacca, {On Exact Proper Time
Wilsonian RG Flows}, Eur.\ Phys.\ J.\ C 80~(3) (2020) 249.
\newblock \href {http://arxiv.org/abs/1912.08135} {\path{arXiv:1912.08135}},
\href {https://doi.org/10.1140/epjc/s10052-020-7798-9}
{\path{doi:10.1140/epjc/s10052-020-7798-9}}.

\bibitem{Falls:2020tmj}
K.~Falls, {Background independent exact renormalisation} (4 2020).
\newblock \href {http://arxiv.org/abs/2004.11409} {\path{arXiv:2004.11409}}.

\bibitem{DAttanasio:1996psq}
M.~D'Attanasio, M.~Pietroni, {Gauge invariant renormalization group at finite
temperature}, Nucl. Phys. B498 (1997) 443--466.
\newblock \href {http://arxiv.org/abs/hep-th/9611038}
{\path{arXiv:hep-th/9611038}}, \href
{https://doi.org/10.1016/S0550-3213(97)00292-7}
{\path{doi:10.1016/S0550-3213(97)00292-7}}.

\bibitem{Comelli:1997ru}
D.~Comelli, M.~Pietroni, {Screening masses in SU(N) from Wilson renormalization
group}, Phys.Lett. B417 (1998) 337--342.
\newblock \href {http://arxiv.org/abs/hep-ph/9708489}
{\path{arXiv:hep-ph/9708489}}, \href
{https://doi.org/10.1016/S0370-2693(97)01372-5}
{\path{doi:10.1016/S0370-2693(97)01372-5}}.

\bibitem{Jungnickel:1997ke}
D.~Jungnickel, C.~Wetterich, {Nonperturbative flow equations, low-energy QCD
and the chiral phase transition}, in: {NATO Advanced Study Institute on
Confinement, Duality and Nonperturbative Aspects of QCD}, 1997, pp. 215--261.
\newblock \href {http://arxiv.org/abs/hep-ph/9710397}
{\path{arXiv:hep-ph/9710397}}.

\bibitem{Litim:1998nf}
D.~F. Litim, J.~M. Pawlowski,
\href{http://alice.cern.ch/format/showfull?sysnb=0302190}{{On gauge invariant
Wilsonian flows}}, in: {The exact renormalization group. Proceedings,
Workshop, Faro, Portugal, September 10-12, 1998}, 1998, pp. 168--185.
\newblock \href {http://arxiv.org/abs/hep-th/9901063}
{\path{arXiv:hep-th/9901063}}.
\newline\urlprefix\url{http://alice.cern.ch/format/showfull?sysnb=0302190}

\bibitem{Schaefer:2006sr}
B.-J. Schaefer, J.~Wambach, {Renormalization group approach towards the QCD
phase diagram}, Phys.Part.Nucl. 39 (2008) 1025--1032.
\newblock \href {http://arxiv.org/abs/hep-ph/0611191}
{\path{arXiv:hep-ph/0611191}}, \href
{https://doi.org/10.1134/S1063779608070083}
{\path{doi:10.1134/S1063779608070083}}.

\bibitem{Gies:2006wv}
H.~Gies, {Introduction to the functional RG and applications to gauge
theories}, Lect.Notes Phys. 852 (2012) 287--348.
\newblock \href {http://arxiv.org/abs/hep-ph/0611146}
{\path{arXiv:hep-ph/0611146}}, \href
{https://doi.org/10.1007/978-3-642-27320-9_6}
{\path{doi:10.1007/978-3-642-27320-9_6}}.

\bibitem{Sonoda:2007av}
H.~Sonoda, {The Exact Renormalization Group: Renormalization theory revisited},
2007.
\newblock \href {http://arxiv.org/abs/0710.1662} {\path{arXiv:0710.1662}}.

\bibitem{Schaefer:2011pn}
B.-J. Schaefer, {Fluctuations and the QCD Phase Diagram}, Phys.Atom.Nucl. 75
(2012) 741--743.
\newblock \href {http://arxiv.org/abs/1102.2772} {\path{arXiv:1102.2772}},
\href {https://doi.org/10.1134/S1063778812060270}
{\path{doi:10.1134/S1063778812060270}}.

\bibitem{Pawlowski:2014aha}
J.~M. Pawlowski, {Equation of state and phase diagram of strongly interacting
matter}, Nucl.Phys. A931 (2014) 113--124.
\newblock \href {https://doi.org/10.1016/j.nuclphysa.2014.09.074}
{\path{doi:10.1016/j.nuclphysa.2014.09.074}}.

\bibitem{Strodthoff:2016dip}
N.~Strodthoff, {Phase Structure and Dynamics of QCD–A Functional
Perspective}, J. Phys. Conf. Ser. 832~(1) (2017) 012040.
\newblock \href {http://arxiv.org/abs/1612.03807} {\path{arXiv:1612.03807}},
\href {https://doi.org/10.1088/1742-6596/832/1/012040}
{\path{doi:10.1088/1742-6596/832/1/012040}}.

\bibitem{Klein:2017shl}
B.~Klein, {Modeling Finite-Volume Effects and Chiral Symmetry Breaking in
Two-Flavor QCD Thermodynamics}, Phys. Rept. 707-708 (2017) 1--51.
\newblock \href {http://arxiv.org/abs/1710.05357} {\path{arXiv:1710.05357}},
\href {https://doi.org/10.1016/j.physrep.2017.09.002}
{\path{doi:10.1016/j.physrep.2017.09.002}}.

\bibitem{Ford:1998bt}
C.~Ford, U.~Mitreuter, T.~Tok, A.~Wipf, J.~Pawlowski, {Monopoles, Polyakov
loops and gauge fixing on the torus}, Annals Phys. 269 (1998) 26--50.
\newblock \href {http://arxiv.org/abs/hep-th/9802191}
{\path{arXiv:hep-th/9802191}}, \href {https://doi.org/10.1006/aphy.1998.5841}
{\path{doi:10.1006/aphy.1998.5841}}.

\bibitem{vanBaal:2000zc}
P.~van Baal, {QCD in a finite volume} (2000).
\newblock \href {http://arxiv.org/abs/hep-ph/0008206}
{\path{arXiv:hep-ph/0008206}}.

\bibitem{Reinosa:2014ooa}
U.~Reinosa, J.~Serreau, M.~Tissier, N.~Wschebor, {Deconfinement transition in
SU($N$) theories from perturbation theory}, Phys. Lett. B 742 (2015) 61--68.
\newblock \href {http://arxiv.org/abs/1407.6469} {\path{arXiv:1407.6469}},
\href {https://doi.org/10.1016/j.physletb.2015.01.006}
{\path{doi:10.1016/j.physletb.2015.01.006}}.

\bibitem{Herbst:2015ona}
T.~K. Herbst, J.~Luecker, J.~M. Pawlowski, {Confinement order parameters and
fluctuations} (2015).
\newblock \href {http://arxiv.org/abs/1510.03830} {\path{arXiv:1510.03830}}.

\bibitem{Gross:1980br}
D.~J. Gross, R.~D. Pisarski, L.~G. Yaffe, {QCD and Instantons at Finite
Temperature}, Rev. Mod. Phys. 53 (1981) 43.
\newblock \href {https://doi.org/10.1103/RevModPhys.53.43}
{\path{doi:10.1103/RevModPhys.53.43}}.

\bibitem{Weiss:1980rj}
N.~Weiss, {The Effective Potential for the Order Parameter of Gauge Theories at
Finite Temperature}, Phys. Rev. D 24 (1981) 475.
\newblock \href {https://doi.org/10.1103/PhysRevD.24.475}
{\path{doi:10.1103/PhysRevD.24.475}}.

\bibitem{Braun:2010cy}
J.~Braun, A.~Eichhorn, H.~Gies, J.~M. Pawlowski, {On the Nature of the Phase
Transition in SU(N), Sp(2) and E(7) Yang-Mills theory}, Eur.Phys.J. C70
(2010) 689--702.
\newblock \href {http://arxiv.org/abs/1007.2619} {\path{arXiv:1007.2619}},
\href {https://doi.org/10.1140/epjc/s10052-010-1485-1}
{\path{doi:10.1140/epjc/s10052-010-1485-1}}.

\bibitem{Braun:2009gm}
J.~Braun, L.~M. Haas, F.~Marhauser, J.~M. Pawlowski, {Phase Structure of
Two-Flavor QCD at Finite Chemical Potential}, Phys.Rev.Lett. 106 (2011)
022002.
\newblock \href {http://arxiv.org/abs/0908.0008} {\path{arXiv:0908.0008}},
\href {https://doi.org/10.1103/PhysRevLett.106.022002}
{\path{doi:10.1103/PhysRevLett.106.022002}}.

\bibitem{Fischer:2013eca}
C.~S. Fischer, L.~Fister, J.~Luecker, J.~M. Pawlowski, {Polyakov loop potential
at finite density}, Phys.Lett. B732 (2014) 273--277.
\newblock \href {http://arxiv.org/abs/1306.6022} {\path{arXiv:1306.6022}},
\href {https://doi.org/10.1016/j.physletb.2014.03.057}
{\path{doi:10.1016/j.physletb.2014.03.057}}.

\bibitem{Fischer:2014vxa}
C.~S. Fischer, J.~Luecker, J.~M. Pawlowski, {Phase structure of QCD for heavy
quarks}, Phys. Rev. D 91~(1) (2015) 014024.
\newblock \href {http://arxiv.org/abs/1409.8462} {\path{arXiv:1409.8462}},
\href {https://doi.org/10.1103/PhysRevD.91.014024}
{\path{doi:10.1103/PhysRevD.91.014024}}.

\bibitem{Fischer:2014ata}
C.~S. Fischer, J.~Luecker, C.~A. Welzbacher, {Phase structure of three and four
flavor QCD}, Phys. Rev. D90~(3) (2014) 034022.
\newblock \href {http://arxiv.org/abs/1405.4762} {\path{arXiv:1405.4762}},
\href {https://doi.org/10.1103/PhysRevD.90.034022}
{\path{doi:10.1103/PhysRevD.90.034022}}.

\bibitem{Reinosa:2014zta}
U.~Reinosa, J.~Serreau, M.~Tissier, N.~Wschebor, {Deconfinement transition in
SU(2) Yang-Mills theory: A two-loop study}, Phys. Rev. D91 (2015) 045035.
\newblock \href {http://arxiv.org/abs/1412.5672} {\path{arXiv:1412.5672}},
\href {https://doi.org/10.1103/PhysRevD.91.045035}
{\path{doi:10.1103/PhysRevD.91.045035}}.

\bibitem{Reinosa:2015oua}
U.~Reinosa, J.~Serreau, M.~Tissier, {Perturbative study of the QCD phase
diagram for heavy quarks at nonzero chemical potential}, Phys. Rev. D 92
(2015) 025021.
\newblock \href {http://arxiv.org/abs/1504.02916} {\path{arXiv:1504.02916}},
\href {https://doi.org/10.1103/PhysRevD.92.025021}
{\path{doi:10.1103/PhysRevD.92.025021}}.

\bibitem{Reinosa:2015gxn}
U.~Reinosa, J.~Serreau, M.~Tissier, N.~Wschebor, {Two-loop study of the
deconfinement transition in Yang-Mills theories: SU(3) and beyond}, Phys.
Rev. D93~(10) (2016) 105002.
\newblock \href {http://arxiv.org/abs/1511.07690} {\path{arXiv:1511.07690}},
\href {https://doi.org/10.1103/PhysRevD.93.105002}
{\path{doi:10.1103/PhysRevD.93.105002}}.

\bibitem{Reinosa:2016iml}
U.~Reinosa, J.~Serreau, M.~Tissier, A.~Tresmontant, {Yang-Mills correlators
across the deconfinement phase transition}, Phys. Rev. D95~(4) (2017) 045014.
\newblock \href {http://arxiv.org/abs/1606.08012} {\path{arXiv:1606.08012}},
\href {https://doi.org/10.1103/PhysRevD.95.045014}
{\path{doi:10.1103/PhysRevD.95.045014}}.

\bibitem{Maelger:2017amh}
J.~Maelger, U.~Reinosa, J.~Serreau, {Perturbative study of the QCD phase
diagram for heavy quarks at nonzero chemical potential: Two-loop
corrections}, Phys. Rev. D 97~(7) (2018) 074027.
\newblock \href {http://arxiv.org/abs/1710.01930} {\path{arXiv:1710.01930}},
\href {https://doi.org/10.1103/PhysRevD.97.074027}
{\path{doi:10.1103/PhysRevD.97.074027}}.

\bibitem{Reinhardt:2012qe}
H.~Reinhardt, J.~Heffner, {The effective potential of the confinement order
parameter in the Hamilton approach}, Phys.Lett. B718 (2012) 672--677.
\newblock \href {http://arxiv.org/abs/1210.1742} {\path{arXiv:1210.1742}},
\href {https://doi.org/10.1016/j.physletb.2012.10.084}
{\path{doi:10.1016/j.physletb.2012.10.084}}.

\bibitem{Reinhardt:2013iia}
H.~Reinhardt, J.~Heffner, {The effective potential of the confinement order
parameter in the Hamiltonian approach} (2013).
\newblock \href {http://arxiv.org/abs/1304.2980} {\path{arXiv:1304.2980}}.

\bibitem{Heffner:2015zna}
J.~Heffner, H.~Reinhardt, {Finite-temperature Yang-Mills theory in the
Hamiltonian approach in Coulomb gauge from a compactified spatial dimension},
Phys. Rev. D 91~(8) (2015) 085022.
\newblock \href {http://arxiv.org/abs/1501.05858} {\path{arXiv:1501.05858}},
\href {https://doi.org/10.1103/PhysRevD.91.085022}
{\path{doi:10.1103/PhysRevD.91.085022}}.

\bibitem{Quandt:2016ykm}
M.~Quandt, H.~Reinhardt, {Covariant variational approach to Yang-Mills Theory:
effective potential of the Polyakov loop}, Phys. Rev. D94~(6) (2016) 065015.
\newblock \href {http://arxiv.org/abs/1603.08058} {\path{arXiv:1603.08058}},
\href {https://doi.org/10.1103/PhysRevD.94.065015}
{\path{doi:10.1103/PhysRevD.94.065015}}.

\bibitem{Gies:2006nz}
H.~Gies, J.~Jaeckel, J.~M. Pawlowski, C.~Wetterich, {Do instantons like a
colorful background?}, Eur.Phys.J. C49 (2007) 997--1010.
\newblock \href {http://arxiv.org/abs/hep-ph/0608171}
{\path{arXiv:hep-ph/0608171}}, \href
{https://doi.org/10.1140/epjc/s10052-006-0178-2}
{\path{doi:10.1140/epjc/s10052-006-0178-2}}.

\bibitem{Pagani:2016pad}
C.~Pagani, {Note on scaling arguments in the effective average action
formalism}, Phys.\ Rev.\ D 94~(4) (2016) 045001.
\newblock \href {http://arxiv.org/abs/1603.07250} {\path{arXiv:1603.07250}},
\href {https://doi.org/10.1103/PhysRevD.94.045001}
{\path{doi:10.1103/PhysRevD.94.045001}}.

\bibitem{Pagani:2016dof}
C.~Pagani, M.~Reuter, {Composite Operators in Asymptotic Safety}, Phys. Rev.
D95~(6) (2017) 066002.
\newblock \href {http://arxiv.org/abs/1611.06522} {\path{arXiv:1611.06522}},
\href {https://doi.org/10.1103/PhysRevD.95.066002}
{\path{doi:10.1103/PhysRevD.95.066002}}.

\bibitem{Becker:2019fhi}
M.~Becker, C.~Pagani, O.~Zanusso, {Fractal geometry of higher derivative
gravity}, Phys. Rev. Lett. 124~(15) (2020) 151302.
\newblock \href {http://arxiv.org/abs/1911.02415} {\path{arXiv:1911.02415}},
\href {https://doi.org/10.1103/PhysRevLett.124.151302}
{\path{doi:10.1103/PhysRevLett.124.151302}}.

\bibitem{Houthoff:2020zqy}
W.~Houthoff, A.~Kurov, F.~Saueressig, {On the scaling of composite operators in
Asymptotic Safety} (2 2020).
\newblock \href {http://arxiv.org/abs/2002.00256} {\path{arXiv:2002.00256}}.

\bibitem{Kurov:2020csd}
A.~Kurov, F.~Saueressig, {On characterizing the Quantum Geometry underlying
Asymptotic Safety} (3 2020).
\newblock \href {http://arxiv.org/abs/2003.07454} {\path{arXiv:2003.07454}}.

\bibitem{Fister:2011uw}
L.~Fister, J.~M. Pawlowski, {Yang-Mills correlation functions at finite
temperature} (2011).
\newblock \href {http://arxiv.org/abs/1112.5440} {\path{arXiv:1112.5440}}.

\bibitem{Fister:2012lug}
L.~Fister, \href{http://www.ub.uni-heidelberg.de/archiv/13877}{{On the Phase
Diagram of QCD with Dynamical Quarks}}, Ph.D. thesis, Heidelberg U. (2012).
\newline\urlprefix\url{http://www.ub.uni-heidelberg.de/archiv/13877}

\bibitem{Boettcher:2012dh}
I.~Boettcher, S.~Diehl, J.~M. Pawlowski, C.~Wetterich, {Tan contact and
universal high momentum behavior of the fermion propagator in the BCS-BEC
crossover}, Phys. Rev. A 87~(2) (2013) 023606.
\newblock \href {http://arxiv.org/abs/1209.5641} {\path{arXiv:1209.5641}},
\href {https://doi.org/10.1103/PhysRevA.87.023606}
{\path{doi:10.1103/PhysRevA.87.023606}}.

\bibitem{Aoki:1996fh}
K.-I. Aoki, K.-i. Morikawa, J.-I. Sumi, H.~Terao, M.~Tomoyose, {Nonperturbative
renormalization group analysis of the chiral critical behaviors in QED},
Prog. Theor. Phys. 97 (1997) 479--490.
\newblock \href {http://arxiv.org/abs/hep-ph/9612459}
{\path{arXiv:hep-ph/9612459}}, \href {https://doi.org/10.1143/PTP.97.479}
{\path{doi:10.1143/PTP.97.479}}.

\bibitem{Aoki:1999dv}
K.-I. Aoki, K.~Morikawa, J.-I. Sumi, H.~Terao, M.~Tomoyose, {Wilson
renormalization group equations for the critical dynamics of chiral
symmetry}, Prog. Theor. Phys. 102 (1999) 1151--1162.
\newblock \href {http://arxiv.org/abs/hep-th/9908042}
{\path{arXiv:hep-th/9908042}}, \href {https://doi.org/10.1143/PTP.102.1151}
{\path{doi:10.1143/PTP.102.1151}}.

\bibitem{Aoki:1999dw}
K.-I. Aoki, K.~Morikawa, J.-I. Sumi, H.~Terao, M.~Tomoyose, {Analysis of the
Wilsonian effective potentials in dynamical chiral symmetry breaking}, Phys.
Rev. D 61 (2000) 045008.
\newblock \href {http://arxiv.org/abs/hep-th/9908043}
{\path{arXiv:hep-th/9908043}}, \href
{https://doi.org/10.1103/PhysRevD.61.045008}
{\path{doi:10.1103/PhysRevD.61.045008}}.

\bibitem{Aoki:2000dh}
K.-I. Aoki, K.~Takagi, H.~Terao, M.~Tomoyose, {Nonladder extended
renormalization group analysis of the dynamical chiral symmetry breaking},
Prog. Theor. Phys. 103 (2000) 815--832.
\newblock \href {http://arxiv.org/abs/hep-th/0002038}
{\path{arXiv:hep-th/0002038}}, \href {https://doi.org/10.1143/PTP.103.815}
{\path{doi:10.1143/PTP.103.815}}.

\bibitem{Meggiolaro:2000kp}
E.~Meggiolaro, C.~Wetterich, {Evolution equations for the effective four quark
interactions in QCD}, Nucl. Phys. B606 (2001) 337--356.
\newblock \href {http://arxiv.org/abs/hep-ph/0012081}
{\path{arXiv:hep-ph/0012081}}, \href
{https://doi.org/10.1016/S0550-3213(01)00130-4}
{\path{doi:10.1016/S0550-3213(01)00130-4}}.

\bibitem{Aoki:2009zza}
K.-I. Aoki, K.~Miyashita, {Evaluation of the spontaneous chiral symmetry
breaking scale in general gauge theories with non-perturbative
renormalization group}, Prog. Theor. Phys. 121 (2009) 875--884.
\newblock \href {https://doi.org/10.1143/PTP.121.875}
{\path{doi:10.1143/PTP.121.875}}.

\bibitem{Aoki:2012mj}
K.-I. Aoki, D.~Sato, {Solving the QCD non-perturbative flow equation as a
partial differential equation and its application to the dynamical chiral
symmetry breaking}, PTEP 2013 (2013) 043B04.
\newblock \href {http://arxiv.org/abs/1212.0063} {\path{arXiv:1212.0063}},
\href {https://doi.org/10.1093/ptep/ptt018} {\path{doi:10.1093/ptep/ptt018}}.

\bibitem{Aoki:2014ola}
K.-I. Aoki, S.-I. Kumamoto, D.~Sato, {Weak solution of the non-perturbative
renormalization group equation to describe dynamical chiral symmetry
breaking}, PTEP 2014~(4) (2014) 043B05.
\newblock \href {http://arxiv.org/abs/1403.0174} {\path{arXiv:1403.0174}},
\href {https://doi.org/10.1093/ptep/ptu039} {\path{doi:10.1093/ptep/ptu039}}.

\bibitem{Braun:2014ata}
J.~Braun, L.~Fister, J.~M. Pawlowski, F.~Rennecke, {From Quarks and Gluons to
Hadrons: Chiral Symmetry Breaking in Dynamical QCD}, Phys. Rev. D94~(3)
(2016) 034016.
\newblock \href {http://arxiv.org/abs/1412.1045} {\path{arXiv:1412.1045}},
\href {https://doi.org/10.1103/PhysRevD.94.034016}
{\path{doi:10.1103/PhysRevD.94.034016}}.

\bibitem{Rennecke:2015eba}
F.~Rennecke, {Vacuum structure of vector mesons in QCD}, Phys. Rev. D92~(7)
(2015) 076012.
\newblock \href {http://arxiv.org/abs/1504.03585} {\path{arXiv:1504.03585}},
\href {https://doi.org/10.1103/PhysRevD.92.076012}
{\path{doi:10.1103/PhysRevD.92.076012}}.

\bibitem{Huber:2011qr}
M.~Q. Huber, J.~Braun, {Algorithmic derivation of functional renormalization
group equations and Dyson-Schwinger equations}, Comput.Phys.Commun. 183
(2012) 1290--1320.
\newblock \href {http://arxiv.org/abs/1102.5307} {\path{arXiv:1102.5307}},
\href {https://doi.org/10.1016/j.cpc.2012.01.014}
{\path{doi:10.1016/j.cpc.2012.01.014}}.

\bibitem{Huber:2019dkb}
M.~Q. Huber, A.~K. Cyrol, J.~M. Pawlowski, {DoFun 3.0: Functional equations in
Mathematica}, Comput. Phys. Commun. 248 (2020) 107058.
\newblock \href {http://arxiv.org/abs/1908.02760} {\path{arXiv:1908.02760}},
\href {https://doi.org/10.1016/j.cpc.2019.107058}
{\path{doi:10.1016/j.cpc.2019.107058}}.

\bibitem{Cyrol:2016zqb}
A.~K. Cyrol, M.~Mitter, N.~Strodthoff, {FormTracer - A Mathematica Tracing
Package Using FORM}, Comput. Phys. Commun. C219 (2017) 346--352.
\newblock \href {http://arxiv.org/abs/1610.09331} {\path{arXiv:1610.09331}},
\href {https://doi.org/10.1016/j.cpc.2017.05.024}
{\path{doi:10.1016/j.cpc.2017.05.024}}.

\bibitem{github:FormTracer}
A.~K. Cyrol, M.~Mitter, J.~M. Pawlowski, N.~Strodthoff, {FormTracer GitHub
Repository}, \url{https://github.com/FormTracer/FormTracer} (2016).

\bibitem{Ellwanger:1993mw}
U.~Ellwanger, {FLow equations for N point functions and bound states}, Z. Phys.
C62 (1994) 503--510, [,206(1993)].
\newblock \href {http://arxiv.org/abs/hep-ph/9308260}
{\path{arXiv:hep-ph/9308260}}, \href {https://doi.org/10.1007/BF01555911}
{\path{doi:10.1007/BF01555911}}.

\bibitem{Ellwanger:1994wy}
U.~Ellwanger, C.~Wetterich, {Evolution equations for the quark - meson
transition}, Nucl.Phys. B423 (1994) 137--170.
\newblock \href {http://arxiv.org/abs/hep-ph/9402221}
{\path{arXiv:hep-ph/9402221}}, \href
{https://doi.org/10.1016/0550-3213(94)90568-1}
{\path{doi:10.1016/0550-3213(94)90568-1}}.

\bibitem{Floerchinger:2009uf}
S.~Floerchinger, C.~Wetterich, {Exact flow equation for composite operators},
Phys.Lett. B680 (2009) 371--376.
\newblock \href {http://arxiv.org/abs/0905.0915} {\path{arXiv:0905.0915}},
\href {https://doi.org/10.1016/j.physletb.2009.09.014}
{\path{doi:10.1016/j.physletb.2009.09.014}}.

\bibitem{Alkofer:2018guy}
R.~Alkofer, A.~Maas, W.~A. Mian, M.~Mitter, J.~París-López, J.~M. Pawlowski,
N.~Wink, {Bound state properties from the functional renormalization group},
Phys. Rev. D99~(5) (2019) 054029.
\newblock \href {http://arxiv.org/abs/1810.07955} {\path{arXiv:1810.07955}},
\href {https://doi.org/10.1103/PhysRevD.99.054029}
{\path{doi:10.1103/PhysRevD.99.054029}}.

\bibitem{Eichmann:2015nra}
G.~Eichmann, C.~S. Fischer, W.~Heupel, {Four-point functions and the
permutation group S4}, Phys. Rev. D 92~(5) (2015) 056006.
\newblock \href {http://arxiv.org/abs/1505.06336} {\path{arXiv:1505.06336}},
\href {https://doi.org/10.1103/PhysRevD.92.056006}
{\path{doi:10.1103/PhysRevD.92.056006}}.

\bibitem{FukushimaPawlowskiStrodthoff}
K.~Fukushima, J.~M. Pawlowski, N.~Strodthoff, {Emergent Hadrons and Diquarks,
and Dense QCD Matter}, in preparation (2020).

\bibitem{Denz:2019ogb}
T.~Denz, M.~Mitter, J.~M. Pawlowski, C.~Wetterich, M.~Yamada, {Partial
bosonisation for the two-dimensional Hubbard model: How well does it work?}
(2019).
\newblock \href {http://arxiv.org/abs/1910.08300} {\path{arXiv:1910.08300}}.

\bibitem{Jakovac:2018dkp}
A.~Jakovac, A.~Patkos, {Bound states in Functional Renormalization Group}, Int.
J. Mod. Phys. A34~(27) (2019) 1950154.
\newblock \href {http://arxiv.org/abs/1811.09418} {\path{arXiv:1811.09418}},
\href {https://doi.org/10.1142/S0217751X19501549}
{\path{doi:10.1142/S0217751X19501549}}.

\bibitem{Jakovac:2019zzw}
A.~Jakovac, A.~Patkos, {Interacting two-particle states in the symmetric phase
of the chiral Nambu-Jona-Lasinio model} (2019).
\newblock \href {http://arxiv.org/abs/1902.06492} {\path{arXiv:1902.06492}}.

\bibitem{Eser:2018jqo}
J.~Eser, F.~Divotgey, M.~Mitter, D.~H. Rischke, {Low-energy limit of the $O(4)$
quark-meson model from the functional renormalization group approach}, Phys.
Rev. D 98~(1) (2018) 014024.
\newblock \href {http://arxiv.org/abs/1804.01787} {\path{arXiv:1804.01787}},
\href {https://doi.org/10.1103/PhysRevD.98.014024}
{\path{doi:10.1103/PhysRevD.98.014024}}.

\bibitem{Divotgey:2019xea}
F.~Divotgey, J.~Eser, M.~Mitter, {Dynamical generation of low-energy couplings
from quark-meson fluctuations}, Phys. Rev. D 99~(5) (2019) 054023.
\newblock \href {http://arxiv.org/abs/1901.02472} {\path{arXiv:1901.02472}},
\href {https://doi.org/10.1103/PhysRevD.99.054023}
{\path{doi:10.1103/PhysRevD.99.054023}}.

\bibitem{Bartels:2018pin}
J.~Bartels, C.~Contreras, G.~P. Vacca, {A functional RG approach for the BFKL
Pomeron}, JHEP 01 (2019) 004.
\newblock \href {http://arxiv.org/abs/1808.07517} {\path{arXiv:1808.07517}},
\href {https://doi.org/10.1007/JHEP01(2019)004}
{\path{doi:10.1007/JHEP01(2019)004}}.

\bibitem{Bartels:2019qho}
J.~Bartels, C.~Contreras, G.~P. Vacca, {The Odderon in QCD with running
coupling}, JHEP 04 (2020) 183.
\newblock \href {http://arxiv.org/abs/1910.04588} {\path{arXiv:1910.04588}},
\href {https://doi.org/10.1007/JHEP04(2020)183}
{\path{doi:10.1007/JHEP04(2020)183}}.

\bibitem{Weyrich:2015hha}
J.~Weyrich, N.~Strodthoff, L.~von Smekal, {Chiral mirror-baryon-meson model and
nuclear matter beyond mean-field approximation}, Phys. Rev. C92~(1) (2015)
015214.
\newblock \href {http://arxiv.org/abs/1504.02697} {\path{arXiv:1504.02697}},
\href {https://doi.org/10.1103/PhysRevC.92.015214}
{\path{doi:10.1103/PhysRevC.92.015214}}.

\bibitem{Eichmann:2016yit}
G.~Eichmann, H.~Sanchis-Alepuz, R.~Williams, R.~Alkofer, C.~S. Fischer,
{Baryons as relativistic three-quark bound states}, Prog. Part. Nucl. Phys.
91 (2016) 1--100.
\newblock \href {http://arxiv.org/abs/1606.09602} {\path{arXiv:1606.09602}},
\href {https://doi.org/10.1016/j.ppnp.2016.07.001}
{\path{doi:10.1016/j.ppnp.2016.07.001}}.

\bibitem{Diehl:2007xz}
S.~Diehl, H.~Krahl, M.~Scherer, {Three-Body Scattering from Nonperturbative
Flow Equations}, Phys. Rev. C 78 (2008) 034001.
\newblock \href {http://arxiv.org/abs/0712.2846} {\path{arXiv:0712.2846}},
\href {https://doi.org/10.1103/PhysRevC.78.034001}
{\path{doi:10.1103/PhysRevC.78.034001}}.

\bibitem{Diehl:2009ma}
S.~Diehl, S.~Floerchinger, H.~Gies, J.~M. Pawlowski, C.~Wetterich, {Functional
renormalization group approach to the BCS-BEC crossover}, Annalen Phys. 522
(2010) 615–656.
\newblock \href {http://arxiv.org/abs/0907.2193} {\path{arXiv:0907.2193}},
\href {https://doi.org/10.1002/andp.201010458}
{\path{doi:10.1002/andp.201010458}}.

\bibitem{Krippa:2009vu}
B.~Krippa, N.~R. Walet, M.~C. Birse, {Renormalization group, dimer-dimer
scattering, and three-body forces}, Phys. Rev. A 81 (2010) 043628.
\newblock \href {http://arxiv.org/abs/0911.4608} {\path{arXiv:0911.4608}},
\href {https://doi.org/10.1103/PhysRevA.81.043628}
{\path{doi:10.1103/PhysRevA.81.043628}}.

\bibitem{Floerchinger:2009pg}
S.~Floerchinger, M.~M. Scherer, C.~Wetterich, {Modified Fermi-sphere, pairing
gap and critical temperature for the BCS-BEC crossover}, Phys. Rev. A81
(2010) 063619.
\newblock \href {http://arxiv.org/abs/0912.4050} {\path{arXiv:0912.4050}},
\href {https://doi.org/10.1103/PhysRevA.81.063619}
{\path{doi:10.1103/PhysRevA.81.063619}}.

\bibitem{Braun:2011uq}
J.~Braun, S.~Diehl, M.~M. Scherer, {Finite-size and Particle-number Effects in
an Ultracold Fermi Gas at Unitarity} (2011).
\newblock \href {http://arxiv.org/abs/1109.1946} {\path{arXiv:1109.1946}}.

\bibitem{Moroz:2009nm}
S.~Moroz, R.~Schmidt, {Nonrelativistic inverse square potential, scale anomaly,
and complex extension}, Annals Phys. 325 (2010) 491--513.
\newblock \href {http://arxiv.org/abs/0909.3477} {\path{arXiv:0909.3477}},
\href {https://doi.org/10.1016/j.aop.2009.10.002}
{\path{doi:10.1016/j.aop.2009.10.002}}.

\bibitem{Floerchinger209}
S.~Floerchinger, R.~Schmidt, S.~Moroz, C.~Wetterich,
\href{http://dx.doi.org/10.1103/PhysRevA.79.013603}{Functional
renormalization for trion formation in ultracold fermion gases}, Physical
Review A 79~(1) (Jan 2009).
\newblock \href {https://doi.org/10.1103/physreva.79.013603}
{\path{doi:10.1103/physreva.79.013603}}.
\newline\urlprefix\url{http://dx.doi.org/10.1103/PhysRevA.79.013603}

\bibitem{Floerchinger109}
S.~Floerchinger, R.~Schmidt, C.~Wetterich,
\href{http://dx.doi.org/10.1103/PhysRevA.79.053633}{Three-body loss in
lithium from functional renormalization}, Physical Review A 79~(5) (May
2009).
\newblock \href {https://doi.org/10.1103/physreva.79.053633}
{\path{doi:10.1103/physreva.79.053633}}.
\newline\urlprefix\url{http://dx.doi.org/10.1103/PhysRevA.79.053633}

\bibitem{Schmidt:2009kq}
R.~Schmidt, S.~Moroz, {Renormalization group study of the four-body problem},
Phys. Rev. A 81 (2010) 052709.
\newblock \href {http://arxiv.org/abs/0910.4586} {\path{arXiv:0910.4586}},
\href {https://doi.org/10.1103/PhysRevA.81.052709}
{\path{doi:10.1103/PhysRevA.81.052709}}.

\bibitem{Schmidt:2012yn}
R.~Schmidt, S.~P. Rath, W.~Zwerger, {Efimov physics beyond universality}, Eur.
Phys. J. B 85 (2012) 386.
\newblock \href {http://arxiv.org/abs/1201.4310} {\path{arXiv:1201.4310}},
\href {https://doi.org/10.1140/epjb/e2012-30841-3}
{\path{doi:10.1140/epjb/e2012-30841-3}}.

\bibitem{Avila:2013rda}
B.~J. Avila, M.~C. Birse, {Universal behavior of four-boson systems from a
functional-renormalization-group analysis}, Phys. Rev. A 88~(4) (2013)
043613.
\newblock \href {http://arxiv.org/abs/1304.5454} {\path{arXiv:1304.5454}},
\href {https://doi.org/10.1103/PhysRevA.88.043613}
{\path{doi:10.1103/PhysRevA.88.043613}}.

\bibitem{Avila:2015iza}
B.~J. Ávila, M.~C. Birse, {Four-boson bound states from a functional
renormalization group}, Phys. Rev. A 92~(2) (2015) 023601.
\newblock \href {http://arxiv.org/abs/1506.04949} {\path{arXiv:1506.04949}},
\href {https://doi.org/10.1103/PhysRevA.92.023601}
{\path{doi:10.1103/PhysRevA.92.023601}}.

\bibitem{Fukushima:2017csk}
K.~Fukushima, V.~Skokov, {Polyakov loop modeling for hot QCD}, Prog. Part.
Nucl. Phys. 96 (2017) 154--199.
\newblock \href {http://arxiv.org/abs/1705.00718} {\path{arXiv:1705.00718}},
\href {https://doi.org/10.1016/j.ppnp.2017.05.002}
{\path{doi:10.1016/j.ppnp.2017.05.002}}.

\bibitem{Springer:2016cji}
P.~Springer, J.~Braun, S.~Rechenberger, F.~Rennecke, {QCD-inspired
determination of NJL model parameters}, EPJ Web Conf. 137 (2017) 03022.
\newblock \href {http://arxiv.org/abs/1611.06020} {\path{arXiv:1611.06020}},
\href {https://doi.org/10.1051/epjconf/201713703022}
{\path{doi:10.1051/epjconf/201713703022}}.

\bibitem{Braun:2018svj}
J.~Braun, M.~Leonhardt, J.~M. Pawlowski, {Renormalization group consistency and
low-energy effective theories}, SciPost Phys. 6~(5) (2019) 056.
\newblock \href {http://arxiv.org/abs/1806.04432} {\path{arXiv:1806.04432}},
\href {https://doi.org/10.21468/SciPostPhys.6.5.056}
{\path{doi:10.21468/SciPostPhys.6.5.056}}.

\bibitem{Resch:2017vjs}
S.~Resch, F.~Rennecke, B.-J. Schaefer, {Mass sensitivity of the three-flavor
chiral phase transition}, Phys. Rev. D99~(7) (2019) 076005.
\newblock \href {http://arxiv.org/abs/1712.07961} {\path{arXiv:1712.07961}},
\href {https://doi.org/10.1103/PhysRevD.99.076005}
{\path{doi:10.1103/PhysRevD.99.076005}}.

\bibitem{Jungnickel:1995fp}
D.~Jungnickel, C.~Wetterich, {Effective action for the chiral quark-meson
model}, Phys.Rev. D53 (1996) 5142--5175.
\newblock \href {http://arxiv.org/abs/hep-ph/9505267}
{\path{arXiv:hep-ph/9505267}}, \href
{https://doi.org/10.1103/PhysRevD.53.5142}
{\path{doi:10.1103/PhysRevD.53.5142}}.

\bibitem{Berges:1997eu}
J.~Berges, D.~Jungnickel, C.~Wetterich, {Two flavor chiral phase transition
from nonperturbative flow equations}, Phys.Rev. D59 (1999) 034010.
\newblock \href {http://arxiv.org/abs/hep-ph/9705474}
{\path{arXiv:hep-ph/9705474}}, \href
{https://doi.org/10.1103/PhysRevD.59.034010}
{\path{doi:10.1103/PhysRevD.59.034010}}.

\bibitem{Berges:1998sd}
J.~Berges, D.-U. Jungnickel, C.~Wetterich, {The Chiral phase transition at high
baryon density from nonperturbative flow equations}, Eur. Phys. J. C13 (2000)
323--329.
\newblock \href {http://arxiv.org/abs/hep-ph/9811347}
{\path{arXiv:hep-ph/9811347}}, \href {https://doi.org/10.1007/s100520000275}
{\path{doi:10.1007/s100520000275}}.

\bibitem{Berges:1998ha}
J.~Berges, D.~U. Jungnickel, C.~Wetterich, {Quark and nuclear matter in the
linear chiral meson model}, Int. J. Mod. Phys. A18 (2003) 3189--3220.
\newblock \href {http://arxiv.org/abs/hep-ph/9811387}
{\path{arXiv:hep-ph/9811387}}, \href
{https://doi.org/10.1142/S0217751X03014034}
{\path{doi:10.1142/S0217751X03014034}}.

\bibitem{Papp:1999he}
G.~Papp, B.-J. Schaefer, H.~Pirner, J.~Wambach, {On the convergence of the
expansion of renormalization group flow equation}, Phys.Rev. D61 (2000)
096002.
\newblock \href {http://arxiv.org/abs/hep-ph/9909246}
{\path{arXiv:hep-ph/9909246}}, \href
{https://doi.org/10.1103/PhysRevD.61.096002}
{\path{doi:10.1103/PhysRevD.61.096002}}.

\bibitem{Bergerhoff:1999hr}
B.~Bergerhoff, J.~Manus, J.~Reingruber, {The Thermal renormalization group for
fermions, universality, and the chiral phase transition}, Phys.Rev. D61
(2000) 125005.
\newblock \href {http://arxiv.org/abs/hep-ph/9912474}
{\path{arXiv:hep-ph/9912474}}, \href
{https://doi.org/10.1103/PhysRevD.61.125005}
{\path{doi:10.1103/PhysRevD.61.125005}}.

\bibitem{Bohr:2000gp}
O.~Bohr, B.-J. Schaefer, J.~Wambach, {Renormalization group flow equations and
the phase transition in O(N) models}, Int.J.Mod.Phys. A16 (2001) 3823--3852.
\newblock \href {http://arxiv.org/abs/hep-ph/0007098}
{\path{arXiv:hep-ph/0007098}}, \href
{https://doi.org/10.1142/S0217751X0100502X}
{\path{doi:10.1142/S0217751X0100502X}}.

\bibitem{Schaefer:2001cn}
B.-J. Schaefer, O.~Bohr, J.~Wambach, {Finite temperature gluon condensate with
renormalization group flow equations}, Phys.Rev. D65 (2002) 105008.
\newblock \href {http://arxiv.org/abs/hep-th/0112087}
{\path{arXiv:hep-th/0112087}}, \href
{https://doi.org/10.1103/PhysRevD.65.105008}
{\path{doi:10.1103/PhysRevD.65.105008}}.

\bibitem{Braun:2009si}
J.~Braun, {Thermodynamics of QCD low-energy models and the derivative expansion
of the effective action}, Phys.Rev. D81 (2010) 016008.
\newblock \href {http://arxiv.org/abs/0908.1543} {\path{arXiv:0908.1543}},
\href {https://doi.org/10.1103/PhysRevD.81.016008}
{\path{doi:10.1103/PhysRevD.81.016008}}.

\bibitem{Fukushima:2010ji}
K.~Fukushima, K.~Kamikado, B.~Klein, {Second-order and Fluctuation-induced
First-order Phase Transitions with Functional Renormalization Group
Equations}, Phys. Rev. D83 (2011) 116005.
\newblock \href {http://arxiv.org/abs/1010.6226} {\path{arXiv:1010.6226}},
\href {https://doi.org/10.1103/PhysRevD.83.116005}
{\path{doi:10.1103/PhysRevD.83.116005}}.

\bibitem{Schaefer:1998my}
B.-J. Schaefer, H.~Pirner, {Nonperturbative flow equations at finite
temperature} (1998).
\newblock \href {http://arxiv.org/abs/nucl-th/9801067}
{\path{arXiv:nucl-th/9801067}}.

\bibitem{Schaefer:1999em}
B.-J. Schaefer, H.-J. Pirner, {Renormalization group flow and equation of state
of quarks and mesons}, Nucl.Phys. A660 (1999) 439--474.
\newblock \href {http://arxiv.org/abs/nucl-th/9903003}
{\path{arXiv:nucl-th/9903003}}, \href
{https://doi.org/10.1016/S0375-9474(99)00409-1}
{\path{doi:10.1016/S0375-9474(99)00409-1}}.

\bibitem{Braun:2003ii}
J.~Braun, K.~Schwenzer, H.-J. Pirner, {Linking the quark meson model with QCD
at high temperature}, Phys.Rev. D70 (2004) 085016.
\newblock \href {http://arxiv.org/abs/hep-ph/0312277}
{\path{arXiv:hep-ph/0312277}}, \href
{https://doi.org/10.1103/PhysRevD.70.085016}
{\path{doi:10.1103/PhysRevD.70.085016}}.

\bibitem{Schaefer:2004en}
B.-J. Schaefer, J.~Wambach, {The Phase diagram of the quark meson model},
Nucl.Phys. A757 (2005) 479--492.
\newblock \href {http://arxiv.org/abs/nucl-th/0403039}
{\path{arXiv:nucl-th/0403039}}, \href
{https://doi.org/10.1016/j.nuclphysa.2005.04.012}
{\path{doi:10.1016/j.nuclphysa.2005.04.012}}.

\bibitem{Braun:2004yk}
J.~Braun, B.~Klein, H.-J. Pirner, {Volume dependence of the pion mass in the
quark-meson-model}, Phys.Rev. D71 (2005) 014032.
\newblock \href {http://arxiv.org/abs/hep-ph/0408116}
{\path{arXiv:hep-ph/0408116}}, \href
{https://doi.org/10.1103/PhysRevD.71.014032}
{\path{doi:10.1103/PhysRevD.71.014032}}.

\bibitem{Braun:2005gy}
J.~Braun, B.~Klein, H.~Pirner, {Influence of quark boundary conditions on the
pion mass in finite volume}, Phys.Rev. D72 (2005) 034017.
\newblock \href {http://arxiv.org/abs/hep-ph/0504127}
{\path{arXiv:hep-ph/0504127}}, \href
{https://doi.org/10.1103/PhysRevD.72.034017}
{\path{doi:10.1103/PhysRevD.72.034017}}.

\bibitem{Braun:2006vd}
J.~Braun, H.-J. Pirner, {Effects of the running of the QCD coupling on the
energy loss in the quark-gluon plasma}, Phys. Rev. D75 (2007) 054031.
\newblock \href {http://arxiv.org/abs/hep-ph/0610331}
{\path{arXiv:hep-ph/0610331}}, \href
{https://doi.org/10.1103/PhysRevD.75.054031}
{\path{doi:10.1103/PhysRevD.75.054031}}.

\bibitem{Braun:2005fj}
J.~Braun, B.~Klein, H.-J. Pirner, A.~Rezaeian, {Volume and quark mass
dependence of the chiral phase transition}, Phys.Rev. D73 (2006) 074010.
\newblock \href {http://arxiv.org/abs/hep-ph/0512274}
{\path{arXiv:hep-ph/0512274}}, \href
{https://doi.org/10.1103/PhysRevD.73.074010}
{\path{doi:10.1103/PhysRevD.73.074010}}.

\bibitem{Braun:2007td}
J.~Braun, B.~Klein, {Scaling functions for the O(4)-model in d=3 dimensions},
Phys. Rev. D77 (2008) 096008.
\newblock \href {http://arxiv.org/abs/0712.3574} {\path{arXiv:0712.3574}},
\href {https://doi.org/10.1103/PhysRevD.77.096008}
{\path{doi:10.1103/PhysRevD.77.096008}}.

\bibitem{Braun:2008sg}
J.~Braun, B.~Klein, {Finite-Size Scaling behavior in the O(4)-Model},
Eur.Phys.J. C63 (2009) 443--460.
\newblock \href {http://arxiv.org/abs/0810.0857} {\path{arXiv:0810.0857}},
\href {https://doi.org/10.1140/epjc/s10052-009-1098-8}
{\path{doi:10.1140/epjc/s10052-009-1098-8}}.

\bibitem{Braun:2010vd}
J.~Braun, B.~Klein, P.~Piasecki, {On the scaling behavior of the chiral phase
transition in QCD in finite and infinite volume}, Eur.Phys.J. C71 (2011)
1576.
\newblock \href {http://arxiv.org/abs/1008.2155} {\path{arXiv:1008.2155}},
\href {https://doi.org/10.1140/epjc/s10052-011-1576-7}
{\path{doi:10.1140/epjc/s10052-011-1576-7}}.

\bibitem{Braun:2011iz}
J.~Braun, B.~Klein, B.-J. Schaefer, {On the Phase Structure of QCD in a Finite
Volume}, Phys.Lett. B713 (2012) 216--223.
\newblock \href {http://arxiv.org/abs/1110.0849} {\path{arXiv:1110.0849}},
\href {https://doi.org/10.1016/j.physletb.2012.05.053}
{\path{doi:10.1016/j.physletb.2012.05.053}}.

\bibitem{Jiang:2012wm}
Y.~Jiang, P.~Zhuang, {Functional Renormalization for Chiral and $U_A(1)$
Symmetries at Finite Temperature}, Phys. Rev. D 86 (2012) 105016.
\newblock \href {http://arxiv.org/abs/1209.0507} {\path{arXiv:1209.0507}},
\href {https://doi.org/10.1103/PhysRevD.86.105016}
{\path{doi:10.1103/PhysRevD.86.105016}}.

\bibitem{Kamikado:2012cp}
K.~Kamikado, T.~Kunihiro, K.~Morita, A.~Ohnishi, {Functional Renormalization
Group Study of Phonon Mode Effects on Chiral Critical Point}, PTEP 2013
(2013) 053D01.
\newblock \href {http://arxiv.org/abs/1210.8347} {\path{arXiv:1210.8347}},
\href {https://doi.org/10.1093/ptep/ptt021} {\path{doi:10.1093/ptep/ptt021}}.

\bibitem{Tripolt:2013zfa}
R.-A. Tripolt, J.~Braun, B.~Klein, B.-J. Schaefer, {Effect of fluctuations on
the QCD critical point in a finite volume}, Phys.Rev. D90~(5) (2014) 054012.
\newblock \href {http://arxiv.org/abs/1308.0164} {\path{arXiv:1308.0164}},
\href {https://doi.org/10.1103/PhysRevD.90.054012}
{\path{doi:10.1103/PhysRevD.90.054012}}.

\bibitem{Mitter:2013fxa}
M.~Mitter, B.-J. Schaefer, {Fluctuations and the axial anomaly with three quark
flavors}, Phys.Rev. D89 (2014) 054027.
\newblock \href {http://arxiv.org/abs/1308.3176} {\path{arXiv:1308.3176}},
\href {https://doi.org/10.1103/PhysRevD.89.054027}
{\path{doi:10.1103/PhysRevD.89.054027}}.

\bibitem{Drews:2013hha}
M.~Drews, T.~Hell, B.~Klein, W.~Weise, {Thermodynamic phases and mesonic
fluctuations in a chiral nucleon-meson model}, Phys. Rev. D 88~(9) (2013)
096011.
\newblock \href {http://arxiv.org/abs/1308.5596} {\path{arXiv:1308.5596}},
\href {https://doi.org/10.1103/PhysRevD.88.096011}
{\path{doi:10.1103/PhysRevD.88.096011}}.

\bibitem{Pawlowski:2014zaa}
J.~M. Pawlowski, F.~Rennecke, {Higher order quark-mesonic scattering processes
and the phase structure of QCD}, Phys. Rev. D90~(7) (2014) 076002.
\newblock \href {http://arxiv.org/abs/1403.1179} {\path{arXiv:1403.1179}},
\href {https://doi.org/10.1103/PhysRevD.90.076002}
{\path{doi:10.1103/PhysRevD.90.076002}}.

\bibitem{Aoki:2015hsa}
K.-I. Aoki, M.~Yamada, {The RG flow of Nambu–Jona-Lasinio model at finite
temperature and density}, Int. J. Mod. Phys. A30~(27) (2015) 1550180.
\newblock \href {http://arxiv.org/abs/1504.00749} {\path{arXiv:1504.00749}},
\href {https://doi.org/10.1142/S0217751X15501808}
{\path{doi:10.1142/S0217751X15501808}}.

\bibitem{Springer:2015kxa}
P.~Springer, B.~Klein, {O(2)-scaling in finite and infinite volume}, Eur. Phys.
J. C 75~(10) (2015) 468.
\newblock \href {http://arxiv.org/abs/1506.00909} {\path{arXiv:1506.00909}},
\href {https://doi.org/10.1140/epjc/s10052-015-3667-3}
{\path{doi:10.1140/epjc/s10052-015-3667-3}}.

\bibitem{Aoki:2015mqa}
K.-I. Aoki, H.~Uoi, M.~Yamada, {Functional renormalization group study of the
Nambu–Jona-Lasinio model at finite temperature and density in an external
magnetic field}, Phys. Lett. B753 (2016) 580--585.
\newblock \href {http://arxiv.org/abs/1507.02527} {\path{arXiv:1507.02527}},
\href {https://doi.org/10.1016/j.physletb.2015.12.063}
{\path{doi:10.1016/j.physletb.2015.12.063}}.

\bibitem{Wang:2015bky}
Z.~Wang, P.~Zhuang, {Critical Behavior and Dimension Crossover of Pion
Superfluidity}, Phys. Rev. D94~(5) (2016) 056012.
\newblock \href {http://arxiv.org/abs/1511.05279} {\path{arXiv:1511.05279}},
\href {https://doi.org/10.1103/PhysRevD.94.056012}
{\path{doi:10.1103/PhysRevD.94.056012}}.

\bibitem{Jiang:2015xqz}
Y.~Jiang, T.~Xia, P.~Zhuang, {Topological Susceptibility in Three-Flavor Quark
Meson Model at Finite Temperature}, Phys. Rev. D 93~(7) (2016) 074006.
\newblock \href {http://arxiv.org/abs/1511.06466} {\path{arXiv:1511.06466}},
\href {https://doi.org/10.1103/PhysRevD.93.074006}
{\path{doi:10.1103/PhysRevD.93.074006}}.

\bibitem{Aoki:2017rjl}
K.-I. Aoki, S.-I. Kumamoto, M.~Yamada, {Phase structure of NJL model with weak
renormalization group}, Nucl. Phys. B931 (2018) 105--131.
\newblock \href {http://arxiv.org/abs/1705.03273} {\path{arXiv:1705.03273}},
\href {https://doi.org/10.1016/j.nuclphysb.2018.04.005}
{\path{doi:10.1016/j.nuclphysb.2018.04.005}}.

\bibitem{Zhang:2017icm}
H.~Zhang, D.~Hou, T.~Kojo, B.~Qin, {Functional renormalization group study of
the quark-meson model with $\omega$ meson}, Phys. Rev. D96~(11) (2017)
114029.
\newblock \href {http://arxiv.org/abs/1709.05654} {\path{arXiv:1709.05654}},
\href {https://doi.org/10.1103/PhysRevD.96.114029}
{\path{doi:10.1103/PhysRevD.96.114029}}.

\bibitem{Tripolt:2017zgc}
R.-A. Tripolt, B.-J. Schaefer, L.~von Smekal, J.~Wambach, {Low-temperature
behavior of the quark-meson model}, Phys. Rev. D97~(3) (2018) 034022.
\newblock \href {http://arxiv.org/abs/1709.05991} {\path{arXiv:1709.05991}},
\href {https://doi.org/10.1103/PhysRevD.97.034022}
{\path{doi:10.1103/PhysRevD.97.034022}}.

\bibitem{Braun:2017srn}
J.~Braun, M.~Leonhardt, M.~Pospiech, {Fierz-complete NJL model study: Fixed
points and phase structure at finite temperature and density}, Phys. Rev.
D96~(7) (2017) 076003.
\newblock \href {http://arxiv.org/abs/1705.00074} {\path{arXiv:1705.00074}},
\href {https://doi.org/10.1103/PhysRevD.96.076003}
{\path{doi:10.1103/PhysRevD.96.076003}}.

\bibitem{Braun:2018bik}
J.~Braun, M.~Leonhardt, M.~Pospiech, {Fierz-complete NJL model study. II.
Toward the fixed-point and phase structure of hot and dense two-flavor QCD},
Phys. Rev. D97~(7) (2018) 076010.
\newblock \href {http://arxiv.org/abs/1801.08338} {\path{arXiv:1801.08338}},
\href {https://doi.org/10.1103/PhysRevD.97.076010}
{\path{doi:10.1103/PhysRevD.97.076010}}.

\bibitem{Yin:2019ebz}
S.~Yin, R.~Wen, W.-j. Fu, {Mesonic dynamics and the QCD phase transition},
Phys. Rev. D 100~(9) (2019) 094029.
\newblock \href {http://arxiv.org/abs/1907.10262} {\path{arXiv:1907.10262}},
\href {https://doi.org/10.1103/PhysRevD.100.094029}
{\path{doi:10.1103/PhysRevD.100.094029}}.

\bibitem{Otto:2019zjy}
K.~Otto, M.~Oertel, B.-J. Schaefer, {Hybrid and quark star matter based on a
non-perturbative equation of state} (10 2019).
\newblock \href {http://arxiv.org/abs/1910.11929} {\path{arXiv:1910.11929}}.

\bibitem{CamaraPereira:2020xla}
R.~Camara~Pereira, R.~Stiele, P.~Costa, {Functional renormalization group study
of the critical region of the quark-meson model with vector interactions} (3
2020).
\newblock \href {http://arxiv.org/abs/2003.12829} {\path{arXiv:2003.12829}}.

\bibitem{Otto:2020hoz}
K.~Otto, M.~Oertel, B.-J. Schaefer, {Nonperturbative quark matter equations of
state with vector interactions} (7 2020).
\newblock \href {http://arxiv.org/abs/2007.07394} {\path{arXiv:2007.07394}}.

\bibitem{Braun:2020bhy}
J.~Braun, T.~D\"ornfeld, B.~Schallmo, S.~T\"opfel, {Renormalization Group
Studies of Dense Relativistic Systems} (8 2020).
\newblock \href {http://arxiv.org/abs/2008.05978} {\path{arXiv:2008.05978}}.

\bibitem{Connelly:2020pno}
A.~Connelly, G.~Johnson, S.~Mukherjee, V.~Skokov, {Universality driven analytic
structure of QCD crossover: radius of convergence and QCD critical point},
in: {28th International Conference on Ultrarelativistic Nucleus-Nucleus
Collisions}, 2020.
\newblock \href {http://arxiv.org/abs/2004.05095} {\path{arXiv:2004.05095}}.

\bibitem{Connelly:2020gwa}
A.~Connelly, G.~Johnson, F.~Rennecke, V.~Skokov, {Universal location of the
Yang-Lee edge singularity in O(N) theories}, Phys. Rev. Lett. 125~(19) (2020)
191602.
\newblock \href {http://arxiv.org/abs/2006.12541} {\path{arXiv:2006.12541}},
\href {https://doi.org/10.1103/PhysRevLett.125.191602}
{\path{doi:10.1103/PhysRevLett.125.191602}}.

\bibitem{Boettcher:2014tfa}
I.~Boettcher, J.~Braun, T.~Herbst, J.~Pawlowski, D.~Roscher, C.~Wetterich,
{Phase structure of spin-imbalanced unitary Fermi gases}, Phys. Rev. A 91~(1)
(2015) 013610.
\newblock \href {http://arxiv.org/abs/1409.5070} {\path{arXiv:1409.5070}},
\href {https://doi.org/10.1103/PhysRevA.91.013610}
{\path{doi:10.1103/PhysRevA.91.013610}}.

\bibitem{Boettcher:2014xna}
I.~Boettcher, T.~Herbst, J.~Pawlowski, N.~Strodthoff, L.~von Smekal,
C.~Wetterich, {Sarma phase in relativistic and non-relativistic systems},
Phys. Lett. B 742 (2015) 86--93.
\newblock \href {http://arxiv.org/abs/1409.5232} {\path{arXiv:1409.5232}},
\href {https://doi.org/10.1016/j.physletb.2015.01.014}
{\path{doi:10.1016/j.physletb.2015.01.014}}.

\bibitem{Skokov:2010wb}
V.~Skokov, B.~Stokic, B.~Friman, K.~Redlich, {Meson fluctuations and
thermodynamics of the Polyakov loop extended quark-meson model}, Phys.Rev.
C82 (2010) 015206.
\newblock \href {http://arxiv.org/abs/1004.2665} {\path{arXiv:1004.2665}},
\href {https://doi.org/10.1103/PhysRevC.82.015206}
{\path{doi:10.1103/PhysRevC.82.015206}}.

\bibitem{Herbst:2010rf}
T.~K. Herbst, J.~M. Pawlowski, B.-J. Schaefer, {The phase structure of the
Polyakov--quark-meson model beyond mean field}, Phys.Lett. B696 (2011)
58--67.
\newblock \href {http://arxiv.org/abs/1008.0081} {\path{arXiv:1008.0081}},
\href {https://doi.org/10.1016/j.physletb.2010.12.003}
{\path{doi:10.1016/j.physletb.2010.12.003}}.

\bibitem{Braun:2011fw}
J.~Braun, A.~Janot, {Dynamical Locking of the Chiral and the Deconfinement
Phase Transition in QCD}, Phys.Rev. D84 (2011) 114022.
\newblock \href {http://arxiv.org/abs/1102.4841} {\path{arXiv:1102.4841}},
\href {https://doi.org/10.1103/PhysRevD.84.114022}
{\path{doi:10.1103/PhysRevD.84.114022}}.

\bibitem{Morita:2011jva}
K.~Morita, V.~Skokov, B.~Friman, K.~Redlich, {Role of mesonic fluctuations in
the Polyakov loop extended quark-meson model at imaginary chemical
potential}, Phys.Rev. D84 (2011) 074020.
\newblock \href {http://arxiv.org/abs/1108.0735} {\path{arXiv:1108.0735}},
\href {https://doi.org/10.1103/PhysRevD.84.074020}
{\path{doi:10.1103/PhysRevD.84.074020}}.

\bibitem{Braun:2012zq}
J.~Braun, T.~K. Herbst, {On the Relation of the Deconfinement and the Chiral
Phase Transition in Gauge Theories with Fundamental and Adjoint Matter}
(2012).
\newblock \href {http://arxiv.org/abs/1205.0779} {\path{arXiv:1205.0779}}.

\bibitem{Herbst:2013ail}
T.~K. Herbst, J.~M. Pawlowski, B.-J. Schaefer, {Phase structure and
thermodynamics of QCD}, Phys.Rev. D88~(1) (2013) 014007.
\newblock \href {http://arxiv.org/abs/1302.1426} {\path{arXiv:1302.1426}},
\href {https://doi.org/10.1103/PhysRevD.88.014007}
{\path{doi:10.1103/PhysRevD.88.014007}}.

\bibitem{Herbst:2013ufa}
T.~K. Herbst, M.~Mitter, J.~M. Pawlowski, B.-J. Schaefer, R.~Stiele,
{Thermodynamics of QCD at vanishing density}, Phys.Lett. B731 (2014)
248--256.
\newblock \href {http://arxiv.org/abs/1308.3621} {\path{arXiv:1308.3621}},
\href {https://doi.org/10.1016/j.physletb.2014.02.045}
{\path{doi:10.1016/j.physletb.2014.02.045}}.

\bibitem{Haas:2013qwp}
L.~M. Haas, R.~Stiele, J.~Braun, J.~M. Pawlowski, J.~Schaffner-Bielich,
{Improved Polyakov-loop potential for effective models from functional
calculations}, Phys.Rev. D87 (2013) 076004.
\newblock \href {http://arxiv.org/abs/1302.1993} {\path{arXiv:1302.1993}},
\href {https://doi.org/10.1103/PhysRevD.87.076004}
{\path{doi:10.1103/PhysRevD.87.076004}}.

\bibitem{Fu:2018wxq}
W.-J. Fu, {Chiral criticality and glue dynamics}, Chin. Phys. C 43~(7) (2019)
074101.
\newblock \href {http://arxiv.org/abs/1801.03213} {\path{arXiv:1801.03213}},
\href {https://doi.org/10.1088/1674-1137/43/7/074101}
{\path{doi:10.1088/1674-1137/43/7/074101}}.

\bibitem{Strodthoff:2011tz}
N.~Strodthoff, B.-J. Schaefer, L.~von Smekal, {Quark-meson-diquark model for
two-color QCD}, Phys.Rev. D85 (2012) 074007.
\newblock \href {http://arxiv.org/abs/1112.5401} {\path{arXiv:1112.5401}},
\href {https://doi.org/10.1103/PhysRevD.85.074007}
{\path{doi:10.1103/PhysRevD.85.074007}}.

\bibitem{Kamikado:2012bt}
K.~Kamikado, N.~Strodthoff, L.~von Smekal, J.~Wambach, {Fluctuations in the
quark-meson model for QCD with isospin chemical potential}, Phys.Lett. B718
(2013) 1044--1053.
\newblock \href {http://arxiv.org/abs/1207.0400} {\path{arXiv:1207.0400}},
\href {https://doi.org/10.1016/j.physletb.2012.11.055}
{\path{doi:10.1016/j.physletb.2012.11.055}}.

\bibitem{Strodthoff:2013cua}
N.~Strodthoff, L.~von Smekal, {Polyakov-Quark-Meson-Diquark Model for two-color
QCD}, Phys. Lett. B731 (2014) 350--357.
\newblock \href {http://arxiv.org/abs/1306.2897} {\path{arXiv:1306.2897}},
\href {https://doi.org/10.1016/j.physletb.2014.03.008}
{\path{doi:10.1016/j.physletb.2014.03.008}}.

\bibitem{Khan:2015puu}
N.~Khan, J.~M. Pawlowski, F.~Rennecke, M.~M. Scherer, {The Phase Diagram of
QC2D from Functional Methods} (2015).
\newblock \href {http://arxiv.org/abs/1512.03673} {\path{arXiv:1512.03673}}.

\bibitem{Schaefer:2006ds}
B.-J. Schaefer, J.~Wambach, {Susceptibilities near the QCD (tri)critical
point}, Phys.Rev. D75 (2007) 085015.
\newblock \href {http://arxiv.org/abs/hep-ph/0603256}
{\path{arXiv:hep-ph/0603256}}, \href
{https://doi.org/10.1103/PhysRevD.75.085015}
{\path{doi:10.1103/PhysRevD.75.085015}}.

\bibitem{Nakano:2009ps}
E.~Nakano, B.-J. Schaefer, B.~Stokic, B.~Friman, K.~Redlich, {Fluctuations and
isentropes near the chiral critical endpoint}, Phys.Lett. B682 (2010)
401--407.
\newblock \href {http://arxiv.org/abs/0907.1344} {\path{arXiv:0907.1344}},
\href {https://doi.org/10.1016/j.physletb.2009.11.027}
{\path{doi:10.1016/j.physletb.2009.11.027}}.

\bibitem{Skokov:2010uh}
V.~Skokov, B.~Friman, K.~Redlich, {Quark number fluctuations in the Polyakov
loop-extended quark-meson model at finite baryon density}, Phys.Rev. C83
(2011) 054904.
\newblock \href {http://arxiv.org/abs/1008.4570} {\path{arXiv:1008.4570}},
\href {https://doi.org/10.1103/PhysRevC.83.054904}
{\path{doi:10.1103/PhysRevC.83.054904}}.

\bibitem{Skokov:2011rq}
V.~Skokov, B.~Friman, K.~Redlich, {Non-perturbative dynamics and charge
fluctuations in effective chiral models}, Phys. Lett. B708 (2012) 179--185.
\newblock \href {http://arxiv.org/abs/1108.3231} {\path{arXiv:1108.3231}},
\href {https://doi.org/10.1016/j.physletb.2012.01.022}
{\path{doi:10.1016/j.physletb.2012.01.022}}.

\bibitem{Skokov:2012ds}
V.~Skokov, B.~Friman, K.~Redlich, {Volume Fluctuations and Higher Order
Cumulants of the Net Baryon Number}, Phys. Rev. C88 (2013) 034911.
\newblock \href {http://arxiv.org/abs/1205.4756} {\path{arXiv:1205.4756}},
\href {https://doi.org/10.1103/PhysRevC.88.034911}
{\path{doi:10.1103/PhysRevC.88.034911}}.

\bibitem{Morita:2013tu}
K.~Morita, B.~Friman, K.~Redlich, V.~Skokov, {Net quark number probability
distribution near the chiral crossover transition}, Phys.Rev. C88~(3) (2013)
034903.
\newblock \href {http://arxiv.org/abs/1301.2873} {\path{arXiv:1301.2873}},
\href {https://doi.org/10.1103/PhysRevC.88.034903}
{\path{doi:10.1103/PhysRevC.88.034903}}.

\bibitem{Fu:2015naa}
W.-J. Fu, J.~M. Pawlowski, {On the relevance of matter and glue dynamics for
baryon number fluctuations}, Phys. Rev. D92~(11) (2015) 116006.
\newblock \href {http://arxiv.org/abs/1508.06504} {\path{arXiv:1508.06504}},
\href {https://doi.org/10.1103/PhysRevD.92.116006}
{\path{doi:10.1103/PhysRevD.92.116006}}.

\bibitem{Fu:2015amv}
W.-J. Fu, J.~M. Pawlowski, {Correlating the skewness and kurtosis of baryon
number distributions}, Phys. Rev. D93~(9) (2016) 091501.
\newblock \href {http://arxiv.org/abs/1512.08461} {\path{arXiv:1512.08461}},
\href {https://doi.org/10.1103/PhysRevD.93.091501}
{\path{doi:10.1103/PhysRevD.93.091501}}.

\bibitem{Fu:2016tey}
W.-j. Fu, J.~M. Pawlowski, F.~Rennecke, B.-J. Schaefer, {Baryon number
fluctuations at finite temperature and density}, Phys. Rev. D94~(11) (2016)
116020.
\newblock \href {http://arxiv.org/abs/1608.04302} {\path{arXiv:1608.04302}},
\href {https://doi.org/10.1103/PhysRevD.94.116020}
{\path{doi:10.1103/PhysRevD.94.116020}}.

\bibitem{Rennecke:2016tkm}
F.~Rennecke, B.-J. Schaefer, {Fluctuation-induced modifications of the phase
structure in (2+1)-flavor QCD}, Phys. Rev. D96~(1) (2017) 016009.
\newblock \href {http://arxiv.org/abs/1610.08748} {\path{arXiv:1610.08748}},
\href {https://doi.org/10.1103/PhysRevD.96.016009}
{\path{doi:10.1103/PhysRevD.96.016009}}.

\bibitem{Almasi:2016zqf}
G.~Almasi, R.~Pisarski, V.~Skokov, {Volume dependence of baryon number
cumulants and their ratios}, Phys. Rev. D95~(5) (2017) 056015.
\newblock \href {http://arxiv.org/abs/1612.04416} {\path{arXiv:1612.04416}},
\href {https://doi.org/10.1103/PhysRevD.95.056015}
{\path{doi:10.1103/PhysRevD.95.056015}}.

\bibitem{Almasi:2017bhq}
G.~A. Almasi, B.~Friman, K.~Redlich, {Baryon number fluctuations in chiral
effective models and their phenomenological implications}, Phys. Rev. D96~(1)
(2017) 014027.
\newblock \href {http://arxiv.org/abs/1703.05947} {\path{arXiv:1703.05947}},
\href {https://doi.org/10.1103/PhysRevD.96.014027}
{\path{doi:10.1103/PhysRevD.96.014027}}.

\bibitem{Sun:2018ozp}
K.-x. Sun, R.~Wen, W.-j. Fu, {Baryon number probability distribution at finite
temperature}, Phys. Rev. D 98~(7) (2018) 074028.
\newblock \href {http://arxiv.org/abs/1805.12025} {\path{arXiv:1805.12025}},
\href {https://doi.org/10.1103/PhysRevD.98.074028}
{\path{doi:10.1103/PhysRevD.98.074028}}.

\bibitem{Fu:2018swz}
W.-j. Fu, J.~M. Pawlowski, F.~Rennecke, {Strangeness neutrality and
baryon-strangeness correlations}, Phys. Rev. D 100~(11) (2019) 111501.
\newblock \href {http://arxiv.org/abs/1809.01594} {\path{arXiv:1809.01594}},
\href {https://doi.org/10.1103/PhysRevD.100.111501}
{\path{doi:10.1103/PhysRevD.100.111501}}.

\bibitem{Fu:2018qsk}
W.-j. Fu, J.~M. Pawlowski, F.~Rennecke, {Strangeness Neutrality and QCD
Thermodynamics} (8 2018).
\newblock \href {http://arxiv.org/abs/1808.00410} {\path{arXiv:1808.00410}},
\href {https://doi.org/10.21468/SciPostPhysCore.2.1.002}
{\path{doi:10.21468/SciPostPhysCore.2.1.002}}.

\bibitem{Wen:2018nkn}
R.~Wen, C.~Huang, W.-J. Fu, {Baryon number fluctuations in the 2+1 flavor low
energy effective model}, Phys. Rev. D 99~(9) (2019) 094019.
\newblock \href {http://arxiv.org/abs/1809.04233} {\path{arXiv:1809.04233}},
\href {https://doi.org/10.1103/PhysRevD.99.094019}
{\path{doi:10.1103/PhysRevD.99.094019}}.

\bibitem{Wen:2019ruz}
R.~Wen, W.-j. Fu, {Correlations of conserved charges and QCD phase structure}
(2019).
\newblock \href {http://arxiv.org/abs/1909.12564} {\path{arXiv:1909.12564}}.

\bibitem{Skokov:2011ib}
V.~Skokov, {Phase diagram in an external magnetic field beyond a mean-field
approximation}, Phys. Rev. D85 (2012) 034026.
\newblock \href {http://arxiv.org/abs/1112.5137} {\path{arXiv:1112.5137}},
\href {https://doi.org/10.1103/PhysRevD.85.034026}
{\path{doi:10.1103/PhysRevD.85.034026}}.

\bibitem{Fukushima:2012xw}
K.~Fukushima, J.~M. Pawlowski, {Magnetic catalysis in hot and dense quark
matter and quantum fluctuations}, Phys. Rev. D86 (2012) 076013.
\newblock \href {http://arxiv.org/abs/1203.4330} {\path{arXiv:1203.4330}},
\href {https://doi.org/10.1103/PhysRevD.86.076013}
{\path{doi:10.1103/PhysRevD.86.076013}}.

\bibitem{Braun:2014fua}
J.~Braun, W.~A. Mian, S.~Rechenberger, {Delayed Magnetic Catalysis}, Phys.
Lett. B755 (2016) 265--269.
\newblock \href {http://arxiv.org/abs/1412.6025} {\path{arXiv:1412.6025}},
\href {https://doi.org/10.1016/j.physletb.2016.02.017}
{\path{doi:10.1016/j.physletb.2016.02.017}}.

\bibitem{Mueller:2015fka}
N.~Mueller, J.~M. Pawlowski, {Magnetic catalysis and inverse magnetic catalysis
in QCD}, Phys. Rev. D91~(11) (2015) 116010.
\newblock \href {http://arxiv.org/abs/1502.08011} {\path{arXiv:1502.08011}},
\href {https://doi.org/10.1103/PhysRevD.91.116010}
{\path{doi:10.1103/PhysRevD.91.116010}}.

\bibitem{Fu:2017vvg}
W.-j. Fu, Y.-x. Liu, {Four-fermion interactions and the chiral symmetry
breaking in an external magnetic field}, Phys. Rev. D 96~(7) (2017) 074019.
\newblock \href {http://arxiv.org/abs/1705.09841} {\path{arXiv:1705.09841}},
\href {https://doi.org/10.1103/PhysRevD.96.074019}
{\path{doi:10.1103/PhysRevD.96.074019}}.

\bibitem{Li:2019nzj}
X.~Li, W.-J. Fu, Y.-X. Liu, {Thermodynamics of 2+1 Flavor Polyakov-Loop
Quark-Meson Model under External Magnetic Field}, Phys. Rev. D 99~(7) (2019)
074029.
\newblock \href {http://arxiv.org/abs/1902.03866} {\path{arXiv:1902.03866}},
\href {https://doi.org/10.1103/PhysRevD.99.074029}
{\path{doi:10.1103/PhysRevD.99.074029}}.

\bibitem{Tripolt:2018jre}
R.-A. Tripolt, C.~Jung, N.~Tanji, L.~von Smekal, J.~Wambach, {In-medium
spectral functions and dilepton rates with the Functional Renormalization
Group}, Nucl. Phys. A982 (2019) 775--778.
\newblock \href {http://arxiv.org/abs/1807.04952} {\path{arXiv:1807.04952}},
\href {https://doi.org/10.1016/j.nuclphysa.2018.08.017}
{\path{doi:10.1016/j.nuclphysa.2018.08.017}}.

\bibitem{Tripolt:2020dac}
R.-A. Tripolt, {Electromagnetic and weak probes: theory}, in: {28th
International Conference on Ultrarelativistic Nucleus-Nucleus Collisions},
2020.
\newblock \href {http://arxiv.org/abs/2001.11232} {\path{arXiv:2001.11232}}.

\bibitem{Bluhm:2018qkf}
M.~Bluhm, Y.~Jiang, M.~Nahrgang, J.~M. Pawlowski, F.~Rennecke, N.~Wink,
{Time-evolution of fluctuations as signal of the phase transition dynamics in
a QCD-assisted transport approach}, Nucl. Phys. A982 (2019) 871--874.
\newblock \href {http://arxiv.org/abs/1808.01377} {\path{arXiv:1808.01377}},
\href {https://doi.org/10.1016/j.nuclphysa.2018.09.058}
{\path{doi:10.1016/j.nuclphysa.2018.09.058}}.

\bibitem{Jung:2016yxl}
C.~Jung, F.~Rennecke, R.-A. Tripolt, L.~von Smekal, J.~Wambach, {In-Medium
Spectral Functions of Vector- and Axial-Vector Mesons from the Functional
Renormalization Group}, Phys. Rev. D95~(3) (2017) 036020.
\newblock \href {http://arxiv.org/abs/1610.08754} {\path{arXiv:1610.08754}},
\href {https://doi.org/10.1103/PhysRevD.95.036020}
{\path{doi:10.1103/PhysRevD.95.036020}}.

\bibitem{Jung:2019nnr}
C.~Jung, L.~von Smekal, {Fluctuating vector mesons in analytically continued
functional RG flow equations}, Phys. Rev. D 100~(11) (2019) 116009.
\newblock \href {http://arxiv.org/abs/1909.13712} {\path{arXiv:1909.13712}},
\href {https://doi.org/10.1103/PhysRevD.100.116009}
{\path{doi:10.1103/PhysRevD.100.116009}}.

\bibitem{Tripolt:2020irx}
R.-A. Tripolt, D.~H. Rischke, L.~von Smekal, J.~Wambach, {Fermionic excitations
at finite temperature and density}, Phys. Rev. D 101 (2020) 094010.
\newblock \href {http://arxiv.org/abs/2003.11871} {\path{arXiv:2003.11871}},
\href {https://doi.org/10.1103/PhysRevD.101.094010}
{\path{doi:10.1103/PhysRevD.101.094010}}.

\bibitem{Tripolt:2018qvi}
R.-A. Tripolt, J.~Weyrich, L.~von Smekal, J.~Wambach, {Fermionic spectral
functions with the Functional Renormalization Group}, Phys. Rev. D98~(9)
(2018) 094002.
\newblock \href {http://arxiv.org/abs/1807.11708} {\path{arXiv:1807.11708}},
\href {https://doi.org/10.1103/PhysRevD.98.094002}
{\path{doi:10.1103/PhysRevD.98.094002}}.

\bibitem{Wang:2018osm}
Z.~Wang, L.~He, {Fermion spectral function in hot strongly interacting matter
from the functional renormalization group}, Phys. Rev. D98~(9) (2018) 094031.
\newblock \href {http://arxiv.org/abs/1808.08535} {\path{arXiv:1808.08535}},
\href {https://doi.org/10.1103/PhysRevD.98.094031}
{\path{doi:10.1103/PhysRevD.98.094031}}.

\bibitem{Binosi:2019ecz}
D.~Binosi, R.-A. Tripolt, {Spectral functions of confined particles}, Phys.
Lett. B801 (2020) 135171.
\newblock \href {http://arxiv.org/abs/1904.08172} {\path{arXiv:1904.08172}},
\href {https://doi.org/10.1016/j.physletb.2019.135171}
{\path{doi:10.1016/j.physletb.2019.135171}}.

\bibitem{Steib:2019xrv}
I.~Steib, S.~Nagy, J.~Polonyi, {Renormalization in Minkowski space-time} (8
2019).
\newblock \href {http://arxiv.org/abs/1908.11311} {\path{arXiv:1908.11311}}.

\bibitem{Helmboldt:2014iya}
A.~J. Helmboldt, J.~M. Pawlowski, N.~Strodthoff, {Towards quantitative
precision in the chiral crossover: masses and fluctuation scales}, Phys.Rev.
D91~(5) (2015) 054010.
\newblock \href {http://arxiv.org/abs/1409.8414} {\path{arXiv:1409.8414}},
\href {https://doi.org/10.1103/PhysRevD.91.054010}
{\path{doi:10.1103/PhysRevD.91.054010}}.

\bibitem{Tripolt:2014wra}
R.-A. Tripolt, L.~von Smekal, J.~Wambach, {Flow equations for spectral
functions at finite external momenta}, Phys.Rev. D90~(7) (2014) 074031.
\newblock \href {http://arxiv.org/abs/1408.3512} {\path{arXiv:1408.3512}},
\href {https://doi.org/10.1103/PhysRevD.90.074031}
{\path{doi:10.1103/PhysRevD.90.074031}}.

\bibitem{Strodthoff:2016pxx}
N.~Strodthoff, {Self-consistent spectral functions in the O(N) model from the
functional renormalization group}, Phys. Rev. D95~(7) (2017) 076002.
\newblock \href {http://arxiv.org/abs/1611.05036} {\path{arXiv:1611.05036}},
\href {https://doi.org/10.1103/PhysRevD.95.076002}
{\path{doi:10.1103/PhysRevD.95.076002}}.

\bibitem{Yokota:2016tip}
T.~Yokota, T.~Kunihiro, K.~Morita, {Functional renormalization group analysis
of the soft mode at the QCD critical point}, PTEP 2016~(7) (2016) 073D01.
\newblock \href {http://arxiv.org/abs/1603.02147} {\path{arXiv:1603.02147}},
\href {https://doi.org/10.1093/ptep/ptw062} {\path{doi:10.1093/ptep/ptw062}}.

\bibitem{Yokota:2017uzu}
T.~Yokota, T.~Kunihiro, K.~Morita, {Tachyonic instability of the scalar mode
prior to the QCD critical point based on the functional renormalization-group
method in the two-flavor case}, Phys. Rev. D96~(7) (2017) 074028.
\newblock \href {http://arxiv.org/abs/1707.05520} {\path{arXiv:1707.05520}},
\href {https://doi.org/10.1103/PhysRevD.96.074028}
{\path{doi:10.1103/PhysRevD.96.074028}}.

\bibitem{Wang:2017vis}
Z.~Wang, P.~Zhuang, {Meson spectral functions at finite temperature and isospin
density with the functional renormalization group}, Phys. Rev. D96~(1) (2017)
014006.
\newblock \href {http://arxiv.org/abs/1703.01035} {\path{arXiv:1703.01035}},
\href {https://doi.org/10.1103/PhysRevD.96.014006}
{\path{doi:10.1103/PhysRevD.96.014006}}.

\bibitem{Gao:2020qsj}
F.~Gao, J.~M. Pawlowski, {QCD phase structure from functional methods}, Phys.
Rev. D 102~(3) (2020) 034027.
\newblock \href {http://arxiv.org/abs/2002.07500} {\path{arXiv:2002.07500}},
\href {https://doi.org/10.1103/PhysRevD.102.034027}
{\path{doi:10.1103/PhysRevD.102.034027}}.

\bibitem{Gao:2020fbl}
F.~Gao, J.~M. Pawlowski, {Chiral phase structure and critical end point in QCD}
(10 2020).
\newblock \href {http://arxiv.org/abs/2010.13705} {\path{arXiv:2010.13705}}.

\bibitem{Bellwied:2015rza}
R.~Bellwied, S.~Borsanyi, Z.~Fodor, J.~G{\"u}nther, S.~D. Katz, C.~Ratti, K.~K.
Szabo, {The QCD phase diagram from analytic continuation}, Phys. Lett. B751
(2015) 559--564.
\newblock \href {http://arxiv.org/abs/1507.07510} {\path{arXiv:1507.07510}},
\href {https://doi.org/10.1016/j.physletb.2015.11.011}
{\path{doi:10.1016/j.physletb.2015.11.011}}.

\bibitem{Bazavov:2018mes}
A.~Bazavov, et~al., {Chiral crossover in QCD at zero and non-zero chemical
potentials}, Phys. Lett. B 795 (2019) 15--21.
\newblock \href {http://arxiv.org/abs/1812.08235} {\path{arXiv:1812.08235}},
\href {https://doi.org/10.1016/j.physletb.2019.05.013}
{\path{doi:10.1016/j.physletb.2019.05.013}}.

\bibitem{Braguta:2019yci}
V.~Braguta, M.~Chernodub, A.~Y. Kotov, A.~Molochkov, A.~Nikolaev,
{Finite-density QCD transition in a magnetic background field}, Phys. Rev. D
100~(11) (2019) 114503.
\newblock \href {http://arxiv.org/abs/1909.09547} {\path{arXiv:1909.09547}},
\href {https://doi.org/10.1103/PhysRevD.100.114503}
{\path{doi:10.1103/PhysRevD.100.114503}}.

\bibitem{Gao:2015kea}
F.~Gao, J.~Chen, Y.-X. Liu, S.-X. Qin, C.~D. Roberts, S.~M. Schmidt, {Phase
diagram and thermal properties of strong-interaction matter}, Phys. Rev. D
93~(9) (2016) 094019.
\newblock \href {http://arxiv.org/abs/1507.00875} {\path{arXiv:1507.00875}},
\href {https://doi.org/10.1103/PhysRevD.93.094019}
{\path{doi:10.1103/PhysRevD.93.094019}}.

\bibitem{Adamczyk:2017iwn}
L.~Adamczyk, et~al., {Bulk Properties of the Medium Produced in Relativistic
Heavy-Ion Collisions from the Beam Energy Scan Program}, Phys. Rev. C96~(4)
(2017) 044904.
\newblock \href {http://arxiv.org/abs/1701.07065} {\path{arXiv:1701.07065}},
\href {https://doi.org/10.1103/PhysRevC.96.044904}
{\path{doi:10.1103/PhysRevC.96.044904}}.

\bibitem{Alba:2014eba}
P.~Alba, W.~Alberico, R.~Bellwied, M.~Bluhm, V.~Mantovani~Sarti, M.~Nahrgang,
C.~Ratti, {Freeze-out conditions from net-proton and net-charge fluctuations
at RHIC}, Phys. Lett. B738 (2014) 305--310.
\newblock \href {http://arxiv.org/abs/1403.4903} {\path{arXiv:1403.4903}},
\href {https://doi.org/10.1016/j.physletb.2014.09.052}
{\path{doi:10.1016/j.physletb.2014.09.052}}.

\bibitem{Andronic:2017pug}
A.~Andronic, P.~Braun-Munzinger, K.~Redlich, J.~Stachel, {Decoding the phase
structure of QCD via particle production at high energy}, Nature 561~(7723)
(2018) 321--330.
\newblock \href {http://arxiv.org/abs/1710.09425} {\path{arXiv:1710.09425}},
\href {https://doi.org/10.1038/s41586-018-0491-6}
{\path{doi:10.1038/s41586-018-0491-6}}.

\bibitem{Becattini:2016xct}
F.~Becattini, J.~Steinheimer, R.~Stock, M.~Bleicher, {Hadronization conditions
in relativistic nuclear collisions and the QCD pseudo-critical line}, Phys.
Lett. B764 (2017) 241--246.
\newblock \href {http://arxiv.org/abs/1605.09694} {\path{arXiv:1605.09694}},
\href {https://doi.org/10.1016/j.physletb.2016.11.033}
{\path{doi:10.1016/j.physletb.2016.11.033}}.

\bibitem{Vovchenko:2015idt}
V.~Vovchenko, V.~V. Begun, M.~I. Gorenstein, {Hadron multiplicities and
chemical freeze-out conditions in proton-proton and nucleus-nucleus
collisions}, Phys. Rev. C93~(6) (2016) 064906.
\newblock \href {http://arxiv.org/abs/1512.08025} {\path{arXiv:1512.08025}},
\href {https://doi.org/10.1103/PhysRevC.93.064906}
{\path{doi:10.1103/PhysRevC.93.064906}}.

\bibitem{Sagun:2017eye}
V.~V. Sagun, K.~A. Bugaev, A.~I. Ivanytskyi, I.~P. Yakimenko, E.~G. Nikonov,
A.~V. Taranenko, C.~Greiner, D.~B. Blaschke, G.~M. Zinovjev, {Hadron
Resonance Gas Model with Induced Surface Tension}, Eur. Phys. J. A54~(6)
(2018) 100.
\newblock \href {http://arxiv.org/abs/1703.00049} {\path{arXiv:1703.00049}},
\href {https://doi.org/10.1140/epja/i2018-12535-1}
{\path{doi:10.1140/epja/i2018-12535-1}}.

\bibitem{Braun:2019aow}
J.~Braun, M.~Leonhardt, M.~Pospiech, {Fierz-complete NJL model study III:
Emergence from quark-gluon dynamics}, Phys. Rev. D 101~(3) (2020) 036004.
\newblock \href {http://arxiv.org/abs/1909.06298} {\path{arXiv:1909.06298}},
\href {https://doi.org/10.1103/PhysRevD.101.036004}
{\path{doi:10.1103/PhysRevD.101.036004}}.

\bibitem{Braun:2008pi}
J.~Braun, {The QCD Phase Boundary from Quark-Gluon Dynamics}, Eur. Phys. J. C64
(2009) 459--482.
\newblock \href {http://arxiv.org/abs/0810.1727} {\path{arXiv:0810.1727}},
\href {https://doi.org/10.1140/epjc/s10052-009-1136-6}
{\path{doi:10.1140/epjc/s10052-009-1136-6}}.

\bibitem{Leonhardt:2019fua}
M.~Leonhardt, M.~Pospiech, B.~Schallmo, J.~Braun, C.~Drischler, K.~Hebeler,
A.~Schwenk, {Symmetric nuclear matter from the strong interaction} (2019).
\newblock \href {http://arxiv.org/abs/1907.05814} {\path{arXiv:1907.05814}}.

\bibitem{Braun:2020ada}
J.~Braun, W.-j. Fu, J.~M. Pawlowski, F.~Rennecke, D.~Rosenbl{\"u}h, S.~Yin,
{Chiral Susceptibility in (2+1)-flavour QCD} (3 2020).
\newblock \href {http://arxiv.org/abs/2003.13112} {\path{arXiv:2003.13112}}.

\bibitem{Borsanyi:2020fev}
S.~Borsanyi, Z.~Fodor, J.~N. Guenther, R.~Kara, S.~D. Katz, P.~Parotto,
A.~Pasztor, C.~Ratti, K.~K. Szabo, {The QCD crossover at finite chemical
potential from lattice simulations} (2 2020).
\newblock \href {http://arxiv.org/abs/2002.02821} {\path{arXiv:2002.02821}}.

\bibitem{Ding:2020rtq}
H.-T. Ding, {New developments in lattice QCD on equilibrium physics and phase
diagram}, in: {28th International Conference on Ultrarelativistic
Nucleus-Nucleus Collisions}, 2020.
\newblock \href {http://arxiv.org/abs/2002.11957} {\path{arXiv:2002.11957}}.

\bibitem{Eichmann:2015kfa}
G.~Eichmann, C.~S. Fischer, C.~A. Welzbacher, {Baryon effects on the location
of QCD's critical end point}, Phys. Rev. D 93~(3) (2016) 034013.
\newblock \href {http://arxiv.org/abs/1509.02082} {\path{arXiv:1509.02082}},
\href {https://doi.org/10.1103/PhysRevD.93.034013}
{\path{doi:10.1103/PhysRevD.93.034013}}.

\bibitem{Isserstedt:2019pgx}
P.~Isserstedt, M.~Buballa, C.~S. Fischer, P.~J. Gunkel, {Baryon number
fluctuations in the QCD phase diagram from Dyson-Schwinger equations}, Phys.
Rev. D 100~(7) (2019) 074011.
\newblock \href {http://arxiv.org/abs/1906.11644} {\path{arXiv:1906.11644}},
\href {https://doi.org/10.1103/PhysRevD.100.074011}
{\path{doi:10.1103/PhysRevD.100.074011}}.

\bibitem{Maelger:2018vow}
J.~Maelger, U.~Reinosa, J.~Serreau, {Universal aspects of the phase diagram of
QCD with heavy quarks}, Phys. Rev. D 98~(9) (2018) 094020.
\newblock \href {http://arxiv.org/abs/1805.10015} {\path{arXiv:1805.10015}},
\href {https://doi.org/10.1103/PhysRevD.98.094020}
{\path{doi:10.1103/PhysRevD.98.094020}}.

\bibitem{Maelger:2019cbk}
J.~Maelger, U.~Reinosa, J.~Serreau, {Localized rainbows in the QCD phase
diagram}, Phys. Rev. D 101~(1) (2020) 014028.
\newblock \href {http://arxiv.org/abs/1903.04184} {\path{arXiv:1903.04184}},
\href {https://doi.org/10.1103/PhysRevD.101.014028}
{\path{doi:10.1103/PhysRevD.101.014028}}.

\bibitem{Fischer:2018sdj}
C.~S. Fischer, {QCD at finite temperature and chemical potential from
Dyson--Schwinger equations}, Prog. Part. Nucl. Phys. 105 (2019) 1--60.
\newblock \href {http://arxiv.org/abs/1810.12938} {\path{arXiv:1810.12938}},
\href {https://doi.org/10.1016/j.ppnp.2019.01.002}
{\path{doi:10.1016/j.ppnp.2019.01.002}}.

\bibitem{Braun:2010qs}
J.~Braun, C.~S. Fischer, H.~Gies, {Beyond Miransky Scaling}, Phys. Rev. D84
(2011) 034045.
\newblock \href {http://arxiv.org/abs/1012.4279} {\path{arXiv:1012.4279}},
\href {https://doi.org/10.1103/PhysRevD.84.034045}
{\path{doi:10.1103/PhysRevD.84.034045}}.

\bibitem{Gies:2005as}
H.~Gies, J.~J\"ackel, {Chiral phase structure of QCD with many flavors},
Eur.Phys.J. C46 (2006) 433--438.
\newblock \href {http://arxiv.org/abs/hep-ph/0507171}
{\path{arXiv:hep-ph/0507171}}, \href
{https://doi.org/10.1140/epjc/s2006-02475-0}
{\path{doi:10.1140/epjc/s2006-02475-0}}.

\bibitem{Braun:2005uj}
J.~Braun, H.~Gies, {Running coupling at finite temperature and chiral symmetry
restoration in QCD}, Phys. Lett. B645 (2007) 53--58.
\newblock \href {http://arxiv.org/abs/hep-ph/0512085}
{\path{arXiv:hep-ph/0512085}}, \href
{https://doi.org/10.1016/j.physletb.2006.11.059}
{\path{doi:10.1016/j.physletb.2006.11.059}}.

\bibitem{Braun:2006jd}
J.~Braun, H.~Gies, {Chiral phase boundary of QCD at finite temperature}, JHEP
0606 (2006) 024.
\newblock \href {http://arxiv.org/abs/hep-ph/0602226}
{\path{arXiv:hep-ph/0602226}}, \href
{https://doi.org/10.1088/1126-6708/2006/06/024}
{\path{doi:10.1088/1126-6708/2006/06/024}}.

\bibitem{Terao:2007jm}
H.~Terao, A.~Tsuchiya, {Conformal dynamics in gauge theories via
non-perturbative renormalization group} (2007).
\newblock \href {http://arxiv.org/abs/0704.3659} {\path{arXiv:0704.3659}}.

\bibitem{Braun:2009ns}
J.~Braun, H.~Gies, {Scaling laws near the conformal window of many-flavor QCD},
JHEP 05 (2010) 060.
\newblock \href {http://arxiv.org/abs/0912.4168} {\path{arXiv:0912.4168}},
\href {https://doi.org/10.1007/JHEP05(2010)060}
{\path{doi:10.1007/JHEP05(2010)060}}.

\bibitem{Braun:2014wja}
J.~Braun, H.~Gies, L.~Janssen, D.~Roscher, {Phase structure of many-flavor
QED$_3$}, Phys. Rev. D90~(3) (2014) 036002.
\newblock \href {http://arxiv.org/abs/1404.1362} {\path{arXiv:1404.1362}},
\href {https://doi.org/10.1103/PhysRevD.90.036002}
{\path{doi:10.1103/PhysRevD.90.036002}}.

\bibitem{Bergerhoff:1994sj}
B.~Bergerhoff, C.~Wetterich, {The Strongly interacting electroweak phase
transition}, Nucl. Phys. B440 (1995) 171--188.
\newblock \href {http://arxiv.org/abs/hep-ph/9409295}
{\path{arXiv:hep-ph/9409295}}, \href
{https://doi.org/10.1016/0550-3213(95)00079-8}
{\path{doi:10.1016/0550-3213(95)00079-8}}.

\bibitem{Eichhorn:2018yfc}
A.~Eichhorn, {An asymptotically safe guide to quantum gravity and matter}
(2018).
\newblock \href {http://arxiv.org/abs/1810.07615} {\path{arXiv:1810.07615}}.

\bibitem{deAlwis:2019aud}
S.~de~Alwis, A.~Eichhorn, A.~Held, J.~M. Pawlowski, M.~Schiffer, F.~Versteegen,
{Asymptotic safety, string theory and the weak gravity conjecture}, Phys.\
Lett.\ B 798 (2019) 134991.
\newblock \href {http://arxiv.org/abs/1907.07894} {\path{arXiv:1907.07894}},
\href {https://doi.org/10.1016/j.physletb.2019.134991}
{\path{doi:10.1016/j.physletb.2019.134991}}.

\bibitem{Held:2020kze}
A.~Held, {Effective asymptotic safety and its predictive power: Gauge-Yukawa
theories} (3 2020).
\newblock \href {http://arxiv.org/abs/2003.13642} {\path{arXiv:2003.13642}}.

\bibitem{Ellwanger:1992us}
U.~Ellwanger, L.~Vergara, {Flow equations for the Higgs top system}, Nucl.
Phys. B398 (1993) 52--68.
\newblock \href {http://arxiv.org/abs/hep-ph/9212265}
{\path{arXiv:hep-ph/9212265}}, \href
{https://doi.org/10.1016/0550-3213(93)90627-2}
{\path{doi:10.1016/0550-3213(93)90627-2}}.

\bibitem{Gies:2003dp}
H.~Gies, J.~Jaeckel, C.~Wetterich, {Towards a renormalizable standard model
without fundamental Higgs scalar}, Phys. Rev. D69 (2004) 105008.
\newblock \href {http://arxiv.org/abs/hep-ph/0312034}
{\path{arXiv:hep-ph/0312034}}, \href
{https://doi.org/10.1103/PhysRevD.69.105008}
{\path{doi:10.1103/PhysRevD.69.105008}}.

\bibitem{Percacci:2009fh}
R.~Percacci, O.~Zanusso, {One loop beta functions and fixed points in Higher
Derivative Sigma Models}, Phys. Rev. D81 (2010) 065012.
\newblock \href {http://arxiv.org/abs/0910.0851} {\path{arXiv:0910.0851}},
\href {https://doi.org/10.1103/PhysRevD.81.065012}
{\path{doi:10.1103/PhysRevD.81.065012}}.

\bibitem{Fabbrichesi:2010xy}
M.~Fabbrichesi, R.~Percacci, A.~Tonero, O.~Zanusso, {Asymptotic safety and the
gauged SU(N) nonlinear sigma- model}, Phys. Rev. D83 (2011) 025016.
\newblock \href {http://arxiv.org/abs/1010.0912} {\path{arXiv:1010.0912}},
\href {https://doi.org/10.1103/PhysRevD.83.025016}
{\path{doi:10.1103/PhysRevD.83.025016}}.

\bibitem{Bazzocchi:2011vr}
F.~Bazzocchi, M.~Fabbrichesi, R.~Percacci, A.~Tonero, L.~Vecchi, {Fermions and
Goldstone bosons in an asymptotically safe model}, Phys. Lett. B 705 (2011)
388--392.
\newblock \href {http://arxiv.org/abs/1105.1968} {\path{arXiv:1105.1968}},
\href {https://doi.org/10.1016/j.physletb.2011.10.029}
{\path{doi:10.1016/j.physletb.2011.10.029}}.

\bibitem{Fabbrichesi:2011bx}
M.~Fabbrichesi, R.~Percacci, A.~Tonero, L.~Vecchi, {The Electroweak S and T
parameters from a fixed point condition}, Phys. Rev. Lett. 107 (2011) 021803.
\newblock \href {http://arxiv.org/abs/1102.2113} {\path{arXiv:1102.2113}},
\href {https://doi.org/10.1103/PhysRevLett.107.021803}
{\path{doi:10.1103/PhysRevLett.107.021803}}.

\bibitem{Gies:2009hq}
H.~Gies, M.~M. Scherer, {Asymptotic safety of simple Yukawa systems}, Eur.
Phys. J. C66 (2010) 387--402.
\newblock \href {http://arxiv.org/abs/0901.2459} {\path{arXiv:0901.2459}},
\href {https://doi.org/10.1140/epjc/s10052-010-1256-z}
{\path{doi:10.1140/epjc/s10052-010-1256-z}}.

\bibitem{Gies:2009sv}
H.~Gies, S.~Rechenberger, M.~M. Scherer, {Towards an Asymptotic-Safety Scenario
for Chiral Yukawa Systems}, Eur. Phys. J. C66 (2010) 403--418.
\newblock \href {http://arxiv.org/abs/0907.0327} {\path{arXiv:0907.0327}},
\href {https://doi.org/10.1140/epjc/s10052-010-1257-y}
{\path{doi:10.1140/epjc/s10052-010-1257-y}}.

\bibitem{Scherer:2009wu}
M.~M. Scherer, H.~Gies, S.~Rechenberger, {An Asymptotic-safety mechanism for
chiral Yukawa systems}, Acta Phys. Polon. Supp. 2 (2009) 541.
\newblock \href {http://arxiv.org/abs/0910.0395} {\path{arXiv:0910.0395}}.

\bibitem{Vacca:2015nta}
G.~P. Vacca, L.~Zambelli, {Multimeson Yukawa interactions at criticality},
Phys. Rev. D91~(12) (2015) 125003.
\newblock \href {http://arxiv.org/abs/1503.09136} {\path{arXiv:1503.09136}},
\href {https://doi.org/10.1103/PhysRevD.91.125003}
{\path{doi:10.1103/PhysRevD.91.125003}}.

\bibitem{Gies:2013pma}
H.~Gies, S.~Rechenberger, M.~M. Scherer, L.~Zambelli, {An asymptotic safety
scenario for gauged chiral Higgs-Yukawa models}, Eur. Phys. J. C73 (2013)
2652.
\newblock \href {http://arxiv.org/abs/1306.6508} {\path{arXiv:1306.6508}},
\href {https://doi.org/10.1140/epjc/s10052-013-2652-y}
{\path{doi:10.1140/epjc/s10052-013-2652-y}}.

\bibitem{Gies:2020xuh}
H.~Gies, J.~Ziebell, {Asymptotically Safe QED} (5 2020).
\newblock \href {http://arxiv.org/abs/2005.07586} {\path{arXiv:2005.07586}}.

\bibitem{Eichhorn:2012uv}
A.~Eichhorn, H.~Gies, D.~Roscher, {Renormalization Flow of Axion
Electrodynamics}, Phys. Rev. D86 (2012) 125014.
\newblock \href {http://arxiv.org/abs/1208.0014} {\path{arXiv:1208.0014}},
\href {https://doi.org/10.1103/PhysRevD.86.125014}
{\path{doi:10.1103/PhysRevD.86.125014}}.

\bibitem{Eichhorn:2018vah}
A.~Eichhorn, A.~Held, P.~Vander~Griend, {Asymptotic safety in the dark}, JHEP
08 (2018) 147.
\newblock \href {http://arxiv.org/abs/1802.08589} {\path{arXiv:1802.08589}},
\href {https://doi.org/10.1007/JHEP08(2018)147}
{\path{doi:10.1007/JHEP08(2018)147}}.

\bibitem{Litim:2014uca}
D.~F. Litim, F.~Sannino, {Asymptotic safety guaranteed}, JHEP 12 (2014) 178.
\newblock \href {http://arxiv.org/abs/1406.2337} {\path{arXiv:1406.2337}},
\href {https://doi.org/10.1007/JHEP12(2014)178}
{\path{doi:10.1007/JHEP12(2014)178}}.

\bibitem{Litim:2015iea}
D.~F. Litim, M.~Mojaza, F.~Sannino, {Vacuum stability of asymptotically safe
gauge-Yukawa theories}, JHEP 01 (2016) 081.
\newblock \href {http://arxiv.org/abs/1501.03061} {\path{arXiv:1501.03061}},
\href {https://doi.org/10.1007/JHEP01(2016)081}
{\path{doi:10.1007/JHEP01(2016)081}}.

\bibitem{Mann:2017wzh}
R.~Mann, J.~Meffe, F.~Sannino, T.~Steele, Z.-W. Wang, C.~Zhang, {Asymptotically
Safe Standard Model via Vectorlike Fermions}, Phys. Rev. Lett. 119~(26)
(2017) 261802.
\newblock \href {http://arxiv.org/abs/1707.02942} {\path{arXiv:1707.02942}},
\href {https://doi.org/10.1103/PhysRevLett.119.261802}
{\path{doi:10.1103/PhysRevLett.119.261802}}.

\bibitem{Bond:2018oco}
A.~D. Bond, D.~F. Litim, {Price of Asymptotic Safety}, Phys. Rev. Lett.
122~(21) (2019) 211601.
\newblock \href {http://arxiv.org/abs/1801.08527} {\path{arXiv:1801.08527}},
\href {https://doi.org/10.1103/PhysRevLett.122.211601}
{\path{doi:10.1103/PhysRevLett.122.211601}}.

\bibitem{Dondi:2019ivp}
N.~A. Dondi, G.~V. Dunne, M.~Reichert, F.~Sannino, {Analytic Coupling Structure
of Large $N_f$ (Super) QED and QCD}, Phys. Rev. D 100~(1) (2019) 015013.
\newblock \href {http://arxiv.org/abs/1903.02568} {\path{arXiv:1903.02568}},
\href {https://doi.org/10.1103/PhysRevD.100.015013}
{\path{doi:10.1103/PhysRevD.100.015013}}.

\bibitem{Bond:2019npq}
A.~D. Bond, D.~F. Litim, T.~Steudtner, {Asymptotic safety with Majorana
fermions and new large $N$ equivalences}, Phys. Rev. D 101~(4) (2020) 045006.
\newblock \href {http://arxiv.org/abs/1911.11168} {\path{arXiv:1911.11168}},
\href {https://doi.org/10.1103/PhysRevD.101.045006}
{\path{doi:10.1103/PhysRevD.101.045006}}.

\bibitem{Dondi:2020qfj}
N.~A. Dondi, G.~V. Dunne, M.~Reichert, F.~Sannino, {Towards the QED beta
function and renormalons at $1/N_f^2$ and $1/N_f^3$} (3 2020).
\newblock \href {http://arxiv.org/abs/2003.08397} {\path{arXiv:2003.08397}}.

\bibitem{Buyukbese:2017ehm}
T.~Buyukbese, D.~F. Litim, {Asymptotic safety of gauge theories beyond marginal
interactions}, PoS LATTICE2016 (2017) 233.
\newblock \href {https://doi.org/10.22323/1.256.0233}
{\path{doi:10.22323/1.256.0233}}.

\bibitem{Gies:2015lia}
H.~Gies, L.~Zambelli, {Asymptotically free scaling solutions in non-Abelian
Higgs models}, Phys. Rev. D92~(2) (2015) 025016.
\newblock \href {http://arxiv.org/abs/1502.05907} {\path{arXiv:1502.05907}},
\href {https://doi.org/10.1103/PhysRevD.92.025016}
{\path{doi:10.1103/PhysRevD.92.025016}}.

\bibitem{Gies:2016kkk}
H.~Gies, L.~Zambelli, {Non-Abelian Higgs models: Paving the way for asymptotic
freedom}, Phys. Rev. D96~(2) (2017) 025003.
\newblock \href {http://arxiv.org/abs/1611.09147} {\path{arXiv:1611.09147}},
\href {https://doi.org/10.1103/PhysRevD.96.025003}
{\path{doi:10.1103/PhysRevD.96.025003}}.

\bibitem{Gies:2018vwk}
H.~Gies, R.~Sondenheimer, A.~Ugolotti, L.~Zambelli, {Asymptotic freedom in
$\mathbb {Z}_2$ -Yukawa-QCD models}, Eur. Phys. J. C 79~(2) (2019) 101.
\newblock \href {http://arxiv.org/abs/1804.09688} {\path{arXiv:1804.09688}},
\href {https://doi.org/10.1140/epjc/s10052-019-6604-z}
{\path{doi:10.1140/epjc/s10052-019-6604-z}}.

\bibitem{Gies:2019nij}
H.~Gies, R.~Sondenheimer, A.~Ugolotti, L.~Zambelli, {Scheme dependence of
asymptotically free solutions}, Eur. Phys. J. C79~(6) (2019) 463.
\newblock \href {http://arxiv.org/abs/1901.08581} {\path{arXiv:1901.08581}},
\href {https://doi.org/10.1140/epjc/s10052-019-6956-4}
{\path{doi:10.1140/epjc/s10052-019-6956-4}}.

\bibitem{Gies:2013fua}
H.~Gies, C.~Gneiting, R.~Sondenheimer, {Higgs Mass Bounds from Renormalization
Flow for a simple Yukawa model}, Phys. Rev. D89~(4) (2014) 045012.
\newblock \href {http://arxiv.org/abs/1308.5075} {\path{arXiv:1308.5075}},
\href {https://doi.org/10.1103/PhysRevD.89.045012}
{\path{doi:10.1103/PhysRevD.89.045012}}.

\bibitem{Jakovac:2015kka}
A.~Jakovac, I.~Kaposvari, A.~Patkos, {Scalar mass stability bound in a simple
Yukawa-theory from renormalization group equations}, Mod. Phys. Lett.
A32~(02) (2016) 1750011.
\newblock \href {http://arxiv.org/abs/1508.06774} {\path{arXiv:1508.06774}},
\href {https://doi.org/10.1142/S0217732317500110}
{\path{doi:10.1142/S0217732317500110}}.

\bibitem{Sondenheimer:2017jin}
R.~Sondenheimer, {Nonpolynomial Higgs interactions and vacuum stability}, Eur.
Phys. J. C 79~(1) (2019) 10.
\newblock \href {http://arxiv.org/abs/1711.00065} {\path{arXiv:1711.00065}},
\href {https://doi.org/10.1140/epjc/s10052-018-6507-4}
{\path{doi:10.1140/epjc/s10052-018-6507-4}}.

\bibitem{Gies:2014xha}
H.~Gies, R.~Sondenheimer, {Higgs Mass Bounds from Renormalization Flow for a
Higgs-top-bottom model}, Eur. Phys. J. C75~(2) (2015) 68.
\newblock \href {http://arxiv.org/abs/1407.8124} {\path{arXiv:1407.8124}},
\href {https://doi.org/10.1140/epjc/s10052-015-3284-1}
{\path{doi:10.1140/epjc/s10052-015-3284-1}}.

\bibitem{Eichhorn:2015kea}
A.~Eichhorn, H.~Gies, J.~Jaeckel, T.~Plehn, M.~M. Scherer, R.~Sondenheimer,
{The Higgs Mass and the Scale of New Physics}, JHEP 04 (2015) 022.
\newblock \href {http://arxiv.org/abs/1501.02812} {\path{arXiv:1501.02812}},
\href {https://doi.org/10.1007/JHEP04(2015)022}
{\path{doi:10.1007/JHEP04(2015)022}}.

\bibitem{Gies:2017zwf}
H.~Gies, R.~Sondenheimer, M.~Warschinke, {Impact of generalized Yukawa
interactions on the lower Higgs mass bound}, Eur. Phys. J. C77~(11) (2017)
743.
\newblock \href {http://arxiv.org/abs/1707.04394} {\path{arXiv:1707.04394}},
\href {https://doi.org/10.1140/epjc/s10052-017-5312-9}
{\path{doi:10.1140/epjc/s10052-017-5312-9}}.

\bibitem{Borchardt:2016xju}
J.~Borchardt, H.~Gies, R.~Sondenheimer, {Global flow of the Higgs potential in
a Yukawa model}, Eur. Phys. J. C76~(8) (2016) 472.
\newblock \href {http://arxiv.org/abs/1603.05861} {\path{arXiv:1603.05861}},
\href {https://doi.org/10.1140/epjc/s10052-016-4300-9}
{\path{doi:10.1140/epjc/s10052-016-4300-9}}.

\bibitem{Eichhorn:2014qka}
A.~Eichhorn, M.~M. Scherer, {Planck scale, Higgs mass, and scalar dark matter},
Phys. Rev. D 90~(2) (2014) 025023.
\newblock \href {http://arxiv.org/abs/1404.5962} {\path{arXiv:1404.5962}},
\href {https://doi.org/10.1103/PhysRevD.90.025023}
{\path{doi:10.1103/PhysRevD.90.025023}}.

\bibitem{Held:2018cxd}
A.~Held, R.~Sondenheimer, {Higgs stability-bound and fermionic dark matter},
JHEP 02 (2019) 166.
\newblock \href {http://arxiv.org/abs/1811.07898} {\path{arXiv:1811.07898}},
\href {https://doi.org/10.1007/JHEP02(2019)166}
{\path{doi:10.1007/JHEP02(2019)166}}.

\bibitem{Gies:2017ajd}
H.~Gies, R.~Sondenheimer, {Renormalization Group Flow of the Higgs Potential},
Phil. Trans. Roy. Soc. Lond. A 376~(2114) (2018) 20170120.
\newblock \href {http://arxiv.org/abs/1708.04305} {\path{arXiv:1708.04305}},
\href {https://doi.org/10.1098/rsta.2017.0120}
{\path{doi:10.1098/rsta.2017.0120}}.

\bibitem{Reichert:2017puo}
M.~Reichert, A.~Eichhorn, H.~Gies, J.~M. Pawlowski, T.~Plehn, M.~M. Scherer,
{Probing baryogenesis through the Higgs boson self-coupling}, Phys. Rev.
D97~(7) (2018) 075008.
\newblock \href {http://arxiv.org/abs/1711.00019} {\path{arXiv:1711.00019}},
\href {https://doi.org/10.1103/PhysRevD.97.075008}
{\path{doi:10.1103/PhysRevD.97.075008}}.

\bibitem{Synatschke:2008pv}
F.~Synatschke, G.~Bergner, H.~Gies, A.~Wipf, {Flow Equation for Supersymmetric
Quantum Mechanics}, JHEP 03 (2009) 028.
\newblock \href {http://arxiv.org/abs/0809.4396} {\path{arXiv:0809.4396}},
\href {https://doi.org/10.1088/1126-6708/2009/03/028}
{\path{doi:10.1088/1126-6708/2009/03/028}}.

\bibitem{Sonoda:2008dz}
H.~Sonoda, K.~Ulker, {Construction of a Wilson action for the Wess-Zumino
model}, Prog. Theor. Phys. 120 (2008) 197--230.
\newblock \href {http://arxiv.org/abs/0804.1072} {\path{arXiv:0804.1072}},
\href {https://doi.org/10.1143/PTP.120.197} {\path{doi:10.1143/PTP.120.197}}.

\bibitem{Rosten:2008ih}
O.~J. Rosten, {On the Renormalization of Theories of a Scalar Chiral
Superfield}, JHEP 03 (2010) 004.
\newblock \href {http://arxiv.org/abs/0808.2150} {\path{arXiv:0808.2150}},
\href {https://doi.org/10.1007/JHEP03(2010)004}
{\path{doi:10.1007/JHEP03(2010)004}}.

\bibitem{Sonoda:2009df}
H.~Sonoda, K.~Ulker, {An Elementary proof of the non-renormalization theorem
for the Wess-Zumino model}, Prog. Theor. Phys. 123 (2010) 989--1002.
\newblock \href {http://arxiv.org/abs/0909.2976} {\path{arXiv:0909.2976}},
\href {https://doi.org/10.1143/PTP.123.989} {\path{doi:10.1143/PTP.123.989}}.

\bibitem{Synatschke:2009nm}
F.~Synatschke, H.~Gies, A.~Wipf, {Phase Diagram and Fixed-Point Structure of
two dimensional N=1 Wess-Zumino Models}, Phys. Rev. D80 (2009) 085007.
\newblock \href {http://arxiv.org/abs/0907.4229} {\path{arXiv:0907.4229}},
\href {https://doi.org/10.1103/PhysRevD.80.085007}
{\path{doi:10.1103/PhysRevD.80.085007}}.

\bibitem{Heilmann:2014iga}
M.~Heilmann, T.~Hellwig, B.~Knorr, M.~Ansorg, A.~Wipf, {Convergence of
Derivative Expansion in Supersymmetric Functional RG Flows}, JHEP 02 (2015)
109.
\newblock \href {http://arxiv.org/abs/1409.5650} {\path{arXiv:1409.5650}},
\href {https://doi.org/10.1007/JHEP02(2015)109}
{\path{doi:10.1007/JHEP02(2015)109}}.

\bibitem{Synatschke:2010ub}
F.~Synatschke, J.~Braun, A.~Wipf, {N=1 Wess Zumino Model in d=3 at zero and
finite temperature}, Phys. Rev. D81 (2010) 125001.
\newblock \href {http://arxiv.org/abs/1001.2399} {\path{arXiv:1001.2399}},
\href {https://doi.org/10.1103/PhysRevD.81.125001}
{\path{doi:10.1103/PhysRevD.81.125001}}.

\bibitem{Granda:1997xk}
L.~N. Granda, S.~D. Odintsov, {Exact renormalization group for O(4) gauged
supergravity}, Phys. Lett. B409 (1997) 206--212.
\newblock \href {http://arxiv.org/abs/hep-th/9706062}
{\path{arXiv:hep-th/9706062}}, \href
{https://doi.org/10.1016/S0370-2693(97)00878-2}
{\path{doi:10.1016/S0370-2693(97)00878-2}}.

\bibitem{Falkenberg:1998bg}
S.~Falkenberg, B.~Geyer, {Effective average action in N=1 superYang-Mills
theory}, Phys. Rev. D58 (1998) 085004.
\newblock \href {http://arxiv.org/abs/hep-th/9802113}
{\path{arXiv:hep-th/9802113}}, \href
{https://doi.org/10.1103/PhysRevD.58.085004}
{\path{doi:10.1103/PhysRevD.58.085004}}.

\bibitem{Arnone:1998zc}
S.~Arnone, C.~Fusi, K.~Yoshida, {Exact renormalization group equation in
presence of rescaling anomaly}, JHEP 02 (1999) 022.
\newblock \href {http://arxiv.org/abs/hep-th/9812022}
{\path{arXiv:hep-th/9812022}}, \href
{https://doi.org/10.1088/1126-6708/1999/02/022}
{\path{doi:10.1088/1126-6708/1999/02/022}}.

\bibitem{Bonini:1998ec}
M.~Bonini, F.~Vian, {Wilson renormalization group for supersymmetric gauge
theories and gauge anomalies}, Nucl. Phys. B532 (1998) 473--497.
\newblock \href {http://arxiv.org/abs/hep-th/9802196}
{\path{arXiv:hep-th/9802196}}, \href
{https://doi.org/10.1016/S0550-3213(98)00458-1}
{\path{doi:10.1016/S0550-3213(98)00458-1}}.

\bibitem{Arnone:2000ij}
S.~Arnone, S.~Chiantese, K.~Yoshida, {Applications of exact renormalization
group techniques to the nonperturbative study of supersymmetric gauge field
theory}, Int. J. Mod. Phys. A16 (2001) 1811.
\newblock \href {http://arxiv.org/abs/hep-th/0012111}
{\path{arXiv:hep-th/0012111}}, \href
{https://doi.org/10.1142/S0217751X01004499}
{\path{doi:10.1142/S0217751X01004499}}.

\bibitem{Litim:2011bf}
D.~F. Litim, M.~C. Mastaler, F.~Synatschke-Czerwonka, A.~Wipf, {Critical
behavior of supersymmetric O(N) models in the large-N limit}, Phys.Rev. D84
(2011) 125009.
\newblock \href {http://arxiv.org/abs/1107.3011} {\path{arXiv:1107.3011}},
\href {https://doi.org/10.1103/PhysRevD.84.125009}
{\path{doi:10.1103/PhysRevD.84.125009}}.

\bibitem{Heilmann:2012yf}
M.~Heilmann, D.~F. Litim, F.~Synatschke-Czerwonka, A.~Wipf, {Phases of
supersymmetric O(N) theories}, Phys. Rev. D86 (2012) 105006.
\newblock \href {http://arxiv.org/abs/1208.5389} {\path{arXiv:1208.5389}},
\href {https://doi.org/10.1103/PhysRevD.86.105006}
{\path{doi:10.1103/PhysRevD.86.105006}}.

\bibitem{Gies:2009az}
H.~Gies, F.~Synatschke, A.~Wipf, {Supersymmetry breaking as a quantum phase
transition}, Phys. Rev. D80 (2009) 101701.
\newblock \href {http://arxiv.org/abs/0906.5492} {\path{arXiv:0906.5492}},
\href {https://doi.org/10.1103/PhysRevD.80.101701}
{\path{doi:10.1103/PhysRevD.80.101701}}.

\bibitem{Gies:2017tod}
H.~Gies, T.~Hellwig, A.~Wipf, O.~Zanusso, {A functional perspective on emergent
supersymmetry}, JHEP 12 (2017) 132.
\newblock \href {http://arxiv.org/abs/1705.08312} {\path{arXiv:1705.08312}},
\href {https://doi.org/10.1007/JHEP12(2017)132}
{\path{doi:10.1007/JHEP12(2017)132}}.

\bibitem{Abbott:2016blz}
B.~P. Abbott, et~al., {Observation of Gravitational Waves from a Binary Black
Hole Merger}, Phys. Rev. Lett. 116~(6) (2016) 061102.
\newblock \href {http://arxiv.org/abs/1602.03837} {\path{arXiv:1602.03837}},
\href {https://doi.org/10.1103/PhysRevLett.116.061102}
{\path{doi:10.1103/PhysRevLett.116.061102}}.

\bibitem{Akiyama:2019cqa}
K.~Akiyama, et~al., {First M87 Event Horizon Telescope Results. I. The Shadow
of the Supermassive Black Hole}, Astrophys. J. 875~(1) (2019) L1.
\newblock \href {http://arxiv.org/abs/1906.11238} {\path{arXiv:1906.11238}},
\href {https://doi.org/10.3847/2041-8213/ab0ec7}
{\path{doi:10.3847/2041-8213/ab0ec7}}.

\bibitem{AmelinoCamelia:1997gz}
G.~Amelino-Camelia, J.~R. Ellis, N.~E. Mavromatos, D.~V. Nanopoulos, S.~Sarkar,
{Tests of quantum gravity from observations of gamma-ray bursts}, Nature 393
(1998) 763--765.
\newblock \href {http://arxiv.org/abs/astro-ph/9712103}
{\path{arXiv:astro-ph/9712103}}, \href {https://doi.org/10.1038/31647}
{\path{doi:10.1038/31647}}.

\bibitem{Abdo:2009zza}
A.~A. Abdo, et~al., {Fermi Observations of High-Energy Gamma-Ray Emission from
GRB 080916C}, Science 323 (2009) 1688--1693.
\newblock \href {https://doi.org/10.1126/science.1169101}
{\path{doi:10.1126/science.1169101}}.

\bibitem{Ackermann:2009aa}
M.~Ackermann, et~al., {A limit on the variation of the speed of light arising
from quantum gravity effects}, Nature 462 (2009) 331--334.
\newblock \href {http://arxiv.org/abs/0908.1832} {\path{arXiv:0908.1832}},
\href {https://doi.org/10.1038/nature08574} {\path{doi:10.1038/nature08574}}.

\bibitem{Vasileiou:2013vra}
V.~Vasileiou, A.~Jacholkowska, F.~Piron, J.~Bolmont, C.~Couturier, J.~Granot,
F.~W. Stecker, J.~Cohen-Tanugi, F.~Longo, {Constraints on Lorentz Invariance
Violation from Fermi-Large Area Telescope Observations of Gamma-Ray Bursts},
Phys. Rev. D87~(12) (2013) 122001.
\newblock \href {http://arxiv.org/abs/1305.3463} {\path{arXiv:1305.3463}},
\href {https://doi.org/10.1103/PhysRevD.87.122001}
{\path{doi:10.1103/PhysRevD.87.122001}}.

\bibitem{Oriti:2009zz}
D.~Oriti,
\href{http://www.cambridge.org/catalogue/catalogue.asp?isbn=9780521860451}{{Approaches
to quantum gravity: Toward a new understanding of space, time and matter}},
Cambridge University Press, 2009.
\newline\urlprefix\url{http://www.cambridge.org/catalogue/catalogue.asp?isbn=9780521860451}

\bibitem{Ashtekar:2014kba}
A.~Ashtekar, M.~Reuter, C.~Rovelli, {From General Relativity to Quantum
Gravity} (2014).
\newblock \href {http://arxiv.org/abs/1408.4336} {\path{arXiv:1408.4336}}.

\bibitem{Carlip:2015asa}
S.~Carlip, D.-W. Chiou, W.-T. Ni, R.~Woodard, {Quantum Gravity: A Brief History
of Ideas and Some Prospects}, Int. J. Mod. Phys. D24~(11) (2015) 1530028.
\newblock \href {http://arxiv.org/abs/1507.08194} {\path{arXiv:1507.08194}},
\href {https://doi.org/10.1142/S0218271815300281}
{\path{doi:10.1142/S0218271815300281}}.

\bibitem{Deser:1974cy}
S.~Deser, P.~van Nieuwenhuizen, {Nonrenormalizability of the Quantized
Dirac-Einstein System}, Phys. Rev. D10 (1974) 411.
\newblock \href {https://doi.org/10.1103/PhysRevD.10.411}
{\path{doi:10.1103/PhysRevD.10.411}}.

\bibitem{Deser:1974cz}
S.~Deser, P.~van Nieuwenhuizen, {One Loop Divergences of Quantized
Einstein-Maxwell Fields}, Phys. Rev. D10 (1974) 401.
\newblock \href {https://doi.org/10.1103/PhysRevD.10.401}
{\path{doi:10.1103/PhysRevD.10.401}}.

\bibitem{tHooft:1974toh}
G.~{'t Hooft}, M.~J.~G. Veltman, {One loop divergencies in the theory of
gravitation}, Ann. Inst. H. Poincare Phys. Theor. A20 (1974) 69--94.

\bibitem{Goroff:1985sz}
M.~H. Goroff, A.~Sagnotti, {Quantum Gravity at two Loops}, Phys. Lett. 160B
(1985) 81--86.
\newblock \href {https://doi.org/10.1016/0370-2693(85)91470-4}
{\path{doi:10.1016/0370-2693(85)91470-4}}.

\bibitem{vandeVen:1991gw}
A.~E.~M. van~de Ven, {Two loop quantum gravity}, Nucl. Phys. B378 (1992)
309--366.
\newblock \href {https://doi.org/10.1016/0550-3213(92)90011-Y}
{\path{doi:10.1016/0550-3213(92)90011-Y}}.

\bibitem{Donoghue:1993eb}
J.~F. Donoghue, {Leading quantum correction to the Newtonian potential}, Phys.
Rev. Lett. 72 (1994) 2996--2999.
\newblock \href {http://arxiv.org/abs/gr-qc/9310024}
{\path{arXiv:gr-qc/9310024}}, \href
{https://doi.org/10.1103/PhysRevLett.72.2996}
{\path{doi:10.1103/PhysRevLett.72.2996}}.

\bibitem{Bern:2018jmv}
Z.~Bern, J.~J. Carrasco, W.-M. Chen, A.~Edison, H.~Johansson,
J.~Parra-Martinez, R.~Roiban, M.~Zeng, {Ultraviolet Properties of $\mathcal N
= 8$ Supergravity at Five Loops}, Phys. Rev. D98~(8) (2018) 086021.
\newblock \href {http://arxiv.org/abs/1804.09311} {\path{arXiv:1804.09311}},
\href {https://doi.org/10.1103/PhysRevD.98.086021}
{\path{doi:10.1103/PhysRevD.98.086021}}.

\bibitem{Machado:2007ea}
P.~F. Machado, F.~Saueressig, {On the renormalization group flow of
f(R)-gravity}, Phys. Rev. D77 (2008) 124045.
\newblock \href {http://arxiv.org/abs/0712.0445} {\path{arXiv:0712.0445}},
\href {https://doi.org/10.1103/PhysRevD.77.124045}
{\path{doi:10.1103/PhysRevD.77.124045}}.

\bibitem{Codello:2010mj}
A.~Codello, {Polyakov Effective Action from Functional Renormalization Group
Equation}, Annals Phys. 325 (2010) 1727--1738.
\newblock \href {http://arxiv.org/abs/1004.2171} {\path{arXiv:1004.2171}},
\href {https://doi.org/10.1016/j.aop.2010.04.013}
{\path{doi:10.1016/j.aop.2010.04.013}}.

\bibitem{Manrique:2008zw}
E.~Manrique, M.~Reuter, {Bare Action and Regularized Functional Integral of
Asymptotically Safe Quantum Gravity}, Phys. Rev. D79 (2009) 025008.
\newblock \href {http://arxiv.org/abs/0811.3888} {\path{arXiv:0811.3888}},
\href {https://doi.org/10.1103/PhysRevD.79.025008}
{\path{doi:10.1103/PhysRevD.79.025008}}.

\bibitem{Morris:2015oca}
T.~R. Morris, Z.~H. Slade, {Solutions to the reconstruction problem in
asymptotic safety}, JHEP 11 (2015) 094.
\newblock \href {http://arxiv.org/abs/1507.08657} {\path{arXiv:1507.08657}},
\href {https://doi.org/10.1007/JHEP11(2015)094}
{\path{doi:10.1007/JHEP11(2015)094}}.

\bibitem{Litim:2012vz}
D.~Litim, A.~Satz, {Limit cycles and quantum gravity} (2012).
\newblock \href {http://arxiv.org/abs/1205.4218} {\path{arXiv:1205.4218}}.

\bibitem{Wetterich:2019qzx}
C.~Wetterich, {Quantum scale symmetry} (2019).
\newblock \href {http://arxiv.org/abs/1901.04741} {\path{arXiv:1901.04741}}.

\bibitem{Weinberg:1980gg}
S.~Weinberg, {ULTRAVIOLET DIVERGENCES IN QUANTUM THEORIES OF GRAVITATION},
General Relativity: An Einstein centenary survey, Eds. Hawking, S.W., Israel,
W; Cambridge University Press (1979) 790--831.

\bibitem{Reuter:1996cp}
M.~Reuter, {Nonperturbative Evolution Equation for Quantum Gravity}, Phys. Rev.
D57 (1998) 971--985.
\newblock \href {http://arxiv.org/abs/hep-th/9605030}
{\path{arXiv:hep-th/9605030}}, \href
{https://doi.org/10.1103/PhysRevD.57.971}
{\path{doi:10.1103/PhysRevD.57.971}}.

\bibitem{Dou:1997fg}
D.~Dou, R.~Percacci, {The running gravitational couplings}, Class.\ Quant.\
Grav. 15 (1998) 3449--3468.
\newblock \href {http://arxiv.org/abs/hep-th/9707239}
{\path{arXiv:hep-th/9707239}}, \href
{https://doi.org/10.1088/0264-9381/15/11/011}
{\path{doi:10.1088/0264-9381/15/11/011}}.

\bibitem{Demmel:2015zfa}
M.~Demmel, A.~Nink, {Connections and geodesics in the space of metrics}, Phys.\
Rev.\ D 92~(10) (2015) 104013.
\newblock \href {http://arxiv.org/abs/1506.03809} {\path{arXiv:1506.03809}},
\href {https://doi.org/10.1103/PhysRevD.92.104013}
{\path{doi:10.1103/PhysRevD.92.104013}}.

\bibitem{Baldazzi:2018mtl}
A.~Baldazzi, R.~Percacci, V.~Skrinjar, {Wicked metrics} (2018).
\newblock \href {http://arxiv.org/abs/1811.03369} {\path{arXiv:1811.03369}}.

\bibitem{Eichhorn:2009ah}
A.~Eichhorn, H.~Gies, M.~M. Scherer, {Asymptotically free scalar
curvature-ghost coupling in Quantum Einstein Gravity}, Phys. Rev. D80 (2009)
104003.
\newblock \href {http://arxiv.org/abs/0907.1828} {\path{arXiv:0907.1828}},
\href {https://doi.org/10.1103/PhysRevD.80.104003}
{\path{doi:10.1103/PhysRevD.80.104003}}.

\bibitem{Benedetti:2012dx}
D.~Benedetti, F.~Caravelli, {The Local potential approximation in quantum
gravity}, JHEP 06 (2012) 017, [Erratum: JHEP10,157(2012)].
\newblock \href {http://arxiv.org/abs/1204.3541} {\path{arXiv:1204.3541}},
\href {https://doi.org/10.1007/JHEP06(2012)017, 10.1007/JHEP10(2012)157}
{\path{doi:10.1007/JHEP06(2012)017, 10.1007/JHEP10(2012)157}}.

\bibitem{Alkofer:2018fxj}
N.~Alkofer, F.~Saueressig, {Asymptotically safe f(R)-gravity coupled to matter
I: the polynomial case} (2018).
\newblock \href {http://arxiv.org/abs/1802.00498} {\path{arXiv:1802.00498}}.

\bibitem{Lauscher:2001ya}
O.~Lauscher, M.~Reuter, {Ultraviolet fixed point and generalized flow equation
of quantum gravity}, Phys. Rev. D65 (2002) 025013.
\newblock \href {http://arxiv.org/abs/hep-th/0108040}
{\path{arXiv:hep-th/0108040}}, \href
{https://doi.org/10.1103/PhysRevD.65.025013}
{\path{doi:10.1103/PhysRevD.65.025013}}.

\bibitem{Reuter:2001ag}
M.~Reuter, F.~Saueressig, {Renormalization group flow of quantum gravity in the
Einstein-Hilbert truncation}, Phys. Rev. D65 (2002) 065016.
\newblock \href {http://arxiv.org/abs/hep-th/0110054}
{\path{arXiv:hep-th/0110054}}, \href
{https://doi.org/10.1103/PhysRevD.65.065016}
{\path{doi:10.1103/PhysRevD.65.065016}}.

\bibitem{Codello:2008vh}
A.~Codello, R.~Percacci, C.~Rahmede, {Investigating the Ultraviolet Properties
of Gravity with a Wilsonian Renormalization Group Equation}, Annals Phys. 324
(2009) 414--469.
\newblock \href {http://arxiv.org/abs/0805.2909} {\path{arXiv:0805.2909}},
\href {https://doi.org/10.1016/j.aop.2008.08.008}
{\path{doi:10.1016/j.aop.2008.08.008}}.

\bibitem{Benedetti:2010nr}
D.~Benedetti, K.~Groh, P.~F. Machado, F.~Saueressig, {The Universal RG
Machine}, JHEP 1106 (2011) 079.
\newblock \href {http://arxiv.org/abs/1012.3081} {\path{arXiv:1012.3081}},
\href {https://doi.org/10.1007/JHEP06(2011)079}
{\path{doi:10.1007/JHEP06(2011)079}}.

\bibitem{Kluth:2019vkg}
Y.~Kluth, D.~F. Litim, {Heat kernel coefficients on the sphere in any
dimension}, Eur.\ Phys.\ J.\ C 80~(3) (2020) 269.
\newblock \href {http://arxiv.org/abs/1910.00543} {\path{arXiv:1910.00543}},
\href {https://doi.org/10.1140/epjc/s10052-020-7784-2}
{\path{doi:10.1140/epjc/s10052-020-7784-2}}.

\bibitem{Gastmans:1977ad}
R.~Gastmans, R.~Kallosh, C.~Truffin, {Quantum Gravity Near Two-Dimensions},
Nucl. Phys. B133 (1978) 417.
\newblock \href {https://doi.org/10.1016/0550-3213(78)90234-1}
{\path{doi:10.1016/0550-3213(78)90234-1}}.

\bibitem{Christensen:1978sc}
S.~M. Christensen, M.~J. Duff, {QUANTUM GRAVITY IN TWO + epsilon DIMENSIONS},
Phys. Lett. B79 (1978) 213.
\newblock \href {https://doi.org/10.1016/0370-2693(78)90225-3}
{\path{doi:10.1016/0370-2693(78)90225-3}}.

\bibitem{Falls:2013bv}
K.~Falls, D.~Litim, K.~Nikolakopoulos, C.~Rahmede, {A bootstrap towards
asymptotic safety} (2013).
\newblock \href {http://arxiv.org/abs/1301.4191} {\path{arXiv:1301.4191}}.

\bibitem{Falls:2014tra}
K.~Falls, D.~F. Litim, K.~Nikolakopoulos, C.~Rahmede, {Further evidence for
asymptotic safety of quantum gravity}, Phys. Rev. D93~(10) (2016) 104022.
\newblock \href {http://arxiv.org/abs/1410.4815} {\path{arXiv:1410.4815}},
\href {https://doi.org/10.1103/PhysRevD.93.104022}
{\path{doi:10.1103/PhysRevD.93.104022}}.

\bibitem{Falls:2018ylp}
K.~G. Falls, D.~F. Litim, J.~Schr{\"o}der, {Aspects of asymptotic safety for
quantum gravity}, Phys.\ Rev.\ D 99~(12) (2019) 126015.
\newblock \href {http://arxiv.org/abs/1810.08550} {\path{arXiv:1810.08550}},
\href {https://doi.org/10.1103/PhysRevD.99.126015}
{\path{doi:10.1103/PhysRevD.99.126015}}.

\bibitem{Lauscher:2002sq}
O.~Lauscher, M.~Reuter, {Flow equation of quantum Einstein gravity in a higher-
derivative truncation}, Phys. Rev. D66 (2002) 025026.
\newblock \href {http://arxiv.org/abs/hep-th/0205062}
{\path{arXiv:hep-th/0205062}}, \href
{https://doi.org/10.1103/PhysRevD.66.025026}
{\path{doi:10.1103/PhysRevD.66.025026}}.

\bibitem{Benedetti:2009rx}
D.~Benedetti, P.~F. Machado, F.~Saueressig, {Asymptotic safety in
higher-derivative gravity}, Mod. Phys. Lett. A24 (2009) 2233--2241.
\newblock \href {http://arxiv.org/abs/0901.2984} {\path{arXiv:0901.2984}},
\href {https://doi.org/10.1142/S0217732309031521}
{\path{doi:10.1142/S0217732309031521}}.

\bibitem{Ohta:2013uca}
N.~Ohta, R.~Percacci, {Higher Derivative Gravity and Asymptotic Safety in
Diverse Dimensions}, Class. Quant. Grav. 31 (2014) 015024.
\newblock \href {http://arxiv.org/abs/1308.3398} {\path{arXiv:1308.3398}},
\href {https://doi.org/10.1088/0264-9381/31/1/015024}
{\path{doi:10.1088/0264-9381/31/1/015024}}.

\bibitem{Ohta:2015fcu}
N.~Ohta, R.~Percacci, G.~P. Vacca, {Renormalization Group Equation and scaling
solutions for f(R) gravity in exponential parametrization}, Eur. Phys. J.
C76~(2) (2016) 46.
\newblock \href {http://arxiv.org/abs/1511.09393} {\path{arXiv:1511.09393}},
\href {https://doi.org/10.1140/epjc/s10052-016-3895-1}
{\path{doi:10.1140/epjc/s10052-016-3895-1}}.

\bibitem{Ohta:2015efa}
N.~Ohta, R.~Percacci, G.~P. Vacca, {Flow equation for $f(R)$ gravity and some
of its exact solutions}, Phys. Rev. D92~(6) (2015) 061501.
\newblock \href {http://arxiv.org/abs/1507.00968} {\path{arXiv:1507.00968}},
\href {https://doi.org/10.1103/PhysRevD.92.061501}
{\path{doi:10.1103/PhysRevD.92.061501}}.

\bibitem{Gies:2016con}
H.~Gies, B.~Knorr, S.~Lippoldt, F.~Saueressig, {Gravitational Two-Loop
Counterterm Is Asymptotically Safe}, Phys. Rev. Lett. 116~(21) (2016) 211302.
\newblock \href {http://arxiv.org/abs/1601.01800} {\path{arXiv:1601.01800}},
\href {https://doi.org/10.1103/PhysRevLett.116.211302}
{\path{doi:10.1103/PhysRevLett.116.211302}}.

\bibitem{Falls:2017lst}
K.~Falls, C.~R. King, D.~F. Litim, K.~Nikolakopoulos, C.~Rahmede, {Asymptotic
safety of quantum gravity beyond Ricci scalars}, Phys. Rev. D97~(8) (2018)
086006.
\newblock \href {http://arxiv.org/abs/1801.00162} {\path{arXiv:1801.00162}},
\href {https://doi.org/10.1103/PhysRevD.97.086006}
{\path{doi:10.1103/PhysRevD.97.086006}}.

\bibitem{deBrito:2018jxt}
G.~P. De~Brito, N.~Ohta, A.~D. Pereira, A.~A. Tomaz, M.~Yamada, {Asymptotic
safety and field parametrization dependence in the $f(R)$ truncation}, Phys.
Rev. D98~(2) (2018) 026027.
\newblock \href {http://arxiv.org/abs/1805.09656} {\path{arXiv:1805.09656}},
\href {https://doi.org/10.1103/PhysRevD.98.026027}
{\path{doi:10.1103/PhysRevD.98.026027}}.

\bibitem{Eichhorn:2018akn}
A.~Eichhorn, P.~Labus, J.~M. Pawlowski, M.~Reichert, {Effective universality in
quantum gravity}, SciPost Phys. 5~(4) (2018) 031.
\newblock \href {http://arxiv.org/abs/1804.00012} {\path{arXiv:1804.00012}},
\href {https://doi.org/10.21468/SciPostPhys.5.4.031}
{\path{doi:10.21468/SciPostPhys.5.4.031}}.

\bibitem{Eichhorn:2018ydy}
A.~Eichhorn, S.~Lippoldt, J.~M. Pawlowski, M.~Reichert, M.~Schiffer, {How
perturbative is quantum gravity?} (2018).
\newblock \href {http://arxiv.org/abs/1810.02828} {\path{arXiv:1810.02828}}.

\bibitem{Eichhorn:2018nda}
A.~Eichhorn, S.~Lippoldt, M.~Schiffer, {Zooming in on fermions and quantum
gravity}, Phys. Rev. D99~(8) (2019) 086002.
\newblock \href {http://arxiv.org/abs/1812.08782} {\path{arXiv:1812.08782}},
\href {https://doi.org/10.1103/PhysRevD.99.086002}
{\path{doi:10.1103/PhysRevD.99.086002}}.

\bibitem{Codello:2006in}
A.~Codello, R.~Percacci, {Fixed Points of Higher Derivative Gravity}, Phys.
Rev. Lett. 97 (2006) 221301.
\newblock \href {http://arxiv.org/abs/hep-th/0607128}
{\path{arXiv:hep-th/0607128}}, \href
{https://doi.org/10.1103/PhysRevLett.97.221301}
{\path{doi:10.1103/PhysRevLett.97.221301}}.

\bibitem{Niedermaier:2009zz}
M.~R. Niedermaier, {Gravitational Fixed Points from Perturbation Theory}, Phys.
Rev. Lett. 103 (2009) 101303.
\newblock \href {https://doi.org/10.1103/PhysRevLett.103.101303}
{\path{doi:10.1103/PhysRevLett.103.101303}}.

\bibitem{Niedermaier:2010zz}
M.~Niedermaier, {Gravitational fixed points and asymptotic safety from
perturbation theory}, Nucl. Phys. B833 (2010) 226--270.
\newblock \href {https://doi.org/10.1016/j.nuclphysb.2010.01.016}
{\path{doi:10.1016/j.nuclphysb.2010.01.016}}.

\bibitem{Eichhorn:2013ug}
A.~Eichhorn, {Faddeev-Popov ghosts in quantum gravity beyond perturbation
theory}, Phys. Rev. D 87~(12) (2013) 124016.
\newblock \href {http://arxiv.org/abs/1301.0632} {\path{arXiv:1301.0632}},
\href {https://doi.org/10.1103/PhysRevD.87.124016}
{\path{doi:10.1103/PhysRevD.87.124016}}.

\bibitem{Eichhorn:2013xr}
A.~Eichhorn, {On unimodular quantum gravity}, Class. Quant. Grav. 30 (2013)
115016.
\newblock \href {http://arxiv.org/abs/1301.0879} {\path{arXiv:1301.0879}},
\href {https://doi.org/10.1088/0264-9381/30/11/115016}
{\path{doi:10.1088/0264-9381/30/11/115016}}.

\bibitem{Eichhorn:2015bna}
A.~Eichhorn, {The Renormalization Group flow of unimodular f(R) gravity}, JHEP
04 (2015) 096.
\newblock \href {http://arxiv.org/abs/1501.05848} {\path{arXiv:1501.05848}},
\href {https://doi.org/10.1007/JHEP04(2015)096}
{\path{doi:10.1007/JHEP04(2015)096}}.

\bibitem{Benedetti:2015zsw}
D.~Benedetti, {Essential nature of Newton's constant in unimodular gravity},
Gen.\ Rel.\ Grav. 48~(5) (2016) 68.
\newblock \href {http://arxiv.org/abs/1511.06560} {\path{arXiv:1511.06560}},
\href {https://doi.org/10.1007/s10714-016-2060-3}
{\path{doi:10.1007/s10714-016-2060-3}}.

\bibitem{Daum:2010qt}
J.~E. Daum, M.~Reuter, {Renormalization Group Flow of the Holst Action}, Phys.
Lett. B710 (2012) 215--218.
\newblock \href {http://arxiv.org/abs/1012.4280} {\path{arXiv:1012.4280}},
\href {https://doi.org/10.1016/j.physletb.2012.01.046}
{\path{doi:10.1016/j.physletb.2012.01.046}}.

\bibitem{Daum:2013fu}
J.~E. Daum, M.~Reuter, {Einstein-Cartan gravity, Asymptotic Safety, and the
running Immirzi parameter}, Annals Phys. 334 (2013) 351--419.
\newblock \href {http://arxiv.org/abs/1301.5135} {\path{arXiv:1301.5135}},
\href {https://doi.org/10.1016/j.aop.2013.04.002}
{\path{doi:10.1016/j.aop.2013.04.002}}.

\bibitem{Harst:2014vca}
U.~Harst, M.~Reuter, {A new functional flow equation for Einstei-Cartan quantum
gravity}, Annals Phys. 354 (2015) 637--704.
\newblock \href {http://arxiv.org/abs/1410.7003} {\path{arXiv:1410.7003}},
\href {https://doi.org/10.1016/j.aop.2015.01.006}
{\path{doi:10.1016/j.aop.2015.01.006}}.

\bibitem{Harst:2015eha}
U.~Harst, M.~Reuter, {On selfdual spin-connections and Asymptotic Safety},
Phys. Lett. B753 (2016) 395--400.
\newblock \href {http://arxiv.org/abs/1509.09122} {\path{arXiv:1509.09122}},
\href {https://doi.org/10.1016/j.physletb.2015.12.016}
{\path{doi:10.1016/j.physletb.2015.12.016}}.

\bibitem{Harst:2012ni}
U.~Harst, M.~Reuter, {The 'Tetrad only' theory space: Nonperturbative
renormalization flow and Asymptotic Safety}, JHEP 05 (2012) 005.
\newblock \href {http://arxiv.org/abs/1203.2158} {\path{arXiv:1203.2158}},
\href {https://doi.org/10.1007/JHEP05(2012)005}
{\path{doi:10.1007/JHEP05(2012)005}}.

\bibitem{Pagani:2015ema}
C.~Pagani, R.~Percacci, {Quantum gravity with torsion and non-metricity},
Class. Quant. Grav. 32~(19) (2015) 195019.
\newblock \href {http://arxiv.org/abs/1506.02882} {\path{arXiv:1506.02882}},
\href {https://doi.org/10.1088/0264-9381/32/19/195019}
{\path{doi:10.1088/0264-9381/32/19/195019}}.

\bibitem{Reuter:2015rta}
M.~Reuter, G.~M. Schollmeyer, {The metric on field space, functional
renormalization, and metric-torsion quantum gravity}, Annals Phys. 367 (2016)
125--191.
\newblock \href {http://arxiv.org/abs/1509.05041} {\path{arXiv:1509.05041}},
\href {https://doi.org/10.1016/j.aop.2015.12.004}
{\path{doi:10.1016/j.aop.2015.12.004}}.

\bibitem{Percacci:2010yk}
R.~Percacci, E.~Sezgin, {One Loop Beta Functions in Topologically Massive
Gravity}, Class. Quant. Grav. 27 (2010) 155009.
\newblock \href {http://arxiv.org/abs/1002.2640} {\path{arXiv:1002.2640}},
\href {https://doi.org/10.1088/0264-9381/27/15/155009}
{\path{doi:10.1088/0264-9381/27/15/155009}}.

\bibitem{Percacci:2013ii}
R.~Percacci, M.~J. Perry, C.~N. Pope, E.~Sezgin, {Beta Functions of
Topologically Massive Supergravity}, JHEP 03 (2014) 083.
\newblock \href {http://arxiv.org/abs/1302.0868} {\path{arXiv:1302.0868}},
\href {https://doi.org/10.1007/JHEP03(2014)083}
{\path{doi:10.1007/JHEP03(2014)083}}.

\bibitem{Binder:2020ifl}
M.~Binder, I.~Schmidt, {Functional Renormalization Group Flow of Massive
Gravity}, Eur.\ Phys.\ J.\ C 80~(3) (2020) 271.
\newblock \href {http://arxiv.org/abs/2003.09030} {\path{arXiv:2003.09030}},
\href {https://doi.org/10.1140/epjc/s10052-020-7835-8}
{\path{doi:10.1140/epjc/s10052-020-7835-8}}.

\bibitem{Ohta:2015zwa}
N.~Ohta, R.~Percacci, {Ultraviolet Fixed Points in Conformal Gravity and
General Quadratic Theories}, Class. Quant. Grav. 33 (2016) 035001.
\newblock \href {http://arxiv.org/abs/1506.05526} {\path{arXiv:1506.05526}},
\href {https://doi.org/10.1088/0264-9381/33/3/035001}
{\path{doi:10.1088/0264-9381/33/3/035001}}.

\bibitem{Reuter:2008qx}
M.~Reuter, H.~Weyer, {Conformal sector of Quantum Einstein Gravity in the local
potential approximation: non-Gaussian fixed point and a phase of
diffeomorphism invariance}, Phys. Rev. D80 (2009) 025001.
\newblock \href {http://arxiv.org/abs/0804.1475} {\path{arXiv:0804.1475}},
\href {https://doi.org/10.1103/PhysRevD.80.025001}
{\path{doi:10.1103/PhysRevD.80.025001}}.

\bibitem{Reuter:2008wj}
M.~Reuter, H.~Weyer, {Background Independence and Asymptotic Safety in
Conformally Reduced Gravity}, Phys. Rev. D79 (2009) 105005.
\newblock \href {http://arxiv.org/abs/0801.3287} {\path{arXiv:0801.3287}},
\href {https://doi.org/10.1103/PhysRevD.79.105005}
{\path{doi:10.1103/PhysRevD.79.105005}}.

\bibitem{Daum:2008gr}
J.-E. Daum, M.~Reuter, {Effective Potential of the Conformal Factor:
Gravitational Average Action and Dynamical Triangulations}, Adv. Sci. Lett. 2
(2009) 255--260.
\newblock \href {http://arxiv.org/abs/0806.3907} {\path{arXiv:0806.3907}},
\href {https://doi.org/10.1166/asl.2009.1033}
{\path{doi:10.1166/asl.2009.1033}}.

\bibitem{Machado:2009ph}
P.~F. Machado, R.~Percacci, {Conformally reduced quantum gravity revisited},
Phys. Rev. D80 (2009) 024020.
\newblock \href {http://arxiv.org/abs/0904.2510} {\path{arXiv:0904.2510}},
\href {https://doi.org/10.1103/PhysRevD.80.024020}
{\path{doi:10.1103/PhysRevD.80.024020}}.

\bibitem{Bonanno:2012dg}
A.~Bonanno, F.~Guarnieri, {Universality and Symmetry Breaking in Conformally
Reduced Quantum Gravity}, Phys. Rev. D 86 (2012) 105027.
\newblock \href {http://arxiv.org/abs/1206.6531} {\path{arXiv:1206.6531}},
\href {https://doi.org/10.1103/PhysRevD.86.105027}
{\path{doi:10.1103/PhysRevD.86.105027}}.

\bibitem{Dietz:2016gzg}
J.~A. Dietz, T.~R. Morris, Z.~H. Slade, {Fixed point structure of the conformal
factor field in quantum gravity}, Phys. Rev. D 94~(12) (2016) 124014.
\newblock \href {http://arxiv.org/abs/1605.07636} {\path{arXiv:1605.07636}},
\href {https://doi.org/10.1103/PhysRevD.94.124014}
{\path{doi:10.1103/PhysRevD.94.124014}}.

\bibitem{Bonanno:2004sy}
A.~Bonanno, M.~Reuter, {Proper time flow equation for gravity}, JHEP 02 (2005)
035.
\newblock \href {http://arxiv.org/abs/hep-th/0410191}
{\path{arXiv:hep-th/0410191}}, \href
{https://doi.org/10.1088/1126-6708/2005/02/035}
{\path{doi:10.1088/1126-6708/2005/02/035}}.

\bibitem{deAlwis:2018azs}
S.~P. De~Alwis, {Higher Derivative Corrections to Lower Order RG Flow
Equations} (2018).
\newblock \href {http://arxiv.org/abs/1809.04671} {\path{arXiv:1809.04671}}.

\bibitem{Christiansen:2014raa}
N.~Christiansen, B.~Knorr, J.~M. Pawlowski, A.~Rodigast, {Global Flows in
Quantum Gravity} (2014).
\newblock \href {http://arxiv.org/abs/1403.1232} {\path{arXiv:1403.1232}}.

\bibitem{Avramidi:1985ki}
I.~Avramidi, A.~Barvinsky, {ASYMPTOTIC FREEDOM IN HIGHER DERIVATIVE QUANTUM
GRAVITY}, Phys. Lett. B 159 (1985) 269--274.
\newblock \href {https://doi.org/10.1016/0370-2693(85)90248-5}
{\path{doi:10.1016/0370-2693(85)90248-5}}.

\bibitem{deBerredoPeixoto:2003pj}
G.~de~Berredo-Peixoto, I.~L. Shapiro, {Conformal quantum gravity with the
Gauss-Bonnet term}, Phys. Rev. D 70 (2004) 044024.
\newblock \href {http://arxiv.org/abs/hep-th/0307030}
{\path{arXiv:hep-th/0307030}}, \href
{https://doi.org/10.1103/PhysRevD.70.044024}
{\path{doi:10.1103/PhysRevD.70.044024}}.

\bibitem{deBerredoPeixoto:2004if}
G.~de~Berredo-Peixoto, I.~L. Shapiro, {Higher derivative quantum gravity with
Gauss-Bonnet term}, Phys. Rev. D 71 (2005) 064005.
\newblock \href {http://arxiv.org/abs/hep-th/0412249}
{\path{arXiv:hep-th/0412249}}, \href
{https://doi.org/10.1103/PhysRevD.71.064005}
{\path{doi:10.1103/PhysRevD.71.064005}}.

\bibitem{DOdorico:2014tyh}
G.~D'Odorico, F.~Saueressig, M.~Schutten, {Asymptotic Freedom in
Horava-Lifshitz Gravity}, Phys. Rev. Lett. 113~(17) (2014) 171101.
\newblock \href {http://arxiv.org/abs/1406.4366} {\path{arXiv:1406.4366}},
\href {https://doi.org/10.1103/PhysRevLett.113.171101}
{\path{doi:10.1103/PhysRevLett.113.171101}}.

\bibitem{Barvinsky:2017kob}
A.~O. Barvinsky, D.~Blas, M.~Herrero-Valea, S.~M. Sibiryakov, C.~F. Steinwachs,
{Ho\v{r}ava Gravity is Asymptotically Free in 2 + 1 Dimensions}, Phys. Rev.
Lett. 119~(21) (2017) 211301.
\newblock \href {http://arxiv.org/abs/1706.06809} {\path{arXiv:1706.06809}},
\href {https://doi.org/10.1103/PhysRevLett.119.211301}
{\path{doi:10.1103/PhysRevLett.119.211301}}.

\bibitem{Souma:1999at}
W.~Souma, {Nontrivial ultraviolet fixed point in quantum gravity},
Prog.Theor.Phys. 102 (1999) 181--195.
\newblock \href {http://arxiv.org/abs/hep-th/9907027}
{\path{arXiv:hep-th/9907027}}, \href {https://doi.org/10.1143/PTP.102.181}
{\path{doi:10.1143/PTP.102.181}}.

\bibitem{Litim:2003vp}
D.~F. Litim, {Fixed points of quantum gravity}, Phys.Rev.Lett. 92 (2004)
201301.
\newblock \href {http://arxiv.org/abs/hep-th/0312114}
{\path{arXiv:hep-th/0312114}}, \href
{https://doi.org/10.1103/PhysRevLett.92.201301}
{\path{doi:10.1103/PhysRevLett.92.201301}}.

\bibitem{Groh:2010ta}
K.~Groh, F.~Saueressig, {Ghost wave-function renormalization in Asymptotically
Safe Quantum Gravity}, J. Phys. A43 (2010) 365403.
\newblock \href {http://arxiv.org/abs/1001.5032} {\path{arXiv:1001.5032}},
\href {https://doi.org/10.1088/1751-8113/43/36/365403}
{\path{doi:10.1088/1751-8113/43/36/365403}}.

\bibitem{Eichhorn:2010tb}
A.~Eichhorn, H.~Gies, {Ghost anomalous dimension in asymptotically safe quantum
gravity}, Phys. Rev. D81 (2010) 104010.
\newblock \href {http://arxiv.org/abs/1001.5033} {\path{arXiv:1001.5033}},
\href {https://doi.org/10.1103/PhysRevD.81.104010}
{\path{doi:10.1103/PhysRevD.81.104010}}.

\bibitem{Nagy:2013hka}
S.~Nagy, B.~Fazekas, L.~Juhasz, K.~Sailer, {Critical exponents in quantum
Einstein gravity}, Phys. Rev. D88~(11) (2013) 116010.
\newblock \href {http://arxiv.org/abs/1307.0765} {\path{arXiv:1307.0765}},
\href {https://doi.org/10.1103/PhysRevD.88.116010}
{\path{doi:10.1103/PhysRevD.88.116010}}.

\bibitem{Falls:2014zba}
K.~Falls, {Asymptotic safety and the cosmological constant}, JHEP 01 (2016)
069.
\newblock \href {http://arxiv.org/abs/1408.0276} {\path{arXiv:1408.0276}},
\href {https://doi.org/10.1007/JHEP01(2016)069}
{\path{doi:10.1007/JHEP01(2016)069}}.

\bibitem{Gies:2015tca}
H.~Gies, B.~Knorr, S.~Lippoldt, {Generalized Parametrization Dependence in
Quantum Gravity}, Phys. Rev. D92~(8) (2015) 084020.
\newblock \href {http://arxiv.org/abs/1507.08859} {\path{arXiv:1507.08859}},
\href {https://doi.org/10.1103/PhysRevD.92.084020}
{\path{doi:10.1103/PhysRevD.92.084020}}.

\bibitem{Nagy:2017zvc}
S.~Nagy, B.~Fazekas, Z.~Peli, K.~Sailer, I.~Steib, {Regulator dependence of
fixed points in quantum Einstein gravity with $R^2$ truncation}, Class.\
Quant.\ Grav. 35~(5) (2018) 055001.
\newblock \href {http://arxiv.org/abs/1707.04934} {\path{arXiv:1707.04934}},
\href {https://doi.org/10.1088/1361-6382/aaa6ee}
{\path{doi:10.1088/1361-6382/aaa6ee}}.

\bibitem{Reuter:2012id}
M.~Reuter, F.~Saueressig, {Quantum Einstein Gravity}, New J. Phys. 14 (2012)
055022.
\newblock \href {http://arxiv.org/abs/1202.2274} {\path{arXiv:1202.2274}},
\href {https://doi.org/10.1088/1367-2630/14/5/055022}
{\path{doi:10.1088/1367-2630/14/5/055022}}.

\bibitem{Percacci:2017fkn}
R.~Percacci, {An Introduction to Covariant Quantum Gravity and Asymptotic
Safety}, Vol.~3 of {100 Years of General Relativity}, World Scientific, 2017.
\newblock \href {https://doi.org/10.1142/10369} {\path{doi:10.1142/10369}}.

\bibitem{Reuter:2019byg}
M.~Reuter, F.~Saueressig,
\href{https://www.cambridge.org/academic/subjects/physics/theoretical-physics-and-mathematical-physics/quantum-gravity-and-functional-renormalization-group-road-towards-asymptotic-safety?format=HB&isbn=9781107107328}{{Quantum
Gravity and the Functional Renormalization Group}}, Cambridge University
Press, 2019.
\newline\urlprefix\url{https://www.cambridge.org/academic/subjects/physics/theoretical-physics-and-mathematical-physics/quantum-gravity-and-functional-renormalization-group-road-towards-asymptotic-safety?format=HB&isbn=9781107107328}

\bibitem{Eichhorn:2020mte}
A.~Eichhorn, {Asymptotically safe gravity}, in: {57th International School of
Subnuclear Physics}: {In Search for the Unexpected}, 2020.
\newblock \href {http://arxiv.org/abs/2003.00044} {\path{arXiv:2003.00044}}.

\bibitem{Reichert:2020mja}
M.~Reichert, {Lecture notes: Functional Renormalisation Group and
Asymptotically Safe Quantum Gravity}, PoS Modave2019 (2020) 005.
\newblock \href {https://doi.org/10.22323/1.384.0005}
{\path{doi:10.22323/1.384.0005}}.

\bibitem{Codello:2007bd}
A.~Codello, R.~Percacci, C.~Rahmede, {Ultraviolet properties of f(R)-gravity},
Int. J. Mod. Phys. A23 (2008) 143--150.
\newblock \href {http://arxiv.org/abs/0705.1769} {\path{arXiv:0705.1769}},
\href {https://doi.org/10.1142/S0217751X08038135}
{\path{doi:10.1142/S0217751X08038135}}.

\bibitem{Demmel:2012ub}
M.~Demmel, F.~Saueressig, O.~Zanusso, {Fixed-Functionals of three-dimensional
Quantum Einstein Gravity}, JHEP 11 (2012) 131.
\newblock \href {http://arxiv.org/abs/1208.2038} {\path{arXiv:1208.2038}},
\href {https://doi.org/10.1007/JHEP11(2012)131}
{\path{doi:10.1007/JHEP11(2012)131}}.

\bibitem{Dietz:2012ic}
J.~A. Dietz, T.~R. Morris, {Asymptotic safety in the f(R) approximation}, JHEP
01 (2013) 108.
\newblock \href {http://arxiv.org/abs/1211.0955} {\path{arXiv:1211.0955}},
\href {https://doi.org/10.1007/JHEP01(2013)108}
{\path{doi:10.1007/JHEP01(2013)108}}.

\bibitem{Dietz:2013sba}
J.~A. Dietz, T.~R. Morris, {Redundant operators in the exact renormalisation
group and in the f(R) approximation to asymptotic safety}, JHEP 07 (2013)
064.
\newblock \href {http://arxiv.org/abs/1306.1223} {\path{arXiv:1306.1223}},
\href {https://doi.org/10.1007/JHEP07(2013)064}
{\path{doi:10.1007/JHEP07(2013)064}}.

\bibitem{Demmel:2015oqa}
M.~Demmel, F.~Saueressig, O.~Zanusso, {A proper fixed functional for
four-dimensional Quantum Einstein Gravity}, JHEP 08 (2015) 113.
\newblock \href {http://arxiv.org/abs/1504.07656} {\path{arXiv:1504.07656}},
\href {https://doi.org/10.1007/JHEP08(2015)113}
{\path{doi:10.1007/JHEP08(2015)113}}.

\bibitem{Gonzalez-Martin:2017gza}
S.~Gonzalez-Martin, T.~R. Morris, Z.~H. Slade, {Asymptotic solutions in
asymptotic safety}, Phys. Rev. D95~(10) (2017) 106010.
\newblock \href {http://arxiv.org/abs/1704.08873} {\path{arXiv:1704.08873}},
\href {https://doi.org/10.1103/PhysRevD.95.106010}
{\path{doi:10.1103/PhysRevD.95.106010}}.

\bibitem{Benedetti:2009gn}
D.~Benedetti, P.~F. Machado, F.~Saueressig, {Taming perturbative divergences in
asymptotically safe gravity}, Nucl. Phys. B824 (2010) 168--191.
\newblock \href {http://arxiv.org/abs/0902.4630} {\path{arXiv:0902.4630}},
\href {https://doi.org/10.1016/j.nuclphysb.2009.08.023}
{\path{doi:10.1016/j.nuclphysb.2009.08.023}}.

\bibitem{Falls:2020qhj}
K.~Falls, N.~Ohta, R.~Percacci, {Towards the determination of the dimension of
the critical surface in asymptotically safe gravity} (4 2020).
\newblock \href {http://arxiv.org/abs/2004.04126} {\path{arXiv:2004.04126}}.

\bibitem{Becker:2012js}
D.~Becker, M.~Reuter, Running boundary actions, asymptotic safety, and black
hole thermodynamics, JHEP 07 (2012) 172.
\newblock \href {http://arxiv.org/abs/1205.3583} {\path{arXiv:1205.3583}},
\href {https://doi.org/10.1007/JHEP07(2012)172}
{\path{doi:10.1007/JHEP07(2012)172}}.

\bibitem{Falls:2017cze}
K.~Falls, {Physical renormalization schemes and asymptotic safety in quantum
gravity}, Phys. Rev. D96~(12) (2017) 126016.
\newblock \href {http://arxiv.org/abs/1702.03577} {\path{arXiv:1702.03577}},
\href {https://doi.org/10.1103/PhysRevD.96.126016}
{\path{doi:10.1103/PhysRevD.96.126016}}.

\bibitem{Fischer:2006fz}
P.~Fischer, D.~F. Litim, {Fixed points of quantum gravity in extra dimensions},
Phys.Lett. B638 (2006) 497--502.
\newblock \href {http://arxiv.org/abs/hep-th/0602203}
{\path{arXiv:hep-th/0602203}}, \href
{https://doi.org/10.1016/j.physletb.2006.05.073}
{\path{doi:10.1016/j.physletb.2006.05.073}}.

\bibitem{Nink:2012vd}
A.~Nink, M.~Reuter, {On the physical mechanism underlying Asymptotic Safety},
JHEP 01 (2013) 062.
\newblock \href {http://arxiv.org/abs/1208.0031} {\path{arXiv:1208.0031}},
\href {https://doi.org/10.1007/JHEP01(2013)062}
{\path{doi:10.1007/JHEP01(2013)062}}.

\bibitem{Falls:2015qga}
K.~Falls, {Renormalization of Newton's constant}, Phys.\ Rev.\ D 92~(12) (2015)
124057.
\newblock \href {http://arxiv.org/abs/1501.05331} {\path{arXiv:1501.05331}},
\href {https://doi.org/10.1103/PhysRevD.92.124057}
{\path{doi:10.1103/PhysRevD.92.124057}}.

\bibitem{Biemans:2016rvp}
J.~Biemans, A.~Platania, F.~Saueressig, {Quantum gravity on foliated
spacetimes: Asymptotically safe and sound}, Phys. Rev. D95~(8) (2017) 086013.
\newblock \href {http://arxiv.org/abs/1609.04813} {\path{arXiv:1609.04813}},
\href {https://doi.org/10.1103/PhysRevD.95.086013}
{\path{doi:10.1103/PhysRevD.95.086013}}.

\bibitem{Ohta:2016npm}
N.~Ohta, R.~Percacci, A.~D. Pereira, {Gauges and functional measures in quantum
gravity I: Einstein theory}, JHEP 06 (2016) 115.
\newblock \href {http://arxiv.org/abs/1605.00454} {\path{arXiv:1605.00454}},
\href {https://doi.org/10.1007/JHEP06(2016)115}
{\path{doi:10.1007/JHEP06(2016)115}}.

\bibitem{Bosma:2019aiu}
L.~Bosma, B.~Knorr, F.~Saueressig, {Resolving Spacetime Singularities within
Asymptotic Safety}, Phys. Rev. Lett. 123~(10) (2019) 101301.
\newblock \href {http://arxiv.org/abs/1904.04845} {\path{arXiv:1904.04845}},
\href {https://doi.org/10.1103/PhysRevLett.123.101301}
{\path{doi:10.1103/PhysRevLett.123.101301}}.

\bibitem{Knorr:2019atm}
B.~Knorr, C.~Ripken, F.~Saueressig, {Form Factors in Asymptotic Safety:
conceptual ideas and computational toolbox}, Class. Quant. Grav. 36~(23)
(2019) 234001.
\newblock \href {http://arxiv.org/abs/1907.02903} {\path{arXiv:1907.02903}},
\href {https://doi.org/10.1088/1361-6382/ab4a53}
{\path{doi:10.1088/1361-6382/ab4a53}}.

\bibitem{Christiansen:2012rx}
N.~Christiansen, D.~F. Litim, J.~M. Pawlowski, A.~Rodigast, {Fixed points and
infrared completion of quantum gravity}, Phys.Lett. B728 (2014) 114--117.
\newblock \href {http://arxiv.org/abs/1209.4038} {\path{arXiv:1209.4038}},
\href {https://doi.org/10.1016/j.physletb.2013.11.025}
{\path{doi:10.1016/j.physletb.2013.11.025}}.

\bibitem{Christiansen:2015rva}
N.~Christiansen, B.~Knorr, J.~Meibohm, J.~M. Pawlowski, M.~Reichert, {Local
Quantum Gravity}, Phys. Rev. D92~(12) (2015) 121501.
\newblock \href {http://arxiv.org/abs/1506.07016} {\path{arXiv:1506.07016}},
\href {https://doi.org/10.1103/PhysRevD.92.121501}
{\path{doi:10.1103/PhysRevD.92.121501}}.

\bibitem{Knorr:2017fus}
B.~Knorr, S.~Lippoldt, {Correlation functions on a curved background}, Phys.
Rev. D96~(6) (2017) 065020.
\newblock \href {http://arxiv.org/abs/1707.01397} {\path{arXiv:1707.01397}},
\href {https://doi.org/10.1103/PhysRevD.96.065020}
{\path{doi:10.1103/PhysRevD.96.065020}}.

\bibitem{Burger:2019upn}
B.~B{\"u}rger, J.~M. Pawlowski, M.~Reichert, B.-J. Schaefer, {Curvature
dependence of quantum gravity with scalars} (2019).
\newblock \href {http://arxiv.org/abs/1912.01624} {\path{arXiv:1912.01624}}.

\bibitem{Morris:2016spn}
T.~R. Morris, {Large curvature and background scale independence in
single-metric approximations to asymptotic safety}, JHEP 11 (2016) 160.
\newblock \href {http://arxiv.org/abs/1610.03081} {\path{arXiv:1610.03081}},
\href {https://doi.org/10.1007/JHEP11(2016)160}
{\path{doi:10.1007/JHEP11(2016)160}}.

\bibitem{Percacci:2016arh}
R.~Percacci, G.~P. Vacca, {The background scale Ward identity in quantum
gravity}, Eur. Phys. J. C77~(1) (2017) 52.
\newblock \href {http://arxiv.org/abs/1611.07005} {\path{arXiv:1611.07005}},
\href {https://doi.org/10.1140/epjc/s10052-017-4619-x}
{\path{doi:10.1140/epjc/s10052-017-4619-x}}.

\bibitem{Nieto:2017ddk}
C.~M. Nieto, R.~Percacci, V.~Skrinjar, {Split Weyl transformations in quantum
gravity}, Phys. Rev. D96~(10) (2017) 106019.
\newblock \href {http://arxiv.org/abs/1708.09760} {\path{arXiv:1708.09760}},
\href {https://doi.org/10.1103/PhysRevD.96.106019}
{\path{doi:10.1103/PhysRevD.96.106019}}.

\bibitem{Ohta:2017dsq}
N.~Ohta, {Background Scale Independence in Quantum Gravity}, PTEP 2017~(3)
(2017) 033E02.
\newblock \href {http://arxiv.org/abs/1701.01506} {\path{arXiv:1701.01506}},
\href {https://doi.org/10.1093/ptep/ptx020} {\path{doi:10.1093/ptep/ptx020}}.

\bibitem{Pagani:2019vfm}
C.~Pagani, M.~Reuter, {Background Independent Quantum Field Theory and
Gravitating Vacuum Fluctuations}, Annals Phys. 411 (2019) 167972.
\newblock \href {http://arxiv.org/abs/1906.02507} {\path{arXiv:1906.02507}},
\href {https://doi.org/10.1016/j.aop.2019.167972}
{\path{doi:10.1016/j.aop.2019.167972}}.

\bibitem{Manrique:2009uh}
E.~Manrique, M.~Reuter, {Bimetric Truncations for Quantum Einstein Gravity and
Asymptotic Safety}, Annals Phys. 325 (2010) 785--815.
\newblock \href {http://arxiv.org/abs/0907.2617} {\path{arXiv:0907.2617}},
\href {https://doi.org/10.1016/j.aop.2009.11.009}
{\path{doi:10.1016/j.aop.2009.11.009}}.

\bibitem{Manrique:2010am}
E.~Manrique, M.~Reuter, F.~Saueressig, {Bimetric Renormalization Group Flows in
Quantum Einstein Gravity}, Annals Phys. 326 (2011) 463--485.
\newblock \href {http://arxiv.org/abs/1006.0099} {\path{arXiv:1006.0099}},
\href {https://doi.org/10.1016/j.aop.2010.11.006}
{\path{doi:10.1016/j.aop.2010.11.006}}.

\bibitem{Becker:2014qya}
D.~Becker, M.~Reuter, {En route to Background Independence: Broken
split-symmetry, and how to restore it with bi-metric average actions}, Annals
Phys. 350 (2014) 225--301.
\newblock \href {http://arxiv.org/abs/1404.4537} {\path{arXiv:1404.4537}},
\href {https://doi.org/10.1016/j.aop.2014.07.023}
{\path{doi:10.1016/j.aop.2014.07.023}}.

\bibitem{Codello:2013fpa}
A.~Codello, G.~D'Odorico, C.~Pagani, {Consistent closure of renormalization
group flow equations in quantum gravity}, Phys.\ Rev.\ D 89~(8) (2014)
081701.
\newblock \href {http://arxiv.org/abs/1304.4777} {\path{arXiv:1304.4777}},
\href {https://doi.org/10.1103/PhysRevD.89.081701}
{\path{doi:10.1103/PhysRevD.89.081701}}.

\bibitem{Knorr:2017mhu}
B.~Knorr, {Infinite order quantum-gravitational correlations}, Class. Quant.
Grav. 35~(11) (2018) 115005.
\newblock \href {http://arxiv.org/abs/1710.07055} {\path{arXiv:1710.07055}},
\href {https://doi.org/10.1088/1361-6382/aabaa0}
{\path{doi:10.1088/1361-6382/aabaa0}}.

\bibitem{Nink:2014yya}
A.~Nink, {Field Parametrization Dependence in Asymptotically Safe Quantum
Gravity}, Phys. Rev. D 91~(4) (2015) 044030.
\newblock \href {http://arxiv.org/abs/1410.7816} {\path{arXiv:1410.7816}},
\href {https://doi.org/10.1103/PhysRevD.91.044030}
{\path{doi:10.1103/PhysRevD.91.044030}}.

\bibitem{Percacci:2015wwa}
R.~Percacci, G.~P. Vacca, {Search of scaling solutions in scalar-tensor
gravity}, Eur. Phys. J. C75~(5) (2015) 188.
\newblock \href {http://arxiv.org/abs/1501.00888} {\path{arXiv:1501.00888}},
\href {https://doi.org/10.1140/epjc/s10052-015-3410-0}
{\path{doi:10.1140/epjc/s10052-015-3410-0}}.

\bibitem{Bonanno:2020bil}
A.~Bonanno, A.~Eichhorn, H.~Gies, J.~M. Pawlowski, R.~Percacci, M.~Reuter,
F.~Saueressig, G.~P. Vacca, {Critical reflections on asymptotically safe
gravity} (4 2020).
\newblock \href {http://arxiv.org/abs/2004.06810} {\path{arXiv:2004.06810}}.

\bibitem{Manrique:2011jc}
E.~Manrique, S.~Rechenberger, F.~Saueressig, {Asymptotically Safe Lorentzian
Gravity}, Phys.Rev.Lett. 106 (2011) 251302.
\newblock \href {http://arxiv.org/abs/1102.5012} {\path{arXiv:1102.5012}},
\href {https://doi.org/10.1103/PhysRevLett.106.251302}
{\path{doi:10.1103/PhysRevLett.106.251302}}.

\bibitem{Visser:2017atf}
M.~Visser, {How to Wick rotate generic curved spacetime} (2 2017).
\newblock \href {http://arxiv.org/abs/1702.05572} {\path{arXiv:1702.05572}}.

\bibitem{Donoghue:2019clr}
J.~F. Donoghue, {A Critique of the Asymptotic Safety Program}, Front.\ in Phys.
8 (2020) 56.
\newblock \href {http://arxiv.org/abs/1911.02967} {\path{arXiv:1911.02967}},
\href {https://doi.org/10.3389/fphy.2020.00056}
{\path{doi:10.3389/fphy.2020.00056}}.

\bibitem{Biemans:2017zca}
J.~Biemans, A.~Platania, F.~Saueressig, {Renormalization group fixed points of
foliated gravity-matter systems}, JHEP 05 (2017) 093.
\newblock \href {http://arxiv.org/abs/1702.06539} {\path{arXiv:1702.06539}},
\href {https://doi.org/10.1007/JHEP05(2017)093}
{\path{doi:10.1007/JHEP05(2017)093}}.

\bibitem{Houthoff:2017oam}
W.~B. Houthoff, A.~Kurov, F.~Saueressig, {Impact of topology in foliated
Quantum Einstein Gravity}, Eur. Phys. J. C77 (2017) 491.
\newblock \href {http://arxiv.org/abs/1705.01848} {\path{arXiv:1705.01848}},
\href {https://doi.org/10.1140/epjc/s10052-017-5046-8}
{\path{doi:10.1140/epjc/s10052-017-5046-8}}.

\bibitem{Ambjorn:2000dv}
J.~Ambjorn, J.~Jurkiewicz, R.~Loll, {A Nonperturbative Lorentzian path integral
for gravity}, Phys. Rev. Lett. 85 (2000) 924--927.
\newblock \href {http://arxiv.org/abs/hep-th/0002050}
{\path{arXiv:hep-th/0002050}}, \href
{https://doi.org/10.1103/PhysRevLett.85.924}
{\path{doi:10.1103/PhysRevLett.85.924}}.

\bibitem{Ambjorn:2004qm}
J.~Ambjorn, J.~Jurkiewicz, R.~Loll, {Emergence of a 4-D world from causal
quantum gravity}, Phys. Rev. Lett. 93 (2004) 131301.
\newblock \href {http://arxiv.org/abs/hep-th/0404156}
{\path{arXiv:hep-th/0404156}}, \href
{https://doi.org/10.1103/PhysRevLett.93.131301}
{\path{doi:10.1103/PhysRevLett.93.131301}}.

\bibitem{Ambjorn:2011cg}
J.~Ambjorn, S.~Jordan, J.~Jurkiewicz, R.~Loll, {A Second-order phase transition
in CDT}, Phys. Rev. Lett. 107 (2011) 211303.
\newblock \href {http://arxiv.org/abs/1108.3932} {\path{arXiv:1108.3932}},
\href {https://doi.org/10.1103/PhysRevLett.107.211303}
{\path{doi:10.1103/PhysRevLett.107.211303}}.

\bibitem{Ambjorn:2012jv}
J.~Ambjorn, A.~Goerlich, J.~Jurkiewicz, R.~Loll, {Nonperturbative Quantum
Gravity}, Phys. Rept. 519 (2012) 127--210.
\newblock \href {http://arxiv.org/abs/1203.3591} {\path{arXiv:1203.3591}},
\href {https://doi.org/10.1016/j.physrep.2012.03.007}
{\path{doi:10.1016/j.physrep.2012.03.007}}.

\bibitem{Bonanno:2015fga}
A.~Bonanno, A.~Platania, {Asymptotically safe inflation from quadratic
gravity}, Phys. Lett. B750 (2015) 638--642.
\newblock \href {http://arxiv.org/abs/1507.03375} {\path{arXiv:1507.03375}},
\href {https://doi.org/10.1016/j.physletb.2015.10.005}
{\path{doi:10.1016/j.physletb.2015.10.005}}.

\bibitem{Arici:2017whq}
F.~Arici, D.~Becker, C.~Ripken, F.~Saueressig, W.~D. van Suijlekom, {Reflection
positivity in higher derivative scalar theories}, J. Math. Phys. 59~(8)
(2018) 082302.
\newblock \href {http://arxiv.org/abs/1712.04308} {\path{arXiv:1712.04308}},
\href {https://doi.org/10.1063/1.5027231} {\path{doi:10.1063/1.5027231}}.

\bibitem{Becker:2017tcx}
D.~Becker, C.~Ripken, F.~Saueressig, {On avoiding Ostrogradski instabilities
within Asymptotic Safety}, JHEP 12 (2017) 121.
\newblock \href {http://arxiv.org/abs/1709.09098} {\path{arXiv:1709.09098}},
\href {https://doi.org/10.1007/JHEP12(2017)121}
{\path{doi:10.1007/JHEP12(2017)121}}.

\bibitem{Nink:2015lmq}
A.~Nink, M.~Reuter, {The unitary conformal field theory behind 2D Asymptotic
Safety}, JHEP 02 (2016) 167.
\newblock \href {http://arxiv.org/abs/1512.06805} {\path{arXiv:1512.06805}},
\href {https://doi.org/10.1007/JHEP02(2016)167}
{\path{doi:10.1007/JHEP02(2016)167}}.

\bibitem{Ambjorn:2001cv}
J.~Ambjorn, J.~Jurkiewicz, R.~Loll, {Dynamically triangulating Lorentzian
quantum gravity}, Nucl. Phys. B 610 (2001) 347--382.
\newblock \href {http://arxiv.org/abs/hep-th/0105267}
{\path{arXiv:hep-th/0105267}}, \href
{https://doi.org/10.1016/S0550-3213(01)00297-8}
{\path{doi:10.1016/S0550-3213(01)00297-8}}.

\bibitem{Anber:2011ut}
M.~M. Anber, J.~F. Donoghue, {On the running of the gravitational constant},
Phys. Rev. D85 (2012) 104016.
\newblock \href {http://arxiv.org/abs/1111.2875} {\path{arXiv:1111.2875}},
\href {https://doi.org/10.1103/PhysRevD.85.104016}
{\path{doi:10.1103/PhysRevD.85.104016}}.

\bibitem{Bonanno:1998ye}
A.~Bonanno, M.~Reuter, {Quantum gravity effects near the null black hole
singularity}, Phys.\ Rev.\ D 60 (1999) 084011.
\newblock \href {http://arxiv.org/abs/gr-qc/9811026}
{\path{arXiv:gr-qc/9811026}}, \href
{https://doi.org/10.1103/PhysRevD.60.084011}
{\path{doi:10.1103/PhysRevD.60.084011}}.

\bibitem{Bonanno:2000ep}
A.~Bonanno, M.~Reuter, {Renormalization group improved black hole spacetimes},
Phys. Rev. D62 (2000) 043008.
\newblock \href {http://arxiv.org/abs/hep-th/0002196}
{\path{arXiv:hep-th/0002196}}, \href
{https://doi.org/10.1103/PhysRevD.62.043008}
{\path{doi:10.1103/PhysRevD.62.043008}}.

\bibitem{Bonanno:2006eu}
A.~Bonanno, M.~Reuter, {Spacetime structure of an evaporating black hole in
quantum gravity}, Phys. Rev. D73 (2006) 083005.
\newblock \href {http://arxiv.org/abs/hep-th/0602159}
{\path{arXiv:hep-th/0602159}}, \href
{https://doi.org/10.1103/PhysRevD.73.083005}
{\path{doi:10.1103/PhysRevD.73.083005}}.

\bibitem{Cai:2010zh}
Y.-F. Cai, D.~A. Easson, {Black holes in an asymptotically safe gravity theory
with higher derivatives}, JCAP 1009 (2010) 002.
\newblock \href {http://arxiv.org/abs/1007.1317} {\path{arXiv:1007.1317}},
\href {https://doi.org/10.1088/1475-7516/2010/09/002}
{\path{doi:10.1088/1475-7516/2010/09/002}}.

\bibitem{Reuter:2010xb}
M.~Reuter, E.~Tuiran, {Quantum Gravity Effects in the Kerr Spacetime}, Phys.
Rev. D83 (2011) 044041.
\newblock \href {http://arxiv.org/abs/1009.3528} {\path{arXiv:1009.3528}},
\href {https://doi.org/10.1103/PhysRevD.83.044041}
{\path{doi:10.1103/PhysRevD.83.044041}}.

\bibitem{Falls:2010he}
K.~Falls, D.~F. Litim, A.~Raghuraman, {Black Holes and Asymptotically Safe
Gravity}, Int.J.Mod.Phys. A27 (2012) 1250019.
\newblock \href {http://arxiv.org/abs/1002.0260} {\path{arXiv:1002.0260}},
\href {https://doi.org/10.1142/S0217751X12500194}
{\path{doi:10.1142/S0217751X12500194}}.

\bibitem{Falls:2012nd}
K.~Falls, D.~F. Litim, {Black hole thermodynamics under the microscope} (2012).
\newblock \href {http://arxiv.org/abs/1212.1821} {\path{arXiv:1212.1821}}.

\bibitem{Torres:2013cya}
R.~Torres, F.~Fayos, O.~Lorente-Espin, {Evaporation of (quantum) black holes
and energy conservation}, Phys.\ Lett.\ B 720 (2013) 198--204.
\newblock \href {http://arxiv.org/abs/1308.4318} {\path{arXiv:1308.4318}},
\href {https://doi.org/10.1016/j.physletb.2013.01.061}
{\path{doi:10.1016/j.physletb.2013.01.061}}.

\bibitem{Litim:2013gga}
D.~F. Litim, K.~Nikolakopoulos, {Quantum gravity effects in Myers-Perry
space-times}, JHEP 04 (2014) 021.
\newblock \href {http://arxiv.org/abs/1308.5630} {\path{arXiv:1308.5630}},
\href {https://doi.org/10.1007/JHEP04(2014)021}
{\path{doi:10.1007/JHEP04(2014)021}}.

\bibitem{Koch:2013owa}
B.~Koch, F.~Saueressig, {Structural aspects of asymptotically safe black
holes}, Class. Quant. Grav. 31 (2014) 015006.
\newblock \href {http://arxiv.org/abs/1306.1546} {\path{arXiv:1306.1546}},
\href {https://doi.org/10.1088/0264-9381/31/1/015006}
{\path{doi:10.1088/0264-9381/31/1/015006}}.

\bibitem{Kofinas:2015sna}
G.~Kofinas, V.~Zarikas, {Avoidance of singularities in asymptotically safe
Quantum Einstein Gravity}, JCAP 1510~(10) (2015) 069.
\newblock \href {http://arxiv.org/abs/1506.02965} {\path{arXiv:1506.02965}},
\href {https://doi.org/10.1088/1475-7516/2015/10/069}
{\path{doi:10.1088/1475-7516/2015/10/069}}.

\bibitem{Torres:2017ygl}
R.~Torres, {Nonsingular black holes, the cosmological constant, and asymptotic
safety}, Phys. Rev. D95~(12) (2017) 124004.
\newblock \href {http://arxiv.org/abs/1703.09997} {\path{arXiv:1703.09997}},
\href {https://doi.org/10.1103/PhysRevD.95.124004}
{\path{doi:10.1103/PhysRevD.95.124004}}.

\bibitem{Pawlowski:2018swz}
J.~M. Pawlowski, D.~Stock, {Quantum-improved Schwarzschild-(A)dS and Kerr-(A)dS
spacetimes}, Phys. Rev. D98~(10) (2018) 106008.
\newblock \href {http://arxiv.org/abs/1807.10512} {\path{arXiv:1807.10512}},
\href {https://doi.org/10.1103/PhysRevD.98.106008}
{\path{doi:10.1103/PhysRevD.98.106008}}.

\bibitem{Adeifeoba:2018ydh}
A.~Adeifeoba, A.~Eichhorn, A.~Platania, {Towards conditions for black-hole
singularity-resolution in asymptotically safe quantum gravity}, Class. Quant.
Grav. 35~(22) (2018) 225007.
\newblock \href {http://arxiv.org/abs/1808.03472} {\path{arXiv:1808.03472}},
\href {https://doi.org/10.1088/1361-6382/aae6ef}
{\path{doi:10.1088/1361-6382/aae6ef}}.

\bibitem{Platania:2019kyx}
A.~Platania, {Dynamical renormalization of black-hole spacetimes} (2019).
\newblock \href {http://arxiv.org/abs/1903.10411} {\path{arXiv:1903.10411}}.

\bibitem{Dymnikova:1992ux}
I.~Dymnikova, {Vacuum nonsingular black hole}, Gen. Rel. Grav. 24 (1992)
235--242.
\newblock \href {https://doi.org/10.1007/BF00760226}
{\path{doi:10.1007/BF00760226}}.

\bibitem{Held:2019xde}
A.~Held, R.~Gold, A.~Eichhorn, Asymptotic safety casts its shadow, JCAP 06
(2019) 029.
\newblock \href {http://arxiv.org/abs/1904.07133} {\path{arXiv:1904.07133}},
\href {https://doi.org/10.1088/1475-7516/2019/06/029}
{\path{doi:10.1088/1475-7516/2019/06/029}}.

\bibitem{Kumar:2019ohr}
R.~Kumar, B.~P. Singh, S.~G. Ghosh, {Rotating black hole shadow in
asymptotically safe gravity} (2019).
\newblock \href {http://arxiv.org/abs/1904.07652} {\path{arXiv:1904.07652}}.

\bibitem{Casadio:2010fw}
R.~Casadio, S.~D.~H. Hsu, B.~Mirza, {Asymptotic Safety, Singularities, and
Gravitational Collapse}, Phys. Lett. B695 (2011) 317--319.
\newblock \href {http://arxiv.org/abs/1008.2768} {\path{arXiv:1008.2768}},
\href {https://doi.org/10.1016/j.physletb.2010.10.060}
{\path{doi:10.1016/j.physletb.2010.10.060}}.

\bibitem{Fayos:2011zza}
F.~Fayos, R.~Torres, {A quantum improvement to the gravitational collapse of
radiating stars}, Class.\ Quant.\ Grav. 28 (2011) 105004.
\newblock \href {https://doi.org/10.1088/0264-9381/28/10/105004}
{\path{doi:10.1088/0264-9381/28/10/105004}}.

\bibitem{Torres:2014gta}
R.~Torres, {Singularity-free gravitational collapse and asymptotic safety},
Phys. Lett. B733 (2014) 21--24.
\newblock \href {http://arxiv.org/abs/1404.7655} {\path{arXiv:1404.7655}},
\href {https://doi.org/10.1016/j.physletb.2014.04.010}
{\path{doi:10.1016/j.physletb.2014.04.010}}.

\bibitem{Torres:2014pea}
R.~Torres, F.~Fayos, {Singularity free gravitational collapse in an effective
dynamical quantum spacetime}, Phys. Lett. B733 (2014) 169--175.
\newblock \href {http://arxiv.org/abs/1405.7922} {\path{arXiv:1405.7922}},
\href {https://doi.org/10.1016/j.physletb.2014.04.038}
{\path{doi:10.1016/j.physletb.2014.04.038}}.

\bibitem{Bonanno:2016dyv}
A.~Bonanno, B.~Koch, A.~Platania, {Cosmic Censorship in Quantum Einstein
Gravity}, Class.\ Quant.\ Grav. 34~(9) (2017) 095012.
\newblock \href {http://arxiv.org/abs/1610.05299} {\path{arXiv:1610.05299}},
\href {https://doi.org/10.1088/1361-6382/aa6788}
{\path{doi:10.1088/1361-6382/aa6788}}.

\bibitem{Bonanno:2017zen}
A.~Bonanno, B.~Koch, A.~Platania, {Gravitational collapse in Quantum Einstein
Gravity}, Found. Phys. 48~(10) (2018) 1393--1406.
\newblock \href {http://arxiv.org/abs/1710.10845} {\path{arXiv:1710.10845}},
\href {https://doi.org/10.1007/s10701-018-0195-7}
{\path{doi:10.1007/s10701-018-0195-7}}.

\bibitem{Bonanno:2017kta}
A.~Bonanno, B.~Koch, A.~Platania, {Asymptotically Safe gravitational collapse:
Kuroda-Papapetrou RG-improved model}, PoS CORFU2016 (2017) 058.
\newblock \href {https://doi.org/10.22323/1.292.0058}
{\path{doi:10.22323/1.292.0058}}.

\bibitem{Bonanno:2019ilz}
A.~Bonanno, R.~Casadio, A.~Platania, {Gravitational antiscreening in stellar
interiors}, JCAP 01~(01) (2020) 022.
\newblock \href {http://arxiv.org/abs/1910.11393} {\path{arXiv:1910.11393}},
\href {https://doi.org/10.1088/1475-7516/2020/01/022}
{\path{doi:10.1088/1475-7516/2020/01/022}}.

\bibitem{Bonanno:2017pkg}
A.~Bonanno, F.~Saueressig, {Asymptotically safe cosmology -- A status report},
Comptes Rendus Physique 18 (2017) 254--264.
\newblock \href {http://arxiv.org/abs/1702.04137} {\path{arXiv:1702.04137}},
\href {https://doi.org/10.1016/j.crhy.2017.02.002}
{\path{doi:10.1016/j.crhy.2017.02.002}}.

\bibitem{Platania:2020lqb}
A.~Platania, {From renormalization group flows to cosmology} (3 2020).
\newblock \href {http://arxiv.org/abs/2003.13656} {\path{arXiv:2003.13656}}.

\bibitem{Lehners:2019ibe}
J.-L. Lehners, K.~Stelle, {A Safe Beginning for the Universe?}, Phys.\ Rev.\ D
100~(8) (2019) 083540.
\newblock \href {http://arxiv.org/abs/1909.01169} {\path{arXiv:1909.01169}},
\href {https://doi.org/10.1103/PhysRevD.100.083540}
{\path{doi:10.1103/PhysRevD.100.083540}}.

\bibitem{Weinberg:2009wa}
S.~Weinberg, {Asymptotically Safe Inflation}, Phys. Rev. D81 (2010) 083535.
\newblock \href {http://arxiv.org/abs/0911.3165} {\path{arXiv:0911.3165}},
\href {https://doi.org/10.1103/PhysRevD.81.083535}
{\path{doi:10.1103/PhysRevD.81.083535}}.

\bibitem{Bonanno:2001xi}
A.~Bonanno, M.~Reuter, {Cosmology of the Planck era from a renormalization
group for quantum gravity}, Phys. Rev. D65 (2002) 043508.
\newblock \href {http://arxiv.org/abs/hep-th/0106133}
{\path{arXiv:hep-th/0106133}}, \href
{https://doi.org/10.1103/PhysRevD.65.043508}
{\path{doi:10.1103/PhysRevD.65.043508}}.

\bibitem{Reuter:2005kb}
M.~Reuter, F.~Saueressig, {From big bang to asymptotic de Sitter: Complete
cosmologies in a quantum gravity framework}, JCAP 0509 (2005) 012.
\newblock \href {http://arxiv.org/abs/hep-th/0507167}
{\path{arXiv:hep-th/0507167}}, \href
{https://doi.org/10.1088/1475-7516/2005/09/012}
{\path{doi:10.1088/1475-7516/2005/09/012}}.

\bibitem{Bonanno:2010bt}
A.~Bonanno, A.~Contillo, R.~Percacci, {Inflationary solutions in asymptotically
safe f(R) theories}, Class. Quant. Grav. 28 (2011) 145026.
\newblock \href {http://arxiv.org/abs/1006.0192} {\path{arXiv:1006.0192}},
\href {https://doi.org/10.1088/0264-9381/28/14/145026}
{\path{doi:10.1088/0264-9381/28/14/145026}}.

\bibitem{Bonanno:2010mk}
A.~Bonanno, M.~Reuter, {Entropy Production during Asymptotically Safe
Inflation} (2010).
\newblock \href {http://arxiv.org/abs/1011.2794} {\path{arXiv:1011.2794}}.

\bibitem{Cai:2012qi}
Y.-F. Cai, D.~A. Easson, {Higgs Boson in RG running Inflationary Cosmology},
Int.\ J.\ Mod.\ Phys.\ D 21 (2013) 1250094.
\newblock \href {http://arxiv.org/abs/1202.1285} {\path{arXiv:1202.1285}},
\href {https://doi.org/10.1142/S0218271812500940}
{\path{doi:10.1142/S0218271812500940}}.

\bibitem{Bonanno:2012jy}
A.~Bonanno, {An effective action for asymptotically safe gravity}, Phys.\ Rev.\
D 85 (2012) 081503.
\newblock \href {http://arxiv.org/abs/1203.1962} {\path{arXiv:1203.1962}},
\href {https://doi.org/10.1103/PhysRevD.85.081503}
{\path{doi:10.1103/PhysRevD.85.081503}}.

\bibitem{Copeland:2013vva}
E.~J. Copeland, C.~Rahmede, I.~D. Saltas, {Asymptotically Safe Starobinsky
Inflation}, Phys.\ Rev.\ D 91~(10) (2015) 103530.
\newblock \href {http://arxiv.org/abs/1311.0881} {\path{arXiv:1311.0881}},
\href {https://doi.org/10.1103/PhysRevD.91.103530}
{\path{doi:10.1103/PhysRevD.91.103530}}.

\bibitem{Tronconi:2017wps}
A.~Tronconi, {Asymptotically Safe Non-Minimal Inflation}, JCAP 07 (2017) 015.
\newblock \href {http://arxiv.org/abs/1704.05312} {\path{arXiv:1704.05312}},
\href {https://doi.org/10.1088/1475-7516/2017/07/015}
{\path{doi:10.1088/1475-7516/2017/07/015}}.

\bibitem{Bonanno:2018gck}
A.~Bonanno, A.~Platania, F.~Saueressig, {Cosmological bounds on the field
content of asymptotically safe gravity--matter models}, Phys.\ Lett.\ B 784
(2018) 229--236.
\newblock \href {http://arxiv.org/abs/1803.02355} {\path{arXiv:1803.02355}},
\href {https://doi.org/10.1016/j.physletb.2018.06.047}
{\path{doi:10.1016/j.physletb.2018.06.047}}.

\bibitem{Liu:2018hno}
L.-H. Liu, T.~Prokopec, A.~A. Starobinsky, {Inflation in an effective
gravitational model and asymptotic safety}, Phys.\ Rev.\ D 98~(4) (2018)
043505.
\newblock \href {http://arxiv.org/abs/1806.05407} {\path{arXiv:1806.05407}},
\href {https://doi.org/10.1103/PhysRevD.98.043505}
{\path{doi:10.1103/PhysRevD.98.043505}}.

\bibitem{Platania:2019qvo}
A.~Platania, {The inflationary mechanism in Asymptotically Safe Gravity},
Universe 5~(8) (2019) 189.
\newblock \href {http://arxiv.org/abs/1908.03897} {\path{arXiv:1908.03897}},
\href {https://doi.org/10.3390/universe5080189}
{\path{doi:10.3390/universe5080189}}.

\bibitem{Hindmarsh:2011hx}
M.~Hindmarsh, D.~Litim, C.~Rahmede, {Asymptotically Safe Cosmology}, JCAP 1107
(2011) 019.
\newblock \href {http://arxiv.org/abs/1101.5401} {\path{arXiv:1101.5401}},
\href {https://doi.org/10.1088/1475-7516/2011/07/019}
{\path{doi:10.1088/1475-7516/2011/07/019}}.

\bibitem{Kofinas:2016lcz}
G.~Kofinas, V.~Zarikas, {Asymptotically Safe gravity and non-singular
inflationary Big Bang with vacuum birth}, Phys.\ Rev.\ D 94~(10) (2016)
103514.
\newblock \href {http://arxiv.org/abs/1605.02241} {\path{arXiv:1605.02241}},
\href {https://doi.org/10.1103/PhysRevD.94.103514}
{\path{doi:10.1103/PhysRevD.94.103514}}.

\bibitem{Bonanno:2002zb}
A.~Bonanno, M.~Reuter, {Cosmological perturbations in renormalization group
derived cosmologies}, Int. J. Mod. Phys. D13 (2004) 107--122.
\newblock \href {http://arxiv.org/abs/astro-ph/0210472}
{\path{arXiv:astro-ph/0210472}}, \href
{https://doi.org/10.1142/S0218271804003809}
{\path{doi:10.1142/S0218271804003809}}.

\bibitem{Contillo:2010ju}
A.~Contillo, {Evolution of cosmological perturbations in an RG-driven
inflationary scenario}, Phys. Rev. D83 (2011) 085016.
\newblock \href {http://arxiv.org/abs/1011.4618} {\path{arXiv:1011.4618}},
\href {https://doi.org/10.1103/PhysRevD.83.085016}
{\path{doi:10.1103/PhysRevD.83.085016}}.

\bibitem{Bonanno:2001hi}
A.~Bonanno, M.~Reuter, {Cosmology with self-adjusting vacuum energy density
from a renormalization group fixed point}, Phys. Lett. B527 (2002) 9--17.
\newblock \href {http://arxiv.org/abs/astro-ph/0106468}
{\path{arXiv:astro-ph/0106468}}, \href
{https://doi.org/10.1016/S0370-2693(01)01522-2}
{\path{doi:10.1016/S0370-2693(01)01522-2}}.

\bibitem{Babic:2004ev}
A.~Babic, B.~Guberina, R.~Horvat, H.~Stefancic, {Renormalization-group running
cosmologies. A Scale-setting procedure}, Phys.\ Rev.\ D 71 (2005) 124041.
\newblock \href {http://arxiv.org/abs/astro-ph/0407572}
{\path{arXiv:astro-ph/0407572}}, \href
{https://doi.org/10.1103/PhysRevD.71.124041}
{\path{doi:10.1103/PhysRevD.71.124041}}.

\bibitem{Ahn:2011qt}
C.~Ahn, C.~Kim, E.~V. Linder, {From Asymptotic Safety to Dark Energy}, Phys.\
Lett.\ B 704 (2011) 10--14.
\newblock \href {http://arxiv.org/abs/1106.1435} {\path{arXiv:1106.1435}},
\href {https://doi.org/10.1016/j.physletb.2011.08.075}
{\path{doi:10.1016/j.physletb.2011.08.075}}.

\bibitem{Bonanno:2011yx}
A.~Bonanno, S.~Carloni, {Dynamical System Analysis of Cosmologies with Running
Cosmological Constant from Quantum Einstein Gravity}, New J.\ Phys. 14 (2012)
025008.
\newblock \href {http://arxiv.org/abs/1112.4613} {\path{arXiv:1112.4613}},
\href {https://doi.org/10.1088/1367-2630/14/2/025008}
{\path{doi:10.1088/1367-2630/14/2/025008}}.

\bibitem{Wetterich:2018qsl}
C.~Wetterich, {Infrared limit of quantum gravity}, Phys.\ Rev.\ D 98~(2) (2018)
026028.
\newblock \href {http://arxiv.org/abs/1802.05947} {\path{arXiv:1802.05947}},
\href {https://doi.org/10.1103/PhysRevD.98.026028}
{\path{doi:10.1103/PhysRevD.98.026028}}.

\bibitem{Anagnostopoulos:2018jdq}
F.~K. Anagnostopoulos, S.~Basilakos, G.~Kofinas, V.~Zarikas, {Constraining the
Asymptotically Safe Cosmology: cosmic acceleration without dark energy}, JCAP
02 (2019) 053.
\newblock \href {http://arxiv.org/abs/1806.10580} {\path{arXiv:1806.10580}},
\href {https://doi.org/10.1088/1475-7516/2019/02/053}
{\path{doi:10.1088/1475-7516/2019/02/053}}.

\bibitem{Gubitosi:2018gsl}
G.~Gubitosi, R.~Ooijer, C.~Ripken, F.~Saueressig, {Consistent early and late
time cosmology from the RG flow of gravity}, JCAP 12 (2018) 004.
\newblock \href {http://arxiv.org/abs/1806.10147} {\path{arXiv:1806.10147}},
\href {https://doi.org/10.1088/1475-7516/2018/12/004}
{\path{doi:10.1088/1475-7516/2018/12/004}}.

\bibitem{Lauscher:2005qz}
O.~Lauscher, M.~Reuter, {Fractal spacetime structure in asymptotically safe
gravity}, JHEP 10 (2005) 050.
\newblock \href {http://arxiv.org/abs/hep-th/0508202}
{\path{arXiv:hep-th/0508202}}, \href
{https://doi.org/10.1088/1126-6708/2005/10/050}
{\path{doi:10.1088/1126-6708/2005/10/050}}.

\bibitem{Reuter:2011ah}
M.~Reuter, F.~Saueressig, {Fractal space-times under the microscope: A
Renormalization Group view on Monte Carlo data}, JHEP 12 (2011) 012.
\newblock \href {http://arxiv.org/abs/1110.5224} {\path{arXiv:1110.5224}},
\href {https://doi.org/10.1007/JHEP12(2011)012}
{\path{doi:10.1007/JHEP12(2011)012}}.

\bibitem{Rechenberger:2012pm}
S.~Rechenberger, F.~Saueressig, {The $R^2$ phase-diagram of QEG and its
spectral dimension}, Phys.Rev. D86 (2012) 024018.
\newblock \href {http://arxiv.org/abs/1206.0657} {\path{arXiv:1206.0657}},
\href {https://doi.org/10.1103/PhysRevD.86.024018}
{\path{doi:10.1103/PhysRevD.86.024018}}.

\bibitem{Calcagni:2013vsa}
G.~Calcagni, A.~Eichhorn, F.~Saueressig, {Probing the quantum nature of
spacetime by diffusion}, Phys. Rev. D87~(12) (2013) 124028.
\newblock \href {http://arxiv.org/abs/1304.7247} {\path{arXiv:1304.7247}},
\href {https://doi.org/10.1103/PhysRevD.87.124028}
{\path{doi:10.1103/PhysRevD.87.124028}}.

\bibitem{Carlip:2017eud}
S.~Carlip, {Dimension and Dimensional Reduction in Quantum Gravity}, Class.
Quant. Grav. 34~(19) (2017) 193001.
\newblock \href {http://arxiv.org/abs/1705.05417} {\path{arXiv:1705.05417}},
\href {https://doi.org/10.1088/1361-6382/aa8535}
{\path{doi:10.1088/1361-6382/aa8535}}.

\bibitem{Becker:2018quq}
M.~Becker, C.~Pagani, {Geometric operators in the asymptotic safety scenario
for quantum gravity}, Phys. Rev. D99~(6) (2019) 066002.
\newblock \href {http://arxiv.org/abs/1810.11816} {\path{arXiv:1810.11816}},
\href {https://doi.org/10.1103/PhysRevD.99.066002}
{\path{doi:10.1103/PhysRevD.99.066002}}.

\bibitem{Percacci:2002ie}
R.~Percacci, D.~Perini, {Constraints on matter from asymptotic safety}, Phys.
Rev. D67 (2003) 081503.
\newblock \href {http://arxiv.org/abs/hep-th/0207033}
{\path{arXiv:hep-th/0207033}}, \href
{https://doi.org/10.1103/PhysRevD.67.081503}
{\path{doi:10.1103/PhysRevD.67.081503}}.

\bibitem{Percacci:2003jz}
R.~Percacci, D.~Perini, {Asymptotic safety of gravity coupled to matter}, Phys.
Rev. D68 (2003) 044018.
\newblock \href {http://arxiv.org/abs/hep-th/0304222}
{\path{arXiv:hep-th/0304222}}, \href
{https://doi.org/10.1103/PhysRevD.68.044018}
{\path{doi:10.1103/PhysRevD.68.044018}}.

\bibitem{Dona:2013qba}
P.~Don{\`a}, A.~Eichhorn, R.~Percacci, {Matter matters in asymptotically safe
quantum gravity}, Phys.Rev. D89 (2014) 084035.
\newblock \href {http://arxiv.org/abs/1311.2898} {\path{arXiv:1311.2898}},
\href {https://doi.org/10.1103/PhysRevD.89.084035}
{\path{doi:10.1103/PhysRevD.89.084035}}.

\bibitem{Dona:2014pla}
P.~Don{\`a}, A.~Eichhorn, R.~Percacci, {Consistency of matter models with
asymptotically safe quantum gravity}, Can. J. Phys. 93~(9) (2015) 988--994.
\newblock \href {http://arxiv.org/abs/1410.4411} {\path{arXiv:1410.4411}},
\href {https://doi.org/10.1139/cjp-2014-0574}
{\path{doi:10.1139/cjp-2014-0574}}.

\bibitem{Dona:2012am}
P.~Don{\`a}, R.~Percacci, {Functional renormalization with fermions and
tetrads}, Phys. Rev. D87~(4) (2013) 045002.
\newblock \href {http://arxiv.org/abs/1209.3649} {\path{arXiv:1209.3649}},
\href {https://doi.org/10.1103/PhysRevD.87.045002}
{\path{doi:10.1103/PhysRevD.87.045002}}.

\bibitem{Gies:2013noa}
H.~Gies, S.~Lippoldt, {Fermions in gravity with local spin-base invariance},
Phys.\ Rev.\ D 89~(6) (2014) 064040.
\newblock \href {http://arxiv.org/abs/1310.2509} {\path{arXiv:1310.2509}},
\href {https://doi.org/10.1103/PhysRevD.89.064040}
{\path{doi:10.1103/PhysRevD.89.064040}}.

\bibitem{Gies:2015cka}
H.~Gies, S.~Lippoldt, {Global surpluses of spin-base invariant fermions},
Phys.\ Lett.\ B 743 (2015) 415--419.
\newblock \href {http://arxiv.org/abs/1502.00918} {\path{arXiv:1502.00918}},
\href {https://doi.org/10.1016/j.physletb.2015.03.014}
{\path{doi:10.1016/j.physletb.2015.03.014}}.

\bibitem{Lippoldt:2015cea}
S.~Lippoldt, {Spin-base invariance of Fermions in arbitrary dimensions}, Phys.\
Rev.\ D 91~(10) (2015) 104006.
\newblock \href {http://arxiv.org/abs/1502.05607} {\path{arXiv:1502.05607}},
\href {https://doi.org/10.1103/PhysRevD.91.104006}
{\path{doi:10.1103/PhysRevD.91.104006}}.

\bibitem{Meibohm:2015twa}
J.~Meibohm, J.~M. Pawlowski, M.~Reichert, {Asymptotic safety of gravity-matter
systems}, Phys. Rev. D93~(8) (2016) 084035.
\newblock \href {http://arxiv.org/abs/1510.07018} {\path{arXiv:1510.07018}},
\href {https://doi.org/10.1103/PhysRevD.93.084035}
{\path{doi:10.1103/PhysRevD.93.084035}}.

\bibitem{Dona:2015tnf}
P.~Don{\`a}, A.~Eichhorn, P.~Labus, R.~Percacci, {Asymptotic safety in an
interacting system of gravity and scalar matter}, Phys. Rev. D93~(4) (2016)
044049, [Erratum: Phys. Rev.D93,no.12,129904(2016)].
\newblock \href {http://arxiv.org/abs/1512.01589} {\path{arXiv:1512.01589}},
\href {https://doi.org/10.1103/PhysRevD.93.129904,
10.1103/PhysRevD.93.044049} {\path{doi:10.1103/PhysRevD.93.129904,
10.1103/PhysRevD.93.044049}}.

\bibitem{Eichhorn:2016vvy}
A.~Eichhorn, S.~Lippoldt, {Quantum gravity and Standard-Model-like fermions},
Phys. Lett. B767 (2017) 142--146.
\newblock \href {http://arxiv.org/abs/1611.05878} {\path{arXiv:1611.05878}},
\href {https://doi.org/10.1016/j.physletb.2017.01.064}
{\path{doi:10.1016/j.physletb.2017.01.064}}.

\bibitem{Eichhorn:2017sok}
A.~Eichhorn, S.~Lippoldt, V.~Skrinjar, {Nonminimal hints for asymptotic
safety}, Phys. Rev. D97~(2) (2018) 026002.
\newblock \href {http://arxiv.org/abs/1710.03005} {\path{arXiv:1710.03005}},
\href {https://doi.org/10.1103/PhysRevD.97.026002}
{\path{doi:10.1103/PhysRevD.97.026002}}.

\bibitem{Christiansen:2017cxa}
N.~Christiansen, D.~F. Litim, J.~M. Pawlowski, M.~Reichert, {Asymptotic safety
of gravity with matter}, Phys. Rev. D97~(10) (2018) 106012.
\newblock \href {http://arxiv.org/abs/1710.04669} {\path{arXiv:1710.04669}},
\href {https://doi.org/10.1103/PhysRevD.97.106012}
{\path{doi:10.1103/PhysRevD.97.106012}}.

\bibitem{Hamada:2017rvn}
Y.~Hamada, M.~Yamada, {Asymptotic safety of higher derivative quantum gravity
non-minimally coupled with a matter system}, JHEP 08 (2017) 070.
\newblock \href {http://arxiv.org/abs/1703.09033} {\path{arXiv:1703.09033}},
\href {https://doi.org/10.1007/JHEP08(2017)070}
{\path{doi:10.1007/JHEP08(2017)070}}.

\bibitem{Bezrukov:2014ina}
F.~Bezrukov, M.~Shaposhnikov, {Why should we care about the top quark Yukawa
coupling?}, J. Exp. Theor. Phys. 120 (2015) 335--343, [Zh. Eksp. Teor.
Fiz.147,389(2015)].
\newblock \href {http://arxiv.org/abs/1411.1923} {\path{arXiv:1411.1923}},
\href {https://doi.org/10.1134/S1063776115030152}
{\path{doi:10.1134/S1063776115030152}}.

\bibitem{Aad:2019ntk}
G.~Aad, et~al., {Measurements of top-quark pair differential and
double-differential cross-sections in the $\ell$+jets channel with $pp$
collisions at $\sqrt{s}=13$ TeV using the ATLAS detector}, Eur. Phys. J.
C79~(12) (2019) 1028.
\newblock \href {http://arxiv.org/abs/1908.07305} {\path{arXiv:1908.07305}},
\href {https://doi.org/10.1140/epjc/s10052-019-7525-6}
{\path{doi:10.1140/epjc/s10052-019-7525-6}}.

\bibitem{Sirunyan:2019zvx}
A.~M. Sirunyan, et~al., {Measurement of $\mathrm{t\bar t}$ normalised
multi-differential cross sections in pp collisions at $\sqrt s=13$ TeV, and
simultaneous determination of the strong coupling strength, top quark pole
mass, and parton distribution functions}, Submitted to: Eur. Phys. J. (2019).
\newblock \href {http://arxiv.org/abs/1904.05237} {\path{arXiv:1904.05237}}.

\bibitem{Eichhorn:2017egq}
A.~Eichhorn, {Status of the asymptotic safety paradigm for quantum gravity and
matter}, Found. Phys. 48~(10) (2018) 1407--1429.
\newblock \href {http://arxiv.org/abs/1709.03696} {\path{arXiv:1709.03696}},
\href {https://doi.org/10.1007/s10701-018-0196-6}
{\path{doi:10.1007/s10701-018-0196-6}}.

\bibitem{Narain:2009fy}
G.~Narain, R.~Percacci, {Renormalization Group Flow in Scalar-Tensor Theories.
I}, Class. Quant. Grav. 27 (2010) 075001.
\newblock \href {http://arxiv.org/abs/0911.0386} {\path{arXiv:0911.0386}},
\href {https://doi.org/10.1088/0264-9381/27/7/075001}
{\path{doi:10.1088/0264-9381/27/7/075001}}.

\bibitem{Narain:2009gb}
G.~Narain, C.~Rahmede, {Renormalization Group Flow in Scalar-Tensor Theories.
II}, Class. Quant. Grav. 27 (2010) 075002.
\newblock \href {http://arxiv.org/abs/0911.0394} {\path{arXiv:0911.0394}},
\href {https://doi.org/10.1088/0264-9381/27/7/075002}
{\path{doi:10.1088/0264-9381/27/7/075002}}.

\bibitem{Eichhorn:2012va}
A.~Eichhorn, {Quantum-gravity-induced matter self-interactions in the
asymptotic-safety scenario}, Phys. Rev. D86 (2012) 105021.
\newblock \href {http://arxiv.org/abs/1204.0965} {\path{arXiv:1204.0965}},
\href {https://doi.org/10.1103/PhysRevD.86.105021}
{\path{doi:10.1103/PhysRevD.86.105021}}.

\bibitem{Henz:2013oxa}
T.~Henz, J.~M. Pawlowski, A.~Rodigast, C.~Wetterich, {Dilaton Quantum Gravity}
(2013).
\newblock \href {http://arxiv.org/abs/1304.7743} {\path{arXiv:1304.7743}}.

\bibitem{Henz:2016aoh}
T.~Henz, J.~M. Pawlowski, C.~Wetterich, {Scaling solutions for Dilaton Quantum
Gravity}, Phys. Lett. B769 (2017) 105--110.
\newblock \href {http://arxiv.org/abs/1605.01858} {\path{arXiv:1605.01858}},
\href {https://doi.org/10.1016/j.physletb.2017.01.057}
{\path{doi:10.1016/j.physletb.2017.01.057}}.

\bibitem{Eichhorn:2017als}
A.~Eichhorn, Y.~Hamada, J.~Lumma, M.~Yamada, {Quantum gravity fluctuations
flatten the Planck-scale Higgs potential}, Phys. Rev. D97~(8) (2018) 086004.
\newblock \href {http://arxiv.org/abs/1712.00319} {\path{arXiv:1712.00319}},
\href {https://doi.org/10.1103/PhysRevD.97.086004}
{\path{doi:10.1103/PhysRevD.97.086004}}.

\bibitem{Wetterich:2019rsn}
C.~Wetterich, {Effective scalar potential in asymptotically safe quantum
gravity} (2019).
\newblock \href {http://arxiv.org/abs/1911.06100} {\path{arXiv:1911.06100}}.

\bibitem{Zanusso:2009bs}
O.~Zanusso, L.~Zambelli, G.~P. Vacca, R.~Percacci, {Gravitational corrections
to Yukawa systems}, Phys. Lett. B689 (2010) 90--94.
\newblock \href {http://arxiv.org/abs/0904.0938} {\path{arXiv:0904.0938}},
\href {https://doi.org/10.1016/j.physletb.2010.04.043}
{\path{doi:10.1016/j.physletb.2010.04.043}}.

\bibitem{Vacca:2010mj}
G.~P. Vacca, O.~Zanusso, {Asymptotic Safety in Einstein Gravity and
Scalar-Fermion Matter}, Phys. Rev. Lett. 105 (2010) 231601.
\newblock \href {http://arxiv.org/abs/1009.1735} {\path{arXiv:1009.1735}},
\href {https://doi.org/10.1103/PhysRevLett.105.231601}
{\path{doi:10.1103/PhysRevLett.105.231601}}.

\bibitem{Oda:2015sma}
K.-y. Oda, M.~Yamada, {Non-minimal coupling in Higgs-Yukawa model with
asymptotically safe gravity}, Class. Quant. Grav. 33~(12) (2016) 125011.
\newblock \href {http://arxiv.org/abs/1510.03734} {\path{arXiv:1510.03734}},
\href {https://doi.org/10.1088/0264-9381/33/12/125011}
{\path{doi:10.1088/0264-9381/33/12/125011}}.

\bibitem{Eichhorn:2016esv}
A.~Eichhorn, A.~Held, J.~M. Pawlowski, {Quantum-gravity effects on a
Higgs-Yukawa model}, Phys. Rev. D94~(10) (2016) 104027.
\newblock \href {http://arxiv.org/abs/1604.02041} {\path{arXiv:1604.02041}},
\href {https://doi.org/10.1103/PhysRevD.94.104027}
{\path{doi:10.1103/PhysRevD.94.104027}}.

\bibitem{Eichhorn:2017eht}
A.~Eichhorn, A.~Held, {Viability of quantum-gravity induced ultraviolet
completions for matter} (2017).
\newblock \href {http://arxiv.org/abs/1705.02342} {\path{arXiv:1705.02342}}.

\bibitem{Eichhorn:2017ylw}
A.~Eichhorn, A.~Held, {Top mass from asymptotic safety}, Phys. Lett. B777
(2018) 217--221.
\newblock \href {http://arxiv.org/abs/1707.01107} {\path{arXiv:1707.01107}},
\href {https://doi.org/10.1016/j.physletb.2017.12.040}
{\path{doi:10.1016/j.physletb.2017.12.040}}.

\bibitem{deBrito:2019epw}
G.~P. De~Brito, Y.~Hamada, A.~D. Pereira, M.~Yamada, {On the impact of Majorana
masses in gravity-matter systems}, JHEP 08 (2019) 142.
\newblock \href {http://arxiv.org/abs/1905.11114} {\path{arXiv:1905.11114}},
\href {https://doi.org/10.1007/JHEP08(2019)142}
{\path{doi:10.1007/JHEP08(2019)142}}.

\bibitem{Classen:2015mar}
L.~Classen, I.~F. Herbut, L.~Janssen, M.~M. Scherer, {Competition of density
waves and quantum multicritical behavior in Dirac materials from functional
renormalization}, Phys. Rev. B 93~(12) (2016) 125119.
\newblock \href {http://arxiv.org/abs/1510.09003} {\path{arXiv:1510.09003}},
\href {https://doi.org/10.1103/PhysRevB.93.125119}
{\path{doi:10.1103/PhysRevB.93.125119}}.

\bibitem{Eichhorn:2011pc}
A.~Eichhorn, H.~Gies, {Light fermions in quantum gravity} (2011).
\newblock \href {http://arxiv.org/abs/1104.5366} {\path{arXiv:1104.5366}}.

\bibitem{Meibohm:2016mkp}
J.~Meibohm, J.~M. Pawlowski, {Chiral fermions in asymptotically safe quantum
gravity}, Eur. Phys. J. C76~(5) (2016) 285.
\newblock \href {http://arxiv.org/abs/1601.04597} {\path{arXiv:1601.04597}},
\href {https://doi.org/10.1140/epjc/s10052-016-4132-7}
{\path{doi:10.1140/epjc/s10052-016-4132-7}}.

\bibitem{Gies:2018jnv}
H.~Gies, R.~Martini, {Curvature bound from gravitational catalysis}, Phys. Rev.
D97~(8) (2018) 085017.
\newblock \href {http://arxiv.org/abs/1802.02865} {\path{arXiv:1802.02865}},
\href {https://doi.org/10.1103/PhysRevD.97.085017}
{\path{doi:10.1103/PhysRevD.97.085017}}.

\bibitem{Daum:2009dn}
J.-E. Daum, U.~Harst, M.~Reuter, {Running Gauge Coupling in Asymptotically Safe
Quantum Gravity}, JHEP 1001 (2010) 084.
\newblock \href {http://arxiv.org/abs/0910.4938} {\path{arXiv:0910.4938}},
\href {https://doi.org/10.1007/JHEP01(2010)084}
{\path{doi:10.1007/JHEP01(2010)084}}.

\bibitem{Daum:2010bc}
J.~E. Daum, U.~Harst, M.~Reuter, {Non-perturbative QEG Corrections to the
Yang-Mills Beta Function} (2010).
\newblock \href {http://arxiv.org/abs/1005.1488} {\path{arXiv:1005.1488}},
\href {https://doi.org/10.1007/s10714-010-1032-2}
{\path{doi:10.1007/s10714-010-1032-2}}.

\bibitem{Harst:2011zx}
U.~Harst, M.~Reuter, {QED coupled to QEG}, JHEP 05 (2011) 119.
\newblock \href {http://arxiv.org/abs/1101.6007} {\path{arXiv:1101.6007}},
\href {https://doi.org/10.1007/JHEP05(2011)119}
{\path{doi:10.1007/JHEP05(2011)119}}.

\bibitem{Christiansen:2017gtg}
N.~Christiansen, A.~Eichhorn, {An asymptotically safe solution to the U(1)
triviality problem}, Phys. Lett. B770 (2017) 154--160.
\newblock \href {http://arxiv.org/abs/1702.07724} {\path{arXiv:1702.07724}},
\href {https://doi.org/10.1016/j.physletb.2017.04.047}
{\path{doi:10.1016/j.physletb.2017.04.047}}.

\bibitem{Eichhorn:2017lry}
A.~Eichhorn, F.~Versteegen, {Upper bound on the Abelian gauge coupling from
asymptotic safety}, JHEP 01 (2018) 030.
\newblock \href {http://arxiv.org/abs/1709.07252} {\path{arXiv:1709.07252}},
\href {https://doi.org/10.1007/JHEP01(2018)030}
{\path{doi:10.1007/JHEP01(2018)030}}.

\bibitem{Eichhorn:2017muy}
A.~Eichhorn, A.~Held, C.~Wetterich, {Quantum-gravity predictions for the
fine-structure constant}, Phys. Lett. B782 (2018) 198--201.
\newblock \href {http://arxiv.org/abs/1711.02949} {\path{arXiv:1711.02949}},
\href {https://doi.org/10.1016/j.physletb.2018.05.016}
{\path{doi:10.1016/j.physletb.2018.05.016}}.

\bibitem{deBrito:2019umw}
G.~P. De~Brito, A.~Eichhorn, A.~D. Pereira, {A link that matters: Towards
phenomenological tests of unimodular asymptotic safety}, JHEP 09 (2019) 100.
\newblock \href {http://arxiv.org/abs/1907.11173} {\path{arXiv:1907.11173}},
\href {https://doi.org/10.1007/JHEP09(2019)100}
{\path{doi:10.1007/JHEP09(2019)100}}.

\bibitem{Wetterich:2016uxm}
C.~Wetterich, M.~Yamada, {Gauge hierarchy problem in asymptotically safe
gravity--the resurgence mechanism}, Phys. Lett. B770 (2017) 268--271.
\newblock \href {http://arxiv.org/abs/1612.03069} {\path{arXiv:1612.03069}},
\href {https://doi.org/10.1016/j.physletb.2017.04.049}
{\path{doi:10.1016/j.physletb.2017.04.049}}.

\bibitem{Shaposhnikov:2009pv}
M.~Shaposhnikov, C.~Wetterich, {Asymptotic safety of gravity and the Higgs
boson mass}, Phys. Lett. B683 (2010) 196--200.
\newblock \href {http://arxiv.org/abs/0912.0208} {\path{arXiv:0912.0208}},
\href {https://doi.org/10.1016/j.physletb.2009.12.022}
{\path{doi:10.1016/j.physletb.2009.12.022}}.

\bibitem{Eichhorn:2018whv}
A.~Eichhorn, A.~Held, {Mass difference for charged quarks from asymptotically
safe quantum gravity}, Phys. Rev. Lett. 121~(15) (2018) 151302.
\newblock \href {http://arxiv.org/abs/1803.04027} {\path{arXiv:1803.04027}},
\href {https://doi.org/10.1103/PhysRevLett.121.151302}
{\path{doi:10.1103/PhysRevLett.121.151302}}.

\bibitem{Eichhorn:2019yzm}
A.~Eichhorn, M.~Schiffer, {$d=4$ as the critical dimensionality of
asymptotically safe interactions}, Phys. Lett. B793 (2019) 383--389.
\newblock \href {http://arxiv.org/abs/1902.06479} {\path{arXiv:1902.06479}},
\href {https://doi.org/10.1016/j.physletb.2019.05.005}
{\path{doi:10.1016/j.physletb.2019.05.005}}.

\bibitem{Reichert:2019car}
M.~Reichert, J.~Smirnov, {Dark Matter meets Quantum Gravity}, Phys. Rev. D
101~(6) (2020) 063015.
\newblock \href {http://arxiv.org/abs/1911.00012} {\path{arXiv:1911.00012}},
\href {https://doi.org/10.1103/PhysRevD.101.063015}
{\path{doi:10.1103/PhysRevD.101.063015}}.

\bibitem{Hamada:2020vnf}
Y.~Hamada, K.~Tsumura, M.~Yamada, {Scalegenesis and fermionic dark matters in
the flatland scenario}, Eur. Phys. J. C 80~(5) (2020) 368.
\newblock \href {http://arxiv.org/abs/2002.03666} {\path{arXiv:2002.03666}},
\href {https://doi.org/10.1140/epjc/s10052-020-7929-3}
{\path{doi:10.1140/epjc/s10052-020-7929-3}}.

\bibitem{Eichhorn:2020kca}
A.~Eichhorn, M.~Pauly, {Safety in darkness: Higgs portal to simple Yukawa
systems} (5 2020).
\newblock \href {http://arxiv.org/abs/2005.03661} {\path{arXiv:2005.03661}}.

\bibitem{Eichhorn:2019dhg}
A.~Eichhorn, A.~Held, C.~Wetterich, {Predictive power of grand unification from
quantum gravity} (2019).
\newblock \href {http://arxiv.org/abs/1909.07318} {\path{arXiv:1909.07318}}.

\bibitem{Kwapisz:2019wrl}
J.~H. Kwapisz, {Asymptotic safety, the Higgs boson mass, and beyond the
standard model physics}, Phys.\ Rev.\ D 100~(11) (2019) 115001.
\newblock \href {http://arxiv.org/abs/1907.12521} {\path{arXiv:1907.12521}},
\href {https://doi.org/10.1103/PhysRevD.100.115001}
{\path{doi:10.1103/PhysRevD.100.115001}}.

\bibitem{Grabowski:2018fjj}
F.~Grabowski, J.~H. Kwapisz, K.~A. Meissner, {Asymptotic safety and Conformal
Standard Model}, Phys.\ Rev.\ D 99~(11) (2019) 115029.
\newblock \href {http://arxiv.org/abs/1810.08461} {\path{arXiv:1810.08461}},
\href {https://doi.org/10.1103/PhysRevD.99.115029}
{\path{doi:10.1103/PhysRevD.99.115029}}.

\bibitem{Percacci:2010af}
R.~Percacci, G.~P. Vacca, {Asymptotic Safety, Emergence and Minimal Length},
Class. Quant. Grav. 27 (2010) 245026.
\newblock \href {http://arxiv.org/abs/1008.3621} {\path{arXiv:1008.3621}},
\href {https://doi.org/10.1088/0264-9381/27/24/245026}
{\path{doi:10.1088/0264-9381/27/24/245026}}.

\bibitem{Estrada:2012te}
C.~Estrada, M.~Marcolli, {Asymptotic safety, hypergeometric functions, and the
Higgs mass in spectral action models}, Int.\ J.\ Geom.\ Meth.\ Mod.\ Phys. 10
(2013) 1350036.
\newblock \href {http://arxiv.org/abs/1208.5023} {\path{arXiv:1208.5023}},
\href {https://doi.org/10.1142/S0219887813500369}
{\path{doi:10.1142/S0219887813500369}}.

\bibitem{Horava:2009uw}
P.~Horava, {Quantum Gravity at a Lifshitz Point}, Phys. Rev. D79 (2009) 084008.
\newblock \href {http://arxiv.org/abs/0901.3775} {\path{arXiv:0901.3775}},
\href {https://doi.org/10.1103/PhysRevD.79.084008}
{\path{doi:10.1103/PhysRevD.79.084008}}.

\bibitem{Yagi:2013qpa}
K.~Yagi, D.~Blas, N.~Yunes, E.~Barausse, {Strong Binary Pulsar Constraints on
Lorentz Violation in Gravity}, Phys. Rev. Lett. 112~(16) (2014) 161101.
\newblock \href {http://arxiv.org/abs/1307.6219} {\path{arXiv:1307.6219}},
\href {https://doi.org/10.1103/PhysRevLett.112.161101}
{\path{doi:10.1103/PhysRevLett.112.161101}}.

\bibitem{Ramos:2018oku}
O.~Ramos, E.~Barausse, {Constraints on Ho\v{r}ava gravity from binary black
hole observations}, Phys. Rev. D99~(2) (2019) 024034.
\newblock \href {http://arxiv.org/abs/1811.07786} {\path{arXiv:1811.07786}},
\href {https://doi.org/10.1103/PhysRevD.99.024034}
{\path{doi:10.1103/PhysRevD.99.024034}}.

\bibitem{Contillo:2013fua}
A.~Contillo, S.~Rechenberger, F.~Saueressig, {Renormalization group flow of
Ho\v{r}ava-Lifshitz gravity at low energies}, JHEP 12 (2013) 017.
\newblock \href {http://arxiv.org/abs/1309.7273} {\path{arXiv:1309.7273}},
\href {https://doi.org/10.1007/JHEP12(2013)017}
{\path{doi:10.1007/JHEP12(2013)017}}.

\bibitem{Rechenberger:2012dt}
S.~Rechenberger, F.~Saueressig, {A functional renormalization group equation
for foliated spacetimes}, JHEP 03 (2013) 010.
\newblock \href {http://arxiv.org/abs/1212.5114} {\path{arXiv:1212.5114}},
\href {https://doi.org/10.1007/JHEP03(2013)010}
{\path{doi:10.1007/JHEP03(2013)010}}.

\bibitem{DOdorico:2015pil}
G.~D'Odorico, J.-W. Goossens, F.~Saueressig, {Covariant computation of
effective actions in Ho{\v{r}}ava-Lifshitz gravity}, JHEP 10 (2015) 126.
\newblock \href {http://arxiv.org/abs/1508.00590} {\path{arXiv:1508.00590}},
\href {https://doi.org/10.1007/JHEP10(2015)126}
{\path{doi:10.1007/JHEP10(2015)126}}.

\bibitem{Knorr:2018fdu}
B.~Knorr, {Lorentz symmetry is relevant}, Phys. Lett. B 792 (2019) 142--148.
\newblock \href {http://arxiv.org/abs/1810.07971} {\path{arXiv:1810.07971}},
\href {https://doi.org/10.1016/j.physletb.2019.01.070}
{\path{doi:10.1016/j.physletb.2019.01.070}}.

\bibitem{Eichhorn:2019ybe}
A.~Eichhorn, A.~Platania, M.~Schiffer, {Lorentz invariance violations in the
interplay of quantum gravity with matter} (2019).
\newblock \href {http://arxiv.org/abs/1911.10066} {\path{arXiv:1911.10066}}.

\bibitem{Eichhorn:2013isa}
A.~Eichhorn, T.~Koslowski, {Continuum limit in matrix models for quantum
gravity from the Functional Renormalization Group}, Phys. Rev. D88 (2013)
084016.
\newblock \href {http://arxiv.org/abs/1309.1690} {\path{arXiv:1309.1690}},
\href {https://doi.org/10.1103/PhysRevD.88.084016}
{\path{doi:10.1103/PhysRevD.88.084016}}.

\bibitem{Sfondrini:2010zm}
A.~Sfondrini, T.~A. Koslowski, {Functional Renormalization of Noncommutative
Scalar Field Theory}, Int. J. Mod. Phys. A26 (2011) 4009--4051.
\newblock \href {http://arxiv.org/abs/1006.5145} {\path{arXiv:1006.5145}},
\href {https://doi.org/10.1142/S0217751X11054048}
{\path{doi:10.1142/S0217751X11054048}}.

\bibitem{Brezin:1992yc}
E.~Brezin, J.~Zinn-Justin, {Renormalization group approach to matrix models},
Phys. Lett. B288 (1992) 54--58.
\newblock \href {http://arxiv.org/abs/hep-th/9206035}
{\path{arXiv:hep-th/9206035}}, \href
{https://doi.org/10.1016/0370-2693(92)91953-7}
{\path{doi:10.1016/0370-2693(92)91953-7}}.

\bibitem{Krajewski:2015clk}
T.~Krajewski, R.~Toriumi, {Polchinski's exact renormalisation group for
tensorial theories: Gaussian universality and power counting}, J. Phys. A
49~(38) (2016) 385401.
\newblock \href {http://arxiv.org/abs/1511.09084} {\path{arXiv:1511.09084}},
\href {https://doi.org/10.1088/1751-8113/49/38/385401}
{\path{doi:10.1088/1751-8113/49/38/385401}}.

\bibitem{Krajewski:2016svb}
T.~Krajewski, R.~Toriumi, {Exact Renormalisation Group Equations and Loop
Equations for Tensor Models}, SIGMA 12 (2016) 068.
\newblock \href {http://arxiv.org/abs/1603.00172} {\path{arXiv:1603.00172}},
\href {https://doi.org/10.3842/SIGMA.2016.068}
{\path{doi:10.3842/SIGMA.2016.068}}.

\bibitem{Eichhorn:2018phj}
A.~Eichhorn, T.~Koslowski, A.~D. Pereira, {Status of background-independent
coarse-graining in tensor models for quantum gravity}, Universe 5~(2) (2019)
53.
\newblock \href {http://arxiv.org/abs/1811.12909} {\path{arXiv:1811.12909}},
\href {https://doi.org/10.3390/universe5020053}
{\path{doi:10.3390/universe5020053}}.

\bibitem{Ambjorn:1990ge}
J.~Ambjorn, B.~Durhuus, T.~Jonsson, {Three-dimensional simplicial quantum
gravity and generalized matrix models}, Mod. Phys. Lett. A6 (1991)
1133--1146.
\newblock \href {https://doi.org/10.1142/S0217732391001184}
{\path{doi:10.1142/S0217732391001184}}.

\bibitem{Sasakura:1990fs}
N.~Sasakura, {Tensor model for gravity and orientability of manifold}, Mod.
Phys. Lett. A6 (1991) 2613--2624.
\newblock \href {https://doi.org/10.1142/S0217732391003055}
{\path{doi:10.1142/S0217732391003055}}.

\bibitem{Godfrey:1990dt}
N.~Godfrey, M.~Gross, {Simplicial quantum gravity in more than two-dimensions},
Phys. Rev. D43 (1991) 1749--1753.
\newblock \href {https://doi.org/10.1103/PhysRevD.43.R1749}
{\path{doi:10.1103/PhysRevD.43.R1749}}.

\bibitem{Gross:1991hx}
M.~Gross, {Tensor models and simplicial quantum gravity in $>$ 2-D}, Nucl.
Phys. B Proc. Suppl. 25A (1992) 144--149.
\newblock \href {https://doi.org/10.1016/S0920-5632(05)80015-5}
{\path{doi:10.1016/S0920-5632(05)80015-5}}.

\bibitem{Gurau:2010ba}
R.~Gurau, {The 1/N expansion of colored tensor models}, Annales Henri Poincare
12 (2011) 829--847.
\newblock \href {http://arxiv.org/abs/1011.2726} {\path{arXiv:1011.2726}},
\href {https://doi.org/10.1007/s00023-011-0101-8}
{\path{doi:10.1007/s00023-011-0101-8}}.

\bibitem{Bonzom:2012hw}
V.~Bonzom, R.~Gurau, V.~Rivasseau, {Random tensor models in the large N limit:
Uncoloring the colored tensor models}, Phys. Rev. D85 (2012) 084037.
\newblock \href {http://arxiv.org/abs/1202.3637} {\path{arXiv:1202.3637}},
\href {https://doi.org/10.1103/PhysRevD.85.084037}
{\path{doi:10.1103/PhysRevD.85.084037}}.

\bibitem{Rivasseau:2011hm}
V.~Rivasseau, {Quantum Gravity and Renormalization: The Tensor Track}, AIP
Conf. Proc. 1444 (2011) 18--29.
\newblock \href {http://arxiv.org/abs/1112.5104} {\path{arXiv:1112.5104}},
\href {https://doi.org/10.1063/1.4715396} {\path{doi:10.1063/1.4715396}}.

\bibitem{Gurau:2011xp}
R.~Gurau, J.~P. Ryan, {Colored Tensor Models - a review}, SIGMA 8 (2012) 020.
\newblock \href {http://arxiv.org/abs/1109.4812} {\path{arXiv:1109.4812}},
\href {https://doi.org/10.3842/SIGMA.2012.020}
{\path{doi:10.3842/SIGMA.2012.020}}.

\bibitem{Rivasseau:2012yp}
V.~Rivasseau, {The Tensor Track: an Update}, in: {29th International Colloquium
on Group-Theoretical Methods in Physics (GROUP 29) Tianjin, China, August
20-26, 2012}, 2012.
\newblock \href {http://arxiv.org/abs/1209.5284} {\path{arXiv:1209.5284}}.

\bibitem{Rivasseau:2013uca}
V.~Rivasseau, {The Tensor Track, III}, Fortsch. Phys. 62 (2014) 81--107.
\newblock \href {http://arxiv.org/abs/1311.1461} {\path{arXiv:1311.1461}},
\href {https://doi.org/10.1002/prop.201300032}
{\path{doi:10.1002/prop.201300032}}.

\bibitem{Rivasseau:2016zco}
V.~Rivasseau, {Random Tensors and Quantum Gravity}, SIGMA 12 (2016) 069.
\newblock \href {http://arxiv.org/abs/1603.07278} {\path{arXiv:1603.07278}},
\href {https://doi.org/10.3842/SIGMA.2016.069}
{\path{doi:10.3842/SIGMA.2016.069}}.

\bibitem{Rivasseau:2016wvy}
V.~Rivasseau, {The Tensor Track, IV}, PoS CORFU2015 (2016) 106.
\newblock \href {http://arxiv.org/abs/1604.07860} {\path{arXiv:1604.07860}},
\href {https://doi.org/10.22323/1.263.0106} {\path{doi:10.22323/1.263.0106}}.

\bibitem{Gurau:2016cjo}
R.~Gurau, {Invitation to Random Tensors}, SIGMA 12 (2016) 094.
\newblock \href {http://arxiv.org/abs/1609.06439} {\path{arXiv:1609.06439}},
\href {https://doi.org/10.3842/SIGMA.2016.094}
{\path{doi:10.3842/SIGMA.2016.094}}.

\bibitem{Eichhorn:2017xhy}
A.~Eichhorn, T.~Koslowski, {Flowing to the continuum in discrete tensor models
for quantum gravity}, Ann. Inst. H. Poincare Comb. Phys. Interact. 5~(2)
(2018) 173--210.
\newblock \href {http://arxiv.org/abs/1701.03029} {\path{arXiv:1701.03029}},
\href {https://doi.org/10.4171/AIHPD/52} {\path{doi:10.4171/AIHPD/52}}.

\bibitem{Eichhorn:2018ylk}
A.~Eichhorn, T.~Koslowski, J.~Lumma, A.~D. Pereira, {Towards background
independent quantum gravity with tensor models} (2018).
\newblock \href {http://arxiv.org/abs/1811.00814} {\path{arXiv:1811.00814}}.

\bibitem{Eichhorn:2019hsa}
A.~Eichhorn, J.~Lumma, A.~D. Pereira, A.~Sikandar, {Universal critical behavior
in tensor models for four-dimensional quantum gravity} (2019).
\newblock \href {http://arxiv.org/abs/1912.05314} {\path{arXiv:1912.05314}}.

\bibitem{Pereira:2019dbn}
A.~D. Pereira, {Quantum spacetime and the renormalization group: Progress and
visions}, in: {Progress and Visions in Quantum Theory in View of Gravity}:
{Bridging foundations of physics and mathematics}, 2019.
\newblock \href {http://arxiv.org/abs/1904.07042} {\path{arXiv:1904.07042}}.

\bibitem{Benedetti:2014qsa}
D.~Benedetti, J.~Ben~Geloun, D.~Oriti, {Functional Renormalisation Group
Approach for Tensorial Group Field Theory: a Rank-3 Model}, JHEP 03 (2015)
084.
\newblock \href {http://arxiv.org/abs/1411.3180} {\path{arXiv:1411.3180}},
\href {https://doi.org/10.1007/JHEP03(2015)084}
{\path{doi:10.1007/JHEP03(2015)084}}.

\bibitem{Geloun:2015qfa}
J.~Ben~Geloun, R.~Martini, D.~Oriti, {Functional Renormalization Group analysis
of a Tensorial Group Field Theory on $\mathbb{R}^3$}, EPL 112~(3) (2015)
31001.
\newblock \href {http://arxiv.org/abs/1508.01855} {\path{arXiv:1508.01855}},
\href {https://doi.org/10.1209/0295-5075/112/31001}
{\path{doi:10.1209/0295-5075/112/31001}}.

\bibitem{Benedetti:2015yaa}
D.~Benedetti, V.~Lahoche, {Functional Renormalization Group Approach for
Tensorial Group Field Theory: A Rank-6 Model with Closure Constraint}, Class.
Quant. Grav. 33~(9) (2016) 095003.
\newblock \href {http://arxiv.org/abs/1508.06384} {\path{arXiv:1508.06384}},
\href {https://doi.org/10.1088/0264-9381/33/9/095003}
{\path{doi:10.1088/0264-9381/33/9/095003}}.

\bibitem{Geloun:2016qyb}
J.~Ben~Geloun, R.~Martini, D.~Oriti, {Functional Renormalisation Group analysis
of Tensorial Group Field Theories on $\mathbb{R}^d$}, Phys. Rev. D94~(2)
(2016) 024017.
\newblock \href {http://arxiv.org/abs/1601.08211} {\path{arXiv:1601.08211}},
\href {https://doi.org/10.1103/PhysRevD.94.024017}
{\path{doi:10.1103/PhysRevD.94.024017}}.

\bibitem{Geloun:2016xep}
J.~Ben~Geloun, T.~A. Koslowski, {Nontrivial UV behavior of rank-4 tensor field
models for quantum gravity} (2016).
\newblock \href {http://arxiv.org/abs/1606.04044} {\path{arXiv:1606.04044}}.

\bibitem{Lahoche:2018vun}
V.~Lahoche, D.~Ousmane~Samary, {Unitary symmetry constraints on tensorial group
field theory renormalization group flow}, Class. Quant. Grav. 35~(19) (2018)
195006.
\newblock \href {http://arxiv.org/abs/1803.09902} {\path{arXiv:1803.09902}},
\href {https://doi.org/10.1088/1361-6382/aad83f}
{\path{doi:10.1088/1361-6382/aad83f}}.

\bibitem{BenGeloun:2018ekd}
J.~Ben~Geloun, T.~A. Koslowski, D.~Oriti, A.~D. Pereira, {Functional
Renormalization Group analysis of rank 3 tensorial group field theory: The
full quartic invariant truncation}, Phys. Rev. D97~(12) (2018) 126018.
\newblock \href {http://arxiv.org/abs/1805.01619} {\path{arXiv:1805.01619}},
\href {https://doi.org/10.1103/PhysRevD.97.126018}
{\path{doi:10.1103/PhysRevD.97.126018}}.

\bibitem{Lahoche:2018oeo}
V.~Lahoche, D.~Ousmane~Samary, {Nonperturbative renormalization group beyond
the melonic sector: The effective vertex expansion method for group fields
theories}, Phys. Rev. D98~(12) (2018) 126010.
\newblock \href {http://arxiv.org/abs/1809.00247} {\path{arXiv:1809.00247}},
\href {https://doi.org/10.1103/PhysRevD.98.126010}
{\path{doi:10.1103/PhysRevD.98.126010}}.

\bibitem{Lahoche:2018ggd}
V.~Lahoche, D.~Ousmane~Samary, {Ward identity violation for melonic
$T^4$-truncation}, Nucl. Phys. B940 (2019) 190--213.
\newblock \href {http://arxiv.org/abs/1809.06081} {\path{arXiv:1809.06081}},
\href {https://doi.org/10.1016/j.nuclphysb.2019.01.005}
{\path{doi:10.1016/j.nuclphysb.2019.01.005}}.

\bibitem{Carrozza:2016tih}
S.~Carrozza, V.~Lahoche, {Asymptotic safety in three-dimensional SU(2) Group
Field Theory: evidence in the local potential approximation}, Class. Quant.
Grav. 34~(11) (2017) 115004.
\newblock \href {http://arxiv.org/abs/1612.02452} {\path{arXiv:1612.02452}},
\href {https://doi.org/10.1088/1361-6382/aa6d90}
{\path{doi:10.1088/1361-6382/aa6d90}}.

\bibitem{Lahoche:2019vzy}
V.~Lahoche, D.~Ousmane~Samary, {Ward-constrained melonic renormalization group
flow}, Phys. Lett. B802 (2020) 135173.
\newblock \href {http://arxiv.org/abs/1904.05655} {\path{arXiv:1904.05655}},
\href {https://doi.org/10.1016/j.physletb.2019.135173}
{\path{doi:10.1016/j.physletb.2019.135173}}.

\bibitem{Eichhorn:2014xaa}
A.~Eichhorn, T.~Koslowski, {Towards phase transitions between discrete and
continuum quantum spacetime from the Renormalization Group}, Phys. Rev.
D90~(10) (2014) 104039.
\newblock \href {http://arxiv.org/abs/1408.4127} {\path{arXiv:1408.4127}},
\href {https://doi.org/10.1103/PhysRevD.90.104039}
{\path{doi:10.1103/PhysRevD.90.104039}}.

\bibitem{Lebellac_book}
M.~Le~Bellac, {Quantum and Statistical Field Theory}, Oxford Science Publ,
Oxford University Press, 1991.

\bibitem{Ma_book}
S.-K. Ma, {Modern theory of critical phenomena}, Advanced book classics,
Perseus Pub, 2000.

\bibitem{Kos14}
F.~Kos, D.~Poland, D.~Simmons-Duffin, {Bootstrapping mixed correlators in the
3D Ising model}, J. High Energ. Phys. 11 (2014) 109.
\newblock \href {https://doi.org/10.1007/JHEP11(2014)109}
{\path{doi:10.1007/JHEP11(2014)109}}.

\end{thebibliography}
